\documentclass[10pt]{article}
\usepackage{amsmath,amssymb}

\DeclareMathOperator\ric{Ric}
\usepackage{empheq}
\usepackage{graphicx}
\usepackage{soul}
\usepackage{float}
\usepackage{slashed}
\usepackage{amsthm}
\usepackage{amsfonts}
\usepackage{amsthm}
\usepackage{amsmath}
\usepackage{amscd}
\usepackage{stmaryrd}
\usepackage{bbm}
\newtheorem{theorem}{Theorem}[section]
\graphicspath{ {Writings/} }
\usepackage{geometry}
\usepackage{enumerate}
\usepackage{pgf,tikz}
\usepackage{csquotes}
\usepackage[bottom]{footmisc}

\usepackage{caption}
\usepackage{subcaption}

\usepackage{color}
\usepackage{xcolor}
\newcommand{\bsy}{\ensuremath{\boldsymbol}}

\DeclareMathOperator{\chih}{\widehat{\chi}}
\DeclareMathOperator{\omegah}{\hat{\omega}}
\newcommand{\ub}{\underbar}
\DeclareMathOperator{\nablasl}{\slashed{\nabla}}

\DeclareMathOperator{\eo}{\bsy{e}_{\bsy{1}}}
\DeclareMathOperator{\etw}{\bsy{e}_{\bsy{2}}}
\DeclareMathOperator{\et}{\bsy{e}_{\bsy{3}}}
\DeclareMathOperator{\ef}{\bsy{e}_{\bsy{4}}}
\DeclareMathOperator{\ea}{\bsy{e}_{\bsy{A}}}
\DeclareMathOperator{\eb}{\bsy{e}_{\bsy{B}}}
\DeclareMathOperator{\chibh}{\widehat{\underline{\chi}}}
\DeclareMathOperator{\omegabh}{\hat{\underline{\omega}}}
\DeclareMathOperator{\etab}{\underline{\eta}}
\DeclareMathOperator{\omegab}{\underline{\omega}}
\DeclareMathOperator{\yb}{\underline{\xi}}

\DeclareMathOperator{\alphab}{\underline{\alpha}}
\DeclareMathOperator{\betab}{\underline{\beta}}
\DeclareMathOperator{\chib}{\underline{\chi}}

\DeclareMathOperator{\etao}{\overset{\text{\scalebox{.6}{$(1)$}}}{\eta}}
\DeclareMathOperator{\zetao}{\overset{\text{\scalebox{.6}{$(1)$}}}{\zeta}}

\DeclareMathOperator{\omegabo}{\overset{\text{\scalebox{.6}{$(1)$}}}{\hat{\omegab}}}
\DeclareMathOperator{\trchio}{\overset{\text{\scalebox{.6}{$(1)$}}}{\text{tr}\chi}}
\DeclareMathOperator{\trchibo}{\overset{\text{\scalebox{.6}{$(1)$}}}{\text{tr}\chib}}
\DeclareMathOperator{\chiho}{\overset{\text{\scalebox{.6}{$(1)$}}}{\chih}}
\DeclareMathOperator{\chibho}{\overset{\text{\scalebox{.6}{$(1)$}}}{\chibh}}

\DeclareMathOperator{\alphao}{\overset{\text{\scalebox{.6}{$(1)$}}}{\alpha}}
\DeclareMathOperator{\alphabo}{\overset{\text{\scalebox{.6}{$(1)$}}}{\alphab}}
\DeclareMathOperator{\betao}{\overset{\text{\scalebox{.6}{$(1)$}}}{\beta}}
\DeclareMathOperator{\betabo}{\overset{\text{\scalebox{.6}{$(1)$}}}{\betab}}
\DeclareMathOperator{\rhoo}{\overset{\text{\scalebox{.6}{$(1)$}}}{\rho}}
\DeclareMathOperator{\sigmao}{\overset{\text{\scalebox{.6}{$(1)$}}}{\sigma}}
\DeclareMathOperator{\ybo}{\overset{\text{\scalebox{.6}{$(1)$}}}{\yb}}
\DeclareMathOperator{\tr}{\text{tr}}
\DeclareMathOperator{\divv}{\text{div}}

\DeclareMathOperator{\htimes}{\widehat{\otimes}}

\newtheorem{corollary}[theorem]{Corollary}

\newtheorem{prop}[theorem]{Proposition}
\newtheorem{remark}[theorem]{Remark}
\newtheorem{definition}[theorem]{Definition}
\newtheorem*{openproblem}{Open Problem}

\newtheorem*{integrprop}{Non-integrable Frames}
\newtheorem*{frobprop}{Frobenius Theorem}

\usepackage[affil-it]{authblk}
\usepackage[linktoc=all]{hyperref}
\hypersetup{
    colorlinks = true,
    linkcolor = {black},
    citecolor = {black}
}

\usepackage{wasysym}
\usepackage{verbatim}
\usepackage{cleveref}
\usepackage{mathtools}
\usepackage{stmaryrd}

\begin{document}

\title{A new gauge for gravitational perturbations of Kerr spacetimes I:\\The linearised theory}
\author{Gabriele Benomio}
\affil{\small Princeton Gravity Initiative, Princeton University\\Washington Road, Princeton NJ 08544, United States of America}

\maketitle

\begin{abstract}
We propose a new geometric framework to address the stability of the Kerr solution to gravitational perturbations in the full sub-extremal range $|a|<M$.~Central to our framework is a new formulation of nonlinear gravitational perturbations of Kerr, whose two novel ingredients are the choice of a geometric gauge and non-integrable null frames both tailored to the \emph{outgoing principal null geodesics} of Kerr.~The vacuum Einstein equations for the perturbations are formulated in our gauge as a system of equations for the connection coefficients and curvature components relative to the chosen frames.~When renormalised with respect to Kerr, the null structure equations with the form of \emph{outgoing} transport equations do \emph{not} possess any derivatives of renormalised connection coefficients on the right hand side.

In this work, we derive the linearised vacuum Einstein equations around Kerr in the new framework.~The system of linearised gravity exhibits two key structural properties.~The first is well-known and consists of the exact decoupling of two \emph{gauge invariant} linearised quantities in the system, satisfying two decoupled spin $\pm 2$ Teukolsky equations.~The second is a \emph{new}, \emph{gauge dependent} structure in the \emph{outgoing} transport equations for the linearised connection coefficients inherited from the nonlinear system:~Unlike previous works, these equations do \emph{not} contain any derivatives of linearised connection coefficients on the right hand side and induce a \emph{new hierarchy} of only outgoing transport equations including all gauge dependent linearised quantities.

Our new framework is designed to effectively capture the stabilising properties of the \emph{red-shifted} transport equations, thereby isolating one of the crucial structures of the problem.~Such a feature is suggestive of future simplifications in the analysis.~As an illustration, our companion work \cite{benomio_schwarzschild_stability} employs the system of linearised gravity and its enhanced red-shifted transport equations, specialised to the $|a|=0$ case, to produce a new simplified proof of linear stability of the Schwarzschild solution.~The full linear stability analysis in the full sub-extremal range $|a|<M$ is deferred to future work.~As already apparent from \cite{benomio_schwarzschild_stability}, our framework will allow to combine the new structure in the transport equations with the elliptic part of the system to establish a linear orbital stability result \emph{without loss of derivatives}, indicating that the framework may be well suited to address nonlinear stability in the full sub-extremal range $|a|<M$.
\end{abstract}

\tableofcontents

\section{Introduction}

The two-parameter family of Kerr spacetimes \cite{Kerr_original_paper} is widely believed to be the unique family of stationary, asymptotically flat black hole solutions to the vacuum Einstein equations
\begin{equation} \label{intro_intro_EVE}
\ric (g) =0 
\end{equation}
and thus to characterise all the possible stationary endstates of vacuum black hole dynamics \cite{Klain_final_state}.

\medskip 

A fundamental open problem in general relativity is to formulate a mathematical proof of nonlinear stability for the Kerr exterior region, in the full sub-extremal range $|a|<M$, as a solution to the Cauchy problem \cite{CB69} for \eqref{intro_intro_EVE}:

\medskip

\begin{openproblem}
For all vacuum initial data sufficiently close to the initial data corresponding to a Kerr exterior solution with parameters $(a_0,M_0)$, $|a_0|<M_0$, the maximal solution to \eqref{intro_intro_EVE} asymptotically settles down to a nearby Kerr exterior solution with parameters $a\approx a_0$ and $M\approx M_0$. 
\end{openproblem}

\medskip

See Conjecture IV.1 in \cite{DHRT} for a more precise formulation.

\medskip

The nonlinear stability of the trivial solution to \eqref{intro_intro_EVE}, Minkowski space, has first been proven in the monumental work \cite{StabMink}.~The breakthrough work \cite{DHRT} is the first nonlinear stability result, without symmetry assumptions, for a black hole solution to \eqref{intro_intro_EVE}, namely the Schwarzschild solution.\footnote{The work \cite{DHRT} characterises the dynamics of all vacuum black holes which arise from initial data in a neighbourhood (in moduli space) of the initial data of a given Schwarzschild solution and asymptote a (possibly different) \emph{Schwarzschild} solution.}~The series of works \cite{Klain_GCM_kerr_1, Klain_GCM_kerr_2, Klain_form_kerr, Klainerman_Szeftel_Kerr_small_a_1, Shen_gcm_spheres_kerr, Giorgi_Klainerman_Szeftel_wave_estimates} prove the nonlinear stability of the Kerr solution in the slowly rotating regime $|a|\ll M$.~Related works are discussed in Section \ref{sec_intro_other_works} of the present introduction.

\subsection{Linear stability as a building block for nonlinear stability}  \label{sec_intro_linear_stability}

Proving nonlinear stability for a solution to \eqref{intro_intro_EVE} poses a number of both conceptual and technical difficulties.~Most of the key conceptual difficulties, however, already appear, and can be understood, in the context of \emph{linear} stability, namely the study of the stability of solutions to the system of linearised vacuum Einstein equations (or, for brevity, system of \emph{linearised gravity}).~This fact has motivated the growing activity around the mathematical study of the linear stability of black holes, although the subject has a long (and, to some extent, independent from the interest in nonlinear problems) tradition in the physics literature (see Section \ref{sec_intro_DHR}).

\medskip

To be employed as a \emph{direct building block in a proof of nonlinear stability}, the mathematical formulation of a linear stability result should include some essential ingredients, or \emph{Steps}:

\begin{itemize}
\item \textbf{Step 1:~Nonlinear gauge}

As a first step, one has to break the general covariance of the equations \eqref{intro_intro_EVE} by choosing a gauge in which the nonlinear vacuum Einstein equations can be formulated as a well-posed initial value problem for a system of nonlinear partial differential equations.

\item \textbf{Step 2:~Linearisation}

To move from the nonlinear to the linear theory, one has to implement a linearisation procedure to linearise the vacuum Einstein equations in the gauge chosen in Step 1.\footnote{Our prescription here should be contrasted to an approach which linearises ``ungauged" equations and later employs so-called ``linear gauges".\label{footnote:linear_gauges}}~The linearised system obtained should come with a natural notion of solutions and initial data and inherit well-posedness from the nonlinear equations.

Ideally, once the linearised system of equations is derived, one wishes to never refer back to the nonlinear theory.~In fact, one should be able to handle the final linearised system of equations without any knowledge about its derivation and phrase all linear stability statements as statements for solutions to the linearised system.

\item \textbf{Step 3:~Orbital stability}

An essential stability result to be proven should be a uniform boundedness statement for solutions, and derivatives of solutions, to the linearised system of equations in terms of an explicit initial data (energy) quantity.\footnote{Orbital stability is a weaker notion of stability than \emph{asymptotic} stability (see Step 4), in that it does not require that solutions to the linearised system decay.~However, we should point out that, in general, one does not expect to be able to achieve an orbital stability result without proving decay for at least some of the quantities in the system.}~To be a true orbital stability result, and be applicable to nonlinear problems, it is important that such a statement \emph{does not lose derivatives} at the top-order energy level.

\item \textbf{Step 4:~Asymptotic stability}

The core of a linear stability result should be a uniform decay statement for all solutions to the linearised system of equations in terms of an explicit initial data (energy) quantity.~The decay statement will necessarily lose derivatives, but the loss should be \emph{quantifiable} and, at least in principle, compatible with nonlinear applications.~One also has to ensure that the decay statement obtained is \emph{quantitative} and again, at least in principle, sufficient for nonlinear applications.~In this sense, the decay rates of certain quantities need to be sufficiently fast.  
\end{itemize}

\subsubsection{The Schwarzschild solution} \label{sec_intro_DHR}

An important milestone in the mathematical study of linear stability is the work \cite{DHR} proving the linear stability of the Schwarzschild solution.\footnote{The work \cite{DHR} is more general than the direct linear analogue of \cite{DHRT}, in that it allows convergence to a linearised \emph{Kerr} solution.}~The linear stability result \cite{DHR} is carried out in line with the prescription of Steps 1-4 and ultimately serves as the fundamental building block in the nonlinear stability result \cite{DHRT}.~Related linear results for the Schwarzschild solution and other solutions to \eqref{intro_intro_EVE}, including the Kerr solution, have recently appeared and are discussed in Section \ref{sec_intro_other_works} of the present introduction.

\medskip

The work \cite{DHR} is centered around long celebrated ideas and contributions from the physics literature on black holes.~Among these, particularly prominent is the realisation that one can derive decoupled equations for some of the quantities in the system of linearised gravity.~Indeed, it has been known since the original work \cite{Bardeen_Press_decoupling_Schw} that the system of linearised gravity on Schwarzschild allows, if suitably formulated, the \emph{exact decoupling} of two \emph{gauge invariant} linearised quantities into two decoupled equations, known as \emph{spin $\pm 2$ Teukolsky equations}.\footnote{The Teukolsky equations can be formulated for general spin $s\in \frac{1}{2}\,\mathbb{Z}$.~For $s=0$, the Teukolsky equations (both) coincide with the wave equation, for $s=\pm 1$ they govern the dynamics of the extremal components of electromagnetic perturbations.~For the sake of the present introduction, we only refer to the case $s=\pm 2$.~We note that some of the results cited hold for general spin.}~These quantities correspond to the linearised extremal curvature components in the Newman--Penrose formalism \cite{Newman_Penrose_formalism}.~Decoupled \emph{wave-like} equations, known as \emph{Regge--Wheeler} \cite{Regge_Wheeler_eqn} and \emph{Zerilli} \cite{Zerilli_eqn} \emph{equations}, have been previously derived following a different approach to linearised gravity, based on metric perturbations (see also \cite{Vishveshwara_stability_schwarzschild, Moncrief_spherically_symmetric_perturbations}).~Work of Chandrasekhar \cite{Chandrasekhar_transformation_theory} developed a transformation theory mapping fixed-frequency \emph{mode} solutions to the spin $\pm 2$ Teukolsky equations to solutions to the Regge--Wheeler equation, connecting the two perspectives.

\medskip

Crucially, the approach to Steps 1 and 2 of the work \cite{DHR} in a \emph{double-null gauge} yields a system of linearised gravity which admits the well-known decoupling of two gauge invariant quantities into Teukolsky equations.~To exploit this decoupling in the stability analysis of Steps 3 and 4, the work \cite{DHR} preliminarily addresses the asymptotic properties of \emph{general} solutions to the spin $\pm 2$ Teukolsky equations on Schwarzschild.~As a first step, work \cite{DHR} generalises Chandrasekhar's theory to a transformation theory for \emph{general} solutions to the Teukolsky equations and establishes uniform boundedness and decay for the transformed Regge--Wheeler quantities.\footnote{The \emph{mode} stability analysis of the Regge--Wheeler equation already appears in the original work \cite{Regge_Wheeler_eqn}.}~The transformation theory of \cite{DHR} is formulated in physical space and allows, as a second step, to translate the decay for the Regge--Wheeler quantities into decay for the original Teukolsky quantities.~As an application to the linear stability problem, uniform decay for the decoupled gauge invariant quantities is established \emph{independently from the other quantities in the system}.~We point out that the asymptotic analysis of the Regge--Wheeler equation and, in turn, of the Teukolsky equations builds upon a series of works, culminating in \cite{FullKerr, new_method}, developing important robust methods in the study of the scalar linear wave equation
\begin{equation*} 
\Box_g \, \varphi=0 
\end{equation*}  
on black hole exterior spacetimes. 

\medskip

The proof of stability for the gauge invariant quantities captures an essential feature of the dynamics of the problem, namely the \emph{hyperbolic} character of the system of linearised gravity, and, in fact, already controls \emph{all} the dynamical degrees of freedom independent of the choice of gauge \cite{Wald_perturbations_kerr}.~In view of this, it has been a common (mis)belief in the physics literature that a stability result\footnote{In fact, a \emph{mode} stability result.} for the gauge invariant quantities constitutes a conclusive linear stability result.~However, to successfully apply the linear theory to the nonlinear stability problem, one needs a \emph{full} linear stability result, namely a stability statement for the \emph{full} system of linearised gravity.~The remaining quantities to control in the system are \emph{gauge dependent}.~Hence, the structure of the remaining part of the system and its stability analysis are very much sensitive to the approach adopted in Step 1 of the problem, where a choice of gauge is made.

\medskip

By adopting a double-null gauge, work \cite{DHR} is the first to resolve the gauge dependent part of the analysis for the system of linearised gravity on Schwarzschild.~After gaining control over the gauge invariant quantities, the proof of \cite{DHR} exploits a \emph{hierarchical structure} in the transport equations in the system to prove decay for \emph{all} the gauge dependent quantities.

\subsubsection{The Kerr problem} \label{sec_intro_teukolsky_to_kerr}

As discovered by the seminal work \cite{Teukolsky_eqn}, a suitable formulation\footnote{We already point out that such a formulation cannot be seen as the Kerr analogue of the approach to Steps 1 and 2 of \cite{DHR}, namely a formulation in a double-null gauge.~See the end of Section \ref{sec_intro_divorce_frames_foliation}.} of the system of linearised gravity on Kerr still admits the exact decoupling of two gauge invariant linearised quantities into decoupled spin $\pm 2$ Teukolsky equations.~The \emph{mode} stability analysis of these equations is, in the case of Kerr, already highly non-trivial.~Relying on the crucial fact that, in analogy with the wave equation \cite{Carter_separability_we_kerr}, the spin $\pm 2$ Teukolsky equations on Kerr fully separate, the celebrated work \cite{Whiting_mode_stability_kerr} proves the non-existence of exponentially growing mode solutions.~As key steps towards addressing the stability of \emph{general} solutions, work \cite{Andersson_Ma_Paganini_Whiting_mode_stability_Teukolsky_real_axis} extends the mode stability result \cite{Whiting_mode_stability_kerr} to the real axis and work \cite{Teixeira_da_Costa_mode_stability} achieves a \emph{quantitative} mode stability statement.~The asymptotic analysis of general solutions is considerably more complicated.~The results \cite{Ma_Teukolsky_slowly_rot_kerr, Teukolsky_DHR} establish uniform boundedness and decay for \emph{general} solutions to the Teukolsky equations in the slowly rotating regime $|a|\ll M$.~The work \cite{Shl_Teix_teuk_1} resolves the frequency space analysis of the equations without any restriction on the spacetime parameters, to be seen as the main missing ingredient towards proving uniform boundedness and decay for the Teukolsky quantities in the full sub-extremal range $|a|<M$.~We remark that all these works address the spin $\pm 2$ Teukolsky equations on Kerr and their asymptotic properties \emph{independently} from the full linear stability problem.~In fact, a system of linearised gravity on Kerr is \emph{not} derived in these works.

\medskip

As in the Schwarzschild case, one desires \textbf{a formulation of Step 1 leading to a system of linearised gravity \emph{both} amenable to the decoupling of the gauge invariant quantities \emph{and} possessing a hierarchical structure sufficient to prove stability for \emph{all} the quantities in the system}.~However, exploiting the decoupling and decay of the gauge invariant quantities in the context of the full linear stability problem for Kerr is more subtle.~Indeed, regardless of the formulation chosen, the equations in the gauge dependent part of the system present a stronger coupling, due to the fact that the Kerr solution is only stationary.~Furthermore, for reasons discussed in Section \ref{sec_intro_divorce_frames_foliation}, a formulation of the problem capturing the decoupling of the gauge invariant quantities is significantly more difficult to implement than in the Schwarzschild case, and necessarily yields a system of linearised gravity with new, more intricate structures in its gauge dependent part.~These additional complications are related to the use of \emph{non-integrable frames} and the appearance of the \emph{antitraces} of the second fundamental forms in the equations.

\medskip

We point out that instances of systems of linearised gravity on Kerr possessing a hierarchical structure compatible with the decoupling of the gauge invariant quantities have been noted, and exploited, in a considerable number of works in the physics literature.~Examples of such works are those concerning metric reconstruction problems \cite{Whiting_Price_metric_reconstruction_review}, in which, starting from the decoupled gauge invariant quantities, one wants to reconstruct the corresponding linearised metric perturbations.~The metric perturbations are typically obtained by descending a hierarchy of transport equations in the system, in which each equation is sourced by one or more of the previously computed quantities \cite{Chrzanowski_metric_reconstruction, Chandrasekhar_book, Loutrel_Ripley_Giorgi_Pretorius_formalism_second_order_perturbations}.~However, it is important to note that employing this type of hierarchies in a proof of linear stability would \emph{not} lead to a full stability analysis \emph{in line with the prescription of Steps 3 and 4},\footnote{We also note that the systems of linearised gravity on Kerr in which these hierarchies appear typically fail to be derived as prescribed by Steps 1 and 2.~See the related footnote \ref{footnote:linear_gauges}.} one typical obstruction being the wasteful loss of derivatives required to establish uniform decay statements.~Finding a hierarchy which is suitable to our objective is a much more delicate task.

\medskip

The purpose of the present work is to develop a \textbf{new geometric framework to address the \emph{full} linear stability of the Kerr solution to gravitational perturbations in the full sub-extremal range $|a|<M$}.~In particular, this work resolves Steps 1 and 2 of the linear stability problem\footnote{The well-posedness of the vacuum Einstein equations in our gauge, which is stated as a feature of Step 1 in Section \ref{sec_intro_linear_stability}, is not addressed in the present work.} by (1) \textbf{introducing a \emph{new} formulation of nonlinear gravitational perturbations of the Kerr solution} and (2) \textbf{deriving a \emph{new} system of linearised gravity on Kerr exterior spacetimes}.~The new linearised system reconciles the two aforementioned aspects of the problem.~Indeed, as a result of the choices made in the formulation of nonlinear gravitational perturbations, the linearised system obtained allows for the well-known decoupling of two gauge invariant linearised quantities and, at the same time, exhibits a \emph{new hierarchical structure} in its gauge dependent part making the stability analysis of the full system tractable.

\medskip

In the future, we intend to perform the full stability analysis of our system of linearised gravity in the full sub-extremal range $|a|<M$ and, \emph{for the first time}, in line with the prescription of Steps 3 and 4.~Our resolution of the linear stability problem will provide the fundamental ingredient towards a proof of nonlinear stability of the Kerr solution in the full sub-extremal range $|a|<M$.~Although the nonlinear stability problem presents a number of additional (and daunting) technical difficulties, the transition from the linear to the nonlinear analysis in our framework is not expected to present any new conceptual elements.

\subsection{The new system of linearised gravity} \label{sec_intro_formulation_problem}

To implement our framework and derive a system of linearised gravity with the features described, some specific \emph{choices} in the formulation of the problem are to be made.~In this section, we give a brief outline of the derivation of our new system of linearised gravity and the choices that it entails.~Further motivation and technical comments on our choices appear in Section \ref{sec_intro_technical_challenges}.

\medskip

As we shall see in the present section, all our choices are made in the formulation of nonlinear perturbations (Section \ref{sec_intro_nonlinear_perturbations}).~The result of our choices is a new nonlinear \emph{geometric} framework tailored to the \emph{outgoing principal null geodesics} of Kerr.~The fact that our framework privileges the outgoing direction reflects in a new convenient form of the outgoing transport equations in both the nonlinear and linearised systems of equations (Section \ref{sec_intro_from_nonlinear_to_linear}) and in the enhanced role played by the event horizon and red-shift effect in the future analysis (see the later Section \ref{sec_intro_elements_analysis}).

\subsubsection{Preliminaries:~The Kerr exterior manifold} \label{sec_intro_preliminaries_kerr}

We start with some preliminaries.~Let $a$ and $M$ be real parameters, with $|a|<M$.~Through a procedure already sketched in Section \ref{sec_overview_definition_kerr_exterior_manifold}, we define the \emph{Kerr exterior manifold} $(\mathcal{M},g_{a,M})$, with ambient manifold
\begin{equation} \label{intro_manifold_fixed_ambient}
\mathcal{M} := (-\infty,\infty)\times [0,\infty) \times \mathbb{S}^2 \, .
\end{equation}
We refer to the boundary 
\begin{equation*}
\mathcal{H}^+ := \partial \mathcal{M}
\end{equation*}
as the \emph{(future) event horizon}.~We define the \emph{algebraically special} (local) null frame $$\mathcal{N}_{\text{as}}=(e_1^{\text{as}},e_2^{\text{as}},e_3^{\text{as}},e_4^{\text{as}})$$ of $(\mathcal{M},g_{a,M})$, with\footnote{For simplicity, the frame vector fields \eqref{intro_intro_as_frame} are given in coordinate form relative to the standard Boyer--Lindquist differentiable structure on $\mathcal{M}\setminus\mathcal{H}^+$.~In the bulk of the paper, the frame vector fields $(e_4^{\text{as}},e_3^{\text{as}})$ are defined relative to a regular differentiable structure on the whole $\mathcal{M}$.~See Section \ref{sec_Kerr_algebr_special_frame}.~We also point out that a choice of the frame vector fields $(e_1^{\text{as}},e_2^{\text{as}})$ is not required.} 
\begin{align} \label{intro_intro_as_frame}
e_4^{\text{as}} &:=\frac{r^2+a^2}{\Sigma}\,\partial_{t}+\frac{\Delta}{\Sigma}\,\partial_{r}+\frac{a}{\Sigma}\,\partial_{\phi}  \, , & e_3^{\text{as}} &:= \frac{r^2+a^2}{\Delta}\,\partial_{t}-\partial_{r}+\frac{a}{\Delta}\,\partial_{\phi}  \, .
\end{align}
The frame vector fields \eqref{intro_intro_as_frame} are global, regular and non-degenerate vector fields on $\mathcal{M}\setminus\mathcal{H}^+$, and smoothly extend to global, regular and non-degenerate vector fields on the whole $\mathcal{M}$, including on $\mathcal{H}^+$.~As we shall explain in Section \ref{sec_intro_divorce_frames_foliation}, the frame vector fields \eqref{intro_intro_as_frame} generate null geodesics of $(\mathcal{M},g_{a,M})$.

\medskip

We refer to any (local) coordinates $(x^1,x^2,\vartheta^A)$ on $\mathcal{M}$, with $(\vartheta^1,\vartheta^2)\in\mathbb{S}^2$, satisfying
\begin{align*}
e_4^{\text{as}}(x^1)&=0 \, , & e_4^{\text{as}}(\vartheta^A)&=0
\end{align*}
as an \emph{outgoing principal differentiable structure} of Kerr.

\medskip

\begin{figure}[H]

\centering

\begin{subfigure}{0.3\textwidth}

\centering

\begin{tikzpicture}[scale=2.2]

\draw (0,2)--(-1.1,0.9);

\node at (-0.8,1.65) {$\mathcal{H}^+\cap \mathcal{M}^*$};
\node at (0.4,1.9) {$\mathcal{I}^+$};
\node at (0,1.3) {$\mathcal{M}^*$};


\draw (0,2)  circle[radius=1pt];




\draw plot [smooth] coordinates { (-0.9,1.1) (-0.35,0.95) (0.3,0.85) };

\draw[dashed] plot [smooth] coordinates {  (0.3,0.85) (0.38,1.3) (0.5,1.5) };


\draw[dashed] (0,2)--(0.7,1.3);


\end{tikzpicture}

\end{subfigure}
\begin{subfigure}{0.3\textwidth}

\centering

\begin{tikzpicture}[scale=2.2]

\draw[color=blue, thick] (0,2)--(-0.9,1.1);
\draw (-1.1,0.9)--(-0.9,1.1);

\node at (-0.8,1.65) {$\mathcal{H}^+\cap \mathcal{M}^*$};
\node at (0.4,1.9) {$\mathcal{I}^+$};


\draw (0,2)  circle[radius=1pt];




\draw plot [smooth] coordinates { (-0.9,1.1) (-0.35,0.95) (0.3,0.85) };

\draw[dashed, color=blue, thick] plot [smooth] coordinates {  (0.3,0.85) (0.38,1.3) (0.5,1.5) };

\draw[color=blue, thick] plot [smooth] coordinates {  (-0.2,0.92) (-0.1,1.4) (0.2,1.8) };


\draw[dashed] (0,2)--(0.7,1.3);


\end{tikzpicture}

\end{subfigure}
\begin{subfigure}{0.3\textwidth}

\centering

\begin{tikzpicture}[scale=2.2]

\draw (0,2)--(-1.1,0.9);

\node at (-0.8,1.65) {$\mathcal{H}^+\cap \mathcal{M}^*$};
\node at (0.4,1.9) {$\mathcal{I}^+$};


\draw (0,2)  circle[radius=1pt];




\draw[color=red, thick] plot [smooth] coordinates { (-0.9,1.1) (-0.35,0.95) (0.3,0.85) };

\draw[color=red, thick] plot [smooth] coordinates { (-0.6,1.4) (-0.2,1.25) (0.34,1.15) };

\draw[dashed] plot [smooth] coordinates {  (0.3,0.85) (0.38,1.3) (0.5,1.5) };


\draw[dashed, color=red, thick] (0,2)--(0.5,1.5);
\draw[dashed] (0.7,1.3)--(0.5,1.5);


\end{tikzpicture}

\end{subfigure}

\caption{The star-normalised, outgoing principal differentiable structure will be defined on a spacetime region $\mathcal{M}^*\subset\mathcal{M}$, on which the coordinates $(s,\underline{s})$ are global.~The figure depicts the Penrose diagram of the manifold $(\mathcal{M}^*,g_{a,M})$ and the foliation of $(\mathcal{M}^*,g_{a,M})$ induced by the coordinates $s$ (in blue) and $\underline{s}$ (in red).~The geometry of the level sets of the coordinates is already discussed in Section \ref{sec_overview_back_diff_structure}.}

\label{fig:intro_foliation}

\end{figure}
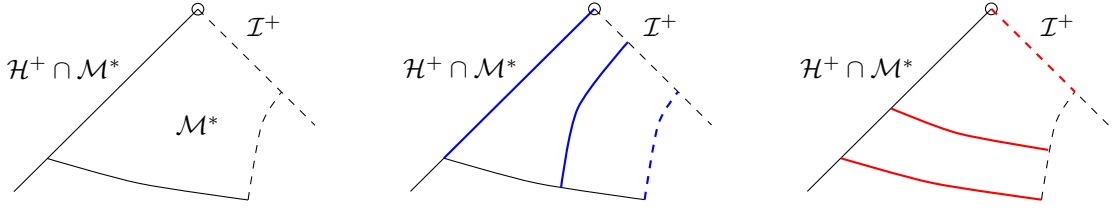

We define the regular differentiable structure 
\begin{equation} \label{intro_s_sbar_diff_structure}
(s,\underline{s},\theta^A)
\end{equation}
on $\mathcal{M}$ (including on $\mathcal{H}^+$),\footnote{Strictly speaking, the differentiable structure $(s,\underline{s},\theta^A)$ will be defined on a spacetime region $\mathcal{M}^*\subset \mathcal{M}$.~See Figure \ref{fig:intro_foliation}.~The distinction between $\mathcal{M}$ and $\mathcal{M}^*$ plays no role in the present introduction, and is thus neglected for simplicity.} with $A=\left\lbrace 1, 2 \right\rbrace$ and 
\begin{align*}
s &\in [r_+,\infty) \, , & \underline{s} &\in[0,\infty) \, , & (\theta^1,\theta^2)&\in\mathbb{S}^2 \, ,
\end{align*}
such that 
\begin{itemize}
\item The coordinates $(s,\underline{s})$ are global on $\mathcal{M}$, whereas the angular coordinates $(\theta^1,\theta^2)$ are subject to the usual degeneration on the spheres $$\mathbb{S}^2_{s,\underline{s}}:=\left\lbrace s,\underline{s} \right\rbrace \times \mathbb{S}^2 \, .$$
\item We have
\begin{equation*}
\underline{s}:=t+\int_{r_0}^r \iota(r^{\prime})\left(\frac{{r^{\prime}}^2+a^2}{\Delta(r^{\prime})} -1 \right) dr^{\prime} 
\end{equation*}
on $\mathcal{M}$, where $(t,r)$ are the standard Boyer--Lindquist time and radial coordinates and $\iota$ is a suitable cut-off function which makes the level sets of $\underline{s}$ everywhere \emph{spacelike}, horizon-penetrating and asymptotically flat (see Section \ref{sec_Kerr_star_coords}).~The coordinate $\underline{s}$ coincides with the time coordinate $t^*$ of the Kerr-star differentiable structure defined in Section \ref{sec_Kerr_star_coords}.
\item For each $s\in [r_+,\infty)$, the $\mathbb{S}^2_{s,0}$-sphere is a sphere of constant Boyer--Lindquist radius.~The coordinates $(\theta^1,\theta^2)$ are arbitrary angular coordinates on each of the $\mathbb{S}^2_{s,0}$-spheres.
\item The null geodesics generated by $e_4^{\text{as}}$ are curves of constant coordinates $(s,\theta^A)$, i.e.
\begin{align} \label{intro_coords_transp_e4}
e_4^{\text{as}}(s)&=0 \, , & e_4^{\text{as}}(\theta^A)&=0 
\end{align}
on $\mathcal{M}$.
\end{itemize}
We refer to the coordinates \eqref{intro_s_sbar_diff_structure} as the \textbf{\emph{star-normalised}, outgoing principal differentiable structure} of Kerr.~See Figure \ref{fig:intro_foliation}.

\medskip

Relative to the differentiable structure \eqref{intro_s_sbar_diff_structure}, the Kerr metric takes the form
\begin{equation} \label{intro_kerr_metric_outgoing_principal}
g_{a,M}=\mathfrak{a}\,ds^2+2\,\Omega^2\,ds\,d\underline{s}+2\,b_{\theta^A}\,ds\,d\theta^A+2\,\underline{b}_{\theta^A}\,d\underline{s}\,d\theta^A+\,\gamma_{\theta^A\theta^B}\,d\theta^A d\theta^B
\end{equation}
on $\mathcal{M}$, with
\begin{align}
\partial_{\underline{s}}\,\Omega^2+\mathfrak{k}_{a,M}\,\Omega^2&=0 \, , & \partial_{\underline{s}}\,\underline{b}_{\theta^A}+\mathfrak{k}_{a,M}\,\underline{b}_{\theta^A}&=0  \label{intro_kerr_gauge_ids}
\end{align}
on $\mathcal{M}$, and the frame vector fields $(e_4^{\text{as}},e_3^{\text{as}})$ take the form
\begin{align} \label{intro_alg_frame_e4_e3_coord}
e_4^{\text{as}}&=k_{a,M}\,\partial_{\underline{s}} \, , & e_3^{\text{as}}&=\underline{j}_{a,M}\,\partial_s+\underline{k}_{a,M}\,\partial_{\underline{s}}+\underline{\Lambda}^{\theta^A}_{a,M}\,\partial_{\theta^A}
\end{align}
on $\mathcal{M}$.~The quantities $\mathfrak{a}$, $\Omega^2$, $\mathfrak{k}_{a,M}$, $k_{a,M}$, $\underline{j}_{a,M}$ and $\underline{k}_{a,M}$ are smooth scalar functions on $\mathcal{M}$, the quantities $b$ and $\underline{b}$ are one-tensors on the $\mathbb{S}^2_{s,\underline{s}}$-spheres, the quantity $\gamma$ is a symmetric two-tensor on the $\mathbb{S}^2_{s,\underline{s}}$-spheres and the quantity $\underline{\Lambda}_{a,M}$ is a vector field tangent to the $\mathbb{S}^2_{s,\underline{s}}$-spheres.~All quantities are regular on the whole $\mathcal{M}$, including on $\mathcal{H}^+$.

\medskip

The Kerr metric coefficients in \eqref{intro_kerr_metric_outgoing_principal} and frame coefficients of $e_3^{\text{as}}$ in \eqref{intro_alg_frame_e4_e3_coord} cannot be made explicit, whereas the functions $\mathfrak{k}_{a,M}$ and $k_{a,M}$ are known explicitly close to the event horizon and null infinity (see Section \ref{sec_Kerr_background_diff_structure}).~In particular, the function $k_{a,M}$ is nowhere vanishing on $\mathcal{M}$.

\medskip

\subsubsection{Nonlinear perturbations}  \label{sec_intro_nonlinear_perturbations}

To formulate nonlinear perturbations of the Kerr exterior manifold, we start by choosing real parameters $a$ and $M$, with $|a|<M$.~We fix the manifold-with-boundary $\mathcal{M}$, which we identify with the Kerr ambient manifold \eqref{intro_manifold_fixed_ambient} with star-normalised, outgoing principal differentiable structure $(s,\underline{s},\theta^A)$.~We introduce the fixed smooth scalar functions $\mathfrak{k}_{a,M}$ and $k_{a,M}$ of the fixed coordinates on $\mathcal{M}$.

\medskip

Once the background structure is fixed, we prescribe a one-parameter family of smooth Lorentzian metrics 
\begin{equation} \label{intro_outline_family_metrics}
\bsy{g}(\epsilon):=\bsy{\mathfrak{a}}(\epsilon)\,ds^2+2\,\bsy{\Omega}^2(\epsilon)\,ds\,d\underline{s}+2\,\bsy{b}_{\theta^A}(\epsilon)\,ds\,d\theta^A+2\,\underline{\bsy{b}}_{\theta^A}(\epsilon)\,d\underline{s}\,d\theta^A+\,\bsy{\gamma}_{\theta^A\theta^B}(\epsilon)\,d\theta^A d\theta^B
\end{equation}
on $\mathcal{M}$, with 
\begin{align}
\partial_{\underline{s}}\,\bsy{\Omega}^2(\epsilon)+\mathfrak{k}_{a,M}\,\bsy{\Omega}^2(\epsilon)&=0 \, , & \partial_{\underline{s}}\,\underline{\bsy{b}}_{\theta^A}(\epsilon)+\mathfrak{k}_{a,M}\,\underline{\bsy{b}}_{\theta^A}(\epsilon)&=0  \label{intro_outline_gauge_ids}
\end{align}
on $\mathcal{M}$ for all $\epsilon$.~For each $\epsilon$, the metric quantities $\bsy{\mathfrak{a}}(\epsilon)$ and $\bsy{\Omega}^2(\epsilon)$ are smooth scalar functions on $\mathcal{M}$, the metric quantities $\bsy{b}(\epsilon)$ and $\underline{\bsy{b}}(\epsilon)$ are one-tensors on the $\mathbb{S}^2_{s,\underline{s}}$-spheres and the metric quantity $\bsy{\gamma}(\epsilon)$ is a symmetric two-tensor on the $\mathbb{S}^2_{s,\underline{s}}$-spheres.~All quantities are regular on the whole $\mathcal{M}$, including on $\mathcal{H}^+$.~We assume that $\bsy{g}(0) = g_{a,M}$ with 
\begin{align*}
\bsy{\mathfrak{a}}(0)&=\mathfrak{a} \, , & \bsy{\Omega}^2(0)&=\Omega^2 \, , & \bsy{b}_{\theta^A}(0)&=b_{\theta^A} \, , & \underline{\bsy{b}}_{\theta^A}(0)&=\underline{b}_{\theta^A} \, , & \bsy{\gamma}_{\theta^A\theta^B}(0)&=\gamma_{\theta^A\theta^B} \, ,
\end{align*}
i.e.~the metric coefficients of $\bsy{g}(0)$ relative to the fixed differentiable structure coincide with the Kerr metric coefficients relative to the star-normalised, outgoing principal differentiable structure of Kerr.

\medskip

The form \eqref{intro_outline_family_metrics} of the metrics $\bsy{g}(\epsilon)$, together with the additional identities \eqref{intro_outline_gauge_ids}, correspond to a choice of gauge for the family of metrics $\bsy{g}(\epsilon)$, called \textbf{outgoing principal gauge}.~See Section \ref{sec_intro_geometric_framework}.

\medskip

To the one-parameter family of metrics $\bsy{g}(\epsilon)$, we associate a one-parameter family of local frames $$\bsy{\mathcal{N}}(\epsilon)=(\eo(\epsilon),\etw(\epsilon),\et(\epsilon),\ef(\epsilon))$$ on $\mathcal{M}$ such that\footnote{A choice of the frame vector fields $(\eo(\epsilon),\etw(\epsilon))$ is not required.}
\begin{align} \label{intro_formulation_problem_e4_e3}
\ef(\epsilon)&:= k_{a,M}\,\partial_{\underline{s}}  & \et(\epsilon)&:=\underline{\bsy{j}}(\epsilon)\,\partial_s+\underline{\bsy{k}}(\epsilon)\,\partial_{\underline{s}}+\underline{\bsy{\Lambda}}^{\theta^A}(\epsilon)\,\partial_{\theta^A} \, .
\end{align}
For each $\epsilon$, the frame quantities $\underline{\bsy{j}}(\epsilon)$ and $\underline{\bsy{k}}(\epsilon)$ are smooth scalar functions on $\mathcal{M}$ and the frame quantity $\underline{\bsy{\Lambda}}(\epsilon)$ is a vector field tangent to the $\mathbb{S}^2_{s,\underline{s}}$-spheres.~All quantities are regular on the whole $\mathcal{M}$, including on $\mathcal{H}^+$.~The frame quantities are \emph{defined} implicitly as solutions to the system of ODEs \eqref{overview_ODE_1}-\eqref{overview_ODE_3} with initial data \eqref{overview_ODE_data_1}-\eqref{overview_ODE_data_3} of Section \ref{sec_overview_family_frames}.

\medskip

As a result of their definition, the frames $\bsy{\mathcal{N}}(\epsilon)$ satisfy the following properties:\footnote{The proof of some of the properties is not immediate from the definition of the frames.~See already the related comments in Section \ref{sec_intro_geometric_framework}.}
\begin{itemize}
\item \textbf{$\bsy{\mathcal{N}}(\epsilon)$ are null frames}

The frames $\bsy{\mathcal{N}}(\epsilon)$ are null relative to the metrics $\bsy{g}(\epsilon)$ for all $\epsilon$.
\item \textbf{$\ef(\epsilon)$ and $\et(\epsilon)$ are global, regular and non-degenerate}

The frame vector fields \eqref{intro_formulation_problem_e4_e3} are global, regular and non-degenerate vector fields on the whole $\mathcal{M}$, including on $\mathcal{H}^+$.
\item \textbf{$\ef(\epsilon)$ is fixed}

The frame vector field $\ef(\epsilon)$ remains fixed on $\mathcal{M}$ for all $\epsilon$ and satisfies the identity 
\begin{equation} \label{intro_fixed_e4}
\ef(\epsilon) = e_4^{\text{as}}
\end{equation}
for all $\epsilon$.
\item \textbf{$\bsy{\mathcal{N}}(0)$ is the algebraically special frame}

We have $$\bsy{\mathcal{N}}(0) = \mathcal{N}_{\text{as}}$$ with 
\begin{align*}
\underline{\bsy{j}}(0)&=\underline{j}_{a,M} \, , & \underline{\bsy{k}}(0)&=\underline{k}_{a,M} \, , & \underline{\bsy{\Lambda}}^{\theta^A}(0)&=\underline{\Lambda}^{\theta^A}_{a,M} \, ,
\end{align*}
i.e.~the frame coefficients of $\bsy{e}_{\bsy{3}}(0)$ relative to the fixed differentiable structure coincide with the frame coefficients of the vector field $e_3^{\text{as}}$ relative to the star-normalised, outgoing principal differentiable structure of Kerr.
\item \textbf{$\bsy{\omegah}(\epsilon)$, $\bsy{\xi}(\epsilon)$ and ``$\bsy{\etab}(\epsilon)$" are fixed}

The perturbed connection coefficients and curvature components are defined as the connection coefficients and curvature components of the metrics $\bsy{g}(\epsilon)$ relative to the frames $\bsy{\mathcal{N}}(\epsilon)$.~For $\epsilon=0$, all the perturbed geometric quantities coincide with the corresponding connection coefficients and curvature components of the Kerr metric $g_{a,M}$ relative to the algebraically special frame $\mathcal{N}_{\text{as}}$.

As a result of our choices, a selected set of perturbed connection coefficients remain fixed on $\mathcal{M}$ for all $\epsilon$ and coincide, in a suitable sense, with their respective Kerr quantities.\footnote{The statement that the perturbed connection coefficient $\underline{\bsy{\eta}}(\epsilon)$ in \eqref{intro_fixed_connection_coeffs} remains fixed for all $\epsilon$ and coincides with the connection coefficient $\underline{\bsy{\eta}}(0)$ can only be understood informally.~See Section \ref{sec_overview_perturbations_kerr}.}~These connection coefficients are
\begin{align}
\bsy{\omegah}(\epsilon)&:=\frac{1}{2}\,\bsy{g}(\bsy{\nabla}_{\bsy{4}}\et(\epsilon),\ef(\epsilon))  \, , &  \bsy{\xi}_{\bsy{A}}(\epsilon)&:=\frac{1}{2}\,\bsy{g}(\bsy{\nabla}_{\bsy{4}}\ef(\epsilon),\ea(\epsilon)) \, , & \underline{\bsy{\eta}}_{\bsy{A}}(\epsilon)&:=\frac{1}{2}\,\bsy{g}(\bsy{\nabla}_{\bsy{4}}\et(\epsilon),\ea(\epsilon)) \label{intro_fixed_connection_coeffs}
\end{align}
and dictate how the frame vector fields $\et(\epsilon)$ and $\ef(\epsilon)$ are transported along the integral curves of $\ef(\epsilon)$.
\end{itemize}
Our choice of frames is \emph{new}.~In particular, the property that the perturbed connection coefficient $\underline{\bsy{\eta}}(\epsilon)$ remains fixed has no analogue in previous works.

\medskip

This concludes the formulation of nonlinear perturbations of the Kerr exterior manifold.~We emphasise that the formulation involves \emph{two choices}, namely the choice of gauge for the metrics $\bsy{g}(\epsilon)$ and the choice of frames $\bsy{\mathcal{N}}(\epsilon)$.~The \textbf{choice of frames is the most important novelty of our framework}, as it is ultimately responsible for the new structural properties of our system of linearised gravity.~See Sections \ref{sec_intro_from_nonlinear_to_linear} and \ref{sec_intro_technical_challenges}.

\subsubsection{From nonlinear perturbations to the system of linearised gravity} \label{sec_intro_from_nonlinear_to_linear}

The derivation of the system of linearised gravity on the Kerr exterior manifold then follows three more steps and does not involve any additional choices:\footnote{Strictly speaking, there is some freedom in setting up the linearisation procedure.~However, all the possible linearisation procedures are equivalent from the point of view of the key structures of the linearised system obtained.}
\begin{enumerate}[(i)]
\item \textbf{A comparison procedure}

We first establish a geometric procedure to compare the perturbed quantities to the corresponding Kerr quantities on the fixed manifold $\mathcal{M}$.\footnote{Such a geometric procedure is needed because the frames $\bsy{\mathcal{N}}(\epsilon)$ induce orthogonal distributions which \emph{vary} with $\epsilon$ (see Appendix \ref{sec_intro_appendix} and Section \ref{sec_intro_divorce_frames_foliation} for the meaning of these words).~For a brief discussion of the geometric procedure, see already Section \ref{sec_overview_identification_horizontal_structures} of the overview.~Although we do not elaborate on this step of the problem in the present introduction, implementing this geometric procedure represents a major technical ingredient of our framework.}

\item \textbf{The vacuum Einstein equations}

Secondly, we assume that the metrics $\bsy{g}(\epsilon)$ solve the vacuum Einstein equations for all $\epsilon$, which are formulated as a nonlinear system of equations for the perturbed connection coefficients and curvature components of $\bsy{g}(\epsilon)$ relative to the frames $\bsy{\mathcal{N}}(\epsilon)$.~Once the vacuum Einstein equations are assumed to hold, we refer to nonlinear perturbations as nonlinear \emph{gravitational} perturbations of the Kerr exterior manifold.~By exploiting our comparison procedure, one can take geometric differences of perturbed and Kerr quantities and \emph{renormalise} the vacuum Einstein equations relative to the Kerr background.

Our choices in the formulation of nonlinear perturbations enforce a \textbf{\emph{new} structure} in the renormalised vacuum Einstein equations.~The new structure characterises the renormalised \emph{null structure equations}.~In our renormalised system, these equations are of three different types:~Outgoing transport equations of the schematic form
\begin{equation} \label{intro_schematic_eqn_1}
\nablasl_{4}(\widetilde{\bsy{\Gamma}}(\epsilon)-\widetilde{\Gamma}_{a,M})=(\bsy{\psi}(\epsilon)-\psi_{a,M}) \, ,
\end{equation} 
ingoing transport equations of the schematic form
\begin{equation} \label{intro_schematic_eqn_2}
\nablasl_{3}(\widetilde{\bsy{\Gamma}}(\epsilon)-\widetilde{\Gamma}_{a,M})=\nablasl(\bsy{\Gamma}(\epsilon)-\Gamma_{a,M})+(\bsy{\psi}(\epsilon)-\psi_{a,M}) 
\end{equation} 
and elliptic equations of the schematic form
\begin{equation} \label{intro_schematic_eqn_3}
\nablasl(\widetilde{\bsy{\Gamma}}(\epsilon)-\widetilde{\Gamma}_{a,M})=\nablasl(\bsy{\Gamma}(\epsilon)-\Gamma_{a,M})+(\bsy{\psi}(\epsilon)-\psi_{a,M}) \, ,
\end{equation} 
where $\widetilde{\bsy{\Gamma}}(\epsilon)$, $\bsy{\Gamma}(\epsilon)$ denote distinct perturbed connection coefficients, $\bsy{\psi}(\epsilon)$ perturbed curvature components and $\widetilde{\Gamma}_{a,M}$, $\Gamma_{a,M}$, $\psi_{a,M}$ their corresponding Kerr quantities.~The schematic equations \eqref{intro_schematic_eqn_1}-\eqref{intro_schematic_eqn_3} only include the top-order terms, with $\nablasl$ denoting the Kerr projected covariant derivative (see Section \ref{sec_intro_divorce_frames_foliation}).~Crucially, as a result of our choice of frames $\bsy{\mathcal{N}}(\epsilon)$, all the \emph{outgoing} transport equations do \emph{not} contain any derivatives of renormalised connection coefficients on the right hand side.~Indeed, all the derivatives of renormalised connection coefficients that would appear on the right hand side of such equations are derivatives of one of the fixed connection coefficients \eqref{intro_fixed_connection_coeffs}.~A similar structure is not present in the ingoing transport and elliptic equations.

\item \textbf{The system of linearised gravity}

Finally, we linearise the system of renormalised vacuum Einstein equations around the Kerr exterior manifold, i.e.~around $\epsilon=0$.~The final system of linearised gravity can be handled without any knowledge about its derivation or reference to the nonlinear formulation of the problem.~The unknowns of the system of linearised gravity are the linearised induced metric, frame coefficients,\footnote{See Section \ref{sec_intro_divorce_frames_foliation} for the meaning of the induced metric and frame coefficients.} connection coefficients and curvature components, denoted by the superscript $\overset{\text{\scalebox{.6}{$(1)$}}}{}$.~The system consists of three sets of equations, namely a set of equations for the linearised frame coefficients, the linearised null structure equations and the linearised Bianchi equations.~The equations for the linearised frame coefficients are (both ingoing and outgoing) transport and elliptic equations.

Two of the linearised curvature components in our system, namely 
\begin{align} \label{intro_alpha_alphab}
&\overset{\text{\scalebox{.6}{$(1)$}}}{\alpha} \, , & &\overset{\text{\scalebox{.6}{$(1)$}}}{\underline{\alpha}} \, ,
\end{align}
are \emph{gauge invariant} linearised quantities and \emph{exactly decouple} into spin $\pm2$ Teukolsky equations.

The new structure of the nonlinear null structure equations directly transfers to the linearised system:~The vanishing of the linearised quantities
\begin{align} \label{intro_vanishing_lin_quantities_2}
&\overset{\text{\scalebox{.6}{$(1)$}}}{\hat{\omega}} \, , & &\overset{\text{\scalebox{.6}{$(1)$}}}{\xi} \, , & &\overset{\text{\scalebox{.6}{$(1)$}}}{\underline{\eta}}
\end{align}
implies that \textbf{\emph{all} the \emph{outgoing} transport equations for the linearised connection coefficients do \emph{not} possess any derivatives of linearised connection coefficients on the right hand side} (see, for instance, the equations \eqref{intro_lin_4_chih}-\eqref{intro_lin_4_atrchi} below).~Furthermore, these equations exhibit an overall \emph{weak coupling} with the rest of the system.~Some examples of weakly coupled equations are the linearised second variational formula
\begin{equation} \label{intro_lin_4_chih}
\nablasl_4 \overset{\text{\scalebox{.6}{$(1)$}}}{\widehat{\chi}} +(\text{tr}\chi)\,\overset{\text{\scalebox{.6}{$(1)$}}}{\widehat{\chi}}-\omegah\, \overset{\text{\scalebox{.6}{$(1)$}}}{\widehat{\chi}} =  -\overset{\text{\scalebox{.6}{$(1)$}}}{\alpha} \, ,
\end{equation}
which only couples with the linearised curvature component $\overset{\text{\scalebox{.6}{$(1)$}}}{\alpha}$, the transport equation\footnote{The linearised equations \eqref{intro_lin_4_chibh}-\eqref{intro_lin_4_atrchi} are obtained by linearising the equations \eqref{intro_intro_4_trchi}-\eqref{intro_intro_4_chibh} appearing in Section \ref{sec_intro_divorce_frames_foliation}.~The reader should refer to Section \ref{sec_intro_divorce_frames_foliation} for a preliminary discussion of the meaning of the antitraces of the second fundamental forms appearing in the equations \eqref{intro_lin_4_chibh}-\eqref{intro_lin_3_trchib}.} 
\begin{align} 
\nablasl_4 \overset{\text{\scalebox{.6}{$(1)$}}}{\widehat{\underline{\chi}}}+\chi\times\overset{\text{\scalebox{.6}{$(1)$}}}{\widehat{\underline{\chi}}}+\omegah\,\overset{\text{\scalebox{.6}{$(1)$}}}{\widehat{\underline{\chi}}} =& \, -\frac{1}{2}\,(\text{tr}\underline{\chi})\,\overset{\text{\scalebox{.6}{$(1)$}}}{\widehat{\chi}}+\frac{1}{2}\,(\slashed{\varepsilon}\cdot\underline{\chi})\,{{}^{\star}\overset{\text{\scalebox{.6}{$(1)$}}}{\widehat{\chi}}}  \label{intro_lin_4_chibh} \\  &  - \widehat{(\nablasl\overset{\text{\scalebox{.6}{$(1)$}}}{\widehat{\slashed{g}}})^{\sharp_3}\cdot\underline{\eta}}+ (\nablasl\overset{\text{\scalebox{.6}{$(1)$}}}{\widehat{\slashed{g}}})^{\sharp_1}\cdot\underline{\eta}  -\frac{1}{2}\,\underline{\eta}\,\widehat{\otimes}\,\nablasl(\textup{tr}\overset{\text{\scalebox{.6}{$(1)$}}}{\slashed{g}})-(\slashed{\textup{div}}\,\underline{\eta} +(\underline{\eta},\underline{\eta}))\,\overset{\text{\scalebox{.6}{$(1)$}}}{\widehat{\slashed{g}}} \nonumber\\
&   -\frac{1}{2}\,(\slashed{\varepsilon}\cdot\chi)\,{}^{\star}\underline{\eta}\,\widehat{\otimes}\,(-2\,\overset{\text{\scalebox{.6}{$(1)$}}}{\mathfrak{\slashed{\mathfrak{f}}}}_{3}+\overset{\text{\scalebox{.6}{$(1)$}}}{\underline{\mathfrak{f}}})  + (\slashed{\varepsilon}\cdot\underline{\chi})\,{}^{\star}\underline{\eta}\,\widehat{\otimes}\, \overset{\text{\scalebox{.6}{$(1)$}}}{\mathfrak{\slashed{\mathfrak{f}}}}_{4} \nonumber\\
&+(\nablasl_3\underline{\eta}-\underline{\chi}^{\sharp_2}\cdot\underline{\eta})\,\widehat{\otimes}\,\overset{\text{\scalebox{.6}{$(1)$}}}{\mathfrak{\slashed{\mathfrak{f}}}}_{4}+(\nablasl_4\underline{\eta}-\chi^{\sharp_2}\cdot\underline{\eta})\,\widehat{\otimes} \, \overset{\text{\scalebox{.6}{$(1)$}}}{\mathfrak{\slashed{\mathfrak{f}}}}_{3} \nonumber
\end{align}
for the linearised ingoing shear, which decouples from any linearised connection coefficient other than the linearised outgoing shear, and the linearised Raychaudhuri equation
\begin{equation} \label{intro_lin_4_trchi}
\nablasl_4 (\overset{\text{\scalebox{.6}{$(1)$}}}{\text{tr}\chi})+ (\text{tr}\chi)(\overset{\text{\scalebox{.6}{$(1)$}}}{\text{tr}\chi})-\omegah \, (\overset{\text{\scalebox{.6}{$(1)$}}}{\text{tr}\chi}) =  (\slashed{\varepsilon}\cdot\chi)(\overset{\text{\scalebox{.6}{$(1)$}}}{\slashed{\varepsilon}\cdot\chi}) 
\end{equation}
and transport equation for the linearised antitrace of the outgoing second fundamental form
\begin{equation} \label{intro_lin_4_atrchi}
\nablasl_4 (\overset{\text{\scalebox{.6}{$(1)$}}}{\slashed{\varepsilon}\cdot\chi})+(\text{tr}\chi)(\overset{\text{\scalebox{.6}{$(1)$}}}{\slashed{\varepsilon}\cdot\chi})-\omegah \, (\overset{\text{\scalebox{.6}{$(1)$}}}{\slashed{\varepsilon}\cdot\chi})   = -(\slashed{\varepsilon}\cdot\chi)(\overset{\text{\scalebox{.6}{$(1)$}}}{\text{tr}\chi})  \, ,
\end{equation}
which form a completely decoupled pair of equations.~The absence of linearised frame coefficients on the right hand side of the equations \eqref{intro_lin_4_chih} and \eqref{intro_lin_4_trchi}-\eqref{intro_lin_4_atrchi} originates from the frame identity \eqref{intro_fixed_e4} and contributes to the weak coupling.

\medskip

Similarly, \emph{all} the \emph{outgoing} transport equations for the linearised induced metric and frame coefficients do \emph{not} possess any derivatives of the linearised induced metric or frame coefficients on the right hand side and are weakly coupled with the rest of the system.

\medskip

Analogous structures are not present in the linearised transport equations in the \emph{ingoing} direction.~In fact, all the ingoing transport equations for the linearised connections coefficients couple with a derivative of a linearised connection coefficient on the right hand side.~An example is the second variational formula
\begin{equation*}
\nablasl_3 \overset{\text{\scalebox{.6}{$(1)$}}}{\widehat{\underline{\chi}}}+(\text{tr}\underline{\chi})\,\overset{\text{\scalebox{.6}{$(1)$}}}{\widehat{\underline{\chi}}} = -2\,\slashed{\mathcal{D}}{}_2^{\star}\, \overset{\text{\scalebox{.6}{$(1)$}}}{\underline{\xi}}+(\underline{\eta}-\eta)\,\widehat{\otimes}\,\overset{\text{\scalebox{.6}{$(1)$}}}{\underline{\xi}}    -\overset{\text{\scalebox{.6}{$(1)$}}}{\underline{\alpha}} \, ,
\end{equation*}
to be contrasted with the equation \eqref{intro_lin_4_chih}.~Similarly, all the ingoing transport equations for the linearised induced metric and frame coefficients couple with a derivative of a linearised induced metric or frame coefficient on the right hand side.~These equations also exhibit a stronger coupling with the rest of the system than their outgoing counterpart.~In this regard, one may compare the equation \eqref{intro_lin_4_trchi} with the corresponding linearised Raychaudhuri equation in the ingoing direction 
\begin{align}
\nablasl_3 (\overset{\text{\scalebox{.6}{$(1)$}}}{\text{tr}\underline{\chi}})+ (\text{tr}\underline{\chi})(\overset{\text{\scalebox{.6}{$(1)$}}}{\text{tr}\underline{\chi}}) =& \,  2\, \slashed{\text{div}} \,\overset{\text{\scalebox{.6}{$(1)$}}}{\underline{\xi}}   +2\,(\underline{\eta}-\eta,\overset{\text{\scalebox{.6}{$(1)$}}}{\underline{\xi}})+(\slashed{\varepsilon}\cdot\underline{\chi})(\overset{\text{\scalebox{.6}{$(1)$}}}{\slashed{\varepsilon}\cdot\underline{\chi}})+(\text{tr}\underline{\chi})\,\overset{\text{\scalebox{.6}{$(1)$}}}{\hat{\underline{\omega}}} \label{intro_lin_3_trchib} \\ &- (\nablasl(\text{tr}\underline{\chi}),\overset{\text{\scalebox{.6}{$(1)$}}}{\underline{\mathfrak{f}}})-\nablasl_4(\text{tr}\underline{\chi})\,\overset{\text{\scalebox{.6}{$(1)$}}}{\underline{\mathfrak{f}}}_3  -\nablasl_3(\text{tr}\underline{\chi})\,\overset{\text{\scalebox{.6}{$(1)$}}}{\underline{\mathfrak{f}}}_4 \, ,  \nonumber
\end{align}
where some extra linearised frame and connection coefficients now appear on the right hand side.

The new structure of the outgoing transport equations in the linearised system allows, crucially, to order these equations into a \textbf{\emph{new} hierarchy} including \emph{all} linearised quantities (see Appendix \ref{sec_hierarchy_system}).~The new hierarchy has the potential to unlock the stability analysis of the full system, which could be carried out in line with the prescription of Steps 3 and 4 of Section \ref{sec_intro_linear_stability}.~See Section \ref{sec_intro_elements_analysis}.
\end{enumerate}

\subsection{Technical features and challenges of our framework}  \label{sec_intro_technical_challenges}

In this section, we elaborate on the choices encountered in the implementation of our framework of Section \ref{sec_intro_formulation_problem} and briefly discuss the technical difficulties that they pose.~As we shall see, the main technical difficulties arise from our choice of frames.

\subsubsection{A geometric framework}  \label{sec_intro_geometric_framework}

We start with a brief comment on the geometric nature of our framework.

\medskip

As outlined in Section \ref{sec_intro_formulation_problem}, our formulation of nonlinear gravitational perturbations involves the choice of a new \emph{geometric} gauge for the vacuum Einstein equations, called \emph{outgoing principal gauge}.~In an outgoing principal gauge, the fixed curves of constant $(s,\theta^A)$ on $\mathcal{M}$ are (i) integral curves of the vector field $e_4^{\text{as}}$ and (ii) \emph{null geodesics} of the metrics $\bsy{g}(\epsilon)$ \emph{for all $\epsilon$}.~Indeed, the gauge identities \eqref{intro_outline_gauge_ids} imply the identity
\begin{equation} \label{intro_christoffel_metrics}
\bsy{\Gamma}^{\mu}_{\underline{s}\,\underline{s}}(\epsilon) = -\mathfrak{k}_{a,M}\, \delta_{\underline{s}}^{\mu}
\end{equation}
for the Christoffel symbols of $\bsy{g}(\epsilon)$ for all $\epsilon$, with $x^{\mu}=(s,\underline{s},\theta^A)$.~By inspecting the geodesic equation, it is an easy check that the identity \eqref{intro_christoffel_metrics} suffices to establish that curves of constant $(s,\theta^A)$ are geodesics of the metrics $\bsy{g}(\epsilon)$ for all $\epsilon$, which are moreover null in view of the metric form \eqref{intro_outline_family_metrics} (in particular, one exploits $\bsy{g}_{\underline{s}\,\underline{s}}(\epsilon)=0$).

\medskip

As part of the fully rigorous implementation of our framework in the bulk of the paper, we will prove that assuming the metric form \eqref{intro_outline_family_metrics} and the identities \eqref{intro_outline_gauge_ids} for all $\epsilon$ can indeed be legitimately viewed as a choice of gauge for the metrics $\bsy{g}(\epsilon)$.~See already Section \ref{sec_overview_gauge_construction}.

\medskip

Other instances of geometric gauges which have been employed to formulate the vacuum Einstein equations are double-null gauges \cite{Christ_Form_BH, DHRT}, constant mean curvature (and maximal) gauges \cite{StabMink}, Bondi gauges \cite{Klainerman_Schw} and radiation gauges \cite{Andersson_Backdahl_Blue_Ma_nonlinear_radiation_gauge}.~Geometric gauges may be contrasted to \emph{wave} gauges \cite{Lindblad_Rodnianski_Stab_Mink_wave_gauge, Lind_Taylor}, for which the coordinates are solutions to a system of wave equations.

\medskip

In our framework, the vacuum Einstein equations are formulated as a nonlinear system of equations for \emph{geometric quantities}, corresponding to the connection coefficients and curvature components of the metrics $\bsy{g}(\epsilon)$ relative to the null frames $\bsy{\mathcal{N}}(\epsilon)$.~These quantities have a manifest geometric meaning and some of the equations in the system can be identified as classical equations from differential geometry (e.g.~the Codazzi equation \eqref{intro_intro_codazzi}).~In addition to the advantages that we shall describe in the present section, the use of frame quantities directly relates with the so-called \emph{tetrad formalisms}, such as the aforementioned Newman-Penrose formalism, which have found extensive application in the treatment of the black hole stability problem in the physics literature.

\medskip

Proving that our frames $\bsy{\mathcal{N}}(\epsilon)$ are well-defined and satisfy the properties listed in Section \ref{sec_intro_nonlinear_perturbations} requires some care.~In particular, an abstract local existence result for the system of ODEs defining the frames will account for their existence, whereas our choice of gauge for the metrics $\bsy{g}(\epsilon)$ will suffice for the frames $\bsy{\mathcal{N}}(\epsilon)$ to satisfy the desired properties.~See already Section \ref{sec_overview_frames_construction}.

\medskip

The geometric character of the nonlinear equations transfers to the linearised system.~Indeed, within our framework, the linearisation procedure is performed \emph{geometrically} and yields a linearised system of equations \emph{in geometric form} for the linearised connection coefficients and curvature components.

\subsubsection{The divorce of frames and spacetime foliation}   \label{sec_intro_divorce_frames_foliation}

The exact decoupling of the gauge invariant quantities \eqref{intro_alpha_alphab} into spin $\pm 2$ Teukolsky equations is a significant advantage guaranteed by our system of linearised gravity, in that it allows to address the analysis of these quantities independently from the other quantities in the system (see Section \ref{sec_intro_elements_analysis}).~However, as we shall now explain, it comes at the price of dealing with new technical difficulties in the formulation of nonlinear perturbations and with new intricacies in the (linearised) system of vacuum Einstein equations. 

\medskip

The gauge invariance and exact decoupling of two linearised curvature quantities are fundamentally tied to some delicate algebraic properties of the Kerr exterior manifold.~In loose terms, these algebraic properties correspond to the existence of exactly two null directions, known as outgoing and ingoing \emph{principal null directions}, which are shear-free geodesic directions ``diagonalising" the Riemann curvature tensor.~For such algebraic properties to become manifest in our problem, a specific choice of frame $\bsy{\mathcal{N}}(0)$ in the formulation of nonlinear perturbations is \emph{necessary}.~Such a choice corresponds to the \emph{algebraically special frame} of Kerr 
\begin{equation} \label{intro_zero_frame}
\bsy{\mathcal{N}}(0) = \mathcal{N}_{\text{as}} \, ,
\end{equation}
as outlined for our frames $\bsy{\mathcal{N}}(\epsilon)$ in Section \ref{sec_intro_formulation_problem}, and is a choice of a null frame whose null frame vector fields are aligned with the two principal null directions of Kerr.~Relative to the frame $\mathcal{N}_{\text{as}}$, the connection coefficients
\begin{align}
&{\widehat{\chi}}_{a,M} \, , & &{\widehat{\underline{\chi}}}_{a,M} \, , \label{intro_chih_chibh_kerr}\\ 
&{\xi}_{a,M} \, , & &{\underline{\xi}}_{a,M} \label{intro_xi_xib_kerr}
\end{align} 
and curvature components
\begin{align} \label{intro_curv_comps_kerr}
&{\alpha}_{a,M} \, , & &{\underline{\alpha}}_{a,M} \, , & &{\beta}_{a,M} \, , & &{\underline{\beta}}_{a,M}
\end{align}
of the Kerr metric identically \emph{vanish}.~The vanishing of these quantities, encoding the shear-free (i.e.~\eqref{intro_chih_chibh_kerr}), geodesic (i.e.~\eqref{intro_xi_xib_kerr}) and ``curvature-diagonalising" (i.e.~\eqref{intro_curv_comps_kerr}) nature of the frame, is what ultimately enforces the gauge invariance and exact decoupling of two linearised curvature quantities.

\medskip

Crucially, \ul{for $|a|>0$}, \textbf{the algebraically special frame of Kerr is a \emph{non-integrable} null frame},\footnote{In stark contrast, the algebraically special frame of the Schwarzschild exterior manifold ($|a|=0$) is \emph{integrable}.~See Section \ref{sec_intro_framework_Schwarzschild}.} in fact forcing our frames $\bsy{\mathcal{N}}(\epsilon)$ to be non-integrable frames \emph{for all $\epsilon$}.\footnote{We think of non-integrability of a frame as an open condition:~Given a one-parameter family of frames $\bsy{\mathcal{N}}(\epsilon)$, if the frame $\bsy{\mathcal{N}}(0)$ is non-integrable, then the frame $\bsy{\mathcal{N}}(\epsilon)$ is non-integrable for $\epsilon$ sufficiently small.}~See Appendix \ref{sec_intro_appendix} for an introduction to the notion of non-integrable frames.

\medskip

Being non-integrable, the frames $\bsy{\mathcal{N}}(\epsilon)$ cannot be possibly \emph{adapted} to \emph{any} foliation of the ambient manifold $\mathcal{M}$, meaning that the induced orthogonal (relative to $\bsy{g}(\epsilon)$) distributions $$\bsy{\mathfrak{D}}_{\bsy{\mathcal{N}}(\epsilon)}:=\left\langle \et(\epsilon),\ef(\epsilon) \right\rangle^{\perp} $$ cannot be possibly identified (even locally) with the tangent bundles to the two-dimensional leaves of any foliation of $\mathcal{M}$.\footnote{This fact follows from Frobenius theorem.~See Appendix \ref{sec_intro_appendix}.}~In particular, \textbf{the frames $\bsy{\mathcal{N}}(\epsilon)$ are \emph{not} adapted to the fixed foliation of $\mathcal{M}$ induced by the fixed $(s,\ub{$s$},\theta^A)$-differentiable structure}.~The frames $\bsy{\mathcal{N}}(\epsilon)$ are nonetheless tailored to the fixed null geodesics of constant $(s,\theta^A)$ (or, equivalently, to the fixed hypersurfaces of constant $s$) on $\mathcal{M}$, which are generated by the frame vector field $\ef(\epsilon)$ for all $\epsilon$.

\medskip

Non-integrable frames are, in general, more complicated to handle than frames adapted to a spacetime foliation.~Some instances of complications, which do not arise for adapted frames, are the following:
\begin{itemize}
\item $\bsy{\mathfrak{D}}_{\bsy{\mathcal{N}}(\epsilon)}$ \textbf{tensors}

The connection coefficients and curvature components of the metrics $\bsy{g}(\epsilon)$ relative to the frames $\bsy{\mathcal{N}}(\epsilon)$ are defined as \emph{$\bsy{\mathfrak{D}}_{\bsy{\mathcal{N}}(\epsilon)}$ tensors}, meaning as smooth sections of the abstract tensor bundle $$\bsy{\mathfrak{D}}_{\bsy{\mathcal{N}}(\epsilon)}\otimes \ldots \otimes \bsy{\mathfrak{D}}_{\bsy{\mathcal{N}}(\epsilon)} \otimes (\bsy{\mathfrak{D}}_{\bsy{\mathcal{N}}(\epsilon)})^{\star} \otimes \ldots \otimes (\bsy{\mathfrak{D}}_{\bsy{\mathcal{N}}(\epsilon)})^{\star} $$ on $\mathcal{M}$.~Being the frames $\bsy{\mathcal{N}}(\epsilon)$ non-integrable, these quantities cannot be interpreted as tensors on a sub-manifold of $\mathcal{M}$, as it is, on the other hand, possible for adapted frames.~One related limitation is that $\bsy{\mathfrak{D}}_{\bsy{\mathcal{N}}(\epsilon)}$ tensors cannot be written in coordinate form, in that non-integrable distributions do not admit coordinate (co-)bases.
\item \textbf{Second fundamental forms}

A crucial role in the problem is played by the outgoing and ingoing second fundamental forms
\begin{align*}
{\bsy{\chi}}_{\bsy{AB}}(\epsilon)&=\bsy{g}(\bsy{\nabla}_{\bsy{A}}\ef(\epsilon),\eb(\epsilon)) \, , & {\underline{\bsy{\chi}}}_{\bsy{AB}}(\epsilon)&=\bsy{g}(\bsy{\nabla}_{\bsy{A}}\et(\epsilon),\eb(\epsilon)) \, . 
\end{align*}
The non-integrability of the frames $\bsy{\mathcal{N}}(\epsilon)$ implies that the second fundamental forms possess non-vanishing outgoing and ingoing \emph{antitraces}
\begin{align*}
&(\bsy{\slashed{\varepsilon}\cdot\chi})(\epsilon) \, , & &(\bsy{\slashed{\varepsilon}\cdot\underline{\chi}})(\epsilon)
\end{align*}
(see already Section \ref{sec_overview_decomp_chi}), which complicate the treatment of these geometric quantities.~In the case of adapted frames, the second fundamental forms are \emph{symmetric}, i.e.~their antitraces identically vanish.
\item \textbf{Induced connection}

The definition of induced geometric quantities over $\bsy{\mathfrak{D}}_{\bsy{\mathcal{N}}(\epsilon)}$ involves new technicalities.~The spacetime metric $\bsy{g}(\epsilon)$ and Levi-Civita connection $\bsy{\nabla}(\epsilon)$ naturally induce a metric $\bsy{\slashed{g}}(\epsilon)$ and a linear connection $$\bsy{\nablasl}(\epsilon)$$ over $\bsy{\mathfrak{D}}_{\bsy{\mathcal{N}}(\epsilon)}$.~For adapted frames, the induced connection $\bsy{\nablasl}(\epsilon)$ is \emph{Levi-Civita} relative to the induced metric $\bsy{\slashed{g}}(\epsilon)$.~In our framework, being $\bsy{\mathfrak{D}}_{\bsy{\mathcal{N}}(\epsilon)}$ a non-integrable distribution, the induced connection $\bsy{\nablasl}(\epsilon)$ has \emph{torsion} (see already Section \ref{sec_overview_connection_torsion}), as it is the case for \emph{any} linear connection over a non-integrable distribution.~Thus, the induced connection $\bsy{\nablasl}(\epsilon)$ is \emph{not} Levi-Civita relative to the induced metric $\bsy{\slashed{g}}(\epsilon)$.~Differential operators acting on $\bsy{\mathfrak{D}}_{\bsy{\mathcal{N}}(\epsilon)}$ tensors are defined relative to $\bsy{\nablasl}(\epsilon)$ and require some extra care to take the torsion of $\bsy{\nablasl}(\epsilon)$ into account.
\item \textbf{Vacuum Einstein equations}

To formulate the vacuum Einstein equations for the connection coefficients and curvature components of $\bsy{g}(\epsilon)$ relative to the frames $\bsy{\mathcal{N}}(\epsilon)$, one has to derive a new system of vacuum Einstein equations.~Our nonlinear system of equations differs from the one in \cite{StabMink}, in that the system of \cite{StabMink} is formulated for connection coefficients and curvature components relative to an adapted frame.\footnote{For $\epsilon =0$, one can easily check that the connection coefficients and curvature components of the Kerr exterior manifold relative to the algebraically special frame do \emph{not} solve the system of vacuum Einstein equations of \cite{StabMink}.}~The appearance of the antitraces of the second fundamental forms in both the null structure and Bianchi equations is the main novelty of our system.~Some examples of equations where the antitraces appear are the outgoing Raychaudhuri equation\footnote{We remove the $\epsilon$ in the equations \eqref{intro_intro_4_trchi}-\eqref{intro_intro_codazzi} to keep the notation lighter.}
\begin{equation}
\bsy{\nablasl_4} (\bsy{\textbf{tr}\chi})+ \frac{1}{2}\,(\bsy{\textbf{tr}\chi})^2-\bsy{\omegah}\,(\bsy{ \textbf{tr}\chi})= -(\widehat{\bsy{\chi}},\widehat{\bsy{\chi}})  +\frac{1}{2}\,(\bsy{\slashed{\varepsilon}\cdot\chi})^2  \, ,   \label{intro_intro_4_trchi}
\end{equation}
the outgoing transport equation for the antitrace of the outgoing second fundamental form
\begin{equation}
\bsy{\nablasl_4} (\bsy{\slashed{\varepsilon}\cdot\chi})+( \bsy{\textbf{tr}\chi})(\bsy{\slashed{\varepsilon}\cdot\chi})-\bsy{\omegah}\, (\bsy{\slashed{\varepsilon}\cdot\chi}) =  0 \, , \label{intro_intro_4_antitrace_chi}
\end{equation}
the outgoing transport equation for the ingoing shear
\begin{equation} \label{intro_intro_4_chibh}
\bsy{\nablasl_4 \underline{\widehat{\chi}}}+\frac{1}{2}\,(\bsy{\textbf{tr}\chi})\,\bsy{\underline{\widehat{\chi}}}+\bsy{\omegah\underline{\widehat{\chi}}}= -2\,\bsy{\slashed{\mathcal{D}}{}_2^{\star}\, \underline{\eta}}+\bsy{\underline{\eta}\,\widehat{\otimes}\,\underline{\eta}} -\frac{1}{2}\,(\bsy{\textbf{tr}\underline{\chi}})\,\bsy{\widehat{\chi}} +\frac{1}{2}\,(\bsy{\slashed{\varepsilon}\cdot\underline{\chi}})\,{}^{\bsy{\star}}\bsy{\widehat{\chi}}  -\frac{1}{2}\,(\bsy{\slashed{\varepsilon}\cdot\chi})\,{}^{\bsy{\star}}\bsy{\underline{\widehat{\chi}}} 
\end{equation}
and the Codazzi equation
\begin{equation} \label{intro_intro_codazzi}
\bsy{\slashed{\textbf{div}}\, \widehat{\chi}}= \frac{1}{2}\,\bsy{\nablasl}( \bsy{\textbf{tr}\chi}) -\frac{1}{2}\,\bsy{{}^{\star}\nablasl}(\bsy{\slashed{\varepsilon}\cdot\chi})-\bsy{\widehat{\chi}^{\sharp}\cdot\zeta}+\frac{1}{2}\,(\bsy{\textbf{tr}\chi})\,\bsy{\zeta} -\frac{1}{2}\,(\bsy{\slashed{\varepsilon}\cdot\chi})\,\bsy{{}^{\star}\zeta}  -(\bsy{\slashed{\varepsilon}\cdot\chi})\,\bsy{{}^{\star}\eta} -\bsy{\beta} \, .
\end{equation}
\end{itemize}

Dealing with the complications coming from the use of the non-integrable frames $\bsy{\mathcal{N}}(\epsilon)$ is one of the main themes and technical challenges of our framework.~Several of these complications transfer to the system of linearised gravity, one example being the appearance of the linearised antitraces of the second fundamental forms (see already the equations \eqref{intro_lin_4_trchi}-\eqref{intro_lin_3_trchib}).

\medskip

We point out that, to enforce the gauge invariance and exact decoupling of two linearised curvature quantities, it is not only necessary, but also \emph{sufficient} that the frames adopted in the formulation of nonlinear perturbations satisfy the property \eqref{intro_zero_frame}.~The other properties satisfied by our frames $\bsy{\mathcal{N}}(\epsilon)$ (see Section \ref{sec_intro_nonlinear_perturbations}) are necessary to produce the new hierarchical structure in the linearised system and mitigate the complications introduced in the equations by non-integrability.~Such properties can only be achieved by exploiting wisely the freedom\footnote{To enforce the gauge invariance and exact decoupling of two linearised curvature quantities, one has the freedom to choose the frames among the infinite one-parameter families of null frames such that $\bsy{\mathcal{N}}(0) = \mathcal{N}_{\text{as}}$.~In particular, being the frames inevitably non-integrable, the freedom in the choice of frames is not limited by any of the constraints coming from possibly adapting the frames to the fixed spacetime foliation.} in the choice of frames, as done by our implicit definition of the frames through the system of ODEs \eqref{overview_ODE_1}-\eqref{overview_ODE_3}.

\medskip

We mention an additional technical difficulty arising from our choice of frames which is not directly ascribable to non-integrability: 
\begin{itemize}
\item \textbf{Frame coefficients}

The frame vector field $\et(\epsilon)$ is only defined implicitly, meaning that the frame quantities $\bsy{\underline{j}}(\epsilon)$, $\bsy{\underline{k}}(\epsilon)$ and $\bsy{\underline{\Lambda}}(\epsilon)$ from \eqref{intro_formulation_problem_e4_e3} are only defined implicitly and cannot be explicitly related to the metric quantities from \eqref{intro_outline_family_metrics}.~It is thus convenient to introduce the \emph{frame coefficients} of $\et(\epsilon)$ 
\begin{align}  \label{intro_frame_coefficients_e3}
\underline{\bsy{\mathfrak{f}}}_{4}(\epsilon) &:= -\frac{1}{2}\, g_{a,M}(\et(\epsilon),e_4^{\text{as}}) \, , & \underline{\bsy{\mathfrak{f}}}_{3}(\epsilon) &:= -\frac{1}{2}\, g_{a,M}(\et(\epsilon),e_3^{\text{as}}) \, , & \underline{\bsy{\mathfrak{f}}}_{A}(\epsilon) &:=   g_{a,M}(\et(\epsilon),e_A^{\text{as}}) \, .
\end{align}
The frame coefficients \eqref{intro_frame_coefficients_e3}, together with additional frame coefficients defined in the bulk of the paper, appear extensively in the problem and replace, in most cases, the use of the metric quantities from \eqref{intro_outline_family_metrics} and frame quantities from \eqref{intro_formulation_problem_e4_e3}.~In fact, the set of all frame coefficients, along with the induced metric $\bsy{\slashed{g}}(\epsilon)$, constitute the full set of perturbed quantities at the level of the metric appearing in the equations.

The linearised frame coefficients are unknowns of the system of linearised gravity (see already the equations \eqref{intro_lin_4_chibh} and \eqref{intro_lin_3_trchib}).~The set of equations for the linearised frame coefficients that we alluded to in Section \ref{sec_intro_from_nonlinear_to_linear} is obtained by suitably linearising the commutators of the $\bsy{\mathcal{N}}(\epsilon)$-frame vector fields.~For the convenient structure of the outgoing transport equations for the linearised frame coefficients discussed in Section \ref{sec_intro_from_nonlinear_to_linear}, it is crucial that the frame vector field $\ef(\epsilon)$ remains fixed for all $\epsilon$.
\end{itemize}

We conclude with a remark about the potential use of adapted frames in the problem.~By Frobenius theorem (see Appendix \ref{sec_intro_appendix}), \emph{all} one-parameter families of frames adapted to a fixed foliation of $\mathcal{M}$ are necessarily families of \emph{integrable} frames, and thus fail at including the algebraically special frame of Kerr as a member of the family.~For this reason, the use of adapted frames is \emph{in}compatible with the gauge invariance and exact decoupling of two linearised curvature quantities, for which the inclusion of the algebraically special frame is necessary.\footnote{The use of adapted frames \emph{is} compatible with the exact decoupling of two gauge invariant linearised quantities in the special case $|a|=0$, in that the algebraically special frame of Schwarzschild is integrable.~In fact, as we shall see in Section \ref{sec_intro_framework_Schwarzschild}, frames adapted to suitable foliations of the Schwarzschild exterior manifold do guarantee the exact decoupling of two gauge invariant linearised quantities.}

\subsubsection{The $|a|=0$ case} \label{sec_intro_framework_Schwarzschild}

We briefly describe our framework in the Schwarzschild $|a|=0$ case.

\medskip

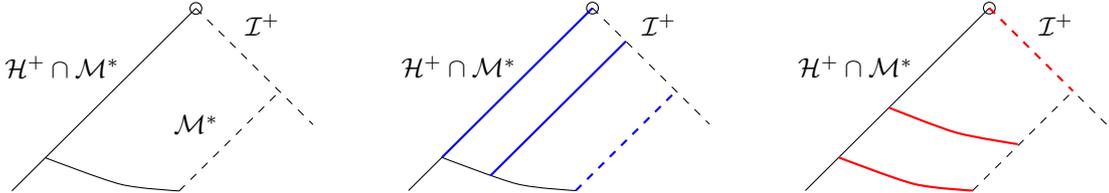
\begin{figure}[H]

\centering

\begin{subfigure}{0.3\textwidth}

\centering

\begin{tikzpicture}[scale=2.2]

\draw (0,2)--(-1.1,0.9);

\node at (-0.8,1.65) {$\mathcal{H}^+\cap \mathcal{M}^*$};
\node at (0.4,1.9) {$\mathcal{I}^+$};
\node at (0,1.3) {$\mathcal{M}^*$};


\draw (0,2)  circle[radius=1pt];




\draw plot [smooth] coordinates { (-0.9,1.1) (-0.45,0.94) (-0.1,0.9) };

\draw[dashed] (-0.1,0.9)--(0.5,1.5);


\draw[dashed] (0,2)--(0.7,1.3);


\end{tikzpicture}

\end{subfigure}
\begin{subfigure}{0.3\textwidth}

\centering

\begin{tikzpicture}[scale=2.2]

\draw[color=blue, thick] (0,2)--(-0.9,1.1);
\draw (-1.1,0.9)--(-0.9,1.1);

\node at (-0.8,1.65) {$\mathcal{H}^+\cap \mathcal{M}^*$};
\node at (0.4,1.9) {$\mathcal{I}^+$};


\draw (0,2)  circle[radius=1pt];




\draw plot [smooth] coordinates { (-0.9,1.1) (-0.45,0.94) (-0.1,0.9) };

\draw[dashed, color=blue, thick] (-0.1,0.9)--(0.5,1.5);

\draw[color=blue, thick] (-0.61,0.99)--(0.2,1.8) ;


\draw[dashed] (0,2)--(0.7,1.3);


\end{tikzpicture}

\end{subfigure}
\begin{subfigure}{0.3\textwidth}

\centering

\begin{tikzpicture}[scale=2.2]

\draw (0,2)--(-1.1,0.9);

\node at (-0.8,1.65) {$\mathcal{H}^+\cap \mathcal{M}^*$};
\node at (0.4,1.9) {$\mathcal{I}^+$};


\draw (0,2)  circle[radius=1pt];




\draw[color=red, thick] plot [smooth] coordinates { (-0.9,1.1) (-0.45,0.94) (-0.1,0.9) };

\draw[color=red, thick] plot [smooth] coordinates { (-0.6,1.4) (-0.2,1.25) (0.17,1.18) };

\draw[dashed] (-0.1,0.9)--(0.5,1.5);


\draw[dashed, color=red, thick] (0,2)--(0.5,1.5);
\draw[dashed] (0.7,1.3)--(0.5,1.5);


\end{tikzpicture}

\end{subfigure}

\caption{The figure depicts the Penrose diagram of the manifold $(\mathcal{M}^*,g_{0,M})$ and the foliation of $(\mathcal{M}^*,g_{0,M})$ induced by the coordinates $s$ (in blue) and $\underline{s}$ (in red).}

\label{fig:intro_foliation_a_0}

\end{figure}

For what concerns the star-normalised, outgoing principal differentiable structure $(s,\underline{s},\theta^A)$ and the algebraically special frame $\mathcal{N}_{\text{as}}$ of the Schwarzschild exterior manifold $(\mathcal{M},g_{0,M})$, two special properties arise.~The first is that the coordinate $s$ is null (see Figure \ref{fig:intro_foliation_a_0}).~The second is that the algebraically special frame $\mathcal{N}_{\text{as}}$ is \emph{integrable} and \emph{adapted} to the foliation induced by the star-normalised, outgoing principal differentiable structure. 

\medskip

\medskip

In the formulation of nonlinear perturbations, \textbf{our framework retains the use of non-integrable frames}.~In fact, the frames $\bsy{\mathcal{N}}(\epsilon)$ are non-integrable \ul{for all $|\epsilon|>0$}.\footnote{The retention of non-integrability follows from choosing frames such that all the perturbed connection coefficients \eqref{intro_fixed_connection_coeffs} remain fixed for all $\epsilon$.~In particular, to achieve the property that the perturbed connection coefficient $\underline{\bsy{\eta}}(\epsilon)$ remains fixed, the frames $\bsy{\mathcal{N}}(\epsilon)$ are, in general, to be non-integrable for $|\epsilon|>0$.}

\medskip

By setting $|a|=0$ in our linearised system of equations on Kerr, one obtains a \textbf{\emph{new} system of linearised gravity on Schwarzschild exterior spacetimes}.~Compared to the Kerr case, many more background quantities in the system identically vanish, including the background antitraces of the second fundamental forms (since $\mathcal{N}_{\text{as}}$ is an integrable frame).~However, as in the Kerr case, the use of non-integrable frames in the formulation of nonlinear perturbations introduces new intricacies in the gauge dependent part of the linearised system, such as the presence of the linearised antitraces of the second fundamental forms.~One instance is the linearised Codazzi equation
\begin{equation*}
\slashed{\text{div}}\,\overset{\text{\scalebox{.6}{$(1)$}}}{\widehat{\chi}}+\ldots = \frac{1}{2}\,\nablasl(\overset{\text{\scalebox{.6}{$(1)$}}}{\text{tr}\chi})-\frac{1}{2}\,{}^{\star}\nablasl(\overset{\text{\scalebox{.6}{$(1)$}}}{\slashed{\varepsilon}\cdot\chi}) -\overset{\text{\scalebox{.6}{$(1)$}}}{\beta} +\ldots \, ,
\end{equation*}
in which the linearised antitrace of the outgoing second fundamental form appears on the right hand side.

\medskip

The system admits the exact decoupling of the gauge invariant quantities \eqref{intro_alpha_alphab} into spin $\pm 2$ Teukolsky equations.~The outgoing transport equations for the linearised induced metric, frame coefficients and connection coefficients present the new structure and weak coupling that we discussed in Section \ref{sec_intro_from_nonlinear_to_linear} for the Kerr case.~In fact, the essence of the new structure of these equations can be more easily appreciated in the Schwarzschild case, where the additional coupling terms arising from linearising around a (only) stationary spacetime are absent.~The equations \eqref{intro_lin_4_chibh}-\eqref{intro_lin_4_atrchi} now assume an even simpler form:~The transport equation
\begin{equation} \label{intro_lin_4_chibh_a_0}
\nablasl_4 \overset{\text{\scalebox{.6}{$(1)$}}}{\widehat{\underline{\chi}}}+\chi\times\overset{\text{\scalebox{.6}{$(1)$}}}{\widehat{\underline{\chi}}}+\omegah\,\overset{\text{\scalebox{.6}{$(1)$}}}{\widehat{\underline{\chi}}} =  -\frac{1}{2}\,(\text{tr}\underline{\chi})\,\overset{\text{\scalebox{.6}{$(1)$}}}{\widehat{\chi}}   
\end{equation}
for the linearised ingoing shear \emph{only} couples with the linearised outgoing shear and the linearised Raychaudhuri equation 
\begin{equation*} 
\nablasl_4 (\overset{\text{\scalebox{.6}{$(1)$}}}{\text{tr}\chi})+ (\text{tr}\chi)(\overset{\text{\scalebox{.6}{$(1)$}}}{\text{tr}\chi})-\omegah \, (\overset{\text{\scalebox{.6}{$(1)$}}}{\text{tr}\chi}) =  0
\end{equation*}
and transport equation for the linearised antitrace of the outgoing second fundamental form
\begin{equation*} 
\nablasl_4 (\overset{\text{\scalebox{.6}{$(1)$}}}{\slashed{\varepsilon}\cdot\chi})+(\text{tr}\chi)(\overset{\text{\scalebox{.6}{$(1)$}}}{\slashed{\varepsilon}\cdot\chi})-\omegah \, (\overset{\text{\scalebox{.6}{$(1)$}}}{\slashed{\varepsilon}\cdot\chi})= 0  
\end{equation*}
are \emph{completely} decoupled.~As in the Kerr case, the outgoing transport equations can be ordered into a \emph{new hierarchy} including \emph{all} linearised quantities in the system.

\medskip

Our new system of linearised gravity on Schwarzschild can be employed in a \emph{new} proof of full linear stability which revisits, and significantly simplifies, some of the key aspects of the analysis from previous works.~See the companion work \cite{benomio_schwarzschild_stability} by the author.

\medskip

To conclude, it is important to point out that, in the Schwarzschild $|a|=0$ case, the use of non-integrable frames in the formulation of nonlinear perturbations is \emph{not} necessary for the gauge invariance and exact decoupling of two linearised curvature quantities.~Indeed, in an outgoing principal gauge, a choice of frames \emph{adapted} to the fixed spacetime foliation \emph{for all $\epsilon$} would still allow the decoupling of two gauge invariant linearised quantities and avoid the complications of non-integrability altogether.~In fact, the use of non-integrable frames is a new element of our framework which is absent in previous works on the linear stability of Schwarzschild.~The work \cite{DHR}, for instance, considers nonlinear perturbations in a double-null gauge and adopts a family of frames adapted to a fixed double-null foliation of the Schwarzschild exterior manifold for all $\epsilon$.

\subsection{Related works}  \label{sec_intro_other_works}

We give a brief account of the most recent developments in the study of both linear and nonlinear stability of black hole solutions to gravitational perturbations.

\medskip

As discussed in Section \ref{sec_intro_DHR}, the breakthrough work \cite{DHR} is the first to establish the linear stability of the Schwarzschild solution to gravitational perturbations.~Subsequent work \cite{Lin_Stab_Schw_wave_gauge} proves the linear stability of Schwarzschild in a generalised harmonic gauge (see also \cite{Wang_Schwarz, Hung_linear_schwarzschild_odd_part, Hung_linear_schwarzschild_even_part}).~Work \cite{Giorgi_full_RN} proves the linear stability of the Reissner--Nordstr\"{o}m solution in the full sub-extremal range. 

\medskip

For what concerns the linear stability of the Kerr solution, two works carry out the analysis of the full system of linearised gravity:

\begin{itemize}
\item The work \cite{Andersson_Blue_Lin_Kerr} establishes a uniform decay statement for solutions to linearised gravity in the full sub-extremal range $|a|<M$.~The result assumes integrated decay estimates for the gauge invariant linearised quantities outside the slowly rotating regime $|a|\ll M$ and requires a sufficiently high loss of derivatives.~An improved control of the regularity loss seems necessary to directly apply the result to the nonlinear problem.

The system of linearised gravity is formulated for spin-weighted (tetrad) quantities in the Geroch--Held--Penrose (GHP) formalism and assumes that the linearised perturbations satisfy a set of ``linear gauge conditions", known as (linear) \emph{outgoing radiation gauge} conditions.~In the follow-up paper \cite{Andersson_Backdahl_Blue_Ma_nonlinear_radiation_gauge}, such gauge conditions are shown to arise from a choice of (nonlinear) outgoing radiation gauge for sufficiently small nonlinear perturbations of Kerr.~The outgoing radiation gauge is tailored to the \emph{ingoing} principal null geodesics of Kerr.~In \cite{Andersson_Backdahl_Blue_Ma_nonlinear_radiation_gauge}, the choice of gauge is combined with a choice of null tetrads similarly tailored to the \emph{ingoing} principal null geodesics of Kerr.~The vacuum Einstein equations for the tetrad quantities in the outgoing radiation gauge are proven to be well-posed.~By linearising the nonlinear system of \cite{Andersson_Backdahl_Blue_Ma_nonlinear_radiation_gauge}, one obtains a system of linearised gravity which, although different from the one presented in \cite{Andersson_Blue_Lin_Kerr}, can be controlled by relating its linearised quantities to the ones in \cite{Andersson_Blue_Lin_Kerr}.

For what concerns the analysis, a key role is played by a hierarchical structure present in the \emph{ingoing} transport equations in the linearised system.

\item The work \cite{Hintz_Vasy_Kerr} establishes a uniform decay statement for solutions to linearised gravity in the slowly rotating regime $|a|\ll M$.~An improved control of the regularity loss seems necessary to directly apply the result to the nonlinear problem.

The system of linearised gravity is formulated for metric perturbations in a harmonic gauge.~The analysis employs tools from microlocal analysis and spectral theory which allow to take a perturbative (relative to the $|a|=0$ case) approach to the proof.~In particular, the decoupling of gauge invariant linearised quantities into spin $\pm 2$ Teukolsky equations for $|a|>0$ is not exploited.

Recent work \cite{Andersson_Hafner_Whiting_mode_linearised_kerr} presents a mode stability analysis of the system of \cite{Hintz_Vasy_Kerr} in the full sub-extremal range $|a|<M$, providing one of the missing ingredients towards the extension of \cite{Hintz_Vasy_Kerr} to the full sub-extremal range $|a|<M$.~Work \cite{Andersson_Hafner_Whiting_mode_linearised_kerr} exploits the seminal mode stability result \cite{Whiting_mode_stability_kerr} for the Teukolsky equations.
\end{itemize}

The work \cite{DHRT} and the series of works \cite{Klain_GCM_kerr_1, Klain_GCM_kerr_2, Klain_form_kerr, Klainerman_Szeftel_Kerr_small_a_1, Shen_gcm_spheres_kerr, Giorgi_Klainerman_Szeftel_wave_estimates} resolve the nonlinear stability problem for Schwarzschild and slowly rotating Kerr black holes respectively.~See also prior work \cite{Klainerman_Schw} for a proof of nonlinear stability of Schwarzschild black holes in polarised axisymmetry and recent work \cite{Lindblad_Tohaneanu_weak_null_kerr} as a step towards proving the nonlinear stability of Kerr black holes in harmonic gauge.~We point out that the work \cite{Klain_form_kerr} presents a geometric framework and a system of vacuum Einstein equations which make use of various non-integrable frames.~The works \cite{Giorgi_Teukolsky_KN, Giorgi_Carter_tensor_KN} initiate the study of the linear stability problem for Kerr--Newman black holes.

\medskip

For positive cosmological constant, the nonlinear stability of slowly rotating Kerr-de Sitter black holes is proven by the work \cite{StabKerrdS}.~See also the recent works \cite{Fang_linear_stability_kerr_dS, Fang_nonlinear_Kerr_deSitter} proving the linear and nonlinear stability of the slowly rotating Kerr-de Sitter solution with a different approach.~Without smallness assumption on the rotational parameter, the work \cite{Casals_Teixeira_mode_stability_kerr_ds} proves a mode stability result for the Teukolsky equations on Kerr-de Sitter black holes in a partial range of spacetime parameters (and, in fact, provides a revisited proof of the original result \cite{Whiting_mode_stability_kerr} for Kerr black holes).

\medskip

For negative cosmological constant, the work \cite{Graf_Holzegel_mode_stability} proves mode stability for the Teukolsky equations on Kerr--anti-de Sitter black holes (satisfying the Hawking--Reall bound), initiating the study of linear stability for these spacetimes.~The nonlinear dynamics of Kerr--anti-de Sitter black holes remains an open problem and may lead, for certain boundary conditions, to instabilities \cite{Holzegel_KG_Slow_KerrAdS}.~See the related breakthrough works \cite{Moschidis_AdS, Moschidis_AdS_Vlasov} for pure AdS space as a solution to the Einstein equations with matter.

\subsection{Outlook}

We describe some open directions related to the present work.

\subsubsection{Linear stability of Kerr}  \label{sec_intro_elements_analysis}

Future work by the author will perform the stability analysis of the system of linearised gravity in the full sub-extremal range $|a|<M$.~As in the $|a|=0$ case of \cite{benomio_schwarzschild_stability}, the analysis will produce uniform boundedness and decay statements for solutions to the linearised system in line with the prescription of Steps 3 and 4 of Section \ref{sec_intro_linear_stability}.~In particular, the linear stability result will include top-order uniform boundedness and (degenerate) integrated decay statements for all linearised curvature components which \emph{do not lose derivatives}, indicating that the framework may be well suited to address nonlinear stability in the full sub-extremal range $|a|<M$.

\medskip

By virtue of the structure of the system of linearised gravity, one may implement a full stability analysis divided into two parts.~The first part of the analysis is already \emph{known} and exploits the exact decoupling of the \emph{gauge invariant} quantities \eqref{intro_alpha_alphab} into decoupled spin $\pm 2$ Teukolsky equations.~Indeed, one can directly transfer the result \cite{Shl_Teix_teuk_1} (and upcoming companion work by the authors) to establish uniform decay for these quantities independently from the other quantities in the system.~For the second part, the new hierarchical structure of the system suggests the possibility to implement a \emph{new scheme} to prove uniform decay for all the remaining \emph{gauge dependent} quantities in the full sub-extremal range $|a|<M$ and, for the first time, in line with the prescription of Steps 3 and 4 of Section \ref{sec_intro_linear_stability}.

\medskip

We expect the outgoing transport equations in the system to play a more prominent role in the gauge dependent analysis than in previous works.~In particular, a crucial role would be played by the \emph{red-shifted} transport equations for the linearised connection coefficients available in the scheme (see Appendix \ref{sec_hierarchy_system}), as for instance the equation \eqref{intro_lin_4_chibh} for the linearised ingoing shear.~The red-shifted transport equations can be employed to prove local integrated decay by direct, forward-in-$\underline{s}$ integration from the initial data.\footnote{One would also need to exploit that the right hand side of the equation suitably decays.}~One such instance, which can already be appreciated in the $|a|=0$ analysis of \cite{benomio_schwarzschild_stability}, would be the proof of local integrated decay for the linearised ingoing shear by direct integration of the equation \eqref{intro_lin_4_chibh}.

\medskip

The red-shift effect at the event horizon has been exploited in a large number of works on the stability analysis of sub-extremal black hole exteriors.~The fact that our new framework effectively isolates this important structure in the analysis is suggestive of potential simplifications in the nonlinear problem in our framework.

\subsubsection{Further directions in linearised gravity}

Beyond its primary application to the linear stability of Kerr, our system of linearised gravity can be employed in a number of related problems.~One interesting example could be the study of conservation laws for solutions to our system of linearised gravity on Kerr.~Work \cite{Holzegel_cons_laws} derives conservation laws and flux bounds for solutions to the system of linearised gravity on Schwarzschild of \cite{DHR}.~For $|a|=0$, our system may be used to present a different approach to deriving conservation laws and flux bounds.

\medskip

From a more general perspective, our framework may be adapted to derive systems of linearised gravity on other black hole exterior spacetimes.~Some natural examples would be \emph{Kerr--(A)dS} exterior spacetimes, as solutions to the vacuum Einstein equations with cosmological constant, and \emph{Kerr--Newman} exterior spacetimes, as solutions to the Einstein--Maxwell equations.~One may expect the systems to enjoy similar features to the ones characterising our system, including the exact decoupling of two gauge invariant linearised quantities and a new hierarchical structure in its gauge dependent part.~Linear stability, a scattering theory and conservation laws for linearised gravity on these spacetimes could then be addressed.~In the particular case of Reissner--Nordstr\"{o}m exterior spacetimes, a new system of linearised gravity may be employed to revisit (and possibly simplify) the proof of linear stability of \cite{Giorgi_full_RN} in the same fashion as described in Section \ref{sec_intro_new_proof_schwarzschild} for the Schwarzschild case.

\subsection{Acknowledgements}

The present work builds upon the framework developed by the author in his doctoral thesis \cite{benomio_thesis}.~The author is indebted to Gustav Holzegel for numerous discussions over the years and for fundamentally contributing to this work with insightful ideas.~The author is also grateful to Mihalis Dafermos, Igor Rodnianski and Martin Taylor for fruitful conversations.~The author's work has been supported by a Roth Scholarship at Imperial College London, by the Royal Society through the Royal Society Tata University Research Fellowship R1-191409 and by the Princeton Gravity Initiative at Princeton University.~The author also acknowledges the University of M\"{u}nster for hospitality during a visit as Young Research Fellow.

\section{Structure of the paper}

The bulk of the present paper is divided into seven sections, namely Sections \ref{sec_nonlinear_EVE}-\ref{sec_system_linearised_gravity}.~The content of each of the sections is summarised in Section \ref{sec_overview}, which serves as an overview.~In this section, we give a brief outline of the paper, which accounts for its overall structure and logic.

\medskip

The first two sections of the bulk of the paper are preliminary sections:
\begin{itemize}
\item In \textbf{Section \ref{sec_nonlinear_EVE}}, we develop the necessary geometric formalism to formulate the vacuum Einstein equations as a system of equations for the connection coefficients and curvature components of a spacetime $(\bsy{\mathcal{M}},\bsy{g})$ relative to a general null frame $\bsy{\mathcal{N}}$.~Crucially, the frame $\bsy{\mathcal{N}}$ is allowed to be a non-integrable frame.~The technical difficulties related to non-integrable frames are all addressed in this section.~The equations are derived in Section \ref{sec_formal_derivation_EVE}.

\medskip

See \textbf{Section \ref{sec_overview_EVE}} for an overview to this section of the paper.

\item In \textbf{Section \ref{sec_Kerr_exterior_manifold}}, we define the Kerr exterior manifold $(\mathcal{M},g_{a,M})$ and the (non-integrable) algebraically special frame $\mathcal{N}_{\text{as}}$ and outgoing principal differentiable structures of $(\mathcal{M},g_{a,M})$.~By making the identification $(\bsy{\mathcal{M}},\bsy{g}) = (\mathcal{M},g_{a,M})$ and $\bsy{\mathcal{N}} = \mathcal{N}_{\text{as}}$, we exploit the formalism of Section \ref{sec_nonlinear_EVE} to treat the connection coefficients and curvature components of $(\mathcal{M},g_{a,M})$ relative to $\mathcal{N}_{\text{as}}$.

\medskip

See \textbf{Section \ref{sec_overview_kerr_exterior_manifold}} for an overview to this section of the paper.
\end{itemize}

At this point, one is ready to start the formulation of the problem proper:

\begin{itemize}
\item In \textbf{Section \ref{sec_perturbations_kerr}}, we formulate nonlinear perturbations of the Kerr exterior manifold $(\mathcal{M},g_{a,M})$.~The formulation employs both the algebraically special frame and an outgoing principal differentiable structure of $(\mathcal{M},g_{a,M})$ from Section \ref{sec_Kerr_exterior_manifold}.~This section is the core of the paper and includes the main elements of our new geometric framework.

\medskip

In Sections \ref{sec_gauge_construction} and \ref{sec_frames_construction}, we prove technical statements concerning our choice of gauge and frames respectively.~These two sections are independent modules and can be skipped by the reader without affecting the overall logic of the paper.

\medskip

See \textbf{Section \ref{sec_overview_perturbations_kerr}} for an overview to this section of the paper.
\end{itemize}

Before proceeding with the formulation of the problem, a first technical digression is required:

\begin{itemize}
\item In \textbf{Section \ref{sec:geom_compare_tensors}}, we prescribe a geometric procedure to compare the perturbed connection coefficients and curvature components with their corresponding Kerr background quantities.~The need for such a procedure is related to certain technical complications arising from our choice of frames.

\medskip

See \textbf{Section \ref{sec_overview_identification_horizontal_structures}} for an overview to this section of the paper.
\end{itemize}

One can now formulate the (renormalised) vacuum Einstein equations for the perturbations:\footnote{The vacuum Einstein equations are nowhere assumed in Sections \ref{sec_perturbations_kerr} and \ref{sec:geom_compare_tensors}.}

\begin{itemize}
\item In \textbf{Section \ref{sec_renormalised_vacuum_Einstein_equations}}, we assume that the perturbations of Section \ref{sec_perturbations_kerr} solve the vacuum Einstein equations.~To formulate the equations for the perturbed connection coefficients and curvature components, one employs the formalism and system of equations of Section \ref{sec_nonlinear_EVE}.~By taking differences of the perturbed and corresponding Kerr quantities as prescribed in Section \ref{sec:geom_compare_tensors}, we renormalise the system of vacuum Einstein equations relative to the Kerr background.\footnote{More conveniently, we will present a slightly different system of equations, dubbed \emph{reduced} vacuum Einstein equations, which can be easily used by the reader to derive the renormalised vacuum Einstein equations.}

\medskip

See \textbf{Section \ref{sec_overview_renormalised_EVE}} for an overview to this section of the paper.
\end{itemize}

Section \ref{sec_renormalised_vacuum_Einstein_equations} concludes the nonlinear formulation of the problem.~To derive the system of linearised gravity, a second (and last) technical digression is required:\footnote{Section \ref{sec_linearisation_procedure} is the first section which departs from the nonlinear theory.~The treatment of Sections \ref{sec_perturbations_kerr}-\ref{sec_renormalised_vacuum_Einstein_equations} is fully nonlinear.}

\begin{itemize}
\item In \textbf{Section \ref{sec_linearisation_procedure}}, we linearise the renormalised quantities obtained by taking geometric differences of perturbed and corresponding Kerr background quantities via the procedure of Section \ref{sec:geom_compare_tensors}.~The linearisation formulae derived in Section \ref{sec_linearisation_procedure} are general and could, in principle, have been derived immediately after Section \ref{sec:geom_compare_tensors}.

\medskip

See \textbf{Section \ref{sec_overview_linearisation}} for an overview to this section of the paper.
\end{itemize}

In the last section, we achieve the formulation of the system of linearised gravity on Kerr exterior spacetimes:

\begin{itemize}
\item In \textbf{Section \ref{sec_system_linearised_gravity}}, we apply the linearisation formulae of Section \ref{sec_linearisation_procedure} to linearise the system of renormalised vacuum Einstein equations and derive the system of linearised gravity on the Kerr exterior manifold.~This section is entirely self-contained and allows to handle the system of linearised equations without knowledge of the rest of the paper. 

\medskip

See \textbf{Section \ref{sec_overview_linearised_equations}} for an overview to this section of the paper.
\end{itemize}

To summarise, the paper presents three systems of equations.~The general vacuum Einstein equations appear in Section \ref{sec_nonlinear_system_eqns}.~The reduced vacuum Einstein equations appear in Section \ref{sec_restricted_vacuum_Einstein_equations}.~The system of linearised gravity on Kerr exterior spacetimes appears in Section \ref{sec_system_linearised_gravity}.

\section{Overview}  \label{sec_overview}

We give an overview of the present work.~Each of the sections in the overview corresponds to a section of the bulk of the paper.

\subsection{The vacuum Einstein equations}  \label{sec_overview_EVE}

In Section \ref{sec_nonlinear_EVE}, we start by considering a $3+1$-dimensional, smooth, time-oriented Lorentzian manifold $(\bsy{\mathcal{M}},\bsy{g})$, endowed with the Levi-Civita connection $\bsy{\nabla}$ of the metric $\bsy{g}$.~We consider a general, possibly \emph{non-integrable}, local null frame on $\bsy{\mathcal{M}}$, meaning a local null frame  
\begin{equation*}
\bsy{\mathcal{N}}=(\eo,\etw,\et,\ef)
\end{equation*}
whose associated local \emph{horizontal distribution} 
\begin{equation*}
\bsy{\mathfrak{D}}_{\bsy{\mathcal{N}}} :=\left\langle \et  ,  \ef \right\rangle^{\bsy{\perp}}
\end{equation*}
is possibly a \emph{non-integrable} distribution on $\bsy{\mathcal{M}}$.~The frame vector fields $(\eo,\etw)$ are assumed to be \emph{orthonormal} relative to $\bsy{g}$ and form a local frame of $\bsy{\mathfrak{D}}_{\bsy{\mathcal{N}}}$.~See Section \ref{sec_preliminary_defns}.

\medskip

In Section \ref{sec_nonlinear_EVE} and throughout the paper, we adopt the bolded notation of \cite{DHR} in the nonlinear setting.

\subsubsection{$\bsy{\mathfrak{D}}_{\bsy{\mathcal{N}}}$ tensors}

In Section \ref{sec_preliminary_defns}, we define \emph{$\bsy{\mathfrak{D}}_{\bsy{\mathcal{N}}}$ $(k,q)$-tensors} on $\bsy{\mathcal{M}}$ as smooth sections of the abstract \emph{horizontal $(k,q)$-tensor bundle} 
\begin{equation*}
\otimes_k \, \bsy{\mathfrak{D}}_{\bsy{\mathcal{N}}}  \otimes_q (\bsy{\mathfrak{D}}_{\bsy{\mathcal{N}}})^{\star}
\end{equation*}
on $\bsy{\mathcal{M}}$.~We refer to $\bsy{\mathfrak{D}}_{\bsy{\mathcal{N}}}$ $(1,0)$-tensors as \emph{$\bsy{\mathfrak{D}}_{\bsy{\mathcal{N}}}$ vector fields} and to $\bsy{\mathfrak{D}}_{\bsy{\mathcal{N}}}$ $(0,q)$-tensors as \emph{$\bsy{\mathfrak{D}}_{\bsy{\mathcal{N}}}$ $q$-tensors}.

\medskip

In Section \ref{sec_ricci_curvature_defns}, the frame connection coefficients and curvature components of the metric $\bsy{g}$ relative to the frame $\bsy{\mathcal{N}}$ are defined as smooth scalar functions and $\bsy{\mathfrak{D}}_{\bsy{\mathcal{N}}}$ $q$-tensors on $\bsy{\mathcal{M}}$.

\medskip

In Section \ref{sec_tensor_contractions_products}, we define contractions and products of $\bsy{\mathfrak{D}}_{\bsy{\mathcal{N}}}$ tensors.

\subsubsection{Decomposition of the second fundamental forms} \label{sec_overview_decomp_chi}

The second fundamental forms 
\begin{align*}
\bsy{\chi}{}_{\bsy{AB}}&= \bsy{g}(\bsy{\nabla}_{\bsy{A}}\ef,\eb) \, , & \underline{\bsy{\chi}}{}_{\bsy{AB}}&= \bsy{g}(\bsy{\nabla}_{\bsy{A}}\et,\eb) 
\end{align*}
are $\bsy{\mathfrak{D}}_{\bsy{\mathcal{N}}}$ two-tensors and can be decomposed into \emph{symmetric traceless}, \emph{pure trace} and \emph{antisymmetric parts} as follows
\begin{align*}
\bsy{\chi} &= \widehat{\bsy{\chi}}+\frac{1}{2}\,(\textbf{tr} \bsy{\chi})\,\bsy{\slashed{g}}+\frac{1}{2}\,(\bsy{\slashed{\varepsilon}\cdot\chi})\,\bsy{\slashed{\varepsilon}} \, , & 
\underline{\bsy{\chi}} &= \underline{\widehat{\bsy{\chi}}}+\frac{1}{2}\,(\textbf{tr} \underline{\bsy{\chi}})\,\bsy{\slashed{g}}+\frac{1}{2}\,(\bsy{\slashed{\varepsilon}\cdot\underline{\chi}})\,\bsy{\slashed{\varepsilon}} \, ,
\end{align*}
where $\bsy{\slashed{g}}$ and $\bsy{\slashed{\varepsilon}}$ are the metric and volume form induced over $\bsy{\mathfrak{D}}_{\bsy{\mathcal{N}}}$ by the spacetime metric $\bsy{g}$ and spacetime standard volume form $\bsy{\varepsilon}$ respectively.~We refer to the smooth scalar functions
\begin{align*}
&(\bsy{\slashed{\varepsilon}\cdot\chi}) \, , & &(\bsy{\slashed{\varepsilon}\cdot\underline{\chi}})
\end{align*}  
as the \emph{antitraces} of the second fundamental forms.~The antitraces of the second fundamental forms both identically vanish if and only if $\bsy{\mathfrak{D}}_{\bsy{\mathcal{N}}}$ is integrable.~Equivalently, the antitraces of the second fundamental forms both identically vanish if and only if the commutator 
\begin{equation} \label{intro_commutator}
[\bsy{e_A},\bsy{e_B}] =  ({}_{\bsy{A}}\bsy{\slashed{H}{}_{B}^C} -{}_{\bsy{B}}\bsy{\slashed{H}{}_{A}^C} )\,\bsy{e_C} +\frac{1}{2}\,(\bsy{\slashed{\varepsilon}\cdot\chi})\,\bsy{\slashed{\varepsilon}}_{\bsy{AB}}\,\bsy{e_3}+\frac{1}{2}\,(\bsy{\slashed{\varepsilon}\cdot\underline{\chi}})\,\bsy{\slashed{\varepsilon}}_{\bsy{AB}}\,\bsy{e_4} 
\end{equation}
of the frame vector fields $(\eo,\etw)$ is generated by the frame vector fields $(\eo,\etw)$.~See Section \ref{sec_shears_decompn}.

\subsubsection{The induced connection and its torsion}  \label{sec_overview_connection_torsion}

The spacetime Levi-Civita connection $\bsy{\nabla}$ induces a canonical linear connection $\bsy{\nablasl}$ over the bundle of $\bsy{\mathfrak{D}}_{\bsy{\mathcal{N}}}$ tensors.~Crucially, the induced connection $\bsy{\nablasl}$ is the \emph{Levi-Civita} connection of the induced metric $\bsy{\slashed{g}}$ if and only if $\bsy{\mathfrak{D}}_{\bsy{\mathcal{N}}}$ is \emph{integrable}.\footnote{As we shall prove in Section \ref{sec_differential_operators}, a stronger statement holds true:~There \emph{exists} a Levi-Civita connection of $\bsy{\slashed{g}}$ over the bundle of $\bsy{\mathfrak{D}}_{\bsy{\mathcal{N}}}$ tensors if and only if $\bsy{\mathfrak{D}}_{\bsy{\mathcal{N}}}$ is integrable.}~In fact, for non-integrable distributions $\bsy{\mathfrak{D}}_{\bsy{\mathcal{N}}}$, the induced connection $\bsy{\nablasl}$ has non-trivial \emph{torsion}.\footnote{We note that, on the other hand, the induced connection $\bsy{\nablasl}$ is always compatible with the induced metric $\bsy{\slashed{g}}$, regardless of the integrability of $\bsy{\mathfrak{D}}_{\bsy{\mathcal{N}}}$.~See Section \ref{sec_differential_operators}.}~The torsion of $\bsy{\nablasl}$ arises from the failure of the commutator \eqref{intro_commutator} to be generated by the frame vector fields $(\eo,\etw)$, which implies the relation 
\begin{equation*}
\bsy{\nablasl}_{\bsy{A}}\eb - \bsy{\nablasl}_{\bsy{B}}\ea \neq [\ea,\eb] \, ,
\end{equation*}
where the left hand side is, by definition of $\bsy{\nablasl}$, a $\bsy{\mathfrak{D}}_{\bsy{\mathcal{N}}}$ vector field.~See Section \ref{sec_differential_operators}.

\medskip

In Section \ref{sec_differential_operators}, we define a set of differential operators acting on $\bsy{\mathfrak{D}}_{\bsy{\mathcal{N}}}$ tensors in terms of the induced connection $\bsy{\nablasl}$.~The formal definition of such operators is, in most part, analogous to that from previous works \cite{DHR, DHRT}.~However, to handle some of these operators, the torsion of $\bsy{\nablasl}$ has to be taken into account.~One relevant example is the second covariant derivative $$\bsy{\nablasl}^2_{\bsy{A},\bsy{B}} \, ,$$ which appears in the definition of the \emph{induced Riemann curvature $\bsy{\slashed{R}}$} of $\bsy{\mathfrak{D}}_{\bsy{\mathcal{N}}}$.~Due to the properties of the second covariant derivative $\bsy{\nablasl}^2$ when the induced connection $\bsy{\nablasl}$ has torsion, it turns out that the induced Riemann curvature $\bsy{\slashed{R}}$ is a \emph{tensorial} $\bsy{\mathfrak{D}}_{\bsy{\mathcal{N}}}$ quantity if and only if $\bsy{\mathfrak{D}}_{\bsy{\mathcal{N}}}$ is \emph{integrable}.~Similarly, the symmetries of $\bsy{\slashed{R}}$ depend on the (non-)integrability of $\bsy{\mathfrak{D}}_{\bsy{\mathcal{N}}}$.~See Section \ref{sec_curvature_horizontal_distribution}.

\subsubsection{The nonlinear system of equations} \label{sec_overview_system_EVE}

In Section \ref{sec_nonlinear_system_eqns}, we assume that the Lorentzian manifold $(\bsy{\mathcal{M}},\bsy{g})$ solves the vacuum Einstein equations
\begin{equation*} 
\bsy{\ric}(\bsy{g})=0 \, .
\end{equation*}
The nonlinear system of equations is formulated as a system of equations for the connection coefficients and curvature components of the metric $\bsy{g}$ relative to the frame $\bsy{\mathcal{N}}$, and thus it is a system for $\bsy{\mathfrak{D}}_{\bsy{\mathcal{N}}}$ tensors.~The complete system of equations appears in Sections \ref{sec_null_structure_equations}-\ref{sec_bianchi_equations}.~The full system of equations is divided into two coupled sets of equations, namely the \emph{null structure equations} and the \emph{Bianchi equations}.~The Bianchi equations encode the \emph{hyperbolic} character of the vacuum Einstein equations.~Our nonlinear system of equations reduces to that of \cite{StabMink} when the frame $\bsy{\mathcal{N}}$ is integrable.

\subsubsection{Formal derivation of the equations}

In Section \ref{sec_formal_derivation_EVE}, we present the formal derivation of the nonlinear system of equations of Section \ref{sec_overview_system_EVE}.~The main new technical ingredient is the presence of the antisymmetric parts of the second fundamental forms in the computations.~Another new technicality arises in the derivation of the \emph{Gauss equation}.~As already mentioned, due to the torsion of the induced connection $\bsy{\nablasl}$, one has to deal with \emph{non-tensorial} $\bsy{\mathfrak{D}}_{\bsy{\mathcal{N}}}$ curvature quantities.~The Gauss equation is written in terms of a \emph{corrected Gauss curvature} $\widetilde{\bsy{\slashed{K}}}$, which we introduce to treat the curvature of non-integrable distributions.~See Section \ref{sec_curvature_horizontal_distribution}.

\subsection{The Kerr exterior manifold}  \label{sec_overview_kerr_exterior_manifold}

In Section \ref{sec_Kerr_exterior_manifold}, we define the \emph{Kerr exterior manifold} and the \emph{algebraically special frame} and \emph{outgoing principal differentiable structures} of the Kerr exterior manifold.

\subsubsection{Definition of the Kerr exterior manifold} \label{sec_overview_definition_kerr_exterior_manifold}

In Section \ref{sec_definition_Kerr_exterior_manifold}, we define the manifold-with-boundary 
\begin{equation}  \label{overview_kerr_ambient_manifold}
\mathcal{M}:= (-\infty,\infty) \times [0,\infty) \times \mathbb{S}^2 
\end{equation}
with coordinates $\bar{t}^*\in (-\infty,\infty)$, $y^*\in [0,\infty)$ and standard spherical coordinates $(\bar{\theta},\bar{\phi}^*)\in\mathbb{S}^2$.~As usual, the coordinates $(\bar{\theta},\bar{\phi}^*)$ only cover a subset of $\mathbb{S}^2$.~We define the \emph{(future) event horizon} as the boundary
\begin{equation*}
\mathcal{H}^+:=\partial\mathcal{M}  
\end{equation*} 
of $\mathcal{M}$ and \emph{(future) null infinity} as the formal hypersurface
\begin{equation*}
\mathcal{I}^+:=\left\lbrace \bar{t}^*=\infty \right\rbrace \, .
\end{equation*}

\medskip

Given real parameters $a$ and $M$, with $|a| <M$, we define two positive constants $r_{\pm}$ and a new regular coordinate $\bar{r}_{a,M}=\bar{r}_{a,M}(y^*)$ (see Section \ref{sec_definition_Kerr_exterior_manifold}).~We will simply denote $\bar{r}_{a,M}$ by $\bar{r}$ and refer to the coordinates $(\bar{t}^*,\bar{r},\bar{\theta},\bar{\phi}^*)$ as \emph{Kerr coordinates}.~We note that $\mathcal{H}^+=\left\lbrace \bar{r}=r_+ \right\rbrace$.

\medskip

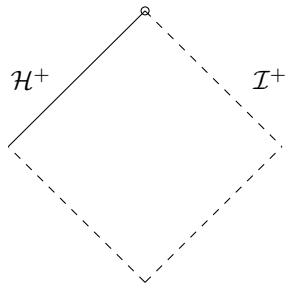
\begin{figure}[H]

\centering
\begin{tikzpicture}[scale=1.5]

\draw (0,2)--(-1.2,0.8);

\node at (-1.0,1.4) {$\mathcal{H}^+$};
\node at (1.1,1.4) {$\mathcal{I}^+$};


\draw (0,2)  circle[radius=1pt];





\draw[dashed] (0,2)--(1.2,0.8);

\draw[dashed] (0,-0.4)--(-1.2,0.8);
\draw[dashed] (0,-0.4)--(1.2,0.8);

\end{tikzpicture}

\caption{Penrose diagram of the Kerr exterior manifold $(\mathcal{M},g_{a,M})$.}

\label{fig:kerr_manifold}

\end{figure}

We define the \emph{Kerr family of metrics} as the two-parameter family of Lorentzian metrics $g_{a,M}$ on $\mathcal{M}$ such that 
\begin{align}
g_{a,M}= &-\left(1-\frac{2\,M\,\bar{r}}{\Sigma}\right){d\bar{t}^*}^2+2\,d\bar{t}^*\, d\bar{r}+\Sigma\, {d\bar{\theta}}^2 +\frac{(\bar{r}^2+a^2)^2-\Delta\, a^2\sin^2\bar{\theta}}{\Sigma}\,\sin^2\bar{\theta} \, {d\bar{\phi}^*}^{2}   \label{intro_def_kerr_metric} \\
&-2\,a\sin^2\bar{\theta} \, d\bar{r} \,d\bar{\phi}^*-\frac{4\,a\,M\,\bar{r}}{\Sigma}\,\sin^2\bar{\theta} \, d\bar{t}^*\, d\bar{\phi}^* \, ,   \nonumber
\end{align}
where the smooth scalar functions $\Delta(\bar{r})$ and $\Sigma(\bar{r},\bar{\theta})$ are defined in Section \ref{sec_definition_Kerr_exterior_manifold}.~The metric \eqref{intro_def_kerr_metric} is manifestly smooth on $\mathcal{M}$, including on $\mathcal{H}^+$.~The event horizon $\mathcal{H}^+$ can be checked to be a null hypersurface relative to $g_{a,M}$.~See Figure \ref{fig:kerr_manifold}.

\medskip

The smooth Lorentzian manifold $(\mathcal{M},g_{a,M})$ will be referred to as the \emph{Kerr exterior manifold}.~One can check that $(\mathcal{M},g_{a,M})$ solves the vacuum Einstein equations.

\medskip

\medskip

In Section \ref{sec_Kerr_star_coords}, we introduce the auxiliary \emph{Kerr-star differentiable structure} $$(t^*,r,\theta,\phi^*)$$ on $\mathcal{M}$.~These coordinates are regular on the whole manifold $\mathcal{M}$, including on $\mathcal{H}^+$, and primarily differ from Kerr coordinates in that the level sets of $t^*$ are everywhere spacelike, horizon-penetrating and asymptotically flat hypersurfaces.~The definition of Kerr-star coordinates requires the use of a radial cut-off function, which prevents an \emph{explicit} form of the Kerr metric in these coordinates.~For this reason, Kerr-star coordinates are not employed in the first place to \emph{define} the Kerr metric.~On the other hand, Kerr-star coordinates are directly employed to \emph{define} the \emph{star-normalised, outgoing principal differentiable structure}.~See Section \ref{sec_overview_back_diff_structure}.

\subsubsection{The algebraically special frame}

In Section \ref{sec_Kerr_algebr_special_frame}, we define the \emph{algebraically special} (local) null frame $$\mathcal{N}_{\text{as}}=(e_1^{\text{as}},e_2^{\text{as}},e_3^{\text{as}},e_4^{\text{as}})$$ of $(\mathcal{M},g_{a,M})$ (see already Section \ref{sec_intro_preliminaries_kerr}).~The associated horizontal distribution
\begin{equation*}
\mathfrak{D}_{\mathcal{N}_{\text{as}}}:=\left\langle e^{\text{as}}_3,e^{\text{as}}_4 \right\rangle^{\perp} 
\end{equation*}
is globally well-defined on $\mathcal{M}$.~For $|a|>0$, the frame $\mathcal{N}_{\text{as}}$ is \emph{non-integrable}, with $\mathfrak{D}_{\mathcal{N}_{\text{as}}}$ a non-integrable distribution.

\medskip

The induced metric, connection coefficients and curvature components of $g_{a,M}$ relative to $\mathcal{N}_{\text{as}}$ are \emph{global} and \emph{regular} $\mathfrak{D}_{\mathcal{N}_{\text{as}}}$ tensors on $\mathcal{M}$ and are explicitly computed in Section \ref{sec_Kerr_connection_coeff_curv_comps}.~We point out the following identities for the connection coefficients 
\begin{align*}
\omegabh_{a,M}&=0 \, , & \eta_{a,M}-\zeta_{a,M}&=0   \, , & \xi_{a,M}=\underline{\xi}_{a,M} &=0 \, , & \widehat{\chi}{}_{a,M}=\underline{\widehat{\chi}}{}_{a,M}&=0  \, ,
\end{align*}
the \emph{non}-vanishing of the antitraces of the second fundamental forms $(\slashed{\varepsilon}\cdot\chi)_{a,M}$ and $(\slashed{\varepsilon}\cdot\underline{\chi})_{a,M}$ and the vanishing of the curvature components
\begin{align*}
\alpha_{a,M}=\underline{\alpha}_{a,M} &=0 \, , & \beta_{a,M}=\underline{\beta}_{a,M} &=0  \, . 
\end{align*}
The connection coefficients and curvature components of $(\mathcal{M},g_{a,M})$ relative to $\mathcal{N}_{\text{as}}$ solve the nonlinear system of null structure and Bianchi equations of Section \ref{sec_overview_system_EVE}.~The formalism developed in Section \ref{sec_overview_EVE} is already needed at this stage in its full generality, in that $\mathcal{N}_{\text{as}}$ is a non-integrable frame.

\subsubsection{Outgoing principal differentiable structures}  \label{sec_overview_back_diff_structure}

In Section \ref{sec_Kerr_background_diff_structure}, we define \emph{outgoing principal differentiable structures} of $(\mathcal{M},g_{a,M})$ as (local) differentiable structures $(x^1,x^2,\vartheta^A)$ on $\mathcal{M}$, with $(\vartheta^1,\vartheta^2)\in\mathbb{S}^2$, such that 
\begin{align*}
e_4^{\text{as}}(x^1)&=0 \, , & e_4^{\text{as}}(\vartheta^A)&=0 \, .
\end{align*}
Relative to an outgoing principal differentiable structure, the Kerr metric satisfies the identities
\begin{align*}
(g_{a,M})_{x^2x^2}&=0 \, , & \partial_{x^2} (g_{a,M})_{x^2\mu}+\mathfrak{h}_{a,M}\, (g_{a,M})_{x^2\mu}&=0
\end{align*}
for some smooth scalar function $\mathfrak{h}_{a,M}$ on $\mathcal{M}$, with $x^{\mu}=(x^1,\vartheta^A)$.

\medskip

\begin{figure}[H]

\centering

\begin{subfigure}{0.45\textwidth}

\centering 

\begin{tikzpicture}[scale=1.5]

\draw (0,2)--(-1.2,0.8);

\node at (-1.0,1.4) {$\mathcal{H}^+$};
\node at (1.1,1.4) {$\mathcal{I}^+$};
\node at (0.2,1.4) {$\Sigma$};


\fill (0.3,0.85)  circle[radius=1pt];

\draw (0,2)  circle[radius=1pt];

\node at (-0.4,0.76) {$\mathcal{S}_{0}$};

\node at (0.45,0.6) {\small{$\mathbb{S}^2_{0,r_+ +1}$}};


\draw plot [smooth] coordinates { (-0.9,1.1) (-0.3,0.92) (0.3,0.85) (1.2,0.8) };

\draw plot [smooth] coordinates {  (0.3,0.85) (0.38,1.3) (0.5,1.5) };


\draw[dashed] (0,2)--(1.2,0.8);

\draw[dashed] (0,-0.4)--(-1.2,0.8);
\draw[dashed] (0,-0.4)--(1.2,0.8);

\end{tikzpicture}

\end{subfigure}
\begin{subfigure}{0.45\textwidth}

\centering

\begin{tikzpicture}[scale=2.2]

\draw (0,2)--(-0.9,1.1);

\node at (-0.8,1.65) {$\mathcal{H}^+\cap \mathcal{M}^*$};
\node at (0.6,1.9) {$\mathcal{I}^+\cap \mathcal{M}^*$};
\node at (0,1.3) {$\mathcal{M}^*$};
\node at (-0.35,0.78) {$\mathcal{S}^*_0$};



\draw (0,2)  circle[radius=1pt];




\draw plot [smooth] coordinates { (-0.9,1.1) (-0.35,0.95) (0.3,0.85) };

\draw[dashed] plot [smooth] coordinates {  (0.3,0.85) (0.38,1.3) (0.5,1.5) };


\draw[dashed] (0,2)--(0.5,1.5);


\end{tikzpicture}

\end{subfigure}

\caption{In the Penrose diagram on the left, the spacelike hypersurface $\mathcal{S}_0$ is defined as the level set $\left\lbrace t^*=0 \right\rbrace$, with $t^*$ the Kerr-star time coordinate.~The timelike hypersurface $\Sigma$ is generated by the integral curves of $e_4^{\text{as}}$ emanating from the Kerr-star two-sphere $\mathbb{S}^2_{t^*=0,r=r_+ +1}$.~The spacetime region $\mathcal{M}^*$ (Penrose diagram on the right) is defined by flowing the hypersurface $\mathcal{S}_0^*:=\mathcal{S}_0\cap\left\lbrace r<r_+ +1 \right\rbrace$ along the integral curves of $e_4^{\text{as}}$.~We note that $\mathcal{S}_0^*\subset\mathcal{M}^*$.}

\label{fig:M_star}

\end{figure}
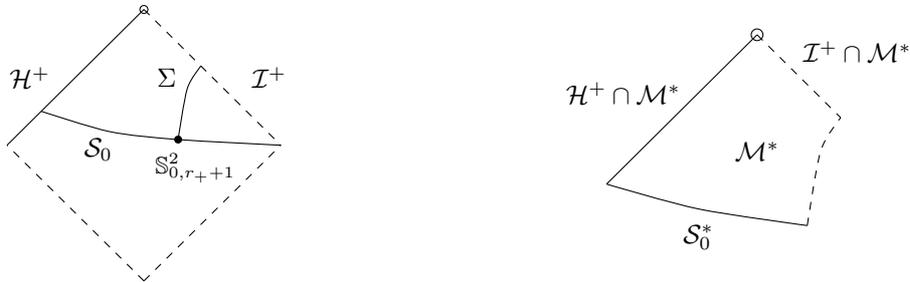

Then, we introduce the \emph{star-normalised}, outgoing principal differentiable structure
\begin{equation} \label{intro_coords_background}
(s,\underline{s},\theta^A)
\end{equation}
on the spacetime region
\begin{equation*}
\mathcal{M}^* \subset \mathcal{M}
\end{equation*}
as in Figure \ref{fig:M_star}, with 
\begin{align*}
 s &\in [r_+,r_+ +1) \, , & \underline{s} &\in  [0,\infty) \, , & (\theta^1,\theta^2) &\in \mathbb{S}^2  \, .
\end{align*}
The coordinates \eqref{intro_coords_background} are \emph{regular} on $\mathcal{M}^*$, including on $\mathcal{H}^+\cap \mathcal{M}^*$.~The coordinates $(s,\underline{s})$ are \emph{global}, whereas the angular coordinates $(\theta^1,\theta^2)$ are subject to the usual degeneration on two-spheres.

\medskip

The coordinate $\underline{s}$ is such that
\begin{equation*}
\underline{s} = t^*
\end{equation*}
on $\mathcal{M}^*$, with $t^*$ the Kerr-star time coordinate on $\mathcal{M}^*$.~To construct the other coordinates, we first consider the hypersurface $$\mathcal{S}^*_0:=\left\lbrace \ub{$s$}=0 \right\rbrace$$ and define a coordinate function $s_0$ on $\mathcal{S}_0^*$ such that $s_0 = r|_{\mathcal{S}^*_0}$, with $r$ the Kerr-star (or, equivalently, Boyer--Lindquist) radius.~We define arbitrary angular coordinates $(\theta^1_0,\theta^2_0)$ on the two-spheres of constant $s_0$ foliating $\mathcal{S}^*_0$.~We then transport the coordinate $s_0$ along the integral curves of the vector field $e_4^{\text{as}}$ on $\mathcal{M}^*$, i.e.~we define the coordinate $s$ on $\mathcal{M}^*$ such that
\begin{equation*}
e_4^{\text{as}}(s)=0  
\end{equation*}
on $\mathcal{M}^*$ and $s|_{\mathcal{S}^*_0} = s_0$.~We define angular coordinates $(\theta^1,\theta^2)$ on the two-spheres
\begin{equation*}
\mathbb{S}^2_{s,\underline{s}}:=\left\lbrace s,\underline{s} \right\rbrace \times \mathbb{S}^2
\end{equation*}
such that 
\begin{align*}
e_4^{\text{as}}(\theta^1)&=0 \, , & e_4^{\text{as}}(\theta^2)&=0
\end{align*}
on $\mathcal{M}^*$ and $(\theta^1,\theta^2)|_{\mathbb{S}^2_{s,0}} = (\theta^1_0,\theta^2_0)|_{\mathbb{S}^2_{s,0}}$.

\medskip

We note that constructing \emph{global} coordinates $(s,\underline{s})$ on $\mathcal{M}^*$ requires to carefully check that the vector field $e_4^{\text{as}}$ is a \emph{complete} vector field on $\mathcal{M}^*$.~See Section \ref{sec_Kerr_background_diff_structure}.

\medskip

The coordinates $(s,\underline{s})$ induce a \emph{global} foliation of $\mathcal{M}^*$.~The level sets of $\ub{$s$}$ are everywhere \emph{spacelike}, horizon-penetrating and asymptotically flat hypersurfaces.~We have the formal \emph{null} hypersurface
\begin{equation*}
 \mathcal{I}^+\cap\mathcal{M}^* := \left\lbrace \ub{$s$}=\infty \right\rbrace  \, .
\end{equation*}
The level sets of $r_+<s< r_+ +1$ are \emph{timelike} hypersurfaces intersecting $\mathcal{I}^+$.~Such hypersurfaces are \emph{asymptotically null} as they approach $\mathcal{I}^+$.~The level set $$\left\lbrace s=r_+ \right\rbrace=\mathcal{H}^+\cap \mathcal{M}^*$$ is \emph{null}.~See already Figure \ref{fig:intro_foliation}.

\medskip

\medskip

In Section \ref{sec_Kerr_background_diff_structure}, we define the smooth scalar functions $k_{a,M}$ and $\mathfrak{k}_{a,M}$ on $\mathcal{M}^*$.~These are regular functions on the whole $\mathcal{M}^*$, including on $\mathcal{H}^+\cap\mathcal{M}^*$, and $k_{a,M}$ is nowhere vanishing.~Relative to the differentiable structure \eqref{intro_coords_background}, the Kerr metric takes the form \eqref{intro_kerr_metric_outgoing_principal} and satisfies the identities \eqref{intro_kerr_gauge_ids} on $\mathcal{M}^*$.~The frame vector fields $(e_4^{\text{as}},e_3^{\text{as}})$ take the form \eqref{intro_alg_frame_e4_e3_coord} on $\mathcal{M}^*$.

\subsection{Nonlinear perturbations of the Kerr exterior manifold}  \label{sec_overview_perturbations_kerr}

In Section \ref{sec_perturbations_kerr}, we formulate nonlinear perturbations of the Kerr exterior manifold.~The formulation is the core of our new geometric framework.

\subsubsection{Family of metrics}  \label{sec_overview_family_metrics}

In Section \ref{sec_family_metrics}, we start by choosing real parameters $a$ and $M$, with $|a|<M$.~We fix the manifold-with-boundary $\mathcal{M}^*$, which we identify with the subset of the Kerr ambient manifold of Figure \ref{fig:M_star} with star-normalised, outgoing principal differentiable structure $(s,\underline{s},\theta^A)$.~We introduce the fixed smooth scalar functions $\mathfrak{k}_{a,M}$, $k_{a,M}$ and $\omegah_{a,M}$ and the fixed one-tensor $\underline{\eta}^{\circ}_{a,M}$ of the fixed coordinates on $\mathcal{M}^*$, with
\begin{equation*}
\omegah_{a,M}:=\frac{1}{2}\,g_{a,M}(\nabla_{e_4^{\text{as}}}e_3^{\text{as}},e_4^{\text{as}})
\end{equation*}
and $\underline{\eta}_{a,M}^{\circ}\in\Gamma((T\mathcal{M}^*)^{\star})$ such that
\begin{align*}
\underline{\eta}^{\circ}_{a,M}(e_A^{\text{as}})&=\frac{1}{2}\,g_{a,M}(\nabla_{e_4^{\text{as}}}e_3^{\text{as}},e_A^{\text{as}}) \, , & \underline{\eta}^{\circ}_{a,M}(e_3^{\text{as}})&=0 \, , & \underline{\eta}^{\circ}_{a,M}(e_4^{\text{as}})&=0 \, .
\end{align*}

\medskip

Once the background structure is fixed, we prescribe a one-parameter family of smooth, non-degenerate Lorentzian metrics $\bsy{g}(\epsilon)$ on $\mathcal{M}^*$ as already outlined in Section \ref{sec_intro_nonlinear_perturbations}.~We further assume that
\begin{equation} \label{intro_uniform_limit_metrics}
\sum_{0\leq |I|\leq 10} |\partial^{I}\bsy{\mathfrak{g}}(\epsilon)-\partial^{I}\bsy{\mathfrak{g}}(0)| \leq \epsilon 
\end{equation}
on $\mathcal{M}^*$, with
\begin{equation*}
\bsy{\mathfrak{g}}(\epsilon)=\left\lbrace \bsy{\mathfrak{a}}(\epsilon)\, , \, \bsy{\Omega}^2(\epsilon) \, , \, \bsy{b}_{\theta^A}(\epsilon) \, , \, \underline{\bsy{b}}_{\theta^A}(\epsilon) \, , \, \bsy{\gamma}_{\theta^A\theta^B}(\epsilon) \right\rbrace
\end{equation*}
and $I$ a multi-index for partial derivatives relative to the fixed differentiable structure.

\medskip

We remark that there exists a sufficiently small constant $\epsilon_0>0$, depending only on the fixed background parameters $a$ and $M$, such that, for all $0\leq |\epsilon|\leq \epsilon_0$, the form \eqref{intro_outline_family_metrics} of the metrics $\bsy{g}(\epsilon)$, together with the additional identities \eqref{intro_outline_gauge_ids}, correspond to a choice of gauge for the family of metrics $\bsy{g}(\epsilon)$, called \emph{outgoing principal gauge}.~The assumption \eqref{intro_uniform_limit_metrics} is relevant for this fact.~See Section \ref{sec_overview_gauge_construction}.

\medskip

From now on, without loss of generality, \ul{we restrict to $0\leq |\epsilon|\leq \epsilon_0$}.~When we write \emph{for all $\epsilon$}, this restriction is implicitly assumed.

\subsubsection{Family of frames}  \label{sec_overview_family_frames}

In Section \ref{sec_family_frames}, to the one-parameter family of metrics $\bsy{g}(\epsilon)$, we associate a one-parameter family of local frames $$\bsy{\mathcal{N}}(\epsilon)=(\eo(\epsilon),\etw(\epsilon),\et(\epsilon),\ef(\epsilon))$$ on $\mathcal{M}^*$ such that
\begin{align} \label{overview_formulation_problem_e4_e3}
\ef(\epsilon)&:= k_{a,M}\,\partial_{\underline{s}}  & \et(\epsilon)&:=\underline{\bsy{j}}(\epsilon)\,\partial_s+\underline{\bsy{k}}(\epsilon)\,\partial_{\underline{s}}+\underline{\bsy{\Lambda}}^{\theta^A}(\epsilon)\,\partial_{\theta^A} \, .
\end{align}
For each $\epsilon$, the frame quantities $\underline{\bsy{j}}(\epsilon)$ and $\underline{\bsy{k}}(\epsilon)$ are smooth scalar functions on $\mathcal{M}^*$ and the frame quantity $\underline{\bsy{\Lambda}}(\epsilon)$ is a vector field tangent to the $\mathbb{S}^2_{s,\underline{s}}$-spheres.~All quantities are regular on the whole $\mathcal{M}^*$, including on $\mathcal{H}^+\cap\mathcal{M}^*$.~For each $\epsilon$ and any $(s,\theta^A)$, the frame quantities are \emph{defined} implicitly as solutions to the system of ODEs\footnote{We note that the only unknown quantities on the right hand side of the system of ODEs \eqref{overview_ODE_1}-\eqref{overview_ODE_3} are the frame quantities $(\underline{\bsy{j}}(\epsilon),\underline{\bsy{k}}(\epsilon),\underline{\bsy{\Lambda}}(\epsilon))$.~All the other quantities are either background quantities or known quantities depending on the given metrics $\bsy{g}(\epsilon)$.~We denote by $\bsy{\Gamma}(\epsilon)$ the Christoffel symbols of the metrics $\bsy{g}(\epsilon)$.}
\begin{align}
k\,\frac{d\,\underline{\bsy{j}}(\epsilon)}{d\underline{s}}=& \,  -2\left(\frac{\underline{\eta}^{\circ}_{s}\,\underline{\bsy{j}}(\epsilon)+\underline{\eta}^{\circ}_{\theta^A}\,\underline{\bsy{\Lambda}}^{\theta^A}(\epsilon)}{\Omega^2\,\underline{\bsy{j}}(\epsilon)+\underline{b}_{\theta^B}\,\underline{\bsy{\Lambda}}^{\theta^B}(\epsilon)}\right)(\Omega^2\bsy{g}^{ss}(\epsilon)+\underline{b}_{\theta^C}\,\bsy{g}^{s\theta^C}(\epsilon)) \label{overview_ODE_1} \\
&-(\omegah+k\,\bsy{\Gamma}_{s\underline{s}}^s(\epsilon))\,\underline{\bsy{j}}(\epsilon)-k\,\bsy{\Gamma}_{\underline{s}\theta^A}^s(\epsilon)\,\underline{\bsy{\Lambda}}^{\theta^A}(\epsilon) \nonumber \\ &+2 \,\underline{\eta}^{\circ}_{s}\,\bsy{g}^{ss}(\epsilon)+2 \,\underline{\eta}^{\circ}_{\theta^A}\,\bsy{g}^{s\theta^A}(\epsilon) \nonumber \\[5pt]
k\,\frac{d\,\underline{\bsy{k}}(\epsilon)}{d\underline{s}}=& \, -2\left(\frac{\underline{\eta}^{\circ}_{s}\,\underline{\bsy{j}}(\epsilon)+\underline{\eta}^{\circ}_{\theta^A}\,\underline{\bsy{\Lambda}}^{\theta^A}(\epsilon)}{\Omega^2\,\underline{\bsy{j}}(\epsilon)+\underline{b}_{\theta^B}\,\underline{\bsy{\Lambda}}^{\theta^B}(\epsilon)}\right)(\Omega^2\bsy{g}^{s\underline{s}}(\epsilon)+\underline{b}_{\theta^C}\,\bsy{g}^{\underline{s}\theta^C}(\epsilon)) \label{overview_ODE_2} \\
&+(k\,\mathfrak{k}-\omegah)\,\underline{\bsy{k}}(\epsilon)-k\,\bsy{\Gamma}_{s\underline{s}}^{\underline{s}}(\epsilon)\,\underline{\bsy{j}}(\epsilon)-k\,\bsy{\Gamma}_{\underline{s}\theta^A}^{\underline{s}}(\epsilon)\,\underline{\bsy{\Lambda}}^{\theta^A}(\epsilon) \nonumber \\ &+2 \,\underline{\eta}^{\circ}_{s}\,\bsy{g}^{s\underline{s}}(\epsilon)+2 \,\underline{\eta}^{\circ}_{\theta^A}\,\bsy{g}^{\underline{s}\theta^A}(\epsilon) \nonumber \\[5pt]
k\,\frac{d\,\underline{\bsy{\Lambda}}^{\theta^A}(\epsilon)}{d\underline{s}} =& \, -2\left(\frac{\underline{\eta}^{\circ}_{s}\,\underline{\bsy{j}}(\epsilon)+\underline{\eta}^{\circ}_{\theta^B}\,\underline{\bsy{\Lambda}}^{\theta^B}(\epsilon)}{\Omega^2\,\underline{\bsy{j}}(\epsilon)+\underline{b}_{\theta^C}\,\underline{\bsy{\Lambda}}^{\theta^C}(\epsilon)}\right)(\Omega^2\bsy{g}^{s\theta^A}(\epsilon)+\underline{b}_{\theta^D}\,\bsy{g}^{\theta^A\theta^D}(\epsilon)) \label{overview_ODE_3} \\
&-\omegah\,\underline{\bsy{\Lambda}}^{\theta^A}(\epsilon)-k\,\bsy{\Gamma}_{\underline{s}\theta^B}^{\theta^A}(\epsilon)\,\underline{\bsy{\Lambda}}^{\theta^B}(\epsilon)-k\,\bsy{\Gamma}_{s\underline{s}}^{\theta^A}(\epsilon)\,\underline{\bsy{j}}(\epsilon) \nonumber \\ &+2 \,\underline{\eta}^{\circ}_{s}\,\bsy{g}^{s\theta^A}(\epsilon)+2 \,\underline{\eta}^{\circ}_{\theta^B}\,\bsy{g}^{\theta^A\theta^B}(\epsilon)  \nonumber
\end{align}
in the variable $\underline{s}$, with initial data
\begin{gather}
\underline{\bsy{j}}(\epsilon)|_{\underline{s}=0}= \left. -\frac{2\,\underline{j}}{k\,(\underline{j}\,\bsy{\Omega}^2(\epsilon)+\underline{\Lambda}^{\theta^A}\underline{\bsy{b}}_{\theta^A}(\epsilon))}\,\right\rvert_{\underline{s}=0} \, , \label{overview_ODE_data_1} \\ 
\underline{\bsy{k}}(\epsilon)|_{\underline{s}=0}= \left.\frac{\underline{j}^2\bsy{\mathfrak{a}}(\epsilon)+2\,\underline{j}\,\underline{k}\,\bsy{\Omega}^2(\epsilon)+2\,\underline{j}\,\underline{\Lambda}^{\theta^A}\bsy{b}_{\theta^A}(\epsilon)+2\,\underline{k}\,\underline{\Lambda}^{\theta^A}\underline{\bsy{b}}_{\theta^A}(\epsilon)+|\underline{\Lambda}|^2_{\bsy{\gamma}(\epsilon)}}{k\,(\underline{j}\,\bsy{\Omega}^2(\epsilon)+\underline{\Lambda}^{\theta^A}\underline{\bsy{b}}_{\theta^A}(\epsilon))^2} -\frac{2\,\underline{k}}{k\,(\underline{j}\,\bsy{\Omega}^2(\epsilon)+\underline{\Lambda}^{\theta^A}\underline{\bsy{b}}_{\theta^A}(\epsilon))} \,\right\rvert_{\underline{s}=0} \, , \label{overview_ODE_data_2} \\ 
\underline{\bsy{\Lambda}}^{\theta^A}(\epsilon)|_{\underline{s}=0}= \left. -\frac{2\,\underline{\Lambda}^{\theta^A}}{k\,(\underline{j}\,\bsy{\Omega}^2(\epsilon)+\underline{\Lambda}^{\theta^A}\underline{\bsy{b}}_{\theta^A}(\epsilon))} \,\right\rvert_{\underline{s}=0} \, . \label{overview_ODE_data_3}
\end{gather}
A choice of the frame vector fields $(\eo(\epsilon),\etw(\epsilon))$ is \emph{not} required.~For further comments on the definition of the frames $\bsy{\mathcal{N}}(\epsilon)$, see Section \ref{sec_overview_frames_construction}.

\medskip

As a result of their definition, the frames $\bsy{\mathcal{N}}(\epsilon)$ satisfy the properties already listed in Section \ref{sec_intro_nonlinear_perturbations} on $\mathcal{M}^*$, whose proof is addressed in Section \ref{sec_overview_frames_construction}.~In particular, the one-parameter family of horizontal distributions
\begin{equation*} 
\bsy{\mathfrak{D}}_{\bsy{\mathcal{N}}(\epsilon)} :=\left\langle \et(\epsilon),\ef(\epsilon)\right\rangle^{\bsy{\perp}_{\bsy{g}(\epsilon)}} 
\end{equation*}
is globally well-defined on $\mathcal{M}^*$ and such that $\bsy{\mathfrak{D}}_{\bsy{\mathcal{N}}(0)} = \mathfrak{D}_{\mathcal{N}_{\text{as}}}$.

\medskip

The connection coefficients and curvature components of the metrics $\bsy{g}(\epsilon)$ relative to the frames $\bsy{\mathcal{N}}(\epsilon)$ are global and regular $\bsy{\mathfrak{D}}_{\bsy{\mathcal{N}}(\epsilon)}$ tensors on $\mathcal{M}^*$.~We have the identities
\begin{align}
\bsy{\omegah}(\epsilon)-\omegah_{a,M} &= 0  \, ,\label{overview_i_2_kerr}\\
\bsy{\xi}(\epsilon) &= 0 \, , \label{overview_i_3_kerr} \\
\underline{\widetilde{\bsy{\eta}}}(\epsilon)-\underline{\eta}{}_{a,M} &= 0 \label{overview_i_4_kerr} 
\end{align} 
on $\mathcal{M}^*$ \emph{for all $\epsilon$}.~The geometric quantity $\underline{\widetilde{\bsy{\eta}}}(\epsilon)$ in the identity \eqref{overview_i_4_kerr} is a global, regular $\mathfrak{D}_{\mathcal{N}_{\text{as}}}$ one-tensor on $\mathcal{M}^*$ that one can uniquely associate to the $\bsy{\mathfrak{D}}_{\bsy{\mathcal{N}}(\epsilon)}$ one-tensor $\underline{\bsy{\eta}}(\epsilon)$ (see Section \ref{sec_overview_identification_horizontal_structures}).\footnote{Consistently, we will have $\underline{\widetilde{\bsy{\eta}}}(0) = \underline{\bsy{\eta}}(0)$, and thus $\underline{\widetilde{\bsy{\eta}}}(0) = \underline{\eta}{}_{a,M}$.}~The identity \eqref{overview_i_4_kerr} is thus an identity for the geometric difference of global, regular $\mathfrak{D}_{\mathcal{N}_{\text{as}}}$ one-tensors on $\mathcal{M}^*$.

\medskip

Two additional properties of the frames $\bsy{\mathcal{N}}(\epsilon)$ are particularly relevant for the problem:
\begin{itemize}
\item \textbf{Non-integrable horizontal distributions}

The frames $\bsy{\mathcal{N}}(\epsilon)$ are non-integrable for all $\epsilon$,\footnote{For $|a|=0$, the frames $\bsy{\mathcal{N}}(\epsilon)$ are non-integrable for all $|\epsilon|>0$, with $\epsilon =0$ excluded.~See already Section \ref{sec_intro_framework_Schwarzschild}.} with $\bsy{\mathfrak{D}}_{\bsy{\mathcal{N}}(\epsilon)}$ non-integrable horizontal distributions.~See already Section \ref{sec_intro_divorce_frames_foliation}.
\item \textbf{Variable horizontal distributions}

The one-parameter family of distributions $\bsy{\mathfrak{D}}_{\bsy{\mathcal{N}}(\epsilon)}$ is a family of \emph{variable distributions} relative to the fixed distribution $\mathfrak{D}_{\mathcal{N}_{\text{as}}}$ on $\mathcal{M}^*$.~This, in particular, means that
\begin{equation} \label{intro_variable_horizontal_structures}
\bsy{\mathfrak{D}}_{\bsy{\mathcal{N}}(\epsilon)} \neq \mathfrak{D}_{\mathcal{N}_{\text{as}}}
\end{equation}
for all $|\epsilon|>0$.~The property persists for $|a|=0$.~For a rigorous definition of variable distributions, see Section \ref{sec_family_frames}.
\end{itemize}
The horizontal distributions $\bsy{\mathfrak{D}}_{\bsy{\mathcal{N}}(\epsilon)}$ are \emph{both} non-integrable \emph{and} variable.~We point out that the two properties are independent and could have, in principle, occurred separately.\footnote{One could, for instance, associate to the metrics $\bsy{g}(\epsilon)$ a one-parameter family of null frames $\bsy{\mathcal{N}}(\epsilon)$ such that $\bsy{\mathfrak{D}}_{\bsy{\mathcal{N}}(\epsilon)} = \mathfrak{D}_{\mathcal{N}_{\text{as}}}$ for all $\epsilon$.~Such frames would not, however, satisfy all the properties listed in Section \ref{sec_intro_nonlinear_perturbations}.}~The variability of the distributions $\bsy{\mathfrak{D}}_{\bsy{\mathcal{N}}(\epsilon)}$ gives rise to some technical difficulties when one wants to renormalise the perturbed quantities relative to their corresponding Kerr background quantities (see Section \ref{sec_overview_identification_horizontal_structures}).~We also point out that a choice of frames \emph{adapted} to the fixed differentiable structure would have induced \emph{fixed}, \emph{integrable} horizontal distributions, and thus neither of these two properties would have occurred.

\medskip

In Section \ref{sec_family_frames}, we also define the \emph{frame coefficients} of $\et(\epsilon)$, together with additional frame coefficients which are convenient for the problem.~See already Section \ref{sec_intro_divorce_frames_foliation}.

\subsubsection{Outgoing principal gauge}  \label{sec_overview_gauge_construction}

In Section \ref{sec_gauge_construction}, we \emph{prove} that the form \eqref{intro_outline_family_metrics} of the metrics $\bsy{g}(\epsilon)$ of Section \ref{sec_overview_family_metrics}, together with the additional identities \eqref{intro_outline_gauge_ids}, correspond to a choice of gauge for the family of metrics $\bsy{g}(\epsilon)$, called \emph{outgoing principal gauge}.~See Proposition \ref{prop_gauge_main}.

\medskip

We emphasise that the gauge statement is \emph{nonlinear} and requires a closeness assumption to the Kerr exterior manifold, whereas the metrics $\bsy{g}(\epsilon)$ are \emph{not} required to satisfy the vacuum Einstein equations or to possess any special Petrov-type property for $|\epsilon|>0$.~We also note that the gauge identities depend on the Kerr parameters.

\medskip

The closeness assumption in the gauge statement involves an analogue of condition \eqref{intro_uniform_limit_metrics} (see the inequality \eqref{bulk_uniform_limit_gauge_constr}).~The condition is employed in various steps of the proof of the statement.~Although a $C^0$-version of the condition (i.e.~$|I|=0$) would suffice for some of these steps, we do need the assumption on derivatives of $\bsy{g}(\epsilon)$ to establish that certain geometric quantities in the proof are sufficiently regular (see Section \ref{sec_auxiliary_family_vf}).~The number of derivatives required by the condition is not sharp.

\subsubsection{Frame construction}  \label{sec_overview_frames_construction}

In Section \ref{sec_frames_construction}, we consider the set-up of Section \ref{sec_overview_family_metrics} and prove that the frames $\bsy{\mathcal{N}}(\epsilon)$ introduced in Section \ref{sec_overview_family_frames} are well-defined and satisfy the properties stated.

\medskip

We emphasise that the frame construction is \emph{nonlinear}.~As for the gauge construction of Section \ref{sec_overview_gauge_construction}, the metrics $\bsy{g}(\epsilon)$ are \emph{not} required to satisfy the vacuum Einstein equations or to possess any special Petrov-type property for $|\epsilon|>0$.~We also point out that the frame construction exploits \emph{all} the gauge identities for the metrics $\bsy{g}(\epsilon)$.

\subsection{A morphism between horizontal tensor bundles}  \label{sec_overview_identification_horizontal_structures}

Let us consider the fixed ambient manifold $\mathcal{M}^*$ and the one-parameter family of metrics $\bsy{g}(\epsilon)$ and associated frames $\bsy{\mathcal{N}}(\epsilon)$ on $\mathcal{M}^*$ introduced in Sections \ref{sec_overview_family_metrics}-\ref{sec_overview_family_frames}.~Since the distributions $\bsy{\mathfrak{D}}_{\bsy{\mathcal{N}}(\epsilon)}$ are variable relative to the fixed distribution $\mathfrak{D}_{\mathcal{N}_{\text{as}}}$, the horizontal tensor bundles
\begin{align} \label{overview_variable_tensor_bundles}
&\otimes_k\bsy{\mathfrak{D}}_{\bsy{\mathcal{N}}(\epsilon)}\otimes_q(\bsy{\mathfrak{D}}_{\bsy{\mathcal{N}}(\epsilon)})^{\star} \, , & &\otimes_k\mathfrak{D}_{\mathcal{N}_{\text{as}}}\otimes_q(\mathfrak{D}_{\mathcal{N}_{\text{as}}})^{\star} 
\end{align}
are \emph{distinct} abstract tensor bundles on $\mathcal{M}^*$ for all $|\epsilon|>0$ (see already the relation \eqref{intro_variable_horizontal_structures}).~In particular, $\bsy{\mathfrak{D}}_{\bsy{\mathcal{N}}(\epsilon)}$ $(k,q)$-tensors and $\mathfrak{D}_{\mathcal{N}_{\text{as}}}$ $(k,q)$-tensors are smooth sections of distinct tensor bundles.

\medskip

In Section \ref{sec:geom_compare_tensors}, we prescribe, for each $\epsilon$, a smooth \emph{bundle morphism}\footnote{We remark that, a priori, there is no canonical bundle morphism between the horizontal tensor bundles \eqref{overview_variable_tensor_bundles}.} 
\begin{align*}
\mathfrak{B}_{q}(\epsilon):\otimes_q(\bsy{\mathfrak{D}}_{\bsy{\mathcal{N}}(\epsilon)})^{\star} &\rightarrow \otimes_q(\mathfrak{D}_{\mathcal{N}_{\text{as}}})^{\star} \\[5pt]
(p,\bsy{\varsigma}_p(\epsilon)) &\mapsto (p,\widetilde{\bsy{\varsigma}}_p(\epsilon)) \, .
\end{align*}
For $\epsilon=0$, the bundle morphism $\mathfrak{B}_{q}(0)$ coincides with the \emph{identity map} over $\otimes_q(\mathfrak{D}_{\mathcal{N}_{\text{as}}})^{\star}$.~For the tensor bundle $\otimes_q(\bsy{\mathfrak{D}}_{\bsy{\mathcal{N}}(\epsilon)})^{\star}$ equipped with the metric $\bsy{\slashed{g}}^{-1}(\epsilon)$, the bundle morphism $\mathfrak{B}_{q}(\epsilon)$ is an \emph{isometry} for all $\epsilon$ (see Section \ref{sec_isometry_horizontal_bundles}).~See Figure \ref{fig:id_horizontal_structures}.

\medskip

\begin{figure}[H]

\centering

\begin{tikzpicture}

\draw[->] (-1,1)--(1,1);

\draw[->] (-2,0.7)--(-0.3,-0.7);

\draw[->] (2,0.7)--(0.3,-0.7);

\node at (-2,1) {$\otimes_q(\bsy{\mathfrak{D}}_{\bsy{\mathcal{N}}(\epsilon)})^{\star}$};

\node at (2,1) {$\otimes_q(\mathfrak{D}_{\mathcal{N}_{\text{as}}})^{\star}$};

\node at (0,-1) {$\mathcal{M}^*$};

\node at (0,1.25) {$\mathfrak{B}_{q}(\epsilon)$};

\node at (-1.8,0) {$\bsy{\pi}(\epsilon)$};

\node at (1.5,0) {$\pi$};

\end{tikzpicture}

\caption{Commutation diagram of the bundle morphism $\mathfrak{B}_{q}(\epsilon)$.~The maps $\bsy{\pi}(\epsilon)$ and $\pi$ are the standard projection maps of the respective tensor bundles over the base manifold $\mathcal{M}^*$.}

\label{fig:id_horizontal_structures}

\end{figure}

\medskip

The bundle morphism $\mathfrak{B}_{q}(\epsilon)$ induces an isometry between \emph{sections} of tensor bundles, i.e.~between $\bsy{\mathfrak{D}}_{\bsy{\mathcal{N}}(\epsilon)}$ $q$-tensors and $\mathfrak{D}_{\mathcal{N}_{\text{as}}}$ $q$-tensors.~In loose terms, the morphism $\mathfrak{B}_{1}(\epsilon)$ maps any $\bsy{\mathfrak{D}}_{\bsy{\mathcal{N}}(\epsilon)}$ one-tensor $\bsy{\varsigma}(\epsilon)$ to a $\mathfrak{D}_{\mathcal{N}_{\text{as}}}$ one-tensor $\widetilde{\bsy{\varsigma}}(\epsilon)$ obtained by first extending $\bsy{\varsigma}(\epsilon)$ to a spacetime one-tensor $\bsy{\varsigma}^{\circ}(\epsilon)\in \Gamma((T\mathcal{M}^*)^{\star})$ such that $\bsy{\varsigma}^{\circ}(\ef(\epsilon))=\bsy{\varsigma}^{\circ}(\et(\epsilon))=0$ and then projecting (and restricting) $\bsy{\varsigma}^{\circ}(\epsilon)$ onto $(\mathfrak{D}_{\mathcal{N}_{\text{as}}})^{\star}$.~Higher order $\bsy{\mathfrak{D}}_{\bsy{\mathcal{N}}(\epsilon)}$ $q$-tensors are mapped accordingly.~We point out that, when applied to the $\bsy{\mathfrak{D}}_{\bsy{\mathcal{N}}(\epsilon)}$ one-tensor $\underline{\bsy{\eta}}(\epsilon)$, the morphism $\mathfrak{B}_{1}(\epsilon)$ yields the $\mathfrak{D}_{\mathcal{N}_{\text{as}}}$ one-tensor $\widetilde{\underline{\bsy{\eta}}}(\epsilon)$ appearing in the identity \eqref{overview_i_4_kerr}.

\medskip

The geometric formalism of Section \ref{sec:geom_compare_tensors} is \emph{nonlinear} and does \emph{not} assume that the metrics $\bsy{g}(\epsilon)$ solve the vacuum Einstein equations.~In fact, in Section \ref{sec:geom_compare_tensors} we only treat general $\bsy{\mathfrak{D}}_{\bsy{\mathcal{N}}(\epsilon)}$ tensors, without identifying them with the connection coefficients and curvature components of the metrics $\bsy{g}(\epsilon)$ relative to the frames $\bsy{\mathcal{N}}(\epsilon)$.~Section \ref{sec:geom_compare_tensors} is the most technical part of the paper and contains an additional overview to its content.~See Section \ref{sec_overview_geom_compare_tensors}.

\subsection{The renormalised vacuum Einstein equations}  \label{sec_overview_renormalised_EVE}

Let us consider the fixed ambient manifold $\mathcal{M}^*$ and the one-parameter family of metrics $\bsy{g}(\epsilon)$ and associated frames $\bsy{\mathcal{N}}(\epsilon)$ on $\mathcal{M}^*$ introduced in Sections \ref{sec_overview_family_metrics}-\ref{sec_overview_family_frames}.~In Section \ref{sec_renormalised_vacuum_Einstein_equations}, we assume that the metrics $\bsy{g}(\epsilon)$ solve the vacuum Einstein equations 
\begin{equation*}
\bsy{\ric}(\bsy{g}(\epsilon))=0
\end{equation*}
for all $\epsilon$.~The nonlinear system of null structure and Bianchi equations for the connection coefficients and curvature components of the metrics $\bsy{g}(\epsilon)$ relative to the frames $\bsy{\mathcal{N}}(\epsilon)$ is a nonlinear system of equations for $\bsy{\mathfrak{D}}_{\bsy{\mathcal{N}}(\epsilon)}$ tensors and coincides, for each $\epsilon$, with the system of equations described in Section \ref{sec_overview_system_EVE}.

\medskip

For later convenience, we \emph{renormalise} the vacuum Einstein equations for the perturbations relative to the Kerr background quantities \emph{in geometric form}.~To this end, one needs to take \emph{geometric differences} of $\bsy{\mathfrak{D}}_{\bsy{\mathcal{N}}(\epsilon)}$ $q$-tensors ($q=1,2$) and $\mathfrak{D}_{\mathcal{N}_{\text{as}}}$ $q$-tensors.~However, since $\bsy{\mathfrak{D}}_{\bsy{\mathcal{N}}(\epsilon)}$ tensors and $\mathfrak{D}_{\mathcal{N}_{\text{as}}}$ tensors are smooth sections of distinct tensor bundles, such differences cannot be taken directly.

\medskip

To circumvent this technical issue, we exploit the formalism discussed in Section \ref{sec_overview_identification_horizontal_structures}.~We first apply the bundle morphism of Section \ref{sec_overview_identification_horizontal_structures} to each of the equations for $\bsy{\mathfrak{D}}_{\bsy{\mathcal{N}}(\epsilon)}$ $q$-tensors in the system, transforming each of the terms in the equations into a $\mathfrak{D}_{\mathcal{N}_{\text{as}}}$ $q$-tensor.~The new nonlinear system of null structure and Bianchi equations for $\mathfrak{D}_{\mathcal{N}_{\text{as}}}$ tensors obtained is dubbed \emph{reduced} system of vacuum Einstein equations and is presented \emph{in geometric form} in Section \ref{sec_restricted_vacuum_Einstein_equations}.

\medskip

With the reduced system of equations at hand, one can now derive the nonlinear system of equations in renormalised form.~One directly takes \emph{geometric differences} between the reduced equations and the corresponding equations for the Kerr background quantities, \emph{all the equations being equations for $\mathfrak{D}_{\mathcal{N}_{\text{as}}}$ tensors}.~As for the reduced system of equations, the renormalised system of vacuum Einstein equations is a system for $\mathfrak{D}_{\mathcal{N}_{\text{as}}}$ tensors, dubbed \emph{renormalised quantities}.~The complete derivation of the renormalised system of equations is left to the reader.

\subsection{The linearisation procedure}  \label{sec_overview_linearisation}

In Section \ref{sec_linearisation_procedure}, we perform the linearisation of general $\bsy{\mathfrak{D}}_{\bsy{\mathcal{N}}(\epsilon)}$ tensors (and their derivatives) around $\epsilon=0$.~The linearisation of a $\bsy{\mathfrak{D}}_{\bsy{\mathcal{N}}(\epsilon)}$ $q$-tensor $\bsy{\varsigma}(\epsilon)$ is the $\mathfrak{D}_{\mathcal{N}_{\text{as}}}$ $q$-tensor $$\overset{\text{\scalebox{.6}{$(1)$}}}{\varsigma}$$ obtained by linearising the geometric difference
\begin{equation*}
\widetilde{\bsy{\varsigma}}(\epsilon)-\bsy{\varsigma}(0)=\epsilon\cdot \overset{\text{\scalebox{.6}{$(1)$}}}{\varsigma}+\mathcal{O}(\epsilon^2)
\end{equation*}
around $\epsilon=0$, with $\widetilde{\bsy{\varsigma}}(\epsilon)$ the $\mathfrak{D}_{\mathcal{N}_{\text{as}}}$ tensor image of $\bsy{\varsigma}(\epsilon)$ via the bundle morphism of Section \ref{sec_overview_identification_horizontal_structures}.~All the linearised quantities are global, \emph{regular} $\mathfrak{D}_{\mathcal{N}_{\text{as}}}$ tensors on the whole $\mathcal{M}^*$, including on $\mathcal{H}^+\cap\mathcal{M}^*$, and all the linearisation formulae are presented for general geometric quantities and \emph{in geometric form}.~Some residual technical difficulties in the linearisation procedure are to be addressed.~These are outlined in Section \ref{sec_overview_linearisation_procedure}, which serves as an overview to the section.

\medskip

The general linearisation formulae are then applied to linearise the connection coefficients and curvature components of the metrics $\bsy{g}(\epsilon)$ relative to the frames $\bsy{\mathcal{N}}(\epsilon)$.~See Section \ref{sec_ids_special_lin_quantities}.

\subsection{The system of linearised gravity} \label{sec_overview_linearised_equations}

In Section \ref{sec_system_linearised_gravity}, we present the system of linearised gravity obtained by applying the linearisation formulae of Section \ref{sec_overview_linearisation} to linearise, in geometric form, the renormalised vacuum Einstein equations.~The final system of linearised equations can be handled without knowledge about its derivation or reference to the nonlinear theory.

\medskip

The system of linearised gravity is a system of equations for global, \emph{regular} $\mathfrak{D}_{\mathcal{N}_{\text{as}}}$ tensors on the whole $\mathcal{M}^*$, including on $\mathcal{H}^+\cap\mathcal{M}^*$, and is presented \emph{in geometric form} in Sections \ref{sec_lin_eqns_frame_coeff}-\ref{sec_lin_eqns_Bianchi}.

\medskip

Appendix \ref{sec_hierarchy_system} presents the key structural properties of the system of linearised gravity.~In addition to the features described therein, we mention that the system exhibits terms reflecting the variability of the distributions $\bsy{\mathfrak{D}}_{\bsy{\mathcal{N}}(\epsilon)}$ employed in its derivation.~This is, for instance, the case of the last two terms in the linearised Codazzi equation
\begin{equation*}
\slashed{\text{div}}\,\overset{\text{\scalebox{.6}{$(1)$}}}{\widehat{\chi}}+\ldots = \frac{1}{2}\,\nablasl(\overset{\text{\scalebox{.6}{$(1)$}}}{\text{tr}\chi})-\frac{1}{2}\,{}^{\star}\nablasl(\overset{\text{\scalebox{.6}{$(1)$}}}{\slashed{\varepsilon}\cdot\chi}) - \overset{\text{\scalebox{.6}{$(1)$}}}{\beta} +\frac{1}{2}\,(\nablasl_3(\text{tr}\chi))\,\overset{\text{\scalebox{.6}{$(1)$}}}{\slashed{\mathfrak{f}}}_4+\frac{1}{2}\,(\nablasl_4(\text{tr}\chi))\,\overset{\text{\scalebox{.6}{$(1)$}}}{\slashed{\mathfrak{f}}}_3 + \ldots \, . \nonumber
\end{equation*}  
Consistently with the fact that the variability of the distributions $\bsy{\mathfrak{D}}_{\bsy{\mathcal{N}}(\epsilon)}$ persists for $|a|=0$, these terms survive for $|a|=0$.

\section{The vacuum Einstein equations} \label{sec_nonlinear_EVE}

Let $(\bsy{\mathcal{M}},\bsy{g})$ be a $3+1$-dimensional, smooth, time-oriented Lorentzian manifold. 

\medskip

In this section and throughout the paper, we adopt the bolded notation of \cite{DHR} in the nonlinear setting.

\subsection{Preliminary definitions} \label{sec_preliminary_defns}

We give some preliminary definitions.

\medskip

\begin{definition}[Null frame] \label{def_null_frame}
A \emph{null frame} $$\bsy{\mathcal{N}}=(\eo,\etw,\et,\ef)$$ on $(\bsy{\mathcal{M}},\bsy{g})$ is a frame on $\bsy{\mathcal{M}}$ such that the identities
\begin{align} 
\bsy{g}(\ea,\eb)&=\delta_{\bsy{AB}}\, , & \bsy{g}(\et,\ef)&=-2 \, , \label{cond_null_frame_1}\\ 
\bsy{g}(\et,\et)=\bsy{g}(\ef,\ef)&= 0 \, , &  \bsy{g}(\ea,\et)=\bsy{g}(\ea,\ef)&=0    \label{cond_null_frame_2}
\end{align}
hold, with $\bsy{A},\bsy{B}=\left\lbrace\bsy{1},\bsy{2}\right\rbrace$.
\end{definition}

\medskip

\begin{remark}
In Definition \ref{def_null_frame}, the frame vector fields $(\eo,\etw)$ are assumed to be \emph{orthonormal}.
\end{remark}

\medskip

\begin{remark}
A \emph{local} null frame on $(\bsy{\mathcal{M}},\bsy{g})$ is a \emph{local} frame on $\bsy{\mathcal{M}}$ such that the identities \eqref{cond_null_frame_1}-\eqref{cond_null_frame_2} hold.
\end{remark}

\medskip

\begin{definition}[Horizontal distribution] \label{def_horizontal_distribution}
The (local) \emph{horizontal distribution} associated to a (local) null frame $\bsy{\mathcal{N}}$ is the (local) distribution $$\bsy{\mathfrak{D}}_{\bsy{\mathcal{N}}}:=\left\langle \et,\ef \right\rangle^{\bsy{\perp}}$$ on $\bsy{\mathcal{M}}$. 
\end{definition}

\medskip

\begin{remark} \label{rmk_domain_horizontal_distribution}
Strictly speaking, the horizontal distribution is associated to the pair of null vector fields $(\et,\ef)$ (as opposed to the full frame $\bsy{\mathcal{N}}$, as phrased for future convenience in Definition \ref{def_horizontal_distribution}).~In fact, null frames that only differ by the choice of orthonormal vector fields $(\eo,\etw)$ induce the same horizontal distribution.~We also emphasise that the domain of $\bsy{\mathcal{M}}$ on which the horizontal distribution is well-defined coincides with that of the null pair $(\et,\ef)$, and may in principle be larger than the domain on which the null pair $(\et,\ef)$ can be completed to a null frame.~Indeed, in Section \ref{sec_Kerr_exterior_manifold} we will deal with a null pair which is \emph{globally} well-defined on the manifold considered, but can only \emph{locally} be completed to a null frame.~In such a scenario, the associated horizontal distribution will be \emph{global}.
\end{remark}

\medskip

\begin{remark}
The horizontal distribution $\bsy{\mathfrak{D}}_{\bsy{\mathcal{N}}}$ can be regarded as an abstract tensor bundle on $\bsy{\mathcal{M}}$ or, alternatively, as a sub-bundle $\bsy{\mathfrak{D}}_{\bsy{\mathcal{N}}}\subset T\bsy{\mathcal{M}}$ on $\bsy{\mathcal{M}}$ via the inclusion map $$\iota: \bsy{\mathfrak{D}}_{\bsy{\mathcal{N}}}\hookrightarrow T\bsy{\mathcal{M}} \, .$$ Such a distinction will be exploited in Definitions \ref{def_canonical_ext_v_field} and \ref{def_canonical_ext_cov_tensor}.
\end{remark}

\medskip

Crucially, the horizontal distribution $\bsy{\mathfrak{D}}_{\bsy{\mathcal{N}}}$ is allowed to be a \emph{non-integrable} distribution.\footnote{We recall that an $n$-dimensional \emph{integrable} distribution is a distribution which admits a (local) basis $\left\lbrace\bsy{X_1},\ldots,\bsy{X_n}\right\rbrace$ spanning all the Lie brackets $[\bsy{X_i},\bsy{X_j}]$, with $1\leq \bsy{i},\bsy{j}\leq n$.~A distribution is \emph{non-integrable} if it fails to be integrable.}

\medskip

\begin{definition}[Non-integrable null frame] \label{def_non_integrable_frame}
A \emph{non-integrable} null frame $\bsy{\mathcal{N}}$ is a null frame whose associated horizontal distribution $\bsy{\mathfrak{D}}_{\bsy{\mathcal{N}}}$ is a non-integrable distribution.
\end{definition}

\medskip

\begin{remark}
The vector fields $(\eo,\etw)$ form a local orthonormal basis of $\bsy{\mathfrak{D}}_{\bsy{\mathcal{N}}}\subset T\bsy{\mathcal{M}}$.~By Definition \ref{def_non_integrable_frame}, the null frame $\bsy{\mathcal{N}}$ is a non-integrable frame if and only if, for any choice of $(\eo,\etw)$, at least one of the relations
\begin{align*}
\bsy{g}([\eo,\etw],\ef)&\neq 0 \, , &  \bsy{g}([\eo,\etw],\et)&\neq 0
\end{align*}
holds.
\end{remark}

\medskip

We now define the main geometric objects of the paper, namely \emph{$\bsy{\mathfrak{D}}_{\bsy{\mathcal{N}}}$ tensors}.~In the following definition, the horizontal distribution $\bsy{\mathfrak{D}}_{\bsy{\mathcal{N}}}$ is regarded as an abstract tensor bundle on $\bsy{\mathcal{M}}$. 

\medskip

\begin{definition}[$\bsy{\mathfrak{D}}_{\bsy{\mathcal{N}}}$ tensors] \label{def_dn_tensor}
A $\bsy{\mathfrak{D}}_{\bsy{\mathcal{N}}}$ \emph{$(k,q)$-tensor field} $\bsy{\varsigma}$ on $\bsy{\mathcal{M}}$, with $k,q\in\mathbb{N}_0$, is a smooth section
\begin{equation*}
\bsy{\varsigma}\in \Gamma(\otimes_k \bsy{\mathfrak{D}}_{\bsy{\mathcal{N}}}\otimes_q(\bsy{\mathfrak{D}}_{\bsy{\mathcal{N}}})^{\bsy{\star}}) \, .
\end{equation*}
In particular, a $\bsy{\mathfrak{D}}_{\bsy{\mathcal{N}}}$ \emph{vector field} $\bsy{X}$ on $\bsy{\mathcal{M}}$ is a smooth section
\begin{equation*}
\bsy{X}\in \Gamma(\bsy{\mathfrak{D}}_{\bsy{\mathcal{N}}}) \, .
\end{equation*}
\end{definition}

\medskip

\begin{remark}
The domain of $\bsy{\mathcal{M}}$ on which a $\bsy{\mathfrak{D}}_{\bsy{\mathcal{N}}}$ tensor is well-defined coincides with that of the horizontal distribution $\bsy{\mathfrak{D}}_{\bsy{\mathcal{N}}}$, and may in principle be larger than the domain on which the null frame $\bsy{\mathcal{N}}$ can be defined.~See the related Remark \ref{rmk_domain_horizontal_distribution}.
\end{remark}

\medskip

\begin{remark}
We often refer to $\bsy{\mathfrak{D}}_{\bsy{\mathcal{N}}}$ $(0,q)$-tensor fields as $\bsy{\mathfrak{D}}_{\bsy{\mathcal{N}}}$ \emph{$q$-tensors}.
\end{remark}

\medskip

Let $\iota_k$ be the \emph{inclusion map}
\begin{equation*}
\iota_k:\otimes_k \bsy{\mathfrak{D}}_{\bsy{\mathcal{N}}} \hookrightarrow \otimes_k T\bsy{\mathcal{M}} \, ,
\end{equation*}
with $k\in\mathbb{N}$.~We denote $\iota_{k=1}$ by $\iota$.

\medskip

For later convenience, we now define how $\bsy{\mathfrak{D}}_{\bsy{\mathcal{N}}}$ tensors can be canonically extended to spacetime tensors.~We start with $\bsy{\mathfrak{D}}_{\bsy{\mathcal{N}}}$ vector fields.

\medskip

\begin{definition}[Canonical extension of $\bsy{\mathfrak{D}}_{\bsy{\mathcal{N}}}$ vector fields] \label{def_canonical_ext_v_field}
Let $\bsy{X}$ be a $\bsy{\mathfrak{D}}_{\bsy{\mathcal{N}}}$ vector field on $\bsy{\mathcal{M}}$.~The \emph{canonical extension of $\bsy{X}$} on $\bsy{\mathcal{M}}$ is the smooth section $$\bsy{X}^{\circ}\in \Gamma(T\bsy{\mathcal{M}})$$ such that $$(p,\bsy{X}_p^{\circ})=\iota(p,\bsy{X}_p)$$ for all $(p,\bsy{X}_p^{\circ})\in T\bsy{\mathcal{M}}$.
\end{definition}

\medskip

\begin{remark} \label{rmk_ids_canonical_ext_vf}
Let $\bsy{X}^{\circ}$ be the canonical extension of a $\bsy{\mathfrak{D}}_{\bsy{\mathcal{N}}}$ vector field $\bsy{X}$ on $\bsy{\mathcal{M}}$.~Then, by Definitions \ref{def_horizontal_distribution} and \ref{def_canonical_ext_v_field}, the identities
\begin{align*}
 \bsy{g}(\bsy{X}^{\circ},\ef)&=0 \, , & \bsy{g}(\bsy{X}^{\circ},\et)&=0  
\end{align*}
hold.
\end{remark}

\medskip

\begin{remark} \label{rmk_abuse_notation_canonical_ext_frame}
With a slight abuse of notation, the frame vector fields $(\eo,\etw)$ will be treated both as $\bsy{\mathfrak{D}}_{\bsy{\mathcal{N}}}$ vector fields on $\bsy{\mathcal{M}}$ and as their canonical extensions.
\end{remark}

\medskip

We now define the canonical extension of higher-rank $\bsy{\mathfrak{D}}_{\bsy{\mathcal{N}}}$ $(k,0)$-tensors.~For $k=1$, the following definition and remark reduce to Definition \ref{def_canonical_ext_v_field} and Remark \ref{rmk_ids_canonical_ext_vf} respectively.

\medskip

\begin{definition}[Canonical extension of $\bsy{\mathfrak{D}}_{\bsy{\mathcal{N}}}$ $(k,0)$-tensors] \label{def_canonical_ext_contrav_tensor}
Let $\bsy{\varsigma}$ be a $\bsy{\mathfrak{D}}_{\bsy{\mathcal{N}}}$ $(k,0)$-tensor on $\bsy{\mathcal{M}}$.~The \emph{canonical extension of $\bsy{\varsigma}$} on $\bsy{\mathcal{M}}$ is the smooth section $$\bsy{\varsigma}^{\circ} \in \Gamma(\otimes_k T\bsy{\mathcal{M}})$$ such that $$(p,\bsy{\varsigma}^{\circ}_p)=\iota_k (p,\bsy{\varsigma}_p)$$ for all $(p,\bsy{\varsigma}^{\circ}_p)\in \otimes_k T\bsy{\mathcal{M}}$.
\end{definition}

\medskip

\begin{remark}
Let $\bsy{\varsigma}^{\circ}$ be the canonical extension of a $\bsy{\mathfrak{D}}_{\bsy{\mathcal{N}}}$ $(k,0)$-tensor $\bsy{\varsigma}$ on $\bsy{\mathcal{M}}$.~Then, by Definitions \ref{def_horizontal_distribution} and \ref{def_canonical_ext_contrav_tensor}, the identities
\begin{align*}
 \bsy{\varsigma}^{\circ}(\ldots,\bsy{g}(\ef,\cdot),\ldots)&=0 \, , & \bsy{\varsigma}^{\circ}(\ldots,\bsy{g}(\et,\cdot),\ldots)&=0
\end{align*}
hold.
\end{remark}

\medskip

We now define the canonical extension of $\bsy{\mathfrak{D}}_{\bsy{\mathcal{N}}}$ $(0,q)$-tensors.

\medskip

\begin{definition}[Canonical extension of $\bsy{\mathfrak{D}}_{\bsy{\mathcal{N}}}$ $(0,q)$-tensors] \label{def_canonical_ext_cov_tensor}
Let $\bsy{\varsigma}$ be a $\bsy{\mathfrak{D}}_{\bsy{\mathcal{N}}}$ $(0,q)$-tensor on $\bsy{\mathcal{M}}$.~The \emph{canonical extension of $\bsy{\varsigma}$} on $\bsy{\mathcal{M}}$ is the smooth section $$\bsy{\varsigma}^{\circ} \in \Gamma(\otimes_q(T\bsy{\mathcal{M}})^{\star})$$ such that, for any $\bsy{\mathfrak{D}}_{\bsy{\mathcal{N}}}$ vector fields $\bsy{X_1},\ldots,\bsy{X_q}$, the identities
\begin{align*}
\bsy{\varsigma}^{\circ}(\bsy{X}_{\bsy{1}}^{\circ},\ldots,\bsy{X}_{\bsy{q}}^{\circ})&=\bsy{\varsigma}(\bsy{X_1},\ldots,\bsy{X_q}) \, , & \bsy{\varsigma}^{\circ}(\ldots,\ef,\ldots)&=\bsy{\varsigma}^{\circ}(\ldots,\et,\ldots)=0 
\end{align*}
hold.
\end{definition}

\medskip

We conclude the section with a series of remarks.

\medskip

\begin{remark}
The canonical extension of a full-rank $\bsy{\mathfrak{D}}_{\bsy{\mathcal{N}}}$ tensor is \emph{not} a full-rank tensor.
\end{remark}

\medskip

\begin{remark} \label{rmk_choice_canonical_extension}
The canonical extension of $\bsy{\mathfrak{D}}_{\bsy{\mathcal{N}}}$ $(0,q)$-tensors, as defined in Definition \ref{def_canonical_ext_cov_tensor}, differs from the extension induced by the inclusion $\otimes_q(\bsy{\mathfrak{D}}_{\bsy{\mathcal{N}}})^{\star}\hookrightarrow \otimes_q(T\bsy{\mathcal{M}})^{\star}$.\footnote{The convenience of our Definition \ref{def_canonical_ext_cov_tensor} of canonical extension of $\bsy{\mathfrak{D}}_{\bsy{\mathcal{N}}}$ $(0,q)$-tensors will become apparent.~See already Remark \ref{rmk_ext_cov_tensor_product} for an example.}~We give a concrete example to elucidate this fact.~We consider the metric $\bsy{g}$ in local double-null form $\bsy{g}=-4\,\bsy{\Omega}^2 d\bsy{u}\,d\bsy{v}+\bsy{\slashed{\gamma}}_{\bsy{\theta^A}\bsy{\theta^B}}(d\bsy{\theta^A}-\bsy{b^{\theta^A}}d\bsy{v})(d\bsy{\theta^B}-\bsy{b^{\theta^B}}d\bsy{v})$ (see \cite{DHRT}) and the local null frame $\bsy{\mathcal{N}}=(\ef,\et,\ea)=(\partial_{\bsy{v}}+\bsy{b^{\theta^A}}\partial_{\bsy{\theta^A}},\bsy{\Omega}^{-2}\partial_{\bsy{u}},\ea)$.~Locally, one has $\bsy{\mathfrak{D}}_{\bsy{\mathcal{N}}}\equiv T\mathbb{S}^2_{\bsy{u},\bsy{v}}$, with $\mathbb{S}^2_{\bsy{u},\bsy{v}}:=\left\lbrace \bsy{u},\bsy{v} \right\rbrace\times\mathbb{S}^2$.~We consider the $\bsy{\mathfrak{D}}_{\bsy{\mathcal{N}}}$ one-tensor $$\bsy{\varsigma}=d\bsy{\theta^A} \, .$$ It is easy to check that the canonical extension of $\bsy{\varsigma}$, as defined in Definition \ref{def_canonical_ext_cov_tensor}, reads $$\bsy{\varsigma}^{\circ}=d\bsy{\theta^A}-\bsy{b^{\theta^A}}d\bsy{v} \, .$$ The extension of $\bsy{\varsigma}$ induced by the inclusion $(\bsy{\mathfrak{D}}_{\bsy{\mathcal{N}}})^{\star}\hookrightarrow (T\bsy{\mathcal{M}})^{\star}$ is $d\bsy{\theta^A}\in\Gamma((T\bsy{\mathcal{M}})^{\star})$, which manifestly differs from $\bsy{\varsigma}^{\circ}$.~In particular, we have $d\bsy{\theta^A}(\ef)=\bsy{b^{\theta^A}}\neq 0$.
\end{remark}

\medskip

\begin{remark} \label{rmk_horizontal_covariant_tensors}
In Section \ref{sec:geom_compare_tensors}, any $\bsy{\varsigma} \in \Gamma(\otimes_q(T\bsy{\mathcal{M}})^{\star})$ such that $\bsy{\varsigma}(\ldots,\ef,\ldots)=\bsy{\varsigma}(\ldots,\et,\ldots)=0$ will be referred to as a \emph{$\bsy{\mathfrak{D}}_{\bsy{\mathcal{N}}}$-horizontal $(0,q)$-tensor}.~Similarly, any $\bsy{\varsigma} \in \Gamma(\otimes_k T\bsy{\mathcal{M}})$ such that $\bsy{\varsigma}(\ldots,\bsy{g}(\ef,\cdot),\ldots)=\bsy{\varsigma}(\ldots,\bsy{g}(\et,\cdot),\ldots)=0$ will be referred to as a \emph{$\bsy{\mathfrak{D}}_{\bsy{\mathcal{N}}}$-horizontal $(k,0)$-tensor}.~Using this terminology, the canonical extension of a $\bsy{\mathfrak{D}}_{\bsy{\mathcal{N}}}$ $(0,q)$-tensor is a $\bsy{\mathfrak{D}}_{\bsy{\mathcal{N}}}$-horizontal $(0,q)$-tensor.~Similarly, the canonical extension of a $\bsy{\mathfrak{D}}_{\bsy{\mathcal{N}}}$ $(k,0)$-tensor is a $\bsy{\mathfrak{D}}_{\bsy{\mathcal{N}}}$-horizontal $(k,0)$-tensor.~Such a terminology is not adopted anywhere else in the paper.
\end{remark}

\medskip

\begin{remark}
All the definitions in this section are manifestly independent of the choice of frame vector fields $(\eo,\etw)$.
\end{remark}

\subsection{Induced metric, connection coefficients and curvature components} \label{sec_ricci_curvature_defns}

Let $\bsy{\nabla}$ be the Levi-Civita connection associated to $\bsy{g}$ on $\bsy{\mathcal{M}}$, $\bsy{R}$ the Riemann curvature tensor on $(\bsy{\mathcal{M}},\bsy{g})$ and $\bsy{\varepsilon}$ the standard volume form associated to $\bsy{g}$ on $\bsy{\mathcal{M}}$.~Let $\bsy{\mathcal{N}}$ be a local null frame on $(\bsy{\mathcal{M}},\bsy{g})$, according to Definition \ref{def_null_frame}.~All the definitions in this section are independent of the choice of frame vector fields $(\eo,\etw)$.~The time-orientation of $(\bsy{\mathcal{M}},\bsy{g})$ is such that $\bsy{\varepsilon}(\eo,\etw,\et,\ef)=2$.

\medskip

We introduce the $\bsy{\mathfrak{D}}_{\bsy{\mathcal{N}}}$ covariant tensor fields
\begin{equation*}
\bsy{\slashed{g}} \, , \,  \bsy{\slashed{\varepsilon}}
\end{equation*}
such that\footnote{Recall the abuse of notation noted in Remark \ref{rmk_abuse_notation_canonical_ext_frame}, which appears here and throughout the section.~To be consistent with the notation of Definition \ref{def_canonical_ext_v_field}, one would consider $(\eo,\etw)$ as $\bsy{\mathfrak{D}}_{\bsy{\mathcal{N}}}$ vector fields and, for instance, write $\bsy{\slashed{g}}(\ea,\eb)=\bsy{g}(\bsy{e}_{\bsy{A}}^{\circ},\bsy{e}_{\bsy{B}}^{\circ})$.}
\begin{align*}
\bsy{\slashed{g}}(\ea,\eb)&=\bsy{g}(\ea,\eb) \, , &  \bsy{\slashed{\varepsilon}}(\ea,\eb)&=\frac{1}{2}\,\bsy{\varepsilon}(\ea,\eb,\et,\ef) \, .
\end{align*}
Informally, we refer to $\bsy{\slashed{g}}$ as the \emph{induced metric} on $\bsy{\mathfrak{D}}_{\bsy{\mathcal{N}}}$ and to $\bsy{\slashed{\varepsilon}}$ as the \emph{induced volume form} on $\bsy{\mathfrak{D}}_{\bsy{\mathcal{N}}}$.

\medskip

\begin{remark}
For any $\bsy{\mathfrak{D}}_{\bsy{\mathcal{N}}}$ vector field $\bsy{X}$, we have the $\bsy{\mathfrak{D}}_{\bsy{\mathcal{N}}}$ one-tensor
\begin{equation*}
 \bsy{\slashed{g}}(\bsy{X},\cdot)   
\end{equation*}
and the identity
\begin{equation*}
(\bsy{\slashed{g}}(\bsy{X},\cdot))^{\circ}=\bsy{g}(\bsy{X}^{\circ},\cdot) \, .
\end{equation*}
\end{remark}

\medskip

\begin{remark}
For any $\bsy{\mathfrak{D}}_{\bsy{\mathcal{N}}}$ vector fields $\bsy{X}, \bsy{Y}$, we have the identity
\begin{equation*}
 \bsy{\slashed{g}}(\bsy{X},\bsy{Y})=\bsy{g}(\bsy{X}^{\circ},\bsy{Y}^{\circ})   \, .
\end{equation*}
\end{remark}

\medskip

We define the $\bsy{\mathfrak{D}}_{\bsy{\mathcal{N}}}$ contravariant tensor $$\bsy{\slashed{g}}^{-1}$$ as the inverses of the metric $\bsy{\slashed{g}}$.~One may view $\bsy{\slashed{g}}^{-1}$ as an induced metric on $(\bsy{\mathfrak{D}}_{\bsy{\mathcal{N}}})^{\star}$.~We will adopt the notation
\begin{equation*}
(\bsy{\slashed{g}}^{-1})^{\bsy{AB}}=\bsy{\slashed{g}}^{-1}(\bsy{\slashed{g}}(\ea,\cdot),\bsy{\slashed{g}}(\eb,\cdot)) \, ,
\end{equation*}
where the left hand side will be often written in the more synthetic form $\bsy{\slashed{g}}^{\bsy{AB}}$.~We note that the $\bsy{\mathfrak{D}}_{\bsy{\mathcal{N}}}$ one-tensors $(\bsy{\slashed{g}}(\eo,\cdot),\bsy{\slashed{g}}(\etw,\cdot))$ are the dual co-frame to $(\eo,\etw)$ and form a local \emph{orthonormal} co-frame of $(\bsy{\mathfrak{D}}_{\bsy{\mathcal{N}}})^{\star}$.

\medskip

\begin{remark}
For any $\bsy{\mathfrak{D}}_{\bsy{\mathcal{N}}}$ one-tensor $\bsy{\varsigma}$, we have the $\bsy{\mathfrak{D}}_{\bsy{\mathcal{N}}}$ vector field
\begin{equation*}
\bsy{\slashed{g}}^{-1}(\bsy{\varsigma},\cdot)  \, . 
\end{equation*}
We will write the expression
\begin{equation}
\bsy{\slashed{g}}^{-1}(\bsy{\varsigma},\cdot)=(\bsy{\slashed{g}}^{-1})^{\bsy{AB}}\bsy{\varsigma}(\ea)\, \eb \, . \label{notation_sum_frame_vectors}
\end{equation}
We have the identity
\begin{equation*}
(\bsy{\slashed{g}}^{-1}(\bsy{\varsigma},\cdot))^{\circ}=\bsy{g}^{-1}(\bsy{\varsigma}^{\circ},\cdot) \, .
\end{equation*}
\end{remark}

\medskip

\begin{remark} \label{rmk_ext_cov_tensor_product}
For any $\bsy{\mathfrak{D}}_{\bsy{\mathcal{N}}}$ one-tensors $\bsy{\varsigma}, \bsy{\theta}$, we have the identity
\begin{equation*}
 \bsy{\slashed{g}}^{-1}(\bsy{\varsigma},\bsy{\theta})=\bsy{g}^{-1}(\bsy{\varsigma}^{\circ},\bsy{\theta}^{\circ})   \, .
\end{equation*}
\end{remark}

\medskip

We define the \emph{connection coefficients} relative to the null frame $\bsy{\mathcal{N}}$ as the smooth scalar functions
\begin{equation*}
\bsy{\omegah} \, , \, \bsy{\omegabh}
\end{equation*}
and the $\bsy{\mathfrak{D}}_{\bsy{\mathcal{N}}}$ covariant tensor fields
\begin{gather*}
\bsy{\eta}\, , \, \bsy{\etab} \, , \, \bsy{\xi} \, , \, \bsy{\yb} \, , \, \bsy{\zeta}\, ,  \\
\bsy{\chi}\, , \, \bsy{\chib}  
\end{gather*}
such that, with the \emph{frame-index notation} $\bsy{\varsigma}(\bsy{e_{A_1}},\ldots,\bsy{e_{A_k}})=\bsy{\varsigma_{A_1\cdots A_k}}$, one has
\begin{align*}
\bsy{\chi}_{\bsy{AB}}&=\bsy{g}(\bsy{\nabla_A e_4}, \bsy{e_B})\, ,  &  \bsy{\chib}_{\bsy{AB}}&=\bsy{g}(\bsy{\nabla_A e_3},\bsy{ e_B})\, ,  \\
\bsy{\eta}_{\bsy{A}} &= \frac{1}{2} \, \bsy{g}(\bsy{\nabla_3 e_4},\bsy{ e_A}) \, , & \bsy{\etab}_{\bsy{A}} &= \frac{1}{2} \, \bsy{g}(\bsy{\nabla_4 e_3},\bsy{ e_A})\, , \\
\bsy{\xi}_{\bsy{A}} &= \frac{1}{2} \, \bsy{g}(\bsy{\nabla_4 e_4},\bsy{ e_A}) \, , & \bsy{\yb}_{\bsy{A}} &= \frac{1}{2} \, \bsy{g}(\bsy{\nabla_3 e_3},\bsy{ e_A})\, ,  \\
\bsy{\omegah} &=\frac{1}{2} \, \bsy{g}(\bsy{\nabla_4 e_3} ,\bsy{e_4})\, , &\bsy{ \omegabh} &=\frac{1}{2} \, \bsy{g}(\bsy{\nabla_3 e_4} ,\bsy{e_3}) \, ,   \\ 
\bsy{\zeta}_{\bsy{A}} &=\frac{1}{2} \, \bsy{g}(\bsy{\nabla_A e_4} , \bsy{e_3}) \, .
\end{align*}

\medskip

We define the \emph{curvature components} relative to the null frame $\bsy{\mathcal{N}}$ as the smooth scalar functions
\begin{equation*}
\bsy{\rho} \, , \, \bsy{\sigma} \, ,  
\end{equation*}
and the $\bsy{\mathfrak{D}}_{\bsy{\mathcal{N}}}$ covariant tensor fields
\begin{gather*}
\bsy{\beta} \, , \, \bsy{\betab} \, ,  \\
\bsy{\alpha}\, , \, \bsy{\alphab} \, , 
\end{gather*}
such that
\begin{align*}
\bsy{\alpha_{AB}} &=\bsy{ R}(\ea,\ef,\eb,\ef) \, ,  &   \bsy{\alphab_{AB}}  &= \bsy{R}(\ea,\et,\eb,\et) \, , \\
\bsy{\beta_{A}} &=  \frac{1}{2}\, \bsy{R}(\ea,\ef,\et,\ef)\, ,  &  \bsy{ \betab_{A}}  &= \frac{1}{2}\, \bsy{R}(\ea,\et,\et,\ef) \, ,\\
\bsy{\rho} &= \frac{1}{4}\, \bsy{ R}(\ef,\et,\ef,\et) \, , &  \bsy{\sigma}&=\frac{1}{4}\, \bsy{{}^{\star} R}(\ef,\et,\ef,\et) \, ,
\end{align*}
where $\bsy{{}^{\star} R}$ is the Hodge dual of $\bsy{R}$ on $(\bsy{\mathcal{M}},\bsy{g})$.~In view of the symmetries of $\bsy{R}$, both $\bsy{\alpha}$ and $\bsy{\alphab}$ are \emph{symmetric} $\bsy{\mathfrak{D}}_{\bsy{\mathcal{N}}}$ tensors.

\subsubsection{Decomposition of the second fundamental forms} \label{sec_shears_decompn}

The following proposition relates the symmetries of the $\bsy{\mathfrak{D}}_{\bsy{\mathcal{N}}}$ tensors $\bsy{\chi}$ and $\bsy{\chib}$ and the integrability of the horizontal distribution $\bsy{\mathfrak{D}}_{\bsy{\mathcal{N}}}$.~The proof is an easy check left to the reader.

\medskip

\begin{prop}
The distribution $\bsy{\mathfrak{D}}_{\bsy{\mathcal{N}}}$ is integrable if and only if there exists a choice of $(\eo,\etw)$ such that the identities
\begin{align*}
\bsy{\chi_{AB}} &= \bsy{\chi_{BA}} \, ,  &  \bsy{\chib_{AB}} &= \bsy{\chib_{BA}} 
\end{align*}
hold. In particular,
\begin{align*}
\bsy{g}([\ea,\eb],\ef)= 0 \quad &\Leftrightarrow \quad \bsy{\chi_{AB}}= \bsy{\chi_{BA}} \, ,  \\
\bsy{g}([\ea,\eb],\et)= 0 \quad &\Leftrightarrow \quad \bsy{\chib_{AB}}= \bsy{\chib_{BA}} \, .
\end{align*}
\end{prop}

\medskip

Since the distribution $\bsy{\mathfrak{D}}_{\bsy{\mathcal{N}}}$ is allowed to be non-integrable, the $\bsy{\mathfrak{D}}_{\bsy{\mathcal{N}}}$ tensors $\bsy{\chi}$ and $\bsy{\chib}$ do \emph{not} possess, in general, any symmetry.

\medskip
 
We define the \emph{symmetric} $\bsy{\mathfrak{D}}_{\bsy{\mathcal{N}}}$ tensors 
\begin{equation*}
\bsy{\chi}_{\tiny{\Circle}}\, ,\,\bsy{\chib_{\tiny{\Circle}}} \, ,
\end{equation*}
the \emph{antisymmetric} $\bsy{\mathfrak{D}}_{\bsy{\mathcal{N}}}$ tensors
\begin{equation*}
\bsy{\chi_{\tiny{\Square}}}\, ,\,\bsy{\chib_{\tiny{\Square}}}
\end{equation*}
and the \emph{symmetric traceless} $\bsy{\mathfrak{D}}_{\bsy{\mathcal{N}}}$ tensors
\begin{equation*}
\bsy{\chih} \, , \, \bsy{\chibh}
\end{equation*}
such that
\begin{align*}
{\bsy{\chi}_{\tiny{\Circle}}}_{\bsy{AB}}&=\frac{1}{2}\,( {\bsy{\chi}}_{\bsy{AB}}+{\bsy{\chi}}_{\bsy{BA}}) \, , & 
{\bsy{\chib_{\tiny{\Circle}}}}_{\bsy{AB}}&=\frac{1}{2}\,({\bsy{\chib}}_{\bsy{AB}}+{\bsy{\chib}}_{\bsy{BA}}) \, ,  \\
{\bsy{\chi}_{\tiny{\Square}}}_{\bsy{AB}}&=\frac{1}{2}\,({\bsy{\chi}}_{\bsy{AB}}-{\bsy{\chi}}_{\bsy{BA}}) \, , & 
{\bsy{\chib_{\tiny{\Square}}}}_{\bsy{AB}}&=\frac{1}{2}\,({\bsy{\chib}}_{\bsy{AB}}-{\bsy{\chib}}_{\bsy{BA}})    \, , \\
\bsy{\chih}&=\bsy{\chi}_{\tiny{\Circle}}- \frac{1}{2}\,(\textbf{tr} \bsy{\chi})\,\bsy{\slashed{g}}  \, , &  \bsy{\chibh}&=\bsy{\chib_{\tiny{\Circle}}}- \frac{1}{2}\,(\textbf{tr} \bsy{\chib})\,\bsy{\slashed{g}} \, ,
\end{align*}
where the traces are taken relative to the inverse metric $\bsy{\slashed{g}}^{-1}$, i.e.
\begin{align*}
(\textbf{tr} \bsy{\chi})&=(\bsy{\slashed{g}}^{-1})^{\bsy{AB}}{\bsy{\chi}}_{\bsy{AB}} \, , & (\textbf{tr} \bsy{\chib})&=(\bsy{\slashed{g}}^{-1})^{\bsy{AB}}\bsy{\chib}_{\bsy{AB}} \, .
\end{align*}

\medskip

We define the \emph{antitraces} of $\bsy{\chi}$ and $\bsy{\chib}$ as the smooth scalar functions\footnote{As we shall clarify in the next section, all the frame indices of $\bsy{\mathfrak{D}}_{\bsy{\mathcal{N}}}$ tensors are raised relative to the inverse metric $\bsy{\slashed{g}}^{-1}$, e.g.~$\bsy{\slashed{\varepsilon}}^{\bsy{AB}}=(\bsy{\slashed{g}}^{-1})^{\bsy{AC}}(\bsy{\slashed{g}}^{-1})^{\bsy{BD}}\bsy{\slashed{\varepsilon}}_{\bsy{CD}}$.}
\begin{align*}
(\bsy{\slashed{\varepsilon}\cdot\chi})&=\bsy{\slashed{\varepsilon}^{AB}\chi{}_{AB}}  \, , &  (\bsy{\slashed{\varepsilon}\cdot\chib})&=\bsy{\slashed{\varepsilon}^{AB}\chib{}_{AB}} 
\end{align*}
respectively.

\medskip

\begin{remark}
The antitraces $(\bsy{\slashed{\varepsilon}\cdot\chi})$ and $(\bsy{\slashed{\varepsilon}\cdot\chib})$ encode the (non-)integrability of the horizontal distribution $\bsy{\mathfrak{D}}_{\bsy{\mathcal{N}}}$, meaning that $\bsy{\mathfrak{D}}_{\bsy{\mathcal{N}}}$ is integrable if and only if both the antitraces identically vanish (see later formula \eqref{frame_commutator_AB}).
\end{remark}

\medskip

We decompose the $\bsy{\mathfrak{D}}_{\bsy{\mathcal{N}}}$ tensors $\bsy{\chi}$ and $\bsy{\chib}$ as
\begin{align*}
\bsy{\chi} &= \bsy{\chih}+\frac{1}{2}\,(\textbf{tr} \bsy{\chi})\,\bsy{\slashed{g}}+\frac{1}{2}\,(\bsy{\slashed{\varepsilon}\cdot\chi})\,\bsy{\slashed{\varepsilon}} \, , \\ 
\bsy{\chib} &= \bsy{\chibh}+\frac{1}{2}\,(\textbf{tr} \bsy{\chib})\,\bsy{\slashed{g}}+\frac{1}{2}\,(\bsy{\slashed{\varepsilon}\cdot\chib})\,\bsy{\slashed{\varepsilon}} \, .
\end{align*}

\subsection{Tensor contractions and products}  \label{sec_tensor_contractions_products}

In this section we define contractions and products of $\bsy{\mathfrak{D}}_{\bsy{\mathcal{N}}}$ tensors.~As a general caveat, we note that the $\bsy{\mathfrak{D}}_{\bsy{\mathcal{N}}}$ tensors involved in the expressions are not assumed to possess any symmetry, and thus the order of the frame indices in the formulae is relevant.

\medskip

Given a $\bsy{\mathfrak{D}}_{\bsy{\mathcal{N}}}$ one-tensor $\bsy{\varsigma}$  and a $\bsy{\mathfrak{D}}_{\bsy{\mathcal{N}}}$ two-tensor $\bsy{\theta}$, we introduce the notation $$\bsy{\varsigma}^{\bsy{\sharp}} \, , \, \bsy{\theta}^{\bsy{\sharp_1}} \, , \, \bsy{\theta}^{\bsy{\sharp_2}}$$ such that
\begin{align*}
{\bsy{\varsigma}^{\bsy{\sharp}}}{}^{\bsy{A}}&=\bsy{\slashed{g}^{AB}}\bsy{\varsigma_B} \, , &
{\bsy{\theta}^{\bsy{\sharp_1}}}{}^{\bsy{B}}_{\bsy{A}} &=\bsy{\slashed{g}}^{\bsy{BC}}\bsy{\theta}_{\bsy{CA}} \, , & 
{\bsy{\theta}^{\bsy{\sharp_2}}}{}^{\bsy{B}}_{\bsy{A}}&=\bsy{\slashed{g}}^{\bsy{BC}}\bsy{\theta}_{\bsy{AC}} \, .
\end{align*}

\medskip

\begin{remark}
The $\bsy{\sharp}_{\bsy{I}}$-notation allows to keep track of the position of the frame indices when we write tensorial expressions involving tensors which do not possess any symmetry.~If $\bsy{\theta}$ is symmetric, then $\bsy{\theta}^{\bsy{\sharp_1}}=\bsy{\theta}^{\bsy{\sharp_2}}$ and we simply write $\bsy{\theta}^{\bsy{\sharp}}$. 
\end{remark}

\medskip

We define the full contraction
\begin{equation*}
\bsy{\slashed{\varepsilon}}\bsy{\cdot}\bsy{\theta}=\bsy{\slashed{\varepsilon}}^{\bsy{AB}}\bsy{\theta_{AB}} 
\end{equation*}
and the duality relations 
\begin{align*}
{\bsy{^{\star}\varsigma}}_{\bsy{A}}&=\bsy{{\slashed{\varepsilon}^{\sharp_2}}{}^B_A}\,\bsy{\varsigma_B} \, , & {\bsy{^{\star}\theta}}_{\bsy{AB}}&=\bsy{{\slashed{\varepsilon}^{\sharp_2}}{}^C_A}\,\bsy{\theta_{BC}} \, .
\end{align*} 
By definition, we have ${\bsy{{^{\star}}{^{\star}}\varsigma}}_{\bsy{A}}=-\bsy{\varsigma_A}$.

\medskip

Given the $\bsy{\mathfrak{D}}_{\bsy{\mathcal{N}}}$ one-tensors $\bsy{\varsigma},\bsy{\tilde{\varsigma}}$ and the $\bsy{\mathfrak{D}}_{\bsy{\mathcal{N}}}$ two-tensors $\bsy{\theta},\bsy{\tilde{\theta}}$, we define
\begin{align*}
(\bsy{\varsigma},\bsy{\tilde{\varsigma}})&=\bsy{\slashed{g}^{AB}}\bsy{\varsigma_A}\,\bsy{\tilde{\varsigma}_B} \, , & \bsy{\varsigma} \bsy{\wedge} \bsy{\tilde{\varsigma}}&=\bsy{\slashed{\varepsilon}^{AB}\varsigma_A\,\tilde{\varsigma}_B} 
\end{align*}
and
\begin{align*}
(\bsy{\theta},\bsy{\tilde{\theta}})&=\bsy{\slashed{g}^{AD}}\bsy{\slashed{g}^{CB}\theta_{AC}\,\tilde{\theta}_{BD}} \, , &
\bsy{\theta} \bsy{\wedge} \bsy{\tilde{\theta}}&=\bsy{\slashed{\varepsilon}^{AD}\slashed{g}^{CB}\theta_{AC}\,\tilde{\theta}_{BD}} \, , &
(\bsy{\theta}\bsy{\times} \bsy{\tilde{\theta}})_{\bsy{AB}}&=\bsy{{\theta^{\sharp_2}}{}_{A}^C\,\tilde{\theta}_{CB}} \, .
\end{align*} 
We also have
\begin{align*}
\bsy{\varsigma} \, \widehat{\otimes} \, \bsy{\tilde{\varsigma}} &=(\bsy{\varsigma} \otimes \bsy{\tilde{\varsigma}})+(\bsy{\tilde{\varsigma}} \otimes \bsy{\varsigma})-(\bsy{\varsigma},\bsy{\tilde{\varsigma}})\,\bsy{\slashed{g}} \, .
\end{align*} 
The $\bsy{\mathfrak{D}}_{\bsy{\mathcal{N}}}$ two-tensor $\bsy{\varsigma} \,\widehat{\otimes} \,\bsy{\tilde{\varsigma}}$ is symmetric and traceless relative to $\bsy{\slashed{g}^{-1}}$.

\subsection{Differential operators}  \label{sec_differential_operators}

This section deals with differential operators acting on $\bsy{\mathfrak{D}}_{\bsy{\mathcal{N}}}$ tensors.~As a general caveat, we note that the $\bsy{\mathfrak{D}}_{\bsy{\mathcal{N}}}$ tensors involved in the expressions are not assumed to possess any symmetry, and thus the order of the frame indices in the formulae is relevant.

\subsubsection{Covariant derivative}

The Levi-Civita connection $\bsy{\nabla}$ of $(\bsy{\mathcal{M}},\bsy{g})$ induces a natural connection $\bsy{\nablasl}$ over the bundle of $\bsy{\mathfrak{D}}_{\bsy{\mathcal{N}}}$ tensors on $\bsy{\mathcal{M}}$.~In this section, we shall define the induced connection $\bsy{\nablasl}$ and discuss its two main properties, namely its compatibility with the (already defined) induced metric $\bsy{\slashed{g}}$ on $\bsy{\mathfrak{D}}_{\bsy{\mathcal{N}}}$ and its torsion.~The latter property is novel and is tied to the non-integrability of $\bsy{\mathfrak{D}}_{\bsy{\mathcal{N}}}$.

\medskip

We start with two definitions.

\medskip

\begin{definition} \label{def_formalism_projected_cov_derivative}
Let $\bsy{X}\in \Gamma(T\bsy{\mathcal{M}})$.~We define the map
\begin{equation*}
\bsy{\nablasl}_{\bsy{X}}
\end{equation*}
between $\bsy{\mathfrak{D}}_{\bsy{\mathcal{N}}}$ tensors such that
\begin{itemize}
\setlength\itemsep{5pt}
\item For any smooth scalar function $\bsy{f}$, we have the smooth scalar function $\bsy{\nablasl_X f}$ such that $$\bsy{\nablasl_X f} =\bsy{X}(\bsy{f}) \, ;$$
\item For any $\bsy{\mathfrak{D}}_{\bsy{\mathcal{N}}}$ vector field $\bsy{Y}$, we have the $\bsy{\mathfrak{D}}_{\bsy{\mathcal{N}}}$ vector field $\bsy{\nablasl_X Y}$ such that\footnote{Recall Definition \ref{def_canonical_ext_v_field} and the abuse of notation noted in Remark \ref{rmk_abuse_notation_canonical_ext_frame}.~To be fully consistent with the notation, one would consider $(\eo,\etw)$ as $\bsy{\mathfrak{D}}_{\bsy{\mathcal{N}}}$ vector fields and write $\bsy{\nablasl_X Y} =\bsy{g}(\bsy{\nabla_X Y^{\circ} },\bsy{e}^{\circ}_{\bsy{1}})\,\bsy{e_1}+\bsy{g}(\bsy{\nabla_X Y^{\circ} },\bsy{e}^{\circ}_{\bsy{2}})\,\bsy{e_2}$.~Note also that one can write $\bsy{\nablasl_X Y}=\bsy{\slashed{g}}^{-1}(\bsy{\varsigma},\cdot)$, with $\bsy{\varsigma}$ the $\bsy{\mathfrak{D}}_{\bsy{\mathcal{N}}}$ one-tensor such that $\bsy{\varsigma}(\ea)=\bsy{g}(\bsy{\nabla_X Y^{\circ} },\bsy{e}^{\circ}_{\bsy{A}})$.~This coincides with the definition \eqref{def_proj_cov_deriv_vector} via the formula \eqref{notation_sum_frame_vectors}.} 
\begin{equation}
\bsy{\nablasl_X Y} =\bsy{\slashed{g}}^{\bsy{AB}}\bsy{g}(\bsy{\nabla_X Y^{\circ} },\bsy{e_A})\,\bsy{e_B} \, ; \label{def_proj_cov_deriv_vector}
\end{equation}
\item For any $\bsy{\mathfrak{D}}_{\bsy{\mathcal{N}}}$ $k$-tensor $\bsy{\varsigma}$ and $\bsy{\mathfrak{D}}_{\bsy{\mathcal{N}}}$ vector fields $\bsy{Y_1},\ldots,\bsy{Y_k}$, we have the $\bsy{\mathfrak{D}}_{\bsy{\mathcal{N}}}$ $k$-tensor $\bsy{\nablasl_X \varsigma}$ such that $$(\bsy{\nablasl_X \varsigma})(\bsy{Y_1},\ldots,\bsy{Y_k}) =\bsy{X}(\bsy{\varsigma}(\bsy{Y_1},\ldots,\bsy{Y_k}))-\bsy{\varsigma}(\bsy{\nablasl_X Y_1},\ldots,\bsy{Y_k})-\cdots-\bsy{\varsigma}(\bsy{Y_1},\ldots,\bsy{\nablasl_X Y_k}) \, .$$ 
\end{itemize}
\end{definition}

\medskip

\begin{definition} \label{def_formalism_projected_connection}
We define the map 
\begin{equation*}
\bsy{\nablasl}
\end{equation*}
between $\bsy{\mathfrak{D}}_{\bsy{\mathcal{N}}}$ tensors such that
\begin{itemize}
\setlength\itemsep{5pt}
\item For any smooth scalar function $\bsy{f}$, we have the $\bsy{\mathfrak{D}}_{\bsy{\mathcal{N}}}$ one-tensor $\bsy{\nablasl f}$ such that $$(\bsy{\nablasl f})(\bsy{X}) =\bsy{\nablasl_X f}$$ for any $\bsy{\mathfrak{D}}_{\bsy{\mathcal{N}}}$ vector field $\bsy{X}$;
\item For any $\bsy{\mathfrak{D}}_{\bsy{\mathcal{N}}}$ vector field $\bsy{Y}$, we have the $\bsy{\mathfrak{D}}_{\bsy{\mathcal{N}}}$ one-tensor $\bsy{\nablasl Y}$ such that $$(\bsy{\nablasl Y})(\bsy{X}) =\bsy{\nablasl_X Y}$$ for any $\bsy{\mathfrak{D}}_{\bsy{\mathcal{N}}}$ vector field $\bsy{X}$;
\item For any $\bsy{\mathfrak{D}}_{\bsy{\mathcal{N}}}$ $k$-tensor $\bsy{\varsigma}$, we have the $\bsy{\mathfrak{D}}_{\bsy{\mathcal{N}}}$ $(k+1)$-tensor $\bsy{\nablasl \varsigma}$ such that $$(\bsy{\nablasl \varsigma})(\bsy{X},\bsy{Y_1},\ldots,\bsy{Y_k}) =(\bsy{\nablasl_X \varsigma})(\bsy{Y_1},\ldots,\bsy{Y_k})$$ for any $\bsy{\mathfrak{D}}_{\bsy{\mathcal{N}}}$ vector fields $\bsy{X},\bsy{Y_1},\ldots,\bsy{Y_k}$.
\end{itemize}
\end{definition}

\medskip

\begin{remark}
It is easy to check that Definition \ref{def_formalism_projected_cov_derivative} defines a \emph{covariant derivative} acting on $\bsy{\mathfrak{D}}_{\bsy{\mathcal{N}}}$ tensors (indeed, the map $\bsy{\nablasl}_{\bsy{X}}$ is linear and satisfies the Leibniz rule).~The map $\bsy{\nablasl}$ is a linear \emph{connection} over the bundle of $\bsy{\mathfrak{D}}_{\bsy{\mathcal{N}}}$ tensors.~We refer to $\bsy{\nablasl}_{\bsy{X}}$ as the \emph{induced covariant derivative} over $\bsy{\mathfrak{D}}_{\bsy{\mathcal{N}}}$ tensors and to $\bsy{\nablasl}$ as the \emph{induced connection} over the bundle of $\bsy{\mathfrak{D}}_{\bsy{\mathcal{N}}}$ tensors.
\end{remark}

\medskip

Let $[\cdot,\cdot]$ denote the standard Lie brackets of $(\bsy{\mathcal{M}},\bsy{g})$, i.e.
\begin{equation*}
[\bsy{X},\bsy{Y}](\bsy{f})=\bsy{X}(\bsy{Y}(\bsy{f}))-\bsy{Y}(\bsy{X}(\bsy{f}))
\end{equation*}
for any $\bsy{X},\bsy{Y}\in\Gamma(T\bsy{\mathcal{M}})$ and $\bsy{f}\in C^{\infty}(\bsy{\mathcal{M}})$.~The two following propositions address two key properties of the covariant derivative $\bsy{\nablasl}$, namely its \emph{torsion} and \emph{compatibility} with the metric $\bsy{\slashed{g}}$.~In particular, Proposition \ref{prop_proj_connection_torsion} captures a new feature of $\bsy{\nablasl}$ arising when non-integrable structures are allowed.~See also Remark \ref{rmk_proj_connection_not_LC}.

\medskip

\begin{prop} \label{prop_proj_connection_torsion}
The horizontal distribution $\bsy{\mathfrak{D}}_{\bsy{\mathcal{N}}}$ is integrable if and only if, for any $\bsy{\mathfrak{D}}_{\bsy{\mathcal{N}}}$ vector fields $\bsy{X},\bsy{Y}$, we have
\begin{equation} \label{torsion_connex_integr_distribution}
(\bsy{\nablasl}_{\bsy{X}}\bsy{Y})^{\circ}-(\bsy{\nablasl}_{\bsy{Y}}\bsy{X})^{\circ} = [\bsy{X}^{\circ},\bsy{Y}^{\circ}]  \, .
\end{equation}
\end{prop}

\medskip

\begin{proof}
We start by noting the identity
\begin{equation} \label{aux_LC_connection_id_torsion_1}
[\bsy{X}^{\circ},\bsy{Y}^{\circ}]=\bsy{\nabla_{X^{\circ}} Y^{\circ}}-\bsy{\nabla_{Y^{\circ}} X^{\circ}} 
\end{equation}
for any $\bsy{\mathfrak{D}}_{\bsy{\mathcal{N}}}$ vector fields $\bsy{X},\bsy{Y}$, with $\bsy{\nabla}$ the Levi-Civita (and thus torsion-free) connection of $\bsy{g}$. 

If $\bsy{\mathfrak{D}}_{\bsy{\mathcal{N}}}$ is integrable, then, for any $\bsy{\mathfrak{D}}_{\bsy{\mathcal{N}}}$ vector fields $\bsy{X},\bsy{Y}$, we have 
\begin{align} 
\bsy{g}([\bsy{X}^{\circ},\bsy{Y}^{\circ}],\ef)&=0 \, , & \bsy{g}([\bsy{X}^{\circ},\bsy{Y}^{\circ}],\et)&=0 \, ,  \label{aux_LC_connection_id_torsion_2}
\end{align}
where the equalities hold by the definition of integrable distribution.~Therefore, we have
\begin{align*}
\bsy{\nabla_{X^{\circ}} Y^{\circ}}-\bsy{\nabla_{Y^{\circ}} X^{\circ}}&=[\bsy{X}^{\circ},\bsy{Y}^{\circ}] \\
&=\bsy{\slashed{g}}^{\bsy{AB}}\bsy{g}([\bsy{X}^{\circ},\bsy{Y}^{\circ}],\ea)\eb \\
&=\bsy{\slashed{g}}^{\bsy{AB}}\bsy{g}(\bsy{\nabla_{X^{\circ}} Y^{\circ}}-\bsy{\nabla_{Y^{\circ}} X^{\circ}},\ea)\eb \\
&=(\bsy{\nablasl}_{\bsy{X}}\bsy{Y})^{\circ}-(\bsy{\nablasl}_{\bsy{Y}}\bsy{X})^{\circ} \, ,
\end{align*}
where the first and third equalities use \eqref{aux_LC_connection_id_torsion_1}, the second uses \eqref{aux_LC_connection_id_torsion_2} and the last equality holds by definition of $\bsy{\nablasl}_{\bsy{X}}\bsy{Y}$.~This proves the identity \eqref{torsion_connex_integr_distribution}. 

To now prove the other direction of Proposition \ref{prop_proj_connection_torsion}, we assume that, for any $\bsy{\mathfrak{D}}_{\bsy{\mathcal{N}}}$ vector fields $\bsy{X},\bsy{Y}$, the identity \eqref{torsion_connex_integr_distribution} holds.~By Definition \ref{def_formalism_projected_cov_derivative}, the quantity $\bsy{\nablasl}_{\bsy{X}}\bsy{Y}-\bsy{\nablasl}_{\bsy{Y}}\bsy{X}$ is a $\bsy{\mathfrak{D}}_{\bsy{\mathcal{N}}}$ vector field.~Thus, the left hand side of \eqref{torsion_connex_integr_distribution} is the canonical extension of a $\bsy{\mathfrak{D}}_{\bsy{\mathcal{N}}}$ vector field, implying $$[\bsy{X}^{\circ},\bsy{Y}^{\circ}]\in \bsy{\mathfrak{D}}_{\bsy{\mathcal{N}}} \, .$$ By definition of integrable distribution, one concludes that $\bsy{\mathfrak{D}}_{\bsy{\mathcal{N}}}$ is integrable.

\end{proof}

\medskip

\begin{prop} \label{prop_proj_connection_compatible}
Let $\bsy{X}\in\Gamma(T\bsy{\mathcal{M}})$.~We have the identities
\begin{align*}
\bsy{\nablasl_X}\,\bsy{\slashed{g}}&=0 \, , & 
\bsy{\nablasl_X}\,\bsy{\slashed{\varepsilon}}&=0 \, .
\end{align*}
\end{prop}

\medskip

\begin{proof}
We prove the first identity, the proof of the second identity is analogous.~We compute
\begin{align*}
(\bsy{\nablasl_X}\,\bsy{\slashed{g}})(\ea,\eb)&=-\bsy{\slashed{g}}(\bsy{\nablasl_X}\ea,\eb)-\bsy{\slashed{g}}(\ea,\bsy{\nablasl_X}\eb) \\
&=-\bsy{\slashed{g}}^{\bsy{CD}}\bsy{g}(\bsy{\nabla_X}\ea,\bsy{e_C})\,\bsy{\slashed{g}}(\bsy{e_D},\eb)-\bsy{\slashed{g}}^{\bsy{CD}}\bsy{g}(\bsy{\nabla_X}\eb,\bsy{e_C})\,\bsy{\slashed{g}}(\bsy{e_D},\ea) \\
&=-\bsy{g}(\bsy{\nabla_X}\ea,\bsy{e_B})-\bsy{g}(\bsy{\nabla_X}\eb,\bsy{e_A}) \\
&=0 \, ,
\end{align*}
where the first equality holds by Definition \ref{def_formalism_projected_cov_derivative} and the equality $\bsy{X}(\bsy{\slashed{g}}(\ea,\eb))=\bsy{X}(\bsy{g}(\ea,\eb))=0$, the second equality by Definition \ref{def_formalism_projected_cov_derivative} and the last equality by the $\bsy{\nabla}$ being the Levi-Civita (and thus metric compatible) connection of $\bsy{g}$.

\end{proof}

\medskip

\begin{remark} \label{rmk_proj_connection_not_LC}
In view of Proposition \ref{prop_proj_connection_compatible}, the induced connection $\bsy{\nablasl}$ over the bundle of $\bsy{\mathfrak{D}}_{\bsy{\mathcal{N}}}$ tensors is \emph{compatible} with the induced metric $\bsy{\slashed{g}}$.~However, Proposition \ref{prop_proj_connection_torsion} implies that the induced connection is \emph{torsion-free} (and thus \emph{Levi-Civita} relative to $\bsy{\slashed{g}}$) if and only if the distribution $\bsy{\mathfrak{D}}_{\bsy{\mathcal{N}}}$ is \emph{integrable}.~Thus, in general, \ul{the induced connection $\bsy{\nablasl}$ is \emph{not} the Levi-Civita connection of the induced metric $\bsy{\slashed{g}}$}.
\end{remark}

\medskip

The following proposition states that, in fact, the failure of $\bsy{\nablasl}$ to be Levi-Civita relative to $\bsy{\slashed{g}}$ noted in Remark \ref{rmk_proj_connection_not_LC} is a manifestation of a more general property:~If the horizontal distribution $\bsy{\mathfrak{D}}_{\bsy{\mathcal{N}}}$ is \emph{non-integrable}, then there exists \emph{no} Levi-Civita connection associated to $\bsy{\slashed{g}}$ over the bundle of $\bsy{\mathfrak{D}}_{\bsy{\mathcal{N}}}$ tensors. 

\medskip

\begin{prop}
The horizontal distribution $\bsy{\mathfrak{D}}_{\bsy{\mathcal{N}}}$ is integrable if and only if there exists a Levi-Civita connection associated to $\bsy{\slashed{g}}$ over the bundle of $\bsy{\mathfrak{D}}_{\bsy{\mathcal{N}}}$ tensors. 
\end{prop}

\medskip

\begin{proof}
Assume that the horizontal distribution $\bsy{\mathfrak{D}}_{\bsy{\mathcal{N}}}$ is integrable.~Then, we consider the connection over the bundle of $\bsy{\mathfrak{D}}_{\bsy{\mathcal{N}}}$ tensors from Definition \ref{def_formalism_projected_connection}.~By Propositions \ref{prop_proj_connection_torsion} and \ref{prop_proj_connection_compatible}, such a connection is Levi-Civita relative to $\bsy{\slashed{g}}$. 

To prove the other direction of the proposition, suppose $\bsy{\nablasl}^{\text{\tiny{LC}}}$ is a Levi-Civita connection associated to $\bsy{\slashed{g}}$ over the bundle of $\bsy{\mathfrak{D}}_{\bsy{\mathcal{N}}}$ tensors.~Then, since $\bsy{\nablasl}^{\text{\tiny{LC}}}$ is torsion-free, we have the identity 
\begin{equation} \label{aux_LC_vs_integr_distrib}
(\bsy{\nablasl}^{\text{\tiny{LC}}}_{\bsy{X}}\bsy{Y})^{\circ}-(\bsy{\nablasl}^{\text{\tiny{LC}}}_{\bsy{Y}}\bsy{X})^{\circ}=[\bsy{X}^{\circ},\bsy{Y}^{\circ}]
\end{equation}
for any $\bsy{\mathfrak{D}}_{\bsy{\mathcal{N}}}$ vector fields $\bsy{X},\bsy{Y}$.~Since $\bsy{\nablasl}^{\text{\tiny{LC}}}$ is a connection over $\bsy{\mathfrak{D}}_{\bsy{\mathcal{N}}}$ tensors, the quantity $\bsy{\nablasl}^{\text{\tiny{LC}}}_{\bsy{X}}\bsy{Y}-\bsy{\nablasl}^{\text{\tiny{LC}}}_{\bsy{Y}}\bsy{X}$ is a $\bsy{\mathfrak{D}}_{\bsy{\mathcal{N}}}$ vector field.~Thus, the left hand side of \eqref{aux_LC_vs_integr_distrib} is the canonical extension of a $\bsy{\mathfrak{D}}_{\bsy{\mathcal{N}}}$ vector field, implying $$[\bsy{X}^{\circ},\bsy{Y}^{\circ}]\in \bsy{\mathfrak{D}}_{\bsy{\mathcal{N}}} \, .$$ By definition of integrable distribution, one concludes that $\bsy{\mathfrak{D}}_{\bsy{\mathcal{N}}}$ is integrable.

\end{proof}

\medskip

In view of Proposition \ref{prop_proj_connection_compatible}, we can define the \emph{antisymmetric} $2\times 2$ matrices $$\bsy{\slashed{M}{}_{A}^B} \, , \, \bsy{\slashed{\underline{M}}{}_{A}^B} \, , \, {}_{\bsy{1}}\bsy{\slashed{H}{}_{A}^B} \, , \, {}_{\bsy{2}}\bsy{\slashed{H}{}_{A}^B}$$ such that
\begin{align*}
\bsy{\nablasl_4}\ea &= \bsy{\slashed{M}{}_{A}^B} \eb \, ,  & \bsy{\nablasl_3}\ea &= \bsy{\slashed{\underline{M}}{}_{A}^B} \eb \, , & \bsy{\nablasl_1}\ea &= {}_{\bsy{1}}\bsy{\slashed{H}{}_{A}^B} \bsy{e_B} \, , & \bsy{\nablasl_2}\ea &= {}_{\bsy{2}}\bsy{\slashed{H}{}_{A}^B} \bsy{e_B} \, .
\end{align*}

\medskip

We shall now define a set of differential operators employing the covariant derivative $\bsy{\nablasl}$.~For any $\bsy{\mathfrak{D}}_{\bsy{\mathcal{N}}}$ one-tensor $\bsy{\varsigma}$ and $\bsy{\mathfrak{D}}_{\bsy{\mathcal{N}}}$ two-tensor $\bsy{\theta}$, we define the divergence operator\footnote{The reader familiar with \cite{DHR} should note that our definition of $\bsy{\slashed{\textbf{div}}}\,\bsy{\theta}$ contracts the frame indices in a different order.} 
\begin{align*}
\bsy{\slashed{\textbf{div}}}\,\bsy{\varsigma}&=\bsy{\slashed{g}^{AB}}(\bsy{\nablasl}_{\bsy{A}}\bsy{\varsigma})_{\bsy{B}}  \, , &
(\bsy{\slashed{\textbf{div}}}\,\bsy{\theta})_{\bsy{A}}&=\bsy{\slashed{g}^{CB}}(\bsy{\nablasl_C}\bsy{\theta})_{\bsy{AB}} \, ,
\end{align*}
the curl operator 
\begin{equation*}
\bsy{\slashed{\textbf{curl}}}\,\bsy{\varsigma}=\bsy{\slashed{\varepsilon}^{AB}}(\bsy{\nablasl_A}\bsy{\varsigma})_{\bsy{B}} 
\end{equation*}
and the differential operator 
\begin{equation*}
\bsy{\slashed{\mathcal{D}}_2^{\star}\,\varsigma}=-\frac{1}{2}((\bsy{\nablasl\varsigma})+(\bsy{\nablasl\varsigma})^{\mathsf{T}}-(\bsy{\slashed{\textbf{div}}}\,\bsy{\varsigma})\,\bsy{\slashed{g}})  \, .
\end{equation*}
The $\bsy{\mathfrak{D}}_{\bsy{\mathcal{N}}}$ two-tensor $\bsy{\slashed{\mathcal{D}}_2^{\star}\,\varsigma}$ is symmetric and traceless relative to $\bsy{\slashed{g}^{-1}}$.

\medskip

For any smooth scalar functions $\bsy{f}$ and $\bsy{h}$, we define the differential operator
\begin{equation*}
\bsy{\slashed{\mathcal{D}}_1^{\star}}(\bsy{f},\bsy{h})=-\bsy{\nablasl}\bsy{f}+{}^{\star}\bsy{\nablasl}\bsy{h} \, .
\end{equation*}

\subsubsection{Lie derivative}

Let $\bsy{X}\in \Gamma(T\bsy{\mathcal{M}})$.~We define the differential operator 
\begin{equation*}
\bsy{\slashed{\mathcal{L}}_{\bsy{X}}}
\end{equation*}
such that, for any smooth scalar function $\bsy{f}$ and $\bsy{\mathfrak{D}}_{\bsy{\mathcal{N}}}$ $k$-tensor $\bsy{\varsigma}$, we have the smooth scalar function $\bsy{\slashed{\mathcal{L}}_{X}}\bsy{f}$ and the $\bsy{\mathfrak{D}}_{\bsy{\mathcal{N}}}$ $k$-tensor $\bsy{\slashed{\mathcal{L}}_{X}}\bsy{\varsigma}$ such that 
\begin{align}
\bsy{\slashed{\mathcal{L}}_{X}}\bsy{f}&= \bsy{\nablasl_X f} \, , &
(\bsy{\slashed{\mathcal{L}}_{X}}\bsy{\varsigma})(\bsy{Y_1},\ldots,\bsy{Y_k})&=(\bsy{\mathcal{L}}_{\bsy{X}}\bsy{\varsigma}^{\circ})(\bsy{Y^{\circ}_1},\ldots,\bsy{Y^{\circ}_k})   \label{def_lie_deriv_tensor}
\end{align}
for any $\bsy{\mathfrak{D}}_{\bsy{\mathcal{N}}}$ vector fields $\bsy{Y_1},\ldots,\bsy{Y_k}$, with $\bsy{\mathcal{L}}_{\bsy{X}}\bsy{\varsigma}^{\circ}$ the spacetime Lie derivative of $\bsy{\varsigma}^{\circ}$ with respect to $\bsy{X}$.

\medskip

The following proposition is useful to relate projected Lie derivatives to projected covariant derivatives.

\medskip

\begin{prop}
For any $\bsy{\mathfrak{D}}_{\bsy{\mathcal{N}}}$ one-tensor $\bsy{\varsigma}$ and $\bsy{\mathfrak{D}}_{\bsy{\mathcal{N}}}$ two-tensor $\bsy{\theta}$, we have the identities
\begin{align}
\bsy{\slashed{\mathcal{L}}_{4}}\,\bsy{\varsigma}&=\bsy{\nablasl_4 \varsigma}+\bsy{\chi^{\sharp_2}\cdot\varsigma} \, , \label{lie_der_formula_1}\\[5pt]
\bsy{\slashed{\mathcal{L}}_{3}}\,\bsy{\varsigma}&=\bsy{\nablasl_3 \varsigma}+\bsy{\chib^{\sharp_2}\cdot\varsigma} \, , \label{lie_der_formula_2}\\[5pt]
\bsy{\slashed{\mathcal{L}}_{4}}\,\bsy{\theta} &=\bsy{\nablasl_4 \theta}+\bsy{\chi_{\tiny{\Circle}}\times\theta}+\bsy{\theta\times \chi_{\tiny{\Circle}}}+\bsy{\chi_{\tiny{\Square}}\times\theta}-\bsy{\theta\times \chi_{\tiny{\Square}}} \, , \label{lie_der_formula_3}\\[5pt]
\bsy{\slashed{\mathcal{L}}_{3}}\,\bsy{\theta} &=\bsy{\nablasl_3 \theta}+\bsy{\chib_{\tiny{\Circle}}\times\,\theta}+\bsy{\theta\times \chib_{\tiny{\Circle}}}+\bsy{\chib_{\tiny{\Square}}\times\,\theta}-\bsy{\theta\times \chib_{\tiny{\Square}}} \, . \label{lie_der_formula_4}
\end{align}
\end{prop}

\medskip

\begin{proof}
We prove the formula \eqref{lie_der_formula_1}, the proof of the formulae \eqref{lie_der_formula_2}, \eqref{lie_der_formula_3} and \eqref{lie_der_formula_4} is analogous.~We compute
\begin{align*}
(\bsy{\slashed{\mathcal{L}}_{4}}\,\bsy{\varsigma})(\ea)&=(\bsy{\mathcal{L}_{4}}\,\bsy{\varsigma}^{\circ})(\ea) \\
&=\ef(\bsy{\varsigma}^{\circ}(\ea))-\bsy{\varsigma}^{\circ}([\ef,\ea]) \\
&=\ef(\bsy{\varsigma}^{\circ}(\ea))-\bsy{\varsigma}^{\circ}((\bsy{\nablasl_4}\ea)^{\circ}-\bsy{\chi^{\sharp_2}}{}^{\bsy{B}}_{\bsy{A}}\eb) \\
&=\ef(\bsy{\varsigma}(\ea))-\bsy{\varsigma}(\bsy{\nablasl_4}\ea)+\bsy{\chi^{\sharp_2}}{}^{\bsy{B}}_{\bsy{A}}\,\bsy{\varsigma}_{\bsy{B}} \\
&=(\bsy{\nablasl_4 \varsigma})(\ea)+(\bsy{\chi^{\sharp_2}\cdot\varsigma})(\ea)  \, ,
\end{align*}
where the first equality holds by the definition of $\bsy{\slashed{\mathcal{L}}_{4}}\,\bsy{\varsigma}$.

\end{proof}

\subsection{Curvature of horizontal distributions} \label{sec_curvature_horizontal_distribution}

In this section, we introduce the notion of \emph{curvature} of the horizontal distribution $\bsy{\mathfrak{D}}_{\bsy{\mathcal{N}}}$.~As we shall see, when $\bsy{\mathfrak{D}}_{\bsy{\mathcal{N}}}$ is allowed to be non-integrable, some of the usual curvature quantities are no longer tensorial and lose their symmetries.~These new features may be seen as yet another manifestation of the torsion of the induced connection $\bsy{\nablasl}$.

\medskip

\begin{definition}  \label{def_proj_second_order_cov_deriv}
For any $\bsy{\mathfrak{D}}_{\bsy{\mathcal{N}}}$ vector fields $\bsy{X},\bsy{Y}$, we define the differential operator 
\begin{equation*}
\bsy{\nablasl}^2_{\bsy{X}, \bsy{Y}} 
\end{equation*}
such that, for any $\bsy{\mathfrak{D}}_{\bsy{\mathcal{N}}}$ $k$-tensor $\bsy{\varsigma}$ and $\bsy{\mathfrak{D}}_{\bsy{\mathcal{N}}}$ vector field $\bsy{Z}$, we have the $\bsy{\mathfrak{D}}_{\bsy{\mathcal{N}}}$ $k$-tensor $\bsy{\nablasl}^2_{\bsy{X}, \bsy{Y}} \bsy{\varsigma}$ and the $\bsy{\mathfrak{D}}_{\bsy{\mathcal{N}}}$ vector field $\bsy{\nablasl}^2_{\bsy{X}, \bsy{Y}} \bsy{Z}$ such that
\begin{align*}
\bsy{\nablasl}^2_{\bsy{X}, \bsy{Y}} \bsy{\varsigma}&= \bsy{\nablasl}_{\bsy{X}}(\bsy{\nablasl_Y}\bsy{\varsigma})-\bsy{\nablasl}_{\bsy{\nablasl_X} \bsy{Y}}\bsy{\varsigma} \, ,  &
\bsy{\nablasl}^2_{\bsy{X}, \bsy{Y}} \bsy{Z}&= \bsy{\nablasl}_{\bsy{X}}(\bsy{\nablasl_Y}\bsy{Z})-\bsy{\nablasl}_{\bsy{\nablasl_X}\bsy{Y}}\bsy{Z} \, . 
\end{align*}
\end{definition}

\medskip

For any point $p\in\bsy{\mathcal{M}}$, we consider the map $$\bsy{\slashed{R}}_p:(\bsy{\mathfrak{D}}_{\bsy{\mathcal{N}}})_p\otimes(\bsy{\mathfrak{D}}_{\bsy{\mathcal{N}}})_p\otimes(\bsy{\mathfrak{D}}_{\bsy{\mathcal{N}}})_p\otimes(\bsy{\mathfrak{D}}_{\bsy{\mathcal{N}}})_p\rightarrow\mathbb{R}$$ such that
\begin{equation}
\bsy{\slashed{R}}_p(\bsy{W},\bsy{Z},\bsy{X},\bsy{Y})=\bsy{\slashed{g}}_p(\bsy{\nablasl}^2_{\bsy{X}, \bsy{Y}} \bsy{Z}-\bsy{\nablasl}^2_{\bsy{Y}, \bsy{X}} \bsy{Z},\bsy{W})
\end{equation}
for any local $\bsy{\mathfrak{D}}_{\bsy{\mathcal{N}}}$ vector fields $\bsy{X},\bsy{Y},\bsy{W}$ and $\bsy{Z}$.

\medskip

\begin{definition}
We define the \emph{Riemann curvature} of the horizontal distribution $\bsy{\mathfrak{D}}_{\bsy{\mathcal{N}}}$ as the (local) smooth map $\bsy{\slashed{R}}$ on $\bsy{\mathcal{M}}$ such that
\begin{equation} \label{curvature_non_integrable_distribution}
\bsy{\slashed{R}}:p\mapsto \bsy{\slashed{R}}_p 
\end{equation}
for any (local) $p\in\bsy{\mathcal{M}}$.  
\end{definition}

\medskip

We have the following proposition, which captures the failure of the Riemann curvature of $\bsy{\mathfrak{D}}_{\bsy{\mathcal{N}}}$ to be tensorial when $\bsy{\mathfrak{D}}_{\bsy{\mathcal{N}}}$ is non-integrable.

\medskip

\begin{prop} \label{prop_tensorial_curvature_distribution}
The Riemann curvature $\bsy{\slashed{R}}$ of the horizontal distribution $\bsy{\mathfrak{D}}_{\bsy{\mathcal{N}}}$ is a (local) $\bsy{\mathfrak{D}}_{\bsy{\mathcal{N}}}$ four-tensor if and only if $\bsy{\mathfrak{D}}_{\bsy{\mathcal{N}}}$ is integrable.
\end{prop}

\medskip

\begin{proof}
For any smooth scalar function $\bsy{f}$, we compute
\begin{align}
\bsy{\slashed{R}}_p(\bsy{W},\bsy{f}\,\bsy{Z},\bsy{X},\bsy{Y}) =& \, \bsy{\slashed{g}}_p(\bsy{\nablasl}^2_{\bsy{X}, \bsy{Y}} (\bsy{f}\,\bsy{Z})-\bsy{\nablasl}^2_{\bsy{Y}, \bsy{X}} (\bsy{f}\,\bsy{Z}),\bsy{W}) \nonumber \\
=& \, \bsy{f} \, \bsy{\slashed{g}}_p(\bsy{\nablasl}^2_{\bsy{X}, \bsy{Y}} \bsy{Z}-\bsy{\nablasl}^2_{\bsy{Y}, \bsy{X}} \bsy{Z},\bsy{W}) \nonumber \\ & + (\bsy{X}(\bsy{Y}(\bsy{f}))-\bsy{Y}(\bsy{X}(\bsy{f}))-(\bsy{\nablasl_X}\bsy{Y})(\bsy{f})+(\bsy{\nablasl_Y}\bsy{X})(\bsy{f}))\,\bsy{\slashed{g}}_p(\bsy{Z},\bsy{W}) \nonumber \\
=& \, \bsy{f}\, \bsy{\slashed{R}}_p(\bsy{W},\bsy{Z},\bsy{X},\bsy{Y}) +([\bsy{X}^{\circ},\bsy{Y}^{\circ}](\bsy{f})-(\bsy{\nablasl_X}\bsy{Y}-\bsy{\nablasl_Y}\bsy{X})(\bsy{f})) \, \bsy{\slashed{g}}_p(\bsy{Z},\bsy{W}) \, , \label{aux_curv_comp}
\end{align}
where in the second identity we used Definition \ref{def_proj_second_order_cov_deriv}.~By applying Proposition \ref{prop_proj_connection_torsion}, we conclude that $\bsy{\slashed{R}}_p$ is a linear map (i.e.~the second term in \eqref{aux_curv_comp} identically vanishes) if and only if $\bsy{\mathfrak{D}}_{\bsy{\mathcal{N}}}$ is integrable.

\end{proof}

\medskip

\begin{remark}
In view of Proposition \ref{prop_tensorial_curvature_distribution}, the Riemann curvature $\bsy{\slashed{R}}$ of the horizontal distribution $\bsy{\mathfrak{D}}_{\bsy{\mathcal{N}}}$ as defined by the map \eqref{curvature_non_integrable_distribution} is \emph{not}, in general, a tensorial quantity, the failure deriving from the torsion associated to the induced connection $\bsy{\nablasl}$ (see Proposition \ref{prop_proj_connection_torsion} and its application in the proof of Proposition \ref{prop_tensorial_curvature_distribution}).~When $\bsy{\mathfrak{D}}_{\bsy{\mathcal{N}}}$ is integrable, the map \eqref{curvature_non_integrable_distribution} coincides with the Riemann curvature tensor of the (local) integral manifold of $\bsy{\mathfrak{D}}_{\bsy{\mathcal{N}}}$.
\end{remark}

\medskip

The two following propositions collect the symmetries of the Riemann curvature $\bsy{\slashed{R}}$.~The proof is easy and is left to the reader.~We note that the symmetries of $\bsy{\slashed{R}}$ collected in Proposition \ref{prop_symm_curvature_2} are conditional to $\bsy{\mathfrak{D}}_{\bsy{\mathcal{N}}}$ being integrable, in that they do not hold when the induced connection $\bsy{\nablasl}$ has torsion.

\medskip

\begin{prop} \label{prop_symm_curvature_1}
For any $\bsy{\mathfrak{D}}_{\bsy{\mathcal{N}}}$ vector fields $\bsy{X},\bsy{Y},\bsy{W}$ and $\bsy{Z}$, we have the identities
\begin{align*}
\bsy{\slashed{R}}_p(\bsy{W},\bsy{Z},\bsy{X},\bsy{Y})&=-\bsy{\slashed{R}}_p(\bsy{Z},\bsy{W},\bsy{X},\bsy{Y}) \, , &
\bsy{\slashed{R}}_p(\bsy{W},\bsy{Z},\bsy{X},\bsy{Y}) &=- \bsy{\slashed{R}}_p(\bsy{W},\bsy{Z},\bsy{Y},\bsy{X}) \, .
\end{align*}
\end{prop}

\medskip

\begin{prop} \label{prop_symm_curvature_2}
We have both the identities
\begin{equation*}
\bsy{\slashed{R}}_p(\bsy{W},\bsy{Z},\bsy{X},\bsy{Y})=\bsy{\slashed{R}}_p(\bsy{X},\bsy{Y},\bsy{W},\bsy{Z}) 
\end{equation*}
and
\begin{equation*}
\bsy{\slashed{R}}_p(\bsy{W},\bsy{Z},\bsy{X},\bsy{Y})+\bsy{\slashed{R}}_p(\bsy{W},\bsy{Y},\bsy{Z},\bsy{X})+\bsy{\slashed{R}}_p(\bsy{W},\bsy{X},\bsy{Y},\bsy{Z})=0
\end{equation*}
for any $\bsy{\mathfrak{D}}_{\bsy{\mathcal{N}}}$ vector fields $\bsy{X},\bsy{Y},\bsy{W}$ and $\bsy{Z}$ if and only if $\bsy{\mathfrak{D}}_{\bsy{\mathcal{N}}}$ is integrable.
\end{prop}

\medskip

\begin{remark} \label{rmk_curvature_volume_form}
It follows immediately from Propositions \ref{prop_symm_curvature_1} and \ref{prop_symm_curvature_2} that $$\bsy{\slashed{R}} \propto \bsy{\slashed{\varepsilon}}\otimes \bsy{\slashed{\varepsilon}}$$ if and only if $\bsy{\mathfrak{D}}_{\bsy{\mathcal{N}}}$ is integrable.
\end{remark}

\medskip

We now introduce the \emph{Gauss curvature} of the horizontal distribution $\bsy{\mathfrak{D}}_{\bsy{\mathcal{N}}}$.~The definition is adapted to the fact that the curvature of $\bsy{\mathfrak{D}}_{\bsy{\mathcal{N}}}$ is, in general, not tensorial.

\medskip

\begin{definition}
We define the \emph{Gauss curvature} of the horizontal distribution $\bsy{\mathfrak{D}}_{\bsy{\mathcal{N}}}$ as the smooth scalar function
\begin{equation} \label{def_gauss_curvature}
\bsy{\slashed{K}}=\frac{1}{2}\,\sum_{\bsy{A},\bsy{B}} \bsy{\slashed{R}}(\ea,\eb,\ea,\eb) \, .
\end{equation}
\end{definition}

\medskip

\begin{remark}
When $\bsy{\mathfrak{D}}_{\bsy{\mathcal{N}}}$ is integrable, the scalar function $\bsy{\slashed{K}}$ coincides with the Gauss curvature of the (local) integral manifold of $\bsy{\mathfrak{D}}_{\bsy{\mathcal{N}}}$.
\end{remark}

\medskip

We have the following proposition.~The proof is easy and left to the reader.~See related Remark \ref{rmk_curvature_volume_form}.

\medskip

\begin{prop}
We have the identity
\begin{equation*}
\bsy{\slashed{R}}= \bsy{\slashed{K}}\, \bsy{\slashed{\varepsilon}}\otimes \bsy{\slashed{\varepsilon}}
\end{equation*}
if and only if $\bsy{\mathfrak{D}}_{\bsy{\mathcal{N}}}$ is integrable.
\end{prop}

\medskip

We define the $\bsy{\mathfrak{D}}_{\bsy{\mathcal{N}}}$ four-\emph{tensor}\footnote{The fact that $\widetilde{\bsy{\slashed{R}}}$ is a tensor can be easily checked.~Crucially, this holds unconditionally, i.e.~without the need to assume that $\bsy{\mathfrak{D}}_{\bsy{\mathcal{N}}}$ is integrable.} $\widetilde{\bsy{\slashed{R}}}$ such that
\begin{align} 
\widetilde{\bsy{\slashed{R}}}(\bsy{e_A},\bsy{e_D},\bsy{e_C},\bsy{e_B})=& \, \bsy{\slashed{R}}(\bsy{e_A},\bsy{e_D},\bsy{e_C},\bsy{e_B}) \label{corrected_curv_tensorial} \\ &+\frac{1}{2}\,(\bsy{\slashed{\varepsilon}\cdot\chi})\,\bsy{\slashed{g}}(\bsy{\nablasl_3e_D},\bsy{e_A})\,\bsy{\slashed{\varepsilon}}(\bsy{e_B},\bsy{e_C}) \nonumber \\ & +\frac{1}{2}\,(\bsy{\slashed{\varepsilon}\cdot\chib})\,\bsy{\slashed{g}}(\bsy{\nablasl_4e_D},\bsy{e_A})\,\bsy{\slashed{\varepsilon}}(\bsy{e_B},\bsy{e_C}) \, . \nonumber
\end{align}
One can check that $\widetilde{\bsy{\slashed{R}}}$ satisfies the symmetries of Proposition \ref{prop_symm_curvature_2} unconditionally, i.e.~without the need to assume that $\bsy{\mathfrak{D}}_{\bsy{\mathcal{N}}}$ is integrable.

\medskip

\begin{remark}
The quantity $\bsy{\slashed{R}}$ is intrinsic to $\bsy{\mathfrak{D}}_{\bsy{\mathcal{N}}}$, whereas he quantity $\widetilde{\bsy{\slashed{R}}}$ contains two terms that are extrinsic with respect to $\bsy{\mathfrak{D}}_{\bsy{\mathcal{N}}}$.~For this reason, it is more natural to consider the former quantity as the Riemann curvature of $\bsy{\mathfrak{D}}_{\bsy{\mathcal{N}}}$.
\end{remark}

\medskip

The tensor $\widetilde{\bsy{\slashed{R}}}$ appears in the formal derivation of the Gauss equation \eqref{eq_nonlinear_Gauss} (see equation \eqref{eq_4} in Section \ref{sec_formal_derivation_EVE}).~In that regard, the following definition will be convenient.

\medskip

\begin{definition}
We define the \emph{corrected Gauss curvature} of the horizontal distribution $\bsy{\mathfrak{D}}_{\bsy{\mathcal{N}}}$ as the smooth scalar function
\begin{equation} \label{def_gauss_curvature_corrected}
\widetilde{\bsy{\slashed{K}}}=\frac{1}{2}\,\bsy{\slashed{g}}^{\bsy{AC}}\bsy{\slashed{g}}^{\bsy{BD}}\,\widetilde{\bsy{\slashed{R}}}(\ea,\eb,\bsy{e_C},\bsy{e_D})  \, .
\end{equation}
\end{definition}

\medskip

We have the identity
\begin{equation}
\widetilde{\bsy{\slashed{K}}}= \bsy{\slashed{K}} +\frac{1}{4}\,(\bsy{\slashed{\varepsilon}\cdot\chi})\,\bsy{\slashed{g}}^{\bsy{BD}} \,\bsy{\slashed{\underline{M}}}{}_{\bsy{D}}^{\bsy{C}}\,\bsy{\slashed{\varepsilon}}_{\bsy{BC}}+\frac{1}{4}\,(\bsy{\slashed{\varepsilon}\cdot\chib})\,\bsy{\slashed{g}}^{\bsy{BD}}\,\bsy{\slashed{M}}{}_{\bsy{D}}^{\bsy{C}}\,\bsy{\slashed{\varepsilon}}_{\bsy{BC}} \, .
\end{equation}

\medskip

\begin{remark}
When $\bsy{\mathfrak{D}}_{\bsy{\mathcal{N}}}$ is integrable, we have $\widetilde{\bsy{\slashed{R}}}\equiv \bsy{\slashed{R}}$ and $\widetilde{\bsy{\slashed{K}}}\equiv \bsy{\slashed{K}}$.
\end{remark}

\subsection{Some useful formulae}

In this section, we collect some useful formulae.

\subsubsection{Covariant derivative}

For the covariant derivative of the frame vector fields, we have the following decomposition formulae 
\begin{align*}
\bsy{\nabla_A e_3}&= \bsy{{\chib^{\bsy{\sharp_2}}}{}_A^B \,e_B}+\bsy{ \zeta_A \,e_3} \, ,  &  \bsy{ \nabla_A e_4}&=\bsy{{\chi^{\bsy{\sharp_2}}}{}_A^B \, e_B}-\bsy{\zeta_A \, e_4 } \, , \\[5pt]
\bsy{\nabla_3  e_3}&= 2\,\bsy{\yb^A e_A}+\bsy{\omegabh e_3} \, , &  \bsy{ \nabla_3  e_4}&=2\,\bsy{\eta^A e_A}-\bsy{\omegabh e_4} \, ,\\[5pt]
\bsy{\nabla_4   e_3 }&=  2\,\bsy{\etab^A e_A}  -\bsy{\omegah e_3}\, ,  &  \bsy{\nabla_4   e_4} &=2 \,\bsy{\xi^A e_A}+\bsy{\omegah e_4} \, ,
\end{align*}
and\footnote{We note the slight abuse of notation in the formulae for the covariant derivatives $\bsy{\nabla_I\ea}$.~To be fully consistent with our notation, one would, for instance, write $\bsy{\nabla_4 e_A }= (\bsy{\nablasl_4 e_A})^{\circ} +\bsy{\xi_A \, e_3} + \bsy{\etab_A  e_4}$.} 
\begin{align*}
\bsy{\nabla_A e_B} &=\bsy{\nablasl_A e_B}+\frac{1}{2}\,\bsy{\chi}_{\bsy{AB}}\bsy{e_3} + \frac{1}{2}\,\bsy{\chib}_{\bsy{AB}} \bsy{e_4}   \, , \\[5pt]
\bsy{\nabla_3 e_A }&= \bsy{\nablasl_3 e_A} +\bsy{\eta_A \,e_3}+ \bsy{\yb_A e_4} \, , \\[5pt]
\bsy{\nabla_4 e_A }&= \bsy{\nablasl_4 e_A} +\bsy{\xi_A \, e_3} + \bsy{\etab_A  e_4} \, .
\end{align*}

\subsubsection{Frame commutators}

The commutators of the frame vector fields read
\begin{align}
[\bsy{e_A},\bsy{e_B}] &=  \left({}_{\bsy{A}}\bsy{\slashed{H}{}_{B}^C} -{}_{\bsy{B}}\bsy{\slashed{H}{}_{A}^C} \right)\bsy{e_C} +\frac{1}{2}\,(\bsy{\slashed{\varepsilon}\cdot\chi})\,\bsy{\slashed{\varepsilon}}_{\bsy{AB}}\,\bsy{e_3}+\frac{1}{2}\,(\bsy{\slashed{\varepsilon}\cdot\chib})\,\bsy{\slashed{\varepsilon}}_{\bsy{AB}}\,\bsy{e_4}  \, , \label{frame_commutator_AB}\\[5pt]
[\bsy{e_3},\bsy{e_A}] &= \left(\bsy{\slashed{\underline{M}}{}_{A}^B}-\bsy{{\chib^{\bsy{\sharp_2}}}{}_A^B}\right)\bsy{e_B}+\left(\bsy{\eta_A}-\bsy{\zeta_A}\right)\bsy{e_3}+\bsy{\yb_A e_4} \, , \label{frame_commutator_3A}\\[5pt]
[\bsy{e_4},\bsy{e_A}] &= \left(\bsy{\slashed{M}{}_{A}^B}-\bsy{{\chi^{\bsy{\sharp_2}}}{}_A^B}\right)\bsy{e_B}+\bsy{\xi_A \, e_3}+\left(\bsy{\etab_A}+\bsy{\zeta_A}\right)\bsy{e_4}  \, , \label{frame_commutator_4A}\\[5pt]
[\bsy{e_3},\bsy{e_4}] &= 2\left(\bsy{\eta^A}-\bsy{\etab^A}\right)\bsy{e_A}+\bsy{\omegah e_3}-\bsy{\omegabh e_4} \, . \label{frame_commutator_34}
\end{align}
The last two terms on the right hand side of \eqref{frame_commutator_AB} encode the (non-)integrability of the horizontal distribution $\bsy{\mathfrak{D}}_{\bsy{\mathcal{N}}}$.

\subsubsection{Commutation formulae} \label{sec_commutation_formulae}

For any $\bsy{\mathfrak{D}}_{\bsy{\mathcal{N}}}$ one-tensor $\bsy{\varsigma}$ and $\bsy{\mathfrak{D}}_{\bsy{\mathcal{N}}}$ two-tensor $\bsy{\theta}$, we have the following commutation formulae
\begin{align}
[\bsy{\nablasl_4},\bsy{\nablasl}]\bsy{\varsigma} =& \, -\bsy{\chi}\bsy{\times}(\bsy{\nablasl\varsigma})+\bsy{\xi}\otimes (\bsy{\nablasl_{\bsy{3}}\varsigma})+(\bsy{\etab}+\bsy{\zeta})\otimes(\bsy{\nablasl_{\bsy{4}}\varsigma})  \label{comm_form_1}\\
&+(\bsy{\xi},\bsy{\varsigma})\,\bsy{\chib}-(\bsy{\chib^{\sharp_2}\cdot\,\varsigma})\otimes \bsy{\xi}+(\bsy{\etab},\bsy{\varsigma})\,\bsy{\chi}-(\bsy{\chi^{\sharp_2}\cdot\varsigma})\otimes \bsy{\etab}+(\bsy{^{\star}\beta})\otimes (\bsy{^{\star}\varsigma}) \, , \nonumber 
\end{align}
\begin{align}
[\bsy{\nablasl_3},\bsy{\nablasl}]\bsy{\varsigma} =& \, -\bsy{\chib}\bsy{\times}(\bsy{\nablasl\varsigma})+\bsy{\yb}\otimes (\bsy{\nablasl_{\bsy{4}}\varsigma})+(\bsy{\eta}-\bsy{\zeta})\otimes(\bsy{\nablasl_{\bsy{3}}\varsigma})  \label{comm_form_2}\\
&+(\bsy{\yb},\bsy{\varsigma})\,\bsy{\chi}-(\bsy{\chi^{\sharp_2}\cdot\varsigma})\otimes \bsy{\yb}+(\bsy{\eta},\bsy{\varsigma})\,\bsy{\chib}-(\bsy{\chib^{\sharp_2}\cdot\,\varsigma})\otimes \bsy{\eta}-(\bsy{^{\star}\betab})\otimes (\bsy{^{\star}\varsigma}) \, , \nonumber 
\end{align}
\begin{align}
[\bsy{\nablasl_3},\bsy{\nablasl_4}]\bsy{\varsigma} =& \, \bsy{\hat{\omega}}\,(\bsy{\nablasl_3}\bsy{\varsigma})-\bsy{\hat{\omegab}}\,(\bsy{\nablasl_4}\bsy{\varsigma})+ 2\,(\bsy{\nablasl\varsigma})^{\bsy{\sharp_1}}\bsy{\cdot}(\bsy{\eta}-\bsy{\etab}) \label{comm_form_3}\\ 
&+2\,(\bsy{\eta},\bsy{\varsigma})\,\bsy{\etab}-2\,(\bsy{\etab},\bsy{\varsigma})\,\bsy{\eta}+2\,(\bsy{\yb},\bsy{\varsigma})\,\bsy{\xi} -2\,(\bsy{\xi},\bsy{\varsigma})\,\bsy{\yb}  +2\,\bsy{\sigma}(\bsy{^{\star}\varsigma})   \nonumber
\end{align}
and 
\begin{align}
([\bsy{\nablasl_4},\bsy{\nablasl}]\bsy{\theta})_{\bsy{ABC}}=& \, -\bsy{\chi^{\sharp_2 }{}_A^D}(\bsy{\nablasl_D\theta})_{\bsy{BC}}+\bsy{\xi_A}(\bsy{\nablasl_{\bsy{3}}\theta})_{\bsy{BC}}+(\bsy{\etab_A}+\bsy{\zeta_A})(\bsy{\nablasl_{\bsy{4}}\theta})_{\bsy{BC}} \label{comm_form_4} \\ & +\bsy{\chib_{AB}\xi^D\theta_{DC}}-\bsy{\xi_B}\bsy{\chib^{\sharp_2}{}_A^D}\bsy{\theta_{DC}}+\bsy{\chi_{AB}\etab^D\theta_{DC}}-\bsy{\etab_B}\bsy{\chi^{\sharp_2}{}_A^D}\bsy{\theta_{DC}} +\bsy{\slashed{\varepsilon}}^{\bsy{\sharp_2}}\bsy{{}_B^D}(\bsy{^{\star}\beta})_{\bsy{A}}\bsy{\theta_{DC}} \nonumber \\ & +\bsy{\chib_{AC}\xi^D\theta_{BD}}-\bsy{\xi_C}\bsy{\chib^{\sharp_2 }{}_A^D}\bsy{\theta_{BD}}+\bsy{\chi_{AC}\etab^D\theta_{BD}}-\bsy{\etab_C}\bsy{\chi^{\sharp_2 }{}_A^D}\bsy{\theta_{BD}} +\bsy{\slashed{\varepsilon}}^{\bsy{\sharp_2}}\bsy{{}_C^D}(\bsy{^{\star}\beta})_{\bsy{A}}\bsy{\theta_{BD}} \, , \nonumber
\end{align}
\begin{align}
([\bsy{\nablasl_3},\bsy{\nablasl}]\bsy{\theta})_{\bsy{ABC}}=& \, -\bsy{\chib^{\sharp_2 }{}_A^D}(\bsy{\nablasl_D\theta})_{\bsy{BC}}+\bsy{\yb_A}(\bsy{\nablasl_{\bsy{4}}\theta})_{\bsy{BC}}+(\bsy{\eta_A}-\bsy{\zeta_A})(\bsy{\nablasl_{\bsy{3}}\theta})_{\bsy{BC}} \label{comm_form_5}\\ & +\bsy{\chi_{AB}\yb^D\theta_{DC}}-\bsy{\yb_B}\bsy{\chi^{\sharp_2 }{}_A^D}\bsy{\theta_{DC}}+\bsy{\chib_{AB}\eta^D\theta_{DC}}-\bsy{\eta_B}\bsy{\chib^{\sharp_2 }{}_A^D}\bsy{\theta_{DC}} -\bsy{\slashed{\varepsilon}}^{\bsy{\sharp_2}}\bsy{{}_B^D}(\bsy{^{\star}\betab})_{\bsy{A}}\bsy{\theta_{DC}} \nonumber\\ & +\bsy{\chi_{AC}\yb^D\theta_{BD}}-\bsy{\yb_C}\bsy{\chi^{\sharp_2 }{}_A^D}\bsy{\theta_{BD}}+\bsy{\chib_{AC}\eta^D\theta_{BD}}-\bsy{\eta_C}\bsy{\chib^{\sharp_2 }{}_A^D}\bsy{\theta_{BD}} -\bsy{\slashed{\varepsilon}}^{\bsy{\sharp_2}}\bsy{{}_C^D}(\bsy{^{\star}\betab})_{\bsy{A}}\bsy{\theta_{BD}} \, ,  \nonumber
\end{align}
\begin{align}
([\bsy{\nablasl_3},\bsy{\nablasl_4}]\bsy{\theta})_{\bsy{AB}} =& \, \bsy{\hat{\omega}}\,(\bsy{\nablasl_3}\bsy{\theta})_{\bsy{AB}}-\bsy{\hat{\omegab}}\,(\bsy{\nablasl_4}\bsy{\theta})_{\bsy{AB}}+ 2\,{(\bsy{\nablasl\theta})^{\bsy{\sharp_1}}}{}^{\bsy{C}}_{\bsy{AB}}(\bsy{\eta}-\bsy{\etab})_{\bsy{C}} \label{comm_form_6}\\
&+2\,\bsy{\eta}^{\bsy{C}}\bsy{\theta}_{\bsy{CB}}\,\bsy{\etab}_{\bsy{A}}-2\,\bsy{\etab}^{\bsy{C}}\bsy{\theta}_{\bsy{CB}}\,\bsy{\eta}_{\bsy{A}}+2\,\bsy{\yb}^{\bsy{C}}\bsy{\theta}_{\bsy{CB}}\,\bsy{\xi}_{\bsy{A}} -2\,\bsy{\xi}^{\bsy{C}}\bsy{\theta}_{\bsy{CB}}\,\bsy{\yb}_{\bsy{A}}  +2\,\bsy{\sigma}\,\bsy{\slashed{\varepsilon}}^{\bsy{\sharp_2}}\bsy{{}_A^C}\bsy{\theta}_{\bsy{CB}} \nonumber\\
&+2\,\bsy{\eta}^{\bsy{C}}\bsy{\theta}_{\bsy{AC}}\,\bsy{\etab}_{\bsy{B}}-2\,\bsy{\etab}^{\bsy{C}}\bsy{\theta}_{\bsy{AC}}\,\bsy{\eta}_{\bsy{B}}+2\,\bsy{\yb}^{\bsy{C}}\bsy{\theta}_{\bsy{AC}}\,\bsy{\xi}_{\bsy{B}} -2\,\bsy{\xi}^{\bsy{C}}\bsy{\theta}_{\bsy{AC}}\,\bsy{\yb}_{\bsy{B}}  +2\,\bsy{\sigma}\,\bsy{\slashed{\varepsilon}}^{\bsy{\sharp_2}}\bsy{{}_B^C}\bsy{\theta}_{\bsy{AC}} \, . \nonumber
\end{align}
The commutation formulae are analogous to the ones derived in \cite{Taylor_thesis}.~Note that, in our case, one has to keep track of the order of the indices of $\bsy{\chi}$ and $\bsy{\chib}$.

\subsection{The nonlinear system of equations} \label{sec_nonlinear_system_eqns}

We assume that $(\bsy{\mathcal{M}},\bsy{g})$ solves the vacuum Einstein equations
\begin{equation*}
\bsy{\ric} (\bsy{g})=0 \, .
\end{equation*}
The nonlinear system of equations for the connections coefficients and curvature components relative to a local null frame $\bsy{\mathcal{N}}$ is presented below.~The formal derivation of the system appears in Section \ref{sec_formal_derivation_EVE}.

\medskip

\begin{remark}
It is crucial, at this stage, that the frame $\bsy{\mathcal{N}}$ is allowed to be non-integrable.~The system of equations of this section differs from the one appearing in \cite{StabMink}, where the frame $\bsy{\mathcal{N}}$ is assumed to be integrable.~The antitraces $(\bsy{\slashed{\varepsilon}\cdot\chi})$ and $(\bsy{\slashed{\varepsilon}\cdot\chib})$ generate new terms in our equations.~Some of the elliptic equations of \cite{StabMink} may now be viewed as transport equations for $(\bsy{\slashed{\varepsilon}\cdot\chi})$ and $(\bsy{\slashed{\varepsilon}\cdot\chib})$.~We also note that, in the equations for symmetric quantities, the second fundamental forms appearing in the equations of \cite{StabMink} are replaced by the symmetric part of the second fundamental forms (e.g.~in the first variational formulae).~The Gauss equation is written in terms of the corrected Gauss curvature \eqref{def_gauss_curvature_corrected}.
\end{remark}

\medskip

\begin{remark}
If the frame $\bsy{\mathcal{N}}$ is assumed to be integrable, then the identities
\begin{align*}
(\bsy{\slashed{\varepsilon}\cdot\chi})&=0 \, , & (\bsy{\slashed{\varepsilon}\cdot\chib})&=0 
\end{align*}
hold and our system of equations reduces to the system of equations appearing in \cite{StabMink}.
\end{remark}

\subsubsection{Null structure equations} \label{sec_null_structure_equations}

We have the first variational formulae
\begin{align}
\bsy{\bsy{\slashed{\mathcal{L}}_{4}}\,\slashed{g}}&=2\,\bsy{\chih}+ (\bsy{\textbf{tr} \bsy{\chi}})\,\bsy{\slashed{g}}\, , &
\bsy{\bsy{\slashed{\mathcal{L}}_{3}}\,\slashed{g}}&=2\,\bsy{\chibh}+ (\textbf{tr} \bsy{\chib})\,\bsy{\slashed{g}}  \, ,
\end{align} 
the second variational formulae
\begin{align}
\bsy{\nablasl_4 \chih}+(\bsy{\textbf{tr}\chi})\,\bsy{\chih}-\bsy{\omegah \chih }=-2\,\bsy{\slashed{\mathcal{D}}_2^{\star}\, \xi} +(\bsy{\eta}+\bsy{\etab}+2\,\bsy{\zeta})\,\bsy{\widehat{\otimes} \, \xi} -\bsy{\alpha } \, ,
\end{align}
\begin{align}
\bsy{\nablasl_3 \chibh}+(\textbf{tr}\bsy{\chib})\,\bsy{\chibh}-\bsy{\omegabh \,\chibh}=-2\,\bsy{\slashed{\mathcal{D}}_2^{\star} \yb}   + (\bsy{\eta}+\bsy{\etab}-2\,\bsy{\zeta})\,\bsy{\widehat{\otimes}\,\yb} -\bsy{\alphab} \, ,
\end{align}
the Raychaudhuri equations
\begin{align}
\bsy{\nablasl_4} (\bsy{\textbf{tr}\chi})+ \frac{1}{2}\,(\bsy{\textbf{tr}\chi})^2-\bsy{\omegah}\,(\bsy{\textbf{tr}\chi})= -(\bsy{\chih},\bsy{\chih})+\frac{1}{2}\,(\bsy{\slashed{\varepsilon}\cdot\chi})^2+2\, \bsy{\slashed{\textbf{div}} \,\xi}  +2\,(\bsy{\eta}+\bsy{\etab}+2\,\bsy{\zeta}, \bsy{\xi})  \, ,
\end{align}
\begin{align}
\bsy{\nablasl_3} (\textbf{tr}\bsy{\chib})+ \frac{1}{2}\,(\textbf{tr}\bsy{\chib})^2-\bsy{\omegabh}\,(\textbf{tr}\bsy{\chib}) = -(\bsy{\chibh},\bsy{\chibh})+\frac{1}{2}\,(\bsy{\slashed{\varepsilon}\cdot\chib})^2+2\, \bsy{\slashed{\textbf{div}} \yb} +2\,(\bsy{\eta}+\bsy{\etab}-2\,\bsy{\zeta}, \bsy{\yb})  \, ,
\end{align}
the mixed transport equations
\begin{align}
\bsy{\nablasl_4 \chibh}+\frac{1}{2}\,(\bsy{\textbf{tr}\chi})\,\bsy{\chibh}+\bsy{\omegah\chibh}=& \, -2\,\bsy{\slashed{\mathcal{D}}_2^{\star} \etab}-\frac{1}{2}\,(\textbf{tr}\bsy{\chib})\,\bsy{\chih}+\frac{1}{2}\,(\bsy{\slashed{\varepsilon}\cdot\chib})(^{\bsy{\star}}\bsy{\chih})-\frac{1}{2}\,(\bsy{\slashed{\varepsilon}\cdot\chi})(^{\bsy{\star}}\bsy{\chibh}) \\ &+\bsy{\etab\,\widehat{\otimes}\,\etab}+\bsy{\xi\,\widehat{\otimes}\, \yb}   \, , \nonumber
\end{align}
\begin{align}
\bsy{\nablasl_3 \chih}+\frac{1}{2}\,(\textbf{tr}\bsy{\chib})\,\bsy{\chih}+\bsy{\omegabh\chih}=& \, -2\,\bsy{\slashed{\mathcal{D}}_2^{\star} \, \eta}-\frac{1}{2}\,(\bsy{\textbf{tr}\chi})\,\bsy{\chibh}+\frac{1}{2}\,(\bsy{\slashed{\varepsilon}\cdot\chi})(^{\bsy{\star}}\bsy{\chibh})-\frac{1}{2}\,(\bsy{\slashed{\varepsilon}\cdot\chib})(^{\bsy{\star}}\bsy{\chih}) \\ &+\bsy{\eta\,\widehat{\otimes}\,\eta}+\bsy{\yb\,\widehat{\otimes} \, \xi} \, , \nonumber
\end{align}
\begin{align}
\bsy{\nablasl_4}( \textbf{tr}\bsy{\chib})+ \frac{1}{2}\,(\bsy{\textbf{tr}\chi})(\textbf{tr}\bsy{\chib})+\bsy{\omegah}\,(\textbf{tr}\bsy{\chib})=& \, -(\bsy{\chih},\bsy{\chibh})+\frac{1}{2}\,(\bsy{\slashed{\varepsilon}\cdot\chi})(\bsy{\slashed{\varepsilon}\cdot\chib}) +2\,(\bsy{\etab} ,\bsy{\etab})+2\,\bsy{\rho} \\ &+2\, \bsy{\slashed{\textbf{div}} \etab} +2\,(\bsy{\xi} , \bsy{\yb})  \, , \nonumber 
\end{align}
\begin{align}
\bsy{\nablasl_3 }(\bsy{\textbf{tr}\chi})+ \frac{1}{2}\,(\textbf{tr}\bsy{\chib})(\bsy{\textbf{tr}\chi})+\bsy{\omegabh}\,(\bsy{\textbf{tr}\chi})=& \, -(\bsy{\chih},\bsy{\chibh})+\frac{1}{2}\,(\bsy{\slashed{\varepsilon}\cdot\chib})(\bsy{\slashed{\varepsilon}\cdot\chi}) +2\,(\bsy{\eta},\bsy{\eta})+2\,\bsy{\rho} \\ & +2\, \bsy{\slashed{\textbf{div}} \,\eta}+2\,(\bsy{\yb}, \bsy{\xi}) \, , \nonumber
\end{align} 
the transport equations for the antitraces of the second fundamental forms
\begin{align}
\bsy{\nablasl_4} (\bsy{\slashed{\varepsilon}\cdot\chi})+( \bsy{\textbf{tr}\chi})(\bsy{\slashed{\varepsilon}\cdot\chi})-\bsy{\omegah}\, (\bsy{\slashed{\varepsilon}\cdot\chi})=  2\,(\bsy{\etab}+2\,\bsy{\zeta})\bsy{\wedge \, \xi} +2\,\bsy{\xi \wedge \eta} +2\, \bsy{\slashed{\textbf{curl}}\, \xi} \, , 
\end{align}
\begin{align}
\bsy{\nablasl_3} (\bsy{\slashed{\varepsilon}\cdot\chib})+( \textbf{tr}\bsy{\chib})(\bsy{\slashed{\varepsilon}\cdot\chib})-\bsy{\omegabh} \,(\bsy{\slashed{\varepsilon}\cdot\chib})=   2\,(\bsy{\eta}-2\,\bsy{\zeta})\bsy{\wedge\yb} +2\,\bsy{\yb\wedge \etab}+2\, \bsy{\slashed{\textbf{curl}} \, \yb} \, ,
\end{align}
\begin{align}
\bsy{\nablasl_4} (\bsy{\slashed{\varepsilon}\cdot\chib})+\frac{1}{2}\,(\bsy{\textbf{tr}\chi})(\bsy{\slashed{\varepsilon}\cdot\chib})+\bsy{\omegah}\,(\bsy{\slashed{\varepsilon}\cdot\chib})=& \, -\bsy{\chih\wedge\chibh} -\frac{1}{2}\,(\textbf{tr}\bsy{\chib})(\bsy{\slashed{\varepsilon}\cdot\chi})+2\,\bsy{\xi \wedge \yb}  \\ &+2\,\bsy{\sigma}+2\,\bsy{\slashed{\textbf{curl}}\, \etab} \, , \nonumber 
\end{align}
\begin{align}
\bsy{\nablasl_3} (\bsy{\slashed{\varepsilon}\cdot\chi})+\frac{1}{2}\,(\textbf{tr}\bsy{\chib})(\bsy{\slashed{\varepsilon}\cdot\chi})+\bsy{\omegabh}\,(\bsy{\slashed{\varepsilon}\cdot\chi})=& \, -\bsy{\chibh\wedge\chih}-\frac{1}{2}\,(\bsy{\textbf{tr}\chi})(\bsy{\slashed{\varepsilon}\cdot\chib}) +2\,\bsy{\yb\wedge \,\xi} \\ &-2\,\bsy{\sigma}+2\,\bsy{\slashed{\textbf{curl}}\, \eta }   \nonumber
\end{align}
and the transport equations
\begin{align} \label{EVE_4_eta}
\bsy{\nablasl_4 \eta   } =\bsy{\nablasl_3 \xi} -\bsy{\chi^{\sharp_1}\cdot}(\bsy{\eta}-\bsy{\etab})+2\,\bsy{\omegabh  \xi }  - \bsy{\beta}  \, ,
\end{align}
\begin{align} \label{EVE_3_etab}
\bsy{\nablasl_3 \etab   } =\bsy{\nablasl_4\yb} +\bsy{\chib^{\sharp_1}\cdot}(\bsy{\eta}-\bsy{\etab})+2\,\bsy{\omegah  \yb }  + \bsy{\betab} \, ,
\end{align}
\begin{align}
\bsy{ \nablasl_4 \omegabh}+\bsy{\nablasl_3 \omegah}=2\,(\bsy{\eta},\bsy{\etab})-2 \,(\bsy{\xi}, \bsy{\yb}) -2\,\bsy{\omegah\omegabh}   -2\,(\bsy{\eta}-\bsy{\etab},\bsy{\zeta})-2\,\bsy{\rho}  
\end{align}
and the transport equations for the torsion
\begin{align} \label{EVE_4_zeta}
\bsy{\nablasl_4\zeta}=-\bsy{\nablasl\omegah}+\bsy{\chi^{\sharp_2}\cdot}(\bsy{\etab}-\bsy{\zeta})-\bsy{\chib^{\sharp_2}\cdot \, \xi}  -\bsy{\omegah}\,(\bsy{\etab}+\bsy{\zeta}) +\bsy{\omegabh \xi}- \bsy{\beta } \, ,
\end{align}
\begin{align} \label{EVE_3_zeta}
\bsy{\nablasl_3\zeta}=\bsy{\nablasl\omegabh}-\bsy{\chib^{\sharp_2}\cdot}(\bsy{\eta}+\bsy{\zeta})+\bsy{\chi^{\sharp_2}\cdot\yb } +\bsy{\omegabh}\,(\bsy{\eta}-\bsy{\zeta}) -\bsy{\omegah\yb}- \bsy{\betab } 
\end{align}
and the elliptic equation for the torsion
\begin{align}
\bsy{\slashed{\textbf{curl}} \, \zeta}=-\frac{1}{2}\,\bsy{\chih\wedge\chibh}+\frac{1}{4}\,(\bsy{\textbf{tr}\chi})(\bsy{\slashed{\varepsilon}\cdot\chib})-\frac{1}{4}\,(\textbf{tr}\bsy{\chib})(\bsy{\slashed{\varepsilon}\cdot\chi})-\frac{1}{2}\,(\bsy{\slashed{\varepsilon}\cdot\chib})\,\bsy{\omegah}+\frac{1}{2}\,(\bsy{\slashed{\varepsilon}\cdot\chi})\,\bsy{\omegabh}+\bsy{\sigma} \, .
\end{align} 
We have the elliptic Codazzi equations
\begin{align}
\bsy{\slashed{\textbf{div}} \chih}=& \, \frac{1}{2}\,\bsy{\nablasl}( \bsy{\textbf{tr}\chi}) -\frac{1}{2}\,{}^{\bsy{\star}}\bsy{\nablasl}(\bsy{\slashed{\varepsilon}\cdot\chi})-\bsy{\chih^{\sharp}\cdot\,\zeta}-\frac{1}{2}\,(\bsy{\slashed{\varepsilon}\cdot\chi})\bsy{{}^{\star}\zeta}+\frac{1}{2}\,(\bsy{\textbf{tr}\chi})\,\bsy{\zeta} -(\bsy{\slashed{\varepsilon}\cdot\chib})\bsy{{}^{\star} \xi} \\ &-(\bsy{\slashed{\varepsilon}\cdot\chi})\bsy{{}^{\star}\eta} -\bsy{\beta} \, , \nonumber
\end{align}
\begin{align}
\bsy{\slashed{\textbf{div}} \chibh}=& \, \frac{1}{2}\,\bsy{\nablasl}( \textbf{tr}\bsy{\chib})-\frac{1}{2}\,{}^{\bsy{\star}}\bsy{\nablasl}(\bsy{\slashed{\varepsilon}\cdot\chib})+\bsy{\chibh^{\sharp}\cdot\,\zeta}+\frac{1}{2}\,(\bsy{\slashed{\varepsilon}\cdot\chib})\bsy{{}^{\star}\zeta}-\frac{1}{2}\,(\textbf{tr}\bsy{\chib})\,\bsy{\zeta} -(\bsy{\slashed{\varepsilon}\cdot\chi})\bsy{{}^{\star} \yb} \\ &-(\bsy{\slashed{\varepsilon}\cdot\chib})\bsy{{}^{\star}\etab} +\bsy{\betab}  \nonumber
\end{align} 
and the elliptic Gauss equation
\begin{align} \label{eq_nonlinear_Gauss}
\widetilde{\bsy{\slashed{K}}}=\frac{1}{2}\,(\bsy{\chih},\bsy{\chibh})-\frac{1}{4}\,(\bsy{\textbf{tr}\chi})(\textbf{tr}\bsy{\chib})-\frac{1}{4}\,(\bsy{\slashed{\varepsilon}\cdot\chi})(\bsy{\slashed{\varepsilon}\cdot\chib})-\bsy{\rho} \, .  
\end{align}
 
\medskip

\begin{remark} \label{rmk_null_structure_lie_derivative}
For future convenience, we note that the null structure equations \eqref{EVE_4_eta}, \eqref{EVE_3_etab}, \eqref{EVE_4_zeta} and \eqref{EVE_3_zeta} can be written in the alternative form
\begin{align*}
\bsy{\slashed{\mathcal{L}}}_{\bsy{4}} \,\bsy{\eta}-(\bsy{\slashed{\varepsilon}\cdot\chi}){}^{\bsy{\star}}\bsy{\eta} =\bsy{\nablasl_3\xi}+\bsy{\chi^{\sharp_1}\cdot}\,\bsy{\etab} +2\,\bsy{\omegabh \xi }  - \bsy{\beta}  \, ,
\end{align*}
\begin{align*}
\bsy{\slashed{\mathcal{L}}}_{\bsy{3}} \,\bsy{\etab}-(\bsy{\slashed{\varepsilon}\cdot\chib}){}^{\bsy{\star}}\bsy{\etab} =\bsy{\nablasl_4\yb} +\bsy{\chib^{\sharp_1}\cdot}\,\bsy{\eta}+2\,\bsy{\omegah  \yb }  + \bsy{\betab} \, ,
\end{align*}
\begin{align*}
\bsy{\slashed{\mathcal{L}}}_{\bsy{4}}\, \bsy{\zeta}=-\bsy{\nablasl\omegah}+\bsy{\chi^{\sharp_2}\cdot}\,\bsy{\etab}-\bsy{\chib^{\sharp_2}\cdot \,\xi}  -\bsy{\omegah}\,(\bsy{\etab}+\bsy{\zeta}) +\bsy{\omegabh \xi}- \bsy{\beta } \, ,
\end{align*}
\begin{align*}
\bsy{\slashed{\mathcal{L}}}_{\bsy{3}}\, \bsy{\zeta}=\bsy{\nablasl\omegabh}-\bsy{\chib^{\sharp_2}\cdot}\,\bsy{\eta}+\bsy{\chi^{\sharp_2}\cdot\yb } +\bsy{\omegabh}\,(\bsy{\eta}-\bsy{\zeta}) -\bsy{\omegah\yb}- \bsy{\betab }  \, .
\end{align*}
\end{remark}

\subsubsection{Bianchi equations}  \label{sec_bianchi_equations}

The Bianchi equations read
\begin{align}
\bsy{\nablasl_3\alpha}+\frac{1}{2}\,(\textbf{tr}\bsy{\chib})\,\bsy{\alpha}+2\,\bsy{\omegabh\alpha} +\frac{1}{2}\,(\bsy{\slashed{\varepsilon}\cdot\chib}) {\bsy{{}^{\star}\alpha}}= -2\,\bsy{\slashed{\mathcal{D}}_2^{\star}\,\beta}-3\,\bsy{\rho\chih}-3\,\bsy{\sigma}{\bsy{{}^{\star} \chih}}   +(4\,\bsy{\eta}+\bsy{\zeta})\,\bsy{\widehat{\otimes}\, \beta} \, ,
\end{align}
\begin{align}
\bsy{\nablasl_4\beta}+2\,(\bsy{\textbf{tr}\chi})\,\bsy{\beta}-\bsy{\omegah\beta}-2\,(\bsy{\slashed{\varepsilon}\cdot\chi})\bsy{{}^{\star}\beta} =\bsy{\slashed{\textbf{div}}\,\alpha}+(\bsy{\etab^{\sharp}}+2\,\bsy{\zeta^{\sharp}})\bsy{\cdot\,\alpha}+3\,\bsy{\rho\, \xi}+3\,\bsy{\sigma {}^{\star}\xi}  \, ,
\end{align}
\begin{align}
\bsy{\nablasl_3\beta}+(\textbf{tr}\bsy{\chib})\,\bsy{\beta}+\bsy{\omegabh\beta}+(\bsy{\slashed{\varepsilon}\cdot\chi})\bsy{{}^{\star}\beta}  =\bsy{\slashed{\mathcal{D}}_1^{\star}}(-\bsy{\rho},\bsy{\sigma})+3\,\bsy{\rho\,\eta}+3\,\bsy{\sigma {}^{\star}\eta}+2\,\bsy{\chih^{\sharp}\cdot\betab}+\bsy{\yb^{\sharp}\cdot\,\alpha }  \, ,
\end{align}
\begin{align}
\bsy{\nablasl_4 \rho}+\frac{3}{2}\,(\bsy{\textbf{tr}\chi})\,\bsy{\rho } =\bsy{\slashed{\textbf{div}}\,\beta}+(2\,\bsy{\etab}+\bsy{\zeta},\bsy{\beta})-\frac{1}{2}\,(\bsy{\chibh},\bsy{\alpha})-2\,(\bsy{\xi},\bsy{\betab})-\frac{3}{2}\,(\bsy{\slashed{\varepsilon}\cdot\chi}) \,\bsy{\sigma} \, , 
\end{align}
\begin{align}
\bsy{\nablasl_4 \sigma}+\frac{3}{2}\,(\bsy{\textbf{tr}\chi})\,\bsy{\sigma} =-\bsy{\slashed{\textbf{curl}}\,\beta}-(2\,\bsy{\etab}+\bsy{\zeta})\bsy{\wedge\,\beta}+\frac{1}{2}\,\bsy{\chibh\wedge\,\alpha}-2\,\bsy{\xi\wedge\betab}+\frac{3}{2}\,(\bsy{\slashed{\varepsilon}\cdot\chi})\, \bsy{\rho}  \, ,
\end{align}
\begin{align}
\bsy{\nablasl_3 \rho}+\frac{3}{2}\,(\textbf{tr}\bsy{\chib})\,\bsy{\rho} =-\bsy{\slashed{\textbf{div}}\betab}-(2\,\bsy{\eta}-\bsy{\zeta},\bsy{\betab})-\frac{1}{2}\,(\bsy{\chih},\bsy{\alphab})+2\,(\bsy{\yb},\bsy{\beta})+\frac{3}{2}\,(\bsy{\slashed{\varepsilon}\cdot\chib})\, \bsy{\sigma } \, , 
\end{align}
\begin{align}
\bsy{\nablasl_3 \sigma}+\frac{3}{2}\,(\textbf{tr}\bsy{\chib})\,\bsy{\sigma}  =-\bsy{\slashed{\textbf{curl}}\betab}-(2\,\bsy{\eta}-\bsy{\zeta})\bsy{\wedge\betab}-\frac{1}{2}\,\bsy{\chih\wedge\alphab}-2\,\bsy{\yb\wedge\,\beta}-\frac{3}{2}\,(\bsy{\slashed{\varepsilon}\cdot\chib})\, \bsy{\rho } \, ,
\end{align}
\begin{align}
\bsy{\nablasl_4\betab}+(\bsy{\textbf{tr} \chi})\,\bsy{\betab}+\bsy{\omegah\betab}+(\bsy{\slashed{\varepsilon}\cdot\chib})\bsy{{}^{\star}\betab}  =\bsy{\slashed{\mathcal{D}}_1^{\star}}(\bsy{\rho},\bsy{\sigma})-3\,\bsy{\rho\,\etab}+3\,\bsy{\sigma {}^{\star}\etab}+2\,\bsy{\chibh^{\sharp}\cdot\,\beta} -\bsy{\xi^{\sharp}\cdot\alphab} \, ,  
\end{align}
\begin{align}
\bsy{\nablasl_3\betab}+2\,(\textbf{tr}\bsy{\chib})\,\bsy{\betab}-\bsy{\omegabh\betab}-2\,(\bsy{\slashed{\varepsilon}\cdot\chib})\bsy{{}^{\star}\betab}=-\bsy{\slashed{\textbf{div}}\alphab}-(\bsy{\eta^{\sharp}}-2\,\bsy{\zeta^{\sharp}})\bsy{\cdot\alphab}-3\,\bsy{\rho \yb}+3\,\bsy{\sigma {}^{\star}\yb }  \, ,
\end{align}
\begin{align}
\bsy{\nablasl_4\alphab}+\frac{1}{2}\,(\bsy{\textbf{tr}\chi})\,\bsy{\alphab}+2\,\bsy{\omegah\alphab}-\frac{1}{2}\,(\bsy{\slashed{\varepsilon}\cdot\chi}) {\bsy{{}^{\star}\alphab}} = 2\,\bsy{\slashed{\mathcal{D}}_2^{\star}\betab}-3\,\bsy{\rho\chibh}+3\,\bsy{\sigma{{}^{\star} \chibh}} -(4\,\bsy{\etab}-\bsy{\zeta})\,\bsy{\widehat{\otimes}\, \betab}  \, .
\end{align}

\subsection{The formal derivation of the equations} \label{sec_formal_derivation_EVE}

We derive the system of vacuum Einstein equations of Section \ref{sec_nonlinear_system_eqns}.~In this section only, we abandon the bolded notation.~In our derivation of the equations, we will box the terms that do not appear in the derivation of the system of equations of \cite{StabMink}.

\subsubsection{Null structure equations}

To derive the null structure equations, one applies the formula 
\begin{equation}
g(\nabla_{e_I}\nabla_{e_J} e_K,e_L)-g(\nabla_{e_J}\nabla_{e_I} e_K,e_L)-g(\nabla_{[e_I,e_J]} e_K,e_L)=R(e_L,e_K,e_I,e_J) \, , \label{appendix_form_null_structure}
\end{equation}
with $I,J,K,L=\left\lbrace 1,2,3,4 \right\rbrace$ and $R$ the Riemann curvature tensor of $(\mathcal{M},g)$.~We recall that, in our notation, $\nabla_{e_I}\nabla_{e_J} e_K=\nabla_{e_I}(\nabla_{e_J} e_K)$.~The reader should already note that, when $I,J=\left\lbrace 1,2 \right\rbrace$, \emph{the commutator in the last term on the left hand side of \eqref{appendix_form_null_structure} captures the non-integrability of the null frame $\mathcal{N}$}.

\medskip

We only derive a subset of the null structure equations.~The remaining equations can be derived by conjugation (see Remark \ref{rmk_conjugate_eqns}).~We also omit the straightforward derivation of the first variational formulae. 

\medskip

Before starting, let us note that, crucially, \emph{in our computations one has to keep track of the order of the frame indices of $\chi$ and $\chib$}.

\medskip

All the equalities in this first part of the computation are equalities between scalar functions and all the connection coefficients and curvature components are intended as $\mathfrak{D}_{\mathcal{N}}$ tensors.~We exploit the spacetime scalar formula \eqref{appendix_form_null_structure} to derive identities for frame components of $\mathfrak{D}_{\mathcal{N}}$ tensors.~The vacuum Einstein equations are not, at this stage, assumed to hold.~We present the computation of the first term in full detail, to clarify the steps.~The logic in computing the following terms is analogous and presented in a more compact fashion. 

\medskip

We compute 
\begin{align*}
g(\nabla_3 \nabla_A e_3, e_B) &= \nabla_3(\chib_{AB})-g(\nabla_Ae_3,\nabla_3 e_B) \\
&= \nabla_3(\chib_{AB})-g(\chib^{\sharp_2}{}_A^Ce_C+\zeta_Ae_3,\slashed{\underline{M}}_B^De_D+\eta_Be_3+\yb_Be_4) \\
&=\nabla_3(\chib_{AB})-\slashed{\underline{M}}_B^C\chib_{AC}+2\zeta_A\yb_B \\
&=\nablasl_3(\chib_{AB})-\slashed{\underline{M}}_B^C\chib_{AC}+2\zeta_A\yb_B \\
&= (\nablasl_3 \chib)_{AB}+\slashed{\underline{M}}_A^C\chib_{CB}+2\zeta_A\yb_B  \, , \\[5pt]
g(\nabla_A \nabla_3 e_3, e_B) &=2(\nablasl \yb)_{AB}+\omegabh \chib_{AB} \, , \\
g(\nabla_{[e_3,e_A]}e_3,e_B) &=(\slashed{\underline{M}}_{A}^C-{\chib^{\sharp_2}}{}_A^C)\chib_{CB}+2(\eta_A-\zeta_A)\yb_B +2\yb_A \etab_B
\end{align*}
and obtain the equation
\begin{equation} \label{eq_1}
 (\nablasl_3 \chib)_{AB}=2(\nablasl \yb)_{AB} +\omegabh \chib_{AB} -{\chib^{\sharp_2}}{}_A^C \chib_{CB}+2(\eta_A-2\zeta_A)\yb_B +2\yb_A \etab_B-\alphab_{AB} \, .
\end{equation}
Note that in the detailed computation of the first term we used the identities $\nabla_3(\chib_{AB})=\nablasl_3(\chib_{AB})$ and $(\nablasl_3 \chib)_{AB}=\nablasl_3(\chib_{AB})-\slashed{\underline{M}}_A^C\chib_{CB}-\slashed{\underline{M}}_B^C\chib_{AC}$, the former holding because we apply the covariant derivative to a scalar function (and thus one can exchange the connection) and the latter by definition of the induced covariant derivative $\nablasl$ applied to a $\mathfrak{D}_{\mathcal{N}}$ covariant tensor.

\medskip

Similarly, we compute
\begin{align*}
g(\nabla_4 \nabla_A e_3, e_B) &=  (\nablasl_4 \chib)_{AB}+\slashed{M}_{A}^C\chib_{CB} +2\zeta_A\etab_B  \, ,\\
g(\nabla_A \nabla_4 e_3, e_B)  &= 2(\nablasl \etab)_{AB}-\omegah\chib_{AB}  \, , \\
g(\nabla_{[e_4,e_A]}e_3,e_B)&=(\slashed{M}_{A}^C-{\chi^{\sharp_2}}{}_A^C)\chib_{CB}+2\xi_A \yb_B+2(\etab_A+\zeta_A)\etab_B     
\end{align*}
and obtain the equation
\begin{equation}  \label{eq_2}
(\nablasl_4 \chib)_{AB}= -2\zeta_A\etab_B +2(\nablasl \etab)_{AB}-\omegah\chib_{AB}-{\chi^{\sharp_2}}{}_A^C\chib_{CB}+2\xi_A \yb_B+2(\etab_A+\zeta_A)\etab_B +R(e_B,e_3,e_4,e_A) \, .
\end{equation}

\medskip

We compute
\begin{align*}
g(\nabla_C \nabla_B e_3, e_A) &= (\nablasl \chib)_{CBA}+{}_{C}\slashed{H}_{B}^D\chib_{DA}+\zeta_B\chib_{CA} \, ,\\
g(\nabla_B \nabla_C e_3, e_A) &= (\nablasl \chib)_{BCA}+{}_{B}\slashed{\underline{M}}_{C}^D\chib_{DA}+\zeta_C\chib_{BA} \, ,\\
g(\nabla_{[e_C,e_B]}  e_3, e_A) &=({}_{C}\slashed{H}_{B}^D-{}_{B}\slashed{\underline{M}}_{C}^D)\chib_{DA}\boxed{+(\chi_{CB}-\chi_{BC})\yb_A} \, \boxed{+(\chib_{CB}-\chib_{BC})\etab_A}
\end{align*}
and obtain the equation
\begin{align}  \label{eq_3}
(\nablasl \chib)_{CBA}+\zeta_B\chib_{CA}=& \, (\nablasl \chib)_{BCA}+\zeta_C\chib_{BA} \boxed{+(\chi_{CB}-\chi_{BC})\yb_A} \, \boxed{+(\chib_{CB}-\chib_{BC})\etab_A}  \\ &+R(e_A,e_3,e_C,e_B) \, . \nonumber
\end{align}

\medskip

We compute
\begin{align*}
g(\nabla_3 \nabla_A e_3, e_4) &=-2(\nablasl_3\zeta)_{A}-2\slashed{\underline{M}}_{A}^B\zeta_B-2{\chib^{\sharp_2}}{}_A^B\eta_B-2\omegabh \zeta_A \, , \\
g(\nabla_A \nabla_3 e_3, e_4) &= -2(\nablasl\omegabh)_A-2{\chi^{\sharp_2}}{}_A^B\yb_B-2\omegabh\zeta_A \, ,\\
g(\nabla_{[e_3,e_A]} e_3, e_4) &=-2(\slashed{\underline{M}}_{A}^B-{\chib^{\sharp_2}}{}_A^B)\zeta_B-2(\eta_A-\zeta_A)\omegabh +2\omegah\yb_A
\end{align*}
and obtain the equation
\begin{equation}  \label{eq_5}
 (\nablasl_3\zeta)_{A}=(\nablasl\omegabh)_A-{\chib^{\sharp_2}}{}_A^B\eta_B+{\chi^{\sharp_2}}{}_A^B\yb_B  -{\chib^{\sharp_2}}{}_A^B\zeta_B+\omegabh(\eta_A-\zeta_A) -\omegah\yb_A- \betab_A \, .
\end{equation}

\medskip

We compute
\begin{align*}
g(\nabla_3 \nabla_4 e_3, e_A) &=2(\nablasl_3 \etab)_A-2\omegah \yb_A \, ,\\
g(\nabla_4 \nabla_3 e_3, e_A) &=2(\nablasl_4\yb)_A +2\omegabh \etab_A \, , \\
g(\nabla_{[e_3,e_4]} e_3, e_A) &=2{\chib^{\sharp_1}}{}^B_A(\eta_B-\etab_B)+2\omegah \yb_A-2\omegabh\etab_A 
\end{align*}
and obtain the equation
\begin{equation}   \label{eq_6}
 (\nablasl_3 \etab)_A -(\nablasl_4\yb)_A    ={\chib^{\sharp_1}}{}^B_A(\eta_B-\etab_B)+2\omegah \yb_A   + \betab_A  \, .
\end{equation}

\medskip

We compute
\begin{align*}
g(\nabla_4 \nabla_3 e_3, e_4) &= -2\nablasl_4 \omegabh -4 \xi^A \yb_A +2\omegah\omegabh \, ,\\
g(\nabla_3 \nabla_4 e_3, e_4) &= 2\nablasl_3 \omegah -4\eta^A\etab_A+2 \omegah\omegabh \, ,\\
g(\nabla_{[e_4,e_3]} e_3, e_4) &= 4(\eta^A-\etab^A)\zeta_A+4\omegah\omegabh 
\end{align*}
and obtain the equation
\begin{equation}  \label{eq_7}
 \nablasl_4 \omegabh+\nablasl_3 \omegah=2\eta^A\etab_A-2 \xi^A \yb_A -2\omegah\omegabh   -2(\eta^A-\etab^A)\zeta_A-2\rho \, .
\end{equation}

\medskip

We compute
\begin{align*}
g(\nabla_A \nabla_B e_4, e_3) &= 2(\nablasl\zeta)_{AB}-\chi_{BD}\chi^{\sharp_2}{}_A^D-2\zeta_A\zeta_B\, , \\
g(\nabla_B \nabla_A e_4, e_3) &=2(\nablasl\zeta)_{BA}-\chi_{AD}\chi^{\sharp_2}{}_B^D-2\zeta_A\zeta_B \, ,\\
g(\nabla_{[e_A,e_B]} e_4, e_3) &=2({}_A\slashed{H}_{B}^C-{}_B\slashed{H}_{A}^C)\zeta_C \boxed{-\omegah(\chib_{AB}-\chib_{BA})} \, \boxed{+\omegabh(\chi_{AB}-\chi_{BA})}
\end{align*}
and obtain the equation
\begin{align}  \label{eq_8}
 2(\nablasl\zeta)_{AB}-2(\nablasl\zeta)_{BA}-\chi_{BD}\chi^{\sharp_2}{}{}_A^D &+\chi_{AD}\chi^{\sharp_2}{}_B^D -2({}_A\slashed{H}_{B}^C-{}_B\slashed{H}_{A}^C)\zeta_C \\ &\boxed{+\omegah(\chib_{AB}-\chib_{BA})} \, \boxed{-\omegabh(\chi_{AB}-\chi_{BA})} =R(e_3,e_4,e_A,e_B) \, . \nonumber
\end{align}

\medskip

Following a slightly different logic, we compute
\begin{align*}
g(\nabla^2_{e_C,e_B} e_D,e_A)=& \, g(\nabla_C(\nabla_{e_B} e_D)-\nabla_{\nabla_C e_B}e_D,e_A) \,  \\
=& \, g(\nabla_C(\nablasl_{B} e_D+\frac{1}{2}\,\chi_{BD}e_3+\frac{1}{2}\,\chib_{BD}e_4),e_A) \\ &-g(\nabla_{\nablasl_{C} e_B+\frac{1}{2}\,\chi_{CB}e_3+\frac{1}{2}\,\chib_{CB}e_4}e_D,e_A) \,   \\
=& \, g(\nabla_C(\nablasl_{B} e_D),e_A)+\frac{1}{2}\,\chi_{BD}\,g(\nabla_C e_3,e_A)+\frac{1}{2}\,\chib_{BD}g(\nabla_C e_4,e_A) \\ &-g (\nabla_{\nablasl_{C} e_B}e_D,e_A) -\frac{1}{2}\,\chi_{CB}\, g(\nabla_3 e_D,e_A)-\frac{1}{2}\,\chib_{CB} g(\nabla_4 e_D,e_A)\\
=& \, g(\nablasl_C(\nablasl_{B} e_D),e_A)+\frac{1}{2}\,\chi_{BD}\,\chib_{CA}+\frac{1}{2}\,\chib_{BD}\,\chi_{CA} \\ &-g (\nablasl_{\nablasl_{C} e_B}e_D,e_A) -\frac{1}{2}\,\chi_{CB}\, g(\nablasl_3 e_D,e_A)-\frac{1}{2}\,\chib_{CB} g(\nablasl_4 e_D,e_A) \\
=& \, \slashed{g}(\nablasl^2_{e_C,e_B} e_D,e_A)+\frac{1}{2}\,\chi_{BD}\chib_{CA}+\frac{1}{2}\,\chi_{CA}\chib_{BD}-\frac{1}{2}\,\chi_{CB}\,\slashed{g}(\nablasl_3e_D,e_A) \\ &-\frac{1}{2}\,\chib_{CB}\slashed{g}(\nablasl_4e_D,e_A) \, ,  \\
g(\nabla^2_{e_B,e_C} e_D,e_A)=& \,  \slashed{g}(\nablasl^2_{e_B,e_C} e_D,e_A)+\frac{1}{2}\,\chi_{CD}\chib_{BA}+\frac{1}{2}\,\chi_{BA}\chib_{CD}-\frac{1}{2}\,\chi_{BC}\,\slashed{g}(\nablasl_3e_D,e_A) \\ &-\frac{1}{2}\,\chib_{BC}\slashed{g}(\nablasl_4e_D,e_A) 
\end{align*}
and obtain the equation
\begin{align}  \label{eq_4}
 \boxed{\slashed{R}(e_A,e_D,e_C,e_B)} \, &+\frac{1}{2}\,(\chi_{BD}\chib_{CA}+\chi_{CA}\chib_{BD})-\frac{1}{2}\,(\chi_{CD}\chib_{BA}+\chi_{BA}\chib_{CD}) \\ & \boxed{-\frac{1}{2}\,(\chi_{CB}-\chi_{BC})\,\slashed{g}(\nablasl_3e_D,e_A)} \, \boxed{-\frac{1}{2}\,(\chib_{CB}-\chib_{BC})\,\slashed{g}(\nablasl_4e_D,e_A)}  \nonumber \\ &= R(e_A,e_D,e_C,e_B) \, . \nonumber
\end{align}
We note that we used the identities $g(\nabla_C(\nablasl_{B} e_D),e_A)=g(\nablasl_C(\nablasl_{B} e_D),e_A)$ and $g (\nabla_{\nablasl_{C} e_B}e_D,e_A)=g (\nablasl_{\nablasl_{C} e_B}e_D,e_A)$, holding by definition of the induced covariant derivative $\nablasl$ applied to a $\mathfrak{D}_{\mathcal{N}}$ vector field, and the identity $g_{AB}=\slashed{g}_{AB}$.~We also recall the definition of the map $\slashed{R}$ from \eqref{curvature_non_integrable_distribution} and the fact that the \emph{sum} of the boxed terms in the equation \eqref{eq_4} is tensorial (although the single terms separately are not).

\medskip

We now assume that $(\mathcal{M},g)$ is a solution to the vacuum Einstein equations 
\begin{equation*}
\ric (g)=0 \, .
\end{equation*}
Then, one has the identities 
\begin{align}
\tr \alpha&=0 \, , & \tr \alphab&=0 \, , \label{appendix_id_0}
\end{align}
where the trace is taken with respect to $\slashed{g}^{-1}$, and the identities
\begin{align}
R_{A34B} &=  \rho \, \slashed{g}_{AB}-\sigma \, \slashed{\varepsilon}_{AB} \, , \label{appendix_id_1}\\
R_{A3CB} &= (^{\star}\betab)_A \slashed{\varepsilon}_{CB} \, , \label{appendix_id_2}\\
R_{34AB} &= 2\,\sigma\,\slashed{\varepsilon}_{AB}  \label{appendix_id_3}
\end{align}
and
\begin{equation}
\slashed{g}^{AC}\slashed{g}^{BD}R_{ABCD} =  -2 \, \rho \, . \label{appendix_id_4}
\end{equation}
The identities \eqref{appendix_id_1}-\eqref{appendix_id_4} rely on the symmetries of the Riemann curvature tensor $R$ and the Bianchi identities for $R$.~The non-integrability of the null frame $\mathcal{N}$ does not enter into the computation.~The proof of the identities \eqref{appendix_id_1}-\eqref{appendix_id_4} is thus analogous to that of \cite{StabMink}.

\medskip

The identities \eqref{appendix_id_1}, \eqref{appendix_id_2} and \eqref{appendix_id_3} can be used to replace the curvature terms in the equations \eqref{eq_2}, \eqref{eq_3} and \eqref{eq_8} respectively.~For instance, one can use the identity \eqref{appendix_id_1} to write the equation \eqref{eq_2} as
\begin{equation*}  
(\nablasl_4 \chib)_{AB}= -2\zeta_A\etab_B +2(\nablasl \etab)_{AB}-\omegah\chib_{AB}-{\chi^{\sharp_2}}{}_A^C\chib_{CB}+2\xi_A \yb_B+2(\etab_A+\zeta_A)\etab_B +\rho \, \slashed{g}_{AB}+\sigma \, \slashed{\varepsilon}_{AB} \, .
\end{equation*}
With these last replacements of the curvature terms, all the equations \eqref{eq_1}-\eqref{eq_8} can now be viewed as equations for frame components of $\mathfrak{D}_{\mathcal{N}}$ tensors, and can thus be written in \emph{geometric form} as equations for $\mathfrak{D}_{\mathcal{N}}$ tensors.~For example, one can write the equation \eqref{eq_2} as
\begin{equation*}  
\nablasl_4 \chib= -2\zeta\otimes \etab +2\nablasl \etab-\omegah\chib-\chi\times\chib +2\xi\otimes \yb+2(\etab+\zeta) \otimes \etab +\rho \, \slashed{g}-\sigma \, \slashed{\varepsilon} \, .
\end{equation*}

\medskip

To obtain the system of equations in its final form, some additional manipulations are needed. 

\medskip

By taking the \emph{trace} (with respect to $\slashed{g}^{-1}$) part, the \emph{antitrace} (with respect to $\slashed{\varepsilon}^{-1}$) part and the \emph{symmetric traceless} part of the equation \eqref{eq_1}, one splits the equation \eqref{eq_1} into the three equations
\begin{gather*}
\nablasl_3 (\text{tr}\chib)+ \frac{1}{2}(\text{tr}\chib)^2-\omegabh (\text{tr}\chib) = -(\chibh,\chibh)\boxed{+\frac{1}{2}(\slashed{\varepsilon}\cdot\chib)^2} +2\, \slashed{\text{div}} \yb +2(\eta+\etab-2\zeta, \yb) \, ,  \\[5pt]
\boxed{\nablasl_3 (\slashed{\varepsilon}\cdot\chib)+(\text{tr}\chib)(\slashed{\varepsilon}\cdot\chib)-\omegabh (\slashed{\varepsilon}\cdot\chib)}=   2(\eta-2\zeta)\wedge\yb +2\yb\wedge \etab+2\, \slashed{\text{curl}} \, \yb \, , \\[5pt]
\nablasl_3 \chibh+(\text{tr}\chib)\chibh-\omegabh \chibh=-2\slashed{\mathcal{D}}_2^{\star} \yb   + (\eta+\etab-2\zeta)\htimes\yb -\alphab
\end{gather*}
respectively.~Note that we used the identities
\begin{align*}
(\chib,\chib)&=(\chibh,\chibh)+\frac{1}{2}\,(\text{tr}\chib)^2\boxed{-\frac{1}{2}\,(\slashed{\varepsilon}\cdot\chib)^2} \\
\chib \wedge \chib&= \boxed{(\text{tr}\chib)(\slashed{\varepsilon}\cdot\chib)} 
\end{align*}
and
\begin{align*}
(\chib\times \chib)-\frac{1}{2}\,(\slashed{g}^{AB}(\chib\times \chib)_{AB})\,\slashed{g}-\frac{1}{2}\,(\slashed{\varepsilon}^{AB}(\chib\times \chib)_{AB})\,\slashed{\varepsilon}=(\text{tr}\chib)\chibh 
\end{align*}
and the second of the identities \eqref{appendix_id_0}.

\medskip

By taking the \emph{trace} part, the \emph{antitrace} part and the \emph{symmetric traceless} part of equation \eqref{eq_2}, one obtains
\begin{gather*}
\nablasl_4( \text{tr}\chib)+ \frac{1}{2}(\text{tr}\chi)(\text{tr}\chib)+\omegah(\text{tr}\chib)= -(\chih,\chibh)\boxed{+\frac{1}{2}\,(\slashed{\varepsilon}\cdot\chi)(\slashed{\varepsilon}\cdot\chib)} +2(\etab ,\etab)+2\rho+2\, \slashed{\text{div}} \etab +2(\xi , \yb) \, ,\\[5pt]
\boxed{\nablasl_4 (\slashed{\varepsilon}\cdot\chib)+\frac{1}{2}\,(\text{tr}\chi)(\slashed{\varepsilon}\cdot\chib)+\omegah(\slashed{\varepsilon}\cdot\chib)}= -\chih  \wedge \chibh\boxed{-\frac{1}{2}\,(\text{tr}\chib)(\slashed{\varepsilon}\cdot\chi)}+2 \xi \wedge \yb +2 \sigma+2\, \slashed{\text{curl}}\, \etab  \, ,\\[5pt]
\nablasl_4 \chibh+\frac{1}{2}(\text{tr}\chi)\chibh+\omegah\chibh = -2 \slashed{\mathcal{D}}_2^{\star} \etab-\frac{1}{2}(\text{tr}\chib)\chih \boxed{ +\frac{1}{2}\,(\slashed{\varepsilon}\cdot\chib)({}^{\star}\chih)} \, \boxed{-\frac{1}{2}\,(\slashed{\varepsilon}\cdot\chi)({}^{\star}\chibh)}  + \etab\htimes\etab + \xi \htimes \yb 
\end{gather*}
respectively.~Note that we used the identities
\begin{align*}
(\chi,\chib)&=(\chih,\chibh)+\frac{1}{2}\,(\text{tr}\chi)(\text{tr}\chib)\boxed{-\frac{1}{2}\,(\slashed{\varepsilon}\cdot\chi)(\slashed{\varepsilon}\cdot\chib)} \, , \\
\chi  \wedge \chib &=\chih  \wedge \chibh \boxed{+\frac{1}{2}\,(\text{tr}\chi)(\slashed{\varepsilon}\cdot\chib)} \, \boxed{+\frac{1}{2}\,(\text{tr}\chib)(\slashed{\varepsilon}\cdot\chi)}
\end{align*}
and
\begin{align*}
(\chi\times \chib)-\frac{1}{2}\,(\slashed{g}^{AB}(\chi\times \chib)_{AB})\,\slashed{g}-\frac{1}{2}\,(\slashed{\varepsilon}^{AB}(\chi\times \chib)_{AB})\,\slashed{\varepsilon}=& \, \frac{1}{2}\,(\text{tr}\chi)\chibh+\frac{1}{2}\,(\text{tr}\chib)\chih \\ & \boxed{-\frac{1}{2}\,(\slashed{\varepsilon}\cdot\chib)({}^{\star}\chih)} \, \boxed{+\frac{1}{2}\,(\slashed{\varepsilon}\cdot\chi)({}^{\star}\chibh)}  \, .
\end{align*}

\medskip

By taking the trace of the equation \eqref{eq_3}, one obtains the Codazzi equation
\begin{align*}
\slashed{\text{div}} \chibh=& \, \boxed{-\frac{1}{2}\,{}^{\star}\nablasl(\slashed{\varepsilon}\cdot\chib)}+\chibh^{\sharp}\cdot\zeta \boxed{+\frac{1}{2}(\slashed{\varepsilon}\cdot\chib){}^{\star}\zeta}-\frac{1}{2}(\text{tr}\chib)\zeta+\frac{1}{2}\nablasl( \text{tr}\chib) \boxed{-(\slashed{\varepsilon}\cdot\chi){}^{\star} \yb} \\ & \boxed{-(\slashed{\varepsilon}\cdot\chib){}^{\star}\etab} + \betab \, .
\end{align*}
Note that one can equivalently take the trace of the equation \eqref{eq_3} relative to either $\slashed{g}^{CA}$ or $\slashed{g}^{BA}$.~The trace relative to $\slashed{g}^{CB}$ vanishes because the equation is antisymmetric in the horizontal indices $B,C$.~Since the equation \eqref{eq_3} is antisymmetric in two of the three horizontal indices, the equation is equivalent to the equation for $\mathfrak{D}_{\mathcal{N}}$ one-tensors obtained by taking any of its non-trivial traces or antitraces.

\medskip

By taking the trace of the equation \eqref{eq_4} relative to $\slashed{g}^{AC}\slashed{g}^{BD}$, one obtains the Gauss equation
\begin{equation*}
\boxed{\widetilde{\slashed{K}}}=\frac{1}{2}(\chih,\chibh)-\frac{1}{4}(\text{tr}\chi)(\text{tr}\chib)\boxed{-\frac{1}{4}(\slashed{\varepsilon}\cdot\chi)(\slashed{\varepsilon}\cdot\chib)}  -\rho  \, ,
\end{equation*}
where we used the identity \eqref{appendix_id_4} and the definition of the corrected Gauss curvature $\widetilde{\slashed{K}}$ from \eqref{def_gauss_curvature_corrected}.

\medskip

By taking the antitrace of equation \eqref{eq_8}, one obtains 
\begin{equation*}
\slashed{\text{curl}} \, \zeta=-\frac{1}{2}\, \chih\wedge\chibh\boxed{+\frac{1}{4}\,( \text{tr}\chi)( \slashed{\varepsilon}\cdot\chib)} \, \boxed{-\frac{1}{4}\,( \text{tr}\chib)(\slashed{\varepsilon}\cdot\chi)} \, \boxed{-\frac{1}{2}\,( \slashed{\varepsilon}\cdot\chib)\, \omegah} \, \boxed{+\frac{1}{2}\,( \slashed{\varepsilon}\cdot\chi)\, \omegabh} + \sigma \, .
\end{equation*}
Since the equation \eqref{eq_8} is antisymmetric in the two horizontal indices, it is equivalent to the scalar equation obtained by taking its antitrace.

\medskip

\begin{remark} \label{rmk_conjugate_eqns}
As already mentioned, the remaining null structure equations can be immediately deduced from the ones derived above.~To do that, one exchanges $\nablasl_4$-derivatives with $\nablasl_3$-derivatives (and vice-versa) and the barred quantities with the unbarred quantities (and vice-versa).~We note that $\rho$ remains unchanged under conjugation, while $\zeta$ and $\sigma$ are to be replaced by $-\zeta$ and $-\sigma$ respectively.~We also note that one substitutes $\beta$ with $-\betab$ and $\betab$ with $-\beta$. 
\end{remark}

\medskip

The derivation of the null structure equations of Section \ref{sec_nonlinear_system_eqns} is now complete.

\subsubsection{Bianchi equations}

We now assume that $(\mathcal{M},g)$ is a solution to the vacuum Einstein equations 
\begin{equation*}
\ric (g)=0 \, .
\end{equation*}
To derive the Bianchi equations, one applies the contracted Bianchi identities\footnote{These are equivalent to the second Bianchi identities $\nabla_{[I} R_{JK]LM}=0$ when $(\mathcal{M},g)$ solves the vacuum Einstein equations.}
\begin{equation}  \label{second_Bianchi_ids}
(\nabla^IR)_{IJKL}=0 \, ,  
\end{equation}
with $I,J,K,L=\left\lbrace 1,2,3,4 \right\rbrace$ and $R$ the Riemann curvature tensor of $(\mathcal{M},g)$.

\medskip

As for the null structure equations, we only derive a subset of the Bianchi equations.~The remaining equations can be derived by conjugation or by Hodge duality (see Remark \ref{rmk_conjugate_eqns_bis}).

\medskip

Before starting, let us note that, crucially, \emph{in our computations one has to keep track of the order of the frame indices of $\chi$ and $\chib$}.

\medskip

All the equalities in this first part of the computation are equalities between scalar functions and all the connection coefficients and curvature components are intended as $\mathfrak{D}_{\mathcal{N}}$ tensors.~We exploit the spacetime scalar identity \eqref{second_Bianchi_ids} to derive identities for frame components of $\mathfrak{D}_{\mathcal{N}}$ tensors.~We present the computation of the first term in full detail, to clarify the steps.~The logic in computing the following terms is analogous and presented in a more compact fashion.

\medskip

We compute
\begin{align*}
(\nabla_3 R)_{A4B4} =& \,  \nabla_3(R_{A4B4})-R(\nabla_3e_A,e_4,e_B,e_4)-R(e_A,\nabla_3e_4,e_B,e_4) \\ &-R(e_A,e_4,\nabla_3e_B,e_4)-R(e_A,e_4,e_B,\nabla_3e_4) \\[5pt]
=& \,  \nabla_3(\alpha_{AB})-\slashed{\underline{M}}_A^C\alpha_{CB}-2\eta_A\beta_B-2\eta^CR_{ACB4}+\omegabh\alpha_{AB} \\ &-\slashed{\underline{M}}_B^C\alpha_{AC}-2\eta_B\beta_A-2\eta^CR_{A4BC}+\omegabh\alpha_{AB} \\[5pt]
=& \, \nablasl_3(\alpha_{AB})-\slashed{\underline{M}}_A^C\alpha_{CB}-\slashed{\underline{M}}_B^C\alpha_{AC}+2\omegabh\alpha_{AB} \\ &-2\eta_A\beta_B-2\eta_B\beta_A+2\eta^C({}^{\star}\beta)_A\slashed{\varepsilon}_{BC}+2\eta^C({}^{\star}\beta)_B\slashed{\varepsilon}_{AC}  \\[5pt]
=& \,(\nablasl_3\alpha)_{AB}+2\omegabh\alpha_{AB}-4(\eta\hat{\otimes}\beta)_{AB} \, , 
\end{align*}
where we used the identities $\nabla_3(\alpha_{AB})=\nablasl_3(\alpha_{AB})$ and $(\nablasl_3 \alpha)_{AB}=\nablasl_3(\alpha_{AB})-\slashed{\underline{M}}_A^C\alpha_{CB}-\slashed{\underline{M}}_B^C\alpha_{AC}$, the former holding because we apply the covariant derivative to a scalar function (and thus one can exchange the connection) and the latter by definition of the induced covariant derivative $\nablasl$ applied to a $\mathfrak{D}_{\mathcal{N}}$ covariant tensor.~We also used the conjugate of the identity \eqref{appendix_id_2} and the identity
\begin{equation*}
2\eta^C({}^{\star}\beta)_A\slashed{\varepsilon}_{BC}=2\eta_B\beta_A-2(\beta\hat{\otimes}\eta)_{AB} \, .
\end{equation*}

\medskip

Similarly, we compute
\begin{align}
(\nabla_4 R)_{A3B3} =& \, (\nablasl_4\alphab)_{AB}+2\omegah\alphab_{AB}+4(\etab\hat{\otimes}\betab)_{AB} \, , \label{Bianchi_term_1}\\[5pt]
(\nabla_4 R)_{A4B3} =& \, -(\nablasl_4\rho)\slashed{g}_{AB}-(\nablasl_4\sigma)\slashed{\varepsilon}_{AB}-4\xi_A\betab_B+2(\betab\hat{\otimes}\xi)_{AB} \label{Bianchi_term_1b} \\ &+4\etab_B \beta_A-2(\beta\hat{\otimes}\etab)_{AB}  \, , \nonumber\\[5pt]
(\nabla_4 R)_{A434} =& \, 2(\nablasl_4\beta)_A-2\omegah\beta_A-2\etab^B\alpha_{AB}-6\xi_A\rho-6({}^{\star}\xi)_A\sigma \, , \label{Bianchi_term_2}\\[5pt]
(\nabla_3 R)_{A434} =& \,  2 (\nablasl_3\beta)_A+2\omegabh\beta_A -2\yb^B\alpha_{AB}-6\eta_A\rho-6({}^{\star}\eta)_A\sigma \, , \label{Bianchi_term_3}\\[5pt]
(\nabla_A R)_{B4C4} =&\,  (\nablasl_A\alpha)_{BC}+2\zeta_A\alpha_{BC}-2(\chi_{AB}\beta_C+\chi_{AC}\beta_B-\chi_{AD}\beta^D\slashed{g}_{BC})  \, , \label{Bianchi_term_4}\\[5pt]
(\nabla_A R)_{B434}=& \,  2(\nablasl_A\beta)_B+2\zeta_A\beta_B-\chib^{\sharp_2}{}_A^C\alpha_{BC}-3\rho\chi_{AB}-3\sigma\slashed{\varepsilon}^{\sharp_2}{}_B^C\chi_{AC}   \label{Bianchi_term_5} \, , \\[5pt]
(\nabla_4R)_{3 434} =& \, 4\nablasl_4\rho-8(\etab,\beta)+8(\xi,\betab) \label{Bianchi_term_5b}
\end{align}
and the contracted terms
\begin{align}
(\nabla^B R)_{A4B4} =& \, (\nablasl^B\alpha)_{AB}+2\zeta^B\alpha_{AB}-2(\tr\chi)\beta_A\boxed{+2(\slashed{\varepsilon}\cdot\chi)({}^{\star}\beta)_A} \, , \label{Bianchi_term_contracted_1}\\[5pt]
(\nabla^B R)_{A3B3} =& \, (\nablasl^B\alphab)_{AB}-2\zeta^B\alphab_{AB}+2(\tr\chib)\betab_A \boxed{-2(\slashed{\varepsilon}\cdot\chib)({}^{\star}\betab)_A} \, , \label{Bianchi_term_contracted_2}
\end{align}
where to compute the identities \eqref{Bianchi_term_contracted_1} and \eqref{Bianchi_term_contracted_2} we used the identity
\begin{equation}
-\chi^{\sharp_1}{}^B_A\beta_B-(\text{tr}\chi)\beta_A+\chi^{BD}({}^{\star}\beta)_B\slashed{\varepsilon}_{AD}+(\slashed{\varepsilon}\cdot\chi)({}^{\star}\beta)_A= -2(\text{tr}\chi)\beta_A+2(\slashed{\varepsilon}\cdot\chi)({}^{\star}\beta)_A \label{Bianchi_aux_comp_term}
\end{equation}
and its conjugate.~Using the identity \eqref{Bianchi_term_5} and its conjugate, we compute the sums
\begin{align}
\frac{1}{2}((\nabla_B R)_{A434}+(\nabla_A R)_{B434}) =& \, \frac{1}{2}\left(2(\nablasl_B\beta)_A+2\zeta_B\beta_A-\chib^{\sharp_2}{}_B^C\alpha_{AC}-3\rho\chi_{BA}-3\sigma\slashed{\varepsilon}^{\sharp_2}{}_A^C\chi_{BC}\right. \label{Bianchi_term_sum_1} \\ &\left. +2(\nablasl_A\beta)_B+2\zeta_A\beta_B-\chib^{\sharp_2}{}_A^C\alpha_{BC}-3\rho\chi_{AB}-3\sigma\slashed{\varepsilon}^{\sharp_2}{}_B^C\chi_{AC}\right) \nonumber
\end{align}
and
\begin{align*}
\frac{1}{2}((\nabla_B R)_{A334}+(\nabla_A R)_{B334}) =& \, \frac{1}{2}\left(2(\nablasl_B\betab)_A-2\zeta_B\betab_A+\chi^{\sharp_2}{}_B^C\alphab_{AC}+3\rho\chib_{BA}-3\sigma\slashed{\varepsilon}^{\sharp_2}{}_A^C\chib_{BC}\right. \\ &\left. +2(\nablasl_A\betab)_B-2\zeta_A\betab_B+\chi^{\sharp_2}{}_A^C\alphab_{BC}+3\rho\chib_{AB}-3\sigma\slashed{\varepsilon}^{\sharp_2}{}_B^C\chib_{AC}\right) 
\end{align*}
respectively.~We compute the term
\begin{align}
(\nabla_C R)_{A3B4} =& \, (\nablasl_C\sigma)\slashed{\varepsilon}_{AB} -(\nablasl_C\rho)\slashed{g}_{AB}+\chib_{CA}\beta_B-\chi_{CB}\betab_A \label{Bianchi_term_6}\\
&+\chib^{\sharp_2}{}_{C}^D({}^{\star}\beta)_B\slashed{\varepsilon}_{AD}-\chi^{\sharp_2}{}_{C}^D({}^{\star}\betab)_A\slashed{\varepsilon}_{BD} \, , \nonumber
\end{align}
where we used the identities \eqref{appendix_id_1} and (the conjugate of) \eqref{appendix_id_2}, and compute the following contractions
\begin{align}
(\nabla^A R)_{A434} =& \, 2\slashed{\divv}\beta+2(\zeta,\beta)-(\chib,\alpha)-3\rho(\tr\chi) \boxed{-3\sigma(\slashed{\varepsilon}\cdot\chi)}  \, , \label{Bianchi_term_7}\\[5pt]
(\nabla^A R)_{A334} =& \,  2\slashed{\divv}\betab-2(\zeta,\betab)+(\chi,\alphab)+3\rho(\tr\chib) \boxed{-3\sigma(\slashed{\varepsilon}\cdot\chib)} \, , \label{Bianchi_term_8}\\[5pt]
(\nabla^B R)_{A3B4} =& \, \slashed{\varepsilon}^{\sharp_2}{}_A^C\nablasl_C\sigma-\nablasl_A\rho +2(\chibh\cdot\beta)_{A} -(\tr \chi)\betab_A \boxed{-(\slashed{\varepsilon}\cdot\chi)({}^{\star}\betab)_A}   \, , \label{Bianchi_term_9} \\[5pt]
(\nabla^B R)_{B3A4} =& \, -\slashed{\varepsilon}^{\sharp_2}{}_A^C\nablasl_C\sigma-\nablasl_A\rho-2(\chih\cdot\betab)_{A}+(\tr \chib)\beta_A \boxed{+(\slashed{\varepsilon}\cdot\chib)({}^{\star}\beta)_A} \label{Bianchi_term_10} 
\end{align}
of the identity \eqref{Bianchi_term_5}, the conjugate of the identity \eqref{Bianchi_term_5}, the identity \eqref{Bianchi_term_6} and the conjugate of the identity \eqref{Bianchi_term_6} respectively.~We note that, in computing \eqref{Bianchi_term_9} and \eqref{Bianchi_term_10}, we used again the identity \eqref{Bianchi_aux_comp_term} and its conjugate.

\medskip

We now apply the contracted Bianchi identities \eqref{second_Bianchi_ids}.~We compute 
\begin{align}
0&=(\nabla^{I}R)_{I 434} \nonumber\\ &=(\nabla^{A}R)_{A 434}+(\nabla^{3}R)_{3 434} \nonumber\\
&=(\nabla^{A}R)_{A 434}-\frac{1}{2}(\nabla_4R)_{3 434} \nonumber\\
&=  2\slashed{\divv}\beta+2(\zeta,\beta)-(\chib,\alpha)-3\rho(\tr\chi) \boxed{-3\sigma(\slashed{\varepsilon}\cdot\chi)} -2\nablasl_4 \rho+4(\etab,\beta)-4(\xi,\betab) \, , \label{Bianchi_rho_computation}
\end{align}
where we used the identities \eqref{Bianchi_term_7} and \eqref{Bianchi_term_5b}.~We compute
\begin{align*}
0=& \, (\nabla^{I}R)_{I 4A4} \\ =& \, (\nabla^BR)_{B 4A4}-\frac{1}{2}(\nabla_4R)_{3 4A4} \\
=& \, (\nablasl^B\alpha)_{AB}+2\zeta^B\alpha_{AB}-2(\tr\chi)\beta_A \boxed{+2(\slashed{\varepsilon}\cdot\chi)({}^{\star}\beta)_A} -(\nablasl_4\beta)_A+\omegah\beta_A+\etab^B\alpha_{AB} \\ & +3\xi_A\rho+3({}^{\star}\xi)_A\sigma \, ,
\end{align*}
where we used the identities \eqref{Bianchi_term_4} and \eqref{Bianchi_term_contracted_1}.~We compute
\begin{align*}
0=& \, (\nabla^{I}R)_{I 3A4} \\ =& \, (\nabla^BR)_{B 3A4}-\frac{1}{2}(\nabla_3R)_{4 3A4} \\
=& \, -\slashed{\varepsilon}^{\sharp_2}{}_A^C\nablasl_C\sigma-\nablasl_A\rho-2(\chih\cdot\betab)_{A}+(\tr \chib)\beta_A \boxed{+(\slashed{\varepsilon}\cdot\chib)({}^{\star}\beta)_A}  + (\nablasl_3\beta)_A+\omegabh\beta_A -\yb^B\alpha_{AB} \\ & -3\eta_A\rho-3({}^{\star}\eta)_A\sigma \, ,
\end{align*}
where we used the identities \eqref{Bianchi_term_3} and \eqref{Bianchi_term_10}.~Using the Bianchi identities, we compute
\begin{align*}
(\nabla_3R)_{A4B4}&= \frac{1}{2}((\nabla_3R)_{A4B4}+(\nabla_3R)_{B4A4})\\
&=\frac{1}{2}\left((\nabla_B R)_{A434}+(\nabla_A R)_{B434}\right)+\frac{1}{2}\left((\nabla_4 R)_{A3B4}+(\nabla_4 R)_{B3A4}\right) 
\end{align*}
and, using the identities \eqref{Bianchi_term_1}, \eqref{Bianchi_term_1b} and \eqref{Bianchi_term_sum_1}, we obtain the equation
\begin{align*}
 (\nablasl_3\alpha)_{AB}+2\omegabh\alpha_{AB}-4(\eta\hat{\otimes}\beta)_{AB} =& \, \frac{1}{2}\left(2(\nablasl_B\beta)_A+2\zeta_B\beta_A-\chib^{\sharp_2}{}_B^C\alpha_{AC}-3\rho\chi_{BA}-3\sigma\slashed{\varepsilon}^{\sharp_2}{}_A^C\chi_{BC}\right. \\ &\left. +2(\nablasl_A\beta)_B+2\zeta_A\beta_B-\chib^{\sharp_2}{}_A^C\alpha_{BC}-3\rho\chi_{AB}-3\sigma\slashed{\varepsilon}^{\sharp_2}{}_B^C\chi_{AC}\right) \\ & +(-\nablasl_4\rho-2(\xi,\betab)+2(\etab,\beta))\slashed{g}_{AB} \, .
\end{align*}
The term $\nablasl_4\rho$ in the last line can be replaced by using the equation \eqref{Bianchi_rho_computation}.

\medskip

\begin{remark} \label{rmk_conjugate_eqns_bis}
The remaining Bianchi equations can be immediately deduced from the ones derived above as explained in Remark \ref{rmk_conjugate_eqns}.~The Bianchi equations for $\sigma$ can be deduced by taking the Hodge dual of the Bianchi equations for $\rho$. 
\end{remark}

\medskip

The derivation of the Bianchi equations of Section \ref{sec_nonlinear_system_eqns} is now complete.

\section{The Kerr exterior manifold}  \label{sec_Kerr_exterior_manifold}

In this section, we identify the Lorentzian manifold $(\bsy{\mathcal{M}},\bsy{g})$ and its Levi-Civita connection $\bsy{\nabla}$ from Section \ref{sec_nonlinear_EVE} with the \emph{Kerr exterior manifold} $(\mathcal{M},g_{a,M})$ and its Levi-Civita connection $\nabla_{a,M}$.~After \emph{defining} the Kerr exterior manifold $(\mathcal{M},g_{a,M})$, we shall discuss a particular choice of differentiable structure and null frame on $(\mathcal{M},g_{a,M})$.~For the latter, we will identify the null frame $\bsy{\mathcal{N}}$ from Section \ref{sec_nonlinear_EVE} with the \emph{algebraically special frame} $\mathcal{N}_{\text{as}}$ of $(\mathcal{M},g_{a,M})$ and the induced horizontal distribution $\bsy{\mathfrak{D}}_{\bsy{\mathcal{N}}}$ with $\mathfrak{D}_{\mathcal{N}_{\text{as}}}$.~The algebraically special frame of $(\mathcal{M},g_{a,M})$ will be defined and, crucially, shown to be a \emph{non-integrable} frame for $|a|>0$.

\medskip

Treating the Kerr geometry relative to the non-integrable frame $\mathcal{N}_{\text{as}}$ requires (and already motivates) the general formalism from Section \ref{sec_nonlinear_EVE}.~The connection coefficients and curvature components of $(\mathcal{M},g_{a,M})$ relative to $\mathcal{N}_{\text{as}}$ solve the nonlinear system of null structure and Bianchi equations of Section \ref{sec_nonlinear_system_eqns}, whereas they do \emph{not} solve the nonlinear system of null structure and Bianchi equations of \cite{StabMink}.

\subsection{Definition of the Kerr exterior manifold} \label{sec_definition_Kerr_exterior_manifold}

We define the manifold-with-boundary 
\begin{equation}  \label{ambient_manifold}
\mathcal{M}:= (-\infty,\infty) \times [0,\infty) \times \mathbb{S}^2 
\end{equation}
with coordinates $\bar{t}^*\in (-\infty,\infty)$, $y^*\in [0,\infty)$ and standard spherical coordinates $(\bar{\theta},\bar{\phi}^*)\in\mathbb{S}^2$.~As usual, the coordinates $(\bar{\theta},\bar{\phi}^*)$ only cover a subset of $\mathbb{S}^2$.~Nonetheless, one can extend $(\bar{\theta},\bar{\phi}^*)$ as \emph{functions} on the whole $\mathbb{S}^2$.~Similarly, the coordinate vector field $\partial_{\bar{\phi}^*}$ can be smoothly extended to a \emph{global} vector field on $\mathbb{S}^2$ which vanishes at the poles of $\mathbb{S}^2$. 

\medskip

We define the vector fields
\begin{align*}
T& :=\partial_{\bar{t}^*} \, ,  &  \Phi& :=\partial_{\bar{\phi}^*}
\end{align*}
on $\mathcal{M}$.~We will denote by $\widetilde{\Phi}$ the smooth extension of $\Phi$ to the whole $\mathbb{S}^2$.~The vector field $\widetilde{\Phi}$ is a \emph{global} vector field on $\mathbb{S}^2$, and thus on $\mathcal{M}$.

\medskip

We define the \emph{(future) event horizon} as the boundary
\begin{equation*}
\mathcal{H}^+:=\partial\mathcal{M}  
\end{equation*} 
of $\mathcal{M}$ and \emph{(future) null infinity} as the formal hypersurface
\begin{equation*}
\mathcal{I}^+ :=\left\lbrace \bar{t}^*=\infty \right\rbrace \, .
\end{equation*}
We note that $\mathcal{H}^+=\left\lbrace y^*=0 \right\rbrace$. 

\medskip

Given real parameters $a$ and $M$, with $|a|<M$, we define the positive constants
\begin{equation*}
r_{\pm}:=M\pm\sqrt{M^2-a^2}
\end{equation*}
and a new coordinate $\bar{r}_{a,M}=\bar{r}_{a,M}(y^*)$ such that
\begin{equation*}
\bar{r}_{a,M}:[0,\infty)\rightarrow [r_+,\infty) \, ,
\end{equation*}
with $\bar{r}_{a,M}(3M)=y^*(3M)$.~We will simply denote $\bar{r}_{a,M}$ by $\bar{r}$.~We define the two-spheres
\begin{equation*}
\mathbb{S}^2_{\bar{t}^*,\bar{r}} :=\left\lbrace \bar{t}^*,\bar{r} \right\rbrace \times \mathbb{S}^2 \, .
\end{equation*} 
The coordinates $$(\bar{t}^*,\bar{r},\bar{\theta},\bar{\phi}^*)$$ are referred to as \emph{Kerr coordinates}.~We note that $\mathcal{H}^+=\left\lbrace \bar{r}=r_+ \right\rbrace$.~We also note that the Kerr coordinates depend on the parameters $a$ and $M$, whereas the definition of the ambient manifold $\mathcal{M}$ and the vector fields $T$ and $\Phi$ are independent of parameters. 

\medskip

For fixed $|a|<M$, we define the smooth scalar functions
\begin{align*}
\Delta(\bar{r}) &:= (\bar{r}-r_+)(\bar{r}-r_-) \, , &
\Sigma(\bar{r},\bar{\theta}) &:= \bar{r}^2+a^2\cos^2\bar{\theta} \, .
\end{align*}
on $\mathcal{M}$.~We note that, strictly speaking, the function $\Sigma$ is well-defined on $\mathbb{S}^2_{\bar{t}^*,\bar{r}}$ only on the domain of the coordinates $(\bar{\theta},\bar{\phi}^*)$.~Nonetheless, one can smoothly extend $\Sigma$ to the whole $\mathbb{S}^2_{\bar{t}^*,\bar{r}}$ and thus understand $\Sigma$ as a \emph{global} smooth function on each of the $\mathbb{S}^2_{\bar{t}^*,\bar{r}}$ (and thus on $\mathcal{M}$).

\medskip

We define the \emph{Kerr family of metrics} as the two-parameter family of Lorentzian metrics $g_{a,M}$ on $\mathcal{M}$ such that 
\begin{align}
g_{a,M}= &-\left(1-\frac{2\,M\,\bar{r}}{\Sigma}\right){d\bar{t}^*}^2+2\,d\bar{t}^*\, d\bar{r}+\Sigma\, {d\bar{\theta}}^2 +\frac{(\bar{r}^2+a^2)^2-\Delta\, a^2\sin^2\bar{\theta}}{\Sigma}\,\sin^2\bar{\theta} \, {d\bar{\phi}^*}^{2}   \label{def_kerr_metric} \\
&-2\,a\sin^2\bar{\theta} \, d\bar{r} \,d\bar{\phi}^*-\frac{4\,a\,M\,\bar{r}}{\Sigma}\,\sin^2\bar{\theta} \, d\bar{t}^*\, d\bar{\phi}^* \, .   \nonumber
\end{align}
The metric $g_{a,M}$ in the form \eqref{def_kerr_metric} is manifestly \emph{smooth} on $\mathcal{M}$, including on $\mathcal{H}^+$.~The event horizon $\mathcal{H}^+$ can be checked to be a \emph{null} hypersurface relative to $g_{a,M}$.~The vector fields $T$ and $\Phi$ are \emph{Killing vector fields} of $g_{a,M}$.

\medskip

The smooth Lorentzian manifold $(\mathcal{M},g_{a,M})$ will be referred to as the \emph{Kerr exterior manifold}.~One can check that $(\mathcal{M},g_{a,M})$ solves the vacuum Einstein equations.

\subsection{Kerr-star coordinates} \label{sec_Kerr_star_coords}

For fixed $|a|<M$, we define the smooth scalar function $\iota_{a,M}=\iota_{a,M}(\bar{r})$ such that $$\iota_{a,M}:[r_+,\infty)\rightarrow [0,1]$$ and
\begin{equation}
\iota(\bar{r})=
\begin{cases}
1 \quad r_+\leq \bar{r} \leq 15\,M/8 \\
0 \quad \bar{r}\geq 9\,M/4 
\end{cases} \, ,
\end{equation}
with
\begin{gather*}
\frac{\bar{r}^2+a^2}{\Delta}-\iota\,\left(\frac{\bar{r}^2+a^2}{\Delta}-1 \right) >0 \, , \\
2-\left( 1-\frac{2\,M\,\bar{r}^2}{\Sigma}\right) \left( \frac{\bar{r}^2+a^2}{\Delta}-\iota\,\left(\frac{\bar{r}^2+a^2}{\Delta}-1 \right) \right) >0 \, .
\end{gather*}

\medskip

We define \emph{Kerr-star coordinates}
\begin{equation} \label{kerr_star_coords}
(t^*,r,\theta,\phi^*)
\end{equation}
on $\mathcal{M}$ such that
\begin{align*}
t^*(\bar{t}^*,\bar{r})&= \bar{t}^*-\int_{r_0}^{\bar{r}}(1-\iota(r^{\prime}))\,\frac{{r^{\prime}}^2+a^2}{\Delta(r^{\prime})}+\iota(r^{\prime})\,dr^{\prime} \, , & r(\bar{r})&=\bar{r} \, , \\
\theta(\bar{\theta})&=\bar{\theta} \, , & \phi^*(\bar{\phi}^*,\bar{r})&=\bar{\phi}^*-\int_{r_0}^{\bar{r}}(1-\iota(r^{\prime}))\,\frac{a}{\Delta(r^{\prime})}\,dr^{\prime} \quad \text{mod} \, 2\pi \, ,
\end{align*}
with $r_0=9\,M/4$.~The coordinates \eqref{kerr_star_coords} are \emph{global} coordinates on $\mathcal{M}$, \ul{including on $\mathcal{H}^+$}.\footnote{Modulo the usual degeneration of the angular coordinates $(\theta,\phi^*)\in\mathbb{S}^2$.}~We define the two-spheres $$\mathbb{S}^2_{t^*,r} :=\left\lbrace t^*,r  \right\rbrace \times \mathbb{S}^2 \, .$$ The $\mathbb{S}^2_{t^*,r}$-spheres are spheres of constant Kerr coordinate functions $(\bar{t}^*,\bar{r})$.~The vector field $\widetilde{\Phi}$ thus remains a global, degenerate (at the poles) vector field on each of the $\mathbb{S}^2_{t^*,r}$ spheres (and thus on $\mathcal{M}$).~Locally, we have $\widetilde{\Phi}=\partial_{\phi^*}$.

\medskip

In Kerr-star coordinates, the Kerr metric $g_{a,M}$ reads
\begin{align}
g_{a,M}= &-\left(1-\frac{2\,M\,r}{\Sigma}\right){dt^*}^2+\iota\,\frac{4\,M\,r}{\Sigma}\,dt^*\, dr  -\frac{4\,a\,M\,r}{\Sigma}\,\sin^2\theta \, dt^*\, d\phi^* \label{kerr_metric_kerr_star}\\
&+\left(\frac{\Sigma}{\Delta}+\iota^2\,\frac{-4\,M^2\,r^2+a^2(r^2+2\,M\,r+a^2)\sin^2\theta-a^4\,\sin^4\theta}{\Delta\,\Sigma}\right)\,dr^2 \nonumber \\
&-\iota\,\frac{2\,a(2\,M\,r+\Sigma)\sin^2\theta}{\Sigma}\,dr\,d\phi^*
+\Sigma\, {d\theta}^2 +\frac{(r^2+a^2)^2-\Delta\, a^2\sin^2\theta}{\Sigma}\,\sin^2\theta \, {d\phi^*}^{2}  \, . \nonumber 
\end{align}
The metric $g_{a,M}$ in the form \eqref{kerr_metric_kerr_star} is manifestly smooth on $\mathcal{M}$, including on $\mathcal{H}^+$.\footnote{Note that $g_{rr}=1+2Mr/\Sigma$ for $r_+\leq r \leq 15\,M/8$.}

\medskip

For any $\tau^*\in\mathbb{R}$, the hypersurface 
\begin{equation} \label{def_S_tau}
\mathcal{S}_{\tau^*} :=\left\lbrace t^*=\tau^* \right\rbrace
\end{equation}
is a \emph{spacelike},\footnote{This can be checked and exploits the careful definition of the function $\iota(\bar{r})$.} asymptotically flat hypersurface intersecting $\mathcal{H}^+$ (see Figure \ref{fig:norm_hyp_kerr}).~The level sets of $r_+<r<\infty$ are \emph{timelike} hypersurfaces.~The level set $\left\lbrace r=r_+ \right\rbrace=\mathcal{H}^+$ is \emph{null} and the formal hypersurface $\left\lbrace t^*=\infty \right\rbrace=\mathcal{I}^+ $ is \emph{null}.

\medskip

\begin{figure}[H]

\centering
\begin{tikzpicture}[scale=1.5]

\draw (0,2)--(-1.2,0.8);

\node at (-1.0,1.4) {$\mathcal{H}^+$};
\node at (1.1,1.4) {$\mathcal{I}^+$};



\node at (0.3,0.8) {$\mathcal{S}_{\tau^*}$};

\draw plot [smooth] coordinates { (-0.6,1.4) (0.3,1) (1.2,0.8) };


\draw[dashed] (0,2)--(1.2,0.8);

\draw[dashed] (0,-0.4)--(-1.2,0.8);
\draw[dashed] (0,-0.4)--(1.2,0.8);

\end{tikzpicture}

\caption{}

\label{fig:norm_hyp_kerr}

\end{figure}

\medskip

\begin{remark}  \label{rmk_transfo_star_coords_to_BL}
The coordinates $(\bar{t}^*,\bar{r},\bar{\theta},\bar{\phi}^*)$ and $(t^*,r,\theta,\phi^*)$ can be related to \emph{Boyer--Lindquist coordinates} $$(t_{\textup{bl}},r_{\textup{bl}},\theta_{\textup{bl}},\phi_{\textup{bl}})$$ on $\mathcal{M}\setminus\mathcal{H}^+$ by the coordinate transformations
\begin{align*}
\bar{t}^*(t_{\textup{bl}},r_{\textup{bl}})&=t_{\textup{bl}}+\int_{r_0}^{r_{\textup{bl}}}\frac{{r^{\prime}}^2+a^2}{\Delta(r^{\prime})}\,dr^{\prime}  \, ,  & \bar{r}(r_{\textup{bl}})&=r_{\textup{bl}} \, ,  \\
\bar{\theta}(\theta_{\textup{bl}})&=\theta_{\textup{bl}}\, , & \bar{\phi}^*(\phi_{\textup{bl}},r_{\textup{bl}})&=\phi_{\textup{bl}}+\int_{r_0}^{r_{\textup{bl}}}\frac{a}{\Delta(r^{\prime})}\,dr^{\prime} \quad \textup{mod} \, 2\pi 
\end{align*}
and
\begin{align*}
t^*(t_{\textup{bl}},r_{\textup{bl}})&= t_{\textup{bl}}+\int_{r_0}^{r_{\textup{bl}}}\iota(r^{\prime})\,\left(\frac{{r^{\prime}}^2+a^2}{\Delta(r^{\prime})}-1\right)\,dr^{\prime} \, , & r(r_{\textup{bl}})&=r_{\textup{bl}} \, , \\
\theta(\theta_{\textup{bl}})&=\theta_{\textup{bl}} \, , & \phi^*(\phi_{\textup{bl}},r_{\textup{bl}})&=\phi_{\textup{bl}}+\int_{r_0}^{r_{\textup{bl}}}\iota(r^{\prime})\, \frac{a}{\Delta(r^{\prime})}\,dr^{\prime} \quad \textup{mod} \, 2\pi \, .
\end{align*}
In Boyer--Lindquist coordinates, the Kerr metric $g_{a,M}$ reads
\begin{equation}
g_{a,M}=-\frac{\Delta}{\Sigma}\,(dt_{\textup{bl}}-a\,\sin^2\theta_{\textup{bl}}\,d\phi_{\textup{bl}})^2+\frac{\Sigma}{\Delta}\,dr_{\textup{bl}}^2+\Sigma \,d\theta_{\textup{bl}}^2+\frac{\sin^2\theta_{\textup{bl}}}{\Sigma}\,(a\, dt_{\textup{bl}}-(r_{\textup{bl}}^2+a^2)\,d\phi_{\textup{bl}})^2 \, . \label{kerr_metric_bl}
\end{equation}
\end{remark}

\medskip

\begin{remark}
The identities
\begin{align*}
(t^*,r,\theta,\phi^*)&\equiv (\bar{t}^*-\bar{r},\bar{r},\bar{\theta},\bar{\phi}^*) & &\textup{for}  \quad r_+\leq r \leq 15\,M/8  \, ,\\
(t^*,r,\theta,\phi^*)&\equiv (t_{\textup{bl}},r_{\textup{bl}},\theta_{\textup{bl}},\phi_{\textup{bl}}) &  &\textup{for} \quad r\geq 9\,M/4 
\end{align*}
hold up to additive constants.~The metric \eqref{kerr_metric_kerr_star} takes the form \eqref{kerr_metric_bl} for $r\geq 9\,M/4$.
\end{remark}

\subsection{The algebraically special frame}   \label{sec_Kerr_algebr_special_frame}

We define the \emph{algebraically special} null vector fields\footnote{Later in the section, we will comment on the algebraically special nature of the vector fields \eqref{as_null_frame_vectors_1}-\eqref{as_null_frame_vectors_2}.}
\begin{align}  
e_4^{\text{as}}&=\frac{r^2+a^2}{\Sigma}\,\left(1+\iota\,\left(1-\frac{\Delta}{r^2+a^2} \right)\right)\,\partial_{t^*} +\frac{\Delta}{\Sigma}\,\partial_r+\frac{a}{\Sigma}(1+\iota)\,\widetilde{\Phi} \, ,  \label{as_null_frame_vectors_1} \\  
e_3^{\text{as}}&=\frac{r^2+a^2}{\Delta}\,\left(1-\iota\,\left(1-\frac{\Delta}{r^2+a^2}\right)\right)\,\partial_{t^*}-\partial_r +\frac{a}{\Delta}\,(1-\iota)\,\widetilde{\Phi} \, . \label{as_null_frame_vectors_2}
\end{align}
The vector fields $e_4^{\text{as}}$ and $e_3^{\text{as}}$ are \emph{global},\footnote{We recall that $\widetilde{\Phi}$ is a global vector field on each of the two-spheres $\mathbb{S}^2_{t^*,r}$ (locally, $\widetilde{\Phi}=\partial_{\phi^*}$).~The vector field coefficients are smooth functions of the coordinates, and can be smoothly extended to global functions on the whole two-spheres $\mathbb{S}^2_{t^*,r}$.} \emph{regular}\footnote{We note that the coordinate components of $e_4^{\text{as}}$ and $e_3^{\text{as}}$ are bounded functions of regular coordinates on $\mathcal{M}$, including on $\mathcal{H}^+$.} and \emph{non-degenerate} vector fields on the whole manifold $\mathcal{M}$, including on $\mathcal{H}^+$.~One can check that the null vector field $e^{\text{as}}_3$ is \emph{geodesic}, i.e.~the identity
\begin{equation*}
\nabla_{e^{\text{as}}_3} e^{\text{as}}_3=0 
\end{equation*} 
holds on $\mathcal{M}$.~We also define the induced \emph{global}, \emph{regular} horizontal distribution
\begin{equation*}
\mathfrak{D}_{\mathcal{N}_{\text{as}}}:=\left\langle e^{\text{as}}_3,e^{\text{as}}_4 \right\rangle^{\perp} 
\end{equation*}
on $\mathcal{M}$.~One can complete the null vector fields \eqref{as_null_frame_vectors_1}-\eqref{as_null_frame_vectors_2} with a \emph{local} orthonormal basis
\begin{equation*}
(e^{\text{as}}_1,e^{\text{as}}_2)
\end{equation*}
of $\mathfrak{D}_{\mathcal{N}_{\text{as}}}$ to form the \emph{local} null frame $$\mathcal{N}_{\text{as}}=(e^{\text{as}}_1,e^{\text{as}}_2,e^{\text{as}}_3,e^{\text{as}}_4) \, ,$$ called the \emph{algebraically special frame} of the Kerr exterior manifold $(\mathcal{M},g_{a,M})$.

\medskip

\begin{remark}
In Boyer-Lindquist coordinates, the algebraically special frame vectors \eqref{as_null_frame_vectors_1} and \eqref{as_null_frame_vectors_2} read
\begin{align*}
e_4^{\textup{as}}&=\frac{r^2+a^2}{\Sigma}\,\partial_{t_{\textup{bl}}}+\frac{\Delta}{\Sigma}\,\partial_{r_{\textup{bl}}}+\frac{a}{\Sigma}\,\partial_{\phi_{\textup{bl}}}  \, , & e_3^{\textup{as}}&= \frac{r^2+a^2}{\Delta}\,\partial_{t_{\textup{bl}}}-\partial_{r_{\textup{bl}}}+\frac{a}{\Delta}\,\partial_{\phi_{\textup{bl}}}  \, .
\end{align*}
\end{remark}

\medskip

Crucially, the frame $\mathcal{N}_{\text{as}}$ is \emph{non-integrable} for $|a|>0$, with $\mathfrak{D}_{\mathcal{N}_{\text{as}}}$ a \emph{non-integrable} distribution.~This fact can be easily checked by inspecting the connection coefficients of $g_{a,M}$ relative to $\mathcal{N}_{\text{as}}$ presented in Section \ref{sec_Kerr_connection_coeff_curv_comps} and noting that both the second fundamental forms possess non-trivial antitraces.~As a consequence, the induced connection $\nablasl_{a,M}$ over the bundle of $\mathfrak{D}_{\mathcal{N}_{\text{as}}}$ tensors is \emph{not} the Levi-Civita connection of the induced metric $\slashed{g}_{a,M}$ (see Remark \ref{rmk_proj_connection_not_LC}).

\medskip

We point out three additional properties of the frame $\mathcal{N}_{\text{as}}$ holding along the event horizon $\mathcal{H}^+$:
\begin{itemize}
\item We have
\begin{equation} \label{intro_e4_horizon_killing}
e_4^{\text{as}}|_{\mathcal{H}^+} = 2\,\frac{r_+^2+a^2}{\Sigma(r_+)}\,\partial_{t^*} +2\,\frac{a}{\Sigma(r_+)}\,\widetilde{\Phi} \, , 
\end{equation}
i.e.~the vector field $e_4^{\text{as}}$ is tangent to $\mathcal{H}^+$ and in the span of the Killing vector fields $\partial_{t^*}$ and $\widetilde{\Phi}$. 
\item We have
\begin{equation*}
\mathfrak{D}_{\mathcal{N}_{\text{as}}}|_{\mathcal{H}^+}\subset T\mathcal{H}^+ \, ,
\end{equation*}
i.e.~the arbitrary local frame $(e^{\text{as}}_1,e^{\text{as}}_2)$ of $\mathfrak{D}_{\mathcal{N}_{\text{as}}}$ is tangent to $\mathcal{H}^+$.\footnote{One way to check this fact is to exhibit an explicit choice of $(e^{\text{as}}_1,e^{\text{as}}_2)$, e.g.~$e_1^{\text{as}}= (a^2 \,\sin(2\theta)/2\,\Sigma)\,\partial_{t^*}+(r/\Sigma)\,\partial_{\theta}+(a\,\cot\theta/\Sigma)\,\partial_{\phi^*}$ and
$e_2^{\text{as}}= (a\,r \,\sin \theta/\Sigma)\,\partial_{t^*}-(a\,\cos\theta/\Sigma)\,\partial_{\theta}+(r\,\csc\theta/\Sigma)\,\partial_{\phi^*}$.~Any local frame vector field of $\mathfrak{D}_{\mathcal{N}_{\text{as}}}$ is a linear combination of such $e_1^{\text{as}}$ and $e_2^{\text{as}}$, which are themselves manifestly tangent to $\mathcal{H}^+$.\label{footnote:e_12_kerr}}~Thus, the vector fields $$(e^{\text{as}}_4,e^{\text{as}}_1,e^{\text{as}}_2)$$ form a local basis of $T\mathcal{H}^+$ (and, in particular, generate the \emph{integrable} distribution $T\mathcal{H}^+\subset T\mathcal{M}$).
\item We have
\begin{equation} \label{eA_kerr_fermi_horizon}
\nablasl_{e_4^{\text{as}}}e_A^{\text{as}}|_{\mathcal{H}^+}=0
\end{equation}
for any local frame $(e^{\text{as}}_1,e^{\text{as}}_2)$ of $\mathfrak{D}_{\mathcal{N}_{\text{as}}}$.\footnote{One can check that $\nablasl_{e_4^{\text{as}}}e_A^{\text{as}}=0$ \emph{on $\mathcal{M}$} for the vector fields $(e^{\text{as}}_1,e^{\text{as}}_2)$ in footnote \ref{footnote:e_12_kerr}.~In view of the identity \eqref{intro_e4_horizon_killing}, any linear combination of such vector fields satisfies the identity \eqref{eA_kerr_fermi_horizon}.}
\end{itemize}

\subsection{Induced metric, connection coefficients and curvature components} \label{sec_Kerr_connection_coeff_curv_comps}

The induced metric, connection coefficients and curvature components of $g_{a,M}$ relative to $\mathcal{N}_{\text{as}}$ are defined as \emph{$\mathfrak{D}_{\mathcal{N}_{\text{as}}}$ tensors} (see Definition \ref{def_dn_tensor} with the identification $\bsy{\mathfrak{D}}_{\bsy{\mathcal{N}}}\equiv \mathfrak{D}_{\mathcal{N}_{\text{as}}}$).~Their explicit expression in coordinate form is given below.\footnote{We note that both their definition and explicit expression are independent of the choice of local vector fields $(e_1^{\text{as}},e_2^{\text{as}})$.~We also note that, for the induced metric and the covariant connection coefficients, we give the explicit expression of their \emph{canonical extension}.~See Remark \ref{rmk_canonical_ext_kerr}.}~Since $\mathfrak{D}_{\mathcal{N}_{\text{as}}}$ is a global, regular distribution on $\mathcal{M}$, the $\mathfrak{D}_{\mathcal{N}_{\text{as}}}$ tensors are defined \emph{globally} on $\mathcal{M}$ and are \emph{regular} quantities on the whole manifold $\mathcal{M}$, including on $\mathcal{H}^+$.

\medskip

In Boyer--Lindquist coordinates, the induced metric $\slashed{g}^{\circ}_{a,M}$ relative to $\mathcal{N}_{\text{as}}$ reads
\begin{equation} \label{kerr_induced_metric}
\slashed{g}^{\circ}_{a,M}=\frac{a^2}{\Sigma}\,\sin^2\theta_{\text{bl}}\,dt_{\text{bl}}^2-2\,\frac{a\,(r_{\text{bl}}^2+a^2)}{\Sigma}\,\sin^2\theta_{\text{bl}}\,dt_{\text{bl}}\,d\phi_{\text{bl}}+\Sigma\,d\theta_{\text{bl}}^2+\frac{(r_{\text{bl}}^2+a^2)^2}{\Sigma}\,\sin^2\theta_{\text{bl}}\,d\phi_{\text{bl}}^2 
\end{equation}
and the connection coefficients $\eta_{a,M}^{\circ}$, $\etab_{a,M}^{\circ}$ and $\zeta_{a,M}^{\circ}$ of $g_{a,M}$ relative to $\mathcal{N}_{\text{as}}$ read
\begin{align}
\eta_{a,M}^{\circ}&=-\frac{a^2r_{\text{bl}}}{\Sigma^2}\,\sin^2\theta_{\text{bl}}\,dt_{\text{bl}}^2-\frac{a^2\sin(2\theta_{\text{bl}})}{2\,\Sigma}\,d\theta_{\text{bl}}+\frac{a\,r_{\text{bl}}(r_{\text{bl}}^2+a^2)}{\Sigma^2}\,\sin^2\theta_{\text{bl}}\,d\phi_{\text{bl}} \, , \label{kerr_eta}\\[5pt]
\etab_{a,M}^{\circ}&= \frac{a^2r_{\text{bl}}}{\Sigma^2}\,\sin^2\theta_{\text{bl}}\,dt_{\text{bl}}^2-\frac{a^2\sin(2\theta_{\text{bl}})}{2\,\Sigma}\,d\theta_{\text{bl}}- \frac{a\,r_{\text{bl}}(r_{\text{bl}}^2+a^2)}{\Sigma^2}\,\sin^2\theta_{\text{bl}}\,d\phi_{\text{bl}} \, , \label{kerr_etab}\\[5pt]
\zeta_{a,M}^{\circ}&=-\etab_{a,M}^{\circ}-\frac{\Sigma}{r_{\text{bl}}^2+a^2}\,\nablasl^{\circ}\left(\frac{r_{\text{bl}}^2+a^2}{\Sigma} \right) \, . \label{kerr_zeta}
\end{align}

\medskip

In Kerr-star coordinates, the remaining connection coefficients of $g_{a,M}$ relative to $\mathcal{N}_{\text{as}}$ read
\begin{align*}
{\chih}_{a,M} &=0 \, , & {\chibh}_{a,M} &= 0 \, ,\\
(\slashed{\varepsilon}\cdot\chi)_{a,M} &= \frac{2\, a\, \Delta\cos \theta }{\Sigma^2} \, ,   & (\slashed{\varepsilon}\cdot\chib)_{a,M}&= \frac{ 2\,a \cos \theta}{\Sigma} \, , \\
(\text{tr}\chi)_{a,M} &= \frac{2\,r\, \Delta}{\Sigma^2} \, ,  &  (\text{tr}\chib)_{a,M} &= -\frac{2 \,r }{\Sigma} 
\end{align*}
and
\begin{align*}
\hat{\omega}_{a,M} &= -\frac{ 2\left(a^2   (M-r)\cos^2  \theta +a^2 r- M r^2\right)}{\Sigma^2} \, , &
\omegabh_{a,M} &= 0 \, , \\
\xi_{a,M} &=0 \, , &  \yb_{a,M} &=0 \, .
\end{align*}
The curvature components of $g_{a,M}$ relative to $\mathcal{N}_{\text{as}}$ read
\begin{align*}
\alpha_{a,M}&=0\, , & \alphab_{a,M}&=0 \, ,\\
\beta_{a,M} &= 0 \, , & \betab_{a,M} &= 0 \, ,\\
\rho_{a,M}&=\frac{ 2\,M\, r \left(3\, a^2 \cos^2 \theta - r^2\right)}{\Sigma^3}  \, , &  \sigma_{a,M} &= \frac{ 2\,a \, M \cos \theta  \left(3 \, r^2-a^2 \cos^2 \theta \right)}{\Sigma^3} \, .
\end{align*}
We note, in particular, the following identities for the connection coefficients\footnote{The second of the identities below is not immediately manifest from the identities \eqref{kerr_eta} and \eqref{kerr_zeta}, but can be easily checked by evaluating the identity relative to the vector fields $(e^{\text{as}}_1,e^{\text{as}}_2)$ in footnote \ref{footnote:e_12_kerr}.~We also note that these identities are geometric, and thus do not depend on our choice of coordinates to write the connection coefficients explicitly.} 
\begin{align*}
\omegabh_{a,M}&=0 \, , & \eta_{a,M}-\zeta_{a,M}&=0   \, , & \xi_{a,M}=\yb_{a,M} &=0 \, , & {\chih}_{a,M}={\chibh}_{a,M}&=0 \, .
\end{align*}
The identities for the curvature components
\begin{align*}
\alpha_{a,M}=\alphab_{a,M} &=0 \, , & \beta_{a,M}=\betab_{a,M} &=0 
\end{align*}
identify the null frame $\mathcal{N}_{\text{as}}$ as the algebraically special frame of the Kerr exterior manifold $(\mathcal{M},g_{a,M})$.

\medskip

\begin{remark} \label{rmk_canonical_ext_kerr}
In the identities \eqref{kerr_induced_metric}, \eqref{kerr_eta}, \eqref{kerr_etab} and \eqref{kerr_zeta}, we expressed the canonical extension (in the sense of Definition \ref{def_canonical_ext_cov_tensor}, with the identification $\bsy{\mathcal{M}}\equiv\mathcal{M}$ and $\bsy{\mathfrak{D}}_{\bsy{\mathcal{N}}}\equiv \mathfrak{D}_{\mathcal{N}_{\textup{as}}}$) of the induced metric $\slashed{g}_{a,M}$ and connection coefficients $\eta_{a,M}$, $\etab_{a,M}$ and $\zeta_{a,M}$, the latter being the actual $\mathfrak{D}_{\mathcal{N}_{\textup{as}}}$ tensors.~The non-integrability of the distribution $\mathfrak{D}_{\mathcal{N}_{\textup{as}}}$ does not allow an explicit expression of $\slashed{g}_{a,M}$, $\eta_{a,M}$, $\etab_{a,M}$ and $\zeta_{a,M}$ in coordinate form, in that $\mathfrak{D}_{\mathcal{N}_{\textup{as}}}$ does not admit a coordinate co-basis.
\end{remark}

\medskip

\begin{remark}
The connection coefficients and curvature components of $(\mathcal{M},g_{a,M})$ relative to $\mathcal{N}_{\textup{as}}$ solve the nonlinear system of null structure and Bianchi equations of Section \ref{sec_nonlinear_system_eqns}.
\end{remark}

\medskip

On the event horizon $\mathcal{H}^+$, the identities \eqref{intro_e4_horizon_killing} and \eqref{eA_kerr_fermi_horizon} imply
\begin{align*}
\nablasl_4\Gamma_{a,M}|_{\mathcal{H}^+}&=0 \, , & \nablasl_4 \psi_{a,M}|_{\mathcal{H}^+}&=0  
\end{align*}
for any connection coefficient $\Gamma_{a,M}$ and curvature component $\psi_{a,M}$ of $g_{a,M}$ relative to $\mathcal{N}_{\text{as}}$.

\medskip

\begin{remark}
In the definition of the algebraically special frame $\mathcal{N}_{\textup{as}}$, we made a choice of scaling for the null vector fields.~Our choice of scaling for $e_3^{\textup{as}}$ and $e_4^{\textup{as}}$ in \eqref{as_null_frame_vectors_1}-\eqref{as_null_frame_vectors_2} is \emph{uniquely} determined by the requirement that the frame extends \emph{regularly} to $\mathcal{H}^+$ and the null vector $e_3^{\textup{as}}$ is \emph{geodesic}.~Alternative scalings that keep the frame regular on $\mathcal{H}^+$ are of course possible, a notable one being the one setting $\nablasl\omegah_{a,M}=0$ and $\etab_{a,M}+\zeta_{a,M}=0$, but now $\omegabh_{a,M}\neq 0$. 
\end{remark}

\subsection{Commutation formulae}

We present the commutation formulae for $\mathfrak{D}_{\mathcal{N}_{\text{as}}}$ tensors.~The formulae are obtained by specialising the general commutation formulae of Section \ref{sec_commutation_formulae}, with the identification $(\bsy{\mathcal{M}},\bsy{g})\equiv (\mathcal{M}^*,g_{a,M})$ and $\bsy{\mathfrak{D}}_{\bsy{\mathcal{N}}}\equiv \mathfrak{D}_{\mathcal{N}_{\text{as}}}$.

\medskip

For any $\mathfrak{D}_{\mathcal{N}_{\text{as}}}$ one-tensor $\varsigma$ and $\mathfrak{D}_{\mathcal{N}_{\text{as}}}$ two-tensor $\theta$, we have the following commutation formulae
\begin{align} 
[\nablasl_4,\nablasl] \, \varsigma =& \,  -\chi_{a,M} \times ( \nablasl\varsigma )+(\etab_{a,M}+\zeta_{a,M})\otimes(\nablasl_{4}\varsigma) \label{kerr_comm_form_1}\\ & +(\etab_{a,M},\varsigma)\, \chi_{a,M}-(\chi^{\sharp_2}_{a,M}\cdot\varsigma)\otimes \etab_{a,M}  \, ,  \nonumber \\[5pt] 
[\nablasl_3,\nablasl]\, \varsigma =& \, -\chib_{a,M} \times (\nablasl\varsigma) \label{kerr_comm_form_2}\\ & +(\eta_{a,M},\varsigma)\,\chib_{a,M}-(\chib^{\sharp_2}_{a,M}\cdot\,\varsigma)\otimes \eta_{a,M} \, ,  \nonumber \\[5pt]
[\nablasl_3,\nablasl_4] \, \varsigma  =& \, \hat{\omega}_{a,M} \,(\nablasl_3 \varsigma)+ 2\,(\nablasl\varsigma)^{\sharp_1} \cdot (\eta_{a,M}-\etab_{a,M}) \label{kerr_comm_form_3}\\
& +2\,(\eta_{a,M},\varsigma)\,\etab_{a,M}-2\,(\etab_{a,M},\varsigma)\,\eta_{a,M}  +2\,\sigma_{a,M}({}^{\star}\varsigma) \nonumber
\end{align}
and 
\begin{align}
([\nablasl_4,\nablasl]\,\theta)_{ABC}=& \, -\chi^{\sharp_2}_{a,M}{}_A^D(\nablasl_D\theta)_{BC}+(\etab_{a,M}+\zeta_{a,M})_A(\nablasl_{4}\theta)_{BC} \label{kerr_comm_form_4} \\ & +\chi_{a,M}{}_{AB}\etab_{a,M}^D\theta_{DC}-\etab_{a,M}{}_B \chi_{a,M}^{\sharp_2}{}_A^D \theta_{DC} \nonumber \\ &+ \chi_{a,M}{}_{AC}\etab_{a,M}^D \theta_{BD} - \etab_{a,M}{}_C  \chi_{a,M}^{\sharp_2 }{}_A^D  \theta_{BD}  \, , \nonumber \\[5pt]
([\nablasl_3,\nablasl]\theta)_{ABC}=& \, -\chib_{a,M}^{\sharp_2 }{}_A^D(\nablasl_D\theta)_{BC} \label{kerr_comm_form_5}\\ & +\chib_{a,M}{}_{AB}\eta_{a,M}^D\theta_{DC}-\eta_{a,M}{}_B \chib^{\sharp_2 }_{a,M}{}_A^D \theta_{DC}  \nonumber\\ & +\chib_{a,M}{}_{AC}\eta_{a,M}^D\theta_{BD}-\eta_{a,M}{}_C \chib_{a,M}^{\sharp_2 }{}_A^D \theta_{BD} \, ,  \nonumber \\[5pt]
([\nablasl_3,\nablasl_4]\,\theta)_{AB} =& \, \hat{\omega}_{a,M}\,(\nablasl_3 \theta)_{AB}+ 2\,{(\nablasl\theta)^{\sharp_1}}{}^{C}_{AB}(\eta_{a,M}-\etab_{a,M})_{C} \label{kerr_comm_form_6}\\
&+2\,\eta_{a,M}^{C} \theta_{CB}\,\etab_{a,M}{}_{A}-2\,\etab_{a,M}^{C} \theta_{CB}\,\eta_{a,M}{}_{A} +2\,\sigma_{a,M}\,\slashed{\varepsilon}^{\sharp_2}{}_A^C \theta_{CB} \nonumber\\
&+2\,\eta_{a,M}^{C} \theta_{AC}\,\etab_{a,M}{}_{B}-2\,\etab_{a,M}^{C}\theta_{AC}\,\eta_{a,M}{}_{B} +2\,\sigma_{a,M}\,\slashed{\varepsilon}^{\sharp_2}{}_B^C\theta_{AC} \, . \nonumber
\end{align}

\subsection{Outgoing principal differentiable structures} \label{sec_Kerr_background_diff_structure}

In this section, we introduce the \emph{star-normalised outgoing principal differentiable structure} on $\mathcal{M}$, which will later serve as the background differentiable structure when formulating nonlinear perturbations of the Kerr exterior manifold $(\mathcal{M},g_{a,M})$ (see Section \ref{sec_family_metrics}).

\medskip

We start with a preliminary definition and a proposition.

\medskip

\begin{definition} \label{def_outgoing_principal_diff_structures}
An \emph{outgoing principal differentiable structure} of $(\mathcal{M},g_{a,M})$ is a (local) differentiable structures $(x^1,x^2,\vartheta^A)$ on $\mathcal{M}$, with $(\vartheta^1,\vartheta^2)\in\mathbb{S}^2$, such that 
\begin{align*}
e_4^{\textup{as}}(x^1)&=0 \, , & e_4^{\textup{as}}(\vartheta^A)&=0 \, .
\end{align*}
\end{definition}

\medskip

\begin{prop}  \label{prop_metric_ids_outgoing_principal_diff_str}
Let $(x^1,x^2,\vartheta^A)$ be an outgoing principal differentiable structure of $(\mathcal{M},g_{a,M})$.~Then, we have the identities
\begin{align*}
(g_{a,M})_{x^2x^2}&=0 \, , & \partial_{x^2} (g_{a,M})_{x^2\mu}+\mathfrak{h}_{a,M}\, (g_{a,M})_{x^2\mu}&=0
\end{align*}
for some smooth scalar function $\mathfrak{h}_{a,M}$ on $\mathcal{M}$, with $x^{\mu}=(x^1,\vartheta^A)$.
\end{prop}

\medskip

\begin{proof}
We have $e_4^{\text{as}}=h_{a,M}\partial_{x^2}$ for some smooth scalar function $h_{a,M}$ on $\mathcal{M}$.~The first identity follows immediately from $e_4^{\text{as}}$ being a null vector field.~We compute
\begin{align*}
\nabla_{e_4^{\text{as}}}(g_{a,M}(e_4^{\text{as}},\partial_{\mu}))&=\omegah_{a,M}\,g_{a,M}(e_4^{\text{as}},\partial_{\mu})+h^2_{a,M}\,g_{a,M}(\partial_{x^2},\nabla_{x^2}\partial_{\mu}) \\
&=\omegah_{a,M}\,g_{a,M}(e_4^{\text{as}},\partial_{\mu})+h^2_{a,M}\,g_{a,M}(\partial_{x^2},\nabla_{\mu}\partial_{x^2}) \\
&=\omegah_{a,M}\,g_{a,M}(e_4^{\text{as}},\partial_{\mu})+\frac{1}{2}\,h^2_{a,M}\,\partial_{\mu}((g_{a,M})_{x^2\,x^2}) \\
&=\omegah_{a,M}\,g_{a,M}(e_4^{\text{as}},\partial_{\mu})
\end{align*}
and then note that the left hand side can be written as follows
\begin{align*}
\nabla_{e_4^{\text{as}}}(g_{a,M}(e_4^{\text{as}},\partial_{\mu}))&=(\nabla_{e_4^{\text{as}}}h_{a,M})\, g_{a,M}(\partial_{x^2},\partial_{\mu})+h_{a,M}\nabla_{e_4^{\text{as}}}(g_{a,M}(\partial_{x^2},\partial_{\mu})) \\
&=h_{a,M}\,(\partial_{x^2}h_{a,M})g_{a,M}(\partial_{x^2},\partial_{\mu})+h^2_{a,M}\,\partial_{x^2}(g_{a,M}(\partial_{x^2},\partial_{\mu})) \, .
\end{align*}
We conclude by setting $\mathfrak{h}_{a,M}=h^{-1}_{a,M}(\partial_{x^2}h_{a,M}-\omegah_{a,M})$.

\end{proof}

\subsubsection{Star-normalised outgoing principal differentiable structure}

In this section, we define the \emph{star-normalised} outgoing principal differentiable structure of $(\mathcal{M},g_{a,M})$.

\medskip

We start by considering the manifold $\mathcal{M}$ with Kerr-star differentiable structure $(t^*,r,\theta,\phi^*)$ and the hypersurface $\mathcal{S}_{0}\subset\mathcal{M}$ (recall the definition \eqref{def_S_tau} and the related Figure \ref{fig:norm_hyp_kerr}).~The hypersurface $\mathcal{S}_{0}$ is foliated by the Kerr-star two-spheres $\mathbb{S}^2_{0,r}$.~For any point $p\in\mathcal{S}_0$, we define the smooth curve 
\begin{gather*}
\gamma_p: [0,\infty)\rightarrow J^+(\mathcal{S}_0)\cap\mathcal{M} \, ,  \\
\gamma_p(\lambda)=(t^*(\lambda),r(\lambda),\theta(\lambda),\phi^*(\lambda))
\end{gather*}
such that $\gamma_p(0)=p$, with $(t^*(0),r(0),\theta(0),\phi^*(0))=(0,r(p),\theta(p),\phi^*(p))$, and
\begin{align}
\frac{d\gamma_p^{t^*}}{d\lambda}(\lambda)&=\frac{\gamma_p^r{}^2(\lambda)+a^2}{\Sigma(\gamma_p^r(\lambda),\gamma_p^\theta(\lambda))}\,\left(1+\iota(\gamma_p^r(\lambda))\,\left(1-\frac{\Delta(\gamma_p^r(\lambda))}{\gamma_p^r{}^2(\lambda)+a^2} \right)\right)  \, , \label{ODE_curves_1}\\[5pt] 
\frac{d\gamma_p^r}{d\lambda}(\lambda)&=\frac{\Delta(\gamma_p^r(\lambda))}{\Sigma(\gamma_p^r(\lambda),\gamma_p^\theta(\lambda))} \, , \label{ODE_curves_2}\\[5pt]
\frac{d\gamma_p^\theta}{d\lambda}(\lambda)&=0 \, , \label{ODE_curves_3}\\[5pt] 
\frac{d\gamma_p^{\phi^*}}{d\lambda}(\lambda)&=  \frac{a}{\Sigma(\gamma_p^r(\lambda),\gamma_p^\theta(\lambda))}(1+\iota(\gamma_p^r(\lambda))) \label{ODE_curves_4}
\end{align}
for all $\lambda\geq 0$, with $\gamma_p^{\mu}(\lambda):=x^{\mu}(\gamma_p(\lambda))$.\footnote{The curves $\gamma_p$ are well-defined in view of the later Proposition \ref{prop_flow_gamma}.}~We note that
\begin{align*}
\gamma_p^{t^*}(\lambda)&\in [0,\infty) \, , & \gamma_p^{r}(\lambda) &\in[r_+,\infty) \, , & (\gamma_p^{\theta}(\lambda),\gamma_p^{\phi^*}(\lambda)))&\in\mathbb{S}^2
\end{align*}
for all $\lambda\geq 0$.~It is an immediate check that the curves $\gamma_p$ are \emph{integral curves} of $e_4^{\text{as}}$.

\medskip

Importantly, the point $p\in\mathcal{S}_0$ can be chosen to coincide with either of the poles of a $\mathbb{S}^2_{0,r}$-sphere.~Although the angular coordinates $(\theta,\phi^*)$ are not defined at the poles (and thus the functions $\gamma_p^\theta(0)$ and $\gamma_p^{\phi^*}(0)$ in the system \eqref{ODE_curves_1}-\eqref{ODE_curves_4} are not well-defined at the poles), one can smoothly extend all the functions of the angular coordinates on the right hand side of \eqref{ODE_curves_1}-\eqref{ODE_curves_4} to the poles.\footnote{We note that being able to extend smoothly to the poles the right hand sides of \eqref{ODE_curves_1}-\eqref{ODE_curves_4} corresponds to the fact that $e_4^{\text{as}}$ is global on the $\mathbb{S}^2_{0,r}$-spheres.}

\medskip

The following propositions prove that the curves $\gamma(\lambda)$ induce a smooth \emph{global} flow $\mathcal{F}_{\lambda\geq 0}$ on $J^+(\mathcal{S}_0)\cap\mathcal{M}$.

\medskip

\begin{prop} \label{prop_flow_gamma}
For any $p\in\mathcal{S}_0$, the curve $\gamma_p$ is unique and future complete.~Furthermore, for any distinct points $p_1,p_2\in\mathcal{S}_0$, we have $\gamma_{p_1}(\lambda)\neq \gamma_{p_2}(\lambda)$ for all $\lambda\geq 0$. 
\end{prop}

\medskip

\begin{proof}
The four ODEs \eqref{ODE_curves_1}-\eqref{ODE_curves_4} form an autonomous system of ODEs of the form
\begin{equation}
\frac{d\gamma_p^{\mu}}{d\lambda}(\lambda) =F^{\mu}(\gamma_p^{\nu}(\lambda)) \, ,
\end{equation}
where each $F^{\mu}$ is a \emph{bounded} smooth function of $\gamma_p^{\nu}(\lambda)=(\gamma_p^{t^*}(\lambda),\gamma_p^{r}(\lambda),\gamma_p^{\theta}(\lambda),\gamma_p^{\phi^*}(\lambda))$ on $\mathcal{M}$.~We note that the right hand side of the ODEs \eqref{ODE_curves_1} and \eqref{ODE_curves_2} are \emph{non-negative} smooth functions on $\mathcal{M}$, and thus $\gamma_p^{t^*}(\lambda)$ and $\gamma_p^r(\lambda)$ are \emph{non-decreasing} functions of $\lambda$ for all $\lambda\geq 0$.~By standard ODE theory, there exists a unique solution $\gamma_p(\lambda)$ for all $\lambda\geq 0$.~Since $\gamma_p^{t^*}(\lambda)$ and $\gamma_p^r(\lambda)$ are non-decreasing functions of $\lambda$, the fact that the domains of $\gamma_p^{t^*}(\lambda)$ and $\gamma_p^r(\lambda)$ are lower bounded poses no issues for the completeness of $\gamma_p$.~The global uniqueness of $\gamma_p$ immediately implies the second part of the proposition.

\end{proof}

\medskip

\begin{prop}  \label{prop_curves_horizon}
Let $p\in \mathcal{S}_0\cap\mathcal{H}^+$.~Then, we have $\gamma_p^r(\lambda)=r_+$ for all $\lambda\geq 0$.
\end{prop}

\medskip

\begin{proof}
One can directly verify that the smooth curve
\begin{equation*}
\gamma_p(\lambda)=\left(2\,\frac{r_+^2+a^2}{\Sigma(r_+,\theta(p))}\,\lambda \, , \, r_+ \, , \, \theta(p) \, , \, \phi^*(p)+2\,\frac{a}{\Sigma(r_+,\theta(p))}\,\lambda\right)
\end{equation*}
solves the system of ODEs \eqref{ODE_curves_1}-\eqref{ODE_curves_4} for all $\lambda\geq 0$.

\end{proof}

\medskip

\begin{prop}  \label{prop_curves_to_scri}
Let $p\in \mathcal{S}_0$ such that $r(p)>r_+$.~Then, we have $\gamma_p^{t^*}(\lambda)\rightarrow \infty$ and $\gamma_p^r(\lambda)\rightarrow \infty$ as $\lambda\rightarrow \infty$.
\end{prop}

\medskip

\begin{proof}
We note that the right hand side of the ODEs \eqref{ODE_curves_1} and \eqref{ODE_curves_2} are positive smooth functions on $\mathcal{M}\setminus\mathcal{H}^+$, and thus $\gamma_p^{t^*}(\lambda)$ and $\gamma_p^r(\lambda)$ are increasing functions of $\lambda$ for all $\lambda\geq 0$.~One concludes by the future completeness of $\gamma_p(\lambda)$ from Proposition \ref{prop_flow_gamma}.
 
\end{proof}

\medskip

\begin{remark}
Let $p\in \mathcal{S}_0$ such that $r(p)>r_+$.~Then, the curve $\gamma_p$ intersects future null infinity $\mathcal{I}^+$.~This follows immediately from Proposition \ref{prop_curves_to_scri}.
\end{remark}

\medskip

We now consider the hypersurface $$\mathcal{S}_0^*:=\mathcal{S}_0\cap\left\lbrace r < r_+ +1 \right\rbrace$$ and the spacetime region $\mathcal{M}^*\subset\mathcal{M}$ such that
\begin{equation} \label{def_M_star}
\mathcal{M}^* :=\mathcal{F}_{\lambda\geq 0}(\mathcal{S}_0^*) \, .
\end{equation}
See the earlier Figure \ref{fig:M_star}.~With a slight abuse of notation, we will denote $\mathcal{H}^+\cap \mathcal{M}^*$ by $\mathcal{H}^+$.~We observe that, by definition, the region $\mathcal{M}^*$ is the spacetime region of all points $q\in J^+(\mathcal{S}_0)\cap\mathcal{M}$ such that there exists an integral curve $\gamma_{p}$ of $e_4^{\text{as}}$ connecting $p$ and $q$, with $p\in\mathcal{S}_0^*$.~For later convenience, we denote such a curve by $\gamma_{p,q}$.

\medskip

We define the smooth scalar function $$s:\mathcal{M}^* \rightarrow [r_+,r_+ +1)$$ such that
\begin{equation*}
s(q)=r(\gamma_{p,q}(0))
\end{equation*}
for any $q\in\mathcal{M}^*$.~Note that $s(q)= r(q)$ for any $q\in\mathcal{S}_0^*$.~We also note that points $q\in\mathcal{M}^*$ with associated $\gamma_{p,q}$ such that $p$ coincides with the pole of a $\mathbb{S}^2_{0,r}$ sphere are allowed.~In view of Propositions \ref{prop_flow_gamma} and \ref{prop_curves_to_scri}, the function $s$ is a \emph{globally} well-defined \emph{coordinate} function on $\mathcal{M}^*$.~One can thus conclude that the intersection of the level sets of $t^*$ and $s$ are (complete) two-spheres
\begin{equation*}
\mathbb{S}^2_{t^*,s}:= \left\lbrace t^*,s \right\rbrace \times \mathbb{S}^2 \, .
\end{equation*}
We note that $\mathbb{S}^2_{0,s}\equiv \mathbb{S}^2_{0,r}$, the latter understood as Kerr-star two-spheres.

\medskip

Let $(\theta^1_0,\theta^2_0)$ be arbitrary angular coordinates on the $\mathbb{S}^2_{0,s}$-spheres.~For each $\mathbb{S}^2_{t^*,s}$-sphere, we define angular coordinates $$(\theta^1,\theta^2)\in\mathbb{S}^2$$ such that $$(\theta^1(q),\theta^2(q))= (\theta^1_0(\gamma_{p,q}(0)),\theta^2_0(\gamma_{p,q}(0)))$$ for any $q\in \mathbb{S}^2_{t^*,s}$.~The coordinates $(\theta^1,\theta^2)$ only cover a subset of each $\mathbb{S}^2_{t^*,s}$-sphere.

\medskip

Finally, we define the smooth scalar function $$\underline{s}:\mathcal{M}^* \rightarrow [0,\infty)$$ such that
\begin{equation*}
\underline{s}\equiv t^*
\end{equation*}
on $\mathcal{M}^*$ and obtain the \emph{star-normalised} outgoing principal differentiable structure
\begin{equation} \label{new_coords}
(s,\underline{s},\theta^A) \, .
\end{equation}
We emphasise that the coordinates $(s,\underline{s})$ are \ul{\emph{global} and \emph{regular} on $\mathcal{M}^*$, including on the event horizon $\mathcal{H}^+$}, whereas the angular coordinates $(\theta^1,\theta^2)$ are regular but only cover a subset of the two-spheres
\begin{equation*}
\mathbb{S}^2_{s,\underline{s}}:= \left\lbrace s,\underline{s} \right\rbrace \times \mathbb{S}^2 \, .
\end{equation*}
We also note that the coordinates \eqref{new_coords} depend on the Kerr parameters $a$ and $M$.

\medskip

It is an immediate check that the differentiable structure \eqref{new_coords} is an outgoing principal differentiable structure in the sense of Definition \ref{def_outgoing_principal_diff_structures}.~In particular, we have 
\begin{align*}
e_4^{\text{as}}(s)&=0 \, , &  e_4^{\text{as}}(\theta^A)&=0 
\end{align*}
on $\mathcal{M}^*$.

\medskip

Let
\begin{align*}
k_{a,M} &:=\frac{r^2+a^2}{\Sigma}\,\left(1+\iota\,\left(1-\frac{\Delta}{r^2+a^2} \right)\right) \, , & \mathfrak{k}_{a,M}&:= k_{a,M}^{-1}\,(\partial_{\underline{s}}k_{a,M}-\omegah_{a,M}) 
\end{align*}
on $\mathcal{M}^*$, with $k_{a,M}$ \emph{nowhere} vanishing on $\mathcal{M}^*$.~Relative to the differentiable structure \eqref{new_coords}, the Kerr metric takes the form
\begin{equation} \label{bulk_kerr_metric_outgoing_principal}
g_{a,M}=\mathfrak{a}\,ds^2+2\,\Omega^2\,ds\,d\underline{s}+2\,b_{\theta^A}\,ds\,d\theta^A+2\,\underline{b}_{\theta^A}\,d\underline{s}\,d\theta^A+\,\gamma_{\theta^A\theta^B}\,d\theta^A d\theta^B
\end{equation}
on $\mathcal{M}^*$, with
\begin{align}
\partial_{\underline{s}}\,\Omega^2+\mathfrak{k}_{a,M}\,\Omega^2&=0 \, , & \partial_{\underline{s}}\,\underline{b}_{\theta^A}+\mathfrak{k}_{a,M}\,\underline{b}_{\theta^A}&=0  \label{bulk_kerr_gauge_ids}
\end{align}
on $\mathcal{M}^*$ (see Proposition \ref{prop_metric_ids_outgoing_principal_diff_str}), and the frame vector fields $(e_4^{\text{as}},e_3^{\text{as}})$ take the form
\begin{align} \label{bulk_alg_frame_e4_e3_coord}
e_4^{\text{as}}&=k_{a,M}\,\partial_{\underline{s}} \, , & e_3^{\text{as}}&=\underline{j}_{a,M}\,\partial_s+\underline{k}_{a,M}\,\partial_{\underline{s}}+\underline{\Lambda}^{\theta^A}_{a,M}\,\partial_{\theta^A}
\end{align}
on $\mathcal{M}^*$.~The quantities $\mathfrak{a}$, $\Omega^2$, $\underline{j}_{a,M}$ and $\underline{k}_{a,M}$ are smooth scalar functions on $\mathcal{M}^*$, the quantities $b$ and $\underline{b}$ are one-tensors on the $\mathbb{S}^2_{s,\underline{s}}$-spheres, the quantity $\gamma$ is a symmetric two-tensor on the $\mathbb{S}^2_{s,\underline{s}}$-spheres and the quantity $\underline{\Lambda}_{a,M}$ is a vector field tangent to the $\mathbb{S}^2_{s,\underline{s}}$-spheres.~All quantities are regular on the whole $\mathcal{M}^*$, including on $\mathcal{H}^+$.~The Kerr metric coefficients in \eqref{bulk_kerr_metric_outgoing_principal} and frame coefficients of $e_3^{\text{as}}$ in \eqref{bulk_alg_frame_e4_e3_coord} cannot be made explicit.

\medskip

We conclude with some remarks.

\medskip

\begin{remark}
In our problem, the ambient manifold $\mathcal{M}^*$ will act as the fixed background manifold and the differentiable structure \eqref{new_coords} will serve as the fixed differentiable structure on $\mathcal{M}^*$.~See Section \ref{sec_family_metrics}. 
\end{remark}

\medskip

\begin{remark} \label{rmk_alternative_coords}
For what follows, it is important for us that the coordinates $(s,\theta^A)$ are transported along the integral curves of the algebraically special null vector $e_4^{\textup{as}}$ and are regular on the event horizon $\mathcal{H}^+$.~In this sense, the choice of these coordinates is somewhat rigid.~On the other hand, our particular choice of the ingoing coordinate $\underline{s}$ will play no fundamental role, and may be replaced.~Note also that one can rescale the coordinate $\underline{s}$ by a regular (on $\mathcal{H}^+$) weight so that $e_4^{\textup{as}}$ becomes a coordinate vector field.
\end{remark}

\subsubsection{The $(s,\protect\underline{s})$-foliation}

The coordinates $(s,\underline{s})$ induce a \emph{global} foliation of $\mathcal{M}^*$.~The level sets of $r_+<s< r_+ +1$ are \emph{timelike} hypersurfaces intersecting the hypersurface $\mathcal{S}_0$ and future null infinity $\mathcal{I}^+$ (see Proposition \ref{prop_curves_to_scri}).~Such hypersurfaces are \emph{asymptotically null} as they approach $\mathcal{I}^+$.~In view of Proposition \ref{prop_curves_horizon}, the event horizon corresponds to the level set $$\mathcal{H}^+=\left\lbrace s=r_+ \right\rbrace$$ and is \emph{null}.~The properties of the level sets of $\underline{s}$ coincide with those of the level sets $\mathcal{S}_{\tau^*}$ of the Kerr-star time coordinate $t^*$ discussed in Section \ref{sec_Kerr_star_coords}.~The reader should refer back to Figure \ref{fig:intro_foliation}.

\subsection{The $|a|=0$ case}

For $|a|=0$, we refer to $(\mathcal{M},g_{0,M})$ as the \emph{Schwarzschild exterior manifold}.~The following properties hold:
\begin{itemize}
\item The algebraically special frame $\mathcal{N}_{\text{as}}$ is integrable and such that
\begin{equation*}
(\mathfrak{D}_{\mathcal{N}_{\text{as}}})_p = T_p\mathbb{S}^2_{s(p),\underline{s}(p)} 
\end{equation*} 
for any $p\in\mathcal{M}$.
\item The induced metric, connection coefficients and curvature components of $g_{0,M}$ relative to $\mathcal{N}_{\text{as}}$ are $\mathbb{S}^2_{s,\underline{s}}$-tensors in the sense of \cite{StabMink, Christ_Form_BH} and such that
\begin{align*}
(\slashed{\varepsilon}\cdot\chi)_{0,M}=(\slashed{\varepsilon}\cdot\chib)_{0,M}&=0 \, , & \eta_{0,M}=\etab_{0,M}=\zeta_{0,M}&=0 \, ,  &  \sigma_{0,M} &=0 
\end{align*}
on $\mathcal{M}$.
\item The outgoing coordinate $s$ is a regular null coordinate, whereas the ingoing coordinate $\underline{s}$ is a spacelike coordinate such that $$\underline{s}=t_{\text{bl}}+2M\log(r_{\text{bl}}-2M)$$ for $2M\leq r_{\text{bl}} \leq 15M/8$ and coincides with the Schwarzschild time coordinate $t_{\text{bl}}$ for $r_{\text{bl}}\geq 9M/4$ (see already Figure \ref{fig:intro_foliation_a_0}).
\end{itemize}

\section{Nonlinear perturbations of the Kerr exterior manifold}  \label{sec_perturbations_kerr}

In this section, we formulate nonlinear perturbations of the Kerr exterior manifold.~We start by introducing a one-parameter family of metrics $\bsy{g}(\epsilon)$ on the fixed ambient manifold $\mathcal{M}^*$ in Section \ref{sec_family_metrics}.~In doing that, we will make a choice of gauge for the metrics $\bsy{g}(\epsilon)$.~To the family of metrics $\bsy{g}(\epsilon)$, we associate a one-parameter family of frames $\bsy{\mathcal{N}}(\epsilon)$ in Section \ref{sec_family_frames}.~Sections \ref{sec_gauge_construction} and \ref{sec_frames_construction} will prove that our choices of gauge and frames are legitimate.

\medskip  

Sections \ref{sec_gauge_construction} and \ref{sec_frames_construction} are independent modules and can be skipped by the reader without affecting the overall logic of the paper.

\subsection{One-parameter family of metrics}  \label{sec_family_metrics}

We choose real parameters $a$ and $M$, with $|a|<M$.~We fix the manifold-with-boundary $\mathcal{M}^*$, which we identify with the subset of the Kerr ambient manifold \eqref{def_M_star} with star-normalised outgoing principal differentiable structure $(s,\underline{s},\theta^A)$.\footnote{We note that the fixed differentiable structure is global (modulo the usual degeneration on two-spheres) and regular on the whole ambient manifold $\mathcal{M}^*$, including on $\mathcal{H}^+$.}~We introduce the fixed smooth scalar functions $\mathfrak{k}_{a,M}$, $k_{a,M}$ and $\omegah_{a,M}$, the fixed one-tensor $\etab^{\circ}_{a,M}$, the fixed vector fields $(e_4^{\text{as}},e_3^{\text{as}})$ and the fixed distribution $\mathfrak{D}_{\mathcal{N}_{\text{as}}}$ of the fixed coordinates on $\mathcal{M}^*$. 

\medskip

We prescribe a one-parameter family of smooth, non-degenerate Lorentzian metrics 
\begin{equation} \label{bulk_outline_family_metrics}
\bsy{g}(\epsilon):=\bsy{\mathfrak{a}}(\epsilon)\,ds^2+2\,\bsy{\Omega}^2(\epsilon)\,ds\,d\underline{s}+2\,\bsy{b}_{\theta^A}(\epsilon)\,ds\,d\theta^A+2\,\underline{\bsy{b}}_{\theta^A}(\epsilon)\,d\underline{s}\,d\theta^A+\,\bsy{\gamma}_{\theta^A\theta^B}(\epsilon)\,d\theta^A d\theta^B
\end{equation}
on $\mathcal{M}^*$, with 
\begin{align}
\partial_{\underline{s}}\,\bsy{\Omega}^2(\epsilon)+\mathfrak{k}_{a,M}\,\bsy{\Omega}^2(\epsilon)&=0 \, , & \partial_{\underline{s}}\,\underline{\bsy{b}}_{\theta^A}(\epsilon)+\mathfrak{k}_{a,M}\,\underline{\bsy{b}}_{\theta^A}(\epsilon)&=0  \label{bulk_outline_gauge_ids}
\end{align}
on $\mathcal{M}^*$ for all $\epsilon$.~For each $\epsilon$, the metric quantities $\bsy{\mathfrak{a}}(\epsilon)$ and $\bsy{\Omega}^2(\epsilon)$ are smooth scalar functions on $\mathcal{M}^*$, the metric quantities $\bsy{b}(\epsilon)$ and $\underline{\bsy{b}}(\epsilon)$ are one-tensors on the $\mathbb{S}^2_{s,\underline{s}}$-spheres and the metric quantity $\bsy{\gamma}(\epsilon)$ is a symmetric two-tensor on the $\mathbb{S}^2_{s,\underline{s}}$-spheres.~All quantities are regular on the whole $\mathcal{M}^*$, including on $\mathcal{H}^+$.~We assume that $\bsy{g}(0)\equiv g_{a,M}$ with 
\begin{align*}
\bsy{\mathfrak{a}}(0)&=\mathfrak{a} \, , & \bsy{\Omega}^2(0)&=\Omega^2 \, , & \bsy{b}_{\theta^A}(0)&=b_{\theta^A} \, , & \underline{\bsy{b}}_{\theta^A}(0)&=\underline{b}_{\theta^A} \, , & \bsy{\gamma}_{\theta^A\theta^B}(0)&=\gamma_{\theta^A\theta^B} \, ,
\end{align*}
i.e.~the metric coefficients of $\bsy{g}(0)$ relative to the fixed differentiable structure coincide with the Kerr metric coefficients relative to the star-normalised, outgoing principal differentiable structure of Kerr from \eqref{bulk_kerr_metric_outgoing_principal} in Section \ref{sec_Kerr_background_diff_structure}.~We further assume that
\begin{equation} \label{bulk_uniform_limit_metrics}
\sum_{0\leq |I|\leq 10} |\partial^{I}\bsy{\mathfrak{g}}(\epsilon)-\partial^{I}\bsy{\mathfrak{g}}(0)| \leq \epsilon 
\end{equation}
on $\mathcal{M}^*$, with
\begin{equation*}
\bsy{\mathfrak{g}}(\epsilon)=\left\lbrace \bsy{\mathfrak{a}}(\epsilon)\, , \, \bsy{\Omega}^2(\epsilon) \, , \, \bsy{b}_{\theta^A}(\epsilon) \, , \, \underline{\bsy{b}}_{\theta^A}(\epsilon) \, , \, \bsy{\gamma}_{\theta^A\theta^B}(\epsilon) \right\rbrace
\end{equation*}
and $I$ a multi-index for partial derivatives relative to the fixed differentiable structure.

\medskip

We remark that there exists a sufficiently small constant $\epsilon_0>0$ (which is chosen once and for all in Section \ref{sec_gauge_construction}), depending only on the fixed background parameters $a$ and $M$, such that, for all $0\leq |\epsilon|\leq \epsilon_0$, the form \eqref{bulk_outline_family_metrics} of the metrics $\bsy{g}(\epsilon)$, together with the additional identities \eqref{bulk_outline_gauge_ids}, correspond to a choice of gauge for the family of metrics $\bsy{g}(\epsilon)$, called \emph{outgoing principal gauge}.~See Section \ref{sec_gauge_construction}.

\medskip

From now on, without loss of generality, \ul{we restrict to $0\leq |\epsilon|\leq \epsilon_0$}.~When we write \emph{for all $\epsilon$}, this restriction is implicitly assumed.

\medskip

\begin{remark}
Since the fixed differentiable structure is regular on the whole ambient manifold $\mathcal{M}^*$ (including on $\mathcal{H}^+$), formulating the smoothness of the metrics $\bsy{g}(\epsilon)$ does not require the slightly more technical characterisation from Section 5.1.1 of \cite{DHR}.
\end{remark}

\subsection{Outgoing principal gauge} \label{sec_gauge_construction}

In this section, we prove the following proposition.~Proposition \ref{prop_gauge_main} implies that the form \eqref{bulk_outline_family_metrics} of the metrics $\bsy{g}(\epsilon)$ introduced in Section \ref{sec_family_metrics}, together with the additional identities \eqref{bulk_outline_gauge_ids}, correspond to a choice of gauge for the family of metrics $\bsy{g}(\epsilon)$ provided that $|\epsilon|$ is sufficiently small.~We refer to such a gauge as an \emph{outgoing principal gauge}.

\medskip

\begin{prop} \label{prop_gauge_main} 
Let $a$ and $M$ be real parameters, with $|a|<M$.~Let $(\mathcal{M},g_{a,M})$ be the Kerr exterior manifold.~Then, there exists a sufficiently small constant $\epsilon_0>0$, depending only on $a$ and $M$, such that the following statement holds true. 

Consider any one-parameter family of smooth, non-degenerate Lorentzian metrics $\bsy{g}(\epsilon)$ on $\mathcal{M}$, with $0\leq |\epsilon|\leq \epsilon_0$ and such that $\bsy{g}(0)\equiv g_{a,M}$ and  
\begin{equation} \label{bulk_uniform_limit_gauge_constr}
\sum_{0\leq |I|\leq 10} |\partial^{I}\bsy{g}_{\mu\nu}(\epsilon)-\partial^{I}g_{a,M}{}_{\mu\nu}| \leq \epsilon 
\end{equation}
on $\mathcal{M}$, where the metric components in \eqref{bulk_uniform_limit_gauge_constr} are relative to the Kerr-star differentiable structure on $\mathcal{M}$.~For any $p\in\mathcal{M}$, there exist an open neighbourhood $\mathcal{R}_p\subset\mathcal{M}$ of $p$ and a one-parameter family of differentiable structures $$(\bsy{x}^1_{\epsilon},\bsy{x}^2_{\epsilon},\bsy{\vartheta}_{\epsilon}^{A})$$ on $\mathcal{R}_p$, with $(\bsy{x}^1_{\epsilon},\bsy{x}^2_{\epsilon})$ global coordinates on $\mathcal{R}_p$ and $(\bsy{\vartheta}_{\epsilon}^{1},\bsy{\vartheta}_{\epsilon}^{2})\in\mathbb{S}^2$ subject to the usual degeneration on two-spheres, such that
\begin{itemize}
\item \textbf{Curves of constant $(\bsy{x}^1_{\epsilon},\bsy{\vartheta}_{\epsilon}^{A})$ are null geodesics of $\bsy{g}(\epsilon)$}

For $0\leq |\epsilon|\leq \epsilon_0$, we have 
\begin{align}
\bsy{g}_{\bsy{x}^2_{\epsilon} \bsy{x}^2_{\epsilon}}(\epsilon)&=0 \, , &
\partial_{\bsy{x}^2_{\epsilon}}\, \bsy{g}_{\bsy{x}^2_{\epsilon}\mu}(\epsilon)+\bsy{\mathfrak{h}}^{\epsilon}_{a,M}\,\bsy{g}_{\bsy{x}^2_{\epsilon}\mu}(\epsilon)&=0 \label{gauge_ids_prop}
\end{align}
on $\mathcal{R}_p$, with $\bsy{\mathfrak{h}}^{\epsilon}_{a,M}$ a one-parameter family of smooth scalar functions such that
\begin{equation*}
\bsy{\mathfrak{h}}^{\epsilon}_{a,M}=h_{a,M}^{-1}(\partial_{\bsy{x}^2_{\epsilon}}h_{a,M}-\omegah_{a,M})
\end{equation*} 
for some smooth, nowhere vanishing scalar function $h_{a,M}$ independent of $\epsilon$ and $x^{\mu}=(\bsy{x}^1_{\epsilon},\bsy{\vartheta}_{\epsilon}^{A})$.
\item \textbf{Curves of constant $(\bsy{x}^1_{0},\bsy{\vartheta}_{0}^{A})$ are outgoing, principal null geodesics of $\bsy{g}(0)$}

For $\epsilon=0$, the coordinates
\begin{equation*} 
(\bsy{x}^1_{0},\bsy{x}^2_{0},\bsy{\vartheta}_{0}^{A})
\end{equation*}
are an outgoing principal differentiable structure of $\bsy{g}(0)$ on $\mathcal{R}_p$.
\end{itemize}
\end{prop}

\medskip

The gauge construction builds upon the definition of an auxiliary one-parameter family of local, null (relative to $\bsy{g}(\epsilon)$) vector fields $\bsy{e}_{\bsy{4}}^{\text{aux}}(\epsilon)$.~The preliminary definition of this family is presented in Section \ref{sec_auxiliary_family_vf}, whereas the gauge construction proper is carried out in Section \ref{sec_local_gauge}.~In Section \ref{sec_gauge_identities}, we prove the gauge identities \eqref{gauge_ids_prop}.

\medskip

\begin{remark}  \label{rmk_extra_assumption_gauge}
The assumption \eqref{bulk_uniform_limit_gauge_constr} is needed in the present section.~In particular, it is exploited in the initial step of the construction of Section \ref{sec_auxiliary_family_vf}.
\end{remark}

\subsubsection{An auxiliary one-parameter family of vector fields}  \label{sec_auxiliary_family_vf}

We consider the set-up stated in Proposition \ref{prop_gauge_main}.

\medskip

In view of the assumption \eqref{bulk_uniform_limit_gauge_constr}, there exists a real constant $\epsilon_0>0$ such that, for all $0\leq |\epsilon| \leq \epsilon_0$, the inequalities 
\begin{align}  \label{ineq_distr_spacelike}
\bsy{g}(e_1^{\text{as}},e_1^{\text{as}})&> 0 \, , & \bsy{g}(e_2^{\text{as}},e_2^{\text{as}})&> 0 
\end{align}
hold \emph{everywhere} on $\mathcal{M}$ for any choice of local frame $(e_1^{\text{as}},e_2^{\text{as}})$ of the fixed distribution $\mathfrak{D}_{\mathcal{N}_{\text{as}}}$ on $\mathcal{M}$.~See the related Remark \ref{rmk_extra_assumption_gauge} and note that, for the existence of the constant $\epsilon_0$, a $C^0$-version of the assumption \eqref{bulk_uniform_limit_gauge_constr} would suffice.~The inequalities \eqref{ineq_distr_spacelike} imply that the distribution $\mathfrak{D}_{\mathcal{N}_{\text{as}}}$ is everywhere \emph{spacelike} \ul{relative to $\bsy{g}(\epsilon)$} on $\mathcal{M}$.\footnote{Meaning that, at each point $p\in \mathcal{M}$, $(\mathfrak{D}_{\mathcal{N}_{\text{as}}},\bsy{g}(\epsilon)|_{\mathfrak{D}_{\mathcal{N}_{\text{as}}}})_p$ is diffeomorphic to the two-dimensional Euclidean space $(\mathbb{R}^2,\delta)_p$.}~For the rest of the section, we restrict to $0\leq |\epsilon|\leq \epsilon_0$.

\medskip

We define the distribution
\begin{equation} \label{aux_def_vertical_space}
\bsy{V}(\epsilon):=(\mathfrak{D}_{\mathcal{N}_{\text{as}}}) ^{\perp_{\bsy{g}(\epsilon)}} 
\end{equation}
on $\mathcal{M}$,\footnote{Recall that both $\mathfrak{D}_{\mathcal{N}_{\text{as}}}$ and $\bsy{g}(\epsilon)$ are global on $\mathcal{M}$, and thus $\bsy{V}(\epsilon)$ is globally well-defined on $\mathcal{M}$.} i.e.~the bundle of all vector fields $\bsy{X}(\epsilon)$ on $\mathcal{M}$ such that $\bsy{g}(\bsy{X}(\epsilon),e_A^{\text{as}})=0$ for any local frame $(e_1^{\text{as}},e_2^{\text{as}})$ of $\mathfrak{D}_{\mathcal{N}_{\text{as}}}$.~For \ul{any} $p\in\mathcal{M}$, by Sylvester's law of inertia,\footnote{By definition, we have $\bsy{V}_p(\epsilon)\oplus(\mathfrak{D}_{\mathcal{N}_{\text{as}}})_p =T_p\mathcal{M}$.~Since the metric $\bsy{g}(\epsilon)$ has Lorentzian signature on $T_p\mathcal{M}$ and is positive definite on $(\mathfrak{D}_{\mathcal{N}_{\text{as}}})_p$, then $\bsy{g}(\epsilon)$ has Lorentzian signature on $\bsy{V}_p(\epsilon)$.} we have that $(\bsy{V}(\epsilon),\bsy{g}(\epsilon)|_{\bsy{V}(\epsilon)})_p $ is diffeomorphic to the two-dimensional Minkowski space $(\mathbb{R}^2,\eta)_p$.~Thus, there exist exactly two distinct \emph{null} (relative to $\bsy{g}(\epsilon)$) directions in $\bsy{V}_p(\epsilon)$.~We choose $(\bsy{e}_{\bsy{4}}^{\text{aux}}(\epsilon))_p$ as the unique vector at $p$ such that
\begin{align*}
(\bsy{e}_{\bsy{4}}^{\text{aux}}(\epsilon))_p &\in \bsy{V}_p(\epsilon)  \, , &
\bsy{g}_p(\bsy{e}_{\bsy{4}}^{\text{aux}}(\epsilon),\bsy{e}_{\bsy{4}}^{\text{aux}}(\epsilon))&=0 \, ,  &
\bsy{g}_p(\bsy{e}_{\bsy{4}}^{\text{aux}}(\epsilon),e_3^{\text{as}})&=-2 \, .
\end{align*}
Note that, by the definition \eqref{aux_def_vertical_space}, we have
\begin{equation*}
\bsy{g}_p(\bsy{e}_{\bsy{4}}^{\text{aux}}(\epsilon),e_A^{\text{as}})=0 
\end{equation*}
for any local frame $(e_1^{\text{as}},e_2^{\text{as}})$ of $\mathfrak{D}_{\mathcal{N}_{\text{as}}}$.

\medskip

From the chosen point $p\in\mathcal{M}$, we propagate $\bsy{e}_{\bsy{4}}^{\text{aux}}(\epsilon)$ \emph{locally} such that
\begin{equation}
\bsy{\nabla}_{\bsy{e}_{\bsy{4}}^{\text{aux}}(\epsilon)}\bsy{e}_{\bsy{4}}^{\text{aux}}(\epsilon)=\omegah_{a,M}\,\bsy{e}_{\bsy{4}}^{\text{aux}}(\epsilon)  \, , \label{aux_geod_e4_aux}
\end{equation}
with $\bsy{\nabla}(\epsilon)$ the Levi-Civita connection of $\bsy{g}(\epsilon)$ and $\omegah_{a,M}=\omegah_{a,M}(r,\theta)$ smooth scalar function of the Kerr-star coordinates on $\mathcal{M}$.~The local existence and uniqueness of $\bsy{e}_{\bsy{4}}^{\text{aux}}(\epsilon)$ away from $p$ is a classical local existence and uniqueness result from ODE theory.~It is an easy check that, propagating via \eqref{aux_geod_e4_aux}, one obtains a local vector field $\bsy{e}_{\bsy{4}}^{\text{aux}}(\epsilon)$ which remains \emph{null} (relative to $\bsy{g}(\epsilon)$) along its integral curve emanating from $p$.

\medskip

Let $$\mathcal{B}_p:=\bigcup_{r(p)-\delta<r<r(p)+\delta}\,\mathbb{S}^2_{t^*(p),r} \, ,$$ with sufficiently small real constant $\delta>0$.~The construction of the local vector field $\bsy{e}_{\bsy{4}}^{\text{aux}}(\epsilon)$ can be repeated starting from \emph{any}\footnote{In particular, the point $q\in\mathcal{B}_p$ can coincide with either of the poles of a $\mathbb{S}^2_{t^*,r}$-sphere foliating $\mathcal{B}_p$.} point $q\in\mathcal{B}_p$.\footnote{For $\delta$ sufficiently small, one can check that, for any $q\in\mathcal{B}_p$, the vector $(\bsy{e}_{\bsy{4}}^{\text{aux}}(\epsilon))_q$ is not tangent to $\mathcal{B}_p$.}~The assumption \eqref{bulk_uniform_limit_gauge_constr} guarantees that the construction yields a null vector field $\bsy{e}_{\bsy{4}}^{\text{aux}}(\epsilon)$ on an open neighbourhood $\mathcal{R}_p\subset \mathcal{M}$ of $p$.~We note that, for this step of the construction, a $C^0$-version of the assumption \eqref{bulk_uniform_limit_gauge_constr} would \emph{not} suffice.~The integral curves of $\bsy{e}_{\bsy{4}}^{\text{aux}}(\epsilon)$ induce a flow on $\mathcal{R}_p$.

\medskip

\begin{remark}
Let $p\in\mathcal{M}$.~For $\epsilon=0$, we have $\bsy{V}(0) \equiv\left\langle e_3^{\textup{as}},e_4^{\textup{as}} \right\rangle$ and
\begin{equation} \label{aux_e4_aux_epsilon_zero}
\bsy{e}_{\bsy{4}}^{\textup{aux}}(0)\equiv e_4^{\textup{as}}
\end{equation}
on $\mathcal{R}_p$.~Indeed, one can check that $e_4^{\textup{as}}$ is the unique null (relative to $g_{a,M}$) vector field on $\mathcal{B}_p$ such that $g_{a,M}(e_4^{\textup{as}},e_A^{\textup{as}})=0$ and $g_{a,M}(e_4^{\textup{as}},e_3^{\textup{as}})=-2$, and thus $\bsy{e}_{\bsy{4}}^{\textup{aux}}(0)\equiv e_4^{\textup{as}}$ on $\mathcal{B}_p$.~Recalling that $\nabla_{e_4^{\textup{as}}}e_4^{\textup{as}}=\omegah_{a,M}\,e_4^{\textup{as}}$ and by uniqueness from ODE theory, one concludes that $\bsy{e}_{\bsy{4}}^{\textup{aux}}(0)\equiv e_4^{\textup{as}}$ on $\mathcal{R}_p$. 
\end{remark}

\medskip

\begin{remark}
As we shall see in the next section, it is crucial that the family of vector fields $\bsy{e}_{\bsy{4}}^{\textup{aux}}(\epsilon)$ satisfies the identities \eqref{aux_geod_e4_aux} and \eqref{aux_e4_aux_epsilon_zero} on $\mathcal{R}_p$.~On the other hand, the choice of the family of vector fields $\bsy{e}_{\bsy{4}}^{\textup{aux}}(\epsilon)$ on $\mathcal{B}_p$ can differ from ours as long as $\bsy{e}_{\bsy{4}}^{\textup{aux}}(0)\equiv e_4^{\textup{as}}$ on $\mathcal{B}_p$.
\end{remark}

\subsubsection{The gauge}  \label{sec_local_gauge}

For any $p\in\mathcal{M}$, we consider the open sets $\mathcal{B}_p$ and $\mathcal{R}_p$ of Section \ref{sec_auxiliary_family_vf}.~We define the scalar functions
\begin{align*}
\bsy{x}^1_{\epsilon} &:\mathcal{R}_p\rightarrow \mathbb{R} \, , & \bsy{x}^2_{\epsilon}:\mathcal{R}_p\rightarrow \mathbb{R} 
\end{align*}
such that $\bsy{x}^1_{\epsilon}\equiv r$ and $\bsy{x}^2_{\epsilon}\equiv t^*$ on $\mathcal{B}_p$ and
\begin{align} 
\bsy{e}_{\bsy{4}}^{\text{aux}}(\bsy{x}^1_{\epsilon})&=0 \, ,  &
\bsy{e}_{\bsy{4}}^{\text{aux}}(\bsy{x}^2_{\epsilon})&=h_{a,M} \label{aux_transport_coords_2}
\end{align}
on $\mathcal{R}_p$, with $h_{a,M}$ an arbitrary smooth, \emph{nowhere vanishing} scalar function of the Kerr-star coordinates on $\mathcal{R}_p$ which is independent of $\epsilon$.~Since the integral curves of $\bsy{e}_{\bsy{4}}^{\text{aux}}(\epsilon)$ induce a flow on $\mathcal{R}_p$, the functions $(\bsy{x}^1_{\epsilon},\bsy{x}^2_{\epsilon})$ are \emph{coordinate} functions on the whole $\mathcal{R}_p$.

\medskip

We define the two-spheres
\begin{equation*}
\mathbb{S}^2_{\bsy{x}^1_{\epsilon},\bsy{x}^2_{\epsilon}} :=\left\lbrace \bsy{x}^1_{\epsilon},\bsy{x}^2_{\epsilon}\right\rbrace \times \mathbb{S}^2
\end{equation*} 
and angular coordinates
\begin{equation*}
(\bsy{\vartheta}^1_{\epsilon},\bsy{\vartheta}^2_{\epsilon})\in\mathbb{S}^2
\end{equation*}
such that $(\bsy{\vartheta}^1_{\epsilon},\bsy{\vartheta}^2_{\epsilon})$ are arbitrary angular coordinates on any two-sphere $\mathbb{S}^2_{\bsy{x}^1_{\epsilon},\bsy{x}^2}\subset\mathcal{B}_p$ and
\begin{equation}
\bsy{e}_{\bsy{4}}^{\text{aux}}(\bsy{\vartheta}^A_{\epsilon})=0  \label{aux_transport_coords_3}
\end{equation}
on $\mathcal{R}_p$, obtaining the differentiable structure $$(\bsy{x}^1_{\epsilon},\bsy{x}^2_{\epsilon},\bsy{\vartheta}^A_{\epsilon})$$ on $\mathcal{R}_p$.~As usual, the angular coordinates $(\bsy{\vartheta}^1_{\epsilon},\bsy{\vartheta}^2_{\epsilon})$ only cover a subset of the two-spheres $\mathbb{S}^2_{\bsy{x}^1_{\epsilon},\bsy{x}^2_{\epsilon}}$.

\subsubsection{The gauge identities}  \label{sec_gauge_identities}

For any $p\in\mathcal{M}$, we consider the differentiable structure $(\bsy{x}^1_{\epsilon},\bsy{x}^2_{\epsilon},\bsy{\vartheta}^A_{\epsilon})$ on $\mathcal{R}_p$ of Section \ref{sec_local_gauge}.~Let 
\begin{equation}  \label{def_k_epsilon}
\bsy{\mathfrak{h}}^{\epsilon}_{a,M}:=h_{a,M}^{-1}(\partial_{\bsy{x}^2_{\epsilon}}h_{a,M}-\omegah_{a,M}) \, .
\end{equation}
The following proposition proves the gauge identities \eqref{gauge_ids_prop}, whereas the second point of Proposition \ref{prop_gauge_main} immediately follows from the identities \eqref{aux_e4_aux_epsilon_zero}, \eqref{aux_transport_coords_2} and \eqref{aux_transport_coords_3}.

\medskip

\begin{prop}
We have 
\begin{align}
\bsy{g}_{\bsy{x}^2_{\epsilon} \bsy{x}^2_{\epsilon}}(\epsilon)&=0 \, , &
\partial_{\bsy{x}^2_{\epsilon}}\, \bsy{g}_{\bsy{x}^2_{\epsilon}\mu}(\epsilon)+\bsy{\mathfrak{h}}^{\epsilon}_{a,M}\,\bsy{g}_{\bsy{x}^2_{\epsilon}\mu}(\epsilon)&=0 \label{gg_1}
\end{align}
on $\mathcal{R}_p$, with $x^{\mu}=(\bsy{x}^1_{\epsilon},\bsy{\vartheta}_{\epsilon}^{A})$. 
\end{prop}

\medskip

\begin{proof}
The first of the identities \eqref{gg_1} is easy to check, in that $\bsy{e}_{\bsy{4}}^{\text{aux}}(\epsilon)$ takes the coordinate form $$\bsy{e}_{\bsy{4}}^{\text{aux}}(\epsilon)=h_{a,M}\partial_{\bsy{x}^2_{\epsilon}}$$ from \eqref{aux_transport_coords_2} and \eqref{aux_transport_coords_3} and is null relative to $\bsy{g}(\epsilon)$ by construction.~Thus, we have 
\begin{align*}
\bsy{g}(\partial_{\bsy{x}^2_{\epsilon}},\partial_{\bsy{x}^2_{\epsilon}})&=h_{a,M}^{-2}\cdot\bsy{g}(\bsy{e}_{\bsy{4}}^{\text{aux}}(\epsilon),\bsy{e}_{\bsy{4}}^{\text{aux}}(\epsilon)) \\
&=0 \, .
\end{align*}
To check the second set of identities \eqref{gg_1}, we compute
\begin{align*}
\bsy{\nabla}_{\bsy{e}_{\bsy{4}}^{\text{aux}}(\epsilon)}(\bsy{g}(\bsy{e}_{\bsy{4}}^{\text{aux}}(\epsilon),\partial_{\mu}))&=\omegah_{a,M}\,\bsy{g}(\bsy{e}_{\bsy{4}}^{\text{aux}}(\epsilon),\partial_{\mu})+\bsy{g}(\bsy{e}_{\bsy{4}}^{\text{aux}}(\epsilon),\bsy{\nabla}_{\bsy{e}_{\bsy{4}}^{\text{aux}}(\epsilon)}\partial_{\mu}) \\
&=\omegah_{a,M}\,\bsy{g}(\bsy{e}_{\bsy{4}}^{\text{aux}}(\epsilon),\partial_{\mu})+h^2_{a,M}\,\bsy{g}(\partial_{\bsy{x}^2_{\epsilon}},\bsy{\nabla}_{\mu}\partial_{\bsy{x}^2_{\epsilon}}) \\
&=\omegah_{a,M}\,\bsy{g}(\bsy{e}_{\bsy{4}}^{\text{aux}}(\epsilon),\partial_{\mu})+\frac{1}{2}\,h^2_{a,M}\,\partial_{\mu}(\bsy{g}(\partial_{\bsy{x}^2_{\epsilon}},\partial_{\bsy{x}^2_{\epsilon}})) \\
&=\omegah_{a,M}\,\bsy{g}(\bsy{e}_{\bsy{4}}^{\text{aux}}(\epsilon),\partial_{\mu}) \, ,
\end{align*}
where in the last equality we used the first of the identities \eqref{gg_1}.~The left hand side can be written as follows
\begin{align*}
\bsy{\nabla}_{\bsy{e}_{\bsy{4}}^{\text{aux}}(\epsilon)}(\bsy{g}(\bsy{e}_{\bsy{4}}^{\text{aux}}(\epsilon),\partial_{\mu}))&=(\bsy{\nabla}_{\bsy{e}_{\bsy{4}}^{\text{aux}}(\epsilon)}h_{a,M})\,\bsy{g}(\partial_{\bsy{x}^2_{\epsilon}},\partial_{\mu})+h_{a,M}\,\bsy{\nabla}_{\bsy{e}_{\bsy{4}}^{\text{aux}}(\epsilon)}(\bsy{g}(\partial_{\bsy{x}^2_{\epsilon}},\partial_{\mu})) \\
&=h_{a,M}\,(\partial_{\bsy{x}^2_{\epsilon}}h_{a,M})\,\bsy{g}(\partial_{\bsy{x}^2_{\epsilon}},\partial_{\mu})+h^2_{a,M}\,\partial_{\bsy{x}^2_{\epsilon}}(\bsy{g}(\partial_{\bsy{x}^2_{\epsilon}},\partial_{\mu})) \, ,
\end{align*}
yielding the second set of identities \eqref{gg_1}.

\end{proof}

\subsection{One-parameter family of frames}  \label{sec_family_frames}

To the one-parameter family of metrics $\bsy{g}(\epsilon)$ introduced in Section \ref{sec_family_metrics}, we associate a one-parameter family of local frames $$\bsy{\mathcal{N}}(\epsilon)=(\eo(\epsilon),\etw(\epsilon),\et(\epsilon),\ef(\epsilon))$$ on $\mathcal{M}^*$ such that
\begin{align} \label{formulation_problem_e4_e3}
\ef(\epsilon)&:= k_{a,M}\,\partial_{\underline{s}}  & \et(\epsilon)&:=\underline{\bsy{j}}(\epsilon)\,\partial_s+\underline{\bsy{k}}(\epsilon)\,\partial_{\underline{s}}+\underline{\bsy{\Lambda}}^{\theta^A}(\epsilon)\,\partial_{\theta^A} \, .
\end{align}
For each $\epsilon$, the frame quantities $\underline{\bsy{j}}(\epsilon)$ and $\underline{\bsy{k}}(\epsilon)$ are smooth scalar functions on $\mathcal{M}^*$ and the frame quantity $\underline{\bsy{\Lambda}}(\epsilon)$ is a vector field tangent to the $\mathbb{S}^2_{s,\underline{s}}$-spheres.~All quantities are regular on the whole $\mathcal{M}^*$, including on $\mathcal{H}^+$.~For each $\epsilon$ and any $(s,\theta^A)$, the frame quantities are \emph{defined} implicitly as solutions to the system of ODEs
\begin{align}
k\,\frac{d\,\underline{\bsy{j}}(\epsilon)}{d\underline{s}}=& \,  -2\left(\frac{\underline{\eta}^{\circ}_{s}\,\underline{\bsy{j}}(\epsilon)+\underline{\eta}^{\circ}_{\theta^A}\,\underline{\bsy{\Lambda}}^{\theta^A}(\epsilon)}{\Omega^2\,\underline{\bsy{j}}(\epsilon)+\underline{b}_{\theta^B}\,\underline{\bsy{\Lambda}}^{\theta^B}(\epsilon)}\right)(\Omega^2\bsy{g}^{ss}(\epsilon)+\underline{b}_{\theta^C}\,\bsy{g}^{s\theta^C}(\epsilon)) \label{ODE_1} \\
&-(\omegah+k\,\bsy{\Gamma}_{s\underline{s}}^s(\epsilon))\,\underline{\bsy{j}}(\epsilon)-k\,\bsy{\Gamma}_{\underline{s}\theta^A}^s(\epsilon)\,\underline{\bsy{\Lambda}}^{\theta^A}(\epsilon) \nonumber \\ &+2 \,\underline{\eta}^{\circ}_{s}\,\bsy{g}^{ss}(\epsilon)+2 \,\underline{\eta}^{\circ}_{\theta^A}\,\bsy{g}^{s\theta^A}(\epsilon) \nonumber \\[5pt]
k\,\frac{d\,\underline{\bsy{k}}(\epsilon)}{d\underline{s}}=& \, -2\left(\frac{\underline{\eta}^{\circ}_{s}\,\underline{\bsy{j}}(\epsilon)+\underline{\eta}^{\circ}_{\theta^A}\,\underline{\bsy{\Lambda}}^{\theta^A}(\epsilon)}{\Omega^2\,\underline{\bsy{j}}(\epsilon)+\underline{b}_{\theta^B}\,\underline{\bsy{\Lambda}}^{\theta^B}(\epsilon)}\right)(\Omega^2\bsy{g}^{s\underline{s}}(\epsilon)+\underline{b}_{\theta^C}\,\bsy{g}^{\underline{s}\theta^C}(\epsilon)) \label{ODE_2} \\
&+(k\,\mathfrak{k}-\omegah)\,\underline{\bsy{k}}(\epsilon)-k\,\bsy{\Gamma}_{s\underline{s}}^{\underline{s}}(\epsilon)\,\underline{\bsy{j}}(\epsilon)-k\,\bsy{\Gamma}_{\underline{s}\theta^A}^{\underline{s}}(\epsilon)\,\underline{\bsy{\Lambda}}^{\theta^A}(\epsilon) \nonumber \\ &+2 \,\underline{\eta}^{\circ}_{s}\,\bsy{g}^{s\underline{s}}(\epsilon)+2 \,\underline{\eta}^{\circ}_{\theta^A}\,\bsy{g}^{\underline{s}\theta^A}(\epsilon) \nonumber \\[5pt]
k\,\frac{d\,\underline{\bsy{\Lambda}}^{\theta^A}(\epsilon)}{d\underline{s}} =& \, -2\left(\frac{\underline{\eta}^{\circ}_{s}\,\underline{\bsy{j}}(\epsilon)+\underline{\eta}^{\circ}_{\theta^B}\,\underline{\bsy{\Lambda}}^{\theta^B}(\epsilon)}{\Omega^2\,\underline{\bsy{j}}(\epsilon)+\underline{b}_{\theta^C}\,\underline{\bsy{\Lambda}}^{\theta^C}(\epsilon)}\right)(\Omega^2\bsy{g}^{s\theta^A}(\epsilon)+\underline{b}_{\theta^D}\,\bsy{g}^{\theta^A\theta^D}(\epsilon)) \label{ODE_3} \\
&-\omegah\,\underline{\bsy{\Lambda}}^{\theta^A}(\epsilon)-k\,\bsy{\Gamma}_{\underline{s}\theta^B}^{\theta^A}(\epsilon)\,\underline{\bsy{\Lambda}}^{\theta^B}(\epsilon)-k\,\bsy{\Gamma}_{s\underline{s}}^{\theta^A}(\epsilon)\,\underline{\bsy{j}}(\epsilon) \nonumber \\ &+2 \,\underline{\eta}^{\circ}_{s}\,\bsy{g}^{s\theta^A}(\epsilon)+2 \,\underline{\eta}^{\circ}_{\theta^B}\,\bsy{g}^{\theta^A\theta^B}(\epsilon)  \nonumber
\end{align}
in the variable $\underline{s}$, with initial data
\begin{gather}
\underline{\bsy{j}}(\epsilon)|_{\underline{s}=0}= \left. -\frac{2\,\underline{j}}{k\,(\underline{j}\,\bsy{\Omega}^2(\epsilon)+\underline{\Lambda}^{\theta^A}\underline{\bsy{b}}_{\theta^A}(\epsilon))}\,\right\rvert_{\underline{s}=0} \, , \label{ODE_data_1} \\ 
\underline{\bsy{k}}(\epsilon)|_{\underline{s}=0}= \left.\frac{\underline{j}^2\bsy{\mathfrak{a}}(\epsilon)+2\,\underline{j}\,\underline{k}\,\bsy{\Omega}^2(\epsilon)+2\,\underline{j}\,\underline{\Lambda}^{\theta^A}\bsy{b}_{\theta^A}(\epsilon)+2\,\underline{k}\,\underline{\Lambda}^{\theta^A}\underline{\bsy{b}}_{\theta^A}(\epsilon)+|\underline{\Lambda}|^2_{\bsy{\gamma}(\epsilon)}}{k\,(\underline{j}\,\bsy{\Omega}^2(\epsilon)+\underline{\Lambda}^{\theta^A}\underline{\bsy{b}}_{\theta^A}(\epsilon))^2} -\frac{2\,\underline{k}}{k\,(\underline{j}\,\bsy{\Omega}^2(\epsilon)+\underline{\Lambda}^{\theta^A}\underline{\bsy{b}}_{\theta^A}(\epsilon))} \,\right\rvert_{\underline{s}=0} \, , \label{ODE_data_2} \\ 
\underline{\bsy{\Lambda}}^{\theta^A}(\epsilon)|_{\underline{s}=0}= \left. -\frac{2\,\underline{\Lambda}^{\theta^A}}{k\,(\underline{j}\,\bsy{\Omega}^2(\epsilon)+\underline{\Lambda}^{\theta^A}\underline{\bsy{b}}_{\theta^A}(\epsilon))} \,\right\rvert_{\underline{s}=0} \, , \label{ODE_data_3}
\end{gather}
where $\bsy{\Gamma}_{\mu\nu}^{\sigma}(\epsilon)$ denote the Christoffel symbols of the metrics $\bsy{g}(\epsilon)$ relative to the fixed differentiable structure on $\mathcal{M}^*$.~A choice of the frame vector fields $(\eo(\epsilon),\etw(\epsilon))$ is not required.~In Section \ref{sec_frames_construction}, we prove that the family of frames $\bsy{\mathcal{N}}(\epsilon)$ is well-defined. 

\medskip

As a result of their definition, the frames $\bsy{\mathcal{N}}(\epsilon)$ satisfy the following properties, whose proof is addressed in Section \ref{sec_frames_construction}.
\begin{itemize}
\item \textbf{$\bsy{\mathcal{N}}(\epsilon)$ are null frames}

The frames $\bsy{\mathcal{N}}(\epsilon)$ are null relative to the metrics $\bsy{g}(\epsilon)$ for all $\epsilon$.
\item \textbf{$\ef(\epsilon)$ and $\et(\epsilon)$ are global, regular and non-degenerate}

The frame vector fields \eqref{formulation_problem_e4_e3} are global, regular and non-degenerate vector fields on the whole $\mathcal{M}^*$, including on $\mathcal{H}^+$.~The one-parameter family of horizontal distributions
\begin{equation*} 
\bsy{\mathfrak{D}}_{\bsy{\mathcal{N}}(\epsilon)} :=\left\langle \et(\epsilon),\ef(\epsilon)\right\rangle^{\bsy{\perp}_{\bsy{g}(\epsilon)}} 
\end{equation*}
is globally well-defined on $\mathcal{M}^*$.
\item \textbf{$\ef(\epsilon)$ is fixed}

The frame vector field $\ef(\epsilon)$ remains fixed on $\mathcal{M}^*$ for all $\epsilon$ and satisfies the identity 
\begin{equation} \label{i_1_kerr}
\ef(\epsilon)\equiv e_4^{\text{as}}
\end{equation}
for all $\epsilon$.
\item \textbf{$\bsy{\mathcal{N}}(0)$ is the algebraically special frame}

We have $$\bsy{\mathcal{N}}(0)\equiv \mathcal{N}_{\text{as}}$$ with 
\begin{align}
\underline{\bsy{j}}(0)&=\underline{j}_{a,M} \, , & \underline{\bsy{k}}(0)&=\underline{k}_{a,M} \, , & \underline{\bsy{\Lambda}}^{\theta^A}(0)&=\underline{\Lambda}^{\theta^A}_{a,M} \, , \label{zero_member_frame_kerr}
\end{align}
i.e.~the frame coefficients of $\bsy{e}_{\bsy{3}}(0)$ relative to the fixed differentiable structure coincide with the frame coefficients of the vector field $e_3^{\text{as}}$ relative to the star-normalised, outgoing principal differentiable structure of Kerr from Section \ref{sec_Kerr_background_diff_structure}.~In particular, we have $$\bsy{\mathfrak{D}}_{\bsy{\mathcal{N}}(0)}\equiv \mathfrak{D}_{\mathcal{N}_{\text{as}}} \, .$$
\item \textbf{$\bsy{\omegah}(\epsilon)$, $\bsy{\xi}(\epsilon)$ and $\underline{\widetilde{\bsy{\eta}}}(\epsilon)$ are fixed}

The connection coefficients and curvature components
\begin{gather*}
\bsy{\omegah}(\epsilon)\, , \, \bsy{\omegabh}(\epsilon)\, , \, \bsy{\eta}(\epsilon)\, , \, \bsy{\etab}(\epsilon) \, , \,\bsy{\xi}(\epsilon) \, , \, \bsy{\yb}(\epsilon) \, , \, \bsy{\zeta}(\epsilon) \, ,  \\
\bsy{\chih}(\epsilon) \, , \, \bsy{\chibh}(\epsilon)\, , \, (\bsy{\textbf{tr}\chi})(\epsilon) \, , \, (\textbf{tr}\bsy{\chib})(\epsilon) \, , \, (\bsy{\slashed{\varepsilon}\cdot\chi})(\epsilon) \, , \, (\bsy{\slashed{\varepsilon}\cdot\chib})(\epsilon) \, , \\
\widetilde{\bsy{\slashed{K}}}(\epsilon) \, , \, \bsy{\rho}(\epsilon)\, , \, \bsy{\sigma}(\epsilon)\, , \, \bsy{\beta}(\epsilon) \, , \, \bsy{\betab}(\epsilon)\, , \, \bsy{\alpha}(\epsilon) \, , \, \bsy{\alphab}(\epsilon) 
\end{gather*}
of the metrics $\bsy{g}(\epsilon)$ relative to the frames $\bsy{\mathcal{N}}(\epsilon)$ are global and regular $\bsy{\mathfrak{D}}_{\bsy{\mathcal{N}}(\epsilon)}$ tensors on $\mathcal{M}^*$.~We have the identities
\begin{align}
\bsy{\omegah}(\epsilon)-\omegah_{a,M} &\equiv 0  \, ,\label{i_2_kerr}\\
\bsy{\xi}(\epsilon) &\equiv 0 \, , \label{i_3_kerr} \\
\underline{\widetilde{\bsy{\eta}}}(\epsilon)-\underline{\eta}{}_{a,M} &\equiv 0 \label{i_4_kerr} 
\end{align} 
on $\mathcal{M}^*$ for all $\epsilon$.~The geometric quantity $\underline{\widetilde{\bsy{\eta}}}(\epsilon)$ in the identity \eqref{i_4_kerr} is a global, regular $\mathfrak{D}_{\mathcal{N}_{\text{as}}}$ one-tensor on $\mathcal{M}^*$ that one can uniquely associate to the $\bsy{\mathfrak{D}}_{\bsy{\mathcal{N}}(\epsilon)}$ one-tensor $\underline{\bsy{\eta}}(\epsilon)$ (see Section \ref{sec:geom_compare_tensors}).

For $\epsilon=0$, all the perturbed geometric quantities coincide with the corresponding connection coefficients and curvature components of the Kerr metric $g_{a,M}$ relative to the algebraically special frame $\mathcal{N}_{\text{as}}$ from Section \ref{sec_Kerr_connection_coeff_curv_comps}.
\end{itemize}

\medskip

\begin{remark}
We recall that the vector fields $(e_3^{\textup{as}},e_4^{\textup{as}})$ are global and regular on the whole $\mathcal{M}^*$ (see Section \ref{sec_Kerr_algebr_special_frame}), which makes the identities \eqref{zero_member_frame_kerr} consistent with the property that the frames $\bsy{\mathcal{N}}(\epsilon)$ are global and regular on the whole $\mathcal{M}^*$.~We also point out that, consistently with the identity \eqref{i_4_kerr}, we will have $\widetilde{\bsy{\etab}}(0)\equiv \bsy{\etab}(0)$ (see Remark \ref{rmk_zero_restriction}), and thus $\widetilde{\bsy{\etab}}(0)\equiv \etab_{a,M}$.
\end{remark}

\medskip

\begin{remark}  
Each of the frame properties is independent of the choice of one-parameter family of local frames $(\eo(\epsilon),\etw(\epsilon))$ of $\bsy{\mathfrak{D}}_{\bsy{\mathcal{N}}(\epsilon)}$.
\end{remark}

\medskip

For $|a|>0$, the one-parameter family of distributions $\bsy{\mathfrak{D}}_{\bsy{\mathcal{N}}(\epsilon)}$ is a family of \emph{non-integrable} distributions for all $\epsilon$ (see Section \ref{sec_frames_construction} for a proof).~Given the Levi-Civita connection $\bsy{\nabla}(\epsilon)$ of $\bsy{g}(\epsilon)$, the induced connection $\bsy{\nablasl}(\epsilon)$ over $\bsy{\mathfrak{D}}_{\bsy{\mathcal{N}}(\epsilon)}$ is \emph{not} the Levi-Civita connection of the induced metric $\bsy{\slashed{g}}(\epsilon)$ over $\bsy{\mathfrak{D}}_{\bsy{\mathcal{N}}(\epsilon)}$ (see the related Remark \ref{rmk_proj_connection_not_LC}).~For $|a|=0$, the distribution $\bsy{\mathfrak{D}}_{\bsy{\mathcal{N}}(0)}$ is integrable and the induced connection $\bsy{\nablasl}(0)$ over $\bsy{\mathfrak{D}}_{\bsy{\mathcal{N}}(0)}$ is the Levi-Civita connection of the induced metric $\bsy{\slashed{g}}(0)$.~See the related comments in Section \ref{sec_intro_framework_Schwarzschild}.

\medskip

We introduce the following definition.

\medskip

\begin{definition}[Variable distributions] \label{def_variable_distribution}
A one-parameter family of \emph{variable distributions} $\bsy{\mathfrak{D}}(\epsilon)$ relative to a fixed distribution $\mathfrak{D}$ on $\mathcal{M}^*$ is a one-parameter family of distributions $\bsy{\mathfrak{D}}(\epsilon)$ on $\mathcal{M}^*$ such that 
\begin{itemize}
\item $\dim (\bsy{\mathfrak{D}}(\epsilon))_p=\dim (\mathfrak{D})_p$ for all $|\epsilon|\geq 0$ and all $p\in\mathcal{M}^*$.
\item $\bsy{\mathfrak{D}}(0)\equiv \mathfrak{D}$.
\item $\bsy{\mathfrak{D}}(\epsilon)\not\equiv \mathfrak{D}$ for all $|\epsilon|>0$.
\end{itemize}
\end{definition}

\medskip

The one-parameter family of distributions $\bsy{\mathfrak{D}}_{\bsy{\mathcal{N}}(\epsilon)}$ on $\mathcal{M}^*$ is a family of \emph{variable} distributions relative to the fixed distribution $\mathfrak{D}_{\mathcal{N}_{\text{as}}}$ on $\mathcal{M}^*$, i.e.~$\bsy{\mathfrak{D}}_{\bsy{\mathcal{N}}(0)} \equiv \mathfrak{D}_{\mathcal{N}_{\text{as}}}$ and, for $|\epsilon|>0$, there exists a choice of local frame $(e_1^{\text{as}},e_2^{\text{as}})$ of $\mathfrak{D}_{\mathcal{N}_{\text{as}}}$ such that at least one of the relations
\begin{align*}
\bsy{g}(\ef(\epsilon),e_A^{\text{as}})&\neq 0 \, , & \bsy{g}(\et(\epsilon),e_A^{\text{as}})&\neq 0
\end{align*} 
holds on $\mathcal{M}^*$ (see Section \ref{sec_frames_construction} for a proof).~The property persists for $|a|=0$.

\medskip

We conclude the section with the following definition.

\medskip

\begin{definition}[Frame coefficients]  \label{def_frame_coefficients}
We define the \emph{frame coefficients} of $\bsy{\mathcal{N}}(\epsilon)$ as the one-parameter families of smooth scalar functions
\begin{equation*}
\bsy{\mathfrak{\mathfrak{f}}}_{3}(\epsilon) \, , \, \bsy{\mathfrak{\mathfrak{f}}}_{4}(\epsilon) \, , \, \bsy{\mathfrak{\underline{\mathfrak{f}}}}_{3}(\epsilon) \, , \, \bsy{\mathfrak{\underline{\mathfrak{f}}}}_{4}(\epsilon) 
\end{equation*}
such that
\begin{align*}
\bsy{\mathfrak{\mathfrak{f}}}_{3}(\epsilon)&=-\frac{1}{2}\,g_{a,M}(\ef(\epsilon),e_3^{\textup{as}}) \, , &
\bsy{\mathfrak{\mathfrak{f}}}_{4}(\epsilon)&=-\frac{1}{2}\,g_{a,M}(\ef(\epsilon),e_4^{\textup{as}}) \, , \\
\bsy{\mathfrak{\underline{\mathfrak{f}}}}_{3}(\epsilon)&=-\frac{1}{2}\,g_{a,M}(\et(\epsilon),e_3^{\textup{as}}) \, , &
\bsy{\mathfrak{\underline{\mathfrak{f}}}}_{4}(\epsilon)&=-\frac{1}{2}\,g_{a,M}(\et(\epsilon),e_4^{\textup{as}})  \, , 
\end{align*}
the one-parameter families of $\bsy{\mathfrak{D}}_{\bsy{\mathcal{N}}(\epsilon)}$ one-tensors
\begin{equation*}
\bsy{\mathfrak{\slashed{\mathfrak{f}}}}_{4}(\epsilon) \, , \, \bsy{\mathfrak{\slashed{\mathfrak{f}}}}_{3}(\epsilon) 
\end{equation*}
such that
\begin{align*}
(\bsy{\mathfrak{\slashed{\mathfrak{f}}}}_{4})_{\bsy{A}}(\epsilon)&=-\frac{1}{2}\,g_{a,M}(e_4^{\textup{as}},\ea(\epsilon)) \, , & (\bsy{\mathfrak{\slashed{\mathfrak{f}}}}_{3})_{\bsy{A}}(\epsilon)&=-\frac{1}{2}\,g_{a,M}(e_3^{\textup{as}},\ea(\epsilon)) 
\end{align*}
and the one-parameter families of $\mathfrak{D}_{\mathcal{N}_{\textup{as}}}$ one-tensors
\begin{equation*}
\bsy{\mathfrak{\mathfrak{f}}}(\epsilon)\, , \,   \bsy{\mathfrak{\underline{\mathfrak{f}}}}(\epsilon) \, , \, \bsy{\mathfrak{\slashed{\mathfrak{f}}}}_{\bsy{1}}(\epsilon) \, , \, \bsy{\mathfrak{\slashed{\mathfrak{f}}}}_{\bsy{2}}(\epsilon)
\end{equation*}
such that
\begin{align*}
\bsy{\mathfrak{\mathfrak{f}}}_{A}(\epsilon)&=g_{a,M}(\ef(\epsilon),e_A^{\textup{as}}) \, , & \bsy{\mathfrak{\underline{\mathfrak{f}}}}_{A}(\epsilon)&=g_{a,M}(\et(\epsilon),e_A^{\textup{as}}) \, , & (\bsy{\mathfrak{\slashed{\mathfrak{f}}}}_{\bsy{A}})_{B}(\epsilon)&=g_{a,M}(\ea(\epsilon),e_B^{\textup{as}}) \, .
\end{align*}
\end{definition}

\medskip

The following proposition is immediate.

\medskip

\begin{prop}
For all $\epsilon$, we have
\begin{align*}
\bsy{\mathfrak{\mathfrak{f}}}_{3}(\epsilon)&=1 \, , & \bsy{\mathfrak{\mathfrak{f}}}_{4}(\epsilon)&=0 \, , & \bsy{\mathfrak{\mathfrak{f}}}(\epsilon)&=0 \, .
\end{align*}
For $\epsilon=0$, we have
\begin{align*}
\bsy{\mathfrak{\underline{\mathfrak{f}}}}_{3}(0)&=0 \, , & \bsy{\mathfrak{\underline{\mathfrak{f}}}}_{4}(0)&=1 \, , \\
 \bsy{\mathfrak{\slashed{\mathfrak{f}}}}_{4}(0)&= \bsy{\mathfrak{\slashed{\mathfrak{f}}}}_{3}(0)=0 \, , & \bsy{\mathfrak{\underline{\mathfrak{f}}}}(0)&=0
\end{align*}
and $(\bsy{\mathfrak{\slashed{\mathfrak{f}}}}_{\bsy{A}})_B(0)=\delta_{AB}$.
\end{prop}

\subsection{Frame construction}  \label{sec_frames_construction}

Consider the set-up of Section \ref{sec_family_metrics}.~Let 
\begin{equation*}
\mathcal{R}_{\underline{s}_0}:=\bigcup_{0\, \leq \, \underline{s} \, \leq \,  \underline{s}_0}\mathbb{S}^2_{s,\underline{s}}  
\end{equation*}
for some constant $\underline{s}_0>0$.~In this section, we prove the following proposition, which implies that the frames introduced in Section \ref{sec_family_frames} are well-defined and satisfy the properties stated.

\medskip

\begin{prop} \label{prop_frame_proposition}
There exists a constant $\underline{s}_0>0$ such that the following statement holds true.~For any $(s,\theta^A)$, the system of ODEs \eqref{ODE_1}-\eqref{ODE_3}, with initial data \eqref{ODE_data_1}-\eqref{ODE_data_3}, admits a unique one-parameter family of smooth (in $\underline{s}$) solutions $$(\underline{\bsy{j}}(\epsilon), \underline{\bsy{k}}(\epsilon), \underline{\bsy{\Lambda}}(\epsilon))$$ for all $0\leq \underline{s} \leq \underline{s}_0$.~By smooth dependence on the initial data, the solutions are smooth in $(s,\theta^A)$.~The resulting one-parameter family of vector fields
\begin{align*}
\ef(\epsilon)&:= k_{a,M}\,\partial_{\underline{s}}  & \et(\epsilon)&:=\underline{\bsy{j}}(\epsilon)\,\partial_s+\underline{\bsy{k}}(\epsilon)\,\partial_{\underline{s}}+\underline{\bsy{\Lambda}}^{\theta^A}(\epsilon)\,\partial_{\theta^A} 
\end{align*}
can be completed to a one-parameter family of frames $\bsy{\mathcal{N}}(\epsilon)$ satisfying all the properties listed in Section \ref{sec_family_frames} on $\mathcal{R}_{\underline{s}_0}$.~Furthermore, the family of distributions $\bsy{\mathfrak{D}}_{\bsy{\mathcal{N}}(\epsilon)}$ is a family of non-integrable and variable distributions.
\end{prop}

\medskip

The section is divided into two parts.~In Section \ref{sec_system_ODE}, we prove the abstract local existence and uniqueness result for solutions to the system of ODEs \eqref{ODE_1}-\eqref{ODE_3} claimed by the first part of Proposition \ref{prop_frame_proposition}.~In Section \ref{sec_properties_frames}, we establish that the family of frames $\bsy{\mathcal{N}}(\epsilon)$ defined in Proposition \ref{prop_frame_proposition} satisfies all the properties listed in Section \ref{sec_family_frames}, thus proving the second part of the proposition.

\subsubsection{The system of ODEs \eqref{ODE_1}-\eqref{ODE_3}}  \label{sec_system_ODE}

In this section, we prove the following proposition.

\medskip

\begin{prop} \label{prop_local_existence_uniqueness_ODE}
There exists a real constant $\underline{s}_0>0$ such that the following statement holds true.~For any $(s,\theta^A)$, the system of ODEs \eqref{ODE_1}-\eqref{ODE_3}, with initial data \eqref{ODE_data_1}-\eqref{ODE_data_3}, admits a unique one-parameter family of smooth (in $\underline{s}$) solutions $$(\underline{\bsy{j}}(\epsilon)\, , \, \underline{\bsy{k}}(\epsilon)\, , \, \underline{\bsy{\Lambda}}(\epsilon))$$ for all $0\leq \underline{s} \leq \underline{s}_0$.~By smooth dependence on the initial data, the solutions are smooth in $(s,\theta^A)$.
\end{prop}

\medskip

\begin{proof}
For each $\epsilon$ and any $(s,\theta^A)$, the strategy of the proof is to apply Picard--Lindel\"{o}f theorem to the system of ODEs \eqref{ODE_1}-\eqref{ODE_3}, with initial data \eqref{ODE_data_1}-\eqref{ODE_data_3}.~To do that, the right hand sides of the equations \eqref{ODE_1}-\eqref{ODE_3} need to be Lipschitz continuous in the unknowns in an open neighbourhood of the initial data \eqref{ODE_data_1}-\eqref{ODE_data_3} (recall that the smooth scalar function $k_{a,M}$ is nowhere vanishing on $\mathcal{M}^*$, so the left hand sides of the equations \eqref{ODE_1}-\eqref{ODE_3} pose no issues).~To check that, it suffices to show that the denominators
\begin{equation*}
\Omega^2\,\underline{\bsy{j}}(\epsilon)+\underline{b}_{\theta^A}\,\underline{\bsy{\Lambda}}^{\theta^A}(\epsilon)
\end{equation*}
are nowhere vanishing for $\underline{s}=0$.~Using the initial data \eqref{ODE_data_1}-\eqref{ODE_data_3}, we compute
\begin{align*}
\Omega^2\,\underline{\bsy{j}}(\epsilon)+\underline{b}_{\theta^A}\,\underline{\bsy{\Lambda}}^{\theta^A}(\epsilon)|_{\underline{s}=0} &= \left. -2\,\frac{\Omega^2\,\underline{j}+\underline{b}_{\theta^B}\,\underline{\Lambda}^{\theta^B}}{k\,(\underline{j}\,\bsy{\Omega}^2(\epsilon)+\underline{\Lambda}^{\theta^C}\underline{\bsy{b}}_{\theta^C}(\epsilon))} \, \right\rvert _{\underline{s}=0} \\
&=\left. -2\,\frac{g_{a,M}(e_4^{\text{as}},e_3^{\text{as}})}{k^2\,(\underline{j}\,\bsy{\Omega}^2(\epsilon)+\underline{\Lambda}^{\theta^C}\underline{\bsy{b}}_{\theta^C}(\epsilon))} \, \right\rvert_{\underline{s}=0}  \\
&= \left. \frac{4}{k^2\,(\underline{j}\,\bsy{\Omega}^2(\epsilon)+\underline{\Lambda}^{\theta^C}\underline{\bsy{b}}_{\theta^C}(\epsilon))} \, \right\rvert_{\underline{s}=0} \\
&\neq 0 \, .
\end{align*}
One can thus conclude by Picard--Lindel\"{o}f theorem.
\end{proof}

\medskip

Proposition \ref{prop_local_existence_uniqueness_ODE} proves the first part of Proposition \ref{prop_frame_proposition}.

\subsubsection{Frame properties}  \label{sec_properties_frames}

Consider the one-parameter family of frames $\bsy{\mathcal{N}}(\epsilon)$ on $\mathcal{R}_{\underline{s}_0}$ defined in Proposition \ref{prop_frame_proposition}.~We prove the following proposition.

\medskip

\begin{prop} \label{prop_properties_frames}
The one-parameter family of frames $\bsy{\mathcal{N}}(\epsilon)$ satisfies all the properties listed in Section \ref{sec_family_frames} on $\mathcal{R}_{\underline{s}_0}$.~Furthermore, the family of distributions $\bsy{\mathfrak{D}}_{\bsy{\mathcal{N}}(\epsilon)}$ is a family of non-integrable and variable distributions.
\end{prop}

\medskip

\begin{proof}
Let $\etab_{a,M}^{\circ}$ be the canonical extension of the $\mathfrak{D}_{\mathcal{N}_{\text{as}}}$ one-tensor $\etab_{a,M}$ (recall Definition \ref{def_canonical_ext_cov_tensor}), i.e.~let $$\etab_{a,M}^{\circ}\in\Gamma((T\mathcal{M}^*)^{\star})$$ such that
\begin{align*}
\etab_{a,M}^{\circ}(e_A^{\text{as}})&=\etab_{a,M}(e_A^{\text{as}})  \, , & \etab_{a,M}^{\circ}(e_4^{\text{as}})&=0 \, , & \etab_{a,M}^{\circ}(e_3^{\text{as}})&=0 \, .
\end{align*}
Here and throughout the proof, the vector fields $(e_4^{\text{as}},e_3^{\text{as}})$ denote the fixed vector fields \eqref{bulk_alg_frame_e4_e3_coord}. 

The central observation of the proof is that $(\underline{\bsy{j}}(\epsilon) ,  \underline{\bsy{k}}(\epsilon) , \underline{\bsy{\Lambda}}(\epsilon))$ is the unique one-parameter family of solutions to the system of ODEs \eqref{ODE_1}-\eqref{ODE_3} for all $0\leq \underline{s} \leq \underline{s}_0$, with initial data \eqref{ODE_data_1}-\eqref{ODE_data_3}, if and only if the one-parameter family of frame vector fields $\et(\epsilon)$ is the unique one-parameter family of solutions to the ODE
\begin{equation} \label{transp_eqn_et_epsilon}
\bsy{\nabla}_{\bsy{e}_{\bsy{4}}(\epsilon)}\bsy{e}_{\bsy{3}}(\epsilon)= \, 2\,\bsy{\Xi}{}^{\bsy{\sharp}}(\epsilon)-\omegah_{a,M} \et(\epsilon)
\end{equation}
on $\mathcal{R}_{\underline{s}_0}$ with initial datum
\begin{equation} \label{initial_datum_vf_et}
\et(\epsilon)=(\bsy{g}(e^{\text{as}}_3,e^{\text{as}}_4))^{-2} \,\bsy{g}(e^{\text{as}}_3,e^{\text{as}}_3)\,e^{\text{as}}_4 -2\, (\bsy{g}(e^{\text{as}}_3,e^{\text{as}}_4))^{-1}\,e^{\text{as}}_3
\end{equation}
on $\mathcal{S}^*_0$, with $\bsy{\Xi}(\epsilon)\in\Gamma((T\mathcal{R}_{\underline{s}_0})^{\star})$ defined as
\begin{equation} \label{def_Xi_proof}
\bsy{\Xi}(\epsilon):=\etab_{a,M}^{\circ}-\frac{\etab_{a,M}^{\circ}(\et(\epsilon))}{g_{a,M}(e_4^{\text{as}},\et(\epsilon))}\,g_{a,M}(e_4^{\text{as}},\cdot)
\end{equation}
and $\bsy{\Xi}{}^{\bsy{\sharp}}(\epsilon)\in\Gamma(T\mathcal{R}_{\underline{s}_0})$ such that
\begin{equation*}
\bsy{\Xi}{}^{\bsy{\sharp}}(\epsilon)=\bsy{g}^{-1}(\bsy{\Xi}(\epsilon),\cdot) \, .
\end{equation*}
This fact is easy (but lengthy) to check by plugging $\et(\epsilon)$ in the form \eqref{bulk_alg_frame_e4_e3_coord} into the ODE \eqref{transp_eqn_et_epsilon} and the initial datum \eqref{initial_datum_vf_et} to obtain the system of ODEs \eqref{ODE_1}-\eqref{ODE_3} with initial data \eqref{ODE_data_1}-\eqref{ODE_data_3} respectively.~We note the identities
\begin{align*}
\bsy{\Xi}(e_A^{\text{as}})&=\etab_{a,M}(e_A^{\text{as}}) \, , & \bsy{\Xi}(\ef(\epsilon))&=0 \, , & \bsy{\Xi}(\et(\epsilon))&=0    
\end{align*}
and
\begin{align}
\bsy{g}(\bsy{\Xi}{}^{\bsy{\sharp}}(\epsilon),\ef(\epsilon))&=0 \, , & \bsy{g}(\bsy{\Xi}{}^{\bsy{\sharp}}(\epsilon),\et(\epsilon))&=0 \, . \label{aux_ids_constr_frames_products}
\end{align}
We also note the identity $\bsy{\Xi}(0)=\etab_{a,M}^{\circ}$.

We now prove the properties satisfied by the frames $\bsy{\mathcal{N}}(\epsilon)$.~The observation that $\et(\epsilon)$ satisfies the ODE \eqref{transp_eqn_et_epsilon} is repeatedly exploited.
\begin{itemize}
\item \ul{$\bsy{\mathcal{N}}(\epsilon)$ are null frames}

The identity $\bsy{g}(\ef(\epsilon),\ef(\epsilon))=0$ on $\mathcal{R}_{\underline{s}_0}$ is immediate to check.~To check the identities
\begin{align*}
\bsy{g}(\ef(\epsilon),\et(\epsilon))&=0 \, , &  \bsy{g}(\et(\epsilon),\et(\epsilon))&=0
\end{align*}
on $\mathcal{R}_{\underline{s}_0}$, we first compute the Christoffel symbols
\begin{align*}
\bsy{\Gamma}_{\underline{s}\,\underline{s}}^{\mu}(\epsilon)&=\bsy{g}^{\mu\sigma}(\epsilon)\,\partial_{\underline{s}}\,\bsy{g}_{\underline{s}\sigma}(\epsilon) \\
&=-\mathfrak{k}_{a,M}\,\delta_{\underline{s}}^{\mu} \, ,
\end{align*}
with $x^{\mu}=(s,\underline{s},\theta^A)$.~We have
\begin{align}
\bsy{\nabla}_{\bsy{e}_{\bsy{4}}(\epsilon)}\bsy{e}_{\bsy{4}}(\epsilon) &=k^2_{a,M} \bsy{\Gamma}_{\underline{s}\,\underline{s}}^{\mu}(\epsilon)\,\partial_{\mu}+k_{a,M}\,(\partial_{\underline{s}}\,k_{a,M})\,\partial_{\underline{s}} \nonumber\\
&= -k_{a,M}\,\mathfrak{k}_{a,M}\,\ef(\epsilon)+(\partial_{\underline{s}}\,k_{a,M})\ef(\epsilon) \nonumber \\
&=\omegah_{a,M}\,\ef(\epsilon) \, .  \label{aux_eqn_omegah_frame}
\end{align}
We then combine \eqref{aux_eqn_omegah_frame} with the equation \eqref{transp_eqn_et_epsilon} to compute
\begin{align*}
\bsy{\nabla}_{\ef(\epsilon)}(\bsy{g}(\ef(\epsilon),\et(\epsilon)))&=  2\,\bsy{g}(\bsy{\Xi}{}^{\bsy{\sharp}}(\epsilon),\ef(\epsilon)) -\omegah_{a,M}\,\bsy{g}(\ef(\epsilon),\et(\epsilon)) +\omegah_{a,M}\,\bsy{g}(\ef(\epsilon),\et(\epsilon)) \nonumber \\
&=0 \, , 
\end{align*}
which, using $\bsy{g}(\ef(\epsilon),\et(\epsilon))=-2$ on $\mathcal{S}^*_{0}$ (easy check left to the reader following from \eqref{initial_datum_vf_et}), allows to conclude 
\begin{equation} \label{inner_prod_relation_constr_frames}
\bsy{g}(\ef(\epsilon),\et(\epsilon))=-2
\end{equation}
on $\mathcal{R}_{\underline{s}_0}$.~We also compute
\begin{align*}
\bsy{\nabla}_{\ef(\epsilon)}(\bsy{g}(\et(\epsilon),\et(\epsilon)))&= 4\,\bsy{g}(\bsy{\Xi}{}^{\bsy{\sharp}}(\epsilon),\et(\epsilon)) -2\omegah_{a,M}\,\bsy{g}(\et(\epsilon),\et(\epsilon)) \\
&=  -2\omegah_{a,M}\,\bsy{g}(\et(\epsilon),\et(\epsilon)) \, , 
\end{align*}
which, using $\bsy{g}(\et(\epsilon),\et(\epsilon))=0$ on $\mathcal{S}^*_{0}$ (easy check left to the reader following from \eqref{initial_datum_vf_et}), allows to conclude $$\bsy{g}(\et(\epsilon),\et(\epsilon))=0$$ on $\mathcal{R}_{\underline{s}_0}$.
\item \ul{$\ef(\epsilon)$ and $\et(\epsilon)$ are global, regular and non-degenerate}

This is immediate for $\ef(\epsilon)$.~The property that $\et(\epsilon)$ is global follows from Proposition \ref{prop_local_existence_uniqueness_ODE} and the fact that the initial data \eqref{ODE_data_1}-\eqref{ODE_data_3} are smooth functions of the coordinates $(s,\theta^A)$ and smoothly extend to global functions on the whole $\mathbb{S}^2_{s,0}$-spheres.~In particular, $\underline{\bsy{\Lambda}}^{\theta^A}(\epsilon)|_{\underline{s}=0}(s,\theta^B)=0$ for some $(s,\theta^B)$ if and only if $\underline{\Lambda}^{\theta^A}|_{\underline{s}=0}(s,\theta^B)=0$, and thus the vector field $\underline{\bsy{\Lambda}}(\epsilon)|_{\mathcal{S}^*_0}$ smoothly extends to a global vector field on the whole $\mathbb{S}^2_{s,0}$-spheres (as $e_3^{\text{as}}|_{\mathcal{S}^*_0}$, and thus $\underline{\Lambda}|_{\mathcal{S}^*_0}$, do).~To prove that $\et(\epsilon)$ is non-degenerate, it suffices to observe that the initial data \eqref{ODE_data_1}-\eqref{ODE_data_3} all vanish for some $(s,\theta^B)$ if and only if the vector field $e_3^{\text{as}}$ degenerates at a point $(s,0,\theta^B)$ (i.e.~$\underline{j}|_{\underline{s}=0}(s,\theta^B)=\underline{k}|_{\underline{s}=0}(s,\theta^B)=\underline{\Lambda}^{\theta^A}|_{\underline{s}=0}(s,\theta^B)=0$).~The initial data \eqref{ODE_data_1}-\eqref{ODE_data_3} are therefore nowhere trivial, giving rise to a nowhere vanishing local solution from Proposition \ref{prop_local_existence_uniqueness_ODE}.
\item \ul{$\ef(\epsilon)$ is fixed}

This immediately follows from the definition of $\ef(\epsilon)$.
\item \ul{$\bsy{\mathcal{N}}(0)$ is the algebraically special frame}

The identity $\ef(0)\equiv e_4^{\text{as}}$ on $\mathcal{R}_{\underline{s}_0}$ immediately follows from the definition of $\ef(\epsilon)$.~The identity $\et(0) \equiv e_3^{\text{as}}$ on $\mathcal{R}_{\underline{s}_0}$ follows from $\et(0)\equiv e_3^{\text{as}}$ on $\mathcal{S}^*_0$ and $\et(0)=e_3^{\text{as}}$ solving the transport equation 
\begin{equation}  \label{transp_eqn_et_epsilon_zero}
\nabla_{e_4^{\text{as}}}\et(0)=2\,(\etab^{\sharp}_{a,M})^{\circ}-\omegah_{a,M}\,\et(0)
\end{equation}
on $\mathcal{R}_{\underline{s}_0}$, obtained from equation \eqref{transp_eqn_et_epsilon} for $\epsilon=0$.\footnote{We recall that $g_{a,M}^{-1}(\etab_{a,M}^{\circ},\cdot)=(\slashed{g}_{a,M}^{-1}(\etab_{a,M},\cdot))^{\circ}=(\etab^{\sharp}_{a,M})^{\circ}$ and $\bsy{\Xi}(0)=\etab_{a,M}^{\circ}$.}~Since $e_3^{\text{as}}$ solves the ODE \eqref{transp_eqn_et_epsilon_zero}, one concludes $\et(0) \equiv e_3^{\text{as}}$ on $\mathcal{R}_{\underline{s}_0}$ by local uniqueness for ODEs.
\item \ul{$\bsy{\omegah}(\epsilon)$, $\bsy{\xi}(\epsilon)$ and} $\widetilde{\underline{\bsy{\eta}}}(\epsilon)$ \ul{are fixed}

The identities \eqref{i_2_kerr} and \eqref{i_3_kerr} are immediate from the equation \eqref{aux_eqn_omegah_frame}.~To check the identity \eqref{i_4_kerr}, we compute
\begin{align*}
\widetilde{\bsy{\etab}}(e_{A}^{\text{as}})&=\bsy{g}((\bsy{\etab}^{\bsy{\sharp}})^{\circ}(\epsilon),e_{A}^{\text{as}})  \\
&=  \bsy{g}(\bsy{\Xi}{}^{\bsy{\sharp}}(\epsilon),e_{A}^{\text{as}})  \\
&= \bsy{\Xi}(e_{A}^{\text{as}}) \\
&= \etab_{a,M}(e_{A}^{\text{as}}) \, ,
\end{align*}
where the first identity follows from the later Definition \ref{def_restriction_tensors} of $\widetilde{\bsy{\etab}}(\epsilon)$ and the identity \eqref{aux_isometry_1} of Section \ref{sec:geom_compare_tensors} and in the second identity we used the transport equation \eqref{transp_eqn_et_epsilon} and the identities \eqref{aux_ids_constr_frames_products}. 
\item \ul{$\bsy{\mathfrak{D}}_{\bsy{\mathcal{N}}(\epsilon)}$ are non-integrable distributions}

It suffices to show that the distributions $\bsy{\mathfrak{D}}_{\bsy{\mathcal{N}}(\epsilon)}$ are non-integrable on $\mathcal{S}_0^*$.~We introduce the local, linearly independent vector fields
\begin{align*}
\bsy{X_1}(\epsilon)&=\left(\frac{\bsy{g}(e_4^{\text{as}},e_3^{\text{as}})\,\bsy{g}(e_4^{\text{as}},e_1^{\text{as}})-\bsy{g}(e_3^{\text{as}},e_3^{\text{as}})\,\bsy{g}(e_4^{\text{as}},e_1^{\text{as}})}{\bsy{g}(e_4^{\text{as}},e_3^{\text{as}})}\right)e_4^{\text{as}}+\bsy{g}(e_4^{\text{as}},e_1^{\text{as}})\,e_3^{\text{as}}-\bsy{g}(e_4^{\text{as}},e_3^{\text{as}})\,e_1^{\text{as}} \, , \\[5pt]
\bsy{X_2}(\epsilon)&=\left(\frac{\bsy{g}(e_4^{\text{as}},e_3^{\text{as}})\,\bsy{g}(e_4^{\text{as}},e_2^{\text{as}})-\bsy{g}(e_3^{\text{as}},e_3^{\text{as}})\,\bsy{g}(e_4^{\text{as}},e_2^{\text{as}})}{\bsy{g}(e_4^{\text{as}},e_3^{\text{as}})}\right)e_4^{\text{as}}+\bsy{g}(e_4^{\text{as}},e_2^{\text{as}})\,e_3^{\text{as}}-\bsy{g}(e_4^{\text{as}},e_3^{\text{as}})\,e_2^{\text{as}} 
\end{align*}
on $\mathcal{S}_0^*$.~By using the identity \eqref{initial_datum_vf_et}, one can easily check that the identities
\begin{align*}
\bsy{g}(\bsy{X_A}(\epsilon),\bsy{e}_{\bsy{4}}(\epsilon))|_{\mathcal{S}_0^*}&=0  \, , & \bsy{g}(\bsy{X_A}(\epsilon),\bsy{e}_{\bsy{3}}(\epsilon))|_{\mathcal{S}_0^*}&=0 
\end{align*}
hold for $\bsy{A}=\left\lbrace \bsy{1},\bsy{2} \right\rbrace$, and thus $(\bsy{X_1}(\epsilon),\bsy{X_2}(\epsilon))$ form a local frame of $\bsy{\mathfrak{D}}_{\bsy{\mathcal{N}}(\epsilon)}$ on $\mathcal{S}_0^*$.~By direct computation, one can also check that both the relations
\begin{align} \label{fail_integrable}
\bsy{g}([\bsy{X_1}(\epsilon),\bsy{X_2}(\epsilon)],\bsy{e}_{\bsy{4}}(\epsilon))&\neq 0 \, , & \bsy{g}([\bsy{X_1}(\epsilon),\bsy{X_2}(\epsilon)],\bsy{e}_{\bsy{3}}(\epsilon))&\neq 0
\end{align}
hold on $\mathcal{S}_0^*$.\footnote{This latter check fails for $\epsilon=0$ when $|a|=0$.}
\item \ul{$\bsy{\mathfrak{D}}_{\bsy{\mathcal{N}}(\epsilon)}$ are variable distributions}

The proof is immediate by recalling Definition \ref{def_variable_distribution} and noting the identity $\bsy{g}(\ef(\epsilon),e_A^{\text{as}})=\bsy{g}(e_4^{\text{as}},e_A^{\text{as}})$ and the relation $\bsy{g}(e_4^{\text{as}},e_A^{\text{as}})\neq 0$.
\end{itemize}

\end{proof}

\medskip

Proposition \ref{prop_properties_frames} proves the second part of Proposition \ref{prop_frame_proposition}, concluding the proof of the proposition.

\section{A morphism between horizontal tensor bundles} \label{sec:geom_compare_tensors}

We consider the fixed ambient manifold $\mathcal{M}^*$ and the one-parameter family of metrics $\bsy{g}(\epsilon)$ and associated frames $\bsy{\mathcal{N}}(\epsilon)$ on $\mathcal{M}^*$ introduced in Sections \ref{sec_family_metrics} and \ref{sec_family_frames}.~Since $\bsy{\mathfrak{D}}_{\bsy{\mathcal{N}}(\epsilon)}$ is a family of \emph{variable} distributions (see Definition \ref{def_variable_distribution} and Proposition \ref{prop_frame_proposition}), $\bsy{\mathfrak{D}}_{\bsy{\mathcal{N}}(\epsilon)}$ covariant tensors and $\mathfrak{D}_{\mathcal{N}_{\text{as}}}$ covariant tensors are smooth sections of \emph{distinct} abstract tensor bundles on $\mathcal{M}^*$ for $|\epsilon|>0$. 

\medskip

In this section, we prescribe a geometric procedure to uniquely identify any $\bsy{\mathfrak{D}}_{\bsy{\mathcal{N}}(\epsilon)}$ covariant tensor $\bsy{\varsigma}(\epsilon)$ with a $\mathfrak{D}_{\mathcal{N}_{\text{as}}}$ covariant tensor, which we shall denote by $\widetilde{\bsy{\varsigma}}(\epsilon)$.~The procedure may be seen as prescribing, for each $\epsilon$, a \emph{bundle morphism} between the bundle of $\bsy{\mathfrak{D}}_{\bsy{\mathcal{N}}(\epsilon)}$ covariant tensors and the bundle of $\mathfrak{D}_{\mathcal{N}_{\text{as}}}$ covariant tensors.\footnote{As we shall see, the procedure also yields, for each $\epsilon$, a bundle morphism between the horizontal distributions $\bsy{\mathfrak{D}}_{\bsy{\mathcal{N}}(\epsilon)}$ and $\mathfrak{D}_{\mathcal{N}_{\text{as}}}$.} 

\medskip

The geometric procedure is nonlinear and, in its abstract formulation of this section, does not assume that the metrics $\bsy{g}(\epsilon)$ solve the vacuum Einstein equations for $|\epsilon|>0$.~The procedure will prove fundamental to formulate the \emph{renormalised} vacuum Einstein equations (see Section \ref{sec_renormalised_vacuum_Einstein_equations}), where $\bsy{\mathfrak{D}}_{\bsy{\mathcal{N}}(\epsilon)}$ covariant tensors and $\mathfrak{D}_{\mathcal{N}_{\text{as}}}$ covariant tensors are to be compared.

\subsection{Overview}  \label{sec_overview_geom_compare_tensors}

Our geometric procedure starts by considering a global, regular $\bsy{\mathfrak{D}}_{\bsy{\mathcal{N}}(\epsilon)}$ covariant tensor $\bsy{\varsigma}(\epsilon)$ on $\mathcal{M}^*$ for some $\epsilon$ and consists of \emph{three} steps, each step producing a new geometric quantity.~The first step, dubbed \emph{extension}, applies Definition \ref{def_canonical_ext_cov_tensor} to obtain the \emph{canonical extension} $\bsy{\varsigma}^{\circ}(\epsilon)$ of $\bsy{\varsigma}(\epsilon)$, which is, by definition, a $\bsy{\mathfrak{D}}_{\bsy{\mathcal{N}}(\epsilon)}$-horizontal tensor.\footnote{We recall Remark \ref{rmk_horizontal_covariant_tensors} for the notion of a $\bsy{\mathfrak{D}}_{\bsy{\mathcal{N}}(\epsilon)}$ (and $\mathfrak{D}_{\mathcal{N}_{\text{as}}}$)-horizontal (covariant) tensor.}~The second step is the \emph{projection} of $\bsy{\varsigma}^{\circ}(\epsilon)$ over the horizontal bundle $\otimes(\mathfrak{D}_{\mathcal{N}_{\text{as}}})^{\star}$, thus obtaining a $\mathfrak{D}_{\mathcal{N}_{\text{as}}}$-horizontal tensor $(\Pi \bsy{\varsigma}^{\circ})(\epsilon)$.~The last step is the \emph{restriction} of $(\Pi \bsy{\varsigma}^{\circ})(\epsilon)$ to a $\mathfrak{D}_{\mathcal{N}_{\text{as}}}$ tensor $\widetilde{\bsy{\varsigma}}(\epsilon)$.~The final quantity $\widetilde{\bsy{\varsigma}}(\epsilon)$ is a \emph{global}, \emph{regular} $\mathfrak{D}_{\mathcal{N}_{\text{as}}}$ tensor on $\mathcal{M}^*$.~For $\epsilon=0$, we have $\widetilde{\bsy{\varsigma}}(0)\equiv \bsy{\varsigma}(0)$, i.e.~our geometric procedure acts as the identity map over the bundle of $\mathfrak{D}_{\mathcal{N}_{\text{as}}}$ covariant tensors.~See Table \ref{table:identification_horiz_structures} for a schematic summary of the procedure.

\medskip

\begin{table}[H]
\centering
\begin{tabular}{ |c||c|c|c| } 
 \hline
 {} & \textbf{Step} & \textbf{Notation} & \textbf{Tensor bundle} \\
 \hline\hline
$\bsy{1}$ & Extension &  $\bsy{\varsigma}^{\circ}(\epsilon)$ & \text{$\bsy{\mathfrak{D}}_{\bsy{\mathcal{N}}(\epsilon)}$-horizontal tensors} \\
\hline
$\bsy{2}$ & Projection & $(\Pi \bsy{\varsigma}^{\circ})(\epsilon)$ & \text{$\mathfrak{D}_{\mathcal{N}_{\text{as}}}$-horizontal tensors}  \\
\hline
$\bsy{3}$ & Restriction & $\widetilde{\bsy{\varsigma}}(\epsilon)$  & \text{$\mathfrak{D}_{\mathcal{N}_{\text{as}}}$ tensors} \\
\hline
\end{tabular}

\caption{}

\label{table:identification_horiz_structures}

\end{table}

\medskip

Each of the Sections \ref{sec_extension}, \ref{sec_projection} and \ref{sec_restriction} corresponds to one of the three steps of the procedure.~The projection step of Section \ref{sec_projection} is the most intricate.~The final result of our procedure is Proposition \ref{prop_final_procedure}.

\medskip

The two main technical difficulties of the procedure are outlined below.~As we shall describe, both these difficulties are overcome by introducing some \emph{auxiliary} geometric quantities induced \emph{directly} by the \emph{spacetime} geometry of $(\mathcal{M}^*,\bsy{g}(\epsilon))$ over $\mathfrak{D}_{\mathcal{N}_{\text{as}}}$.~As a result, one has to deal with a delicate mix between the $\mathfrak{D}_{\mathcal{N}_{\text{as}}}$ quantities arising from applying our geometric procedure to $\bsy{\mathfrak{D}}_{\bsy{\mathcal{N}}(\epsilon)}$ quantities and the auxiliary $\mathfrak{D}_{\mathcal{N}_{\text{as}}}$ quantities. 
\begin{itemize}
\item For later convenience, one needs to apply our geometric procedure to \emph{covariant} and \emph{Lie derivatives} of $\bsy{\mathfrak{D}}_{\bsy{\mathcal{N}}(\epsilon)}$ tensors,\footnote{Products and contractions of $\bsy{\mathfrak{D}}_{\bsy{\mathcal{N}}(\epsilon)}$ tensors are also considered, but they pose less severe technical issues.} i.e.~to $\bsy{\mathfrak{D}}_{\bsy{\mathcal{N}}(\epsilon)}$ tensors of the form
\begin{align*}
&(\bsy{\nablasl}_{\bsy{X}}\,\bsy{\varsigma})(\epsilon) \, , & &(\bsy{\slashed{\mathcal{L}}}_{\bsy{X}}\,\bsy{\varsigma})(\epsilon)
\end{align*}
for some $\bsy{X}(\epsilon)\in \Gamma(T\mathcal{M}^*)$, and be able to relate the $\mathfrak{D}_{\mathcal{N}_{\text{as}}}$ tensors 
\begin{align} \label{rough_product_procedure}
&(\widetilde{\bsy{\nablasl}_{\bsy{X}}\,\bsy{\varsigma}})(\epsilon) \, , & &(\widetilde{\bsy{\slashed{\mathcal{L}}}_{\bsy{X}}\,\bsy{\varsigma}})(\epsilon)
\end{align}
produced by our procedure to $\mathfrak{D}_{\mathcal{N}_{\text{as}}}$ tensors of the form
\begin{align} \label{convenient_product_procedure}
&(\widetilde{\bsy{\nablasl}}_{\bsy{X}}\,\widetilde{\bsy{\varsigma}})(\epsilon) \, , & &(\widetilde{\bsy{\slashed{\mathcal{L}}}}_{\bsy{X}}\,\widetilde{\bsy{\varsigma}})(\epsilon) \, ,
\end{align}
with $\widetilde{\bsy{\nablasl}}(\epsilon)$ a suitable linear connection over the bundle of $\mathfrak{D}_{\mathcal{N}_{\text{as}}}$ tensors and $\widetilde{\bsy{\slashed{\mathcal{L}}}}_{\bsy{X}}(\epsilon)$ a suitable differential operator over the bundle of $\mathfrak{D}_{\mathcal{N}_{\text{as}}}$ tensors.~The definition of the connection $\widetilde{\bsy{\nablasl}}(\epsilon)$ and the derivation of the formulae relating the $\mathfrak{D}_{\mathcal{N}_{\text{as}}}$ tensors \eqref{rough_product_procedure} and \eqref{convenient_product_procedure} will be rather technical.~These are mainly addressed at the level of the \emph{projection} step of the procedure (see Section \ref{sec_proj_cov_lie_deriv}), to then obtain the correct formulae in the final restriction step (see, for instance, the formulae in Proposition \ref{prop_final_procedure}).

It is already worth pointing out that the \emph{auxiliary} connection $\widetilde{\bsy{\nablasl}}(\epsilon)$ may be thought as the connection induced by the \emph{spacetime} Levi-Civita connection $\bsy{\nabla}(\epsilon)$ of $\bsy{g}(\epsilon)$ over the bundle of $\mathfrak{D}_{\mathcal{N}_{\text{as}}}$ tensors.~The connection $\widetilde{\bsy{\nablasl}}(\epsilon)$ is such that $\widetilde{\bsy{\nablasl}}(0)\equiv\nablasl_{a,M}$.

Some of the terms in the final formulae relating the geometric quantities \eqref{rough_product_procedure} and \eqref{convenient_product_procedure} will be referred to as \emph{nonlinear terms} and will not be explicitly computed (see Proposition \ref{prop_final_procedure}).~These terms contain products of two or more quantities which identically vanish for $\epsilon=0$.\footnote{We note that such a notion of nonlinear terms depends, of course, on our choice of gauge and frames from Section \ref{sec_perturbations_kerr}.}~We will comment on the structure of the omitted nonlinear terms when relevant.
\item We will have to deal with two distinct metric quantities on $\mathfrak{D}_{\mathcal{N}_{\text{as}}}$.~One will be the $\mathfrak{D}_{\mathcal{N}_{\text{as}}}$ tensor\footnote{Note the slight abuse of notation in \eqref{rough_metric_procedure} and \eqref{convenient_metric_procedure}, where the tilded quantity \eqref{convenient_metric_procedure} is an auxiliary quantity and \emph{not} the product of our procedure (which is double-tilded instead).} 
\begin{equation} \label{rough_metric_procedure}
\widetilde{\widetilde{\bsy{\slashed{g}}}}(\epsilon)
\end{equation} 
obtained by applying our procedure to the $\bsy{\mathfrak{D}}_{\bsy{\mathcal{N}}(\epsilon)}$ metric tensor $\bsy{\slashed{g}}(\epsilon)$, the other will be an \emph{auxiliary} symmetric $\mathfrak{D}_{\mathcal{N}_{\text{as}}}$ two-tensor 
\begin{equation} \label{convenient_metric_procedure}
\widetilde{\bsy{\slashed{g}}}(\epsilon) \, ,
\end{equation}
which may be thought as the metric induced by the \emph{spacetime} metric $\bsy{g}(\epsilon)$ over $\mathfrak{D}_{\mathcal{N}_{\text{as}}}$.~The quantity \eqref{convenient_metric_procedure} is such that $\widetilde{\bsy{\slashed{g}}}(0)\equiv \slashed{g}_{a,M}$.~The auxiliary metric quantity is introduced at the level of the \emph{projection} step of the procedure (see Section \ref{sec_proj_metric_quantities}, in particular Definition \ref{def_auxiliary_proj_metric}), to then obtain the correct metric quantity \eqref{convenient_metric_procedure} in the final step of the procedure (see Definition \ref{def_restricted_metrics}).

The quantity \eqref{convenient_metric_procedure} comes with a crucial computational advantage, in that it can be \emph{explicitly} related to the aforementioned connection $\widetilde{\bsy{\nablasl}}(\epsilon)$ over $\mathfrak{D}_{\mathcal{N}_{\text{as}}}$.~Indeed, although $\widetilde{\bsy{\nablasl}}(\epsilon)$ is \emph{not} the Levi-Civita connection of either $\widetilde{\widetilde{\bsy{\slashed{g}}}}(\epsilon)$ or $\widetilde{\bsy{\slashed{g}}}(\epsilon)$,\footnote{We recall that, being $\mathfrak{D}_{\mathcal{N}_{\text{as}}}$ a non-integrable distribution, any given metric on $\mathfrak{D}_{\mathcal{N}_{\text{as}}}$ admits no Levi-Civita connection, in that all connections over $\mathfrak{D}_{\mathcal{N}_{\text{as}}}$ fail to be torsion-free.~However, the obstruction is, in this case, even more severe.~Indeed, the connection $\widetilde{\bsy{\nablasl}}(\epsilon)$ is not even \emph{compatible} with the metric $\widetilde{\bsy{\slashed{g}}}(\epsilon)$ (see Remark \ref{rmk_incompatibility_proj_metric_connection}).} the connection $\widetilde{\bsy{\nablasl}}(\epsilon)$ and the metric $\widetilde{\bsy{\slashed{g}}}(\epsilon)$ are induced by the spacetime connection $\bsy{\nabla}(\epsilon)$ and metric $\bsy{g}(\epsilon)$ respectively, the former being \emph{Levi-Civita} of the latter.~This gives the hope that the induced quantities $\widetilde{\bsy{\nablasl}}(\epsilon)$ and $\widetilde{\bsy{\slashed{g}}}(\epsilon)$ inherit some induced version of the well-known explicit formulae relating $\bsy{\nabla}(\epsilon)$ and $\bsy{g}(\epsilon)$.\footnote{One instance of these formulae is the explicit expression for the Christoffel symbols of $\bsy{\nabla}(\epsilon)$ in terms of $\bsy{g}(\epsilon)$.}~As we shall see, this is indeed the case.

Another fundamental observation will be that the two quantities \eqref{rough_metric_procedure} and \eqref{convenient_metric_procedure} only differ by \emph{nonlinear terms} (see Proposition \ref{prop_restricted_metrics_coincide}, and Section \ref{sec_preliminary_tech_remarks} for the notion of nonlinear terms), and thus the latter quantity can replace the former in all the formulae \emph{up to nonlinear terms}.~The only metric quantity over $\mathfrak{D}_{\mathcal{N}_{\text{as}}}$ appearing in our final formulae will be $\widetilde{\bsy{\slashed{g}}}(\epsilon)$ (see Proposition \ref{prop_final_procedure}).
\end{itemize}

\medskip

Before moving to some more technical remarks, we point out that, in principle, there is no canonical bundle morphism between $\bsy{\mathfrak{D}}_{\bsy{\mathcal{N}}(\epsilon)}$ covariant tensors and $\mathfrak{D}_{\mathcal{N}_{\text{as}}}$ covariant tensors.~Our geometric procedure represents one of the many possible ways to construct a morphism between the two abstract tensor bundles on $\mathcal{M}^*$.\footnote{There are, for instance, ways to extend $\bsy{\mathfrak{D}}_{\bsy{\mathcal{N}}(\epsilon)}$ tensors to $\bsy{\mathfrak{D}}_{\bsy{\mathcal{N}}(\epsilon)}$-horizontal tensors which differ from the one implemented in our procedure.~See the related Remark \ref{rmk_choice_canonical_extension}.}~The fundamental aspect underpinning our particular choice of construction is that our bundle morphism
\begin{align*}
(\mathfrak{B}_q(\epsilon))_p \, : \, (\otimes_q(\bsy{\mathfrak{D}}_{\bsy{\mathcal{N}}(\epsilon)})^{\star},|\cdot|_{\bsy{\slashed{g}}(\epsilon)})_p &\rightarrow (\otimes_q(\mathfrak{D}_{\mathcal{N}_{\text{as}}})^{\star},|\cdot|_{\widetilde{\bsy{\slashed{g}}}(\epsilon)})_p \, , \\
\bsy{\varsigma}_p(\epsilon) &\mapsto \widetilde{\bsy{\varsigma}}_p(\epsilon)
\end{align*}
is an \emph{isometry} between normed vector spaces at each $p\in\mathcal{M}^*$, with $|\cdot|_{\bsy{\slashed{g}}(\epsilon)}$ and $|\cdot|_{\widetilde{\bsy{\slashed{g}}}(\epsilon)}$ the natural pointwise norms induced by $\bsy{\slashed{g}}(\epsilon)$ and $\widetilde{\bsy{\slashed{g}}}(\epsilon)$ over the respective tensor bundles (see Proposition \ref{prop_final_procedure} and, in particular, the identity \eqref{isometry_product}).~This fact, discussed in Section \ref{sec_isometry_horizontal_bundles}, guarantees that the later application of the geometric procedure to our problem is legitimate.

\subsubsection{Preliminary technical remarks}  \label{sec_preliminary_tech_remarks}

Some technical remarks before we start:
\begin{itemize}
\item The steps of our geometric procedure apply to global, regular, totally covariant $\bsy{\mathfrak{D}}_{\bsy{\mathcal{N}}(\epsilon)}$ tensors.~The mixed covariant-contravariant tensors appearing in the section (e.g.~$\bsy{\chi}^{\bsy{\sharp_2}}(\epsilon)$) should be always thought as suitable contractions between the inverse metric $\bsy{\slashed{g}}^{-1}(\epsilon)$ and totally covariant tensors.~This observation includes the volume form $\bsy{\slashed{\varepsilon}}(\epsilon)$, which is always treated as a totally covariant tensor.
\item Both bolded and unbolded horizontal frame indices will appear, i.e.~$\bsy{A},\bsy{B},\ldots$ and $A,B,\ldots$.~Bold indices refer to the vector fields $(\eo,\etw)$ and are raised/lowered by $\bsy{\slashed{g}}(\epsilon)$ (and its inverse $\bsy{\slashed{g}}^{-1}(\epsilon)$), unbolded indices refer to the vector fields $(e_1^{\text{as}},e_2^{\text{as}})$ and are raised/lowered by $\slashed{g}_{a,M}$ (and its inverse $\slashed{g}^{-1}_{a,M}$).~When repeated, indices are either both bolded or both unbolded.~The bolded flat symbol $\bsy{\flat}$ refers to lowering relative to $\bsy{\slashed{g}}(\epsilon)$, the unbolded flat symbol $\flat$ refers to lowering relative to $\slashed{g}_{a,M}$.~The choice of the vector fields $(\eo(\epsilon),\etw(\epsilon))$ and $(e_1^{\text{as}},e_2^{\text{as}})$ is arbitrary throughout the section.
\item All the bolded quantities depend on $\epsilon$ and may be interpreted as one-parameter families of geometric quantities.~All the identities involving $\epsilon$ hold for all $\epsilon$.~To keep the notation lighter, the $\epsilon$ will be often removed.
\end{itemize}

\subsection{Extension}  \label{sec_extension}

For any $\bsy{\mathfrak{D}}_{\bsy{\mathcal{N}}(\epsilon)}$ tensor $\bsy{\varsigma}(\epsilon)$, we canonically extend $\bsy{\varsigma}(\epsilon)$ to a $\bsy{\mathfrak{D}}_{\bsy{\mathcal{N}}(\epsilon)}$-horizontal tensor $\bsy{\varsigma}^{\circ}(\epsilon)$.~The canonical extension is intended as defined in Definition \ref{def_canonical_ext_cov_tensor}. 

\medskip

We define the $\bsy{\mathfrak{D}}_{\bsy{\mathcal{N}}(\epsilon)}$-horizontal tensor $$\bsy{\slashed{g}}^{\circ}(\epsilon)$$ on $\mathcal{M}^*$ as the canonical extension of the $\bsy{\mathfrak{D}}_{\bsy{\mathcal{N}}(\epsilon)}$ tensor $\bsy{\slashed{g}}(\epsilon)$.~We have the identity
\begin{equation} \label{decomp_spacetime_metric}
\bsy{g}(\epsilon)=-\frac{1}{2}\,(\bsy{\ef})_{\bsy{\flat}}(\epsilon)\otimes (\bsy{\et})_{\bsy{\flat}}(\epsilon)-\frac{1}{2}\,(\bsy{\et})_{\bsy{\flat}}(\epsilon)\otimes (\bsy{\ef})_{\bsy{\flat}}(\epsilon)+\bsy{\slashed{g}}^{\circ}(\epsilon) \, .
\end{equation} 

\medskip

\begin{definition} \label{def_extension_cov_derivative}
For any $\bsy{f}\in C^{\infty}(\mathcal{M}^*)$, any $\bsy{X}\in \Gamma(T\mathcal{M}^*)$ and any $\bsy{\mathfrak{D}}_{\bsy{\mathcal{N}}(\epsilon)}$ $k$-tensor $\bsy{\varsigma}$, we define the smooth scalar function $\bsy{\nablasl}^{\circ}_{\bsy{X}} \,\bsy{f}$ such that
\begin{equation*}
\bsy{\nablasl}^{\circ}_{\bsy{X}}\, \bsy{f}=\bsy{\nabla}_{\bsy{X}}\,\bsy{f} \, ,
\end{equation*}
the $\bsy{\mathfrak{D}}_{\bsy{\mathcal{N}}(\epsilon)}$-horizontal one-tensor $\bsy{\nablasl}^{\circ} \bsy{f}$ such that
\begin{align*}
(\bsy{\nablasl}^{\circ} \bsy{f})(\ea)&= (\bsy{\nabla} \bsy{f})(\ea) \, ,
\end{align*}
the $\bsy{\mathfrak{D}}_{\bsy{\mathcal{N}}(\epsilon)}$-horizontal $k$-tensors $\bsy{\nablasl^{\circ}_{\bsy{X}} \varsigma^{\circ}}$ and $\bsy{\slashed{\mathcal{L}}^{\circ}_{\bsy{X}} \varsigma^{\circ}}$ such that
\begin{align*}
(\bsy{\nablasl^{\circ}_{\bsy{X}} \varsigma^{\circ}})(\bsy{e_{A_1}},\ldots,\bsy{e_{A_K}})&= (\bsy{\nabla_{\bsy{X}} \,\varsigma^{\circ}})(\bsy{e_{A_1}},\ldots,\bsy{e_{A_K}}) \, ,  \\
(\bsy{\slashed{\mathcal{L}}^{\circ}_{\bsy{X}} \,\varsigma^{\circ}})(\bsy{e_{A_1}},\ldots,\bsy{e_{A_K}}) &= (\bsy{\mathcal{L}_{\bsy{X}}\, \varsigma^{\circ}})(\bsy{e_{A_1}},\ldots,\bsy{e_{A_K}})  
\end{align*}
and the $\bsy{\mathfrak{D}}_{\bsy{\mathcal{N}}(\epsilon)}$-horizontal $(k+1)$-tensor $\bsy{\nablasl}^{\circ} \bsy{\varsigma^{\circ}}$ such that
\begin{align*}
(\bsy{\nablasl}^{\circ} \bsy{\varsigma^{\circ}})(\bsy{e_{A_1}},\ldots,\bsy{e_{A_{K+1}}}) &= (\bsy{\nabla} \,\bsy{\varsigma^{\circ}})(\bsy{e_{A_1}},\ldots,\bsy{e_{A_{K+1}}})  \, .
\end{align*} 
\end{definition}

\medskip

We have the following proposition.

\medskip

\begin{prop} \label{prop_non_trivial_ext}
For any $\bsy{f}\in C^{\infty}(\mathcal{M}^*)$, any $\bsy{X}\in \Gamma(T\mathcal{M}^*)$ and any $\bsy{\mathfrak{D}}_{\bsy{\mathcal{N}}(\epsilon)}$ $k$-tensor $\bsy{\varsigma}$, we have
\begin{align*}
(\bsy{\varsigma}\otimes\bsy{\theta})^{\bsy{\circ}}&=\bsy{\varsigma}^{\bsy{\circ}}\otimes\bsy{\theta}^{\bsy{\circ}} 
\end{align*}
and
\begin{align*}
(\bsy{\nablasl_{\bsy{X}}}\, \bsy{f})^{\bsy{\circ}}&=\bsy{\nablasl^{\circ}_{\bsy{X}}} \, \bsy{f} \, , &  (\bsy{\nablasl}\, \bsy{f})^{\bsy{\circ}}&=\bsy{\nablasl^{\circ}} \, \bsy{f} \, , \\
(\bsy{\nablasl_{\bsy{X}} \varsigma})^{\bsy{\circ}}&=\bsy{\nablasl^{\circ}_{\bsy{X}} \varsigma^{\circ}} \, , & (\bsy{\nablasl \varsigma})^{\bsy{\circ}}&=\bsy{\nablasl}^{\circ} \bsy{\varsigma^{\circ}}   \, , \\  (\bsy{\slashed{\mathcal{L}}_{\bsy{X}} \varsigma})^{\circ} &= \bsy{\slashed{\mathcal{L}}^{\circ}_{\bsy{X}} \varsigma^{\circ}}
\end{align*}
and similarly for the extension of all differential operators defined in Section \ref{sec_nonlinear_EVE}.
\end{prop}

\medskip

\begin{proof}
The proof is immediate.

\end{proof}

\medskip

The following is an immediate corollary.

\medskip

\begin{corollary}  \label{corollary_extension}
For any $\bsy{f}\in C^{\infty}(\mathcal{M}^*)$, any $\bsy{X}\in \Gamma(T\mathcal{M}^*)$ and any $\bsy{\mathfrak{D}}_{\bsy{\mathcal{N}}(\epsilon)}$ $k$-tensor $\bsy{\varsigma}$, we have
\begin{align*}
(\bsy{\nablasl}^{\circ} \bsy{f})(\ea)&= (\bsy{\nablasl} \bsy{f})(\ea) 
\end{align*}
and
\begin{align*}
(\bsy{\nablasl^{\circ}_{\bsy{X}} \varsigma^{\circ}})(\bsy{e_{A_1}},\ldots,\bsy{e_{A_K}})&= (\bsy{\nablasl_{\bsy{X}} \varsigma})(\bsy{e_{A_1}},\ldots,\bsy{e_{A_K}}) \, ,  \\
(\bsy{\slashed{\mathcal{L}}^{\circ}_{\bsy{X}} \,\varsigma^{\circ}})(\bsy{e_{A_1}},\ldots,\bsy{e_{A_K}}) &= (\bsy{\slashed{\mathcal{L}}_{\bsy{X}} \varsigma})(\bsy{e_{A_1}},\ldots,\bsy{e_{A_K}})  \, , \\
(\bsy{\nablasl}^{\circ} \bsy{\varsigma^{\circ}})(\bsy{e_{A_1}},\ldots,\bsy{e_{A_{K+1}}}) &= (\bsy{\nablasl} \,\bsy{\varsigma})(\bsy{e_{A_1}},\ldots,\bsy{e_{A_{K+1}}})  \, .
\end{align*}
\end{corollary}

\medskip

\begin{proof}
This follows from simply computing, for instance,
\begin{align*}
(\bsy{\nablasl^{\circ}_{\bsy{X}} \varsigma^{\circ}})(\bsy{e_{A_1}},\ldots,\bsy{e_{A_K}})&= (\bsy{\nablasl_{\bsy{X}} \varsigma})^{\circ}(\bsy{e_{A_1}},\ldots,\bsy{e_{A_K}}) \\
&=(\bsy{\nablasl_{\bsy{X}} \varsigma})(\bsy{e_{A_1}},\ldots,\bsy{e_{A_K}}) \, ,
\end{align*}
where the first equality follows from Proposition \ref{prop_non_trivial_ext} and the second equality from Definition \ref{def_canonical_ext_cov_tensor}. 

\end{proof}

\subsection{Projection}  \label{sec_projection}

We start with the main definition of this step of our geometric procedure.

\medskip

\begin{definition} \label{def_proj_cov_tensors}
For any $\bsy{\mathfrak{D}}_{\bsy{\mathcal{N}}(\epsilon)}$ $k$-tensor $\bsy{\varsigma}$, we define the $\mathfrak{D}_{\mathcal{N}_{\textup{as}}}$-horizontal $k$-tensor $\Pi\bsy{\varsigma}^{\circ}$ such that
\begin{equation*}
(\Pi\bsy{\varsigma}^{\circ})(e_{A_1}^{\textup{as}},\ldots,e_{A_k}^{\textup{as}})=\bsy{\varsigma}^{\circ}(e_{A_1}^{\textup{as}},\ldots,e_{A_k}^{\textup{as}}) \, .
\end{equation*}
\end{definition}

\medskip

For later convenience, we note that any local frame $(e_{1}^{\text{as}},e_{2}^{\text{as}})$ of $\mathfrak{D}_{\mathcal{N}_{\text{as}}}$ induces a local frame $(\bsy{G}_{1}(\epsilon),\bsy{G}_{2}(\epsilon))$ of $\bsy{\mathfrak{D}}_{\bsy{\mathcal{N}}(\epsilon)}$, with
\begin{equation} \label{matrix_G}
\bsy{G}_{A}(\epsilon)=\bsy{\slashed{g}}^{-1}(\bsy{\varsigma}_A(\epsilon),\cdot) \, ,
\end{equation}
where $\bsy{\varsigma}_A(\epsilon)$ is the $\bsy{\mathfrak{D}}_{\bsy{\mathcal{N}}(\epsilon)}$ one-tensor such that $\bsy{\varsigma}_A(\eb(\epsilon))=\bsy{g}(e_A^{\text{as}},\bsy{e_B})$.~Note that $\bsy{G}_{A}(0)=e_{A}^{\text{as}}$.

\medskip

We state the following propositions.

\medskip

\begin{prop} \label{prop_rewrite_proj}
For any $\bsy{\mathfrak{D}}_{\bsy{\mathcal{N}}(\epsilon)}$ $k$-tensor $\bsy{\varsigma}$, we have
\begin{equation*}
(\Pi\bsy{\varsigma}^{\circ})(e_{A_1}^{\textup{as}},\ldots,e_{A_k}^{\textup{as}})=\bsy{\varsigma}(\bsy{G}_{A_1},\ldots,\bsy{G}_{A_k}) \, .
\end{equation*}
\end{prop}

\medskip

\begin{proof}
Immediate from Definition \ref{def_proj_cov_tensors}.

\end{proof}

\medskip

\begin{prop} \label{prop_geometric_form_horizontal_tensor}
For any $\bsy{\mathfrak{D}}_{\bsy{\mathcal{N}}(\epsilon)}$ one-tensor $\bsy{\varsigma}$, we have
\begin{equation*}
(\Pi\bsy{\varsigma}^{\circ})(\epsilon)=\bsy{\varsigma}^{\circ}(\epsilon)+\frac{1}{2}\,\bsy{\varsigma}^{\circ}(e_3^{\textup{as}})(e_4^{\textup{as}})_{\flat} \, ,
\end{equation*}
with $(e_I^{\textup{as}})_{\flat}=g_{a,M}(e_I^{\textup{as}},\cdot )$.
\end{prop}

\medskip

\begin{proof}
It is easy to check that 
\begin{equation*}
(\Pi\bsy{\varsigma}^{\circ})(\epsilon)=\bsy{\varsigma}^{\circ}(\epsilon)+\frac{1}{2}\,\bsy{\varsigma}^{\circ}(e_3^{\text{as}})(e_4^{\text{as}})_{\flat}+\frac{1}{2}\,\bsy{\varsigma}^{\circ}(e_4^{\text{as}})(e_3^{\text{as}})_{\flat} \, .
\end{equation*}
Crucially, for our choice of frames, we have $\bsy{\varsigma}^{\circ}(e_4^{\text{as}})=\bsy{\varsigma}^{\circ}(\ef(\epsilon))=0$.

\end{proof}

\medskip

\begin{prop}  \label{prop_alt_decomp_proj_metric}
We have
\begin{align*}
(\Pi\bsy{\slashed{g}}^{\circ})(\epsilon)= \,\bsy{\slashed{g}}^{\circ}(\epsilon)&+\frac{1}{2}\,\bsy{\slashed{g}}^{\circ}(e^{\textup{as}}_3,e^{\textup{as}}_A)(e_4^{\textup{as}})_{\flat}\otimes(e_A^{\textup{as}})_{\flat} \\ &+\frac{1}{2}\,\bsy{\slashed{g}}^{\circ}(e^{\textup{as}}_A,e^{\textup{as}}_3)(e_A^{\textup{as}})_{\flat}\otimes(e_4^{\textup{as}})_{\flat} \\ &-\frac{1}{4}\,\bsy{\slashed{g}}^{\circ}(e^{\textup{as}}_3,e^{\textup{as}}_3)(e_4^{\textup{as}})_{\flat}\otimes(e_4^{\textup{as}})_{\flat} \, .
\end{align*}
\end{prop}

\medskip

\begin{proof}
Easy check, analogous to the proof of Proposition \ref{prop_geometric_form_horizontal_tensor}.

\end{proof}

\medskip

The following definition will be useful to write some of the upcoming formulae in a compact fashion.

\medskip

\begin{definition} \label{def_proj_frame}
We define the $\mathfrak{D}_{\mathcal{N}_{\text{as}}}$-horizontal one-tensors $\Pi(\ef)_{\bsy{\flat}}$, $\Pi(\et)_{\bsy{\flat}}$ and $\Pi(\ea)_{\bsy{\flat}}$ such that 
\begin{align*}
\Pi(\ef)_{\bsy{\flat}}(e_A^{\textup{as}})&=\bsy{g}(\ef,e_A^{\textup{as}})  \, , & \Pi(\et)_{\bsy{\flat}}(e_A^{\textup{as}})&= \bsy{g}(\et,e_A^{\textup{as}})  \, , & \Pi(\ea)_{\bsy{\flat}}(e_B^{\textup{as}})&=\bsy{g}(\ea,e_B^{\textup{as}})  \, .
\end{align*}
\end{definition}

\subsubsection{Metric quantities} \label{sec_proj_metric_quantities}

We start with the definition of an \ul{\emph{auxiliary} metric quantity}, which arises from projecting the spacetime metric $\bsy{g}(\epsilon)$ onto $\otimes_2(\mathfrak{D}_{\mathcal{N}_{\text{as}}})^{\star}$.~For some preliminary comments on this part of the section, see the overview in Section \ref{sec_overview_geom_compare_tensors}.

\medskip

\begin{definition}  \label{def_auxiliary_proj_metric}
We define the $\mathfrak{D}_{\mathcal{N}_{\textup{as}}}$-horizontal two-tensor $(\Pi\bsy{g})(\epsilon)$ such that
\begin{equation*}
(\Pi\bsy{g})(e_A^{\textup{as}},e_B^{\textup{as}})=\bsy{g}(e_A^{\textup{as}},e_B^{\textup{as}}) \, .
\end{equation*}
\end{definition}

\medskip

We have the following proposition.

\medskip

\begin{prop}
We have
\begin{align*}
(\Pi\bsy{g})(\epsilon)= \, \bsy{g}(\epsilon)&+\frac{1}{2}\,\bsy{g}(e^{\textup{as}}_3,e^{\textup{as}}_A)(e_4^{\textup{as}})_{\flat}\otimes(e_A^{\textup{as}})_{\flat}+\frac{1}{2}\,\bsy{g}(e^{\textup{as}}_A,e^{\textup{as}}_3)(e_A^{\textup{as}})_{\flat}\otimes(e_4^{\textup{as}})_{\flat}-\frac{1}{4}\,\bsy{g}(e^{\textup{as}}_3,e^{\textup{as}}_3)(e_4^{\textup{as}})_{\flat}\otimes(e_4^{\textup{as}})_{\flat} \\
 &+\frac{1}{2}\,\bsy{g}(e^{\textup{as}}_4,e^{\textup{as}}_A)(e_3^{\textup{as}})_{\flat}\otimes(e_A^{\textup{as}})_{\flat}+\frac{1}{2}\,\bsy{g}(e^{\textup{as}}_A,e^{\textup{as}}_4)(e_A^{\textup{as}})_{\flat}\otimes(e_3^{\textup{as}})_{\flat}-\frac{1}{4}\,\bsy{g}(e^{\textup{as}}_3,e^{\textup{as}}_4)(e_4^{\textup{as}})_{\flat}\otimes(e_3^{\textup{as}})_{\flat} \\
& -\frac{1}{4}\,\bsy{g}(e^{\textup{as}}_4,e^{\textup{as}}_3)(e_3^{\textup{as}})_{\flat}\otimes(e_4^{\textup{as}})_{\flat} \, .
\end{align*}
\end{prop}

\medskip

Note that we have the four distinct spacetime tensors  
\begin{align*}
&\bsy{g}(\epsilon)\, , & &(\Pi\bsy{g})(\epsilon) \, , & &\bsy{\slashed{g}}^{\circ}(\epsilon) \,  , & &(\Pi\bsy{\slashed{g}}^{\circ})(\epsilon) \, , 
\end{align*}
with 
\begin{align*}
(\Pi\bsy{g})(0)&\equiv\bsy{\slashed{g}}^{\circ}(0)\equiv(\Pi\bsy{\slashed{g}}^{\circ})(0) \equiv \slashed{g}_{a,M}^{\circ}  \, .
\end{align*}
Note that $(\slashed{g}_{a,M}^{\circ})_{AB}= (g_{a,M})_{AB}$, but $\slashed{g}_{a,M}^{\circ} \not\equiv  g_{a,M}$.

\medskip

We have the following proposition, which will be crucial.

\medskip

\begin{prop} \label{m_1}
We have 
\begin{equation} 
(\Pi\bsy{\slashed{g}}^{\circ})(\epsilon)=(\Pi\bsy{g})(\epsilon)+ \textup{nonlinear terms} \, .
\end{equation}
\end{prop}

\medskip

\begin{proof}
It suffices to show that the two quantities agree, up to nonlinear terms, when evaluated relative to an arbitrary choice of $(e_A^{\text{as}},e_B^{\text{as}})$.~Using the identity \eqref{decomp_spacetime_metric}, we have
\begin{align*}
(\Pi\bsy{g})(e_A^{\text{as}},e_B^{\text{as}}) &= \bsy{g}(e_A^{\text{as}},e_B^{\text{as}}) \\
&= \bsy{\slashed{g}}^{\circ}(e_A^{\text{as}},e_B^{\text{as}}) -\frac{1}{2}\,\bsy{g}(e_A^{\text{as}},\bsy{e_3})\,\bsy{g}(e_B^{\text{as}},\bsy{e_4})-\frac{1}{2}\,\bsy{g}(e_A^{\text{as}},\bsy{e_4})\,\bsy{g}(e_B^{\text{as}},\bsy{e_3}) \\
&=(\Pi\bsy{\slashed{g}}^{\circ})(e_A^{\text{as}},e_B^{\text{as}}) -\frac{1}{2}\,\bsy{g}(e_A^{\text{as}},\bsy{e_3})\,\bsy{g}(e_B^{\text{as}},\bsy{e_4})-\frac{1}{2}\,\bsy{g}(e_A^{\text{as}},\bsy{e_4})\,\bsy{g}(e_B^{\text{as}},\bsy{e_3}) \, ,
\end{align*}
where the last two terms in the last line are nonlinear.

\end{proof}

\medskip

We shall now define a key contravariant tensor.~To do that, let $(\widehat{\bsy{G}}{}^1(\epsilon),\widehat{\bsy{G}}{}^2(\epsilon))$ be the local co-frame of $(\bsy{\mathfrak{D}}_{\bsy{\mathcal{N}}(\epsilon)})^{\star}$ relative to $(\bsy{G}_1(\epsilon),\bsy{G}_2(\epsilon))$ from \eqref{matrix_G}, i.e.
\begin{align*}
\widehat{\bsy{G}}{}^B(\bsy{G}_A) &=\delta_A^B \, , &  \bsy{G}_C^{\bsy{B}}\,\widehat{\bsy{G}}{}_{\bsy{A}}^C &=\delta_{\bsy{A}}^{\bsy{B}} \, .
\end{align*}
Note that $\widehat{\bsy{G}}{}^A(0)=\slashed{g}_{a,M}(e_A^{\text{as}},\cdot)$.

\medskip

\begin{definition} \label{def_inverse_proj_metric}
We define the $\mathfrak{D}_{\mathcal{N}_{\textup{as}}}$-horizontal (contravariant) tensor
\begin{align} \label{inverse_proj_metric}
(\Pi(\bsy{\slashed{g}}^{-1})^{\circ})(\epsilon) 
\end{align}
such that 
\begin{gather*}
\Pi(\bsy{\slashed{g}}^{-1})^{\circ} = \bsy{\slashed{g}}^{-1}(\widehat{\bsy{G}}{}^A,\widehat{\bsy{G}}{}^B)\, e_A^{\textup{as}}\otimes e_B^{\textup{as}} \\
(\Pi(\bsy{\slashed{g}}^{-1})^{\circ})(g_{a,M}(e_A^{\textup{as}},\cdot),g_{a,M}(e_B^{\textup{as}},\cdot))=\bsy{\slashed{g}}^{-1}(\widehat{\bsy{G}}{}^A,\widehat{\bsy{G}}{}^B) \, .
\end{gather*}
\end{definition}

\medskip

We adopt the notation
\begin{align*}
(\Pi(\bsy{\slashed{g}}^{-1})^{\circ})(g_{a,M}(e_A^{\text{as}},\cdot),g_{a,M}(e_B^{\text{as}},\cdot))&= (\Pi(\bsy{\slashed{g}}^{-1})^{\circ})^{AB} \, .
\end{align*}

\medskip

We have the following proposition, which allows to interpret the tensor \eqref{inverse_proj_metric} as the inverse of $(\Pi\bsy{\slashed{g}}^{\circ})(\epsilon)$ in a suitable sense.\footnote{Strictly speaking, the tensor $(\Pi\bsy{\slashed{g}}^{\circ})(\epsilon)$ is not invertible.}~Moreover, the tensor \eqref{inverse_proj_metric} may be interpreted as the inverse of $(\Pi\bsy{g})(\epsilon)$ in a suitable sense, up to nonlinear terms.

\medskip

\begin{prop} \label{prop_inverse_metric_proj}
We have
\begin{align*}
(\Pi(\bsy{\slashed{g}}^{-1})^{\circ})^{BC}(\Pi\bsy{\slashed{g}}^{\circ})_{AC}&= \delta_A^B \, , & (\Pi(\bsy{\slashed{g}}^{-1})^{\circ})^{BC}(\Pi\bsy{g})_{AC}&= \delta_A^B+\textup{nonlinear terms} \, .
\end{align*}
\end{prop}

\medskip

\begin{proof}
From Proposition \ref{prop_rewrite_proj}, we note that
\begin{align*}
(\Pi\bsy{\slashed{g}}^{\circ})(e_A^{\text{as}},e_B^{\text{as}})&=\bsy{\slashed{g}}(\bsy{G}_A,\bsy{G}_B) \, .
\end{align*}
We compute 
\begin{align*}
(\Pi(\bsy{\slashed{g}}^{-1})^{\circ})^{BC}(\Pi\bsy{\slashed{g}}^{\circ})_{AC}&=\bsy{\slashed{g}}^{-1}(\widehat{\bsy{G}}{}^B,\widehat{\bsy{G}}{}^C) \,\bsy{\slashed{g}}(\bsy{G}_A,\bsy{G}_C)\\
&=\widehat{\bsy{G}}{}^B(\bsy{G}_A) \\
&=\delta_A^B \, .
\end{align*}
This proves the first identity.~For the second identity, we apply Proposition \ref{m_1}.

\end{proof}

\medskip

The following proposition is useful to treat the projected volume form.

\medskip

\begin{prop} \label{prop_proj_volume_form}
We have
\begin{align*}
(\Pi\bsy{\slashed{\varepsilon}}^{\bsy{\circ}})(e_A^{\textup{as}},e_B^{\textup{as}})=& \, (\Pi\bsy{\slashed{g}}^{\circ})(e^{\textup{as}}_A,\bsy{e_1})\cdot(\Pi\bsy{\slashed{g}}^{\circ})(e^{\textup{as}}_B,\bsy{e_2})-(\Pi\bsy{\slashed{g}}^{\circ})(e^{\textup{as}}_A,\bsy{e_2})\cdot(\Pi\bsy{\slashed{g}}^{\circ})(e^{\textup{as}}_B,\bsy{e_1}) \\ &+ \textup{nonlinear terms} \, .
\end{align*}
\end{prop}

\medskip

\begin{proof}
We compute
\begin{align*}
(\Pi\bsy{\slashed{\varepsilon}}^{\bsy{\circ}})(e_A^{\textup{as}},e_B^{\textup{as}})&=\bsy{\slashed{\varepsilon}}^{\bsy{\circ}}(e_A^{\textup{as}},e_B^{\textup{as}}) \\
&=\bsy{\slashed{g}}^{\circ}(e^{\textup{as}}_A,\bsy{e_1})\,\bsy{\slashed{g}}^{\circ}(e^{\textup{as}}_B,\bsy{e_2})-\bsy{\slashed{g}}^{\circ}(e^{\textup{as}}_A,\bsy{e_2})\,\bsy{\slashed{g}}^{\circ}(e^{\textup{as}}_B,\bsy{e_1}) \\
&=(\Pi\bsy{\slashed{g}}^{\circ})(e^{\textup{as}}_A,\bsy{e_1})\,(\Pi\bsy{\slashed{g}}^{\circ})(e^{\textup{as}}_B,\bsy{e_2})-(\Pi\bsy{\slashed{g}}^{\circ})(e^{\textup{as}}_A,\bsy{e_2})\,(\Pi\bsy{\slashed{g}}^{\circ})(e^{\textup{as}}_B,\bsy{e_1})+ \text{nonlinear terms} \, .
\end{align*}

\end{proof}

\subsubsection{Projection of products and contractions of tensors} \label{sec_proj_products_contractions}

We start with the following proposition.

\medskip

\begin{prop}
For any $\bsy{\mathfrak{D}}_{\bsy{\mathcal{N}}(\epsilon)}$ $k_1$-tensor $\bsy{\varsigma}(\epsilon)$ and any $\bsy{\mathfrak{D}}_{\bsy{\mathcal{N}}(\epsilon)}$ $k_2$-tensor $\bsy{\theta}(\epsilon)$, we have
\begin{equation*}
\Pi(\bsy{\varsigma}\otimes \bsy{\theta})^{\circ}=(\Pi\bsy{\varsigma}^{\circ})\otimes (\Pi\bsy{\theta}^{\circ}) \, .
\end{equation*}
\end{prop}

\medskip

\begin{proof}
This is immediate by using the identity for tensor products from Proposition \ref{prop_non_trivial_ext}.

\end{proof}

\medskip

We give two definitions for products of $\mathfrak{D}_{\mathcal{N}_{\text{as}}}$-horizontal tensors.~The first one allows for some immediate considerations (see Propositions \ref{claim_contractions} and \ref{claim_contractions_bis}), the second one will be applied later on (see, for instance, Proposition \ref{prop_proj_D2star}).~Note that, in Definition \ref{def_proj_product_frame_vectors}, the replacement of $(\Pi\bsy{\slashed{g}}^{\circ})(\epsilon)$ by $(\Pi\,\bsy{g})(\epsilon)$ is intentional and will be convenient in the sequel.

\medskip

\begin{definition}
For any $\bsy{\mathfrak{D}}_{\bsy{\mathcal{N}}(\epsilon)}$ one-tensors $\bsy{\varsigma}(\epsilon)$ and $\bsy{\theta}(\epsilon)$, we define the product
\begin{align*}
(\Pi\,\bsy{\varsigma}^{\circ},\Pi\,\bsy{\theta}^{\circ})_{\Pi\bsy{\slashed{g}}^{\circ}}&:=(\Pi(\bsy{\slashed{g}}^{-1})^{\circ})^{AB}(\Pi\,\bsy{\varsigma}^{\circ})_A(\Pi\,\bsy{\theta}^{\circ})_B  \, .
\end{align*}
\end{definition}

\medskip

\begin{definition} \label{def_proj_product_frame_vectors}
We define the product
\begin{equation*}
\Pi(\bsy{e_I})_{\bsy{\flat}}\,\widehat{\otimes}_{\Pi\,\bsy{g}}\,\Pi(\bsy{e_J})_{\bsy{\flat}} :=\Pi(\bsy{e_I})_{\bsy{\flat}}\otimes\Pi(\bsy{e_J})_{\bsy{\flat}}+\Pi(\bsy{e_J})_{\bsy{\flat}}\otimes\Pi(\bsy{e_I})_{\bsy{\flat}}-(\Pi(\bsy{e_I})_{\bsy{\flat}},\Pi(\bsy{e_J})_{\bsy{\flat}})_{\Pi\bsy{\slashed{g}}^{\circ}}\,(\Pi\,\bsy{g}) \, .
\end{equation*}
\end{definition}

\medskip

The following convenient proposition holds.

\medskip

\begin{prop} \label{claim_contractions}
For any $\bsy{\mathfrak{D}}_{\bsy{\mathcal{N}}(\epsilon)}$ one-tensors $\bsy{\varsigma}(\epsilon)$ and $\bsy{\theta}(\epsilon)$ and $\bsy{\mathfrak{D}}_{\bsy{\mathcal{N}}(\epsilon)}$ two-tensor $\bsy{\omega}(\epsilon)$, we have the identity between smooth scalar functions 
\begin{align*}
(\Pi(\bsy{\slashed{g}}^{-1})^{\circ})^{AB}(\Pi\,\bsy{\varsigma}^{\circ})_A(\Pi\,\bsy{\theta}^{\circ})_B &=\bsy{\slashed{g}}^{\bsy{AB}}\bsy{\varsigma}_{\bsy{A}}\,\bsy{\theta}_{\bsy{B}}  \, , \\
(\Pi(\bsy{\slashed{g}}^{-1})^{\circ})^{AB}(\Pi\bsy{\omega}^{\circ})_{AB} &=\bsy{\slashed{g}}^{\bsy{AB}}\bsy{\omega}_{\bsy{AB}} \, .
\end{align*}
In particular,
\begin{align}
(\Pi\,\bsy{\varsigma}^{\circ},\Pi\,\bsy{\theta}^{\circ})_{\Pi\bsy{\slashed{g}}^{\circ}}&=(\bsy{\varsigma},\bsy{\theta})_{\bsy{\slashed{g}}} \, ,  & |\Pi\,\bsy{\varsigma}^{\circ}|^2_{\Pi\bsy{\slashed{g}}^{\circ}}&= |\bsy{\varsigma}|^2_{\bsy{\slashed{g}}} \, .
\end{align}
Analogous formulae hold for inner products of higher rank tensors and for more general full contractions.
\end{prop}

\medskip

\begin{proof}
We use Proposition \ref{prop_rewrite_proj} and Definition \ref{def_inverse_proj_metric} to compute
\begin{align*}
(\Pi(\bsy{\slashed{g}}^{-1})^{\circ})^{AB}(\Pi\,\bsy{\varsigma}^{\circ})_A(\Pi\,\bsy{\theta}^{\circ})_B &= \bsy{\slashed{g}}^{-1}(\widehat{\bsy{G}}{}^A,\widehat{\bsy{G}}{}^B)\, \bsy{\varsigma}(\bsy{G}_A)\,\bsy{\theta}(\bsy{G}_B) \\
&=\bsy{\slashed{g}}^{\bsy{AB}} \,\bsy{\varsigma}_{\bsy{A}}\,\bsy{\theta}_{\bsy{B}} \, ,
\end{align*}
where the second identity holds because tensor contractions are independent of the choice of basis.

\end{proof}

\medskip

We have the following proposition for partial contractions.

\medskip

\begin{prop} \label{claim_contractions_bis}
For any $\bsy{\mathfrak{D}}_{\bsy{\mathcal{N}}(\epsilon)}$ one-tensor $\bsy{\varsigma}(\epsilon)$ and $\bsy{\mathfrak{D}}_{\bsy{\mathcal{N}}(\epsilon)}$ two-tensor $\bsy{\theta}(\epsilon)$, we have
\begin{align*}
(\Pi(\bsy{\slashed{g}}^{-1})^{\bsy{\circ}})^{BC}(\Pi\bsy{\theta}^{\circ})_{AC}(\Pi\,\bsy{\varsigma}^{\circ})_B &=(\Pi (\bsy{\theta}{}^{\bsy{\sharp_2}} \bsy{\cdot} \bsy{\varsigma})^{\circ})_A  \, .
\end{align*}
\end{prop}

\medskip

\begin{proof}
We compute
\begin{align*}
(\Pi(\bsy{\slashed{g}}^{-1})^{\bsy{\circ}})^{BC}(\Pi\bsy{\theta}^{\circ})_{AC}(\Pi\,\bsy{\varsigma}^{\circ})_B &= \bsy{\slashed{g}}^{-1}(\widehat{\bsy{G}}{}^B,\widehat{\bsy{G}}{}^C) \,\bsy{\theta}(\bsy{G}_A,\bsy{G}_C)\,\bsy{\varsigma}(\bsy{G}_B) \\
&=\bsy{\slashed{g}}^{\bsy{BC}} \,\bsy{\theta}(\bsy{G}_A,\bsy{e_C})\,\bsy{\varsigma}_{\bsy{B}} \, ,
\end{align*}
where the second identity holds because tensor contractions are independent of the choice of basis.~By Proposition \ref{prop_rewrite_proj}, we have
\begin{equation*}
(\Pi (\bsy{\theta}{}^{\bsy{\sharp_2}} \bsy{\cdot} \bsy{\varsigma})^{\circ})_A =\bsy{\slashed{g}}^{\bsy{BC}} \,\bsy{\theta}(\bsy{G}_A,\bsy{e_C})\,\bsy{\varsigma}_{\bsy{B}} \, .
\end{equation*}

\end{proof}

\medskip

\begin{remark}  \label{rmk_contraction_volume_form}
For any $\bsy{\mathfrak{D}}_{\bsy{\mathcal{N}}(\epsilon)}$ one-tensors $\bsy{\varsigma}(\epsilon)$ and $\bsy{\theta}(\epsilon)$, Propositions \ref{claim_contractions} and \ref{claim_contractions_bis} allow to treat quantities of the form
\begin{align*}
&\bsy{\varsigma}\bsy{\wedge}\bsy{\theta} \, , & &{}^{\bsy{\star}}\bsy{\varsigma} \, .
\end{align*}
Indeed, one first writes
\begin{align*}
\bsy{\varsigma}\bsy{\wedge}\bsy{\theta}&=\bsy{\slashed{g}}^{\bsy{AC}}\bsy{\slashed{g}}^{\bsy{BD}}\bsy{\slashed{\varepsilon}}_{\bsy{CD}}\bsy{\varsigma}{}_{\bsy{A}}\,\bsy{\theta}{}_{\bsy{B}} \, , &  {}^{\bsy{\star}}\bsy{\varsigma}&= \bsy{\slashed{\varepsilon}}^{\bsy{\sharp_2}}\bsy{\cdot}\bsy{\varsigma}
\end{align*}
and then, by applying Propositions \ref{claim_contractions} and \ref{claim_contractions_bis}, gets
\begin{align*}
(\Pi(\bsy{\slashed{g}}^{-1})^{\circ}){}^{AC}(\Pi(\bsy{\slashed{g}}^{-1})^{\circ}){}^{BD}(\Pi\bsy{\slashed{\varepsilon}^{\circ}})_{CD}(\Pi\bsy{\varsigma}^{\circ}){}_{A}\,(\Pi\bsy{\theta}^{\circ}){}_{B}&=\bsy{\varsigma}\bsy{\wedge}\bsy{\theta} \\
(\Pi(\bsy{\slashed{g}}^{-1})^{\circ}){}^{BC}(\Pi\bsy{\slashed{\varepsilon}^{\circ}})_{AC}(\Pi\bsy{\varsigma}^{\circ}){}_{B}&=(\Pi({}^{\bsy{\star}}\bsy{\varsigma})^{\circ})_A \, .
\end{align*}
\end{remark}

\subsubsection{Projection of covariant and Lie derivatives}  \label{sec_proj_cov_lie_deriv}

In this section, we present some technical propositions to treat the projection of covariant and Lie derivatives.~The formulae in the propositions relate the raw projection of the $\bsy{\mathfrak{D}}_{\bsy{\mathcal{N}}(\epsilon)}$-horizontal tensors obtained in the extension step of the procedure, e.g.~the $\mathfrak{D}_{\mathcal{N}_{\text{as}}}$-horizontal tensor
\begin{align*}
&\Pi(\bsy{\nablasl}_{\bsy{X}}^{\circ}\,\bsy{\varsigma}^{\circ}) \, ,  & &\Pi(\bsy{\slashed{\mathcal{L}}}^{\circ}_X\bsy{\varsigma}^{\circ}) \, ,
\end{align*}
to more convenient $\mathfrak{D}_{\mathcal{N}_{\text{as}}}$-horizontal tensors, e.g.~$\mathfrak{D}_{\mathcal{N}_{\text{as}}}$-horizontal tensors of the form
\begin{align*}
&\Pi(\bsy{\nabla}_{\bsy{4}}(\Pi\bsy{\varsigma}^{\circ})) \, , & &\Pi(\bsy{\mathcal{L}}_X(\Pi\bsy{\varsigma}^{\circ})) \, ,
\end{align*}
with $\bsy{\nabla}(\epsilon)$ the spacetime Levi-Civita connection of $\bsy{g}(\epsilon)$ and $\bsy{\mathcal{L}}_X(\epsilon)$ the spacetime Lie derivative.~We recall that all the formulae are independent of the choice of the one-parameter family $(\eo(\epsilon),\etw(\epsilon))$.~For some preliminary comments on this part of the section, see the overview in Section \ref{sec_overview_geom_compare_tensors}.

\medskip

We give a brief outline of the propositions addressing the projection of the \emph{covariant derivatives}.~Propositions \ref{prop_proj_cov_deriv_function} and \ref{prop_proj_cov_derivatives} state the main formulae for the covariant derivatives of scalar functions and one-tensors respectively.~The proof of Proposition \ref{prop_proj_cov_derivatives} is the most technical part of the section.~Proposition \ref{prop_proj_cov_derivative_schw} states the formulae for the covariant derivative of one-tensors in a special case.~Proposition \ref{prop_proj_cov_derivatives_shears} states the relevant formulae for the covariant derivative of higher-rank tensors.~The set of propositions of the present section can be combined with those from Sections \ref{sec_proj_metric_quantities} and \ref{sec_proj_products_contractions} to write formulae for the projection of all the differential operators defined in Section \ref{sec_nonlinear_EVE}.~We give an example for the projection of a more complicated differential operator at the end of this section, in Proposition \ref{prop_proj_D2star}.~The treatment of the projection of the \emph{Lie derivatives} is somewhat less involved and is limited to Proposition \ref{prop_proj_lie_derivative_tensors}.

\medskip

We are now ready to start.

\medskip

\begin{prop} \label{prop_proj_cov_deriv_function}
For any $f\in C^{\infty}(\mathcal{M}^*)$, we have
\begin{align}
\Pi(\bsy{\nabla}f) =& \, \Pi(\bsy{\nablasl}^{\circ}f)-\frac{1}{2}\,(\bsy{\nablasl}^{\circ}_{\bsy{3}}f)\, \Pi(\ef)_{\bsy{\flat}}-\frac{1}{2}\,(\bsy{\nablasl}^{\circ}_{\bsy{4}}f) \,\Pi(\et)_{\bsy{\flat}} \, , 
\end{align}
where we recall Definition \ref{def_proj_frame} of the $\mathfrak{D}_{\mathcal{N}_{\textup{as}}}$-horizontal one-tensors $\Pi(\bsy{e_I})_{\bsy{\flat}}(\epsilon)$.
\end{prop}

\medskip

\begin{proof}
We compute
\begin{align*} 
\bsy{\nabla}f &=  \bsy{\nablasl}^{\circ}f-\frac{1}{2}\,(\bsy{\nabla}_{\bsy{3}}f) (\ef)_{\bsy{\flat}}-\frac{1}{2}\,(\bsy{\nabla}_{\bsy{4}}f) (\et)_{\bsy{\flat}} \\
&= \bsy{\nablasl}^{\circ}f-\frac{1}{2}\,(\bsy{\nablasl}^{\circ}_{\bsy{3}}f) (\ef)_{\bsy{\flat}}-\frac{1}{2}\,(\bsy{\nablasl}^{\circ}_{\bsy{4}}f) (\et)_{\bsy{\flat}} \, ,
\end{align*}
where the second identity uses Definition \ref{def_extension_cov_derivative}.~We then take the projection of both the left and right hand sides.  

\end{proof}

\medskip

\begin{prop}  \label{prop_proj_cov_derivatives}
For any $\bsy{\mathfrak{D}}_{\bsy{\mathcal{N}}(\epsilon)}$ one-tensor $\bsy{\varsigma}(\epsilon)$, we have
\begin{align}
\Pi(\bsy{\nabla}_{\bsy{4}}(\Pi\bsy{\varsigma}^{\circ}))=& \, \Pi(\bsy{\nablasl}_{\bsy{4}}^{\circ}\,\bsy{\varsigma}^{\circ})+(\Pi\bsy{\varsigma}^{\circ},\Pi\bsy{\etab}^{\circ})_{\Pi\bsy{\slashed{g}}^{\circ}}\cdot \Pi(\ef)_{\bsy{\flat}} \label{proj_cov_derivatives_4}\\ &+ \textup{nonlinear terms} \, , \nonumber\\[5pt]
\Pi(\bsy{\nabla}_{\bsy{3}}(\Pi\bsy{\varsigma}^{\circ}))=& \, \Pi(\bsy{\nablasl}_{\bsy{3}}^{\circ}\,\bsy{\varsigma}^{\circ})+(\Pi\bsy{\varsigma}^{\circ},\Pi\bsy{\eta}^{\circ})_{\Pi\bsy{\slashed{g}}^{\circ}}\cdot\Pi(\et)_{\bsy{\flat}} -g_{a,M}^{-1}(\Pi\bsy{\varsigma}^{\circ},\bsy{\underline{\mathfrak{f}}}^{\circ})\cdot(\Pi\bsy{\eta}^{\circ}) \label{proj_cov_derivatives_3}\\ &+ \textup{nonlinear terms} \, , \nonumber\\[5pt]
\Pi(\bsy{\nabla}_{\bsy{A}}(\Pi\bsy{\varsigma}^{\circ}))=& \, \Pi(\bsy{\nablasl}_{\bsy{A}}^{\circ}\,\bsy{\varsigma}^{\circ})+\frac{1}{2}\left({\bsy{\chib}^{\bsy{\sharp_2}}}{}_{\bsy{A}}^{\bsy{B}}\,\bsy{\varsigma}_{\bsy{B}}\right)\Pi(\ef)_{\bsy{\flat}}+\frac{1}{2}\left({\bsy{\chi}^{\bsy{\sharp_2}}}{}_{\bsy{A}}^{\bsy{B}}\,\bsy{\varsigma}_{\bsy{B}}\right)\Pi(\et)_{\bsy{\flat}} \label{proj_cov_derivatives_A}\\ &-\frac{1}{2}\,g_{a,M}^{-1}(\Pi\bsy{\varsigma}^{\circ},\bsy{\underline{\mathfrak{f}}}^{\circ})\cdot{\bsy{\chi}^{\bsy{\sharp_2}}}{}_{\bsy{A}}^{\bsy{B}}\,\Pi(\bsy{e_B})_{\bsy{\flat}} \nonumber \\ &+ \textup{nonlinear terms}  \nonumber
\end{align}
and
\begin{align} \label{ang_deriv_formula}
\Pi(\bsy{\nabla}(\Pi\bsy{\varsigma}^{\circ}))=& \,\Pi(\bsy{\nablasl}^{\circ}\bsy{\varsigma}^{\circ}) +\frac{1}{2}\left({\bsy{\chib}}{}^{\bsy{AB}}\,\bsy{\varsigma}_{\bsy{B}}\right)\Pi(\bsy{e_A})_{\bsy{\flat}}\otimes\Pi(\ef)_{\bsy{\flat}}+\frac{1}{2}\left({\bsy{\chi}}{}^{\bsy{AB}}\,\bsy{\varsigma}_{\bsy{B}}\right)\Pi(\bsy{e_A})_{\bsy{\flat}}\otimes \Pi(\et)_{\bsy{\flat}} \\&-\frac{1}{2}\,(\bsy{\nablasl}_{\bsy{4}}\,\bsy{\varsigma})^{\bsy{A}}\,\Pi(\et)_{\bsy{\flat}}\otimes\Pi(\ea)_{\bsy{\flat}}-\frac{1}{2}\,(\bsy{\nablasl}_{\bsy{3}}\,\bsy{\varsigma})^{\bsy{A}}\,\Pi(\ef)_{\bsy{\flat}}\otimes\Pi(\ea)_{\bsy{\flat}}  \nonumber \\ &-\frac{1}{2}\,g_{a,M}^{-1}(\Pi\bsy{\varsigma}^{\circ},\bsy{\underline{\mathfrak{f}}}^{\circ})\cdot {\bsy{\chi}}{}^{\bsy{AB}}\,\Pi(\bsy{e_A})_{\bsy{\flat}}\otimes\Pi(\bsy{e_B})_{\bsy{\flat}} \nonumber \\ &+ \textup{nonlinear terms} \, , \nonumber
\end{align}
where we recall Definition \ref{def_proj_frame} of the $\mathfrak{D}_{\mathcal{N}_{\textup{as}}}$-horizontal one-tensors $\Pi(\bsy{e_I})_{\bsy{\flat}}(\epsilon)$ and Definition \ref{def_frame_coefficients} of the $\mathfrak{D}_{\mathcal{N}_{\textup{as}}}$ one-tensor $\bsy{\underline{\mathfrak{f}}}(\epsilon)$.~The $\mathfrak{D}_{\mathcal{N}_{\textup{as}}}$-horizontal one-tensor $\bsy{\underline{\mathfrak{f}}}^{\circ}(\epsilon)$ is the canonical extension of $\bsy{\underline{\mathfrak{f}}}(\epsilon)$.
\end{prop}

\medskip

Before presenting the proof of Proposition \ref{prop_proj_cov_derivatives}, we make some remarks.

\medskip

\begin{remark} \label{rmk_proj_cov_deriv_structure_nonlinear_terms}
It will be manifest from the proof of Proposition \ref{prop_proj_cov_derivatives} that \ul{none of the \emph{nonlinear terms} in the formulae} \eqref{proj_cov_derivatives_4}, \eqref{proj_cov_derivatives_3}, \eqref{proj_cov_derivatives_A} \ul{and} \eqref{ang_deriv_formula} \ul{contain any \emph{derivatives} of $\bsy{\varsigma}^{\circ}(\epsilon)$ or $(\Pi\bsy{\varsigma}^{\circ})(\epsilon)$ (or any derivatives of connection coefficients or curvature components)}.~This remains true for the \emph{nonlinear terms} appearing in the formulae \eqref{proj_cov_derivative_schw_1}, \eqref{proj_cov_derivative_schw_2} and \eqref{proj_cov_derivative_schw_3} of Proposition \ref{prop_proj_cov_derivative_schw}.
\end{remark}

\medskip

\begin{remark}
In the formulae of Proposition \ref{prop_proj_cov_derivatives}, we allow for terms that involve both $\bsy{\mathfrak{D}}_{\bsy{\mathcal{N}}(\epsilon)}$ tensors and $\mathfrak{D}_{\mathcal{N}_{\textup{as}}}$-horizontal tensors, e.g.~the tensors $(\bsy{\nablasl}_{\bsy{4}}\,\bsy{\varsigma})(\epsilon)$ and $(\Pi(\et)_{\bsy{\flat}})(\epsilon)$ in the term
\begin{equation} \label{example_mixed_term}
-\frac{1}{2}\,(\bsy{\nablasl}_{\bsy{4}}\,\bsy{\varsigma})^{\bsy{A}}\,\Pi(\et)_{\bsy{\flat}}\otimes\Pi(\ea)_{\bsy{\flat}}
\end{equation}
from the formula \eqref{ang_deriv_formula}.~This is consistent, in that a term like \eqref{example_mixed_term} only involves \emph{components} of $\bsy{\mathfrak{D}}_{\bsy{\mathcal{N}}(\epsilon)}$ tensors, while the geometric part of the term is a $\mathfrak{D}_{\mathcal{N}_{\textup{as}}}$-horizontal tensor.~Recall that the frame indices of $\bsy{\mathfrak{D}}_{\bsy{\mathcal{N}}(\epsilon)}$ tensors are always raised relative to $\bsy{\slashed{g}}^{-1}(\epsilon)$.
\end{remark}

\medskip

\begin{remark}
Using Proposition \ref{claim_contractions}, the formulae \eqref{proj_cov_derivatives_4}-\eqref{proj_cov_derivatives_3} can be re-written as
\begin{align*}
\Pi(\bsy{\nabla}_{\bsy{4}}(\Pi\bsy{\varsigma}^{\circ}))=& \, \Pi(\bsy{\nablasl}_{\bsy{4}}^{\circ}\,\bsy{\varsigma}^{\circ})+(\bsy{\varsigma},\bsy{\etab})_{\bsy{\slashed{g}}}\cdot \Pi(\ef)_{\bsy{\flat}} \\ &+ \textup{nonlinear terms} \, ,\\[5pt]
\Pi(\bsy{\nabla}_{\bsy{3}}(\Pi\bsy{\varsigma}^{\circ}))=& \, \Pi(\bsy{\nablasl}_{\bsy{3}}^{\circ}\,\bsy{\varsigma}^{\circ})+(\bsy{\varsigma},\bsy{\eta})_{\bsy{\slashed{g}}}\cdot\Pi(\et)_{\bsy{\flat}} -g_{a,M}^{-1}(\Pi\bsy{\varsigma}^{\circ},\bsy{\underline{\mathfrak{f}}}^{\circ})\cdot(\Pi\bsy{\eta}^{\circ}) \\ &+ \textup{nonlinear terms} \, ,
\end{align*}
where we replaced the inner products.
\end{remark}

\medskip

\begin{proof}
For any $X\in \Gamma(T\mathcal{M}^*)$, we compute
\begin{align}
\bsy{\nabla}_X(\Pi\,\bsy{\varsigma}^{\circ})=& \,  \bsy{\nabla}_X \bsy{\varsigma}^{\circ} +\frac{1}{2}\,\bsy{\nabla}_X(\bsy{\varsigma}^{\circ}(e_3^{\text{as}})(e_4^{\text{as}})_{\flat}) \nonumber \\
=& \, \bsy{\nablasl}_X^{\circ}\bsy{\varsigma}^{\circ} -\frac{1}{2}((\bsy{\nabla}_X\bsy{\varsigma}^{\circ})(\et))(\ef)_{\bsy{\flat}}-\frac{1}{2}((\bsy{\nabla}_X\bsy{\varsigma}^{\circ})(\ef))(\et)_{\bsy{\flat}}+\frac{1}{2}\,\bsy{\nabla}_X(\bsy{\varsigma}^{\circ}(e_3^{\text{as}})(e_4^{\text{as}})_{\flat}) \nonumber \\
=& \, \bsy{\nablasl}_X^{\circ}\bsy{\varsigma}^{\circ}+\frac{1}{2}\,\bsy{\varsigma}^{\circ}(\bsy{\nabla}_X\et)(\ef)_{\bsy{\flat}}+\frac{1}{2}\,\bsy{\varsigma}^{\circ}(\bsy{\nabla}_X\ef)(\et)_{\bsy{\flat}}+\frac{1}{2}\,X(\bsy{\varsigma}^{\circ}(e_3^{\text{as}}))(e_4^{\text{as}})_{\flat} 
\nonumber\\ &+\frac{1}{2}\,\bsy{\varsigma}^{\circ}(e_3^{\text{as}})\bsy{\nabla}_X((e_4^{\text{as}})_{\flat}) \, , \label{proof_proj_cov_deriv_1}
\end{align}
where we recall $(e_I^{\text{as}})_{\flat}=g_{a,M}(e_I^{\text{as}},\cdot)$ and $(\bsy{e_I})_{\bsy{\flat}}=\bsy{g}(\bsy{e_I},\cdot)$ and note that
\begin{align}
\bsy{\varsigma}^{\circ}(e_3^{\text{as}})(\bsy{\nabla}_X((e_4^{\text{as}})_{\flat})(e_A^{\text{as}})=& \, -\bsy{\varsigma}^{\circ}(e_3^{\text{as}})\,(e_4^{\text{as}})_{\flat}(\bsy{\nabla}_X e_A^{\text{as}}) \nonumber\\
=& \,-\bsy{\varsigma}^{\circ}(e_3^{\text{as}})\,g_{a,M}(e_4^{\text{as}},\bsy{\nabla}_X e_A^{\text{as}}) \nonumber\\
=& \, -\bsy{\varsigma}^{\circ}(e_3^{\text{as}})\,\bsy{\slashed{g}}^{\bsy{BC}}\,g_{a,M}(e_4^{\text{as}},\bsy{\nabla}_X (\bsy{g}(e_A^{\text{as}},\eb)\bsy{e_C}))+\frac{1}{2}\,\bsy{\varsigma}^{\circ}(e_3^{\text{as}})\,g_{a,M}(e_4^{\text{as}},\bsy{\nabla}_X (\bsy{g}(e_A^{\text{as}},\et)\ef)) \nonumber\\
&+\frac{1}{2}\,\bsy{\varsigma}^{\circ}(e_3^{\text{as}})\,g_{a,M}(e_4^{\text{as}},\bsy{\nabla}_X (\bsy{g}(e_A^{\text{as}},\ef)\et)) \nonumber\\
=& -\bsy{\varsigma}^{\circ}(e_3^{\text{as}})\,\bsy{\slashed{g}}^{\bsy{BC}}\,g_{a,M}(e_4^{\text{as}},\bsy{e_C})\bsy{\nabla}_X (\bsy{g}(e_A^{\text{as}},\eb))\boxed{-\bsy{\varsigma}^{\circ}(e_3^{\text{as}})\,\bsy{\slashed{g}}^{\bsy{BC}}\,\bsy{g}(e_A^{\text{as}},\eb)\,g_{a,M}(e_4^{\text{as}},\bsy{\nabla}_X \bsy{e_C})} \nonumber\\
&+\frac{1}{2}\,\bsy{\varsigma}^{\circ}(e_3^{\text{as}})\bsy{g}(e_A^{\text{as}},\et)\,g_{a,M}(e_4^{\text{as}},\bsy{\nabla}_X \ef)+\frac{1}{2}\,\bsy{\varsigma}^{\circ}(e_3^{\text{as}})\,g_{a,M}(e_4^{\text{as}},\et)\bsy{\nabla}_X (\bsy{g}(e_A^{\text{as}},\ef)) \nonumber\\ &+\frac{1}{2}\,\bsy{\varsigma}^{\circ}(e_3^{\text{as}})\bsy{g}(e_A^{\text{as}},\ef)\,g_{a,M}(e_4^{\text{as}},\bsy{\nabla}_X \et)   \, . \label{proof_proj_cov_deriv_2}
\end{align}
The boxed term is the only term on the right hand side of \eqref{proof_proj_cov_deriv_2} which is potentially (depending on $X$) \emph{linear}, all the other terms being \emph{nonlinear} for any $X$.~We also compute
\begin{align}
\bsy{\nabla}(\Pi\,\bsy{\varsigma}^{\circ})=& \,\bsy{\nabla}\bsy{\varsigma}^{\circ}+\frac{1}{2}\bsy{\nabla}(\bsy{\varsigma}^{\circ}(e_3)(e_4^{\text{as}})_{\flat}) \nonumber\\
=& \, \bsy{\nablasl}^{\circ}\bsy{\varsigma}^{\circ}-\frac{1}{2}\,(\bsy{\nablasl}_{\bsy{4}}\,\bsy{\varsigma})^{\bsy{A}}(\et)_{\bsy{\flat}}\otimes(\ea)_{\bsy{\flat}}-\frac{1}{2}\,(\bsy{\nablasl}_{\bsy{3}}\,\bsy{\varsigma})^{\bsy{A}}(\ef)_{\bsy{\flat}}\otimes(\ea)_{\bsy{\flat}} \nonumber\\ &+\frac{1}{2}\,\bsy{\chi}^{\bsy{AB}}\,\bsy{\varsigma}_{\bsy{B}}(\ea)_{\bsy{\flat}}\otimes(\et)_{\bsy{\flat}}+\frac{1}{2}\,\bsy{\chib}^{\bsy{AB}}\,\bsy{\varsigma}_{\bsy{B}}(\ea)_{\bsy{\flat}}\otimes(\ef)_{\bsy{\flat}} \nonumber\\ &+\frac{1}{4}\,(\bsy{\nabla}\bsy{\varsigma}^{\circ})(\ef,\ef)(\et)_{\bsy{\flat}}\otimes(\et)_{\bsy{\flat}}+\frac{1}{4}\,(\bsy{\nabla}\bsy{\varsigma}^{\circ})(\et,\et)(\ef)_{\bsy{\flat}}\otimes(\ef)_{\bsy{\flat}} \nonumber\\ &+\frac{1}{4}\,(\bsy{\nabla}\bsy{\varsigma}^{\circ})(\et,\ef)(\ef)_{\bsy{\flat}}\otimes(\et)_{\bsy{\flat}}+\frac{1}{4}\,(\bsy{\nabla}\bsy{\varsigma}^{\circ})(\ef,\et)(\et)_{\bsy{\flat}}\otimes(\ef)_{\bsy{\flat}} \nonumber\\ &+\frac{1}{2}\,\bsy{\nabla}(\bsy{\varsigma}^{\circ}(e_3))\otimes(e_4^{\text{as}})_{\flat}+\frac{1}{2}\,\bsy{\varsigma}^{\circ}(e_3)\bsy{\nabla}(e_4^{\text{as}})_{\flat} \, , \label{proof_proj_cov_deriv_3}
\end{align}
where we used the equality
\begin{align*}
-\frac{1}{2}\,\bsy{\slashed{g}}^{\bsy{AB}}(\bsy{\nabla}_{\bsy{4}}\bsy{\varsigma}^{\circ})_{\bsy{A}}(\et)_{\bsy{\flat}}\otimes(\eb)_{\bsy{\flat}}&=-\frac{1}{2}\,\bsy{\slashed{g}}^{\bsy{AB}}(\bsy{\nablasl}_{\bsy{4}}\bsy{\varsigma})_{\bsy{A}}(\et)_{\bsy{\flat}}\otimes(\eb)_{\bsy{\flat}} 
\end{align*}
from Corollary \ref{corollary_extension} and analogous ones to treat the second to fifth terms on the right hand side.~The last term on the right hand side of \eqref{proof_proj_cov_deriv_3} yields
\begin{align}
\bsy{\varsigma}^{\circ}(e_3)(\bsy{\nabla}(e_4^{\text{as}})_{\flat})(e_B^{\text{as}},e_A^{\text{as}})=& \, -\bsy{\varsigma}^{\circ}(e_3^{\text{as}})\,(e_4^{\text{as}})_{\flat}(\bsy{\nabla}_{e_B^{\text{as}}} e_A^{\text{as}}) \nonumber\\
=& \, -\bsy{\varsigma}^{\circ}(e_3^{\text{as}})\,\bsy{\slashed{g}}^{\bsy{CD}}\,g_{a,M}(e_4^{\text{as}},\bsy{e_C})\,\bsy{\nabla}_{e_B^{\text{as}}} (\bsy{g}(e_A^{\text{as}},\bsy{e_D})) \nonumber \\ &-\bsy{\varsigma}^{\circ}(e_3^{\text{as}})\,\bsy{\slashed{g}}^{\bsy{CD}}\,\bsy{g}(e_A^{\text{as}},\bsy{e_D})\,g_{a,M}(e_4^{\text{as}},\bsy{\nabla}_{e_B^{\text{as}}} \bsy{e_C}) \nonumber\\
&+\frac{1}{2}\,\bsy{\varsigma}^{\circ}(e_3^{\text{as}})\,\bsy{g}(e_A^{\text{as}},\et)\,g_{a,M}(e_4^{\text{as}},\bsy{\nabla}_{e_B^{\text{as}}} \ef) \nonumber \\ &+\frac{1}{2}\,\bsy{\varsigma}^{\circ}(e_3^{\text{as}})\,g_{a,M}(e_4^{\text{as}},\et)\,\bsy{\nabla}_{e_B^{\text{as}}} (\bsy{g}(e_A^{\text{as}},\ef)) \nonumber\\ &+\frac{1}{2}\,\bsy{\varsigma}^{\circ}(e_3^{\text{as}})\,\bsy{g}(e_A^{\text{as}},\ef)\,g_{a,M}(e_4^{\text{as}},\bsy{\nabla}_{e_B^{\text{as}}} \et) \nonumber\\
=& \, -\bsy{\varsigma}^{\circ}(e_3^{\text{as}})\,\bsy{\slashed{g}}^{\bsy{CD}}\,g_{a,M}(e_4^{\text{as}},\bsy{e_C})\,\bsy{\nabla}_{e_B^{\text{as}}} (\bsy{g}(e_A^{\text{as}},\bsy{e_D})) \nonumber \\
&\boxed{-\bsy{\varsigma}^{\circ}(e_3^{\text{as}})\,\bsy{\slashed{g}}^{\bsy{CD}}\,\bsy{\slashed{g}}^{\bsy{EF}}\,\bsy{g}(e_A^{\text{as}},\bsy{e_D})\,\bsy{g}(e_B^{\text{as}},\bsy{e_E})\,g_{a,M}(e_4^{\text{as}},\bsy{\nabla}_{\bsy{e_F}} \bsy{e_C})} \nonumber\\
&+\frac{1}{2}\,\bsy{\varsigma}^{\circ}(e_3^{\text{as}})\,\bsy{\slashed{g}}^{\bsy{CD}}\,\bsy{g}(e_A^{\text{as}},\bsy{e_D})\,\bsy{g}(e_B^{\text{as}},\et)\,g_{a,M}(e_4^{\text{as}},\bsy{\nabla}_{\ef} \bsy{e_C}) \nonumber \\
&+\frac{1}{2}\,\bsy{\varsigma}^{\circ}(e_3^{\text{as}})\,\bsy{\slashed{g}}^{\bsy{CD}}\,\bsy{g}(e_A^{\text{as}},\bsy{e_D})\,\bsy{g}(e_B^{\text{as}},\ef)\,g_{a,M}(e_4^{\text{as}},\bsy{\nabla}_{\et} \bsy{e_C}) \nonumber\\
&+\frac{1}{2}\,\bsy{\varsigma}^{\circ}(e_3^{\text{as}})\,\bsy{g}(e_A^{\text{as}},\et)\,g_{a,M}(e_4^{\text{as}},\bsy{\nabla}_{e_B^{\text{as}}} \ef) \nonumber \\ &+\frac{1}{2}\,\bsy{\varsigma}^{\circ}(e_3^{\text{as}})\,g_{a,M}(e_4^{\text{as}},\et)\,\bsy{\nabla}_{e_B^{\text{as}}} (\bsy{g}(e_A^{\text{as}},\ef)) \nonumber\\ &+\frac{1}{2}\,\bsy{\varsigma}^{\circ}(e_3^{\text{as}})\,\bsy{g}(e_A^{\text{as}},\ef)\,g_{a,M}(e_4^{\text{as}},\bsy{\nabla}_{e_B^{\text{as}}} \et) \, , \label{proof_proj_cov_deriv_4}
\end{align}
where the second equality follows from the identity \eqref{proof_proj_cov_deriv_2} with $X\equiv e_B^{\text{as}}$.~The boxed term is the only term on the right hand side of \eqref{proof_proj_cov_deriv_4} which is potentially (depending on $X$) \emph{linear}, all the other terms being \emph{nonlinear} for any $X$.~To obtain the final formulae, we project the formulae \eqref{proof_proj_cov_deriv_1} and \eqref{proof_proj_cov_deriv_3} and use the formulae \eqref{proof_proj_cov_deriv_2} and \eqref{proof_proj_cov_deriv_4} to treat the last term on the right hand side of the formulae \eqref{proof_proj_cov_deriv_1} and \eqref{proof_proj_cov_deriv_3} respectively.~For the boxed term in \eqref{proof_proj_cov_deriv_2}, we note that
\begin{align*}
-\bsy{\varsigma}^{\circ}(e_3^{\text{as}})\,\bsy{\slashed{g}}^{\bsy{BC}}\,\bsy{g}(e_A^{\text{as}},\eb)\,g_{a,M}(e_4^{\text{as}},\bsy{\nabla}_{\ef} \bsy{e_C}) =& \, \text{nonlinear terms}  
\end{align*}
for $X=\ef$,
\begin{align*}
-\bsy{\varsigma}^{\circ}(e_3^{\text{as}})\, \bsy{\slashed{g}}^{\bsy{BC}}\,\bsy{g}(e_A^{\text{as}},\eb)\,g_{a,M}(e_4^{\text{as}},\bsy{\nabla}_{\et} \bsy{e_C}) =& \, -\bsy{\varsigma}^{\circ}(e_3^{\text{as}})\,\bsy{g}(e_A^{\text{as}},\eb)\,g_{a,M}(e_4^{\text{as}},\et)\,\bsy{\eta}^{\bsy{B}} \\ &+\text{nonlinear terms} \\
=& \, -\bsy{\varsigma}^{\circ}(e_3^{\text{as}})\,g_{a,M}(e_4^{\text{as}},\et)\,(\Pi\bsy{\eta^{\circ}})_{A} \\ &+\text{nonlinear terms} \\
=& \,  2\,\bsy{\varsigma}^{\circ}\left(-\frac{1}{2}\,g_{a,M}(e_4^{\text{as}},\et)\,e_3^{\text{as}}\right)\,(\Pi\bsy{\eta^{\circ}})_{A} \\ &+\text{nonlinear terms} \\
=& \,  2\,\bsy{\varsigma}^{\circ}\left(\et+\frac{1}{2}\,g_{a,M}(e_3^{\text{as}},\et)\,e_4^{\text{as}}-\slashed{g}_{a,M}^{BC}\,g_{a,M}(e_B^{\text{as}},\et)\,e_C^{\text{as}}\right)\,(\Pi\bsy{\eta^{\circ}})_{A} \\ &+\text{nonlinear terms} \\
=& \,  -2\,\slashed{g}_{a,M}^{BC}\,g_{a,M}(e_B^{\text{as}},\et)\,(\Pi\bsy{\varsigma}^{\circ})_C\,(\Pi\bsy{\eta^{\circ}})_{A} \\ &+\text{nonlinear terms} 
\end{align*}
for $X=\et$, where we used Proposition \ref{prop_rewrite_proj} in the second equality, and
\begin{align*}
-\bsy{\varsigma}^{\circ}(e_3^{\text{as}})\,\bsy{\slashed{g}}^{\bsy{BD}}\,\bsy{g}(e_A^{\text{as}},\eb)\,g_{a,M}(e_4^{\text{as}},\bsy{\nabla}_{\bsy{e_C}} \bsy{e_D}) =& \, -\frac{1}{2}\,\bsy{\varsigma}^{\circ}(e_3^{\text{as}})\,\bsy{\slashed{g}}^{\bsy{BD}}\,\bsy{g}(e_A^{\text{as}},\eb)\,g_{a,M}(e_4^{\text{as}},\et)\,\bsy{\chi}_{\bsy{CD}} \\ &+\text{nonlinear terms} \\
=& \, -\slashed{g}_{a,M}^{EF}\,\bsy{\slashed{g}}^{\bsy{BD}}\,g_{a,M}(e_E^{\text{as}},\et)\,\bsy{g}(e_A^{\text{as}},\eb)\,(\Pi\bsy{\varsigma}^{\circ})_F\,\bsy{\chi}_{\bsy{CD}} \\ &+\text{nonlinear terms} 
\end{align*}
for $X=\bsy{e_C}$.~For the boxed term in \eqref{proof_proj_cov_deriv_4}, we note that
\begin{align*}
-\bsy{\varsigma}^{\circ}(e_3^{\text{as}})\,\bsy{\slashed{g}}^{\bsy{CD}}\,\bsy{\slashed{g}}^{\bsy{EF}}\,\bsy{g}(e_A^{\text{as}},\bsy{e_D})\,\bsy{g}(e_B^{\text{as}},&\bsy{e_E})\,g_{a,M}(e_4^{\text{as}},\bsy{\nabla}_{\bsy{e_F}} \bsy{e_C}) \\ =& \, -\frac{1}{2}\,\bsy{\varsigma}^{\circ}(e_3^{\text{as}})\,\bsy{\slashed{g}}^{\bsy{CD}}\,\bsy{\slashed{g}}^{\bsy{EF}}\,\bsy{g}(e_A^{\text{as}},\bsy{e_D})\,\bsy{g}(e_B^{\text{as}},\bsy{e_E})\,g_{a,M}(e_4^{\text{as}},\et)\,\bsy{\chi}_{\bsy{FC}} \\ &+\text{nonlinear terms} \\
=& \, -\slashed{g}_{a,M}^{HL}\,\bsy{\slashed{g}}^{\bsy{CD}}\,\bsy{\slashed{g}}^{\bsy{EF}}\,g_{a,M}(e_H^{\text{as}},\et)\,\bsy{g}(e_A^{\text{as}},\bsy{e_D})\,\bsy{g}(e_B^{\text{as}},\bsy{e_E})\,(\Pi\bsy{\varsigma}^{\circ})_L \, \bsy{\chi}_{\bsy{FC}} \\ &+\text{nonlinear terms} \, .
\end{align*}
This concludes the proof.

\end{proof}

\medskip

\begin{remark}
In relation to Remark \ref{rmk_proj_cov_deriv_structure_nonlinear_terms}, we note that the fourth term on the right hand side of \eqref{proof_proj_cov_deriv_1} identically vanishes upon projection.~All the terms in the third and fourth lines on the right hand side of \eqref{proof_proj_cov_deriv_3} can be unpacked and seen not to depend on derivatives of $\bsy{\varsigma}^{\circ}(\epsilon)$, whereas the first term on the last line identically vanishes upon projection.
\end{remark}

\medskip

\begin{prop}  \label{prop_proj_cov_derivative_schw}
For any $\bsy{\mathfrak{D}}_{\bsy{\mathcal{N}}(\epsilon)}$ one-tensor $\bsy{\varsigma}(\epsilon)$ such that $\bsy{\varsigma}(0)\equiv 0$, we have
\begin{align}
\Pi(\bsy{\nabla}_{{\bsy{4}}}(\Pi\bsy{\varsigma}))=& \, \Pi(\bsy{\nablasl}_{\bsy{4}}^{\circ}\,\bsy{\varsigma})+ \textup{nonlinear terms} \, , \label{proj_cov_derivative_schw_1}\\[5pt]
\Pi(\bsy{\nabla}_{\bsy{3}}(\Pi\bsy{\varsigma}))=& \, \Pi(\bsy{\nablasl}_{\bsy{3}}^{\circ}\,\bsy{\varsigma})+ \textup{nonlinear terms} \, , \label{proj_cov_derivative_schw_2}\\[5pt]
\Pi(\bsy{\nabla}_{\bsy{A}}(\Pi\bsy{\varsigma}))=& \, \Pi(\bsy{\nablasl}_{\bsy{A}}^{\circ}\,\bsy{\varsigma})+ \textup{nonlinear terms} \, , \label{proj_cov_derivative_schw_3}\\[5pt]
\Pi(\bsy{\nabla}(\Pi\bsy{\varsigma}))=& \,\Pi(\bsy{\nablasl}^{\circ}\bsy{\varsigma}) +\textup{nonlinear terms} \, . \label{proj_cov_derivative_schw_4}
\end{align}
\end{prop}

\medskip

\begin{proof}
This is immediate from the formulae of Proposition \ref{prop_proj_cov_derivatives}.

\end{proof}

\medskip

We will not need formulae analogous to those of Proposition \ref{prop_proj_cov_derivatives} for general higher rank tensors.~It suffices to have the following propositions.

\medskip

\begin{prop} \label{prop_proj_cov_derivatives_shears}
For any $X\in \Gamma(T\mathcal{M}^*)$, we have
\begin{align}
\Pi(\bsy{\nabla}_{X}(\Pi\bsy{\chih^{\circ}}))&= \Pi(\bsy{\nablasl}_{X}^{\circ}\,\bsy{\chih^{\circ}})+ \textup{nonlinear terms}  \, , \label{proj_cov_deriv_outgoing_shear} \\[5pt]
\Pi(\bsy{\nabla}_{X}(\Pi\bsy{\chibh^{\circ}}))&= \Pi(\bsy{\nablasl}_{X}^{\circ}\,\bsy{\chibh^{\circ}})+ \textup{nonlinear terms} \label{proj_cov_deriv_ingoing_shear}
\end{align}
and
\begin{align} 
\Pi(\bsy{\nabla}(\Pi\bsy{\chih^{\circ}}))&= \Pi(\bsy{\nablasl}^{\circ}\bsy{\chih^{\circ}})+ \textup{nonlinear terms}  \, , \\[5pt]
\Pi(\bsy{\nabla}(\Pi\bsy{\chibh^{\circ}}))&= \Pi(\bsy{\nablasl}^{\circ}\bsy{\chibh^{\circ}})+ \textup{nonlinear terms} \, .
\end{align}
The same identities hold for $\bsy{\alpha}$ and $\bsy{\alphab}$ replacing $\bsy{\chih}$ and $\bsy{\chibh}$.
\end{prop}

\medskip

\begin{proof}
We start by noting that we have $$\Pi\bsy{\chih^{\circ}}=\bsy{\chih^{\circ}}+\text{nonlinear terms} \, .$$ We compute
\begin{align*}
\Pi(\bsy{\nabla}_{X}(\Pi\bsy{\chih^{\circ}}))&=\Pi(\bsy{\nabla}_{X}\bsy{\chih^{\circ}})+\text{nonlinear terms} \\
&=\Pi(\bsy{\nablasl}_{X}^{\circ}\,\bsy{\chih^{\circ}})+ \textup{nonlinear terms} \, ,
\end{align*}
where the second equality is easy to check and follows from the schematic formula
\begin{align*}
\Pi(\bsy{\nabla}_{X}\bsy{\chih^{\circ}})
=& \, \Pi(\bsy{\nablasl}_{X}^{\circ}\,\bsy{\chih^{\circ}}) \\ &+(\bsy{\nabla}_X\,\bsy{\chih^{\circ}})(\ef,\bsy{e_I}) \cdot \Pi(\bsy{e_3})_{\bsy{\flat}}\otimes \Pi(\bsy{e_I})_{\bsy{\flat}} \\ &+(\bsy{\nabla}_X\,\bsy{\chih^{\circ}})(\et,\bsy{e_I}) \cdot \Pi(\bsy{e_4})_{\bsy{\flat}}\otimes \Pi(\bsy{e_I})_{\bsy{\flat}} \, ,
\end{align*}
for which the last two terms are nonlinear.~We also compute 
\begin{align*}
\Pi(\bsy{\nabla}(\Pi\bsy{\chih^{\circ}}))&=\Pi(\bsy{\nabla}\bsy{\chih^{\circ}})+\text{nonlinear terms} \\
&=\Pi(\bsy{\nablasl}^{\circ}\bsy{\chih^{\circ}})+ \textup{nonlinear terms} \, ,
\end{align*}
where the second equality is easy to check and follows from the schematic formula
\begin{align*}
\Pi(\bsy{\nabla}\bsy{\chih^{\circ}})
=& \, \Pi(\bsy{\nablasl}^{\circ} \bsy{\chih^{\circ}}) \\ 
&+(\bsy{\nabla}_{\bsy{4}}\,\bsy{\chih^{\circ}})(\bsy{e_I},\bsy{e_J}) \cdot \Pi(\bsy{e_3})_{\bsy{\flat}}\otimes \Pi(\bsy{e_I})_{\bsy{\flat}}\otimes \Pi(\bsy{e_J})_{\bsy{\flat}} \\ &+(\bsy{\nabla}_{\bsy{3}}\,\bsy{\chih^{\circ}})(\bsy{e_I},\bsy{e_J}) \cdot \Pi(\bsy{e_4})_{\bsy{\flat}}\otimes \Pi(\bsy{e_I})_{\bsy{\flat}}\otimes \Pi(\bsy{e_J})_{\bsy{\flat}} \\
&+(\bsy{\nabla}_{\bsy{I}}\,\bsy{\chih^{\circ}})(\ef,\bsy{e_J}) \cdot \Pi(\bsy{e_I})_{\bsy{\flat}}\otimes\Pi(\bsy{e_3})_{\bsy{\flat}}\otimes \Pi(\bsy{e_J})_{\bsy{\flat}} \\ &+(\bsy{\nabla}_{\bsy{I}}\,\bsy{\chih^{\circ}})(\et,\bsy{e_J}) \cdot \Pi(\bsy{e_I})_{\bsy{\flat}}\otimes \Pi(\bsy{e_4})_{\bsy{\flat}}\otimes \Pi(\bsy{e_J})_{\bsy{\flat}} \, ,
\end{align*}
for which the last four terms are nonlinear.~Analogous considerations apply to $\bsy{\chibh^{\circ}}$. 

\end{proof}

\medskip

\begin{remark} \label{rmk_proj_cov_deriv_shears_structure_nonlinear_terms}
It is manifest from the proof of Proposition \ref{prop_proj_cov_derivatives_shears} that \ul{none of the \emph{nonlinear terms} in the formulae} \eqref{proj_cov_deriv_outgoing_shear} \ul{and} \eqref{proj_cov_deriv_ingoing_shear} \ul{contain any \emph{derivatives} of $\bsy{\chih^{\circ}}(\epsilon)$, $(\Pi\bsy{\chih^{\circ}})(\epsilon)$ or $\bsy{\chibh^{\circ}}(\epsilon)$, $(\Pi\bsy{\chibh^{\circ}})(\epsilon)$ respectively (or any derivatives of connection coefficients or curvature components)}.
\end{remark}

\medskip

We now state the relevant formulae for the Lie derivatives.~Proposition \ref{prop_proj_lie_derivative_tensors} states the formula for the Lie derivative of one-tensors.~Proposition \ref{prop_proj_lie_derivative_metric} states the formula for the Lie derivative of the metric, together with a useful formula for the covariant derivative of the metric.

\medskip

\begin{prop} \label{prop_proj_lie_derivative_tensors}
For any $\bsy{\mathfrak{D}}_{\bsy{\mathcal{N}}(\epsilon)}$ one-tensor $\bsy{\varsigma}(\epsilon)$ and any $X\in \Gamma(T\mathcal{M}^*)$, we have
\begin{align}
\Pi(\bsy{\mathcal{L}}_X(\Pi\bsy{\varsigma}^{\circ}))=& \, \Pi(\bsy{\slashed{\mathcal{L}}}^{\circ}_X\bsy{\varsigma}^{\circ})+\frac{1}{2}(\bsy{\varsigma}^{\circ}([X,\et]))\Pi(\ef)_{\bsy{\flat}}+\frac{1}{2}(\bsy{\varsigma}^{\circ}([X,\ef]))\Pi(\et)_{\bsy{\flat}} \\ &+\frac{1}{2}\,\bsy{\varsigma}^{\circ}(e_3^{\text{as}})\Pi(\bsy{\mathcal{L}}_X(e_4^{\textup{as}})_{\flat})   \, . \nonumber 
\end{align}
\end{prop}

\medskip

\begin{proof}
We compute
\begin{align*}
\bsy{\mathcal{L}}_X(\Pi\bsy{\varsigma}^{\circ})=& \, \bsy{\mathcal{L}}_X\bsy{\varsigma}^{\circ}+\frac{1}{2}\,\bsy{\mathcal{L}}_X(\bsy{\varsigma}^{\circ}(e_3^{\text{as}})(e_4^{\text{as}})_{\flat}) \\
=& \, \bsy{\slashed{\mathcal{L}}}^{\circ}_X\bsy{\varsigma}^{\circ}-\frac{1}{2}((\bsy{\mathcal{L}}_X\bsy{\varsigma}^{\circ})(\et))(\ef)_{\bsy{\flat}}-\frac{1}{2}((\bsy{\mathcal{L}}_X\bsy{\varsigma}^{\circ})(\ef))(\et)_{\bsy{\flat}}+\frac{1}{2}\,\bsy{\mathcal{L}}_X(\bsy{\varsigma}^{\circ}(e_3^{\text{as}})(e_4^{\text{as}})_{\flat}) \\
=& \, \bsy{\slashed{\mathcal{L}}}^{\circ}_X\bsy{\varsigma}^{\circ}+\frac{1}{2}(\bsy{\varsigma}^{\circ}([X,\et]))(\ef)_{\bsy{\flat}}+\frac{1}{2}(\bsy{\varsigma}^{\circ}([X,\ef]))(\et)_{\bsy{\flat}} \\ &+\frac{1}{2}\,X(\bsy{\varsigma}^{\circ}(e_3^{\text{as}}))(e_4^{\text{as}})_{\flat}+\frac{1}{2}\,\bsy{\varsigma}^{\circ}(e_3^{\text{as}})\bsy{\mathcal{L}}_X((e_4^{\text{as}})_{\flat}) \, .
\end{align*}

\end{proof}

\medskip

The following proposition states two important identities.

\medskip

\begin{prop} \label{prop_proj_lie_derivative_metric}
For any $X\in \Gamma(T\mathcal{M}^*)$, we have
\begin{align}
\Pi(\bsy{\nabla}_X(\Pi\bsy{g}))=& \,\frac{1}{2}\,\slashed{g}_{a,M}^{AB}\,\bsy{g}(e^{\textup{as}}_3,e^{\textup{as}}_A)\,\left(\Pi(\bsy{\nabla}_X(e_4^{\textup{as}})_{\flat})\otimes (e_B^{\textup{as}})_{\flat}+(e_B^{\textup{as}})_{\flat}\otimes \Pi(\bsy{\nabla}_X(e_4^{\textup{as}})_{\flat}) \right) \label{deriv_metric_formula}\\
 &+\frac{1}{2}\,\slashed{g}_{a,M}^{AB}\,\bsy{g}(e^{\textup{as}}_4,e^{\textup{as}}_A)\,\left(\Pi(\bsy{\nabla}_X(e_3^{\textup{as}})_{\flat})\otimes(e_B^{\textup{as}})_{\flat}+(e_B^{\textup{as}})_{\flat}\otimes\Pi(\bsy{\nabla}_X(e_3^{\textup{as}})_{\flat})\right) \nonumber\\
 & +\textup{nonlinear terms} \nonumber 
\end{align}
and
\begin{align}
\Pi(\bsy{\mathcal{L}}_X(\Pi\bsy{g})) =&\, \Pi(\bsy{\slashed{\mathcal{L}}}^{\circ}_X\,\bsy{\slashed{g}}^{\circ})+\frac{1}{2}\,\bsy{\slashed{g}}^{\bsy{AB}}\,\bsy{\slashed{g}}^{\circ}([X,\et],\ea)\left(\Pi(\ef)_{\bsy{\flat}}\otimes \Pi(\eb)_{\bsy{\flat}}+\Pi(\eb)_{\bsy{\flat}}\otimes \Pi(\ef)_{\bsy{\flat}}\right) \label{Lie_deriv_metric_formula}\\ & +\frac{1}{2}\,\bsy{\slashed{g}}^{\bsy{AB}}\,\bsy{\slashed{g}}^{\circ}([X,\ef],\ea)\left(\Pi(\et)_{\bsy{\flat}}\otimes \Pi(\eb)_{\bsy{\flat}} +\Pi(\eb)_{\bsy{\flat}}\otimes \Pi(\et)_{\bsy{\flat}}\right) \nonumber\\
&+\frac{1}{2}\,\slashed{g}_{a,M}^{AB}\,\bsy{\slashed{g}}^{\circ}(e^{\textup{as}}_3,e^{\textup{as}}_A)\left(\Pi(\bsy{\mathcal{L}}_X(e_4^{\textup{as}})_{\flat})\otimes(e_B^{\textup{as}})_{\flat}+(e_B^{\textup{as}})_{\flat}\otimes\Pi(\bsy{\mathcal{L}}_X(e_4^{\textup{as}})_{\flat})\right) \nonumber\\ &+\textup{nonlinear terms} \, . \nonumber
\end{align}
\end{prop}

\medskip

\begin{proof}
To prove the formula \eqref{deriv_metric_formula}, one simply applies the Definition \ref{def_auxiliary_proj_metric} and the compatibility identity $\bsy{\nabla}_X\bsy{g}=0$.~To prove the formula \eqref{Lie_deriv_metric_formula}, we first observe that
\begin{align*}
\Pi(\bsy{\mathcal{L}}_X(\Pi\bsy{g}))=& \, \Pi(\bsy{\mathcal{L}}_X(\Pi\bsy{\slashed{g}}^{\circ})) +\text{nonlinear terms}   \\
=& \, \Pi(\bsy{\mathcal{L}}_X\,\bsy{\slashed{g}}^{\circ})+\frac{1}{2}\,\slashed{g}_{a,M}^{AB}\,\bsy{\slashed{g}}^{\circ}(e^{\text{as}}_3,e^{\text{as}}_A)\left(\Pi(\bsy{\mathcal{L}}_X(e_4^{\text{as}})_{\flat})\otimes(e_B^{\text{as}})_{\flat}+(e_B^{\text{as}})_{\flat}\otimes\Pi(\bsy{\mathcal{L}}_X(e_4^{\text{as}})_{\flat})\right) \\ &+\text{nonlinear terms} \, ,
\end{align*} 
where the first equality follows from Proposition \ref{m_1}, the second equality from Proposition \ref{prop_alt_decomp_proj_metric}.~We also have
\begin{align*}
\Pi(\bsy{\mathcal{L}}_X\,\bsy{\slashed{g}}^{\circ})=& \, \Pi(\bsy{\slashed{\mathcal{L}}}^{\circ}_X\,\bsy{\slashed{g}}^{\circ})-\frac{1}{2}\,\bsy{\slashed{g}}^{\bsy{AB}}\,(\bsy{\mathcal{L}}_X\,\bsy{\slashed{g}}^{\circ})(\et,\ea)\cdot\Pi(\ef)_{\bsy{\flat}}\otimes\Pi(\eb)_{\bsy{\flat}} \\
&  -\frac{1}{2}\,\bsy{\slashed{g}}^{\bsy{AB}}\,(\bsy{\mathcal{L}}_X\,\bsy{\slashed{g}}^{\circ})(\ea,\et)\cdot \Pi(\eb)_{\bsy{\flat}}\otimes \Pi(\ef)_{\bsy{\flat}} \\ &-\frac{1}{2}\,\bsy{\slashed{g}}^{\bsy{AB}}\,(\bsy{\mathcal{L}}_X\,\bsy{\slashed{g}}^{\circ})(\ea,\ef)\cdot \Pi(\eb)_{\bsy{\flat}}\otimes \Pi(\et)_{\bsy{\flat}}\\ &-\frac{1}{2}\,\bsy{\slashed{g}}^{\bsy{AB}}\,(\bsy{\mathcal{L}}_X\,\bsy{\slashed{g}}^{\circ})(\ef,\ea)\cdot \Pi(\et)_{\bsy{\flat}}\otimes \Pi(\eb)_{\bsy{\flat}} \\ &+\text{nonlinear terms} \\
=& \, \Pi(\bsy{\slashed{\mathcal{L}}}^{\circ}_X\,\bsy{\slashed{g}}^{\circ})+\frac{1}{2}\,\bsy{\slashed{g}}^{\bsy{AB}}\,\bsy{\slashed{g}}^{\circ}([X,\et],\ea)\cdot\Pi(\ef)_{\bsy{\flat}}\otimes \Pi(\eb)_{\bsy{\flat}} \\ & +\frac{1}{2}\,\bsy{\slashed{g}}^{\bsy{AB}}\,\bsy{\slashed{g}}^{\circ}([X,\et],\ea)\cdot\Pi(\eb)_{\bsy{\flat}}\otimes \Pi(\ef)_{\bsy{\flat}} \\ &+\frac{1}{2}\,\bsy{\slashed{g}}^{\bsy{AB}}\,\bsy{\slashed{g}}^{\circ}([X,\ef],\ea)\cdot\Pi(\eb)_{\bsy{\flat}}\otimes \Pi(\et)_{\bsy{\flat}} \\ &+\frac{1}{2}\,\bsy{\slashed{g}}^{\bsy{AB}}\,\bsy{\slashed{g}}^{\circ}([X,\ef],\ea)\cdot\Pi(\et)_{\bsy{\flat}}\otimes \Pi(\eb)_{\bsy{\flat}} \\ &+\text{nonlinear terms} \, .
\end{align*}
Combining the two computations yields the formula \eqref{Lie_deriv_metric_formula}.

\end{proof}

\medskip

\begin{remark}  \label{rmk_incompatibility_proj_metric_connection}
The reader should note that the identity \eqref{deriv_metric_formula} implies the incompatibility relation $$\Pi(\bsy{\nabla}_{e_A^{\textup{as}}}(\Pi\bsy{g})) \neq 0$$ when $|\epsilon|>0$.~The incompatibility arises because our geometric projection is performed relative to a frame (i.e.~$\mathcal{N}_{\textup{as}}$) that is \emph{not} null relative to $\bsy{g}(\epsilon)$ when $|\epsilon|>0$.
\end{remark}

\medskip

We conclude the section by noting that the projection formulae of Section \ref{sec_proj_products_contractions} and the formulae derived in the present section can be combined together to treat the projection of all the differential operators defined in Section \ref{sec_nonlinear_EVE}.~For instance, using Propositions \ref{m_1} and \ref{claim_contractions}, one can state and prove the following proposition.

\medskip

\begin{prop} \label{prop_proj_D2star}
For any $\bsy{\mathfrak{D}}_{\bsy{\mathcal{N}}(\epsilon)}$ one-tensor $\bsy{\varsigma}(\epsilon)$, we have
\begin{align}
-2\,\Pi(\bsy{\slashed{\mathcal{D}}_2^{\star}}^{\circ}\bsy{\varsigma}^{\circ})=&\, \Pi(\bsy{\nabla}(\Pi\bsy{\varsigma}^{\circ}))+\Pi(\bsy{\nabla}(\Pi\bsy{\varsigma}^{\circ}))^{\mathsf{T}}-(\Pi(\bsy{\slashed{g}}^{-1})^{\circ})^{AB}(\Pi(\bsy{\nabla}(\Pi\bsy{\varsigma}^{\circ})))_{AB}\,(\Pi\,\bsy{g}) \label{formula_proj_D2star} \\
&+\frac{1}{2}\,(\bsy{\nablasl}_{\bsy{3}}\,\bsy{\varsigma}-{\bsy{\chib}}{}^{\bsy{\sharp_2}}\bsy{\cdot}\bsy{\varsigma})^{\bsy{A}}\,\Pi(\ef)_{\bsy{\flat}}\,\widehat{\otimes}_{\Pi\bsy{g}}\,\Pi(\ea)_{\bsy{\flat}} \nonumber\\ & +\frac{1}{2}\,(\bsy{\nablasl}_{\bsy{4}}\,\bsy{\varsigma}-{\bsy{\chi}}{}^{\bsy{\sharp_2}}\bsy{\cdot}\bsy{\varsigma})^{\bsy{A}}\,\Pi(\et)_{\bsy{\flat}}\,\widehat{\otimes}_{\Pi\bsy{g}}\,\Pi(\ea)_{\bsy{\flat}} \nonumber \\ &+ \textup{nonlinear terms} \, . \nonumber
\end{align}
\end{prop}

\medskip

\begin{proof}
We compute
\begin{equation*}
-2\,\Pi(\bsy{\slashed{\mathcal{D}}_2^{\star}}{}^{\circ}\bsy{\varsigma}^{\circ})=\Pi(\bsy{\nablasl}^{\circ}\bsy{\varsigma}^{\circ})+\Pi(\bsy{\nablasl}^{\circ}\bsy{\varsigma}^{\circ})^{\mathsf{T}}-(\Pi(\bsy{\slashed{g}}^{-1})^{\circ})^{AB}(\Pi(\bsy{\nablasl}^{\circ}\bsy{\varsigma}^{\circ}))_{AB}\,(\Pi\,\bsy{g})+\textup{nonlinear terms} \, .
\end{equation*}
Using then the formula \eqref{ang_deriv_formula}, we get
\begin{align}
-2\,\Pi(\bsy{\slashed{\mathcal{D}}_2^{\star}}{}^{\circ}\bsy{\varsigma}^{\circ})=& \, \Pi(\bsy{\nabla}(\Pi\bsy{\varsigma}^{\circ}))-\frac{1}{2}\left({\bsy{\chib}}{}^{\bsy{AB}}\,\bsy{\varsigma}_{\bsy{B}}\right)\Pi(\bsy{e_A})_{\bsy{\flat}}\otimes\Pi(\ef)_{\bsy{\flat}}-\frac{1}{2}\left({\bsy{\chi}}{}^{\bsy{AB}}\,\bsy{\varsigma}_{\bsy{B}}\right)\Pi(\bsy{e_A})_{\bsy{\flat}}\otimes \Pi(\et)_{\bsy{\flat}} \label{ex_eq_line_1} \\ 
&+\frac{1}{2}\,(\bsy{\nablasl}_{\bsy{4}}\,\bsy{\varsigma})^{\bsy{A}}\,\Pi(\et)_{\bsy{\flat}}\otimes\Pi(\ea)_{\bsy{\flat}}+\frac{1}{2}\,(\bsy{\nablasl}_{\bsy{3}}\,\bsy{\varsigma})^{\bsy{A}}\,\Pi(\ef)_{\bsy{\flat}}\otimes\Pi(\ea)_{\bsy{\flat}}  \label{ex_eq_line_2} \\ 
&+\frac{1}{2}\,g_{a,M}^{-1}(\Pi\bsy{\varsigma}^{\circ},\bsy{\underline{\mathfrak{f}}}^{\circ})\cdot{\bsy{\chi}}{}^{\bsy{AB}}\, \Pi(\bsy{e_A})_{\bsy{\flat}}\otimes\Pi(\bsy{e_B})_{\bsy{\flat}} \label{ex_eq_line_3}\\ 
&+(\text{line} \,\eqref{ex_eq_line_1}+ \text{line} \,\eqref{ex_eq_line_2}+\text{line} \,\eqref{ex_eq_line_3})^{\mathsf{T}} \nonumber\\
&-(\Pi(\bsy{\slashed{g}}^{-1})^{\circ})^{AB}(\text{line} \,\eqref{ex_eq_line_1}+\text{line} \,\eqref{ex_eq_line_2}+\text{line} \,\eqref{ex_eq_line_3})_{AB} (\Pi\bsy{g}) \nonumber\\
&+\text{nonlinear terms} \, . \nonumber
\end{align}
Note that in the formula \eqref{formula_proj_D2star} we dropped the nonlinear term
\begin{equation*}
\frac{1}{2}\,g_{a,M}^{-1}(\Pi\bsy{\varsigma}^{\circ},\bsy{\underline{\mathfrak{f}}}^{\circ})\cdot{\bsy{\chi}}{}^{\bsy{AB}}\, \Pi(\bsy{e_A})_{\bsy{\flat}}\,\widehat{\otimes}_{\Pi\bsy{g}}\,\Pi(\bsy{e_B})_{\bsy{\flat}} \, .
\end{equation*}

\end{proof}

\medskip

\begin{remark}  \label{rmk_proj_cov_deriv_D2star_structure_nonlinear_terms}
The \emph{nonlinear terms} in the formula \eqref{formula_proj_D2star} contain derivatives of $\bsy{\varsigma}(\epsilon)$ and $(\Pi\bsy{\varsigma}^{\circ})(\epsilon)$.~These terms originate from replacing $(\Pi\bsy{\slashed{g}}^{\circ})(\epsilon)$ by $(\Pi\bsy{g})(\epsilon)$ in the formula.
\end{remark}

\subsection{Restriction}  \label{sec_restriction}

We start by defining the restriction of $\mathfrak{D}_{\mathcal{N}_{\text{as}}}$-horizontal $k$-tensors to $\mathfrak{D}_{\mathcal{N}_{\text{as}}}$ $k$-tensors.

\medskip

\begin{definition} \label{def_restriction_tensors}
For any $\bsy{\mathfrak{D}}_{\bsy{\mathcal{N}}(\epsilon)}$ $k$-tensor $\bsy{\varsigma}(\epsilon)$, we define the $\mathfrak{D}_{\mathcal{N}_{\textup{as}}}$ $k$-tensor $\widetilde{\bsy{\varsigma}}(\epsilon)$ such that
\begin{equation*}
\widetilde{\bsy{\varsigma}}(e_{A_1}^{\textup{as}},\ldots,e_{A_k}^{\textup{as}})=(\Pi\bsy{\varsigma}^{\circ})(e_{A_1}^{\textup{as}},\ldots,e_{A_k}^{\textup{as}}) \, .
\end{equation*}
\end{definition}

\medskip

\begin{remark}  \label{rmk_zero_restriction}
For any $\bsy{\mathfrak{D}}_{\bsy{\mathcal{N}}(\epsilon)}$ $k$-tensor $\bsy{\varsigma}(\epsilon)$, we have
\begin{equation*}
\widetilde{\bsy{\varsigma}}(0) \equiv \bsy{\varsigma}(0) \, .
\end{equation*}
\end{remark}

\medskip

With a slight abuse of notation (see later Remark \ref{rmk_abuse_notation_restriction_metric}), we give the following definition.

\medskip

\begin{definition} \label{def_restricted_metrics}
We define the $\mathfrak{D}_{\mathcal{N}_{\textup{as}}}$ two-tensors $\widetilde{\bsy{\slashed{g}}}(\epsilon)$ and $\widetilde{\widetilde{\bsy{\slashed{g}}}}(\epsilon)$ such that
\begin{align*}
\widetilde{\bsy{\slashed{g}}}(e_A^{\textup{as}},e_B^{\textup{as}})&=(\Pi\,\bsy{g})(e_A^{\textup{as}},e_B^{\textup{as}}) \, ,  &
\widetilde{\widetilde{\bsy{\slashed{g}}}}(e_A^{\textup{as}},e_B^{\textup{as}})&=(\Pi\,\bsy{\slashed{g}}^{\circ})(e_A^{\textup{as}},e_B^{\textup{as}}) \, .
\end{align*}
\end{definition}

\medskip

\begin{remark} \label{rmk_abuse_notation_restriction_metric}
We remark that $\widetilde{\bsy{\slashed{g}}}(\epsilon)$ is the restriction to $\mathfrak{D}_{\mathcal{N}_{\textup{as}}}$ of $(\Pi\bsy{g})(\epsilon)$, instead of $(\Pi\bsy{\slashed{g}}^{\circ})(\epsilon)$.~For a preliminary discussion of why this slight abuse of notation will be later convenient, see the overview in Section \ref{sec_overview_geom_compare_tensors}.
\end{remark}

\medskip

\begin{remark}
In general, the $\mathfrak{D}_{\mathcal{N}_{\textup{as}}}$ two-tensor $\widetilde{\bsy{\slashed{\varepsilon}}}(\epsilon)$ (defined according to Definition \ref{def_restriction_tensors}) differs from both the standard volume form associated to $\widetilde{\bsy{\slashed{g}}}(\epsilon)$ and the standard volume form associated to $\widetilde{\widetilde{\bsy{\slashed{g}}}}(\epsilon)$.
\end{remark}

\medskip

\begin{definition}
We define the $\mathfrak{D}_{\mathcal{N}_{\textup{as}}}$ contravariant tensor $\widetilde{\bsy{\slashed{g}}^{-1}}(\epsilon)$ such that
\begin{equation*}
\widetilde{\bsy{\slashed{g}}^{-1}}(\slashed{g}_{a,M}(e_A^{\textup{as}},\cdot),\slashed{g}_{a,M}(e_B^{\textup{as}},\cdot))=(\Pi(\bsy{\slashed{g}}^{-1})^{\circ})(g_{a,M}(e_A^{\textup{as}},\cdot),g_{a,M}(e_B^{\textup{as}},\cdot))  \, .
\end{equation*}
\end{definition}

\medskip

We adopt the notation
\begin{equation*}
\widetilde{\bsy{\slashed{g}}^{-1}}{}^{AB}=\widetilde{\bsy{\slashed{g}}^{-1}}(\slashed{g}_{a,M}(e_A^{\text{as}},\cdot),\slashed{g}_{a,M}(e_B^{\text{as}},\cdot)) \, .
\end{equation*}

\medskip

In view of Propositions \ref{m_1} and \ref{prop_inverse_metric_proj}, we have the following proposition.

\medskip

\begin{prop} \label{prop_restricted_metrics_coincide}
We have the identities
\begin{equation}
\widetilde{\widetilde{\bsy{\slashed{g}}}}(\epsilon)=\widetilde{\bsy{\slashed{g}}}(\epsilon)+\textup{nonlinear terms} \label{restriction_two_metrics_relation}
\end{equation}
and
\begin{equation}
\widetilde{\bsy{\slashed{g}}^{-1}}{}^{BC}\,\widetilde{\bsy{\slashed{g}}}_{AC}=\delta_{A}^B  + \textup{nonlinear terms} \, . \label{restriction_inverse_metric_relation}
\end{equation}
\end{prop}

\medskip

Let the $\mathfrak{D}_{\mathcal{N}_{\text{as}}}$ one-tensor $$\widetilde{(\bsy{e_I})_{\bsy{\flat}}}(\epsilon)$$ be the restriction of $\Pi(\bsy{e_I})_{\bsy{\flat}}(\epsilon)$ to $\mathfrak{D}_{\mathcal{N}_{\text{as}}}$.~We have the following definition for the restriction of covariant and Lie derivatives.

\medskip

\begin{definition} \label{def_restriction_derivatives}
For any $\bsy{f}\in C^{\infty}(\mathcal{M}^*)$, any $\bsy{X}\in \Gamma(T\mathcal{M}^*)$ and any $\bsy{\mathfrak{D}}_{\bsy{\mathcal{N}}(\epsilon)}$ $k$-tensor $\bsy{\varsigma}$, we define the smooth scalar function $\widetilde{\bsy{\nablasl}}_{\bsy{X}} \,\bsy{f}$ such that
\begin{equation*}
\widetilde{\bsy{\nablasl}}_{\bsy{X}} \,\bsy{f}=\Pi(\bsy{\nabla}_{\bsy{X}}\,\bsy{f}) \, ,
\end{equation*}
the $\mathfrak{D}_{\mathcal{N}_{\textup{as}}}$ one-tensor $\widetilde{\bsy{\nablasl}} \bsy{f}$ such that
\begin{align*}
(\widetilde{\bsy{\nablasl}} \bsy{f})(e_A^{\textup{as}})&= (\Pi(\bsy{\nabla} \bsy{f}))(e_A^{\textup{as}}) \, ,
\end{align*}
the $\mathfrak{D}_{\mathcal{N}_{\textup{as}}}$ $k$-tensors $\bsy{\widetilde{\nablasl}_{\bsy{X}} \widetilde{\varsigma}}$ and $\bsy{\widetilde{\slashed{\mathcal{L}}}_{\bsy{X}} \widetilde{\varsigma}}$ such that
\begin{align*}
(\bsy{\widetilde{\nablasl}_{\bsy{X}} \widetilde{\varsigma}})(e_{A_1}^{\textup{as}},\ldots,e_{A_k}^{\textup{as}})&= (\Pi(\bsy{\nabla_{\bsy{X}}} \,(\Pi\bsy{\varsigma^{\circ}})))(e_{A_1}^{\textup{as}},\ldots,e_{A_k}^{\textup{as}}) \, ,  \\
(\bsy{\widetilde{\slashed{\mathcal{L}}}_{\bsy{X}} \,\widetilde{\varsigma}})(e_{A_1}^{\textup{as}},\ldots,e_{A_k}^{\textup{as}}) &= (\Pi(\bsy{\mathcal{L}_{\bsy{X}}}\, (\Pi\bsy{\varsigma^{\circ}})))(e_{A_1}^{\textup{as}},\ldots,e_{A_k}^{\textup{as}})  
\end{align*}
and the $\mathfrak{D}_{\mathcal{N}_{\textup{as}}}$ $(k+1)$-tensor $\widetilde{\bsy{\nablasl}}\, \widetilde{\bsy{\varsigma}}$ such that
\begin{align*}
(\widetilde{\bsy{\nablasl}} \,\widetilde{\bsy{\varsigma}})(e_{A_1}^{\textup{as}},\ldots,e_{A_k}^{\textup{as}}) &= (\Pi(\bsy{\nabla} \,(\Pi\bsy{\varsigma^{\circ}})))(e_{A_1}^{\textup{as}},\ldots,e_{A_k}^{\textup{as}})  \, .
\end{align*}
\end{definition}

\medskip

\begin{remark}
The $\mathfrak{D}_{\mathcal{N}_{\textup{as}}}$ tensor $\widetilde{\bsy{\slashed{g}}}(\epsilon)$ is a metric tensor on $\mathfrak{D}_{\mathcal{N}_{\textup{as}}}$ and $\widetilde{\bsy{\nablasl}}(\epsilon)$ is a linear connection over the bundle of $\mathfrak{D}_{\mathcal{N}_{\textup{as}}}$ tensors.~The latter does \emph{not} coincide with the Levi-Civita connection of $\widetilde{\bsy{\slashed{g}}}(\epsilon)$ \ul{for all $|\epsilon|\geq 0$}.~In particular,
\begin{equation}
(\widetilde{\bsy{\nablasl}}_{e_A^{\textup{as}}}\,\widetilde{\bsy{\slashed{g}}})(\epsilon) \not\equiv 0  \label{incomp_metric_connection_restricted}
\end{equation}
\ul{when $|\epsilon|>0$} (see already Remark \ref{rmk_incompatibility_proj_metric_connection}).~Although the incompatibility relation \eqref{incomp_metric_connection_restricted} originates from our particular geometric procedure, the connection $\widetilde{\bsy{\nablasl}}(\epsilon)$ also inherits the torsion of the original connection $\bsy{\nablasl}(\epsilon)$ on $\bsy{\mathfrak{D}}_{\bsy{\mathcal{N}}(\epsilon)}$ \ul{for all $|\epsilon|\geq 0$} (see Remark \ref{rmk_proj_connection_not_LC}), which is a general aspect when one treats non-integrable structures.
\end{remark}

\medskip

\begin{remark} \label{rmk_zero_restriction_connection}
We have the identity 
\begin{equation*}
\widetilde{\bsy{\slashed{g}}}(0)\equiv \slashed{g}_{a,M}
\end{equation*}
between metric tensors on $\mathfrak{D}_{\mathcal{N}_{\textup{as}}}$ and the identity
\begin{equation*}
\widetilde{\bsy{\nablasl}}(0)\equiv \nablasl_{a,M} 
\end{equation*}
between linear connections over the bundle of $\mathfrak{D}_{\mathcal{N}_{\textup{as}}}$ tensors.
\end{remark}

\medskip

We need one last definition in order to state our final formulae in Proposition \ref{prop_final_procedure}.

\medskip

\begin{definition}
For any $\bsy{\mathfrak{D}}_{\bsy{\mathcal{N}}(\epsilon)}$ one-tensors $\bsy{\varsigma}(\epsilon)$ and $\bsy{\theta}(\epsilon)$, we define the products
\begin{align*}
(\widetilde{\bsy{\varsigma}},\widetilde{\bsy{\theta}})_{\widetilde{\bsy{\slashed{g}}}}&:=(\widetilde{\bsy{\slashed{g}}^{-1}})^{AB}\,\widetilde{\bsy{\varsigma}}_A\, \widetilde{\bsy{\theta}}_B  \, ,  \\
\widetilde{\bsy{\varsigma}}\,\bsy{\wedge}_{\widetilde{\bsy{\slashed{g}}}}\,\widetilde{\bsy{\theta}} &:= (\widetilde{\bsy{\slashed{g}}^{-1}}){}^{AC}(\widetilde{\bsy{\slashed{g}}^{-1}}){}^{BD}\widetilde{\bsy{\slashed{\varepsilon}}}_{CD}\widetilde{\bsy{\varsigma}}_{A}\,\widetilde{\bsy{\theta}}_{B} 
\end{align*}
and
\begin{align*}
\widetilde{\bsy{\varsigma}}\,\widehat{\otimes}_{\widetilde{\bsy{\slashed{g}}}}\,\widetilde{\bsy{\theta}} &:=\widetilde{\bsy{\varsigma}}\otimes \widetilde{\bsy{\theta}}+\widetilde{\bsy{\theta}}\otimes \widetilde{\bsy{\varsigma}}-(\widetilde{\bsy{\varsigma}},\widetilde{\bsy{\theta}})_{\widetilde{\bsy{\slashed{g}}}}\,\widetilde{\bsy{\slashed{g}}}  
\end{align*}
and the differential operators
\begin{align*}
\bsy{\widetilde{\slashed{\textbf{div}}} \,\widetilde{\varsigma}}&:= (\widetilde{\bsy{\slashed{g}}^{-1}})^{AB} (\widetilde{\bsy{\nablasl}}\,\widetilde{\bsy{\varsigma}})_{AB} \, , \\
\bsy{\widetilde{\slashed{\textbf{curl}}} \,\widetilde{\varsigma}}&:=  (\widetilde{\bsy{\slashed{g}}^{-1}}){}^{AC}(\widetilde{\bsy{\slashed{g}}^{-1}}){}^{BD}\widetilde{\bsy{\slashed{\varepsilon}}}_{CD} (\widetilde{\bsy{\nablasl}}\,\widetilde{\bsy{\varsigma}})_{AB} \, , \\
\widetilde{\bsy{\slashed{\mathcal{D}}}_{\bsy{2}}^{\star}}\,\widetilde{\bsy{\varsigma}}&:=-\frac{1}{2}\,(\widetilde{\bsy{\nablasl}}\,\widetilde{\bsy{\varsigma}}+(\widetilde{\bsy{\nablasl}}\,\widetilde{\bsy{\varsigma}})^{\mathsf{T}}-(\widetilde{\bsy{\slashed{g}}^{-1}})^{AB}(\widetilde{\bsy{\nablasl}}\,\widetilde{\bsy{\varsigma}})_{AB}\,\widetilde{\bsy{\slashed{g}}}) \, .
\end{align*}
For any $\bsy{\mathfrak{D}}_{\bsy{\mathcal{N}}(\epsilon)}$ two-tensors $\bsy{\theta}(\epsilon)$ and $\bsy{\omega}(\epsilon)$, we define
\begin{align*}
\textbf{tr}_{\widetilde{\bsy{\slashed{g}}}}\,\widetilde{\bsy{\theta}}&:= (\widetilde{\bsy{\slashed{g}}^{-1}})^{AB}\widetilde{\bsy{\theta}}_{AB} \, , \\
(\widetilde{\bsy{\theta}}{}^{\bsy{\sharp_2}}\bsy{\cdot}_{\widetilde{\bsy{\slashed{g}}}}\,\widetilde{\bsy{\xi}})_A &:=(\widetilde{\bsy{\slashed{g}}^{-1}})^{BC}\widetilde{\bsy{\theta}}_{AC}\widetilde{\bsy{\xi}}_B \, , \\
(\widetilde{\bsy{\theta}},\widetilde{\bsy{\omega}})_{\widetilde{\bsy{\slashed{g}}}}&:=(\widetilde{\bsy{\slashed{g}}^{-1}})^{AD}(\widetilde{\bsy{\slashed{g}}^{-1}})^{CB}\widetilde{\bsy{\theta}}_{AC} \widetilde{\bsy{\omega}}_{BD} \, , \\
\widetilde{\bsy{\theta}}\,\bsy{\wedge}_{\widetilde{\bsy{\slashed{g}}}}\,\widetilde{\bsy{\omega}} &:= (\widetilde{\bsy{\slashed{g}}^{-1}})^{AE}(\widetilde{\bsy{\slashed{g}}^{-1}})^{DF}(\widetilde{\bsy{\slashed{g}}^{-1}})^{CB}\widetilde{\bsy{\slashed{\varepsilon}}}_{EF}\widetilde{\bsy{\theta}}_{AC} \widetilde{\bsy{\omega}}_{BD}  \, , \\
(\bsy{\widetilde{\slashed{\textbf{div}}} \,\widetilde{\theta}})_A &:=(\widetilde{\bsy{\slashed{g}}^{-1}})^{CB}(\widetilde{\bsy{\nablasl}}_C\,\widetilde{\bsy{\theta}})_{AB} \, .
\end{align*}
\end{definition}

\medskip

\begin{remark}
In view of the identity \eqref{restriction_inverse_metric_relation}, the $\mathfrak{D}_{\mathcal{N}_{\textup{as}}}$ tensors $\widetilde{\bsy{\varsigma}}\,\widehat{\otimes}_{\widetilde{\bsy{\slashed{g}}}}\,\widetilde{\bsy{\theta}}$ and $\widetilde{\bsy{\slashed{\mathcal{D}}}_{\bsy{2}}^{\star}}\,\widetilde{\bsy{\varsigma}}$ are \emph{traceless} relative to $\widetilde{\bsy{\slashed{g}}^{-1}}$ up to nonlinear terms.
\end{remark}

\medskip

The following proposition combines all the formulae derived in the previous sections and may be seen as the final product of our geometric procedure.~The proof is immediate.

\medskip 

\begin{prop} \label{prop_final_procedure}
For any $\bsy{\mathfrak{D}}_{\bsy{\mathcal{N}}(\epsilon)}$ one-tensors $\bsy{\varsigma}(\epsilon)$ and $\bsy{\theta}(\epsilon)$ and $\bsy{\mathfrak{D}}_{\bsy{\mathcal{N}}(\epsilon)}$ two-tensor $\bsy{\omega}(\epsilon)$, we have 
\begin{equation*}
\widetilde{\bsy{\varsigma}}\otimes \widetilde{\bsy{\theta}}=\widetilde{\bsy{\varsigma}\otimes \bsy{\theta}}
\end{equation*}
and the identities between smooth scalar functions 
\begin{align*}
(\widetilde{\bsy{\varsigma}},\widetilde{\bsy{\theta}})_{\widetilde{\bsy{\slashed{g}}}}&=(\bsy{\varsigma},\bsy{\theta})_{\bsy{\slashed{g}}}  \, , &
\textbf{tr}_{\widetilde{\bsy{\slashed{g}}}}\,\widetilde{\bsy{\omega}}  &=\textbf{tr}_{\bsy{\slashed{g}}}\,\bsy{\omega} \, , &
 \widetilde{\bsy{\varsigma}}\,\bsy{\wedge}_{\widetilde{\bsy{\slashed{g}}}}\,\widetilde{\bsy{\theta}}&=\bsy{\varsigma}\,\bsy{\wedge}_{\bsy{\slashed{g}}}\,\bsy{\theta} \, .
\end{align*}
In particular, we have
\begin{equation} \label{isometry_product}
 |\widetilde{\bsy{\varsigma}}|^2_{\widetilde{\bsy{\slashed{g}}}} = |\bsy{\varsigma}|^2_{\bsy{\slashed{g}}} 
\end{equation}
and
\begin{equation*}
\widetilde{\bsy{\varsigma}}\,\widehat{\otimes}_{\widetilde{\bsy{\slashed{g}}}}\,\widetilde{\bsy{\theta}}=\widetilde{\bsy{\varsigma}\,\widehat{\otimes} \, \bsy{\theta}} + \textup{nonlinear terms} \, .
\end{equation*}
For any $\bsy{\mathfrak{D}}_{\bsy{\mathcal{N}}(\epsilon)}$ one-tensor $\bsy{\varsigma}(\epsilon)$ and $\bsy{\mathfrak{D}}_{\bsy{\mathcal{N}}(\epsilon)}$ two-tensor $\bsy{\theta}(\epsilon)$, we have
\begin{align*}
(\widetilde{\bsy{\theta}}{}^{\bsy{\sharp_2}}\bsy{\cdot}_{\widetilde{\bsy{\slashed{g}}}}\,\widetilde{\bsy{\varsigma}})_A &=(\widetilde{\bsy{\theta}{}^{\bsy{\sharp_2}} \bsy{\cdot}_{\bsy{\slashed{g}}}\, \bsy{\varsigma}})_A  \, , &
(\widetilde{\bsy{\slashed{g}}^{-1}}){}^{BC}\widetilde{\bsy{\slashed{\varepsilon}}}_{AC}\widetilde{\bsy{\varsigma}}_{B}&=(\widetilde{{}^{\bsy{\star}}\bsy{\varsigma}})_A \, .
\end{align*}
For any $f\in C^{\infty}(\mathcal{M}^*)$, we have
\begin{align}
\widetilde{\bsy{\nablasl}}f =& \, \widetilde{\bsy{\nablasl}f}-\frac{1}{2}\,(\widetilde{\bsy{\nablasl}}_{\bsy{3}}f)\, \widetilde{(\ef)_{\bsy{\flat}}}-\frac{1}{2}\,(\widetilde{\bsy{\nablasl}}_{\bsy{4}}f) \,\widetilde{(\et)_{\bsy{\flat}}} \, . 
\end{align}
For any $\bsy{\mathfrak{D}}_{\bsy{\mathcal{N}}(\epsilon)}$ one-tensor $\bsy{\varsigma}(\epsilon)$, we have
\begin{align}
\widetilde{\bsy{\nablasl}}_{\bsy{4}}\,\widetilde{\bsy{\varsigma}}=& \, \widetilde{\bsy{\nablasl}_{\bsy{4}}\,\bsy{\varsigma}}+(\widetilde{\bsy{\varsigma}},\widetilde{\bsy{\etab}})_{\widetilde{\bsy{\slashed{g}}}}\cdot \widetilde{(\ef)_{\bsy{\flat}}} \label{restr_proj_cov_derivatives_4}\\ &+ \textup{nonlinear terms} \, , \nonumber\\[5pt]
\widetilde{\bsy{\nablasl}}_{\bsy{3}}\,\widetilde{\bsy{\varsigma}}=& \, \widetilde{\bsy{\nablasl}_{\bsy{3}}\,\bsy{\varsigma}}+(\widetilde{\bsy{\varsigma}},\widetilde{\bsy{\eta}})_{\widetilde{\bsy{\slashed{g}}}}\cdot\widetilde{(\et)_{\bsy{\flat}}} -(\bsy{\underline{\mathfrak{f}}},\widetilde{\bsy{\varsigma}})_{\slashed{g}_{a,M}}\cdot\widetilde{\bsy{\eta}} \label{restr_proj_cov_derivatives_3}\\ &+ \textup{nonlinear terms} \, , \nonumber\\[5pt]
\widetilde{\bsy{\nablasl}}_{\bsy{A}}\,\widetilde{\bsy{\varsigma}}=& \, \widetilde{\bsy{\nablasl}_{\bsy{A}}\,\bsy{\varsigma}}+\frac{1}{2}\left({\bsy{\chib}^{\bsy{\sharp_2}}}{}_{\bsy{A}}^{\bsy{B}}\,\bsy{\varsigma}_{\bsy{B}}\right)\widetilde{(\ef)_{\bsy{\flat}}}+\frac{1}{2}\left({\bsy{\chi}^{\bsy{\sharp_2}}}{}_{\bsy{A}}^{\bsy{B}}\,\bsy{\varsigma}_{\bsy{B}}\right)\widetilde{(\et)_{\bsy{\flat}}} \label{restr_proj_cov_derivatives_A}\\ &-\frac{1}{2}\,(\bsy{\underline{\mathfrak{f}}},\widetilde{\bsy{\varsigma}})_{\slashed{g}_{a,M}}\cdot{\bsy{\chi}^{\bsy{\sharp_2}}}{}_{\bsy{A}}^{\bsy{B}} \,\widetilde{(\bsy{e_B})_{\bsy{\flat}}} \nonumber \\ &+ \textup{nonlinear terms}  \nonumber
\end{align}
and
\begin{align} \label{restr_ang_deriv_formula}
\widetilde{\bsy{\nablasl}}\,\widetilde{\bsy{\varsigma}}=& \,\widetilde{\bsy{\nablasl}\,\bsy{\varsigma}} +\frac{1}{2}\left({\bsy{\chib}}{}^{\bsy{AB}}\,\bsy{\varsigma}_{\bsy{B}}\right)\widetilde{(\bsy{e_A})_{\bsy{\flat}}}\otimes\widetilde{(\ef)_{\bsy{\flat}}}+\frac{1}{2}\left({\bsy{\chi}}{}^{\bsy{AB}}\,\bsy{\varsigma}_{\bsy{B}}\right)\widetilde{(\bsy{e_A})_{\bsy{\flat}}}\otimes \widetilde{(\et)_{\bsy{\flat}}} \\&-\frac{1}{2}\,(\bsy{\nablasl}_{\bsy{4}}\,\bsy{\varsigma})^{\bsy{A}}\,\widetilde{(\et)_{\bsy{\flat}}}\otimes\widetilde{(\ea)_{\bsy{\flat}}}-\frac{1}{2}\,(\bsy{\nablasl}_{\bsy{3}}\,\bsy{\varsigma})^{\bsy{A}}\,\widetilde{(\ef)_{\bsy{\flat}}}\otimes\widetilde{(\ea)_{\bsy{\flat}}}  \nonumber \\ &-\frac{1}{2}\,(\bsy{\underline{\mathfrak{f}}},\widetilde{\bsy{\varsigma}})_{\slashed{g}_{a,M}}\cdot{\bsy{\chi}}{}^{\bsy{AB}}\, \widetilde{(\bsy{e_A})_{\bsy{\flat}}}\otimes\widetilde{(\bsy{e_B})_{\bsy{\flat}}} \nonumber \\ &+ \textup{nonlinear terms} \, . \nonumber
\end{align}
We also have the formulae
\begin{align}
 \slashed{\textbf{div}}\,\bsy{\varsigma}=& \,\widetilde{\slashed{\textbf{div}}}\,\widetilde{\bsy{\varsigma}} +\frac{1}{2}\,(\bsy{\nablasl}_{\bsy{3}}\,\bsy{\varsigma}-{\bsy{\chib}}{}^{\bsy{\sharp_2}}\bsy{\cdot}\bsy{\varsigma})^{\bsy{A}}\,(\widetilde{(\ef)_{\bsy{\flat}}},\widetilde{(\ea)_{\bsy{\flat}}})_{\widetilde{\bsy{\slashed{g}}}} \\ & +\frac{1}{2}\,(\bsy{\nablasl}_{\bsy{4}}\,\bsy{\varsigma}-{\bsy{\chi}}{}^{\bsy{\sharp_2}}\bsy{\cdot}\bsy{\varsigma})^{\bsy{A}}(\widetilde{(\et)_{\bsy{\flat}}},\widetilde{(\ea)_{\bsy{\flat}}})_{\widetilde{\bsy{\slashed{g}}}}  \nonumber \\
 &+ \frac{1}{2}\,(\bsy{\underline{\mathfrak{f}}},\widetilde{\bsy{\varsigma}})_{\slashed{g}_{a,M}}\cdot{\bsy{\chi}}{}^{\bsy{AB}}\, (\widetilde{(\bsy{e_A})_{\bsy{\flat}}},\widetilde{(\bsy{e_B})_{\bsy{\flat}}})_{\widetilde{\bsy{\slashed{g}}}} \nonumber \\ &+ \textup{nonlinear terms}     \, , \nonumber \\[5pt]
\slashed{\textbf{curl}}\,\bsy{\varsigma}=& \,\widetilde{\slashed{\textbf{curl}}}\,\widetilde{\bsy{\varsigma}} +\frac{1}{2}\,(\bsy{\nablasl}_{\bsy{3}}\,\bsy{\varsigma}-{\bsy{\chib}}{}^{\bsy{\sharp_2}}\bsy{\cdot}\bsy{\varsigma})^{\bsy{A}}\,\widetilde{(\ef)_{\bsy{\flat}}}\,\bsy{\wedge}_{\widetilde{\bsy{\slashed{g}}}}\,\widetilde{(\ea)_{\bsy{\flat}}} \\ & +\frac{1}{2}\,(\bsy{\nablasl}_{\bsy{4}}\,\bsy{\varsigma}-{\bsy{\chi}}{}^{\bsy{\sharp_2}}\bsy{\cdot}\bsy{\varsigma})^{\bsy{A}} \widetilde{(\et)_{\bsy{\flat}}}\,\bsy{\wedge}_{\widetilde{\bsy{\slashed{g}}}}\,\widetilde{(\ea)_{\bsy{\flat}}}  \nonumber \\
 &+ \frac{1}{2}\,(\bsy{\underline{\mathfrak{f}}},\widetilde{\bsy{\varsigma}})_{\slashed{g}_{a,M}}\cdot{\bsy{\chi}}{}^{\bsy{AB}}\, \widetilde{(\bsy{e_A})_{\bsy{\flat}}}\,\bsy{\wedge}_{\widetilde{\bsy{\slashed{g}}}}\,\widetilde{(\bsy{e_B})_{\bsy{\flat}}} \nonumber \\ &+ \textup{nonlinear terms}     \, , \nonumber \\[5pt]
-2\,\widetilde{\bsy{\slashed{\mathcal{D}}_2^{\star}}\,\bsy{\varsigma}}=&\, -2 \, \widetilde{\bsy{\slashed{\mathcal{D}}}_{\bsy{2}}^{\star}}\,\widetilde{\bsy{\varsigma}}  \label{restr_formula_proj_D2star} \\
&+\frac{1}{2}\,(\bsy{\nablasl}_{\bsy{3}}\,\bsy{\varsigma}-{\bsy{\chib}}{}^{\bsy{\sharp_2}}\bsy{\cdot}\bsy{\varsigma})^{\bsy{A}}\,\widetilde{(\ef)_{\bsy{\flat}}}\,\widehat{\otimes}_{\widetilde{\bsy{\slashed{g}}}}\,\widetilde{(\ea)_{\bsy{\flat}}} \nonumber\\ & +\frac{1}{2}\,(\bsy{\nablasl}_{\bsy{4}}\,\bsy{\varsigma}-{\bsy{\chi}}{}^{\bsy{\sharp_2}}\bsy{\cdot}\bsy{\varsigma})^{\bsy{A}}\,\widetilde{(\et)_{\bsy{\flat}}}\,\widehat{\otimes}_{\widetilde{\bsy{\slashed{g}}}}\,\widetilde{(\ea)_{\bsy{\flat}}} \nonumber \\ &+ \textup{nonlinear terms} \, . \nonumber
\end{align}
For any $X\in \Gamma(T\mathcal{M}^*)$, we have
\begin{align}
\widetilde{\bsy{\nablasl}}_{X}\widetilde{\bsy{\chih}}&= \widetilde{\bsy{\nablasl}_{X}\,\bsy{\chih}}+ \textup{nonlinear terms} \, , & \widetilde{\bsy{\nablasl}}_{X}\widetilde{\bsy{\alpha}}&= \widetilde{\bsy{\nablasl}_{X}\,\bsy{\alpha}}+ \textup{nonlinear terms} \, , \label{restr_proj_cov_deriv_outgoing_shear}\\[5pt]
\widetilde{\bsy{\nablasl}}_{X}\widetilde{\bsy{\chibh}}&= \widetilde{\bsy{\nablasl}_{X}\,\bsy{\chibh}}+ \textup{nonlinear terms} \, , & \widetilde{\bsy{\nablasl}}_{X}\widetilde{\bsy{\alphab}}&= \widetilde{\bsy{\nablasl}_{X}\,\bsy{\alphab}}+ \textup{nonlinear terms} \label{restr_proj_cov_deriv_ingoing_shear}
\end{align}
and
\begin{align} 
\widetilde{\bsy{\nablasl}}\,\widetilde{\bsy{\chih}}&= \widetilde{\bsy{\nablasl}\,\bsy{\chih}}+ \textup{nonlinear terms}  \, , & \widetilde{\bsy{\nablasl}}\,\widetilde{\bsy{\alpha}}&= \widetilde{\bsy{\nablasl}\,\bsy{\alpha}}+ \textup{nonlinear terms} \, , \\[5pt]
\widetilde{\bsy{\nablasl}} \, \widetilde{\bsy{\chibh}}&= \widetilde{\bsy{\nablasl} \,\bsy{\chibh}}+ \textup{nonlinear terms} \, , & \widetilde{\bsy{\nablasl}}\,\widetilde{\bsy{\alphab}}&= \widetilde{\bsy{\nablasl}\,\bsy{\alphab}}+ \textup{nonlinear terms} \, , \\[5pt]
(\widetilde{\slashed{\textbf{div}}}\,\widetilde{\bsy{\chih}})_A &= (\widetilde{\slashed{\textbf{div}}\,\bsy{\chih}})_A + \textup{nonlinear terms} \, , & (\widetilde{\slashed{\textbf{div}}}\,\widetilde{\bsy{\alpha}})_A &= (\widetilde{\slashed{\textbf{div}}\,\bsy{\alpha}})_A + \textup{nonlinear terms} \, , \\[5pt]
(\widetilde{\slashed{\textbf{div}}}\,\widetilde{\bsy{\chibh}})_A &=  (\widetilde{\slashed{\textbf{div}}\,\bsy{\chibh}})_A + \textup{nonlinear terms} \, , & (\widetilde{\slashed{\textbf{div}}}\,\widetilde{\bsy{\alphab}})_A &= (\widetilde{\slashed{\textbf{div}}\,\bsy{\alphab}})_A + \textup{nonlinear terms} \, .
\end{align}
We have the formulae
\begin{align}
\widetilde{\bsy{\slashed{\mathcal{L}}}}_{\bsy{4}}\,\widetilde{\bsy{\varsigma}}=& \, \widetilde{\bsy{\slashed{\mathcal{L}}}_{\bsy{4}}\,\bsy{\varsigma}}+(\bsy{\etab}-\bsy{\eta})^{\bsy{A}}\bsy{\varsigma}_{\bsy{A}}\,\widetilde{(\ef)_{\bsy{\flat}}} +\frac{1}{2}\,\slashed{\bsy{g}}^{\bsy{AB}}\bsy{g}(e_3^{\textup{as}},\ea)\,\bsy{\varsigma}_{\bsy{B}}\,\widetilde{\bsy{\mathcal{L}}_{\bsy{4}}(e_4^{\textup{as}})_{\flat}}   \, ,  \\[5pt]
\widetilde{\bsy{\slashed{\mathcal{L}}}}_{\bsy{3}}\,\widetilde{\bsy{\varsigma}}=& \, \widetilde{\bsy{\slashed{\mathcal{L}}}_{\bsy{3}}\,\bsy{\varsigma}}+(\bsy{\eta}-\bsy{\etab})^{\bsy{A}}\bsy{\varsigma}_{\bsy{A}}\,\widetilde{(\et)_{\bsy{\flat}}} +\frac{1}{2}\,\slashed{\bsy{g}}^{\bsy{AB}}\bsy{g}(e_3^{\textup{as}},\ea)\,\bsy{\varsigma}_{\bsy{B}}\,\widetilde{\bsy{\mathcal{L}}_{\bsy{3}}(e_4^{\textup{as}})_{\flat}}  \, ,  \\[5pt]
\widetilde{\bsy{\slashed{\mathcal{L}}}}_{\bsy{4}}\,\widetilde{\bsy{\slashed{g}}} =&\, \widetilde{\bsy{\slashed{\mathcal{L}}}_{\bsy{4}}\,\bsy{\slashed{g}}}+(\bsy{\etab}-\bsy{\eta})^{\bsy{A}}\left(\widetilde{(\ef)_{\bsy{\flat}}}\otimes \widetilde{(\ea)_{\bsy{\flat}}}+\widetilde{(\ea)_{\bsy{\flat}}}\otimes \widetilde{(\ef)_{\bsy{\flat}}}\right) \label{restr_Lie_deriv_4_metric_formula}\\ 
&+\frac{1}{2}\,\slashed{g}_{a,M}^{AB}\,\bsy{\slashed{g}}{}^{\bsy{CD}}\,\bsy{g}(e^{\textup{as}}_3,\bsy{e_C})\,\bsy{g}(e^{\textup{as}}_A,\bsy{e_D})\left(\widetilde{\bsy{\mathcal{L}}_{\bsy{4}}(e_4^{\textup{as}})_{\flat}}\otimes\widetilde{(e_B^{\textup{as}})_{\flat}}+\widetilde{(e_B^{\textup{as}})_{\flat}}\otimes\widetilde{\bsy{\mathcal{L}}_{\bsy{4}}(e_4^{\textup{as}})_{\flat}}\right) \nonumber\\ &+\textup{nonlinear terms} \, , \nonumber \\[5pt]
\widetilde{\bsy{\slashed{\mathcal{L}}}}_{\bsy{3}}\,\widetilde{\bsy{\slashed{g}}}  =&\, \widetilde{\bsy{\slashed{\mathcal{L}}}_{\bsy{3}}\,\bsy{\slashed{g}}}  -(\bsy{\etab}-\bsy{\eta})^{\bsy{A}}\left(\widetilde{(\et)_{\bsy{\flat}}}\otimes \widetilde{(\ea)_{\bsy{\flat}}}+\widetilde{(\ea)_{\bsy{\flat}}}\otimes \widetilde{(\et)_{\bsy{\flat}}}\right) \label{restr_Lie_deriv_3_metric_formula}\\
&+\frac{1}{2}\,\slashed{g}_{a,M}^{AB}\,\bsy{\slashed{g}}{}^{\bsy{CD}}\,\bsy{g}(e^{\textup{as}}_3,\bsy{e_C})\,\bsy{g}(e^{\textup{as}}_A,\bsy{e_D})\left(\widetilde{\bsy{\mathcal{L}}_{\bsy{3}}(e_4^{\textup{as}})_{\flat}}\otimes\widetilde{(e_B^{\textup{as}})_{\flat}}+\widetilde{(e_B^{\textup{as}})_{\flat}}\otimes\widetilde{\bsy{\mathcal{L}}_{\bsy{3}}(e_4^{\textup{as}})_{\flat}}\right) \nonumber\\ &+\textup{nonlinear terms} \, . \nonumber
\end{align}
For any $\bsy{\mathfrak{D}}_{\bsy{\mathcal{N}}(\epsilon)}$ one-tensor $\bsy{\varsigma}(\epsilon)$ such that $\bsy{\varsigma}(0)\equiv 0$, we have
\begin{align}
\widetilde{\bsy{\nablasl}}_{\bsy{4}}\,\widetilde{\bsy{\varsigma}}=& \, \widetilde{\bsy{\nablasl}_{\bsy{4}}\,\bsy{\varsigma}}+ \textup{nonlinear terms} \, , \label{restr_proj_cov_deriv_schw_1}\\[5pt]
\widetilde{\bsy{\nablasl}}_{\bsy{3}}\,\widetilde{\bsy{\varsigma}}=& \, \widetilde{\bsy{\nablasl}_{\bsy{3}}\,\bsy{\varsigma}}+ \textup{nonlinear terms} \, , \label{restr_proj_cov_deriv_schw_2}\\[5pt]
\widetilde{\bsy{\nablasl}}_{\bsy{A}}\,\widetilde{\bsy{\varsigma}}=& \, \widetilde{\bsy{\nablasl}_{\bsy{A}}\,\bsy{\varsigma}}+ \textup{nonlinear terms} \, , \label{restr_proj_cov_deriv_schw_3}\\[5pt]
\widetilde{\bsy{\nablasl}}\,\widetilde{\bsy{\varsigma}}=& \,\widetilde{\bsy{\nablasl}\,\bsy{\varsigma}} +\textup{nonlinear terms} \, . 
\end{align}
\end{prop}

\medskip

\begin{remark}  \label{rmk_final_structure_nonlinear_terms}
We recall the Remarks \ref{rmk_proj_cov_deriv_structure_nonlinear_terms} and \ref{rmk_proj_cov_deriv_shears_structure_nonlinear_terms}.~\ul{The \emph{nonlinear terms} in the formulae} \eqref{restr_proj_cov_derivatives_4}, \eqref{restr_proj_cov_derivatives_3}, \eqref{restr_proj_cov_derivatives_A}, \eqref{restr_ang_deriv_formula}, \eqref{restr_proj_cov_deriv_outgoing_shear}, \eqref{restr_proj_cov_deriv_ingoing_shear}, \eqref{restr_proj_cov_deriv_schw_1}, \eqref{restr_proj_cov_deriv_schw_2} \ul{and} \eqref{restr_proj_cov_deriv_schw_3} \ul{do \emph{not} contain any \emph{derivatives} of the respective quantities (or any derivatives of connection coefficients or curvature components)}.~On the other hand, some of the \emph{nonlinear terms} in the formulae for the angular derivatives of tensors \emph{do} involve derivatives of the respective quantities.~An example is the formula \eqref{restr_formula_proj_D2star}.~See the related Remark \ref{rmk_proj_cov_deriv_D2star_structure_nonlinear_terms}.
\end{remark}

\subsubsection{Auxiliary identities}  \label{sec_auxiliary_ids}

In this section, we collect some auxiliary propositions.~These propositions will not be directly needed to formulate the renormalised system of vacuum Einstein equations (see Section \ref{sec_renormalised_vacuum_Einstein_equations}), but will later allow for convenient manipulations of some of the terms appearing in the system.~Immediate proofs are omitted.

\medskip

\begin{prop} \label{prop_frame_commutators_ricci_1}
For any $\bsy{\mathfrak{D}}_{\bsy{\mathcal{N}}(\epsilon)}$ $k$-tensor $\bsy{\varsigma}(\epsilon)$, we have the identity
\begin{equation}
\bsy{\varsigma}_{\bsy{A}_1\cdots \bsy{A}_k}(\epsilon)=(\bsy{\mathfrak{\slashed{\mathfrak{f}}}}_{\bsy{A}_1})^{B_1}(\epsilon)\cdots (\bsy{\mathfrak{\slashed{\mathfrak{f}}}}_{\bsy{A}_k})^{B_k}(\epsilon)\,\widetilde{\bsy{\varsigma}}_{B_1\cdots B_k}(\epsilon)+\textup{nonlinear terms}\, . \label{aux_eqn_frame_coeff_nonlinear}
\end{equation}
\end{prop}

\medskip

\begin{prop} \label{prop_frame_commutators_ricci_2}
We have the identities
\begin{align*}
(\bsy{\mathfrak{\slashed{\mathfrak{f}}}}_{\bsy{A}}(\epsilon),\widetilde{\bsy{\mathfrak{\slashed{\mathfrak{f}}}}}_{4}(\epsilon))_{\slashed{g}_{a,M}}\cdot\bsy{g}(e_4^{\textup{as}}, e_3^{\textup{as}})+(\bsy{\mathfrak{\slashed{\mathfrak{f}}}}_{\bsy{A}}(\epsilon),\widetilde{(\bsy{e_4})_{\bsy{\flat}}}(\epsilon))_{\slashed{g}_{a,M}}&=\textup{nonlinear terms} \, , \\[5pt]
-2\,(\bsy{\mathfrak{\slashed{\mathfrak{f}}}}_{\bsy{A}}(\epsilon),\widetilde{\bsy{\mathfrak{\slashed{\mathfrak{f}}}}}_{3}(\epsilon))_{\slashed{g}_{a,M}}+(\bsy{\mathfrak{\slashed{\mathfrak{f}}}}_{\bsy{A}}(\epsilon),\widetilde{(\bsy{e_3})_{\bsy{\flat}}}(\epsilon))_{\slashed{g}_{a,M}}&=\textup{nonlinear terms} \, , \\[5pt]
\bsy{\underline{\mathfrak{f}}}_4(\epsilon)\cdot\bsy{g}(e^{\textup{as}}_4,e^{\textup{as}}_3)+2 &=   \textup{nonlinear terms} \, , \\[5pt]
(\bsy{\underline{\mathfrak{f}}}_4(\epsilon))^2\cdot\bsy{g}(e^{\textup{as}}_3,e^{\textup{as}}_3)+2\,\bsy{\underline{\mathfrak{f}}}_4(\epsilon)\cdot\bsy{\underline{\mathfrak{f}}}_3(\epsilon)\cdot\bsy{g}(e^{\textup{as}}_4,e^{\textup{as}}_3)&= \textup{nonlinear terms} \, , \\[5pt]
\widetilde{\bsy{\slashed{g}}}((\bsy{\slashed{\mathfrak{f}}}_{\bsy{A}})(\epsilon),(\bsy{\slashed{\mathfrak{f}}}_{\bsy{B}})(\epsilon))-\delta_{\bsy{AB}}&=\textup{nonlinear terms} 
\end{align*}
and
\begin{align*}
(\widetilde{(\bsy{e_3})_{\bsy{\flat}}})_A(\epsilon)&=\bsy{\underline{\mathfrak{f}}}_4(\epsilon)\cdot\bsy{g}(e^{\textup{as}}_3,e_A^{\textup{as}})+(\bsy{\underline{\mathfrak{f}}}^{\sharp}(\epsilon)\cdot\widetilde{\bsy{\slashed{g}}}(\epsilon))_A+\textup{nonlinear terms} \, .
\end{align*}
All the inner products are products of $\mathfrak{D}_{\mathcal{N}_{\textup{as}}}$ tensors relative to $\slashed{g}_{a,M}$ and indices are raised by $\slashed{g}^{-1}_{a,M}$.
\end{prop}

\medskip

\begin{proof}
We compute
\begin{align*}
0&=\bsy{g}(\ef(\epsilon),\ea(\epsilon)) \\
&=\bsy{g}(e^{\text{as}}_4,(\bsy{\slashed{\mathfrak{f}}}_4)_{\bsy{A}}(\epsilon)e^{\text{as}}_3+\slashed{g}_{a,M}^{CB}(\bsy{\slashed{\mathfrak{f}}}_{\bsy{A}})_{C}(\epsilon)e^{\text{as}}_B)+\text{nonlinear terms} \\
&=(\bsy{\slashed{\mathfrak{f}}}_4)_{\bsy{A}}(\epsilon)\cdot\bsy{g}(e^{\text{as}}_4,e^{\text{as}}_3)+\slashed{g}_{a,M}^{CB}(\bsy{\slashed{\mathfrak{f}}}_{\bsy{A}})_{C}(\epsilon)\bsy{g}(e^{\text{as}}_4,e^{\text{as}}_B)+\text{nonlinear terms} \\
&=(\bsy{\mathfrak{\slashed{\mathfrak{f}}}}_{\bsy{A}}(\epsilon),\widetilde{\bsy{\mathfrak{\slashed{\mathfrak{f}}}}}_{4}(\epsilon))\cdot\bsy{g}(e^{\text{as}}_4,e^{\text{as}}_3)+\slashed{g}_{a,M}^{CB}(\bsy{\slashed{\mathfrak{f}}}_{\bsy{A}})_{C}(\epsilon)\bsy{g}(\ef(\epsilon),e^{\text{as}}_B)+\text{nonlinear terms} \, ,
\end{align*}
where in the last identity we used the identity \eqref{aux_eqn_frame_coeff_nonlinear}.~We compute
\begin{align*}
0&=\bsy{g}(\et(\epsilon),\ea(\epsilon)) \\
&=\bsy{g}(\et(\epsilon),(\bsy{\slashed{\mathfrak{f}}}_4)_{\bsy{A}}(\epsilon)e^{\text{as}}_3+(\bsy{\slashed{\mathfrak{f}}}_3)_{\bsy{A}}(\epsilon)e^{\text{as}}_4+\slashed{g}_{a,M}^{CB}(\bsy{\slashed{\mathfrak{f}}}_{\bsy{A}})_{C}(\epsilon)e^{\text{as}}_B) \\
&= (\bsy{\slashed{\mathfrak{f}}}_3)_{\bsy{A}}(\epsilon)\cdot \bsy{g}(\et(\epsilon),e^{\text{as}}_4)+\slashed{g}_{a,M}^{CB}(\bsy{\slashed{\mathfrak{f}}}_{\bsy{A}})_{C}(\epsilon)\bsy{g}(\et(\epsilon),e^{\text{as}}_B)+\text{nonlinear terms} \\
&=-2\,(\bsy{\mathfrak{\slashed{\mathfrak{f}}}}_{\bsy{A}}(\epsilon),\widetilde{\bsy{\mathfrak{\slashed{\mathfrak{f}}}}}_{3}(\epsilon))+\slashed{g}_{a,M}^{CB}(\bsy{\slashed{\mathfrak{f}}}_{\bsy{A}})_{C}(\epsilon)\bsy{g}(\et(\epsilon),e^{\text{as}}_B)+\text{nonlinear terms} \, ,
\end{align*}
where in the last identity we used the identity \eqref{aux_eqn_frame_coeff_nonlinear}.~We compute
\begin{align*}
-2 &=\bsy{g}(\ef(\epsilon),\et(\epsilon)) \\
&=\bsy{g}(e_4^{\text{as}},\bsy{\underline{\mathfrak{f}}}_4(\epsilon)e_3^{\text{as}}+\bsy{\underline{\mathfrak{f}}}_3(\epsilon)e_4^{\text{as}})+\text{nonlinear terms} \\
&=\bsy{\underline{\mathfrak{f}}}_4(\epsilon) \cdot\bsy{g}(e^{\text{as}}_4,e^{\text{as}}_3) +\text{nonlinear terms} \, .
\end{align*}
We compute
\begin{align*}
0 &=\bsy{g}(\et(\epsilon),\et(\epsilon)) \\
&=\bsy{g}(\bsy{\underline{\mathfrak{f}}}_4(\epsilon)e_3^{\text{as}}+\bsy{\ub{$\mathfrak{f}$}}_3(\epsilon)e_4^{\text{as}},\bsy{\underline{\mathfrak{f}}}_4(\epsilon)e_3^{\text{as}}+\bsy{\underline{\mathfrak{f}}}_3(\epsilon)e_4^{\text{as}}) +\text{nonlinear terms} \\
&=(\bsy{\underline{\mathfrak{f}}}_4(\epsilon))^2\cdot\bsy{g}(e_3^{\text{as}},e_3^{\text{as}})+2\,\bsy{\underline{\mathfrak{f}}}_3(\epsilon)\,\bsy{\underline{\mathfrak{f}}}_4(\epsilon)\cdot\bsy{g}(e_3^{\text{as}},e_4^{\text{as}}) +\text{nonlinear terms} \, .
\end{align*}
We compute
\begin{align*}
\delta_{\bsy{AB}} &=\bsy{g}(\ea(\epsilon),\eb(\epsilon)) \\
&=\bsy{g}(\slashed{g}_{a,M}^{CD}(\bsy{\slashed{\mathfrak{f}}}_{\bsy{A}})_{C}(\epsilon)e^{\text{as}}_D,\slashed{g}_{a,M}^{EF}(\bsy{\slashed{\mathfrak{f}}}_{\bsy{B}})_{E}(\epsilon)e^{\text{as}}_F)+\text{nonlinear terms} \\
&=\widetilde{\bsy{\slashed{g}}}((\bsy{\slashed{\mathfrak{f}}}_{\bsy{A}})(\epsilon),(\bsy{\slashed{\mathfrak{f}}}_{\bsy{B}})(\epsilon))+\text{nonlinear terms} \, .
\end{align*}
This concludes the proof.

\end{proof}

\medskip

\begin{prop}  \label{prop_restricted_volume_form}
We have the identity
\begin{align*}
\widetilde{\bsy{\slashed{\varepsilon}}}(\epsilon)=& \, ((\bsy{\mathfrak{\slashed{\mathfrak{f}}}}_{\bsy{1}})^{\sharp}(\epsilon)\cdot\widetilde{\bsy{\slashed{g}}}(\epsilon))\otimes((\bsy{\mathfrak{\slashed{\mathfrak{f}}}}_{\bsy{2}})^{\sharp}(\epsilon)\cdot\widetilde{\bsy{\slashed{g}}}(\epsilon))-((\bsy{\mathfrak{\slashed{\mathfrak{f}}}}_{\bsy{2}})^{\sharp}(\epsilon)\cdot\widetilde{\bsy{\slashed{g}}}(\epsilon))\otimes((\bsy{\mathfrak{\slashed{\mathfrak{f}}}}_{\bsy{1}})^{\sharp}(\epsilon)\cdot\widetilde{\bsy{\slashed{g}}}(\epsilon)) \\ &+ \textup{nonlinear terms} \, ,
\end{align*}
where all the quantities are $\mathfrak{D}_{\mathcal{N}_{\textup{as}}}$ tensors and indices are raised by $\slashed{g}^{-1}_{a,M}$.
\end{prop}

\medskip

\begin{prop}   \label{prop_aux_restriction_lie_derivative}
For any $\bsy{\mathfrak{D}}_{\bsy{\mathcal{N}}(\epsilon)}$ one-tensor $\bsy{\varsigma}(\epsilon)$, we have the identity
\begin{align*}
(\widetilde{\bsy{\slashed{\mathcal{L}}}}_{\bsy{3}}\,\widetilde{\bsy{\varsigma}})(\epsilon) =& \bsy{\mathfrak{\underline{\mathfrak{f}}}}_{3}(\epsilon)\, (\nablasl^{a,M}_4\,\widetilde{\bsy{\varsigma}})(\epsilon)+\bsy{\mathfrak{\underline{\mathfrak{f}}}}_{4}(\epsilon)\, (\nablasl^{a,M}_3\,\widetilde{\bsy{\varsigma}})(\epsilon)+((\nablasl_{a,M}\,\widetilde{\bsy{\varsigma}})^{\sharp_1} \cdot \bsy{\mathfrak{\underline{\mathfrak{f}}}})(\epsilon) \\
& +((\nablasl_{a,M}\,\bsy{\mathfrak{\underline{\mathfrak{f}}}})^{\sharp_2}\cdot\widetilde{\bsy{\varsigma}})(\epsilon) +\bsy{\mathfrak{\underline{\mathfrak{f}}}}_{3}(\epsilon)\,(\chi^{\sharp_2}_{a,M}\cdot\widetilde{\bsy{\varsigma}})(\epsilon)+\bsy{\mathfrak{\underline{\mathfrak{f}}}}_{4}(\epsilon)\,(\chib^{\sharp_2}_{a,M}\cdot\,\widetilde{\bsy{\varsigma}})(\epsilon) \, .
\end{align*}
We also have the identity
\begin{align*}
(\widetilde{\bsy{\slashed{\mathcal{L}}}}_{\bsy{3}}\,\widetilde{\bsy{\slashed{g}}})(\epsilon) =& \, \bsy{\mathfrak{\underline{\mathfrak{f}}}}_{4}(\epsilon)\, (\nablasl^{a,M}_3\,\widetilde{\bsy{\slashed{g}}})(\epsilon) \\
&+\bsy{\mathfrak{\underline{\mathfrak{f}}}}_{3}(\epsilon)\,(\chi_{a,M}\times\widetilde{\bsy{\slashed{g}}})(\epsilon)+\bsy{\mathfrak{\underline{\mathfrak{f}}}}_{4}(\epsilon)\,(\chib_{a,M}\times\,\widetilde{\bsy{\slashed{g}}})(\epsilon)+(\nablasl_{a,M}\bsy{\mathfrak{\underline{\mathfrak{f}}}}\times\widetilde{\bsy{\slashed{g}}})(\epsilon) \\
&+\bsy{\mathfrak{\underline{\mathfrak{f}}}}_{3}(\epsilon)\,(\chi_{a,M}\times\widetilde{\bsy{\slashed{g}}})^{\mathsf{T}}(\epsilon)+\bsy{\mathfrak{\underline{\mathfrak{f}}}}_{4}(\epsilon)\,(\chib_{a,M}\times\,\widetilde{\bsy{\slashed{g}}})^{\mathsf{T}}(\epsilon)+(\nablasl_{a,M}\bsy{\mathfrak{\underline{\mathfrak{f}}}}\times\widetilde{\bsy{\slashed{g}}})^{\mathsf{T}}(\epsilon) \\
&+\textup{nonlinear terms} \, .
\end{align*}
\end{prop}

\medskip

\begin{proof}
We compute
\begin{align*}
(\widetilde{\bsy{\slashed{\mathcal{L}}}}_{\bsy{3}}\,\widetilde{\bsy{\varsigma}})_{A}(\epsilon) =& \, (\bsy{\mathcal{L}}_{\bsy{3}}\,(\Pi\bsy{\varsigma}^{\circ}))_{A}(\epsilon) \\[5pt]
=& \, \et((\Pi\bsy{\varsigma}^{\circ})_{A}(\epsilon))-(\Pi\bsy{\varsigma}^{\circ})([\et(\epsilon),e^{\text{as}}_A]) \\[5pt]
=& \, \bsy{\mathfrak{\underline{\mathfrak{f}}}}_{3}(\epsilon)\, e_4^{\text{as}}((\Pi\bsy{\varsigma}^{\circ})_{A}(\epsilon))+\bsy{\mathfrak{\underline{\mathfrak{f}}}}_{4}(\epsilon)\, e_3^{\text{as}}((\Pi\bsy{\varsigma}^{\circ})_{A}(\epsilon))+\bsy{\mathfrak{\underline{\mathfrak{f}}}}^{C}(\epsilon)\, e_C^{\text{as}}((\Pi\bsy{\varsigma}^{\circ})_{A}(\epsilon)) \\
&-(\Pi\bsy{\varsigma}^{\circ})([\bsy{\mathfrak{\underline{\mathfrak{f}}}}_{3}(\epsilon)\, e_4^{\text{as}},e^{\text{as}}_A])-(\Pi\bsy{\varsigma}^{\circ})([\bsy{\mathfrak{\underline{\mathfrak{f}}}}_{4}(\epsilon)\, e_3^{\text{as}},e^{\text{as}}_A])-(\Pi\bsy{\varsigma}^{\circ})([\bsy{\mathfrak{\underline{\mathfrak{f}}}}^{C}(\epsilon)\, e_C^{\text{as}},e^{\text{as}}_A]) \\[5pt]
=& \, \bsy{\mathfrak{\underline{\mathfrak{f}}}}_{3}(\epsilon)\, e_4^{\text{as}}((\Pi\bsy{\varsigma}^{\circ})_{A}(\epsilon))+\bsy{\mathfrak{\underline{\mathfrak{f}}}}_{4}(\epsilon)\, e_3^{\text{as}}((\Pi\bsy{\varsigma}^{\circ})_{A}(\epsilon))+\bsy{\mathfrak{\underline{\mathfrak{f}}}}^{C}(\epsilon)\, e_C^{\text{as}}((\Pi\bsy{\varsigma}^{\circ})_{A}(\epsilon)) \\
&-\bsy{\mathfrak{\underline{\mathfrak{f}}}}_{3}(\epsilon)\,(\Pi\bsy{\varsigma}^{\circ})([ e_4^{\text{as}},e^{\text{as}}_A])-\bsy{\mathfrak{\underline{\mathfrak{f}}}}_{4}(\epsilon)\,(\Pi\bsy{\varsigma}^{\circ})([ e_3^{\text{as}},e^{\text{as}}_A]) \\ &-\bsy{\mathfrak{\underline{\mathfrak{f}}}}^{C}(\epsilon)\,(\Pi\bsy{\varsigma}^{\circ})([ e_C^{\text{as}},e^{\text{as}}_A]) +e^{\text{as}}_A(\bsy{\mathfrak{\underline{\mathfrak{f}}}}^{C}(\epsilon))(\Pi\bsy{\varsigma}^{\circ})( e_C^{\text{as}}) \, ,
\end{align*} 
where the first equality holds by Definition \ref{def_restriction_derivatives}.~We compute
\begin{align*}
(\widetilde{\bsy{\slashed{\mathcal{L}}}}_{\bsy{3}}\,\widetilde{\bsy{\slashed{g}}})_{AB}(\epsilon) =& \, (\bsy{\mathcal{L}}_{\bsy{3}}\,(\Pi\bsy{g}))_{AB}(\epsilon) \\[5pt]
=& \, \et((\Pi\bsy{g})_{AB}(\epsilon))-(\Pi\bsy{g})([\et(\epsilon),e^{\text{as}}_A],e^{\text{as}}_B)-(\Pi\bsy{g})([\et(\epsilon),e^{\text{as}}_B],e^{\text{as}}_A) \\[5pt]
=& \, \bsy{\mathfrak{\underline{\mathfrak{f}}}}_{3}(\epsilon)\, e_4^{\text{as}}((\Pi\bsy{g})_{AB}(\epsilon))+\bsy{\mathfrak{\underline{\mathfrak{f}}}}_{4}(\epsilon)\, e_3^{\text{as}}((\Pi\bsy{g})_{AB}(\epsilon))+\bsy{\mathfrak{\underline{\mathfrak{f}}}}^{C}(\epsilon)\, e_C^{\text{as}}((\Pi\bsy{g})_{AB}(\epsilon)) \\
&-(\Pi\bsy{g})([\bsy{\mathfrak{\underline{\mathfrak{f}}}}_{3}(\epsilon)\, e_4^{\text{as}},e^{\text{as}}_A],e^{\text{as}}_B)-(\Pi\bsy{g})([\bsy{\mathfrak{\underline{\mathfrak{f}}}}_{4}(\epsilon)\, e_3^{\text{as}},e^{\text{as}}_A],e^{\text{as}}_B)-(\Pi\bsy{g})([\bsy{\mathfrak{\underline{\mathfrak{f}}}}^{C}(\epsilon)\, e_C^{\text{as}},e^{\text{as}}_A],e^{\text{as}}_B) \\
&-(\Pi\bsy{g})([\bsy{\mathfrak{\underline{\mathfrak{f}}}}_{3}(\epsilon)\, e_4^{\text{as}},e^{\text{as}}_B],e^{\text{as}}_A)-(\Pi\bsy{g})([\bsy{\mathfrak{\underline{\mathfrak{f}}}}_{4}(\epsilon)\, e_3^{\text{as}},e^{\text{as}}_B],e^{\text{as}}_A)-(\Pi\bsy{g})([\bsy{\mathfrak{\underline{\mathfrak{f}}}}^{C}(\epsilon)\, e_C^{\text{as}},e^{\text{as}}_B],e^{\text{as}}_A) \\[5pt]
=& \, \bsy{\mathfrak{\underline{\mathfrak{f}}}}_{4}(\epsilon)\, e_3^{\text{as}}((\Pi\bsy{g})_{AB}(\epsilon)) \\
&-\bsy{\mathfrak{\underline{\mathfrak{f}}}}_{3}(\epsilon)\,(\Pi\bsy{g})([ e_4^{\text{as}},e^{\text{as}}_A],e^{\text{as}}_B)-\bsy{\mathfrak{\underline{\mathfrak{f}}}}_{4}(\epsilon)\,(\Pi\bsy{g})([ e_3^{\text{as}},e^{\text{as}}_A],e^{\text{as}}_B)-\bsy{\mathfrak{\underline{\mathfrak{f}}}}^{C}(\epsilon)\,(\Pi\bsy{g})([ e_C^{\text{as}},e^{\text{as}}_A],e^{\text{as}}_B) \\
&-\bsy{\mathfrak{\underline{\mathfrak{f}}}}_{3}(\epsilon)\,(\Pi\bsy{g})([ e_4^{\text{as}},e^{\text{as}}_B],e^{\text{as}}_A)-\bsy{\mathfrak{\underline{\mathfrak{f}}}}_{4}(\epsilon)\,(\Pi\bsy{g})([ e_3^{\text{as}},e^{\text{as}}_B],e^{\text{as}}_A)-\bsy{\mathfrak{\underline{\mathfrak{f}}}}^{C}(\epsilon)\,(\Pi\bsy{g})([ e_C^{\text{as}},e^{\text{as}}_B],e^{\text{as}}_A) \\
&+(e^{\text{as}}_A(\bsy{\mathfrak{\underline{\mathfrak{f}}}}^{C}(\epsilon)))\,(\Pi\bsy{g})( e_C^{\text{as}},e^{\text{as}}_B) +(e^{\text{as}}_B(\bsy{\mathfrak{\underline{\mathfrak{f}}}}^{C}(\epsilon)))\,(\Pi\bsy{g})( e_C^{\text{as}},e^{\text{as}}_A) \\
&+\text{nonlinear terms}   \, ,
\end{align*}
where the first equality holds by Definition \ref{def_restriction_derivatives}.

\end{proof}

\medskip

\begin{prop} \label{prop_decomp_commutators_nonlinear}
We have the identities
\begin{align*}
[\ef(\epsilon),\et(\epsilon)]=& \, (\nablasl^{a,M}_{e_4^{\textup{as}}}\bsy{\mathfrak{\underline{\mathfrak{f}}}}_{3}(\epsilon)) e_4^{\textup{as}}+(\nablasl^{a,M}_{e_4^{\textup{as}}}\bsy{\mathfrak{\underline{\mathfrak{f}}}}_{4}(\epsilon)) e_3^{\textup{as}}+(\nablasl^{a,M}_{e_4^{\textup{as}}}\bsy{\mathfrak{\underline{\mathfrak{f}}}}^{A}(\epsilon)) e_A^{\textup{as}}  \\
&+\bsy{\mathfrak{\underline{\mathfrak{f}}}}_{4}(\epsilon)\, [e_4^{\textup{as}},e_3^{\textup{as}}]+\bsy{\mathfrak{\underline{\mathfrak{f}}}}^{A}(\epsilon)\, [e_4^{\textup{as}}, e_A^{\textup{as}}]  \, , 
\end{align*}
\begin{align*}
[\ef(\epsilon),\ea(\epsilon)]=& \, (\nabla^{a,M}_{e_4^{\textup{as}}}(\bsy{\mathfrak{\slashed{\mathfrak{f}}}}_{3})_{\bsy{A}}(\epsilon))\, e_4^{\textup{as}}+(\nabla^{a,M}_{e_4^{\textup{as}}}(\bsy{\mathfrak{\slashed{\mathfrak{f}}}}_{4})_{\bsy{A}}(\epsilon))\, e_3^{\textup{as}} +\slashed{g}_{a,M}^{BC}(\nabla^{a,M}_{e_4^{\textup{as}}}(\bsy{\mathfrak{\slashed{\mathfrak{f}}}}_{\bsy{A}})_{B}(\epsilon))\, e_C^{\textup{as}} \\
&+(\bsy{\mathfrak{\slashed{\mathfrak{f}}}}_{4})_{\bsy{A}}(\epsilon)\, [e_4^{\textup{as}},e_3^{\textup{as}}] +\slashed{g}_{a,M}^{BC}(\bsy{\mathfrak{\slashed{\mathfrak{f}}}}_{\bsy{A}})_{B}(\epsilon)\,[e_4^{\textup{as}}, e_C^{\textup{as}}] \, ,  
\end{align*}
\begin{align*}
[\et(\epsilon),\ea(\epsilon)]=& \left( \bsy{\mathfrak{\underline{\mathfrak{f}}}}_{3}(\epsilon)(\nabla^{a,M}_{e_4^{\textup{as}}}  (\bsy{\mathfrak{\slashed{\mathfrak{f}}}}_{3})_{\bsy{A}}(\epsilon))+\bsy{\mathfrak{\underline{\mathfrak{f}}}}_{4}(\epsilon)(\nabla^{a,M}_{e_3^{\textup{as}}} (\bsy{\mathfrak{\slashed{\mathfrak{f}}}}_{3})_{\bsy{A}}(\epsilon))
+\bsy{\mathfrak{\underline{\mathfrak{f}}}}^{B}(\epsilon)(\nabla^{a,M}_{e_B^{\textup{as}}} (\bsy{\mathfrak{\slashed{\mathfrak{f}}}}_{3})_{\bsy{A}}(\epsilon)) \right. \\ &\left. -(\bsy{\mathfrak{\slashed{\mathfrak{f}}}}_{3})_{\bsy{A}}(\epsilon)\, (\nabla^{a,M}_{e_4^{\textup{as}}} \bsy{\mathfrak{\underline{\mathfrak{f}}}}_{3}(\epsilon))-(\bsy{\mathfrak{\slashed{\mathfrak{f}}}}_{4})_{\bsy{A}}(\epsilon)\, (\nabla^{a,M}_{e_3^{\textup{as}}} \bsy{\mathfrak{\underline{\mathfrak{f}}}}_{3}(\epsilon))-\slashed{g}_{a,M}^{BC}(\bsy{\mathfrak{\slashed{\mathfrak{f}}}}_{\bsy{A}})_{B}(\epsilon)\, (\nabla^{a,M}_{e_C^{\textup{as}}} \bsy{\mathfrak{\underline{\mathfrak{f}}}}_{3}(\epsilon)) \right) e_4^{\textup{as}}\\
&+\left( \bsy{\mathfrak{\underline{\mathfrak{f}}}}_{3}(\epsilon)(\nabla^{a,M}_{e_4^{\textup{as}}} (\bsy{\mathfrak{\slashed{\mathfrak{f}}}}_{4})_{\bsy{A}}(\epsilon))+\bsy{\mathfrak{\underline{\mathfrak{f}}}}_{4}(\epsilon)(\nabla^{a,M}_{e_3^{\textup{as}}}(\bsy{\mathfrak{\slashed{\mathfrak{f}}}}_{4})_{\bsy{A}}(\epsilon))+\bsy{\mathfrak{\underline{\mathfrak{f}}}}^{B}(\epsilon)(\nabla^{a,M}_{e_B^{\textup{as}}}(\bsy{\mathfrak{\slashed{\mathfrak{f}}}}_{4})_{\bsy{A}}(\epsilon)) \right. \\ & \left. -(\bsy{\mathfrak{\slashed{\mathfrak{f}}}}_{3})_{\bsy{A}}(\epsilon)\, (\nabla^{a,M}_{e_4^{\textup{as}}}\bsy{\mathfrak{\underline{\mathfrak{f}}}}_{4}(\epsilon))-(\bsy{\mathfrak{\slashed{\mathfrak{f}}}}_{4})_{\bsy{A}}(\epsilon)\, (\nabla^{a,M}_{e_3^{\textup{as}}}\bsy{\mathfrak{\underline{\mathfrak{f}}}}_{4}(\epsilon))-\slashed{g}_{a,M}^{BC}(\bsy{\mathfrak{\slashed{\mathfrak{f}}}}_{\bsy{A}})_{B}(\epsilon)\,( \nabla^{a,M}_{e_C^{\textup{as}}}\bsy{\mathfrak{\underline{\mathfrak{f}}}}_{4}(\epsilon)) \right) e_3^{\textup{as}} \\
&+\left( \bsy{\mathfrak{\underline{\mathfrak{f}}}}_{3}(\epsilon)(\nabla^{a,M}_{e_4^{\textup{as}}} \slashed{g}_{a,M}^{BC}(\bsy{\mathfrak{\slashed{\mathfrak{f}}}}_{\bsy{A}})_{C}(\epsilon))+\bsy{\mathfrak{\underline{\mathfrak{f}}}}_{4}(\epsilon)(\nabla^{a,M}_{e_3^{\textup{as}}}\slashed{g}_{a,M}^{BC}(\bsy{\mathfrak{\slashed{\mathfrak{f}}}}_{\bsy{A}})_{C}(\epsilon))\,  +\bsy{\mathfrak{\underline{\mathfrak{f}}}}^{C}(\epsilon)(\nabla^{a,M}_{e_C^{\textup{as}}}\slashed{g}_{a,M}^{DB}(\bsy{\mathfrak{\slashed{\mathfrak{f}}}}_{\bsy{A}})_{D}(\epsilon)) \right. \\ & \left. -(\bsy{\mathfrak{\slashed{\mathfrak{f}}}}_{3})_{\bsy{A}}(\epsilon)\, (\nabla^{a,M}_{e_4^{\textup{as}}}\bsy{\mathfrak{\underline{\mathfrak{f}}}}^{B}(\epsilon))-(\bsy{\mathfrak{\slashed{\mathfrak{f}}}}_{4})_{\bsy{A}}(\epsilon)\, (\nabla^{a,M}_{e_3^{\textup{as}}}\bsy{\mathfrak{\underline{\mathfrak{f}}}}^{B}(\epsilon))-\slashed{g}_{a,M}^{DC}(\bsy{\mathfrak{\slashed{\mathfrak{f}}}}_{\bsy{A}})_{D}(\epsilon)\, (\nabla^{a,M}_{e_C^{\textup{as}}}\bsy{\mathfrak{\underline{\mathfrak{f}}}}^{B}(\epsilon))  \right) e_B^{\textup{as}} \\
&  +\left((\bsy{\mathfrak{\slashed{\mathfrak{f}}}}_{4})_{\bsy{A}}(\epsilon)\, \bsy{\mathfrak{\underline{\mathfrak{f}}}}_{3}(\epsilon)-(\bsy{\mathfrak{\slashed{\mathfrak{f}}}}_{3})_{\bsy{A}}(\epsilon)\, \bsy{\mathfrak{\underline{\mathfrak{f}}}}_{4}(\epsilon)\right)\,[e_4^{\textup{as}},e_3^{\textup{as}}]  \\
&+\left(\slashed{g}_{a,M}^{BC}(\bsy{\mathfrak{\slashed{\mathfrak{f}}}}_{\bsy{A}})_{B}(\epsilon)\, \bsy{\mathfrak{\underline{\mathfrak{f}}}}_{3}(\epsilon)-(\bsy{\mathfrak{\slashed{\mathfrak{f}}}}_{3})_{\bsy{A}}(\epsilon)\, \bsy{\mathfrak{\underline{\mathfrak{f}}}}^{C}(\epsilon)\right)\, [e_4^{\textup{as}},e_C^{\textup{as}}]   \\
& +\left(\slashed{g}_{a,M}^{BC}(\bsy{\mathfrak{\slashed{\mathfrak{f}}}}_{\bsy{A}})_{B}(\epsilon)\, \bsy{\mathfrak{\underline{\mathfrak{f}}}}_{4}(\epsilon)-(\bsy{\mathfrak{\slashed{\mathfrak{f}}}}_{4})_{\bsy{A}}(\epsilon)\, \bsy{\mathfrak{\underline{\mathfrak{f}}}}^{C}(\epsilon)\right)\, [e_3^{\textup{as}}, e_C^{\textup{as}}]   \\
&+\slashed{g}_{a,M}^{BD}(\bsy{\mathfrak{\slashed{\mathfrak{f}}}}_{\bsy{A}})_{B}(\epsilon)\, \bsy{\mathfrak{\underline{\mathfrak{f}}}}^{C}(\epsilon) \, [e_C^{\textup{as}} , e_D^{\textup{as}} ]   \, .
\end{align*}
\end{prop}

\medskip

\begin{proof}
We first decompose the frame vector fields as follows
\begin{align*}
\bsy{e_3}(\epsilon)&=\bsy{\mathfrak{\underline{\mathfrak{f}}}}_{3}(\epsilon)\, e_4^{\text{as}}+\bsy{\mathfrak{\underline{\mathfrak{f}}}}_{4}(\epsilon)\, e_3^{\text{as}}+\bsy{\mathfrak{\underline{\mathfrak{f}}}}^{A}(\epsilon)\, e_A^{\text{as}}  \, , \\
\bsy{e_A}(\epsilon)&=(\bsy{\mathfrak{\slashed{\mathfrak{f}}}}_{3})_{\bsy{A}}(\epsilon)\, e_4^{\text{as}}+(\bsy{\mathfrak{\slashed{\mathfrak{f}}}}_{4})_{\bsy{A}}(\epsilon)\, e_3^{\text{as}} +\slashed{g}_{a,M}^{BC}(\bsy{\mathfrak{\slashed{\mathfrak{f}}}}_{\bsy{A}})_{B}(\epsilon)\, e_C^{\text{as}} 
\end{align*} 
and then compute
\begin{align*}
[\ef,\et]=& \, \nabla^{a,M}_{\ef} \et-\nabla^{a,M}_{\et} \ef \\
=& \,\nabla^{a,M}_{e_4^{\text{as}}} \et-\nabla^{a,M}_{\et} e_4^{\text{as}} \\
=& \, \nabla^{a,M}_{e_4^{\text{as}}}\left(\bsy{\mathfrak{\underline{\mathfrak{f}}}}_{3}(\epsilon)\, e_4^{\text{as}}+\bsy{\mathfrak{\underline{\mathfrak{f}}}}_{4}(\epsilon)\, e_3^{\text{as}}+\bsy{\mathfrak{\underline{\mathfrak{f}}}}^{A}(\epsilon)\, e_A^{\text{as}} \right) \\
&-\bsy{\mathfrak{\underline{\mathfrak{f}}}}_{3}(\epsilon)\nabla^{a,M}_{e_4^{\text{as}}} e_4^{\text{as}}-\bsy{\mathfrak{\underline{\mathfrak{f}}}}_{4}(\epsilon)\nabla^{a,M}_{e_3^{\text{as}}} e_4^{\text{as}}-\bsy{\mathfrak{\underline{\mathfrak{f}}}}^{A}(\epsilon)\nabla^{a,M}_{e_A^{\text{as}}} e_4^{\text{as}} \\
=& \, (\nabla^{a,M}_{e_4^{\text{as}}}\bsy{\mathfrak{\underline{\mathfrak{f}}}}_{3}(\epsilon)) e_4^{\text{as}}+(\nabla^{a,M}_{e_4^{\text{as}}}\bsy{\mathfrak{\underline{\mathfrak{f}}}}_{4}(\epsilon)) e_3^{\text{as}}+(\nabla^{a,M}_{e_4^{\text{as}}}\bsy{\mathfrak{\underline{\mathfrak{f}}}}^{A}(\epsilon)) e_A^{\text{as}}  \\
&+\bsy{\mathfrak{\underline{\mathfrak{f}}}}_{4}(\epsilon)\, \nabla^{a,M}_{e_4^{\text{as}}}e_3^{\text{as}}+\bsy{\mathfrak{\underline{\mathfrak{f}}}}^{A}(\epsilon)\, \nabla^{a,M}_{e_4^{\text{as}}}e_A^{\text{as}} \\
&-\bsy{\mathfrak{\underline{\mathfrak{f}}}}_{4}(\epsilon)\nabla^{a,M}_{e_3^{\text{as}}} e_4^{\text{as}}-\bsy{\mathfrak{\underline{\mathfrak{f}}}}^{A}(\epsilon)\nabla^{a,M}_{e_A^{\text{as}}} e_4^{\text{as}} \, ,
\end{align*}
\begin{align*}
[\ef,\ea]=& \, \nabla^{a,M}_{\ef} \ea-\nabla^{a,M}_{\et} \ea \\
=& \,\nabla^{a,M}_{e_4^{\text{as}}} \ea-\nabla^{a,M}_{\ea} e_4^{\text{as}} \\
=& \,\nabla^{a,M}_{e_4^{\text{as}}}\left( (\bsy{\mathfrak{\slashed{\mathfrak{f}}}}_{3})_{\bsy{A}}(\epsilon)\, e_4^{\text{as}}+(\bsy{\mathfrak{\slashed{\mathfrak{f}}}}_{4})_{\bsy{A}}(\epsilon)\, e_3^{\text{as}} +\slashed{g}_{a,M}^{BC}(\bsy{\mathfrak{\slashed{\mathfrak{f}}}}_{\bsy{A}})_{B}(\epsilon)\, e_C^{\text{as}}\right) \\
&-(\bsy{\mathfrak{\slashed{\mathfrak{f}}}}_{3})_{\bsy{A}}(\epsilon)\, \nabla^{a,M}_{e_4^{\text{as}}}e_4^{\text{as}}-(\bsy{\mathfrak{\slashed{\mathfrak{f}}}}_{4})_{\bsy{A}}(\epsilon)\, \nabla^{a,M}_{e_3^{\text{as}}}e_4^{\text{as}} -\slashed{g}_{a,M}^{BC}(\bsy{\mathfrak{\slashed{\mathfrak{f}}}}_{\bsy{A}})_{B}(\epsilon)\, \nabla^{a,M}_{e_C^{\text{as}}}e_4^{\text{as}} \\
=& \,(\nabla^{a,M}_{e_4^{\text{as}}}(\bsy{\mathfrak{\slashed{\mathfrak{f}}}}_{3})_{\bsy{A}}(\epsilon))\, e_4^{\text{as}}+(\nabla^{a,M}_{e_4^{\text{as}}}(\bsy{\mathfrak{\slashed{\mathfrak{f}}}}_{4})_{\bsy{A}}(\epsilon))\, e_3^{\text{as}} +(\nabla^{a,M}_{e_4^{\text{as}}}\slashed{g}_{a,M}^{BC}(\bsy{\mathfrak{\slashed{\mathfrak{f}}}}_{\bsy{A}})_{B}(\epsilon))\, e_C^{\text{as}} \\
&+(\bsy{\mathfrak{\slashed{\mathfrak{f}}}}_{4})_{\bsy{A}}(\epsilon)\, \nabla^{a,M}_{e_4^{\text{as}}}e_3^{\text{as}} +\slashed{g}_{a,M}^{BC}(\bsy{\mathfrak{\slashed{\mathfrak{f}}}}_{\bsy{A}})_{B}(\epsilon)\,\nabla^{a,M}_{e_4^{\text{as}}} e_C^{\text{as}}  \\
&-(\bsy{\mathfrak{\slashed{\mathfrak{f}}}}_{4})_{\bsy{A}}(\epsilon)\, \nabla^{a,M}_{e_3^{\text{as}}}e_4^{\text{as}} -\slashed{g}_{a,M}^{BC}(\bsy{\mathfrak{\slashed{\mathfrak{f}}}}_{\bsy{A}})_{B}(\epsilon)\, \nabla^{a,M}_{e_C^{\text{as}}}e_4^{\text{as}} \, ,
\end{align*}
\begin{align*}
[\et,\ea]=& \,\nabla^{a,M}_{\et} \ea-\nabla^{a,M}_{\ea} \et \\[5pt]
=& \,\bsy{\mathfrak{\underline{\mathfrak{f}}}}_{3}(\epsilon)\nabla^{a,M}_{e_4^{\text{as}}} \ea+\bsy{\mathfrak{\underline{\mathfrak{f}}}}_{4}(\epsilon)\nabla^{a,M}_{e_3^{\text{as}}} \ea+\bsy{\mathfrak{\underline{\mathfrak{f}}}}^{B}(\epsilon)\nabla^{a,M}_{e_B^{\text{as}}} \ea \\
&-(\bsy{\mathfrak{\slashed{\mathfrak{f}}}}_{3})_{\bsy{A}}(\epsilon)\, \nabla^{a,M}_{e_4^{\text{as}}}\et-(\bsy{\mathfrak{\slashed{\mathfrak{f}}}}_{4})_{\bsy{A}}(\epsilon)\, \nabla^{a,M}_{e_3^{\text{as}}}\et -\slashed{g}_{a,M}^{BC}(\bsy{\mathfrak{\slashed{\mathfrak{f}}}}_{\bsy{A}})_{B}(\epsilon)\, \nabla^{a,M}_{e_C^{\text{as}}} \et \\[5pt]
=& \,\bsy{\mathfrak{\underline{\mathfrak{f}}}}_{3}(\epsilon)\nabla^{a,M}_{e_4^{\text{as}}} \left( (\bsy{\mathfrak{\slashed{\mathfrak{f}}}}_{3})_{\bsy{A}}(\epsilon)\, e_4^{\text{as}}+(\bsy{\mathfrak{\slashed{\mathfrak{f}}}}_{4})_{\bsy{A}}(\epsilon)\, e_3^{\text{as}} +\slashed{g}_{a,M}^{BC}(\bsy{\mathfrak{\slashed{\mathfrak{f}}}}_{\bsy{A}})_{B}(\epsilon)\, e_C^{\text{as}}  \right) \\
&+\bsy{\mathfrak{\underline{\mathfrak{f}}}}_{4}(\epsilon)\nabla^{a,M}_{e_3^{\text{as}}} \left( (\bsy{\mathfrak{\slashed{\mathfrak{f}}}}_{3})_{\bsy{A}}(\epsilon)\, e_4^{\text{as}}+(\bsy{\mathfrak{\slashed{\mathfrak{f}}}}_{4})_{\bsy{A}}(\epsilon)\, e_3^{\text{as}} +\slashed{g}_{a,M}^{BC}(\bsy{\mathfrak{\slashed{\mathfrak{f}}}}_{\bsy{A}})_{B}(\epsilon)\, e_C^{\text{as}}  \right) \\
&+\bsy{\mathfrak{\underline{\mathfrak{f}}}}^{B}(\epsilon)\nabla^{a,M}_{e_B^{\text{as}}} \left( (\bsy{\mathfrak{\slashed{\mathfrak{f}}}}_{3})_{\bsy{A}}(\epsilon)\, e_4^{\text{as}}+(\bsy{\mathfrak{\slashed{\mathfrak{f}}}}_{4})_{\bsy{A}}(\epsilon)\, e_3^{\text{as}} +\slashed{g}_{a,M}^{DC}(\bsy{\mathfrak{\slashed{\mathfrak{f}}}}_{\bsy{A}})_{D}(\epsilon)\, e_C^{\text{as}}  \right) \\
&-(\bsy{\mathfrak{\slashed{\mathfrak{f}}}}_{3})_{\bsy{A}}(\epsilon)\, \nabla^{a,M}_{e_4^{\text{as}}} \left(\bsy{\mathfrak{\underline{\mathfrak{f}}}}_{3}(\epsilon)\, e_4^{\text{as}}+\bsy{\mathfrak{\underline{\mathfrak{f}}}}_{4}(\epsilon)\, e_3^{\text{as}}+\bsy{\mathfrak{\underline{\mathfrak{f}}}}^{B}(\epsilon)\, e_B^{\text{as}} \right) \\
&-(\bsy{\mathfrak{\slashed{\mathfrak{f}}}}_{4})_{\bsy{A}}(\epsilon)\, \nabla^{a,M}_{e_3^{\text{as}}} \left(\bsy{\mathfrak{\underline{\mathfrak{f}}}}_{3}(\epsilon)\, e_4^{\text{as}}+\bsy{\mathfrak{\underline{\mathfrak{f}}}}_{4}(\epsilon)\, e_3^{\text{as}}+\bsy{\mathfrak{\underline{\mathfrak{f}}}}^{B}(\epsilon)\, e_B^{\text{as}} \right) \\
&-\slashed{g}_{a,M}^{BC}(\bsy{\mathfrak{\slashed{\mathfrak{f}}}}_{\bsy{A}})_{B}(\epsilon)\, \nabla^{a,M}_{e_C^{\text{as}}} \left(\bsy{\mathfrak{\underline{\mathfrak{f}}}}_{3}(\epsilon)\, e_4^{\text{as}}+\bsy{\mathfrak{\underline{\mathfrak{f}}}}_{4}(\epsilon)\, e_3^{\text{as}}+\bsy{\mathfrak{\underline{\mathfrak{f}}}}^{D}(\epsilon)\, e_D^{\text{as}} \right) \\[5pt]
=& \,\bsy{\mathfrak{\underline{\mathfrak{f}}}}_{3}(\epsilon)(\nabla^{a,M}_{e_4^{\text{as}}}  (\bsy{\mathfrak{\slashed{\mathfrak{f}}}}_{3})_{\bsy{A}}(\epsilon))\, e_4^{\text{as}}+\bsy{\mathfrak{\underline{\mathfrak{f}}}}_{3}(\epsilon)(\nabla^{a,M}_{e_4^{\text{as}}} (\bsy{\mathfrak{\slashed{\mathfrak{f}}}}_{4})_{\bsy{A}}(\epsilon))\, e_3^{\text{as}} +\bsy{\mathfrak{\underline{\mathfrak{f}}}}_{3}(\epsilon)(\nabla^{a,M}_{e_4^{\text{as}}} \slashed{g}_{a,M}^{BC}(\bsy{\mathfrak{\slashed{\mathfrak{f}}}}_{\bsy{A}})_{B}(\epsilon))\, e_C^{\text{as}}   \\
&+\bsy{\mathfrak{\underline{\mathfrak{f}}}}_{4}(\epsilon)(\nabla^{a,M}_{e_3^{\text{as}}} (\bsy{\mathfrak{\slashed{\mathfrak{f}}}}_{3})_{\bsy{A}}(\epsilon))\, e_4^{\text{as}}+\bsy{\mathfrak{\underline{\mathfrak{f}}}}_{4}(\epsilon)(\nabla^{a,M}_{e_3^{\text{as}}}(\bsy{\mathfrak{\slashed{\mathfrak{f}}}}_{4})_{\bsy{A}}(\epsilon))\, e_3^{\text{as}} +\bsy{\mathfrak{\underline{\mathfrak{f}}}}_{4}(\epsilon)(\nabla^{a,M}_{e_3^{\text{as}}}\slashed{g}_{a,M}^{BC}(\bsy{\mathfrak{\slashed{\mathfrak{f}}}}_{\bsy{A}})_{B}(\epsilon))\, e_C^{\text{as}}   \\
&+\bsy{\mathfrak{\underline{\mathfrak{f}}}}^{B}(\epsilon)(\nabla^{a,M}_{e_B^{\text{as}}} (\bsy{\mathfrak{\slashed{\mathfrak{f}}}}_{3})_{\bsy{A}}(\epsilon))\, e_4^{\text{as}}+\bsy{\mathfrak{\underline{\mathfrak{f}}}}^{B}(\epsilon)(\nabla^{a,M}_{e_B^{\text{as}}}(\bsy{\mathfrak{\slashed{\mathfrak{f}}}}_{4})_{\bsy{A}}(\epsilon))\, e_3^{\text{as}} +\bsy{\mathfrak{\underline{\mathfrak{f}}}}^{B}(\epsilon)(\nabla^{a,M}_{e_B^{\text{as}}}\slashed{g}_{a,M}^{DC}(\bsy{\mathfrak{\slashed{\mathfrak{f}}}}_{\bsy{A}})_{D}(\epsilon))\, e_C^{\text{as}}   \\
&-(\bsy{\mathfrak{\slashed{\mathfrak{f}}}}_{3})_{\bsy{A}}(\epsilon)\, (\nabla^{a,M}_{e_4^{\text{as}}} \bsy{\mathfrak{\underline{\mathfrak{f}}}}_{3}(\epsilon))\, e_4^{\text{as}}-(\bsy{\mathfrak{\slashed{\mathfrak{f}}}}_{3})_{\bsy{A}}(\epsilon)\, (\nabla^{a,M}_{e_4^{\text{as}}}\bsy{\mathfrak{\underline{\mathfrak{f}}}}_{4}(\epsilon))\, e_3^{\text{as}}-(\bsy{\mathfrak{\slashed{\mathfrak{f}}}}_{3})_{\bsy{A}}(\epsilon)\, (\nabla^{a,M}_{e_4^{\text{as}}}\bsy{\mathfrak{\underline{\mathfrak{f}}}}^{B}(\epsilon))\, e_B^{\text{as}}  \\
&-(\bsy{\mathfrak{\slashed{\mathfrak{f}}}}_{4})_{\bsy{A}}(\epsilon)\, (\nabla^{a,M}_{e_3^{\text{as}}} \bsy{\mathfrak{\underline{\mathfrak{f}}}}_{3}(\epsilon))\, e_4^{\text{as}}-(\bsy{\mathfrak{\slashed{\mathfrak{f}}}}_{4})_{\bsy{A}}(\epsilon)\, (\nabla^{a,M}_{e_3^{\text{as}}}\bsy{\mathfrak{\underline{\mathfrak{f}}}}_{4}(\epsilon))\, e_3^{\text{as}}-(\bsy{\mathfrak{\slashed{\mathfrak{f}}}}_{4})_{\bsy{A}}(\epsilon)\, (\nabla^{a,M}_{e_3^{\text{as}}}\bsy{\mathfrak{\underline{\mathfrak{f}}}}^{B}(\epsilon))\, e_B^{\text{as}}  \\
&-\slashed{g}_{a,M}^{BC}(\bsy{\mathfrak{\slashed{\mathfrak{f}}}}_{\bsy{A}})_{B}(\epsilon)\, (\nabla^{a,M}_{e_C^{\text{as}}} \bsy{\mathfrak{\underline{\mathfrak{f}}}}_{3}(\epsilon))\, e_4^{\text{as}}-\slashed{g}_{a,M}^{BC}(\bsy{\mathfrak{\slashed{\mathfrak{f}}}}_{\bsy{A}})_{B}(\epsilon)\,( \nabla^{a,M}_{e_C^{\text{as}}}\bsy{\mathfrak{\underline{\mathfrak{f}}}}_{4}(\epsilon))\, e_3^{\text{as}}-\slashed{g}_{a,M}^{BC}(\bsy{\mathfrak{\slashed{\mathfrak{f}}}}_{\bsy{A}})_{B}(\epsilon)\, (\nabla^{a,M}_{e_C^{\text{as}}}\bsy{\mathfrak{\underline{\mathfrak{f}}}}^{D}(\epsilon))\, e_D^{\text{as}}  \\
&+ (\bsy{\mathfrak{\slashed{\mathfrak{f}}}}_{3})_{\bsy{A}}(\epsilon)\, \bsy{\mathfrak{\underline{\mathfrak{f}}}}_{3}(\epsilon)\nabla^{a,M}_{e_4^{\text{as}}}e_4^{\text{as}}+(\bsy{\mathfrak{\slashed{\mathfrak{f}}}}_{4})_{\bsy{A}}(\epsilon)\, \bsy{\mathfrak{\underline{\mathfrak{f}}}}_{3}(\epsilon)\nabla^{a,M}_{e_4^{\text{as}}}e_3^{\text{as}} +\slashed{g}_{a,M}^{BC}(\bsy{\mathfrak{\slashed{\mathfrak{f}}}}_{\bsy{A}})_{B}(\epsilon)\, \bsy{\mathfrak{\underline{\mathfrak{f}}}}_{3}(\epsilon)\nabla^{a,M}_{e_4^{\text{as}}}e_C^{\text{as}}   \\
&+ (\bsy{\mathfrak{\slashed{\mathfrak{f}}}}_{3})_{\bsy{A}}(\epsilon)\, \bsy{\mathfrak{\underline{\mathfrak{f}}}}_{4}(\epsilon)\nabla^{a,M}_{e_3^{\text{as}}} e_4^{\text{as}}+(\bsy{\mathfrak{\slashed{\mathfrak{f}}}}_{4})_{\bsy{A}}(\epsilon)\, \bsy{\mathfrak{\underline{\mathfrak{f}}}}_{4}(\epsilon)\nabla^{a,M}_{e_3^{\text{as}}} e_3^{\text{as}} +\slashed{g}_{a,M}^{BC}(\bsy{\mathfrak{\slashed{\mathfrak{f}}}}_{\bsy{A}})_{B}(\epsilon)\, \bsy{\mathfrak{\underline{\mathfrak{f}}}}_{4}(\epsilon)\nabla^{a,M}_{e_3^{\text{as}}} e_C^{\text{as}}   \\
&+ (\bsy{\mathfrak{\slashed{\mathfrak{f}}}}_{3})_{\bsy{A}}(\epsilon)\, \bsy{\mathfrak{\underline{\mathfrak{f}}}}^{B}(\epsilon)\nabla^{a,M}_{e_B^{\text{as}}}e_4^{\text{as}}+(\bsy{\mathfrak{\slashed{\mathfrak{f}}}}_{4})_{\bsy{A}}(\epsilon)\, \bsy{\mathfrak{\underline{\mathfrak{f}}}}^{B}(\epsilon)\nabla^{a,M}_{e_B^{\text{as}}}e_3^{\text{as}} +\slashed{g}_{a,M}^{DC}(\bsy{\mathfrak{\slashed{\mathfrak{f}}}}_{\bsy{A}})_{D}(\epsilon)\, \bsy{\mathfrak{\underline{\mathfrak{f}}}}^{B}(\epsilon)\nabla^{a,M}_{e_B^{\text{as}}}e_C^{\text{as}}   \\
&-\bsy{\mathfrak{\underline{\mathfrak{f}}}}_{3}(\epsilon)\,(\bsy{\mathfrak{\slashed{\mathfrak{f}}}}_{3})_{\bsy{A}}(\epsilon)\, \nabla^{a,M}_{e_4^{\text{as}}}  e_4^{\text{as}}-\bsy{\mathfrak{\underline{\mathfrak{f}}}}_{4}(\epsilon)\, (\bsy{\mathfrak{\slashed{\mathfrak{f}}}}_{3})_{\bsy{A}}(\epsilon)\, \nabla^{a,M}_{e_4^{\text{as}}} e_3^{\text{as}}-\bsy{\mathfrak{\underline{\mathfrak{f}}}}^{B}(\epsilon)\, (\bsy{\mathfrak{\slashed{\mathfrak{f}}}}_{3})_{\bsy{A}}(\epsilon)\, \nabla^{a,M}_{e_4^{\text{as}}} e_B^{\text{as}}  \\
&- \bsy{\mathfrak{\underline{\mathfrak{f}}}}_{3}(\epsilon)\, (\bsy{\mathfrak{\slashed{\mathfrak{f}}}}_{4})_{\bsy{A}}(\epsilon)\, \nabla^{a,M}_{e_3^{\text{as}}} e_4^{\text{as}}-\bsy{\mathfrak{\underline{\mathfrak{f}}}}_{4}(\epsilon)\, (\bsy{\mathfrak{\slashed{\mathfrak{f}}}}_{4})_{\bsy{A}}(\epsilon)\, \nabla^{a,M}_{e_3^{\text{as}}}e_3^{\text{as}}-\bsy{\mathfrak{\underline{\mathfrak{f}}}}^{B}(\epsilon)\, (\bsy{\mathfrak{\slashed{\mathfrak{f}}}}_{4})_{\bsy{A}}(\epsilon)\, \nabla^{a,M}_{e_3^{\text{as}}} e_B^{\text{as}}  \\
&- \bsy{\mathfrak{\underline{\mathfrak{f}}}}_{3}(\epsilon)\, \slashed{g}_{a,M}^{BC}(\bsy{\mathfrak{\slashed{\mathfrak{f}}}}_{\bsy{A}})_{B}(\epsilon)\, \nabla^{a,M}_{e_C^{\text{as}}}e_4^{\text{as}}-\bsy{\mathfrak{\underline{\mathfrak{f}}}}_{4}(\epsilon)\, \slashed{g}_{a,M}^{BC}(\bsy{\mathfrak{\slashed{\mathfrak{f}}}}_{\bsy{A}})_{B}(\epsilon)\, \nabla^{a,M}_{e_C^{\text{as}}}e_3^{\text{as}}-\bsy{\mathfrak{\underline{\mathfrak{f}}}}^{D}(\epsilon)\, \slashed{g}_{a,M}^{BC}(\bsy{\mathfrak{\slashed{\mathfrak{f}}}}_{\bsy{A}})_{B}(\epsilon)\, \nabla^{a,M}_{e_C^{\text{as}}}e_D^{\text{as}} \, .
\end{align*}
One concludes the proof by rearranging the terms.

\end{proof}

\medskip

\begin{prop}  \label{prop_M_Mbar_nonlinear}
We have the identities
\begin{align*}
\bsy{\slashed{M}{}_A^B}(\epsilon) =& \, \bsy{g}^{\bsy{BC}}(\epsilon)\,g_{a,M}^{DE}\,g_{a,M}([\ef(\epsilon),\ea(\epsilon)],e^{\textup{as}}_D)\,\bsy{g}(e^{\textup{as}}_E,\bsy{e_C}(\epsilon))   \\ &-\frac{1}{2}\,g_{a,M}([\ef(\epsilon),\ea(\epsilon)],e^{\textup{as}}_4)\,\bsy{g}^{\bsy{BC}}(\epsilon)\,\bsy{g}(e^{\textup{as}}_3,\bsy{e_C}(\epsilon)) \\ &+\bsy{\slashed{g}}^{\bsy{BC}}(\epsilon)\,\bsy{\chi}_{\bsy{AC}}(\epsilon) \, , \\[5pt]
\bsy{\slashed{\underline{M}}{}_A^B}(\epsilon) =& \, \bsy{g}^{\bsy{BC}}(\epsilon)\,g_{a,M}^{DE}\,g_{a,M}([\et(\epsilon),\ea(\epsilon)],e^{\textup{as}}_D)\,\bsy{g}(e^{\textup{as}}_E,\bsy{e_C}(\epsilon))   \\  &-\frac{1}{2}\,g_{a,M}([\et(\epsilon),\ea(\epsilon)],e^{\textup{as}}_4)\,\bsy{g}^{\bsy{BC}}(\epsilon)\,\bsy{g}(e^{\textup{as}}_3,\bsy{e_C}(\epsilon)) \\ &+\bsy{\slashed{g}}^{\bsy{BC}}(\epsilon)\,\bsy{\chib}_{\bsy{AC}}(\epsilon) \, .
\end{align*}
\end{prop}

\medskip

\begin{proof}
We compute
\begin{align*}
\bsy{\slashed{M}{}_A^B}  =& \, \slashed{\bsy{g}}^{\bsy{BC}}\slashed{\bsy{g}}(\bsy{\nablasl}_{\bsy{4}}\ea,\bsy{e_C}) \\
=& \, \bsy{g}^{\bsy{BC}}\bsy{g}(\bsy{\nabla}_{\bsy{4}}\ea,\bsy{e_C}) \\
=& \,  \bsy{g}^{\bsy{BC}}\bsy{g}([\ef,\ea],\bsy{e_C})+\bsy{g}^{\bsy{BC}}\bsy{\chi}_{\bsy{AC}}  \\
=& \, \bsy{g}^{\bsy{BC}}g_{a,M}^{DE}\,g_{a,M}([\ef,\ea],e^{\text{as}}_D)\,\bsy{g}(e^{\text{as}}_E,\bsy{e_C})-\frac{1}{2}\,g_{a,M}([\ef,\ea],e^{\text{as}}_3)\,\bsy{g}^{\bsy{BC}}\bsy{g}(e^{\text{as}}_4,\bsy{e_C}) \\ & -\frac{1}{2}\,g_{a,M}([\ef,\ea],e^{\text{as}}_4)\,\bsy{g}^{\bsy{BC}}\bsy{g}(e^{\text{as}}_3,\bsy{e_C}) +\bsy{\slashed{g}}^{\bsy{BC}}\bsy{\chi}_{\bsy{AC}} \, .
\end{align*}
The identity for $\bsy{\slashed{\underline{M}}{}_A^B}(\epsilon)$ follows from an analogous computation.

\end{proof}

\subsection{An isometry between horizontal tensor bundles} \label{sec_isometry_horizontal_bundles}

We summarise our geometric procedure as follows.~For any $\bsy{\mathfrak{D}}_{\bsy{\mathcal{N}}(\epsilon)}$ $k$-tensor $\bsy{\varsigma}(\epsilon)$, our procedure yields the $\mathfrak{D}_{\mathcal{N}_{\text{as}}}$ $k$-tensor $\widetilde{\bsy{\varsigma}}(\epsilon)$ such that
\begin{equation} \label{aux_isometry_1}
\widetilde{\bsy{\varsigma}}(e_{A_1}^{\text{as}},\ldots,e_{A_k}^{\text{as}})=\bsy{\varsigma}(\bsy{G}_{A_1},\ldots,\bsy{G}_{A_k})
\end{equation}
and
\begin{equation} \label{aux_isometry_2}
|\widetilde{\bsy{\varsigma}}|^2_{\widetilde{\bsy{\slashed{g}}}}= |\bsy{\varsigma}|^2_{\bsy{\slashed{g}}} \, ,
\end{equation}
where the identity \eqref{aux_isometry_1} follows from Proposition \ref{prop_rewrite_proj} and the identity \eqref{aux_isometry_2} from Proposition \ref{prop_final_procedure} (see, in particular, the identity \eqref{isometry_product}).~For $\epsilon=0$, we have $\bsy{G}_{A_i}(0)=e_{A_i}^{\text{as}}$ from the identity \eqref{matrix_G}, and thus $$\widetilde{\bsy{\varsigma}}(0)=\bsy{\varsigma}(0) \, .$$ 

\medskip

From a more abstract perspective, our geometric procedure provides, for each $\epsilon$, the bundle morphism 
\begin{align*}
\mathfrak{G}(\epsilon)  :  \mathfrak{D}_{\mathcal{N}_{\text{as}}} &\rightarrow \bsy{\mathfrak{D}}_{\bsy{\mathcal{N}}(\epsilon)}   \\[5pt]
(p,(e_A^{\text{as}})_p) &\mapsto (p,(\bsy{G}_{A}(\epsilon))_p)
\end{align*}
between horizontal distributions on $\mathcal{M}^*$ and the bundle morphism
\begin{align*}
\mathfrak{B}_k(\epsilon)  :  (\otimes_k(\bsy{\mathfrak{D}}_{\bsy{\mathcal{N}}(\epsilon)})^{\star},|\cdot|_{\bsy{\slashed{g}}(\epsilon)}) &\rightarrow (\otimes_k(\mathfrak{D}_{\mathcal{N}_{\text{as}}})^{\star},|\cdot|_{\widetilde{\bsy{\slashed{g}}}(\epsilon)})  \\[5pt]
(p,\bsy{\varsigma}_p(\epsilon)) &\mapsto (p,\widetilde{\bsy{\varsigma}}_p(\epsilon))
\end{align*}
between normed horizontal tensor bundles on $\mathcal{M}^*$.~In fact, the bundle morphism $\mathfrak{B}_k(\epsilon)$ is an \emph{isometry}.~For $\epsilon=0$, the bundle morphism $\mathfrak{B}_k(0)$ coincides with the identity map over $\otimes_k(\mathfrak{D}_{\mathcal{N}_{\text{as}}})^{\star}$.

\section{The renormalised vacuum Einstein equations} \label{sec_renormalised_vacuum_Einstein_equations}

Consider the fixed ambient manifold $\mathcal{M}^*$ and the one-parameter family of metrics $\bsy{g}(\epsilon)$ and associated frames $\bsy{\mathcal{N}}(\epsilon)$ on $\mathcal{M}^*$ introduced in Section \ref{sec_perturbations_kerr}.~We assume that the metrics $\bsy{g}(\epsilon)$ solve the vacuum Einstein equations 
\begin{equation*}
\bsy{\ric}(\bsy{g}(\epsilon))=0 
\end{equation*}
for all $\epsilon$.~The nonlinear system of null structure and Bianchi equations for the connection coefficients and curvature components of the metrics $\bsy{g}(\epsilon)$ relative to the frames $\bsy{\mathcal{N}}(\epsilon)$ is a system of equations for smooth scalar functions and $\bsy{\mathfrak{D}}_{\bsy{\mathcal{N}}(\epsilon)}$ covariant tensors and coincides, for each $\epsilon$, with the system of equations presented in Section \ref{sec_nonlinear_system_eqns}.

\medskip

\begin{remark}
Since $\bsy{\mathcal{N}}(\epsilon)$ is a family of \emph{non-integrable} frames (see Section \ref{sec_family_frames}), the formulation of the nonlinear system of null structure and Bianchi equations for the metrics $\bsy{g}(\epsilon)$ relative to the frames $\bsy{\mathcal{N}}(\epsilon)$ employs (and ultimately motivates) the geometric formalism from Section \ref{sec_nonlinear_EVE} \emph{in its full generality}.
\end{remark}

\medskip

We now want to \emph{renormalise} the nonlinear system of equations relative to the Kerr background quantities, maintaining the equations in \emph{geometric} form.~To this end, the formalism developed in Section \ref{sec:geom_compare_tensors} becomes necessary.~Indeed, when dealing with equations for covariant tensors, one needs to take differences of $\bsy{\mathfrak{D}}_{\bsy{\mathcal{N}}(\epsilon)}$ covariant tensors and $\mathfrak{D}_{\mathcal{N}_{\text{as}}}$ covariant tensors for $|\epsilon|>0$.

\medskip

We apply the geometric procedure of Section \ref{sec:geom_compare_tensors} to each of the equations for $\bsy{\mathfrak{D}}_{\bsy{\mathcal{N}}(\epsilon)}$ covariant tensors in the system, transforming each of the terms in these equations into a $\mathfrak{D}_{\mathcal{N}_{\text{as}}}$ covariant tensor.~Proposition \ref{prop_final_procedure} is essential to treat the various terms and give to the final $\mathfrak{D}_{\mathcal{N}_{\text{as}}}$ covariant tensors a convenient form.~The new nonlinear system of null structure and Bianchi equations for $\mathfrak{D}_{\mathcal{N}_{\text{as}}}$ tensors that one obtains is dubbed \emph{reduced} system of equations and is presented in Section \ref{sec_restricted_vacuum_Einstein_equations}.

\medskip

\begin{remark}  \label{rmk_extra_terms_restricted_system}
In view of Proposition \ref{prop_final_procedure}, the reduced system of equations differs from the system that one would obtain by simply replacing the $\bsy{\mathfrak{D}}_{\bsy{\mathcal{N}}(\epsilon)}$ covariant tensors (and differential operators) in the original system with their corresponding restricted (i.e.~tilded) versions.~The formulae from Proposition \ref{prop_final_procedure} account for the \emph{extra terms} appearing in the reduced system of equations.
\end{remark}

\medskip

With the reduced system of equations at hand, one can derive the nonlinear system of equations in \emph{renormalised} form.~In the case of covariant tensors, one can indeed directly, and \emph{geometrically}, subtract the equations for the Kerr background quantities to the corresponding reduced equations, all the equations being equations for $\mathfrak{D}_{\mathcal{N}_{\text{as}}}$ covariant tensors (see Section \ref{sec_renormalised_quantities}).~The derivation of the renormalised system of equations is a straightforward (but lengthy) computation left to the reader.~Nonetheless, we will comment on some remarkable structures of the system, which can be already deduced from the reduced equations of Section \ref{sec_restricted_vacuum_Einstein_equations}.

\subsection{The reduced system of equations} \label{sec_restricted_vacuum_Einstein_equations}

In this section, we present the \emph{reduced} system of null structure and Bianchi equations.

\subsubsection{Preliminary remarks}  \label{sec_preliminary_rmks_restricted_eqns}

Some brief remarks before we present the equations:
\begin{itemize}
\item We adopt the notation $\equiv$ to denote an equality up to nonlinear terms.~Among the nonlinear terms appearing in the equations, we only drop those arising from applying the identities of Proposition \ref{prop_final_procedure} and referred to as \emph{nonlinear terms} in the identities of Proposition \ref{prop_final_procedure}.
\item In view of Remark \ref{rmk_final_structure_nonlinear_terms}, the dropped nonlinear terms in the equations contain derivatives of the restricted $\mathfrak{D}_{\mathcal{N}_{\text{as}}}$ tensor $\widetilde{\bsy{\varsigma}}(\epsilon)$ when arising from applying the identities of Proposition \ref{prop_final_procedure} to the restricted \emph{angular} covariant derivative $(\widetilde{\bsy{\nablasl}\bsy{\varsigma}})(\epsilon)$.~On the other hand, the dropped nonlinear terms arising from applying the identities of Proposition \ref{prop_final_procedure} to the restriction of the covariant derivatives $(\widetilde{\bsy{\nablasl_4}\,\bsy{\varsigma}})(\epsilon)$ and $(\widetilde{\bsy{\nablasl_3}\,\bsy{\varsigma}})(\epsilon)$ and Lie derivatives $(\widetilde{\bsy{\slashed{\mathcal{L}}_4}\,\bsy{\varsigma}})(\epsilon)$ and $(\widetilde{\bsy{\slashed{\mathcal{L}}_3}\,\bsy{\varsigma}})(\epsilon)$ do \emph{not} contain derivatives of the restricted $\mathfrak{D}_{\mathcal{N}_{\text{as}}}$ tensor $\widetilde{\bsy{\varsigma}}(\epsilon)$.~Thus, \ul{for the reduced \emph{null structure equations} with the form of \emph{transport equations}, the dropped nonlinear terms containing derivatives of (restricted) connection coefficients are \emph{only} generated by the restriction of the angular covariant derivatives on the right hand side}.~In particular, for the reduced null structure equations with the form of transport equations in the \emph{outgoing} direction, the dropped nonlinear terms containing derivatives of (restricted) connection coefficients \emph{only} contain derivatives of $\bsy{\omegah}(\epsilon)$ and $\widetilde{\bsy{\etab}}(\epsilon)$.
\item In writing the reduced system of equations, we box the \emph{extra terms}\footnote{The \emph{extra terms} are understood in the sense of Remark \ref{rmk_extra_terms_restricted_system}.} generated by the identities of Proposition \ref{prop_final_procedure}.~For example, we write the formula \eqref{restr_formula_proj_D2star} as 
\begin{align}
-2\,\widetilde{\bsy{\slashed{\mathcal{D}}_2^{\star}}\,\bsy{\varsigma}} \equiv &\, -2 \, \widetilde{\bsy{\slashed{\mathcal{D}}}_{\bsy{2}}^{\star}}\,\widetilde{\bsy{\varsigma}} \label{formula_restr_D2star_boxed} \\
&\boxed{+\frac{1}{2}\,(\bsy{\nablasl}_{\bsy{3}}\,\bsy{\varsigma}-{\bsy{\chib}}{}^{\bsy{\sharp_2}}\bsy{\cdot}\bsy{\varsigma})^{\bsy{A}}\,\widetilde{(\ef)_{\bsy{\flat}}}\,\widehat{\otimes}_{\widetilde{\bsy{\slashed{g}}}}\,\widetilde{(\ea)_{\bsy{\flat}}}}  \nonumber \\ 
& \boxed{+\frac{1}{2}\,(\bsy{\nablasl}_{\bsy{4}}\,\bsy{\varsigma}-{\bsy{\chi}}{}^{\bsy{\sharp_2}}\bsy{\cdot}\bsy{\varsigma})^{\bsy{A}}\,\widetilde{(\et)_{\bsy{\flat}}}\,\widehat{\otimes}_{\widetilde{\bsy{\slashed{g}}}}\,\widetilde{(\ea)_{\bsy{\flat}}}}  \, . \nonumber
\end{align}
Note that some of the boxed terms may be nonlinear when specialised to particular connection coefficients or curvature components, but will \emph{not} be dropped.
\item When appearing on the right hand side of the reduced null structure equations with the form of transport equations in the \emph{outgoing} direction, the boxed terms containing derivatives of connection coefficients will be re-written in a more convenient form.~For example, the identity \eqref{formula_restr_D2star_boxed} will be, in such instances, re-written as
\begin{align}
-2\,\widetilde{\bsy{\slashed{\mathcal{D}}_2^{\star}}\,\bsy{\varsigma}} \equiv &\, -2 \, \widetilde{\bsy{\slashed{\mathcal{D}}}_{\bsy{2}}^{\star}}\,\widetilde{\bsy{\varsigma}}  \label{formula_restr_D2star_boxed_bis}\\
&\boxed{+\frac{1}{2}\,(\widetilde{\bsy{\nablasl}}_{\bsy{3}}\,\widetilde{\bsy{\varsigma}}-(\widetilde{\bsy{\varsigma}},\widetilde{\bsy{\eta}})_{\widetilde{\bsy{\slashed{g}}}}\cdot\widetilde{(\et)_{\bsy{\flat}}}+(\bsy{\underline{\mathfrak{f}}},\widetilde{\bsy{\varsigma}})_{\slashed{g}}\cdot\widetilde{\bsy{\eta}}-\widetilde{\bsy{\chib}}{}^{\bsy{\sharp_2}}\bsy{\cdot}_{\widetilde{\bsy{\slashed{g}}}}\,\widetilde{\bsy{\varsigma}},\bsy{\mathfrak{\slashed{\mathfrak{f}}}}^{\bsy{A}})_{\slashed{g}}\,\widetilde{(\ef)_{\bsy{\flat}}}\,\widehat{\otimes}_{\widetilde{\bsy{\slashed{g}}}}\,\widetilde{(\ea)_{\bsy{\flat}}}} \nonumber  \\ 
& \boxed{+\frac{1}{2}\,(\widetilde{\bsy{\nablasl}}_{\bsy{4}}\,\widetilde{\bsy{\varsigma}}-(\widetilde{\bsy{\varsigma}},\widetilde{\bsy{\etab}})_{\widetilde{\bsy{\slashed{g}}}}\cdot \widetilde{(\ef)_{\bsy{\flat}}}-\widetilde{\bsy{\chi}}{}^{\bsy{\sharp_2}}\bsy{\cdot}_{\widetilde{\bsy{\slashed{g}}}}\,\widetilde{\bsy{\varsigma}},\bsy{\mathfrak{\slashed{\mathfrak{f}}}}^{\bsy{A}})_{\slashed{g}}\,\widetilde{(\et)_{\bsy{\flat}}}\,\widehat{\otimes}_{\widetilde{\bsy{\slashed{g}}}}\,\widetilde{(\ea)_{\bsy{\flat}}}}  \, , \nonumber
\end{align}
where we used Propositions \ref{prop_final_procedure} and \ref{prop_frame_commutators_ricci_1}.~Relative to the identity \eqref{formula_restr_D2star_boxed}, the identity \eqref{formula_restr_D2star_boxed_bis} is more convenient because the $\mathfrak{D}_{\mathcal{N}_{\text{as}}}$ tensor $\widetilde{\bsy{\varsigma}}(\epsilon)$ replaces the $\bsy{\mathfrak{D}}_{\bsy{\mathcal{N}}(\epsilon)}$ tensor $\bsy{\varsigma}(\epsilon)$.~This allows to apply the identities \eqref{i_2_kerr}, \eqref{i_3_kerr} and \eqref{i_4_kerr} for the connection coefficients and conclude that the reduced null structure equations with the form of transport equations in the \emph{outgoing} direction contain \emph{no} derivatives of (restricted) connection coefficients on the right hand side other than derivatives of the \emph{background} connection coefficients $\omegah_{a,M}$ and $\etab_{a,M}$.~This fact remains true when one takes into account the dropped nonlinear terms.
\item The identities \eqref{i_2_kerr}, \eqref{i_3_kerr} and \eqref{i_4_kerr} for the connection coefficients are repeatedly applied throughout the reduced system of equations and allow for numerous simplifications.
\item Each of the quantities in the reduced system of equations depends on $\epsilon$.~To keep the notation lighter, the $\epsilon$ will be dropped throughout. 
\item For future convenience, the reduced null structure equations corresponding to the equations \eqref{EVE_4_eta}, \eqref{EVE_3_etab}, \eqref{EVE_4_zeta} and \eqref{EVE_3_zeta} are presented in two alternative forms.~See the related Remark \ref{rmk_null_structure_lie_derivative}.
\end{itemize}

\subsubsection{Reduced null structure equations}

We have the first variational formulae
\begin{align}
\widetilde{\bsy{\slashed{\mathcal{L}}}}_{\bsy{4}}\,\widetilde{\bsy{\slashed{g}}}  \equiv & \, 2\,\widetilde{\bsy{\chih}}+ (\bsy{\textbf{tr} \bsy{\chi}})\,\widetilde{\bsy{\slashed{g}}} \,\boxed{+(\bsy{\etab}-\bsy{\eta})^{\bsy{A}}\left(\widetilde{(\ef)_{\bsy{\flat}}}\otimes \widetilde{(\ea)_{\bsy{\flat}}}+\widetilde{(\ea)_{\bsy{\flat}}}\otimes \widetilde{(\ef)_{\bsy{\flat}}}\right)} \\ 
&\boxed{+\frac{1}{2}\,\slashed{g}_{a,M}^{AB}\,\bsy{\slashed{g}}{}^{\bsy{CD}}\,\bsy{g}(e^{\textup{as}}_3,\bsy{e_C})\,\bsy{g}(e^{\textup{as}}_A,\bsy{e_D})\left(\widetilde{\bsy{\mathcal{L}}_{\bsy{4}}(e_4^{\textup{as}})_{\flat}}\otimes\widetilde{(e_B^{\textup{as}})_{\flat}}+\widetilde{(e_B^{\textup{as}})_{\flat}}\otimes\widetilde{\bsy{\mathcal{L}}_{\bsy{4}}(e_4^{\textup{as}})_{\flat}}\right)}   \, ,\nonumber 
\end{align}
\begin{align}
\widetilde{\bsy{\slashed{\mathcal{L}}}}_{\bsy{3}}\,\widetilde{\bsy{\slashed{g}}}   \equiv & \, 2\, \widetilde{\bsy{\chibh}}+ (\textbf{tr} \bsy{\chib})\,\widetilde{\bsy{\slashed{g}}} \,\boxed{-(\bsy{\etab}-\bsy{\eta})^{\bsy{A}}\left(\widetilde{(\et)_{\bsy{\flat}}}\otimes \widetilde{(\ea)_{\bsy{\flat}}}+\widetilde{(\ea)_{\bsy{\flat}}}\otimes \widetilde{(\et)_{\bsy{\flat}}}\right)} \\
&\boxed{+\frac{1}{2}\,\slashed{g}_{a,M}^{AB}\,\bsy{\slashed{g}}{}^{\bsy{CD}}\,\bsy{g}(e^{\textup{as}}_3,\bsy{e_C})\,\bsy{g}(e^{\textup{as}}_A,\bsy{e_D})\left(\widetilde{\bsy{\mathcal{L}}_{\bsy{3}}(e_4^{\textup{as}})_{\flat}}\otimes\widetilde{(e_B^{\textup{as}})_{\flat}}+\widetilde{(e_B^{\textup{as}})_{\flat}}\otimes\widetilde{\bsy{\mathcal{L}}_{\bsy{3}}(e_4^{\textup{as}})_{\flat}}\right)} \, , \nonumber 
\end{align} 
the second variational formulae
\begin{align}
\bsy{\widetilde{\nablasl}_4 \widetilde{\chih}}+(\bsy{\textbf{tr}\chi})\widetilde{\bsy{\chih}}-\omegah_{a,M}\bsy{ \widetilde{\chih} } \equiv & \,  -\widetilde{\bsy{\alpha }} \, ,
\end{align}
\begin{align}
\bsy{\widetilde{\nablasl}_3 \widetilde{\chibh}}+(\bsy{\textbf{tr}\chib})\widetilde{\bsy{\chibh}}-\bsy{\omegabh \widetilde{\chibh}} \equiv & \,  -2\,\widetilde{\bsy{\slashed{\mathcal{D}}}_{\bsy{2}}^{\star}}\,\widetilde{\bsy{\yb}}   + (\widetilde{\bsy{\eta}}+\etab_{a,M}-2\,\widetilde{\bsy{\zeta}})\widehat{\otimes}_{\widetilde{\bsy{\slashed{g}}}}\,\widetilde{\bsy{\yb}} -\widetilde{\bsy{\alphab}} \\
&\boxed{+\frac{1}{2}\,(\bsy{\nablasl}_{\bsy{3}}\,\bsy{\yb}-{\bsy{\chib}}{}^{\bsy{\sharp_2}}\bsy{\cdot}\bsy{\yb})^{\bsy{A}}\,\widetilde{(\ef)_{\bsy{\flat}}}\,\widehat{\otimes}_{\widetilde{\bsy{\slashed{g}}}}\,\widetilde{(\ea)_{\bsy{\flat}}}} \nonumber\\  &\boxed{+\frac{1}{2}\,(\bsy{\nablasl}_{\bsy{4}}\,\bsy{\yb}-{\bsy{\chi}}{}^{\bsy{\sharp_2}}\bsy{\cdot}\bsy{\yb})^{\bsy{A}}\,\widetilde{(\et)_{\bsy{\flat}}}\,\widehat{\otimes}_{\widetilde{\bsy{\slashed{g}}}}\,\widetilde{(\ea)_{\bsy{\flat}}}} \, , \nonumber
\end{align}
the Raychaudhuri equations
\begin{align}
\bsy{\widetilde{\nablasl}_4} (\bsy{\textbf{tr}\chi})+ \frac{1}{2}(\bsy{\textbf{tr}\chi})^2-\omegah_{a,M}(\bsy{ \textbf{tr}\chi}) \equiv & \, -|\widetilde{\bsy{\chih}}|^2_{\widetilde{\bsy{\slashed{g}}}}+\frac{1}{2}(\bsy{\slashed{\varepsilon}\cdot\chi})^2  \, ,
\end{align}
\begin{align}
\bsy{\widetilde{\nablasl}_3} (\bsy{\textbf{tr}\chib})+ \frac{1}{2}(\bsy{\textbf{tr}\chib})^2-\bsy{\omegabh}(\bsy{\textbf{tr}\chib}) \equiv & \, -|\widetilde{\bsy{\chibh}}|^2_{\widetilde{\bsy{\slashed{g}}}}+\frac{1}{2}(\bsy{\slashed{\varepsilon}\cdot\chib})^2+2\, \bsy{\widetilde{\slashed{\textbf{div}}}\, \widetilde{\yb}} +2(\widetilde{\bsy{\eta}}+\etab_{a,M}-2\,\widetilde{\bsy{\zeta}}, \widetilde{\bsy{\yb}})_{\widetilde{\bsy{\slashed{g}}}} \\
&\boxed{+(\bsy{\nablasl}_{\bsy{3}}\,\bsy{\yb}-{\bsy{\chib}}{}^{\bsy{\sharp_2}}\bsy{\cdot}\bsy{\yb})^{\bsy{A}}\,(\widetilde{(\ef)_{\bsy{\flat}}},\widetilde{(\ea)_{\bsy{\flat}}})_{\widetilde{\bsy{\slashed{g}}}}} \nonumber\\ & \boxed{+(\bsy{\nablasl}_{\bsy{4}}\,\bsy{\yb}-{\bsy{\chi}}{}^{\bsy{\sharp_2}}\bsy{\cdot}\bsy{\yb})^{\bsy{A}}(\widetilde{(\et)_{\bsy{\flat}}},\widetilde{(\ea)_{\bsy{\flat}}})_{\widetilde{\bsy{\slashed{g}}}}}  \nonumber \\
 & \boxed{+ (\bsy{\underline{\mathfrak{f}}},\widetilde{\bsy{\yb}})_{\slashed{g}}\,{\bsy{\chi}}{}^{\bsy{AB}}(\widetilde{(\bsy{e_A})_{\bsy{\flat}}},\widetilde{(\bsy{e_B})_{\bsy{\flat}}})_{\widetilde{\bsy{\slashed{g}}}}} \, , \nonumber
\end{align}
the mixed transport equations
\begin{align}
\bsy{\widetilde{\nablasl}_4 \widetilde{\chibh}}+\frac{1}{2}(\bsy{\textbf{tr}\chi})\widetilde{\bsy{\chibh}}+\omegah_{a,M}\widetilde{\bsy{\chibh}} \equiv & \, -2\bsy{\widetilde{\slashed{\mathcal{D}}_2^{\star}}} \etab_{a,M}-\frac{1}{2}(\bsy{\textbf{tr}\chib})\widetilde{\bsy{\chih}}+\frac{1}{2}(\bsy{\slashed{\varepsilon}\cdot\chib})(\widetilde{{}^{\bsy{\star}}\bsy{\chih}})-\frac{1}{2}(\bsy{\slashed{\varepsilon}\cdot\chi})(\widetilde{{}^{\bsy{\star}}\bsy{\chibh}}) \\ &+\etab_{a,M}\widehat{\otimes}_{\widetilde{\bsy{\slashed{g}}}}\,\etab_{a,M} \nonumber \\ & \boxed{+\frac{1}{2}\,(\widetilde{\bsy{\nablasl}}_{\bsy{4}}\,\etab_{a,M},\bsy{\mathfrak{\slashed{\mathfrak{f}}}}^{\bsy{A}})_{\slashed{g}}\,\widetilde{(\et)_{\bsy{\flat}}}\,\widehat{\otimes}_{\widetilde{\bsy{\slashed{g}}}}\,\widetilde{(\ea)_{\bsy{\flat}}}}  \, \boxed{+\frac{1}{2}\,(\widetilde{\bsy{\nablasl}}_{\bsy{3}}\,\etab_{a,M},\bsy{\mathfrak{\slashed{\mathfrak{f}}}}^{\bsy{A}})_{\slashed{g}}\,\widetilde{(\ef)_{\bsy{\flat}}}\,\widehat{\otimes}_{\widetilde{\bsy{\slashed{g}}}}\,\widetilde{(\ea)_{\bsy{\flat}}}} \nonumber \\ &\boxed{+\frac{1}{2}\,(-(\etab_{a,M},\widetilde{\bsy{\eta}})_{\widetilde{\bsy{\slashed{g}}}}\cdot\widetilde{(\et)_{\bsy{\flat}}}+(\bsy{\underline{\mathfrak{f}}},\etab_{a,M})_{\slashed{g}}\cdot\widetilde{\bsy{\eta}}-\widetilde{\bsy{\chib}}{}^{\bsy{\sharp_2}}\bsy{\cdot}_{\widetilde{\bsy{\slashed{g}}}}\,\etab_{a,M},\bsy{\mathfrak{\slashed{\mathfrak{f}}}}^{\bsy{A}})_{\slashed{g}}\,\widetilde{(\ef)_{\bsy{\flat}}}\,\widehat{\otimes}_{\widetilde{\bsy{\slashed{g}}}}\,\widetilde{(\ea)_{\bsy{\flat}}}} \nonumber\\ &\boxed{+\frac{1}{2}\,(-(\etab_{a,M},\etab_{a,M})_{\widetilde{\bsy{\slashed{g}}}}\cdot \widetilde{(\ef)_{\bsy{\flat}}}-\widetilde{\bsy{\chi}}{}^{\bsy{\sharp_2}}\bsy{\cdot}_{\widetilde{\bsy{\slashed{g}}}}\,\etab_{a,M},\bsy{\mathfrak{\slashed{\mathfrak{f}}}}^{\bsy{A}})_{\slashed{g}}\,\widetilde{(\et)_{\bsy{\flat}}}\,\widehat{\otimes}_{\widetilde{\bsy{\slashed{g}}}}\,\widetilde{(\ea)_{\bsy{\flat}}}}  \, , \nonumber 
\end{align}
\begin{align}
\bsy{\widetilde{\nablasl}_3 \widetilde{\chih}}+\frac{1}{2}(\bsy{\textbf{tr}\chib})\widetilde{\bsy{\chih}}+\bsy{\omegabh}\,\widetilde{\bsy{\chih}} \equiv & \, -2\bsy{\widetilde{\slashed{\mathcal{D}}_2^{\star}} \,\widetilde{\eta}}-\frac{1}{2}(\bsy{\textbf{tr}\chi})\widetilde{\bsy{\chibh}}+\frac{1}{2}(\bsy{\slashed{\varepsilon}\cdot\chi})(\widetilde{{}^{\bsy{\star}}\bsy{\chibh}})-\frac{1}{2}(\bsy{\slashed{\varepsilon}\cdot\chib})(\widetilde{{}^{\bsy{\star}}\bsy{\chih}}) \\ &+\widetilde{\bsy{\eta}}\widehat{\otimes}_{\widetilde{\bsy{\slashed{g}}}}\,\widetilde{\bsy{\eta}} \,\boxed{+\frac{1}{2}\,(\bsy{\nablasl}_{\bsy{3}}\,\bsy{\eta}-{\bsy{\chib}}{}^{\bsy{\sharp_2}}\bsy{\cdot}\bsy{\eta})^{\bsy{A}}\,\widetilde{(\ef)_{\bsy{\flat}}}\,\widehat{\otimes}_{\widetilde{\bsy{\slashed{g}}}}\,\widetilde{(\ea)_{\bsy{\flat}}}} \nonumber\\  &\boxed{+\frac{1}{2}\,(\bsy{\nablasl}_{\bsy{4}}\,\bsy{\eta}-{\bsy{\chi}}{}^{\bsy{\sharp_2}}\bsy{\cdot}\bsy{\eta})^{\bsy{A}}\,\widetilde{(\et)_{\bsy{\flat}}}\,\widehat{\otimes}_{\widetilde{\bsy{\slashed{g}}}}\,\widetilde{(\ea)_{\bsy{\flat}}}} \, , \nonumber
\end{align}
\begin{align}
\bsy{\widetilde{\nablasl}_4}( \bsy{\textbf{tr}\chib})+ \frac{1}{2}(\bsy{\textbf{tr}\chi})(\bsy{\textbf{tr}\chib})+\omegah_{a,M}(\bsy{\textbf{tr}\chib}) \equiv & \, -(\widetilde{\bsy{\chih}},\widetilde{\bsy{\chibh}})_{\widetilde{\bsy{\slashed{g}}}}+\frac{1}{2}(\bsy{\slashed{\varepsilon}\cdot\chi})(\bsy{\slashed{\varepsilon}\cdot\chib}) +2(\etab_{a,M} ,\etab_{a,M})_{\widetilde{\bsy{\slashed{g}}}}+2\bsy{\rho} \\ &+2\, \widetilde{\bsy{\slashed{\textbf{div}}}}\, \etab_{a,M} \nonumber \\ & \boxed{+(\widetilde{\bsy{\nablasl}}_{\bsy{4}}\,\etab_{a,M},\bsy{\mathfrak{\slashed{\mathfrak{f}}}}^{\bsy{A}})_{\slashed{g}}\,(\widetilde{(\et)_{\bsy{\flat}}},\widetilde{(\ea)_{\bsy{\flat}}})_{\widetilde{\bsy{\slashed{g}}}}} \, \boxed{+(\widetilde{\bsy{\nablasl}}_{\bsy{3}}\,\etab_{a,M},\bsy{\mathfrak{\slashed{\mathfrak{f}}}}^{\bsy{A}})_{\slashed{g}}\,(\widetilde{(\ef)_{\bsy{\flat}}},\widetilde{(\ea)_{\bsy{\flat}}})_{\widetilde{\bsy{\slashed{g}}}}} \nonumber\\ &\boxed{+(-(\etab_{a,M},\widetilde{\bsy{\eta}})_{\widetilde{\bsy{\slashed{g}}}}\cdot\widetilde{(\et)_{\bsy{\flat}}}+(\bsy{\underline{\mathfrak{f}}},\etab_{a,M})_{\slashed{g}}\cdot\widetilde{\bsy{\eta}}-\widetilde{\bsy{\chib}}{}^{\bsy{\sharp_2}}\bsy{\cdot}_{\widetilde{\bsy{\slashed{g}}}}\,\etab_{a,M},\bsy{\mathfrak{\slashed{\mathfrak{f}}}}^{\bsy{A}})_{\slashed{g}}\,(\widetilde{(\ef)_{\bsy{\flat}}},\widetilde{(\ea)_{\bsy{\flat}}})_{\widetilde{\bsy{\slashed{g}}}}} \nonumber\\  & \boxed{+(-(\etab_{a,M},\etab_{a,M})_{\widetilde{\bsy{\slashed{g}}}}\cdot \widetilde{(\ef)_{\bsy{\flat}}}-\widetilde{\bsy{\chi}}{}^{\bsy{\sharp_2}}\bsy{\cdot}_{\widetilde{\bsy{\slashed{g}}}}\,\etab_{a,M},\bsy{\mathfrak{\slashed{\mathfrak{f}}}}^{\bsy{A}})_{\slashed{g}}\,(\widetilde{(\et)_{\bsy{\flat}}},\widetilde{(\ea)_{\bsy{\flat}}})_{\widetilde{\bsy{\slashed{g}}}}}  \nonumber \\
 & \boxed{+ (\bsy{\underline{\mathfrak{f}}},\etab_{a,M})_{\slashed{g}}\,{\bsy{\chi}}{}^{\bsy{AB}}(\widetilde{(\bsy{e_A})_{\bsy{\flat}}},\widetilde{(\bsy{e_B})_{\bsy{\flat}}})_{\widetilde{\bsy{\slashed{g}}}}}  \, , \nonumber 
\end{align}
\begin{align}
\bsy{\widetilde{\nablasl}_3 }(\bsy{\textbf{tr}\chi})+ \frac{1}{2}(\bsy{\textbf{tr}\chib})(\bsy{\textbf{tr}\chi})+\bsy{\omegabh}(\bsy{\textbf{tr}\chi}) \equiv & \, -(\widetilde{\bsy{\chih}},\widetilde{\bsy{\chibh}})_{\widetilde{\bsy{\slashed{g}}}}+\frac{1}{2}(\bsy{\slashed{\varepsilon}\cdot\chib})(\bsy{\slashed{\varepsilon}\cdot\chi}) +2(\widetilde{\bsy{\eta}},\widetilde{\bsy{\eta}})_{\widetilde{\bsy{\slashed{g}}}}+2\bsy{\rho} \\ & +2\, \widetilde{\bsy{\slashed{\textbf{div}}}}\,\widetilde{\bsy{ \eta}} \, \boxed{+(\bsy{\nablasl}_{\bsy{3}}\,\bsy{\eta}-{\bsy{\chib}}{}^{\bsy{\sharp_2}}\bsy{\cdot}\bsy{\eta})^{\bsy{A}}\,(\widetilde{(\ef)_{\bsy{\flat}}},\widetilde{(\ea)_{\bsy{\flat}}})_{\widetilde{\bsy{\slashed{g}}}}} \nonumber\\ & \boxed{+(\bsy{\nablasl}_{\bsy{4}}\,\bsy{\eta}-{\bsy{\chi}}{}^{\bsy{\sharp_2}}\bsy{\cdot}\bsy{\eta})^{\bsy{A}}(\widetilde{(\et)_{\bsy{\flat}}},\widetilde{(\ea)_{\bsy{\flat}}})_{\widetilde{\bsy{\slashed{g}}}}}  \nonumber \\
 & \boxed{+ (\bsy{\underline{\mathfrak{f}}},\widetilde{\bsy{\eta}})_{\slashed{g}}\,{\bsy{\chi}}{}^{\bsy{AB}}(\widetilde{(\bsy{e_A})_{\bsy{\flat}}},\widetilde{(\bsy{e_B})_{\bsy{\flat}}})_{\widetilde{\bsy{\slashed{g}}}}} \, , \nonumber
\end{align} 
the transport equations for the antitraces of the second fundamental forms
\begin{align}
\bsy{\widetilde{\nablasl}_4} (\bsy{\slashed{\varepsilon}\cdot\chi})+( \bsy{\textbf{tr}\chi})(\bsy{\slashed{\varepsilon}\cdot\chi})-\omegah_{a,M} (\bsy{\slashed{\varepsilon}\cdot\chi}) \equiv & \, 0 \, , 
\end{align}
\begin{align}
\bsy{\widetilde{\nablasl}_3} (\bsy{\slashed{\varepsilon}\cdot\chib})+( \bsy{\textbf{tr}\chib})(\bsy{\slashed{\varepsilon}\cdot\chib})-\bsy{\omegabh} (\bsy{\slashed{\varepsilon}\cdot\chib}) \equiv & \,  2(\widetilde{\bsy{\eta}}-2\,\widetilde{\bsy{\zeta}})\bsy{\wedge}_{\widetilde{\bsy{\slashed{g}}}}\,\widetilde{\bsy{\yb}} +2\,\widetilde{\bsy{\yb}}\bsy{\wedge}_{\widetilde{\bsy{\slashed{g}}}}\,\etab_{a,M}+2\, \bsy{\widetilde{\slashed{\textbf{curl}}} \, \widetilde{\yb}} \\
& \boxed{+(\bsy{\nablasl}_{\bsy{3}}\,\bsy{\yb}-{\bsy{\chib}}{}^{\bsy{\sharp_2}}\bsy{\cdot}\bsy{\yb})^{\bsy{A}}\,\widetilde{(\ef)_{\bsy{\flat}}}\,\bsy{\wedge}_{\widetilde{\bsy{\slashed{g}}}}\,\widetilde{(\ea)_{\bsy{\flat}}}} \nonumber\\ & \boxed{+(\bsy{\nablasl}_{\bsy{4}}\,\bsy{\yb}-{\bsy{\chi}}{}^{\bsy{\sharp_2}}\bsy{\cdot}\bsy{\yb})^{\bsy{A}} \widetilde{(\et)_{\bsy{\flat}}}\,\bsy{\wedge}_{\widetilde{\bsy{\slashed{g}}}}\,\widetilde{(\ea)_{\bsy{\flat}}}}  \nonumber \\
 &\boxed{+ (\bsy{\underline{\mathfrak{f}}},\widetilde{\bsy{\yb}})_{\slashed{g}}\,{\bsy{\chi}}{}^{\bsy{AB}}\widetilde{(\bsy{e_A})_{\bsy{\flat}}}\,\bsy{\wedge}_{\widetilde{\bsy{\slashed{g}}}}\,\widetilde{(\bsy{e_B})_{\bsy{\flat}}}} \, , \nonumber
\end{align}
\begin{align}
\bsy{\widetilde{\nablasl}_4} (\bsy{\slashed{\varepsilon}\cdot\chib})+\frac{1}{2}(\bsy{\textbf{tr}\chi})(\bsy{\slashed{\varepsilon}\cdot\chib})+\omegah_{a,M}(\bsy{\slashed{\varepsilon}\cdot\chib}) \equiv & \, -\widetilde{\bsy{\chih}}\bsy{\wedge}_{\widetilde{\bsy{\slashed{g}}}}\,\widetilde{\bsy{\chibh}} -\frac{1}{2}(\bsy{\textbf{tr}\chib})(\bsy{\slashed{\varepsilon}\cdot\chi})  +2\bsy{\sigma}+2\,\bsy{\widetilde{\slashed{\textbf{curl}}}}\, \etab_{a,M}   \\
& \boxed{+(\widetilde{\bsy{\nablasl}}_{\bsy{4}}\,\etab_{a,M},\bsy{\mathfrak{\slashed{\mathfrak{f}}}}^{\bsy{A}})_{\slashed{g}}\, \widetilde{(\et)_{\bsy{\flat}}}\,\bsy{\wedge}_{\widetilde{\bsy{\slashed{g}}}}\,\widetilde{(\ea)_{\bsy{\flat}}}} \, \boxed{+(\widetilde{\bsy{\nablasl}}_{\bsy{3}}\,\etab_{a,M},\bsy{\mathfrak{\slashed{\mathfrak{f}}}}^{\bsy{A}})_{\slashed{g}}\,\widetilde{(\ef)_{\bsy{\flat}}}\,\bsy{\wedge}_{\widetilde{\bsy{\slashed{g}}}}\,\widetilde{(\ea)_{\bsy{\flat}}}} \nonumber\\ & \boxed{+(-(\etab_{a,M},\widetilde{\bsy{\eta}})_{\widetilde{\bsy{\slashed{g}}}}\cdot\widetilde{(\et)_{\bsy{\flat}}}+(\bsy{\underline{\mathfrak{f}}},\etab_{a,M})_{\slashed{g}}\cdot\widetilde{\bsy{\eta}}-\widetilde{\bsy{\chib}}{}^{\bsy{\sharp_2}}\bsy{\cdot}_{\widetilde{\bsy{\slashed{g}}}}\,\etab_{a,M},\bsy{\mathfrak{\slashed{\mathfrak{f}}}}^{\bsy{A}})_{\slashed{g}}\,\widetilde{(\ef)_{\bsy{\flat}}}\,\bsy{\wedge}_{\widetilde{\bsy{\slashed{g}}}}\,\widetilde{(\ea)_{\bsy{\flat}}}} \nonumber\\  & \boxed{+(-(\etab_{a,M},\etab_{a,M})_{\widetilde{\bsy{\slashed{g}}}}\cdot \widetilde{(\ef)_{\bsy{\flat}}}-\widetilde{\bsy{\chi}}{}^{\bsy{\sharp_2}}\bsy{\cdot}_{\widetilde{\bsy{\slashed{g}}}}\,\etab_{a,M},\bsy{\mathfrak{\slashed{\mathfrak{f}}}}^{\bsy{A}})_{\slashed{g}}\, \widetilde{(\et)_{\bsy{\flat}}}\,\bsy{\wedge}_{\widetilde{\bsy{\slashed{g}}}}\,\widetilde{(\ea)_{\bsy{\flat}}}}  \nonumber \\
 &\boxed{+ (\bsy{\underline{\mathfrak{f}}},\etab_{a,M})_{\slashed{g}}\,{\bsy{\chi}}{}^{\bsy{AB}}\widetilde{(\bsy{e_A})_{\bsy{\flat}}}\,\bsy{\wedge}_{\widetilde{\bsy{\slashed{g}}}}\,\widetilde{(\bsy{e_B})_{\bsy{\flat}}}} \, , \nonumber 
\end{align}
\begin{align}
\bsy{\widetilde{\nablasl}_3} (\bsy{\slashed{\varepsilon}\cdot\chi})+\frac{1}{2}(\bsy{\textbf{tr}\chib})(\bsy{\slashed{\varepsilon}\cdot\chi})+\bsy{\omegabh}(\bsy{\slashed{\varepsilon}\cdot\chi}) \equiv & \, -\bsy{\widetilde{\chibh}}\bsy{\wedge}_{\widetilde{\bsy{\slashed{g}}}}\,\widetilde{\bsy{\chih}}-\frac{1}{2}(\bsy{\textbf{tr}\chi})(\bsy{\slashed{\varepsilon}\cdot\chib})  -2\bsy{\sigma}+2\,\bsy{\widetilde{\slashed{\textbf{curl}}}\, \widetilde{\eta} }    \\
& \boxed{+(\bsy{\nablasl}_{\bsy{3}}\,\bsy{\eta}-{\bsy{\chib}}{}^{\bsy{\sharp_2}}\bsy{\cdot}\bsy{\eta})^{\bsy{A}}\,\widetilde{(\ef)_{\bsy{\flat}}}\,\bsy{\wedge}_{\widetilde{\bsy{\slashed{g}}}}\,\widetilde{(\ea)_{\bsy{\flat}}}} \nonumber\\ & \boxed{+(\bsy{\nablasl}_{\bsy{4}}\,\bsy{\eta}-{\bsy{\chi}}{}^{\bsy{\sharp_2}}\bsy{\cdot}\bsy{\eta})^{\bsy{A}} \widetilde{(\et)_{\bsy{\flat}}}\,\bsy{\wedge}_{\widetilde{\bsy{\slashed{g}}}}\,\widetilde{(\ea)_{\bsy{\flat}}}}  \nonumber \\
 &\boxed{+ (\bsy{\underline{\mathfrak{f}}},\widetilde{\bsy{\eta}})_{\slashed{g}}\,{\bsy{\chi}}{}^{\bsy{AB}} \widetilde{(\bsy{e_A})_{\bsy{\flat}}}\,\bsy{\wedge}_{\widetilde{\bsy{\slashed{g}}}}\,\widetilde{(\bsy{e_B})_{\bsy{\flat}}}}  \nonumber
\end{align}
and the transport equations
\begin{align}
\widetilde{\bsy{\nablasl}}_{\bsy{4}}\,\widetilde{\bsy{\eta}} \equiv & \,  -\bsy{\widetilde{\chih}{}^{\sharp}\cdot_{\widetilde{\bsy{\slashed{g}}}}}\,(\widetilde{\bsy{\eta}}-\etab_{a,M}) -\frac{1}{2}(\textbf{tr}\bsy{\chi})(\widetilde{\bsy{\eta}}-\etab_{a,M})+\frac{1}{2}(\bsy{\slashed{\varepsilon}\cdot\chi})(\widetilde{{}^{\star}\bsy{\eta}}-\widetilde{{}^{\star}\bsy{\etab}}) - \widetilde{\bsy{\beta}} \\
&\boxed{+(\widetilde{\bsy{\eta}},\etab_{a,M})_{\widetilde{\bsy{\slashed{g}}}}\cdot \widetilde{(\ef)_{\bsy{\flat}}}}  \, , \nonumber 
\end{align}
\begin{align}
\widetilde{\bsy{\slashed{\mathcal{L}}}}_{\bsy{4}}\,\widetilde{\bsy{\eta}} -(\bsy{\slashed{\varepsilon}\cdot\chi})(\widetilde{{}^{\star}\bsy{\eta}}) =&\, \bsy{\widetilde{\chih}{}^{\sharp}\cdot_{\widetilde{\bsy{\slashed{g}}}}}\,\etab_{a,M} +\frac{1}{2}(\textbf{tr}\bsy{\chi})\,\etab_{a,M}-\frac{1}{2}(\bsy{\slashed{\varepsilon}\cdot\chi})\,\widetilde{{}^{\star}\bsy{\etab}}  - \widetilde{\bsy{\beta}} \\
&\boxed{+(\bsy{\etab}-\bsy{\eta})^{\bsy{A}}\bsy{\eta}_{\bsy{A}}\,\widetilde{(\ef)_{\bsy{\flat}}}} \, \boxed{ +\frac{1}{2}\,\slashed{\bsy{g}}^{\bsy{AB}}\bsy{g}(e_3^{\text{as}},\ea)\,\bsy{\eta}_{\bsy{B}}\,\widetilde{\bsy{\mathcal{L}}_{\bsy{4}}(e_4^{\text{as}})_{\flat}}} \, , \nonumber
\end{align}
\begin{align}
\widetilde{\bsy{\nablasl}}_{\bsy{3}}\,\etab_{a,M} \equiv & \, \widetilde{\bsy{\nablasl}}_{\bsy{4}}\,\widetilde{\bsy{\yb}} +\bsy{\widetilde{\chibh}{}^{\sharp}\cdot_{\widetilde{\bsy{\slashed{g}}}}}\,(\widetilde{\bsy{\eta}}-\etab_{a,M}) +\frac{1}{2}(\textbf{tr}\bsy{\chib})(\widetilde{\bsy{\eta}}-\etab_{a,M})-\frac{1}{2}(\bsy{\slashed{\varepsilon}\cdot\chib})(\widetilde{{}^{\star}\bsy{\eta}}-\widetilde{{}^{\star}\bsy{\etab}})+2\,\omegah_{a,M}\, \widetilde{\bsy{\yb}}  + \widetilde{\bsy{\betab}} \\
&\boxed{+(\etab_{a,M},\widetilde{\bsy{\eta}})_{\widetilde{\bsy{\slashed{g}}}}\cdot\widetilde{(\et)_{\bsy{\flat}}}} \, \boxed{-(\widetilde{\bsy{\yb}},\etab_{a,M})_{\widetilde{\bsy{\slashed{g}}}}\cdot \widetilde{(\ef)_{\bsy{\flat}}}} \, \boxed{ -(\bsy{\underline{\mathfrak{f}}},\etab_{a,M})_{\slashed{g}} \widetilde{\bsy{\eta}}} \, , \nonumber
\end{align}
\begin{align}
\widetilde{\bsy{\slashed{\mathcal{L}}}}_{\bsy{3}}\,\etab_{a,M}-(\bsy{\slashed{\varepsilon}\cdot\chib})(\widetilde{{}^{\star}\bsy{\etab}}) =& \,\widetilde{\bsy{\nablasl}}_{\bsy{4}}\,\widetilde{\bsy{\yb}} +\bsy{\widetilde{\chibh}{}^{\sharp}\cdot_{\widetilde{\bsy{\slashed{g}}}}}\,\widetilde{\bsy{\eta}} +\frac{1}{2}(\textbf{tr}\bsy{\chib})\,\widetilde{\bsy{\eta}}-\frac{1}{2}(\bsy{\slashed{\varepsilon}\cdot\chib})(\widetilde{{}^{\star}\bsy{\eta}})+2\,\omegah_{a,M}\, \widetilde{\bsy{\yb}}  + \widetilde{\bsy{\betab}} \\
&\boxed{+(\bsy{\eta}-\bsy{\etab})^{\bsy{A}}\bsy{\etab}_{\bsy{A}}\,\widetilde{(\et)_{\bsy{\flat}}}} \, \boxed{ +\frac{1}{2}\,\slashed{\bsy{g}}^{\bsy{AB}}\bsy{g}(e_3^{\text{as}},\ea)\,\bsy{\etab}_{\bsy{B}}\,\widetilde{\bsy{\mathcal{L}}_{\bsy{3}}(e_4^{\text{as}})_{\flat}}} \, \boxed{-(\widetilde{\bsy{\yb}},\etab_{a,M})_{\widetilde{\bsy{\slashed{g}}}}\cdot \widetilde{(\ef)_{\bsy{\flat}}}} \, , \nonumber
\end{align}
\begin{align}
\bsy{ \widetilde{\nablasl}_4 \omegabh}+\bsy{\widetilde{\nablasl}_3} \omegah_{a,M} \equiv 2(\widetilde{\bsy{\eta}},\etab_{a,M})_{\widetilde{\bsy{\slashed{g}}}} -2\,\omegah_{a,M}\bsy{\omegabh}   -2(\widetilde{\bsy{\eta}}-\etab_{a,M},\widetilde{\bsy{\zeta}})_{\widetilde{\bsy{\slashed{g}}}}-2\bsy{\rho}  
\end{align}
and the transport equations for the torsion
\begin{align}
\widetilde{\bsy{\nablasl}}_{\bsy{4}}\,\widetilde{\bsy{\zeta}} \equiv & \, -\widetilde{\bsy{\nablasl}}\omegah_{a,M} +\bsy{\widetilde{\chih}{}^{\sharp}\cdot_{\widetilde{\bsy{\slashed{g}}}}}\,(\etab_{a,M}-\widetilde{\bsy{\zeta}})+\frac{1}{2}(\textbf{tr}\bsy{\chi})(\etab_{a,M}-\widetilde{\bsy{\zeta}})+\frac{1}{2}(\bsy{\slashed{\varepsilon}\cdot\chi})(\widetilde{{}^{\star}\bsy{\etab}}-\widetilde{{}^{\star}\bsy{\zeta}})  -\omegah_{a,M}(\etab_{a,M}+\widetilde{\bsy{\zeta}}) - \widetilde{\bsy{\beta}}  \\
&\boxed{+(\widetilde{\bsy{\zeta}},\etab_{a,M})_{\widetilde{\bsy{\slashed{g}}}}\cdot \widetilde{(\ef)_{\bsy{\flat}}}} \, \boxed{ -\frac{1}{2}\,(\widetilde{\bsy{\nablasl}}_{\bsy{3}}\omegah_{a,M})\, \widetilde{(\ef)_{\bsy{\flat}}}} \, \boxed{-\frac{1}{2}\,(\widetilde{\bsy{\nablasl}}_{\bsy{4}}\omegah_{a,M}) \,\widetilde{(\et)_{\bsy{\flat}}}}   \, ,  \nonumber 
\end{align}
\begin{align}
\widetilde{\bsy{\slashed{\mathcal{L}}}}_{\bsy{4}}\,\widetilde{\bsy{\zeta}} \equiv & \, -\widetilde{\bsy{\nablasl}}\omegah_{a,M} +\bsy{\widetilde{\chih}{}^{\sharp}\cdot_{\widetilde{\bsy{\slashed{g}}}}}\,\etab_{a,M}+\frac{1}{2}(\textbf{tr}\bsy{\chi})\,\etab_{a,M}+\frac{1}{2}(\bsy{\slashed{\varepsilon}\cdot\chi})(\widetilde{{}^{\star}\bsy{\etab}})  -\omegah_{a,M}(\etab_{a,M}+\widetilde{\bsy{\zeta}}) - \widetilde{\bsy{\beta}}  \\
& \boxed{+(\bsy{\etab}-\bsy{\eta})^{\bsy{A}}\bsy{\zeta}_{\bsy{A}}\,\widetilde{(\ef)_{\bsy{\flat}}}} \, \boxed{ +\frac{1}{2}\,\slashed{\bsy{g}}^{\bsy{AB}}\bsy{g}(e_3^{\text{as}},\ea)\,\bsy{\zeta}_{\bsy{B}}\,\widetilde{\bsy{\mathcal{L}}_{\bsy{4}}(e_4^{\text{as}})_{\flat}}}\, \boxed{ -\frac{1}{2}\,(\widetilde{\bsy{\nablasl}}_{\bsy{3}}\omegah_{a,M})\, \widetilde{(\ef)_{\bsy{\flat}}}} \, \boxed{-\frac{1}{2}\,(\widetilde{\bsy{\nablasl}}_{\bsy{4}}\omegah_{a,M}) \,\widetilde{(\et)_{\bsy{\flat}}}}   \, ,  \nonumber 
\end{align}
\begin{align}
\widetilde{\bsy{\nablasl}}_{\bsy{3}}\,\widetilde{\bsy{\zeta}}\equiv & \, \widetilde{\bsy{\nablasl}}\bsy{\omegabh} +\bsy{\widetilde{\chih}{}^{\sharp}\cdot_{\widetilde{\bsy{\slashed{g}}}}}\,\widetilde{\bsy{\yb}}+\frac{1}{2}(\textbf{tr}\bsy{\chi})\widetilde{\bsy{\yb}}+\frac{1}{2}(\bsy{\slashed{\varepsilon}\cdot\chi})(\widetilde{{}^{\star}\bsy{\yb}})  -\omegah_{a,M}\widetilde{\bsy{\yb}}+\bsy{\omegabh}(\widetilde{\bsy{\eta}}-\widetilde{\bsy{\zeta}}) - \widetilde{\bsy{\betab}}  \\
&-\bsy{\widetilde{\chibh}{}^{\sharp}\cdot_{\widetilde{\bsy{\slashed{g}}}}}\,(\widetilde{\bsy{\eta}}+\widetilde{\bsy{\zeta}})-\frac{1}{2}(\textbf{tr}\bsy{\chib})(\widetilde{\bsy{\eta}}+\widetilde{\bsy{\zeta}})-\frac{1}{2}(\bsy{\slashed{\varepsilon}\cdot\chib})(\widetilde{{}^{\star}\bsy{\eta}}+\widetilde{{}^{\star}\bsy{\zeta}}) \nonumber \\
& \boxed{+(\widetilde{\bsy{\zeta}},\widetilde{\bsy{\eta}})_{\widetilde{\bsy{\slashed{g}}}}\cdot\widetilde{(\et)_{\bsy{\flat}}}} \, \boxed{-(\bsy{\underline{\mathfrak{f}}},\widetilde{\bsy{\zeta}})_{\slashed{g}} \widetilde{\bsy{\eta}}} \, \boxed{ +\frac{1}{2}\,(\widetilde{\bsy{\nablasl}}_{\bsy{3}}\,\bsy{\omegabh})\, \widetilde{(\ef)_{\bsy{\flat}}}} \, \boxed{+\frac{1}{2}\,(\widetilde{\bsy{\nablasl}}_{\bsy{4}}\,\bsy{\omegabh}) \,\widetilde{(\et)_{\bsy{\flat}}}}  \, ,  \nonumber 
\end{align}
\begin{align}
\widetilde{\bsy{\slashed{\mathcal{L}}}}_{\bsy{3}}\,\widetilde{\bsy{\zeta}}\equiv & \, \widetilde{\bsy{\nablasl}}\bsy{\omegabh} +\bsy{\widetilde{\chih}{}^{\sharp}\cdot_{\widetilde{\bsy{\slashed{g}}}}}\,\widetilde{\bsy{\yb}}+\frac{1}{2}(\textbf{tr}\bsy{\chi})\widetilde{\bsy{\yb}}+\frac{1}{2}(\bsy{\slashed{\varepsilon}\cdot\chi})(\widetilde{{}^{\star}\bsy{\yb}})  -\omegah_{a,M}\widetilde{\bsy{\yb}}+\bsy{\omegabh}(\widetilde{\bsy{\eta}}-\widetilde{\bsy{\zeta}}) - \widetilde{\bsy{\betab}}  \\
&-\bsy{\widetilde{\chibh}{}^{\sharp}\cdot_{\widetilde{\bsy{\slashed{g}}}}}\,\widetilde{\bsy{\eta}}-\frac{1}{2}(\textbf{tr}\bsy{\chib})\,\widetilde{\bsy{\eta}}-\frac{1}{2}(\bsy{\slashed{\varepsilon}\cdot\chib})\,(\widetilde{{}^{\star}\bsy{\eta}}) \nonumber \\
& \boxed{+(\bsy{\eta}-\bsy{\etab})^{\bsy{A}}\bsy{\zeta}_{\bsy{A}}\,\widetilde{(\et)_{\bsy{\flat}}}} \, \boxed{ +\frac{1}{2}\,\slashed{\bsy{g}}^{\bsy{AB}}\bsy{g}(e_3^{\text{as}},\ea)\,\bsy{\zeta}_{\bsy{B}}\,\widetilde{\bsy{\mathcal{L}}_{\bsy{3}}(e_4^{\text{as}})_{\flat}}} \, \boxed{ +\frac{1}{2}\,(\widetilde{\bsy{\nablasl}}_{\bsy{3}}\,\bsy{\omegabh})\, \widetilde{(\ef)_{\bsy{\flat}}}} \, \boxed{+\frac{1}{2}\,(\widetilde{\bsy{\nablasl}}_{\bsy{4}}\,\bsy{\omegabh}) \,\widetilde{(\et)_{\bsy{\flat}}}}   \nonumber 
\end{align}
and the elliptic equation for the torsion
\begin{align}
\widetilde{\slashed{\textbf{curl}}}\,\widetilde{\bsy{\zeta}}  \equiv & \,-\frac{1}{2}\,\widetilde{\bsy{\chih}}\,\bsy{\wedge}_{\widetilde{\bsy{\slashed{g}}}}\,\widetilde{\bsy{\chibh}}+\frac{1}{4}\,(\bsy{\textbf{tr}\chi})(\bsy{\slashed{\varepsilon}\cdot\chib})-\frac{1}{4}\,(\bsy{\textbf{tr}\chib})(\bsy{\slashed{\varepsilon}\cdot\chi})-\frac{1}{2}\,(\bsy{\slashed{\varepsilon}\cdot\chib})\,\omegah_{a,M}+\frac{1}{2}\,(\bsy{\slashed{\varepsilon}\cdot\chi})\,\bsy{\omegabh}+\bsy{\sigma} \\
&\boxed{-\frac{1}{2}\,(\bsy{\nablasl}_{\bsy{3}}\,\bsy{\zeta}-{\bsy{\chib}}{}^{\bsy{\sharp_2}}\bsy{\cdot}\bsy{\zeta})^{\bsy{A}}\,\widetilde{(\ef)_{\bsy{\flat}}}\,\bsy{\wedge}_{\widetilde{\bsy{\slashed{g}}}}\,\widetilde{(\ea)_{\bsy{\flat}}}} \, \boxed{-\frac{1}{2}\,(\bsy{\nablasl}_{\bsy{4}}\,\bsy{\zeta}-{\bsy{\chi}}{}^{\bsy{\sharp_2}}\bsy{\cdot}\bsy{\zeta})^{\bsy{A}} \widetilde{(\et)_{\bsy{\flat}}}\,\bsy{\wedge}_{\widetilde{\bsy{\slashed{g}}}}\,\widetilde{(\ea)_{\bsy{\flat}}} } \nonumber \\
 & \boxed{- \frac{1}{2}(\bsy{\underline{\mathfrak{f}}},\widetilde{\bsy{\zeta}})_{\slashed{g}}\,{\bsy{\chi}}{}^{\bsy{AB}}\widetilde{(\bsy{e_A})_{\bsy{\flat}}}\,\bsy{\wedge}_{\widetilde{\bsy{\slashed{g}}}}\,\widetilde{(\bsy{e_B})_{\bsy{\flat}}} } \, . \nonumber
\end{align} 
We have the Codazzi equations
\begin{align}
\widetilde{\slashed{\textbf{div}}}\,\widetilde{\bsy{\chih}} \equiv & \, -\bsy{\widetilde{\chih}{}^{\sharp}\cdot_{\widetilde{\bsy{\slashed{g}}}}\,\widetilde{\zeta}}-\frac{1}{2}(\bsy{\slashed{\varepsilon}\cdot\chi})(\bsy{\widetilde{{}^{\star}\zeta}})+\frac{1}{2}(\bsy{\textbf{tr}\chi})\widetilde{\bsy{\zeta}}  -(\bsy{\slashed{\varepsilon}\cdot\chi})(\widetilde{\bsy{{}^{\star}\eta}}) -\widetilde{\bsy{\beta}}   \\
& +\frac{1}{2}\widetilde{\bsy{\nablasl}}( \bsy{\textbf{tr}\chi}) \,\boxed{+\frac{1}{4}\,(\widetilde{\bsy{\nablasl}}_{\bsy{3}}( \bsy{\textbf{tr}\chi}))\, \widetilde{(\ef)_{\bsy{\flat}}}} \, \boxed{+\frac{1}{4}\,(\widetilde{\bsy{\nablasl}}_{\bsy{4}}( \bsy{\textbf{tr}\chi})) \,\widetilde{(\et)_{\bsy{\flat}}}}  \nonumber\\
&-\frac{1}{2}\bsy{\widetilde{\slashed{\varepsilon}}{}^{\sharp_2}\cdot_{\widetilde{\bsy{\slashed{g}}}}}\,(\widetilde{\bsy{\nablasl}}(\bsy{\slashed{\varepsilon}\cdot\chi})) \, \boxed{ -\frac{1}{4}\,(\widetilde{\bsy{\nablasl}}_{\bsy{3}}(\bsy{\slashed{\varepsilon}\cdot\chi}))\, \widetilde{\bsy{\slashed{\varepsilon}}}{}^{\bsy{\sharp_2}}\bsy{\cdot}_{\widetilde{\bsy{\slashed{g}}}} \, \widetilde{(\ef)_{\bsy{\flat}}}} \, \boxed{-\frac{1}{4}\,(\widetilde{\bsy{\nablasl}}_{\bsy{4}}(\bsy{\slashed{\varepsilon}\cdot\chi})) \,\widetilde{\bsy{\slashed{\varepsilon}}}{}^{\bsy{\sharp_2}}\bsy{\cdot}_{\widetilde{\bsy{\slashed{g}}}}\,\widetilde{(\et)_{\bsy{\flat}}}} \, , \nonumber 
\end{align}
\begin{align}
\widetilde{\slashed{\textbf{div}}}\,\widetilde{\bsy{\chibh}} \equiv & \, \bsy{\widetilde{\chibh}{}^{\sharp}\cdot_{\widetilde{\bsy{\slashed{g}}}}\,\widetilde{\zeta}}+\frac{1}{2}(\bsy{\slashed{\varepsilon}\cdot\chib})(\widetilde{\bsy{{}^{\star}\zeta}})-\frac{1}{2}(\bsy{\textbf{tr}\chib})\widetilde{\bsy{\zeta}} -(\bsy{\slashed{\varepsilon}\cdot\chi})(\widetilde{\bsy{{}^{\star} \yb}}) -(\bsy{\slashed{\varepsilon}\cdot\chib})(\widetilde{\bsy{^{\star}\etab}}) +\widetilde{\bsy{\betab}}   \\
& +\frac{1}{2}\widetilde{\bsy{\nablasl}}(\bsy{\textbf{tr}\chib}) \,\boxed{+\frac{1}{4}\,(\widetilde{\bsy{\nablasl}}_{\bsy{3}}( \bsy{\textbf{tr}\chib}))\, \widetilde{(\ef)_{\bsy{\flat}}}} \, \boxed{+\frac{1}{4}\,(\widetilde{\bsy{\nablasl}}_{\bsy{4}}( \bsy{\textbf{tr}\chib})) \,\widetilde{(\et)_{\bsy{\flat}}}}  \nonumber\\
&-\frac{1}{2}\bsy{\widetilde{\slashed{\varepsilon}}{}^{\sharp_2}\cdot_{\widetilde{\bsy{\slashed{g}}}}}\,(\widetilde{\bsy{\nablasl}}(\bsy{\slashed{\varepsilon}\cdot\chib})) \, \boxed{ -\frac{1}{4}\,(\widetilde{\bsy{\nablasl}}_{\bsy{3}}(\bsy{\slashed{\varepsilon}\cdot\chib}))\, \widetilde{\bsy{\slashed{\varepsilon}}}{}^{\bsy{\sharp_2}}\bsy{\cdot}_{\widetilde{\bsy{\slashed{g}}}} \, \widetilde{(\ef)_{\bsy{\flat}}}} \, \boxed{-\frac{1}{4}\,(\widetilde{\bsy{\nablasl}}_{\bsy{4}}(\bsy{\slashed{\varepsilon}\cdot\chib})) \,\widetilde{\bsy{\slashed{\varepsilon}}}{}^{\bsy{\sharp_2}}\bsy{\cdot}_{\widetilde{\bsy{\slashed{g}}}}\,\widetilde{(\et)_{\bsy{\flat}}}}  \nonumber
\end{align} 
and the Gauss equation
\begin{align}
\widetilde{\bsy{\slashed{K}}} \equiv \frac{1}{2}(\widetilde{\bsy{\chih}},\widetilde{\bsy{\chibh}})_{\widetilde{\bsy{\slashed{g}}}}-\frac{1}{4}(\bsy{\textbf{tr}\chi})(\bsy{\textbf{tr}\chib})-\frac{1}{4}(\bsy{\slashed{\varepsilon}\cdot\chi})(\bsy{\slashed{\varepsilon}\cdot\chib})-\bsy{\rho} \, .  
\end{align}

\subsubsection{Reduced Bianchi equations}

The reduced Bianchi equations read
\begin{align}
\bsy{\widetilde{\nablasl}_3\widetilde{\alpha}}+\frac{1}{2}(\bsy{\textbf{tr}\chib})\widetilde{\bsy{\alpha}}+2\bsy{\omegabh\widetilde{\alpha}} +\frac{1}{2}(\bsy{\slashed{\varepsilon}\cdot\chib}) (\widetilde{{\bsy{{}^{\star}\alpha}}})\equiv & \,  -2\bsy{\widetilde{\slashed{\mathcal{D}}_2^{\star}}\,\widetilde{\beta}}-3\,\bsy{\rho}\,\widetilde{\bsy{\chih}}-3\bsy{\sigma}(\widetilde{{{}\bsy{^{\star} \chih}}})   +(4\widetilde{\bsy{\eta}}+\widetilde{\bsy{\zeta}})\bsy{\widehat{\otimes}}_{\widetilde{\bsy{\slashed{g}}}}\,\widetilde{\bsy{ \beta}} \\
&\boxed{+\frac{1}{2}\,(\bsy{\nablasl}_{\bsy{3}}\,\bsy{\beta}-{\bsy{\chib}}{}^{\bsy{\sharp_2}}\bsy{\cdot}\bsy{\beta})^{\bsy{A}}\,\widetilde{(\ef)_{\bsy{\flat}}}\,\widehat{\otimes}_{\widetilde{\bsy{\slashed{g}}}}\,\widetilde{(\ea)_{\bsy{\flat}}}} \nonumber\\  &\boxed{+\frac{1}{2}\,(\bsy{\nablasl}_{\bsy{4}}\,\bsy{\beta}-{\bsy{\chi}}{}^{\bsy{\sharp_2}}\bsy{\cdot}\bsy{\beta})^{\bsy{A}}\,\widetilde{(\et)_{\bsy{\flat}}}\,\widehat{\otimes}_{\widetilde{\bsy{\slashed{g}}}}\,\widetilde{(\ea)_{\bsy{\flat}}}}  \, , \nonumber
\end{align}
\begin{align}
\bsy{\widetilde{\nablasl}_4\widetilde{\beta}}+2(\bsy{\textbf{tr}\chi})\widetilde{\bsy{\beta}}-\omegah_{a,M}\widetilde{\bsy{\beta}}-2(\bsy{\slashed{\varepsilon}\cdot\chi})(\widetilde{\bsy{^{\star}\beta}}) \equiv & \, \bsy{\widetilde{\slashed{\textbf{div}}}\,\widetilde{\alpha}}+(\bsy{\widetilde{\etab}{}^{\sharp}}+2\bsy{\widetilde{\zeta}{}^{\sharp}})\bsy{\cdot_{\widetilde{\bsy{\slashed{g}}}}\,\widetilde{\alpha}}
\,\boxed{+(\widetilde{\bsy{\beta}},\etab_{a,M})_{\widetilde{\bsy{\slashed{g}}}}\cdot \widetilde{(\ef)_{\bsy{\flat}}}} \, , 
\end{align}
\begin{align}
\bsy{\widetilde{\nablasl}_3\widetilde{\beta}}+(\bsy{\textbf{tr} \chib})\widetilde{\bsy{\beta}}+\bsy{\omegabh}\widetilde{\bsy{\beta}}+(\bsy{\slashed{\varepsilon}\cdot\chi})(\widetilde{\bsy{{}^{\star}\beta}})  \equiv & \, \widetilde{\bsy{\nablasl}}\bsy{\rho}+\bsy{\widetilde{\slashed{\varepsilon}}{}^{\sharp_2}\cdot_{\widetilde{\bsy{\slashed{g}}}}}\,(\widetilde{\bsy{\nablasl}}\bsy{\sigma}) +3\,\bsy{\rho}\,\widetilde{\bsy{\eta}}+3\,\bsy{\sigma} (\widetilde{\bsy{ {}^{\star}\eta}})+2\,\bsy{\widetilde{\chih}{}^{\sharp}\cdot_{\widetilde{\bsy{\slashed{g}}}}\,\widetilde{\betab}}+\bsy{\widetilde{\yb}{}^{\sharp}\cdot_{\widetilde{\bsy{\slashed{g}}}}\widetilde{\alpha} }  \\
& \boxed{+\frac{1}{2}\,(\widetilde{\bsy{\nablasl}}_{\bsy{3}}\bsy{\rho})\, \widetilde{(\ef)_{\bsy{\flat}}}} \, \boxed{+\frac{1}{2}\,(\widetilde{\bsy{\nablasl}}_{\bsy{4}}\bsy{\rho}) \,\widetilde{(\et)_{\bsy{\flat}}}}  \nonumber\\
& \boxed{ +\frac{1}{2}\,(\widetilde{\bsy{\nablasl}}_{\bsy{3}}\bsy{\sigma})\, \widetilde{\bsy{\slashed{\varepsilon}}}{}^{\bsy{\sharp_2}}\bsy{\cdot}_{\widetilde{\bsy{\slashed{g}}}} \, \widetilde{(\ef)_{\bsy{\flat}}}} \, \boxed{+\frac{1}{2}\,(\widetilde{\bsy{\nablasl}}_{\bsy{4}}\bsy{\sigma}) \,\widetilde{\bsy{\slashed{\varepsilon}}}{}^{\bsy{\sharp_2}}\bsy{\cdot}_{\widetilde{\bsy{\slashed{g}}}}\,\widetilde{(\et)_{\bsy{\flat}}}} \nonumber \\
&\boxed{+(\widetilde{\bsy{\beta}},\widetilde{\bsy{\eta}})_{\widetilde{\bsy{\slashed{g}}}}\cdot\widetilde{(\et)_{\bsy{\flat}}}} \, \boxed{-(\bsy{\underline{\mathfrak{f}}},\widetilde{\bsy{\beta}})_{\slashed{g}} \widetilde{\bsy{\eta}}} \, , \nonumber
\end{align}
\begin{align}
\bsy{\widetilde{\nablasl}_4 \rho}+\frac{3}{2}(\bsy{\textbf{tr}\chi})\bsy{\rho } \equiv & \, \bsy{\widetilde{\slashed{\textbf{div}}}\,\widetilde{\beta}}+(2\etab_{a,M}+\widetilde{\bsy{\zeta}},\widetilde{\bsy{\beta}})_{\widetilde{\bsy{\slashed{g}}}}-\frac{1}{2}(\widetilde{\bsy{\chibh}},\widetilde{\bsy{\alpha}})_{\widetilde{\bsy{\slashed{g}}}}-\frac{3}{2}(\bsy{\slashed{\varepsilon}\cdot\chi}) \bsy{\sigma} \\
&\boxed{+\frac{1}{2}(\bsy{\nablasl}_{\bsy{3}}\,\bsy{\beta}-{\bsy{\chib}}{}^{\bsy{\sharp_2}}\bsy{\cdot}\bsy{\beta})^{\bsy{A}}\,(\widetilde{(\ef)_{\bsy{\flat}}},\widetilde{(\ea)_{\bsy{\flat}}})_{\widetilde{\bsy{\slashed{g}}}}} \nonumber\\ & \boxed{+\frac{1}{2}(\bsy{\nablasl}_{\bsy{4}}\,\bsy{\beta}-{\bsy{\chi}}{}^{\bsy{\sharp_2}}\bsy{\cdot}\bsy{\beta})^{\bsy{A}}(\widetilde{(\et)_{\bsy{\flat}}},\widetilde{(\ea)_{\bsy{\flat}}})_{\widetilde{\bsy{\slashed{g}}}}}  \nonumber \\
 & \boxed{+ \frac{1}{2}(\bsy{\underline{\mathfrak{f}}},\widetilde{\bsy{\beta}})_{\slashed{g}}\,{\bsy{\chi}}{}^{\bsy{AB}}(\widetilde{(\bsy{e_A})_{\bsy{\flat}}},\widetilde{(\bsy{e_B})_{\bsy{\flat}}})_{\widetilde{\bsy{\slashed{g}}}}}\, , \nonumber 
\end{align}
\begin{align}
\bsy{\widetilde{\nablasl}_4 \sigma}+\frac{3}{2}(\bsy{\textbf{tr}\chi})\bsy{\sigma} \equiv & \, -\bsy{\widetilde{\slashed{\textbf{curl}}}\,\widetilde{\beta}}-(2\etab_{a,M}+\widetilde{\bsy{\zeta}})\bsy{\wedge}_{\widetilde{\bsy{\slashed{g}}}}\widetilde{\bsy{\beta}}+\frac{1}{2}\bsy{\widetilde{\chibh}}\bsy{\wedge}_{\widetilde{\bsy{\slashed{g}}}}\widetilde{\bsy{\alpha}}+\frac{3}{2}(\bsy{\slashed{\varepsilon}\cdot\chi}) \bsy{\rho} \\
&\boxed{-\frac{1}{2}\,(\bsy{\nablasl}_{\bsy{3}}\,\bsy{\beta}-{\bsy{\chib}}{}^{\bsy{\sharp_2}}\bsy{\cdot}\bsy{\beta})^{\bsy{A}}\,\widetilde{(\ef)_{\bsy{\flat}}}\,\bsy{\wedge}_{\widetilde{\bsy{\slashed{g}}}}\,\widetilde{(\ea)_{\bsy{\flat}}}} \nonumber \\ & \boxed{-\frac{1}{2}\,(\bsy{\nablasl}_{\bsy{4}}\,\bsy{\beta}-{\bsy{\chi}}{}^{\bsy{\sharp_2}}\bsy{\cdot}\bsy{\beta})^{\bsy{A}} \widetilde{(\et)_{\bsy{\flat}}}\,\bsy{\wedge}_{\widetilde{\bsy{\slashed{g}}}}\,\widetilde{(\ea)_{\bsy{\flat}}} } \nonumber \\
 & \boxed{- \frac{1}{2}(\bsy{\underline{\mathfrak{f}}},\widetilde{\bsy{\beta}})_{\slashed{g}}\,{\bsy{\chi}}{}^{\bsy{AB}} \widetilde{(\bsy{e_A})_{\bsy{\flat}}}\,\bsy{\wedge}_{\widetilde{\bsy{\slashed{g}}}}\,\widetilde{(\bsy{e_B})_{\bsy{\flat}}} } \, , \nonumber
\end{align}
\begin{align}
\bsy{\widetilde{\nablasl}_3 \rho}+\frac{3}{2}(\bsy{\textbf{tr}\chib})\bsy{\rho} \equiv & \, -\bsy{\widetilde{\slashed{\textbf{div}}}\,\widetilde{\betab}}-(2\widetilde{\bsy{\eta}}-\widetilde{\bsy{\zeta}},\widetilde{\bsy{\betab}})_{\widetilde{\bsy{\slashed{g}}}}-\frac{1}{2}(\widetilde{\bsy{\chih}},\widetilde{\bsy{\alphab}})_{\widetilde{\bsy{\slashed{g}}}}+2(\widetilde{\bsy{\yb}},\widetilde{\bsy{\beta}})_{\widetilde{\bsy{\slashed{g}}}}+\frac{3}{2}(\bsy{\slashed{\varepsilon}\cdot\chib}) \bsy{\sigma } \\
&\boxed{-\frac{1}{2}(\bsy{\nablasl}_{\bsy{3}}\,\bsy{\betab}-{\bsy{\chib}}{}^{\bsy{\sharp_2}}\bsy{\cdot}\bsy{\betab})^{\bsy{A}}\,(\widetilde{(\ef)_{\bsy{\flat}}},\widetilde{(\ea)_{\bsy{\flat}}})_{\widetilde{\bsy{\slashed{g}}}}} \nonumber\\ & \boxed{-\frac{1}{2}(\bsy{\nablasl}_{\bsy{4}}\,\bsy{\betab}-{\bsy{\chi}}{}^{\bsy{\sharp_2}}\bsy{\cdot}\bsy{\betab})^{\bsy{A}}(\widetilde{(\et)_{\bsy{\flat}}},\widetilde{(\ea)_{\bsy{\flat}}})_{\widetilde{\bsy{\slashed{g}}}}}  \nonumber \\
 & \boxed{- \frac{1}{2}(\bsy{\underline{\mathfrak{f}}},\widetilde{\bsy{\betab}})_{\slashed{g}}\,{\bsy{\chi}}{}^{\bsy{AB}} (\widetilde{(\bsy{e_A})_{\bsy{\flat}}},\widetilde{(\bsy{e_B})_{\bsy{\flat}}})_{\widetilde{\bsy{\slashed{g}}}}}\, , \nonumber 
\end{align}
\begin{align}
\bsy{\widetilde{\nablasl}_3 \sigma}+\frac{3}{2}(\bsy{\textbf{tr}\chib})\bsy{\sigma}  \equiv & \, -\bsy{\widetilde{\slashed{\textbf{curl}}}\,\widetilde{\betab}}-(2\widetilde{\bsy{\eta}}-\widetilde{\bsy{\zeta}})\bsy{\wedge}_{\widetilde{\bsy{\slashed{g}}}}\widetilde{\bsy{\betab}}-\frac{1}{2}\bsy{\widetilde{\chih}}\bsy{\wedge}_{\widetilde{\bsy{\slashed{g}}}}\widetilde{\bsy{\alphab}}-2\widetilde{\bsy{\yb}}\bsy{\wedge}_{\widetilde{\bsy{\slashed{g}}}}\widetilde{\bsy{\beta}}-\frac{3}{2}(\bsy{\slashed{\varepsilon}\cdot\chib}) \bsy{\rho } \\
&\boxed{-\frac{1}{2}\,(\bsy{\nablasl}_{\bsy{3}}\,\bsy{\betab}-{\bsy{\chib}}{}^{\bsy{\sharp_2}}\bsy{\cdot}\bsy{\betab})^{\bsy{A}}\,\widetilde{(\ef)_{\bsy{\flat}}}\,\bsy{\wedge}_{\widetilde{\bsy{\slashed{g}}}}\,\widetilde{(\ea)_{\bsy{\flat}}}} \nonumber \\ & \boxed{-\frac{1}{2}\,(\bsy{\nablasl}_{\bsy{4}}\,\bsy{\betab}-{\bsy{\chi}}{}^{\bsy{\sharp_2}}\bsy{\cdot}\bsy{\betab})^{\bsy{A}} \widetilde{(\et)_{\bsy{\flat}}}\,\bsy{\wedge}_{\widetilde{\bsy{\slashed{g}}}}\,\widetilde{(\ea)_{\bsy{\flat}}} } \nonumber \\
 & \boxed{- \frac{1}{2}(\bsy{\underline{\mathfrak{f}}},\widetilde{\bsy{\betab}})_{\slashed{g}}\,{\bsy{\chi}}{}^{\bsy{AB}} \widetilde{(\bsy{e_A})_{\bsy{\flat}}}\,\bsy{\wedge}_{\widetilde{\bsy{\slashed{g}}}}\,\widetilde{(\bsy{e_B})_{\bsy{\flat}}} } \, , \nonumber
\end{align}
\begin{align}
\bsy{\widetilde{\nablasl}_4\widetilde{\betab}}+(\bsy{\textbf{tr} \chi})\widetilde{\bsy{\betab}}+\omegah_{a,M}\widetilde{\bsy{\betab}}+(\bsy{\slashed{\varepsilon}\cdot\chib})(\widetilde{\bsy{{}^{\star}\betab}})  \equiv & \, -\widetilde{\bsy{\nablasl}}\bsy{\rho}+\bsy{\widetilde{\slashed{\varepsilon}}{}^{\sharp_2}\cdot_{\widetilde{\bsy{\slashed{g}}}}}\,(\widetilde{\bsy{\nablasl}}\bsy{\sigma})-3\bsy{\rho}\etab_{a,M}+3\bsy{\sigma}(\widetilde{\bsy{ {}^{\star}\etab}})+2\bsy{\widetilde{\chibh}{}^{\sharp}\cdot_{\widetilde{\bsy{\slashed{g}}}} \widetilde{\beta}}  \\
& \boxed{-\frac{1}{2}\,(\widetilde{\bsy{\nablasl}}_{\bsy{3}}\bsy{\rho})\, \widetilde{(\ef)_{\bsy{\flat}}}} \, \boxed{-\frac{1}{2}\,(\widetilde{\bsy{\nablasl}}_{\bsy{4}}\bsy{\rho}) \,\widetilde{(\et)_{\bsy{\flat}}}}  \nonumber\\
& \boxed{ +\frac{1}{2}\,(\widetilde{\bsy{\nablasl}}_{\bsy{3}}\bsy{\sigma})\, \widetilde{\bsy{\slashed{\varepsilon}}}{}^{\bsy{\sharp_2}}\bsy{\cdot}_{\widetilde{\bsy{\slashed{g}}}} \, \widetilde{(\ef)_{\bsy{\flat}}}} \, \boxed{+\frac{1}{2}\,(\widetilde{\bsy{\nablasl}}_{\bsy{4}}\bsy{\sigma}) \,\widetilde{\bsy{\slashed{\varepsilon}}}{}^{\bsy{\sharp_2}}\bsy{\cdot}_{\widetilde{\bsy{\slashed{g}}}}\,\widetilde{(\et)_{\bsy{\flat}}}} \nonumber \\
&\boxed{+(\widetilde{\bsy{\betab}},\etab_{a,M})_{\widetilde{\bsy{\slashed{g}}}}\cdot \widetilde{(\ef)_{\bsy{\flat}}}} \, ,  \nonumber
\end{align}
\begin{align}
\bsy{\widetilde{\nablasl}_3\widetilde{\betab}}+2(\bsy{\textbf{tr}\chib})\widetilde{\bsy{\betab}}-\bsy{\omegabh\widetilde{\betab}}-2(\bsy{\slashed{\varepsilon}\cdot\chib})(\widetilde{\bsy{{}^{\star}\betab}}) \equiv & \, -\bsy{\widetilde{\slashed{\textbf{div}}}\,\widetilde{\alphab}}-(\bsy{\widetilde{\eta}{}^{\sharp}}-2\bsy{\widetilde{\zeta}{}^{\sharp}})\bsy{\cdot_{\widetilde{\bsy{\slashed{g}}}}\,\widetilde{\alphab}}-3\bsy{\rho}\widetilde{\bsy{\yb}}+3\bsy{\sigma} (\widetilde{\bsy{{}^{\star}\yb}})  \\
&\boxed{+(\widetilde{\bsy{\betab}},\widetilde{\bsy{\eta}})_{\widetilde{\bsy{\slashed{g}}}}\cdot\widetilde{(\et)_{\bsy{\flat}}}} \, \boxed{-(\bsy{\underline{\mathfrak{f}}},\widetilde{\bsy{\betab}})_{\slashed{g}} \widetilde{\bsy{\eta}}} \, , \nonumber
\end{align}
\begin{align}
\bsy{\widetilde{\nablasl}_4\widetilde{\alphab}}+\frac{1}{2}(\bsy{\textbf{tr}\chi})\widetilde{\bsy{\alphab}}+2\omegah_{a,M}\widetilde{\bsy{\alphab}}-\frac{1}{2}(\bsy{\slashed{\varepsilon}\cdot\chi}) (\widetilde{\bsy{{}^{\star}\alphab}}) \equiv & \, 2\bsy{\widetilde{\slashed{\mathcal{D}}_2^{\star}}\,\widetilde{\betab}}-3\,\bsy{\rho}\,\widetilde{\bsy{\chibh}}+3\bsy{\sigma}(\widetilde{\bsy{{{}^{\star} \chibh}}}) -(4\etab_{a,M}-\widetilde{\bsy{\zeta}})\bsy{\widehat{\otimes}}_{\widetilde{\bsy{\slashed{g}}}}\, \widetilde{\bsy{ \betab}} \\
&\boxed{-\frac{1}{2}\,(\bsy{\nablasl}_{\bsy{3}}\,\bsy{\betab}-{\bsy{\chib}}{}^{\bsy{\sharp_2}}\bsy{\cdot}\bsy{\betab})^{\bsy{A}}\,\widetilde{(\ef)_{\bsy{\flat}}}\,\widehat{\otimes}_{\widetilde{\bsy{\slashed{g}}}}\,\widetilde{(\ea)_{\bsy{\flat}}}} \nonumber\\  &\boxed{-\frac{1}{2}\,(\bsy{\nablasl}_{\bsy{4}}\,\bsy{\betab}-{\bsy{\chi}}{}^{\bsy{\sharp_2}}\bsy{\cdot}\bsy{\betab})^{\bsy{A}}\,\widetilde{(\et)_{\bsy{\flat}}}\,\widehat{\otimes}_{\widetilde{\bsy{\slashed{g}}}}\,\widetilde{(\ea)_{\bsy{\flat}}}}  \, . \nonumber
\end{align}

\subsection{The system for the renormalised quantities} \label{sec_renormalised_quantities}

Using the reduced nonlinear system of equations of Section \ref{sec_restricted_vacuum_Einstein_equations}, one can derive the \emph{renormalised} system of equations.

\medskip

For any smooth scalar function $\bsy{f}(\epsilon)$ and $\mathfrak{D}_{\mathcal{N}_{\text{as}}}$ covariant tensor $\widetilde{\bsy{\varsigma}}(\epsilon)$ in the reduced system of equations, we define the \emph{renormalised quantities}
\begin{align} \label{proto_diff}
\delta\bsy{f}(\epsilon) :&= \bsy{f}(\epsilon)-\bsy{f}(0) \, , & \delta\widetilde{\bsy{\varsigma}}(\epsilon) :&= \widetilde{\bsy{\varsigma}}(\epsilon)-\bsy{\varsigma}(0) \, .
\end{align}
We emphasise that $\delta\widetilde{\bsy{\varsigma}}(\epsilon)$ is a difference of $\mathfrak{D}_{\mathcal{N}_{\text{as}}}$ covariant tensors for all $\epsilon$, and thus a $\mathfrak{D}_{\mathcal{N}_{\text{as}}}$ covariant tensor for all $\epsilon$.~In view of Remark \ref{rmk_zero_restriction}, we have $$\delta\widetilde{\bsy{\varsigma}}(0)=0 \, .$$ 

\medskip

Applying the definition \eqref{proto_diff} to the induced metric $\bsy{\slashed{g}}(\epsilon)$ yields the renormalised quantity
\begin{equation*} 
\delta\widetilde{\bsy{\slashed{g}}}(\epsilon) = \widetilde{\bsy{\slashed{g}}}(\epsilon)-\slashed{g}_{a,M} \, .
\end{equation*} 
Similarly, the definition \eqref{proto_diff} can be applied to any connection coefficient and curvature component in the reduced system of equations to define the corresponding renormalised quantity.~The connection coefficient $\bsy{\eta}(\epsilon)$, for instance, yields the renormalised quantity 
\begin{equation} \label{proto_diff_eta}
\delta\widetilde{\bsy{\eta}}(\epsilon) = \widetilde{\bsy{\eta}}(\epsilon)-\eta_{a,M} 
\end{equation} 
and the corrected Gauss curvature $\widetilde{\bsy{\slashed{K}}}(\epsilon)$ yields the renormalised quantity 
\begin{equation*} 
\delta\widetilde{\bsy{\slashed{K}}}(\epsilon) = \widetilde{\bsy{\slashed{K}}}(\epsilon)-\widetilde{\slashed{K}}_{a,M}  \, .
\end{equation*} 

\medskip

We also define the \emph{renormalised connection} $\delta\widetilde{\bsy{\nablasl}}(\epsilon)$ on $\mathfrak{D}_{\mathcal{N}_{\text{as}}}$ as the $\mathfrak{D}_{\mathcal{N}_{\text{as}}}$ $(1,2)$-\ul{tensor}
\begin{equation*}
\delta\widetilde{\bsy{\nablasl}}(\epsilon) := \widetilde{\bsy{\nablasl}}(\epsilon)-\nablasl_{a,M}  \, .
\end{equation*}
The renormalised connection $\delta\widetilde{\bsy{\nablasl}}(\epsilon)$ on $\mathfrak{D}_{\mathcal{N}_{\text{as}}}$ is a $\mathfrak{D}_{\mathcal{N}_{\text{as}}}$ \emph{tensor} since it is the difference of two linear connections on $\mathfrak{D}_{\mathcal{N}_{\text{as}}}$.~Moreover, in view of Remark \ref{rmk_zero_restriction_connection}, we have the identically vanishing $\mathfrak{D}_{\mathcal{N}_{\text{as}}}$ tensor
\begin{equation*}
\delta\widetilde{\bsy{\nablasl}}(0)=0 \, .
\end{equation*} 

\medskip

\begin{remark}
The fact that the difference between two connections is a tensor only relies on both the connections being \emph{linear}.~As already pointed out, the connection $\widetilde{\bsy{\nablasl}}(\epsilon)$ on $\mathfrak{D}_{\mathcal{N}_{\textup{as}}}$ has \emph{torsion} for all $\epsilon$, but this does \emph{not} prevent $\delta\widetilde{\bsy{\nablasl}}(\epsilon)$ from being a $\mathfrak{D}_{\mathcal{N}_{\textup{as}}}$ \emph{tensor}.
\end{remark}

\medskip

For future convenience, we define the \emph{renormalised frame commutators}
\begin{equation*}
\delta([\bsy{e_I},\bsy{e_J}])(\epsilon):=[\bsy{e_I}(\epsilon),\bsy{e_J}(\epsilon)]-[e_I^{\text{as}},e_J^{\text{as}}] 
\end{equation*}
for $\bsy{I},\bsy{J}=\left\lbrace \bsy{1},\bsy{2},\bsy{3},\bsy{4} \right\rbrace$.

\medskip

\begin{remark}
All the renormalised quantities are \emph{global}, \emph{regular} $\mathfrak{D}_{\mathcal{N}_{\textup{as}}}$ tensors on the Kerr exterior manifold $(\mathcal{M}^*,g_{a,M})$.
\end{remark}

\medskip

The explicit derivation of the renormalised system of equations is omitted, but can be easily carried out by the reader.~As for the reduced system of equations, the renormalised system of equations is a system of equations for $\mathfrak{D}_{\mathcal{N}_{\text{as}}}$ tensors. 

\medskip

Crucially, the renormalised null structure equations with the form of transport equations in the \emph{outgoing} direction do \emph{not} contain any derivatives of renormalised connection coefficients on the right hand side.~This structure is already apparent from the reduced system of equations of Section \ref{sec_restricted_vacuum_Einstein_equations}.~In this regard, see the preliminary discussion in Section \ref{sec_preliminary_rmks_restricted_eqns}.

\medskip

A similar structure does \emph{not} appear for the corresponding transport equations in the \emph{ingoing} direction.

\section{The linearisation procedure}  \label{sec_linearisation_procedure}

In this section, we derive general linearisation formulae around $\epsilon=0$ for the renormalised quantities defined in Section \ref{sec_renormalised_quantities}.~We already remark that, exploiting the geometric formulation of the renormalised quantities achieved in Section \ref{sec:geom_compare_tensors}, we will be able to linearise all renormalised quantities \emph{geometrically} and write all linearisation formulae \emph{in geometric form}.~All the linearised quantities are \emph{global}, \emph{regular} $\mathfrak{D}_{\mathcal{N}_{\text{as}}}$ tensors on the Kerr exterior manifold $(\mathcal{M},g_{a,M})$.

\medskip

In Section \ref{sec_system_linearised_gravity}, the formulae will be applied to linearise the renormalised vacuum Einstein equations around the Kerr exterior manifold $(\mathcal{M},g_{a,M})$ relative to the frame $\mathcal{N}_{\text{as}}$, i.e.~around $\epsilon=0$.

\subsection{Overview}  \label{sec_overview_linearisation_procedure}

The formal linearisation of the renormalised quantities is straightforward.~However, some more work is needed if one wants to write certain linearised quantities in a convenient form.~We comment on two such instances.

\begin{itemize}
\item Let $\widetilde{\bsy{\varsigma}}(\epsilon)$ be a $\mathfrak{D}_{\mathcal{N}_{\text{as}}}$ covariant tensor.~The linearisation of the renormalised covariant derivative $$\delta(\widetilde{\bsy{\nablasl}}\,\widetilde{\bsy{\varsigma}})(\epsilon)$$ is the most technical part of the section.~Crucially, we will be able to relate the linearisation of the renormalised connection $\delta\widetilde{\bsy{\nablasl}}(\epsilon)$ (which we recall is a $\mathfrak{D}_{\mathcal{N}_{\text{as}}}$ covariant tensor) to the linearisation of the renormalised metric $\delta\widetilde{\bsy{\slashed{g}}}(\epsilon)$ (see Proposition \ref{prop_linearisation_cov_derivatives}).~For this to be possible, the choice of induced connection $\widetilde{\bsy{\nablasl}}(\epsilon)$ over $\mathfrak{D}_{\mathcal{N}_{\text{as}}}$ made in Section \ref{sec:geom_compare_tensors} plays a fundamental role.~See the related preliminary comments in Section \ref{sec_overview_geom_compare_tensors}.

\item Let $\widetilde{\bsy{\varsigma}}(\epsilon)$ be a $\mathfrak{D}_{\mathcal{N}_{\text{as}}}$ covariant tensor.~We will \emph{not} need general linearisation formulae for the renormalised covariant derivatives 
\begin{align} 
&\delta(\widetilde{\bsy{\nablasl}}_{\ef}\,\widetilde{\bsy{\varsigma}})(\epsilon) \, , &  &\delta(\widetilde{\bsy{\nablasl}}_{\et}\,\widetilde{\bsy{\varsigma}})(\epsilon) \, . \label{overview_cov_derivs_ef_et}
\end{align}
In fact, when we later apply the linearisation formulae of the present section to linearise the renormalised vacuum Einstein equations, we \emph{never} linearise the covariant derivatives \eqref{overview_cov_derivs_ef_et} \emph{unless} $\widetilde{\bsy{\varsigma}}(0)\equiv 0$.~The renormalised versions of the equations \eqref{EVE_4_eta}, \eqref{EVE_3_etab}, \eqref{EVE_4_zeta} and \eqref{EVE_3_zeta} will be considered in their alternative form of Remark \ref{rmk_null_structure_lie_derivative}, where one linearises the renormalised Lie derivatives 
\begin{align}
&\delta (\widetilde{\bsy{\slashed{\mathcal{L}}}}_{\ef}\,\widetilde{\bsy{\varsigma}})(\epsilon) \, , & &\delta (\widetilde{\bsy{\slashed{\mathcal{L}}}}_{\ef}\,\widetilde{\bsy{\varsigma}})(\epsilon) \label{overview_lie_derivs_ef_et}
\end{align}
instead (see Proposition \ref{prop_linearisation_lie_derivatives}).~The renormalised Lie derivatives \eqref{overview_lie_derivs_ef_et} are more convenient to linearise than the renormalised covariant derivatives \eqref{overview_cov_derivs_ef_et}.~This computational advantage is already manifest from the nonlinear formulae for $(\widetilde{\bsy{\slashed{\mathcal{L}}}}_{\et}\,\widetilde{\bsy{\varsigma}})(\epsilon)$ in Proposition \ref{prop_aux_restriction_lie_derivative}, where the differential operators on the right hand side are \emph{independent of $\epsilon$}.\footnote{The reader should realise that the dependence on $\epsilon$ of the
Lie derivative $\widetilde{\bsy{\slashed{\mathcal{L}}}}_{\bsy{X}(\epsilon)}$ only originates from the dependence on $\epsilon$ of the vector field $\bsy{X}(\epsilon)$ relative to which the derivative is taken.~In this regard, we recall that the Lie derivative $\widetilde{\bsy{\slashed{\mathcal{L}}}}_{\bsy{X}(\epsilon)}$ arises from the spacetime Lie derivative $\bsy{\mathcal{L}}_{\bsy{X}(\epsilon)}$ (whose dependence on $\epsilon$ only originates from $\bsy{X}(\epsilon)$) projected (and then restricted) over $\mathfrak{D}_{\mathcal{N}_{\text{as}}}$, the projection being independent of $\epsilon$.}~For $(\widetilde{\bsy{\slashed{\mathcal{L}}}}_{\ef}\,\widetilde{\bsy{\varsigma}})(\epsilon)$, one has the very convenient identity $$(\widetilde{\bsy{\slashed{\mathcal{L}}}}_{\ef}\,\widetilde{\bsy{\varsigma}})(\epsilon)=(\slashed{\mathcal{L}}^{a,M}_{e_4^{\text{as}}}\, \widetilde{\bsy{\varsigma}})(\epsilon) \, .$$
\end{itemize} 

\medskip

To keep the notation lighter, all the quantities corresponding to $|\epsilon|=0$ are simply denoted as de-bolded, de-tilded quantities.~For instance, we write the $\mathfrak{D}_{\mathcal{N}_{\text{as}}}$ covariant tensor $\widetilde{\bsy{\varsigma}}(0)\equiv \bsy{\varsigma}(0)$ as $\varsigma$.~We also often remove the reference to the Kerr parameters.~For instance, we write the geometric quantities $\slashed{g}_{a,M}$, $\nablasl^{a,M}$ and $\slashed{\mathcal{L}}^{a,M}$ as $\slashed{g}$, $\nablasl$ and $\slashed{\mathcal{L}}$ respectively.

\subsection{Scalar functions}  \label{sec_lin_scalar_functions}

We consider the scalar renormalised quantities $\delta\bsy{f}(\epsilon)$ from Section \ref{sec_renormalised_quantities}.~The one-parameter family $\delta\bsy{f}(\epsilon)$ is regarded as a smooth map in the $\epsilon$ variable
\begin{equation*}
\delta\bsy{f}(\epsilon) :\mathbb{R}\rightarrow C^{\infty}(\mathcal{M}) \, .
\end{equation*} 
We consider the Taylor expansion of $\delta\bsy{f}(\epsilon)$ around $\epsilon=0$ as follows
\begin{equation*}
\delta\bsy{f}(\epsilon)=\epsilon \cdot \overset{\text{\scalebox{.6}{$(1)$}}}{f} +\mathcal{O}(\epsilon^2)  \, .
\end{equation*}
The linearisation of $\delta\bsy{f}(\epsilon)$ consists in dropping the $\mathcal{O}(\epsilon^2)$-terms in the expansion.

\subsection{Covariant tensors}   \label{sec_lin_cov_tensors}

We consider the covariant renormalised quantities $\delta\widetilde{\bsy{\varsigma}}(\epsilon)$ from Section \ref{sec_renormalised_quantities}.~The one-parameter family $\delta\widetilde{\bsy{\varsigma}}(\epsilon)$ is regarded as a smooth map in the $\epsilon$ variable
\begin{equation} \label{map_diff_tens}
\delta\widetilde{\bsy{\varsigma}}(\epsilon):\mathbb{R}\rightarrow (\mathfrak{D}_{\mathcal{N}_{\text{as}}})^{\star} \, .
\end{equation} 
We consider the Taylor expansion of $\delta\widetilde{\bsy{\varsigma}}(\epsilon)$ around $\epsilon=0$ as follows
\begin{equation*}
\delta\widetilde{\bsy{\varsigma}}(\epsilon)=\epsilon \cdot \overset{\text{\scalebox{.6}{$(1)$}}}{\varsigma} +\mathcal{O}(\epsilon^2)  \, .
\end{equation*}
The quantity $\overset{\text{\scalebox{.6}{$(1)$}}}{\varsigma}$ is a $\mathfrak{D}_{\mathcal{N}_{\text{as}}}$ one-tensor.\footnote{One has the identity $\overset{\text{\scalebox{.6}{$(1)$}}}{\varsigma}=\partial_{\epsilon}\delta\widetilde{\bsy{\varsigma}}(\epsilon)|_{\epsilon=0}$, with $\delta\widetilde{\bsy{\varsigma}}(\epsilon)$ a smooth function of $\epsilon$ as in \eqref{map_diff_tens}.}~The linearisation of $\delta\widetilde{\bsy{\varsigma}}(\epsilon)$ consists in dropping the $\mathcal{O}(\epsilon^2)$-terms in the expansion.

\medskip

For the renormalised metrics $\delta\widetilde{\bsy{\slashed{g}}}(\epsilon)$, we can expand as follows
\begin{equation*}
\delta\widetilde{\bsy{\slashed{g}}}(\epsilon)=\epsilon \cdot \overset{\text{\scalebox{.6}{$(1)$}}}{\slashed{g}} +\mathcal{O}(\epsilon^2)  \, .
\end{equation*}
For the renormalised $\mathfrak{D}_{\mathcal{N}_{\text{as}}}$ one-tensors $\delta\widetilde{(\bsy{e_I})_{\bsy{\flat}}}(\epsilon)$, we can expand as follows $$\delta\widetilde{(\bsy{e_I})_{\bsy{\flat}}}(\epsilon)=\epsilon\cdot \overset{\text{\scalebox{.6}{$(1)$}}}{(e_I)_{\flat}}+\mathcal{O}(\epsilon^2)$$ for all $\bsy{I},\bsy{J}=\left\lbrace \bsy{1},\bsy{2},\bsy{3},\bsy{4} \right\rbrace$.

\medskip

\begin{remark}
The linearisation of a \emph{symmetric} covariant renormalised quantity $\delta\widetilde{\bsy{\varsigma}}(\epsilon)$ is a \emph{symmetric} $\mathfrak{D}_{\mathcal{N}_{\text{as}}}$ covariant tensor. 
\end{remark}

\medskip

\begin{remark}
Let $\widetilde{\bsy{\varsigma}}(\epsilon)$ be a symmetric \emph{traceless} (relative to $\widetilde{\bsy{\slashed{g}}^{-1}}(\epsilon)$) $\mathfrak{D}_{\mathcal{N}_{\text{as}}}$ covariant tensor.~The linearisation of the covariant renormalised quantity $\delta\widetilde{\bsy{\varsigma}}(\epsilon)$ is \emph{not}, in general, a symmetric \emph{traceless} (relative to $\slashed{g}^{-1}_{a,M}$) $\mathfrak{D}_{\mathcal{N}_{\text{as}}}$ tensor.~However, if $\delta\widetilde{\bsy{\varsigma}}(\epsilon) \equiv \widetilde{\bsy{\varsigma}}(\epsilon)$ (or, equivalently, if $\widetilde{\bsy{\varsigma}}(0)\equiv 0$), then the linearised quantity \emph{is} symmetric traceless.~The linearised quantities
\begin{align*}
&\overset{\text{\scalebox{.6}{$(1)$}}}{\chih} \, , & &\overset{\text{\scalebox{.6}{$(1)$}}}{\chibh} \, , & &\overset{\text{\scalebox{.6}{$(1)$}}}{\alpha} \, , & &\overset{\text{\scalebox{.6}{$(1)$}}}{\alphab}
\end{align*}
\emph{are} symmetric traceless $\mathfrak{D}_{\mathcal{N}_{\text{as}}}$ tensors, whereas the linearisations of the renormalised quantities
\begin{align*}
&\delta(\widetilde{\bsy{\varsigma}}\,\widehat{\otimes}_{\widetilde{\bsy{\slashed{g}}}}\, \widetilde{\bsy{\theta}})(\epsilon) \, , & &\delta(\widetilde{\bsy{\slashed{\mathcal{D}}}_{\bsy{2}}^{\star}}\,\widetilde{\bsy{\varsigma}})(\epsilon)
\end{align*}
are \emph{not}, in general, symmetric traceless $\mathfrak{D}_{\mathcal{N}_{\text{as}}}$ tensors (see the later general formulae \eqref{linearisation_otimes_hat_tensors} and \eqref{linearisation_D2star_tensor}).~The linearisation of special combinations of renormalised quantities, like for instance 
\begin{align*}
&\delta(-2\,\widetilde{\bsy{\slashed{\mathcal{D}}}_{\bsy{2}}^{\star}}\,\widetilde{\bsy{\eta}}+\widetilde{\bsy{\eta}}\,\widehat{\otimes}_{\widetilde{\bsy{\slashed{g}}}}\, \widetilde{\bsy{\eta}})(\epsilon) \, , & &\delta(-2\,\widetilde{\bsy{\slashed{\mathcal{D}}}_{\bsy{2}}^{\star}}\,\widetilde{\bsy{\etab}}+\widetilde{\bsy{\etab}}\,\widehat{\otimes}_{\widetilde{\bsy{\slashed{g}}}}\, \widetilde{\bsy{\etab}})(\epsilon) \, ,
\end{align*}
\emph{are} symmetric traceless $\mathfrak{D}_{\mathcal{N}_{\text{as}}}$ tensors (although the linearisation of the single terms separately are not).~This latter fact follows from the identities 
\begin{align*}
(-2\,\widetilde{\bsy{\slashed{\mathcal{D}}}_{\bsy{2}}^{\star}}\,\widetilde{\bsy{\eta}}+\widetilde{\bsy{\eta}}\,\widehat{\otimes}_{\widetilde{\bsy{\slashed{g}}}}\, \widetilde{\bsy{\eta}})(0)&\equiv 0 \, , & (-2\,\widetilde{\bsy{\slashed{\mathcal{D}}}_{\bsy{2}}^{\star}}\,\widetilde{\bsy{\etab}}+\widetilde{\bsy{\etab}}\,\widehat{\otimes}_{\widetilde{\bsy{\slashed{g}}}}\, \widetilde{\bsy{\etab}})(0)&\equiv 0
\end{align*}
and can be checked via the general formula \eqref{linearisation_D2star_plus_otimes_hat_tensor}. 
\end{remark}

\medskip

We define the smooth scalar function
\begin{equation*}
\text{tr}\overset{\text{\scalebox{.6}{$(1)$}}}{\slashed{g}} :=(\slashed{g}^{-1}){}^{AB}\overset{\text{\scalebox{.6}{$(1)$}}}{\slashed{g}}_{AB} 
\end{equation*}
and the \emph{symmetric traceless} $\mathfrak{D}_{\mathcal{N}_{\text{as}}}$ two-tensor
\begin{equation*}
\overset{\text{\scalebox{.6}{$(1)$}}}{\widehat{\slashed{g}}} :=\overset{\text{\scalebox{.6}{$(1)$}}}{\slashed{g}}-\frac{1}{2}\,(\text{tr}\overset{\text{\scalebox{.6}{$(1)$}}}{\slashed{g}})\,\slashed{g} \, .
\end{equation*}

\medskip

We remark that we always linearise fully covariant renormalised quantities, with the exception of the \emph{renormalised inverse metric} $$\delta\widetilde{\bsy{\slashed{g}}^{-1}}(\epsilon):=\widetilde{\bsy{\slashed{g}}^{-1}}(\epsilon)-\slashed{g}^{-1}_{a,M} \, ,$$ whose Taylor expansion around $\epsilon=0$ reads
\begin{equation*}
\delta\widetilde{\bsy{\slashed{g}}^{-1}}(\epsilon)=\epsilon \cdot \overset{\text{\scalebox{.6}{$(1)$}}}{\slashed{g}^{-1}} +\mathcal{O}(\epsilon^2)  \, .
\end{equation*}

\subsection{Frame coefficients}  \label{sec_lin_frame_vectors}

We Taylor-expand and linearise the renormalised quantities
\begin{equation*}
\delta\bsy{\mathfrak{\underline{\mathfrak{f}}}}_{3}(\epsilon) \, , \, \delta\bsy{\mathfrak{\underline{\mathfrak{f}}}}_{4}(\epsilon)   
\end{equation*}
as outlined in Section \ref{sec_lin_scalar_functions}.

\medskip

The one-parameter families of renormalised $\mathfrak{D}_{\mathcal{N}_{\text{as}}}$ one-tensors $\delta\bsy{\mathfrak{\underline{\mathfrak{f}}}}(\epsilon)$ and $\delta\bsy{\mathfrak{\slashed{\mathfrak{f}}}}_{\bsy{A}}(\epsilon
)$ are regarded as smooth maps in the $\epsilon$ variable
\begin{align}  
\delta\bsy{\mathfrak{\underline{\mathfrak{f}}}}(\epsilon) &:\mathbb{R}\rightarrow (\mathfrak{D}_{\mathcal{N}_{\text{as}}})^* \, , &  \delta \bsy{\mathfrak{\slashed{\mathfrak{f}}}}_{\bsy{A}}(\epsilon
) &:\mathbb{R}\rightarrow (\mathfrak{D}_{\mathcal{N}_{\text{as}}})^* \, . \label{fb_map}
\end{align} 
We consider the Taylor expansion of $\delta\bsy{\mathfrak{\underline{\mathfrak{f}}}}(\epsilon)$ and $\delta\bsy{\mathfrak{\slashed{\mathfrak{f}}}}_{\bsy{A}}(\epsilon
)$ around $\epsilon=0$ as follows
\begin{align*}
\delta\bsy{\mathfrak{\underline{\mathfrak{f}}}}(\epsilon)&=\epsilon \cdot \overset{\text{\scalebox{.6}{$(1)$}}}{\mathfrak{\underline{\mathfrak{f}}}} +\mathcal{O}(\epsilon^2)  \, , \\
\delta\bsy{\mathfrak{\slashed{\mathfrak{f}}}}_{\bsy{A}}(\epsilon
) &=\epsilon \cdot \overset{\text{\scalebox{.6}{$(1)$}}}{\mathfrak{\slashed{\mathfrak{f}}}}_A +\mathcal{O}(\epsilon^2)  \, .
\end{align*}
The quantities $\overset{\text{\scalebox{.6}{$(1)$}}}{\mathfrak{\underline{\mathfrak{f}}}}$ and $\overset{\text{\scalebox{.6}{$(1)$}}}{\mathfrak{\slashed{\mathfrak{f}}}}_A$ are $\mathfrak{D}_{\mathcal{N}_{\text{as}}}$ one-tensors.~The linearisation of $\delta\bsy{\mathfrak{\underline{\mathfrak{f}}}}(\epsilon)$ and $\delta\bsy{\mathfrak{\slashed{\mathfrak{f}}}}_{\bsy{A}}(\epsilon
)$ consists in dropping the $\mathcal{O}(\epsilon^2)$-terms in the expansion.

\medskip

We Taylor-expand and linearise the renormalised quantities
\begin{equation*}
\delta\widetilde{\bsy{\mathfrak{\slashed{\mathfrak{f}}}}_{4}}(\epsilon) \, , \, \delta\widetilde{\bsy{\mathfrak{\slashed{\mathfrak{f}}}}_{3}}(\epsilon)  
\end{equation*}
as outlined in Section \ref{sec_lin_cov_tensors}.

\subsection{Identities for linearised quantities}  \label{sec_ids_special_lin_quantities}

The following proposition is crucial.

\begin{prop}
We have the identities for smooth scalar functions 
\begin{align}
\overset{\text{\scalebox{.6}{$(1)$}}}{\mathfrak{f}}_3&=0 \, , & \overset{\text{\scalebox{.6}{$(1)$}}}{\mathfrak{f}}_4&=0 \, , &  \overset{\text{\scalebox{.6}{$(1)$}}}{\omegah}&=0 
\end{align}
and for $\mathfrak{D}_{\mathcal{N}_{\text{as}}}$ one-tensors
\begin{align}
\overset{\text{\scalebox{.6}{$(1)$}}}{\mathfrak{f}}&=0 \, , & \overset{\text{\scalebox{.6}{$(1)$}}}{\xi}&=0 \, , & \overset{\text{\scalebox{.6}{$(1)$}}}{\etab}&=0 \, .
\end{align}
\end{prop}

\medskip

\begin{proof}
The identities \eqref{i_1_kerr}, \eqref{i_2_kerr}, \eqref{i_3_kerr} and \eqref{i_4_kerr} imply the identities
\begin{align*}
\delta\,\bsy{\mathfrak{f}_3}(\epsilon)&=0 \, , & \delta\,\bsy{\mathfrak{f}_4}(\epsilon)&=0 \, ,   & \delta\,\bsy{\mathfrak{f}}(\epsilon)&=0 \, ,\\
\delta\,\bsy{\omegah}(\epsilon)&=0 \, , & \delta\,\widetilde{\bsy{\xi}}(\epsilon)&=0 \, , & \delta\,\widetilde{\bsy{\etab}}(\epsilon)&=0 
\end{align*}
for the respective renormalised quantities.~The identities for the linearised quantities from the proposition follow immediately.

\end{proof}

\medskip

The two following propositions provide useful relations between linearised quantities.~The identities stated are repeatedly used in the remaining part of the section.

\medskip

\begin{prop} \label{prop_exchange_lin_frame_coeff}
We have the identities 
\begin{align}
\overset{\text{\scalebox{.6}{$(1)$}}}{(e_4)_{\flat}}-2\overset{\text{\scalebox{.6}{$(1)$}}}{\slashed{\mathfrak{f}}}_4&=0 \, ,  & \overset{\text{\scalebox{.6}{$(1)$}}}{(e_3)_{\flat}}-2\overset{\text{\scalebox{.6}{$(1)$}}}{\slashed{\mathfrak{f}}}_3&=0 \, .
\end{align}
and
\begin{align}  \label{eq_exchange_lin_frame_coeff_2}
\overset{\text{\scalebox{.6}{$(1)$}}}{\bsy{g}(e_3,e_A)}-2\,(\overset{\text{\scalebox{.6}{$(1)$}}}{\mathfrak{\slashed{\mathfrak{f}}}}_3)_{A}+\overset{\text{\scalebox{.6}{$(1)$}}}{\mathfrak{\underline{\mathfrak{f}}}}_{A} &=0 \, , & \overset{\text{\scalebox{.6}{$(1)$}}}{\slashed{g}}_{AB}+(\overset{\text{\scalebox{.6}{$(1)$}}}{\mathfrak{\slashed{\mathfrak{f}}}}_{A})_{B}+(\overset{\text{\scalebox{.6}{$(1)$}}}{\mathfrak{\slashed{\mathfrak{f}}}}_{B})_{A} &=0 \, .
\end{align}
\end{prop} 

\medskip

\begin{proof}
We linearise the identities from Proposition \ref{prop_frame_commutators_ricci_2}.

\end{proof}

\medskip

\begin{prop}  \label{prop_identities_linearised_metric_quantities}
We have the identities
\begin{equation}  \label{eq_lin_inverse_metric_contracted}
(\slashed{g}^{-1}_{a,M})^{BC}\overset{\text{\scalebox{.6}{$(1)$}}}{\slashed{g}}_{AC}=-(\overset{\text{\scalebox{.6}{$(1)$}}}{\slashed{g}^{-1}})^{BC}(\slashed{g}_{a,M})_{AC}
\end{equation}
and
\begin{equation} \label{eq_lin_volume_form}
\overset{\text{\scalebox{.6}{$(1)$}}}{\slashed{\varepsilon}}=\frac{1}{2}\,(\textup{tr}\overset{\text{\scalebox{.6}{$(1)$}}}{\slashed{g}})\,\slashed{\varepsilon}_{a,M} \, .
\end{equation}
In particular, we have the identity
\begin{equation}  \label{eq_lin_inverse_metric}
\overset{\text{\scalebox{.6}{$(1)$}}}{\slashed{g}^{-1}}=-\overset{\text{\scalebox{.6}{$(1)$}}}{\widehat{\slashed{g}}}{}^{\sharp\sharp}-\frac{1}{2}\,(\textup{tr}\overset{\text{\scalebox{.6}{$(1)$}}}{\slashed{g}})\,\slashed{g}_{a,M}^{-1} \, .
\end{equation}
\end{prop}

\medskip

\begin{proof}
The first identity is immediate from linearising the identity \eqref{restriction_inverse_metric_relation}.~For the second identity, we linearise the identity from Proposition \ref{prop_restricted_volume_form} and obtain
\begin{align*}
\overset{\text{\scalebox{.6}{$(1)$}}}{\slashed{\varepsilon}}=& \, (\overset{\text{\scalebox{.6}{$(1)$}}}{\mathfrak{\slashed{\mathfrak{f}}}}_1+\mathfrak{\slashed{\mathfrak{f}}}_1^{\sharp}\cdot \overset{\text{\scalebox{.6}{$(1)$}}}{\slashed{g}})\otimes \mathfrak{\slashed{\mathfrak{f}}}_2+\mathfrak{\slashed{\mathfrak{f}}}_1\otimes (\overset{\text{\scalebox{.6}{$(1)$}}}{\mathfrak{\slashed{\mathfrak{f}}}}_2+\mathfrak{\slashed{\mathfrak{f}}}_2^{\sharp}\cdot \overset{\text{\scalebox{.6}{$(1)$}}}{\slashed{g}}) \\
&-(\overset{\text{\scalebox{.6}{$(1)$}}}{\mathfrak{\slashed{\mathfrak{f}}}}_2+\mathfrak{\slashed{\mathfrak{f}}}_2^{\sharp}\cdot \overset{\text{\scalebox{.6}{$(1)$}}}{\slashed{g}})\otimes \mathfrak{\slashed{\mathfrak{f}}}_1-\mathfrak{\slashed{\mathfrak{f}}}_2\otimes (\overset{\text{\scalebox{.6}{$(1)$}}}{\mathfrak{\slashed{\mathfrak{f}}}}_1+\mathfrak{\slashed{\mathfrak{f}}}_1^{\sharp}\cdot \overset{\text{\scalebox{.6}{$(1)$}}}{\slashed{g}}) \\
=& \, ((\overset{\text{\scalebox{.6}{$(1)$}}}{\mathfrak{\slashed{\mathfrak{f}}}}_1)_1+ \overset{\text{\scalebox{.6}{$(1)$}}}{\slashed{g}}_{11}+(\overset{\text{\scalebox{.6}{$(1)$}}}{\mathfrak{\slashed{\mathfrak{f}}}}_2)_2+\overset{\text{\scalebox{.6}{$(1)$}}}{\slashed{g}}_{22})\,\slashed{\varepsilon}_{a,M} \\
=& \, (-\frac{1}{2}\,\overset{\text{\scalebox{.6}{$(1)$}}}{\slashed{g}}_{11}+ \overset{\text{\scalebox{.6}{$(1)$}}}{\slashed{g}}_{11}-\frac{1}{2}\,\overset{\text{\scalebox{.6}{$(1)$}}}{\slashed{g}}_{22}+\overset{\text{\scalebox{.6}{$(1)$}}}{\slashed{g}}_{22})\,\slashed{\varepsilon}_{a,M} \\
=& \, \frac{1}{2}\,(\overset{\text{\scalebox{.6}{$(1)$}}}{\slashed{g}}_{11}+\overset{\text{\scalebox{.6}{$(1)$}}}{\slashed{g}}_{22})\,\slashed{\varepsilon}_{a,M} \, , 
\end{align*}
where in the penultimate equality we used the second of the identities \eqref{eq_exchange_lin_frame_coeff_2}.~The last identity is obtained by observing that the identity \eqref{eq_lin_inverse_metric_contracted} implies the identity 
\begin{equation*}
\overset{\text{\scalebox{.6}{$(1)$}}}{\slashed{g}^{-1}}=-\overset{\text{\scalebox{.6}{$(1)$}}}{\slashed{g}}{}^{\sharp\sharp} 
\end{equation*}
and by decomposing the linearised metric $\overset{\text{\scalebox{.6}{$(1)$}}}{\slashed{g}}$ on the right hand side into pure trace and symmetric traceless parts.

\end{proof}

\subsection{Frame commutators}

We have the following proposition for the linearisation of the renormalised frame commutators.~We recall the definition of the renormalised frame commutators from Section \ref{sec_renormalised_quantities}.

\medskip

\begin{prop} \label{prop_linearisation_frame_commutators_1}
We have the identities
\begin{align}
\delta([\ef,\et])(\epsilon)=& \, \epsilon\cdot\left( (\nablasl^{a,M}_{e_4^{\text{as}}} \overset{\text{\scalebox{.6}{$(1)$}}}{\mathfrak{\underline{\mathfrak{f}}}}_{3})\, e_4^{\text{as}}+(\nablasl^{a,M}_{e_4^{\text{as}}}\overset{\text{\scalebox{.6}{$(1)$}}}{\mathfrak{\underline{\mathfrak{f}}}}_{4})\,  e_3^{\text{as}}+\slashed{g}_{a,M}^{AB}(\nablasl^{a,M}_{e_4^{\text{as}}}\overset{\text{\scalebox{.6}{$(1)$}}}{\mathfrak{\underline{\mathfrak{f}}}}_{A}) \, e_B^{\text{as}}  \right.\\
&\left. +\overset{\text{\scalebox{.6}{$(1)$}}}{\mathfrak{\underline{\mathfrak{f}}}}_{4}\, [e_4^{\text{as}},e_3^{\text{as}}]+\overset{\text{\scalebox{.6}{$(1)$}}}{\mathfrak{\underline{\mathfrak{f}}}}{}^{A}\, [e_4^{\text{as}}, e_A^{\text{as}}] \right) + \mathcal{O}(\epsilon^2) \, ,  \nonumber 
\end{align}
\begin{align}
\delta([\ef,\ea])(\epsilon)=& \, \epsilon\cdot\left((\nablasl^{a,M}_{e_4^{\text{as}}}(\overset{\text{\scalebox{.6}{$(1)$}}}{\mathfrak{\slashed{\mathfrak{f}}}}_{3})_A)\, e_4^{\text{as}}+(\nablasl^{a,M}_{e_4^{\text{as}}}(\overset{\text{\scalebox{.6}{$(1)$}}}{\mathfrak{\slashed{\mathfrak{f}}}}_{4})_A)\, e_3^{\text{as}} +\slashed{g}_{a,M}^{BC}(\nablasl^{a,M}_{e_4^{\text{as}}}(\overset{\text{\scalebox{.6}{$(1)$}}}{\mathfrak{\slashed{\mathfrak{f}}}}_{A})_{B})\, e_C^{\text{as}} \right.\\
&\left. +(\overset{\text{\scalebox{.6}{$(1)$}}}{\mathfrak{\slashed{\mathfrak{f}}}}_{4})_A\, [e_4^{\text{as}},e_3^{\text{as}}] +\slashed{g}_{a,M}^{BC}(\overset{\text{\scalebox{.6}{$(1)$}}}{\mathfrak{\slashed{\mathfrak{f}}}}_{A})_{B}\,[e_4^{\text{as}}, e_C^{\text{as}}]\right)  + \mathcal{O}(\epsilon^2) \, , \nonumber 
\end{align}
\begin{align}
\delta([\et,\ea])(\epsilon)=& \, \epsilon\cdot\left(\left( (\nablasl^{a,M}_{e_3^{\text{as}}}(\overset{\text{\scalebox{.6}{$(1)$}}}{\mathfrak{\slashed{\mathfrak{f}}}}_{3})_A)
  - (\nablasl^{a,M}_{e_A^{\text{as}}} \overset{\text{\scalebox{.6}{$(1)$}}}{\mathfrak{\underline{\mathfrak{f}}}}_{3}) \right) e_4^{\text{as}} \right.\\
&+\left( (\nablasl^{a,M}_{e_3^{\text{as}}}(\overset{\text{\scalebox{.6}{$(1)$}}}{\mathfrak{\slashed{\mathfrak{f}}}}_{4})_A)  -( \nablasl^{a,M}_{e_A^{\text{as}}}\overset{\text{\scalebox{.6}{$(1)$}}}{\mathfrak{\underline{\mathfrak{f}}}}_{4}) \right) e_3^{\text{as}} \nonumber \\
&+\slashed{g}_{a,M}^{BC}\left( (\nablasl^{a,M}_{e_3^{\text{as}}}(\overset{\text{\scalebox{.6}{$(1)$}}}{\mathfrak{\slashed{\mathfrak{f}}}}_{A})_C) - (\nablasl^{a,M}_{e_A^{\text{as}}}\overset{\text{\scalebox{.6}{$(1)$}}}{\mathfrak{\underline{\mathfrak{f}}}}_C)  \right) e_B^{\text{as}} \nonumber \\
&  + (\overset{\text{\scalebox{.6}{$(1)$}}}{\mathfrak{\slashed{\mathfrak{f}}}}_{3})_A\, [e_3^{\text{as}},e_4^{\text{as}}]  + \overset{\text{\scalebox{.6}{$(1)$}}}{\mathfrak{\underline{\mathfrak{f}}}}_{3} \, [e_4^{\text{as}},e_A^{\text{as}}] +\slashed{g}_{a,M}^{BC}\overset{\text{\scalebox{.6}{$(1)$}}}{\mathfrak{\underline{\mathfrak{f}}}}_B \, [e_C^{\text{as}} , e_A^{\text{as}} ]  \nonumber \\
& \left. +\left( \overset{\text{\scalebox{.6}{$(1)$}}}{\mathfrak{\underline{\mathfrak{f}}}}_{4}\,\delta_A^C+\slashed{g}_{a,M}^{BC}(\overset{\text{\scalebox{.6}{$(1)$}}}{\mathfrak{\slashed{\mathfrak{f}}}}_{A})_B \right)\, [e_3^{\text{as}}, e_C^{\text{as}}]  \right)+ \mathcal{O}(\epsilon^2)   \, .  \nonumber
\end{align}
\end{prop} 

\medskip

\begin{proof}
We linearise the formulae of Proposition \ref{prop_decomp_commutators_nonlinear}. Before linearising, we use Proposition \ref{prop_frame_commutators_ricci_1} to write
\begin{align*}
(\bsy{\mathfrak{\slashed{\mathfrak{f}}}}_{4})_{\bsy{A}}(\epsilon)&=((\bsy{\mathfrak{\slashed{\mathfrak{f}}}}_{\bsy{A}})(\epsilon),(\widetilde{\bsy{\mathfrak{\slashed{\mathfrak{f}}}}_{4}})(\epsilon))_{\slashed{g}_{a,M}}+\textup{nonlinear terms} \, , \\
(\bsy{\mathfrak{\slashed{\mathfrak{f}}}}_{3})_{\bsy{A}}(\epsilon)&=((\bsy{\mathfrak{\slashed{\mathfrak{f}}}}_{\bsy{A}})(\epsilon),(\widetilde{\bsy{\mathfrak{\slashed{\mathfrak{f}}}}_{3}})(\epsilon))_{\slashed{g}_{a,M}}+\textup{nonlinear terms} 
\end{align*}
in the formulae of Proposition \ref{prop_decomp_commutators_nonlinear}.

\end{proof}

\medskip

By linearising the right hand side of the frame commutators \eqref{frame_commutator_3A}, \eqref{frame_commutator_4A} and \eqref{frame_commutator_34}, one can easily prove the following proposition.~The proof is an easy computation that employs Proposition \ref{prop_frame_commutators_ricci_1} in the same way as it is used in the proof of Proposition \ref{prop_linearisation_frame_commutators_1}.

\medskip

\begin{prop} \label{prop_linearisation_frame_commutators_2}
We have the identities
\begin{align}
\delta([\ef,\et])(\epsilon)=& \, \epsilon\cdot\left(  -\left(2\,\slashed{g}_{a,M}^{AB}(\overset{\text{\scalebox{.6}{$(1)$}}}{\slashed{\mathfrak{f}}}_B,\eta-\etab) +2\overset{\text{\scalebox{.6}{$(1)$}}}{\eta}{}^A  +2\,(\eta-\etab)^C\slashed{g}_{a,M}^{BA}(\overset{\text{\scalebox{.6}{$(1)$}}}{\mathfrak{\slashed{\mathfrak{f}}}}_{C})_{B} +\omegah\overset{\text{\scalebox{.6}{$(1)$}}}{\mathfrak{\underline{\mathfrak{f}}}}{}^{A}\right) e_A^{\text{as}} \right. \\
&-\left(\omegah\overset{\text{\scalebox{.6}{$(1)$}}}{\mathfrak{\underline{\mathfrak{f}}}}_{4}+2\,(\eta-\etab,\overset{\text{\scalebox{.6}{$(1)$}}}{\mathfrak{\slashed{\mathfrak{f}}}}_{4})\right) e_3^{\text{as}}   \nonumber\\ 
& \left.+\left( \overset{\text{\scalebox{.6}{$(1)$}}}{\omegab} -\omegah \overset{\text{\scalebox{.6}{$(1)$}}}{\mathfrak{\underline{\mathfrak{f}}}}_{3}  -2\,(\eta-\etab,\overset{\text{\scalebox{.6}{$(1)$}}}{\mathfrak{\slashed{\mathfrak{f}}}}_{3}) \right) e_4^{\text{as}} \right) + \mathcal{O}(\epsilon^2) \, ,  \nonumber 
\end{align}
\begin{align}
\delta([\ef,\ea])(\epsilon)=& \, \epsilon\cdot\left( \left(\overset{\text{\scalebox{.6}{$(1)$}}}{\slashed{M}}{}_A^B-\overset{\text{\scalebox{.6}{$(1)$}}}{\chih}{}_A^B- \frac{1}{2}\,(\overset{\text{\scalebox{.6}{$(1)$}}}{\textup{tr}\chi})\,\delta_A^B-\frac{1}{2}\,(\overset{\text{\scalebox{.6}{$(1)$}}}{\slashed{\varepsilon}\cdot\chi})\,\slashed{\varepsilon}^{\sharp_2}{}_A^B+(\slashed{M}{}_A^D-\chi^{\sharp_2}{}_A^D)\,\slashed{g}_{a,M}^{CB}(\overset{\text{\scalebox{.6}{$(1)$}}}{\mathfrak{\slashed{\mathfrak{f}}}}_{D})_{C} \right)e_B^{\text{as}} \right. \\
&+(\slashed{M}{}_A^B-\chi^{\sharp_2}{}_A^B)(\overset{\text{\scalebox{.6}{$(1)$}}}{\mathfrak{\slashed{\mathfrak{f}}}}_{4})_{B}\, e_3^{\text{as}}   \nonumber\\
&\left. +\left((\slashed{M}{}_A^B-\chi^{\sharp_2}{}_A^B)(\overset{\text{\scalebox{.6}{$(1)$}}}{\mathfrak{\slashed{\mathfrak{f}}}}_{3})_{B} +  (\overset{\text{\scalebox{.6}{$(1)$}}}{\slashed{\mathfrak{f}}}_A,\etab+\zeta)+\overset{\text{\scalebox{.6}{$(1)$}}}{\zeta}_A \right)e^{\text{as}}_4 \right)  + \mathcal{O}(\epsilon^2) \, , \nonumber 
\end{align}
\begin{align}
\delta([\et,\ea])(\epsilon)=& \, \epsilon\cdot\left(  \left(\overset{\text{\scalebox{.6}{$(1)$}}}{\underline{\slashed{M}}}{}_A^B-\overset{\text{\scalebox{.6}{$(1)$}}}{\chibh}{}_A^B- \frac{1}{2}\,(\overset{\text{\scalebox{.6}{$(1)$}}}{\textup{tr}\chib})\,\delta_A^B-\frac{1}{2}\,(\overset{\text{\scalebox{.6}{$(1)$}}}{\slashed{\varepsilon}\cdot\chib})\,\slashed{\varepsilon}^{\sharp_2}{}_A^B +(\underline{\slashed{M}}{}_A^D-\chib^{\sharp_2}{}_A^D)\slashed{g}_{a,M}^{CB}(\overset{\text{\scalebox{.6}{$(1)$}}}{\mathfrak{\slashed{\mathfrak{f}}}}_{D})_{C} \right)e_B^{\text{as}} \right. \\
&+\left( (\underline{\slashed{M}}{}_A^B-\chib^{\sharp_2}{}_A^B)(\overset{\text{\scalebox{.6}{$(1)$}}}{\mathfrak{\slashed{\mathfrak{f}}}}_{4})_{B}  + \overset{\text{\scalebox{.6}{$(1)$}}}{\eta}_A-\overset{\text{\scalebox{.6}{$(1)$}}}{\zeta}_A \right) e^{\text{as}}_3 \nonumber\\
&\left. +\left((\underline{\slashed{M}}{}_A^B-\chib^{\sharp_2}{}_A^B)(\overset{\text{\scalebox{.6}{$(1)$}}}{\mathfrak{\slashed{\mathfrak{f}}}}_{3})_{B}+\overset{\text{\scalebox{.6}{$(1)$}}}{\yb}_A \right) e_4^{\text{as}}   \right)+ \mathcal{O}(\epsilon^2)   \, .  \nonumber
\end{align}
\end{prop}

\subsection{Contractions and products}

We have the following proposition for the linearisation of the renormalised contractions and products of $\mathfrak{D}_{\mathcal{N}_{\text{as}}}$ tensors.~The proof is a computation which, in particular, exploits Proposition \ref{prop_identities_linearised_metric_quantities} to rearrange several of the terms in the identities.

\medskip

\begin{prop}
For any $\bsy{\mathfrak{D}}_{\bsy{\mathcal{N}}(\epsilon)}$ one-tensors $\bsy{\varsigma}(\epsilon)$ and $\bsy{\theta}(\epsilon)$ and $\bsy{\mathfrak{D}}_{\bsy{\mathcal{N}}(\epsilon)}$ two-tensor $\bsy{\omega}(\epsilon)$, we have 
\begin{equation}
\delta(\widetilde{\bsy{\varsigma}}\otimes \widetilde{\bsy{\theta}})(\epsilon)=\epsilon\cdot(\overset{\text{\scalebox{.6}{$(1)$}}}{\varsigma}\otimes \theta+\varsigma\otimes\overset{\text{\scalebox{.6}{$(1)$}}}{\theta})+\mathcal{O}(\epsilon^2)
\end{equation}
and the identity between smooth scalar functions 
\begin{align}
\delta(\textbf{tr}\,{}_{\widetilde{\bsy{\slashed{g}}}}\,\widetilde{\bsy{\omega}})(\epsilon)  =& \, \epsilon\cdot \left(\textup{tr}\,\overset{\text{\scalebox{.6}{$(1)$}}}{\omega}- (\omega,\overset{\text{\scalebox{.6}{$(1)$}}}{\widehat{\slashed{g}}}) - \frac{1}{2}\,(\textup{tr}\,\omega)(\textup{tr}\overset{\text{\scalebox{.6}{$(1)$}}}{\slashed{g}}) \right)+\mathcal{O}(\epsilon^2)  \, , \\
\delta((\widetilde{\bsy{\varsigma}},\widetilde{\bsy{\theta}})_{\widetilde{\bsy{\slashed{g}}}})(\epsilon)=& \, \epsilon\cdot\left((\overset{\text{\scalebox{.6}{$(1)$}}}{\varsigma},\theta) +(\varsigma,\overset{\text{\scalebox{.6}{$(1)$}}}{\theta}) -\frac{1}{2}\,(\varsigma\,\widehat{\otimes}\,\theta, \overset{\text{\scalebox{.6}{$(1)$}}}{\widehat{\slashed{g}}}) -\frac{1}{2}\,(\textup{tr}\overset{\text{\scalebox{.6}{$(1)$}}}{\slashed{g}})\,(\varsigma,\theta) \right)+\mathcal{O}(\epsilon^2) \, ,  \\
\delta(\widetilde{\bsy{\varsigma}}\bsy{\wedge}_{\widetilde{\bsy{\slashed{g}}}}\,\widetilde{\bsy{\theta}})(\epsilon)=& \, \epsilon\cdot \left( \overset{\text{\scalebox{.6}{$(1)$}}}{\varsigma}\wedge\theta+\varsigma\wedge\overset{\text{\scalebox{.6}{$(1)$}}}{\theta} +\frac{1}{2}\,({}^{\star}\varsigma\,\widehat{\otimes}\,\theta,\overset{\text{\scalebox{.6}{$(1)$}}}{\widehat{\slashed{g}}}) -\frac{1}{2}\,(\varsigma\,\widehat{\otimes}\,{}^{\star}\theta,\overset{\text{\scalebox{.6}{$(1)$}}}{\widehat{\slashed{g}}}) -\frac{1}{2}\,(\textup{tr}\overset{\text{\scalebox{.6}{$(1)$}}}{\slashed{g}})\,\varsigma \wedge \theta \right)+\mathcal{O}(\epsilon^2) 
\end{align}
and the identity
\begin{align}
\delta(\widetilde{\bsy{\varsigma}}\,\widehat{\otimes}_{\widetilde{\bsy{\slashed{g}}}}\, \widetilde{\bsy{\theta}})(\epsilon)=& \, \epsilon\cdot\left( \overset{\text{\scalebox{.6}{$(1)$}}}{\varsigma}\widehat{\otimes}\,\theta+\varsigma \,\widehat{\otimes}\overset{\text{\scalebox{.6}{$(1)$}}}{\theta}-(\varsigma,\theta) \overset{\text{\scalebox{.6}{$(1)$}}}{\widehat{\slashed{g}}} + \frac{1}{2}\,(\varsigma\,\widehat{\otimes}\,\theta, \overset{\text{\scalebox{.6}{$(1)$}}}{\widehat{\slashed{g}}})\,\slashed{g}_{a,M} \right)+\mathcal{O}(\epsilon^2) \, .  \label{linearisation_otimes_hat_tensors}
\end{align}
For any $\bsy{\mathfrak{D}}_{\bsy{\mathcal{N}}(\epsilon)}$ one-tensor $\bsy{\varsigma}(\epsilon)$ and $\bsy{\mathfrak{D}}_{\bsy{\mathcal{N}}(\epsilon)}$ two-tensor $\bsy{\theta}(\epsilon)$, we have
\begin{align}
\delta(\widetilde{\bsy{\theta}}{}^{\bsy{\sharp_2}} \bsy{\cdot}_{\widetilde{\bsy{\slashed{g}}}} \widetilde{\bsy{\varsigma}})(\epsilon)&=\epsilon\cdot \left( \theta^{\sharp_2}\cdot\overset{\text{\scalebox{.6}{$(1)$}}}{\varsigma} +\overset{\text{\scalebox{.6}{$(1)$}}}{\theta}{}^{\sharp_2}\cdot\varsigma-(\theta\times\overset{\text{\scalebox{.6}{$(1)$}}}{\widehat{\slashed{g}}})^{\sharp_2}\cdot\varsigma-\frac{1}{2}\,(\textup{tr}\overset{\text{\scalebox{.6}{$(1)$}}}{\slashed{g}})\,\theta^{\sharp_2}\cdot\varsigma  \right) + \mathcal{O}(\epsilon^2)  \, , \\
\delta(\widetilde{{}^{\bsy{\star}}\bsy{\varsigma}})(\epsilon)&= \epsilon\cdot \left( {}^{\star}\overset{\text{\scalebox{.6}{$(1)$}}}{\varsigma} -({}^{\star}\overset{\text{\scalebox{.6}{$(1)$}}}{\widehat{\slashed{g}}})^{\sharp_1}\cdot\varsigma  \right) +\mathcal{O}(\epsilon^2) \, .
\end{align}
\end{prop}

\subsection{Covariant and Lie derivatives}

We have the following propositions for the linearisation of the renormalised Lie derivatives and covariant derivatives of scalar functions and $\mathfrak{D}_{\mathcal{N}_{\text{as}}}$ covariant tensors.

\medskip

\begin{prop}  \label{prop_linearisation_lie_derivatives}
For any $\bsy{\mathfrak{D}}_{\bsy{\mathcal{N}}(\epsilon)}$ one-tensor $\bsy{\varsigma}(\epsilon)$, we have
\begin{align}
\delta(\widetilde{\bsy{\slashed{\mathcal{L}}}}_{\bsy{4}}\,\widetilde{\bsy{\varsigma}})(\epsilon)=& \,  \epsilon\cdot \slashed{\mathcal{L}}_4\,\overset{\text{\scalebox{.6}{$(1)$}}}{\varsigma} +\mathcal{O}(\epsilon^2)  \, ,  \label{lin_lie_deriv_4_cov_tensor}\\
\delta(\widetilde{\bsy{\slashed{\mathcal{L}}}}_{\bsy{3}}\,\widetilde{\bsy{\varsigma}})(\epsilon)=& \, \epsilon\cdot \left(\nablasl_3\overset{\text{\scalebox{.6}{$(1)$}}}{\varsigma}+\chib^{\sharp_2}\cdot\overset{\text{\scalebox{.6}{$(1)$}}}{\varsigma} +(\nablasl\varsigma)^{\sharp_1}\cdot \overset{\text{\scalebox{.6}{$(1)$}}}{\underline{\mathfrak{f}}} \right. \\ 
&-(\slashed{\mathcal{D}}_2^{\star}\overset{\text{\scalebox{.6}{$(1)$}}}{\underline{\mathfrak{f}}})^{\sharp_2}\cdot \varsigma+\frac{1}{2}\,(\slashed{\textup{div}}\overset{\text{\scalebox{.6}{$(1)$}}}{\underline{\mathfrak{f}}})\,\varsigma+\frac{1}{2}\,(\slashed{\textup{curl}}\overset{\text{\scalebox{.6}{$(1)$}}}{\underline{\mathfrak{f}}})\,{}^{\star}\varsigma  \nonumber \\
& \left. +(\nablasl_4\varsigma+\chi^{\sharp_2}\cdot\varsigma)\overset{\text{\scalebox{.6}{$(1)$}}}{\underline{\mathfrak{f}}}_3+ (\nablasl_3\varsigma+\chib^{\sharp_2}\cdot\,\varsigma)\overset{\text{\scalebox{.6}{$(1)$}}}{\underline{\mathfrak{f}}}_4 \right)+\mathcal{O}(\epsilon^2) \, . \nonumber
\end{align}
We have the identities
\begin{align}
\delta(\widetilde{\bsy{\slashed{\mathcal{L}}}}_{\bsy{4}}\,\widetilde{\bsy{\slashed{g}}})(\epsilon) =& \, \epsilon\cdot \left(\slashed{\mathcal{L}}_4\,\overset{\text{\scalebox{.6}{$(1)$}}}{\widehat{\slashed{g}}}+\frac{1}{2}\,(\slashed{\mathcal{L}}_4\,(\textup{tr}\overset{\text{\scalebox{.6}{$(1)$}}}{\slashed{g}}))\,\slashed{g}_{a,M}+\frac{1}{2}\,(\textup{tr}\chi)(\textup{tr}\overset{\text{\scalebox{.6}{$(1)$}}}{\slashed{g}})\,\slashed{g}_{a,M} \right)+\mathcal{O}(\epsilon^2)  \, ,  \label{lin_lie_deriv_4_metric}\\
\delta(\widetilde{\bsy{\slashed{\mathcal{L}}}}_{\bsy{3}}\,\widetilde{\bsy{\slashed{g}}})(\epsilon)  =& \, \epsilon\cdot \left(\nablasl_3\overset{\text{\scalebox{.6}{$(1)$}}}{\widehat{\slashed{g}}}+\frac{1}{2}\,(\nablasl_3(\textup{tr}\overset{\text{\scalebox{.6}{$(1)$}}}{\slashed{g}}))\,\slashed{g}_{a,M} \right. \\ & +(\textup{tr}\chib)\overset{\text{\scalebox{.6}{$(1)$}}}{\widehat{\slashed{g}}}+(\slashed{\varepsilon}\cdot\chib)\,{}^{\star}\overset{\text{\scalebox{.6}{$(1)$}}}{\widehat{\slashed{g}}}+\frac{1}{2}\,(\textup{tr}\chib)(\textup{tr}\overset{\text{\scalebox{.6}{$(1)$}}}{\slashed{g}})\,\slashed{g}_{a,M} \nonumber\\ & \left. -2\,(\slashed{\mathcal{D}}_2^{\star}\overset{\text{\scalebox{.6}{$(1)$}}}{\underline{\mathfrak{f}}})+(\slashed{\textup{div}}\overset{\text{\scalebox{.6}{$(1)$}}}{\underline{\mathfrak{f}}})\,\slashed{g}_{a,M}   +\overset{\text{\scalebox{.6}{$(1)$}}}{\underline{\mathfrak{f}}}_3(\textup{tr}\chi)\,\slashed{g}_{a,M}+\overset{\text{\scalebox{.6}{$(1)$}}}{\underline{\mathfrak{f}}}_4(\textup{tr}\chib)\,\slashed{g}_{a,M} \right)+\mathcal{O}(\epsilon^2) \nonumber \, . 
\end{align}
\end{prop}

\medskip

\begin{proof}
The proof is immediate by linearising the identities of Proposition \ref{prop_aux_restriction_lie_derivative}.~Note that the differential operator $\widetilde{\bsy{\slashed{\mathcal{L}}}}_{\bsy{4}}$ does \emph{not} depend on $\epsilon$.

\end{proof}

\medskip

\begin{remark}
One can re-write the identities \eqref{lin_lie_deriv_4_cov_tensor} and \eqref{lin_lie_deriv_4_metric} as
\begin{align}
\delta(\widetilde{\bsy{\slashed{\mathcal{L}}}}_{\bsy{4}}\,\widetilde{\bsy{\varsigma}})(\epsilon)=& \,  \epsilon\cdot \left(\nablasl_4\,\overset{\text{\scalebox{.6}{$(1)$}}}{\varsigma}+\chi^{\sharp_2}\cdot\overset{\text{\scalebox{.6}{$(1)$}}}{\varsigma} \right)+\mathcal{O}(\epsilon^2) \, , \\
\delta(\widetilde{\bsy{\slashed{\mathcal{L}}}}_{\bsy{4}}\,\widetilde{\bsy{\slashed{g}}})(\epsilon) =& \, \epsilon\cdot \left(\nablasl_4\overset{\text{\scalebox{.6}{$(1)$}}}{\widehat{\slashed{g}}}+\frac{1}{2}\,(\nablasl_4(\textup{tr}\overset{\text{\scalebox{.6}{$(1)$}}}{\slashed{g}}))\,\slashed{g}_{a,M}  \right. \\
& \left. +(\textup{tr}\chi)\overset{\text{\scalebox{.6}{$(1)$}}}{\widehat{\slashed{g}}}+(\slashed{\varepsilon}\cdot\chi)\,{}^{\star}\overset{\text{\scalebox{.6}{$(1)$}}}{\widehat{\slashed{g}}}+\frac{1}{2}\,(\textup{tr}\chi)(\textup{tr}\overset{\text{\scalebox{.6}{$(1)$}}}{\slashed{g}})\,\slashed{g}_{a,M} \right)+\mathcal{O}(\epsilon^2) \, . \nonumber 
\end{align}
\end{remark}

\medskip

\begin{prop}  \label{prop_linearisation_cov_derivatives}
For any smooth scalar function $\bsy{f}(\epsilon)$, we have
\begin{equation} \label{linearisation_ang_cov_deriv_function}
\delta(\widetilde{\bsy{\nablasl}}\,\bsy{f})(\epsilon)=\epsilon\cdot(\nablasl\,\overset{\text{\scalebox{.6}{$(1)$}}}{f})+\mathcal{O}(\epsilon^2) \, .
\end{equation}
For any $\bsy{\mathfrak{D}}_{\bsy{\mathcal{N}}(\epsilon)}$ one-tensor $\bsy{\varsigma}(\epsilon)$, we have
\begin{align}
\delta(\widetilde{\bsy{\nablasl}}\,\widetilde{\bsy{\varsigma}})(\epsilon)=& \, \epsilon\cdot \left(-\slashed{\mathcal{D}}_2^{\star}\overset{\text{\scalebox{.6}{$(1)$}}}{\varsigma}+\frac{1}{2}\,(\slashed{\textup{div}}\overset{\text{\scalebox{.6}{$(1)$}}}{\varsigma})\,\slashed{g}_{a,M}+\frac{1}{2}\,(\slashed{\textup{curl}}\overset{\text{\scalebox{.6}{$(1)$}}}{\varsigma})\,\slashed{\varepsilon}_{a,M}  \right. \label{linearisation_ang_cov_deriv_tensor}\\ &- \frac{1}{2}\,\widehat{(\nablasl\overset{\text{\scalebox{.6}{$(1)$}}}{\widehat{\slashed{g}}})^{\sharp_3}\cdot\varsigma} - \frac{1}{2}\,((\slashed{\textup{div}}\overset{\text{\scalebox{.6}{$(1)$}}}{\widehat{\slashed{g}}})^{\sharp}\cdot\varsigma)\,\slashed{g}_{a,M}+\frac{1}{2}\,(\nablasl\overset{\text{\scalebox{.6}{$(1)$}}}{\widehat{\slashed{g}}})^{\sharp_1}\cdot\varsigma  -\frac{1}{4}\,\varsigma\,\widehat{\otimes}\,(\nablasl(\textup{tr}\overset{\text{\scalebox{.6}{$(1)$}}}{\slashed{g}})) \nonumber\\
&-\frac{1}{4}\,(\textup{tr}\chi)\,(-2\overset{\text{\scalebox{.6}{$(1)$}}}{\mathfrak{\slashed{\mathfrak{f}}}}_{3}+\overset{\text{\scalebox{.6}{$(1)$}}}{\mathfrak{\underline{\mathfrak{f}}}},\varsigma)\,\slashed{g}_{a,M}   
-\frac{1}{4}\,(\slashed{\varepsilon}\cdot\chi)\,(-2\overset{\text{\scalebox{.6}{$(1)$}}}{\mathfrak{\slashed{\mathfrak{f}}}}_{3}+\overset{\text{\scalebox{.6}{$(1)$}}}{\mathfrak{\underline{\mathfrak{f}}}},{}^{\star}\varsigma)\,\slashed{g}_{a,M}  \nonumber\\
&-\frac{1}{4}\,(\slashed{\varepsilon}\cdot\chi)\,({}^{\star}\varsigma)\,\widehat{\otimes}\,(-2\overset{\text{\scalebox{.6}{$(1)$}}}{\mathfrak{\slashed{\mathfrak{f}}}}_{3}+\overset{\text{\scalebox{.6}{$(1)$}}}{\mathfrak{\underline{\mathfrak{f}}}})  +\frac{1}{2}\,(\slashed{\varepsilon}\cdot\chib)\,({}^{\star}\varsigma)\,\widehat{\otimes}\, \overset{\text{\scalebox{.6}{$(1)$}}}{\mathfrak{\slashed{\mathfrak{f}}}}_{4} \nonumber\\
& \left. +\frac{1}{2}\,(\textup{tr}\chib)(\overset{\text{\scalebox{.6}{$(1)$}}}{\mathfrak{\slashed{\mathfrak{f}}}}_{4},\varsigma)\,\slashed{g}_{a,M}  
+\frac{1}{2}\,(\slashed{\varepsilon}\cdot\chib)(\overset{\text{\scalebox{.6}{$(1)$}}}{\mathfrak{\slashed{\mathfrak{f}}}}_{4}, {}^{\star}\varsigma)\,\slashed{g}_{a,M}  \right) +\mathcal{O}(\epsilon^2) \, . \nonumber 
\end{align}
\end{prop}

\medskip

\begin{proof}
To prove the identity \eqref{linearisation_ang_cov_deriv_function}, it suffices to observe that the difference of two linear connections over $\mathfrak{D}_{\mathcal{N}_{\text{as}}}$ vanishes when applied to a scalar function.~To prove the identity \eqref{linearisation_ang_cov_deriv_tensor}, we compute
\begin{align}
(\widetilde{\bsy{\nablasl}}\,\widetilde{\bsy{\varsigma}})_{AB}&=(\bsy{\nabla}(\Pi\bsy{\varsigma}^{\circ}))_{AB} \nonumber\\
&=\bsy{\nabla}_A((\Pi\bsy{\varsigma}^{\circ})_B)-(\Pi\bsy{\varsigma}^{\circ})(\bsy{\nabla}_Ae_B) \nonumber\\
&=\bsy{\nabla}_A((\Pi\bsy{\varsigma}^{\circ})_B)-\slashed{g}_{a,M}^{CD}\,g_{a,M}(\bsy{\nabla}_Ae_B,e_C)\,(\Pi\bsy{\varsigma}^{\circ})_D  \nonumber\\
&=\bsy{\nabla}_A((\Pi\bsy{\varsigma}^{\circ})_B)-\slashed{g}_{a,M}^{CD}\,g_{a,M}((\bsy{\nabla}-\nabla_{a,M})_Ae_B,e_C)\,(\Pi\bsy{\varsigma}^{\circ})_D-\slashed{g}_{a,M}^{CD}\,g_{a,M}(\nabla_Ae_B,e_C)\,(\Pi\bsy{\varsigma}^{\circ})_D \nonumber\\
&=(\nablasl\,\widetilde{\bsy{\varsigma}})_{AB}-\slashed{g}_{a,M}^{CD}\,g_{a,M}((\bsy{\nabla}-\nabla_{a,M})_Ae_B,e_C)\,\widetilde{\bsy{\varsigma}}_D  \, . \label{aux_lin_connection}
\end{align}
Note that the second term in \eqref{aux_lin_connection} is tensorial in the indices $A$ and $B$, as one expects for the difference of two linear connections.~We have
\begin{align*}
g_{a,M}((\bsy{\nabla}-\nabla_{a,M})_Ae_B,e_C) =&\, \epsilon\cdot\left(\frac{1}{2}\,(\nablasl_{B}\overset{\text{\scalebox{.6}{$(1)$}}}{\widehat{\slashed{g}}})_{AC}+\frac{1}{2}\,(\nablasl_{A}\overset{\text{\scalebox{.6}{$(1)$}}}{\widehat{\slashed{g}}})_{BC}-\frac{1}{2}\,(\nablasl_{C}\overset{\text{\scalebox{.6}{$(1)$}}}{\widehat{\slashed{g}}})_{AB} \right. \\
&+\frac{1}{4}\,(\nablasl_{B}(\text{tr}\overset{\text{\scalebox{.6}{$(1)$}}}{\slashed{g}}))(\slashed{g}_{a,M})_{AC}+\frac{1}{4}\,(\nablasl_{A}(\text{tr}\overset{\text{\scalebox{.6}{$(1)$}}}{\slashed{g}}))(\slashed{g}_{a,M})_{BC}-\frac{1}{4}\,(\nablasl_{C}(\text{tr}\overset{\text{\scalebox{.6}{$(1)$}}}{\slashed{g}}))(\slashed{g}_{a,M})_{AB} \\
&+\frac{1}{4}\,(\text{tr}\chi)\,(-2(\overset{\text{\scalebox{.6}{$(1)$}}}{\mathfrak{\slashed{\mathfrak{f}}}}_{3})_C+\overset{\text{\scalebox{.6}{$(1)$}}}{\mathfrak{\underline{\mathfrak{f}}}}_{C})(\slashed{g}_{a,M})_{AB}  \\ 
&+\frac{1}{4}\,(\slashed{\varepsilon}\cdot\chi)\,(-2(\overset{\text{\scalebox{.6}{$(1)$}}}{\mathfrak{\slashed{\mathfrak{f}}}}_{3})_A+\overset{\text{\scalebox{.6}{$(1)$}}}{\mathfrak{\underline{\mathfrak{f}}}}_{A})(\slashed{\varepsilon}_{a,M})_{BC}  \\
&+\frac{1}{4}\,(\slashed{\varepsilon}\cdot\chi)\,(-2(\overset{\text{\scalebox{.6}{$(1)$}}}{\mathfrak{\slashed{\mathfrak{f}}}}_{3})_B+\overset{\text{\scalebox{.6}{$(1)$}}}{\mathfrak{\underline{\mathfrak{f}}}}_{B})(\slashed{\varepsilon}_{a,M})_{AC} \\
&-\frac{1}{2}\,(\text{tr}\chib)(\overset{\text{\scalebox{.6}{$(1)$}}}{\mathfrak{\slashed{\mathfrak{f}}}}_{4})_C(\slashed{g}_{a,M})_{AB}\\ 
&-\frac{1}{2}\,(\slashed{\varepsilon}\cdot\chib)(\overset{\text{\scalebox{.6}{$(1)$}}}{\mathfrak{\slashed{\mathfrak{f}}}}_{4})_A(\slashed{\varepsilon}_{a,M})_{BC}   \\
&\left. -\frac{1}{2}\,(\slashed{\varepsilon}\cdot\chib)\,(\overset{\text{\scalebox{.6}{$(1)$}}}{\mathfrak{\slashed{\mathfrak{f}}}}_{4})_B(\slashed{\varepsilon}_{a,M})_{AC}\right)+\mathcal{O}(\epsilon^2) \, .
\end{align*}
The identity \eqref{linearisation_ang_cov_deriv_tensor} follows by a simple rearrangement.

\end{proof}

\medskip

\begin{remark}
Each of the terms in the formula \eqref{linearisation_ang_cov_deriv_tensor} is either a symmetric traceless, pure trace (relative to $\slashed{g}_{a,M}^{-1}$) or antisymmetric $\mathfrak{D}_{\mathcal{N}_{\text{as}}}$ covariant tensor.
\end{remark}

\medskip

Proposition \ref{prop_linearisation_cov_derivatives} can be used to compute the linearisation formulae for \emph{all} the renormalised differential operators.~In the following proposition, we collect the ones that will be necessary in the sequel.

\medskip

\begin{prop}
For any $\bsy{\mathfrak{D}}_{\bsy{\mathcal{N}}(\epsilon)}$ one-tensor $\bsy{\varsigma}(\epsilon)$, we have
\begin{align}
\delta(\widetilde{\slashed{\textbf{div}}}\,\widetilde{\bsy{\varsigma}})(\epsilon) =& \, \epsilon\cdot \left(\slashed{\textup{div}}\overset{\text{\scalebox{.6}{$(1)$}}}{\varsigma} + (\slashed{\mathcal{D}}_2^{\star}\,\varsigma,\overset{\text{\scalebox{.6}{$(1)$}}}{\widehat{\slashed{g}}})   - (\slashed{\textup{div}}\overset{\text{\scalebox{.6}{$(1)$}}}{\widehat{\slashed{g}}})^{\sharp}\cdot\varsigma -\frac{1}{2}\,(\slashed{\textup{div}}\,\varsigma)(\textup{tr}\overset{\text{\scalebox{.6}{$(1)$}}}{\slashed{g}}) \right. \label{linearisation_div_tensor}\\ & 
-\frac{1}{2}\,(\textup{tr}\chi)\,(-2\overset{\text{\scalebox{.6}{$(1)$}}}{\mathfrak{\slashed{\mathfrak{f}}}}_{3}+\overset{\text{\scalebox{.6}{$(1)$}}}{\mathfrak{\underline{\mathfrak{f}}}},\varsigma) -\frac{1}{2}\,(\slashed{\varepsilon}\cdot\chi)\,(-2\overset{\text{\scalebox{.6}{$(1)$}}}{\mathfrak{\slashed{\mathfrak{f}}}}_{3}+\overset{\text{\scalebox{.6}{$(1)$}}}{\mathfrak{\underline{\mathfrak{f}}}},{}^{\star}\varsigma)   \nonumber\\
& \left. +(\textup{tr}\chib)(\overset{\text{\scalebox{.6}{$(1)$}}}{\mathfrak{\slashed{\mathfrak{f}}}}_{4},\varsigma)  
+(\slashed{\varepsilon}\cdot\chib)(\overset{\text{\scalebox{.6}{$(1)$}}}{\mathfrak{\slashed{\mathfrak{f}}}}_{4}, {}^{\star}\varsigma) \right) +\mathcal{O}(\epsilon^2) \, , \nonumber \\
\delta(\widetilde{\slashed{\textbf{curl}}}\,\widetilde{\bsy{\varsigma}})(\epsilon) =& \, \epsilon\cdot \left(  \slashed{\textup{curl}}\overset{\text{\scalebox{.6}{$(1)$}}}{\varsigma}  -\frac{1}{2}\,(\textup{tr}\overset{\text{\scalebox{.6}{$(1)$}}}{\slashed{g}})(\slashed{\textup{curl}}\,\varsigma)   \right)+\mathcal{O}(\epsilon^2)   \, ,  \label{linearisation_curl_tensor}\\
-2\,\delta(\widetilde{\bsy{\slashed{\mathcal{D}}}_{\bsy{2}}^{\star}}\,\widetilde{\bsy{\varsigma}})(\epsilon) =& \, \epsilon\cdot \left(-2\,(\slashed{\mathcal{D}}_2^{\star}\overset{\text{\scalebox{.6}{$(1)$}}}{\varsigma}) -(\slashed{\textup{div}}\,\varsigma)\overset{\text{\scalebox{.6}{$(1)$}}}{\widehat{\slashed{g}}}  - (\slashed{\mathcal{D}}_2^{\star}\,\varsigma ,\overset{\text{\scalebox{.6}{$(1)$}}}{\widehat{\slashed{g}}})\,\slashed{g}_{a,M}  \label{linearisation_D2star_tensor} \right. \\ & - \widehat{(\nablasl\overset{\text{\scalebox{.6}{$(1)$}}}{\widehat{\slashed{g}}})^{\sharp_3}\cdot\varsigma}+ (\nablasl\overset{\text{\scalebox{.6}{$(1)$}}}{\widehat{\slashed{g}}})^{\sharp_1}\cdot\varsigma  -\frac{1}{2}\,\varsigma\,\widehat{\otimes}\,(\nablasl(\textup{tr}\overset{\text{\scalebox{.6}{$(1)$}}}{\slashed{g}})) \nonumber\\
& \left. -\frac{1}{2}\,(\slashed{\varepsilon}\cdot\chi)\,({}^{\star}\varsigma)\,\widehat{\otimes}\,(-2\overset{\text{\scalebox{.6}{$(1)$}}}{\mathfrak{\slashed{\mathfrak{f}}}}_{3}+\overset{\text{\scalebox{.6}{$(1)$}}}{\mathfrak{\underline{\mathfrak{f}}}})  + (\slashed{\varepsilon}\cdot\chib)\,({}^{\star}\varsigma)\,\widehat{\otimes}\, \overset{\text{\scalebox{.6}{$(1)$}}}{\mathfrak{\slashed{\mathfrak{f}}}}_{4} \right) +\mathcal{O}(\epsilon^2) \nonumber \, .
\end{align}
\end{prop}

\medskip

\begin{remark}
For future convenience, we report the identity
\begin{align}
\delta(-2\,\widetilde{\bsy{\slashed{\mathcal{D}}}_{\bsy{2}}^{\star}}\,\widetilde{\bsy{\varsigma}}+\widetilde{\bsy{\varsigma}}\,\widehat{\otimes}_{\widetilde{\bsy{\slashed{g}}}}\, \widetilde{\bsy{\varsigma}})(\epsilon) =& \, \epsilon\cdot \left(-2\,(\slashed{\mathcal{D}}_2^{\star}\overset{\text{\scalebox{.6}{$(1)$}}}{\varsigma})+2\, \varsigma \,\widehat{\otimes}\overset{\text{\scalebox{.6}{$(1)$}}}{\varsigma}   \right. \label{linearisation_D2star_plus_otimes_hat_tensor}\\ 
&-(\slashed{\textup{div}}\,\varsigma +(\varsigma,\varsigma))\overset{\text{\scalebox{.6}{$(1)$}}}{\widehat{\slashed{g}}}  +\frac{1}{2}\, (-2\,\slashed{\mathcal{D}}_2^{\star}\,\varsigma+\varsigma\,\widehat{\otimes}\,\varsigma ,\overset{\text{\scalebox{.6}{$(1)$}}}{\widehat{\slashed{g}}})\,\slashed{g}_{a,M} \nonumber \\
& - \widehat{(\nablasl\overset{\text{\scalebox{.6}{$(1)$}}}{\widehat{\slashed{g}}})^{\sharp_3}\cdot\varsigma}+ (\nablasl\overset{\text{\scalebox{.6}{$(1)$}}}{\widehat{\slashed{g}}})^{\sharp_1}\cdot\varsigma  -\frac{1}{2}\,\varsigma\,\widehat{\otimes}\,(\nablasl(\textup{tr}\overset{\text{\scalebox{.6}{$(1)$}}}{\slashed{g}})) \nonumber\\
& \left.  -\frac{1}{2}\,(\slashed{\varepsilon}\cdot\chi)\,({}^{\star}\varsigma)\,\widehat{\otimes}\,(-2\overset{\text{\scalebox{.6}{$(1)$}}}{\mathfrak{\slashed{\mathfrak{f}}}}_{3}+\overset{\text{\scalebox{.6}{$(1)$}}}{\mathfrak{\underline{\mathfrak{f}}}})  + (\slashed{\varepsilon}\cdot\chib)\,({}^{\star}\varsigma)\,\widehat{\otimes}\, \overset{\text{\scalebox{.6}{$(1)$}}}{\mathfrak{\slashed{\mathfrak{f}}}}_{4}  \right) +\mathcal{O}(\epsilon^2) \, , \nonumber
\end{align}
obtained by simply combining the identities \eqref{linearisation_otimes_hat_tensors} and \eqref{linearisation_D2star_tensor}. 
\end{remark}

\section{The system of linearised gravity}  \label{sec_system_linearised_gravity}

With the linearisation formulae of Section \ref{sec_linearisation_procedure} at hand, one can linearise the renormalised vacuum Einstein equations and derive the full system of linearised vacuum Einstein equations on the Kerr exterior manifold $(\mathcal{M}^*,g_{a,M})$.

\medskip

In this section, we present the full system of linearised gravity on $(\mathcal{M}^*,g_{a,M})$.~The system is written \emph{in geometric form}.~The unknowns of the system are \emph{global}, \emph{regular} $\mathfrak{D}_{\mathcal{N}_{\text{as}}}$ tensors on the Kerr exterior manifold $(\mathcal{M}^*,g_{a,M})$.~A complete list of the unknowns can be found in Table \ref{table:linearised_unknowns}. 

\medskip

A \emph{solution} to the system of linearised gravity is a collection of all the linearised quantities in Table \ref{table:linearised_unknowns}.

\medskip

\begin{table}[H]
\centering
\begin{tabular}{ |c||c|c|c| } 
 \hline
 {} & \textbf{Scalar function} & $\mathfrak{D}_{\mathcal{N}_{\text{as}}}$ \textbf{one-tensor} & $\mathfrak{D}_{\mathcal{N}_{\text{as}}}$ \textbf{two-tensor} \\
 \hline\hline
\textbf{Metric/Frame} & $\overset{\text{\scalebox{.6}{$(1)$}}}{\mathfrak{\underline{\mathfrak{f}}}}_4,\overset{\text{\scalebox{.6}{$(1)$}}}{\mathfrak{\underline{\mathfrak{f}}}}_3,(\text{tr}\overset{\text{\scalebox{.6}{$(1)$}}}{\slashed{g}})$ &  $\overset{\text{\scalebox{.6}{$(1)$}}}{\mathfrak{\underline{\mathfrak{f}}}},\overset{\text{\scalebox{.6}{$(1)$}}}{\mathfrak{\slashed{\mathfrak{f}}}}_{3},\overset{\text{\scalebox{.6}{$(1)$}}}{\mathfrak{\slashed{\mathfrak{f}}}}_{4}$ & $\overset{\text{\scalebox{.6}{$(1)$}}}{\widehat{\slashed{g}}}$ \\
\hline
\textbf{Connection} & $(\overset{\text{\scalebox{.6}{$(1)$}}}{\text{tr}\chi}),(\overset{\text{\scalebox{.6}{$(1)$}}}{\slashed{\varepsilon}\cdot\chi}),(\overset{\text{\scalebox{.6}{$(1)$}}}{\text{tr}\chib}),(\overset{\text{\scalebox{.6}{$(1)$}}}{\slashed{\varepsilon}\cdot\chib}),\overset{\text{\scalebox{.6}{$(1)$}}}{\omegab}$ & $\overset{\text{\scalebox{.6}{$(1)$}}}{\yb},\overset{\text{\scalebox{.6}{$(1)$}}}{\eta},\overset{\text{\scalebox{.6}{$(1)$}}}{\zeta}$ & $\overset{\text{\scalebox{.6}{$(1)$}}}{\chih},\overset{\text{\scalebox{.6}{$(1)$}}}{\chibh}$  \\
\hline
\textbf{Curvature} & $\overset{\text{\scalebox{.6}{$(1)$}}}{\widetilde{\slashed{K}}},\overset{\text{\scalebox{.6}{$(1)$}}}{\rho},\overset{\text{\scalebox{.6}{$(1)$}}}{\sigma}$ & $\overset{\text{\scalebox{.6}{$(1)$}}}{\beta},\overset{\text{\scalebox{.6}{$(1)$}}}{\betab}$  & $\overset{\text{\scalebox{.6}{$(1)$}}}{\alpha},\overset{\text{\scalebox{.6}{$(1)$}}}{\alphab}$ \\
\hline
\end{tabular}
\caption{Unknowns of the system of linearised gravity.}
\label{table:linearised_unknowns}
\end{table}

\medskip 

The system of linearised null structure and Bianchi equations is complemented by a set of equation for the linearised frame coefficients.~This latter set of equations appears in Section \ref{sec_lin_eqns_frame_coeff} and is obtained by combining Propositions \ref{prop_linearisation_frame_commutators_1} and \ref{prop_linearisation_frame_commutators_2}.

\subsection{The equations for the linearised frame coefficients}  \label{sec_lin_eqns_frame_coeff}

The equations for the linearised frame coefficients read as follows.~We have the transport equations
\begin{align}
\nablasl_4 \overset{\text{\scalebox{.6}{$(1)$}}}{\mathfrak{\slashed{\mathfrak{f}}}}_{4}+\chi^{\sharp_2}\cdot \overset{\text{\scalebox{.6}{$(1)$}}}{\mathfrak{\slashed{\mathfrak{f}}}}_{4} -\omegah \overset{\text{\scalebox{.6}{$(1)$}}}{\mathfrak{\slashed{\mathfrak{f}}}}_{4}  &= 0 \, , \\[6pt]
\nablasl_3 \overset{\text{\scalebox{.6}{$(1)$}}}{\mathfrak{\slashed{\mathfrak{f}}}}_{3} +\chib^{\sharp_2} \cdot\, \overset{\text{\scalebox{.6}{$(1)$}}}{\mathfrak{\slashed{\mathfrak{f}}}}_{3} &= \nablasl \overset{\text{\scalebox{.6}{$(1)$}}}{\mathfrak{\underline{\mathfrak{f}}}}_{3}- \overset{\text{\scalebox{.6}{$(1)$}}}{\mathfrak{\underline{\mathfrak{f}}}}_{3}\,(\eta+\etab)+\frac{1}{2}\,(\slashed{\varepsilon}\cdot\chib)\,{}^{\star}\overset{\text{\scalebox{.6}{$(1)$}}}{\mathfrak{\underline{\mathfrak{f}}}} + \overset{\text{\scalebox{.6}{$(1)$}}}{\yb}  \, , 
\end{align}
the mixed transport equations
\begin{align}
\nablasl_4 \overset{\text{\scalebox{.6}{$(1)$}}}{\mathfrak{\slashed{\mathfrak{f}}}}_{3} +\chi^{\sharp_2} \cdot \overset{\text{\scalebox{.6}{$(1)$}}}{\mathfrak{\slashed{\mathfrak{f}}}}_{3}  &=   \zetao \, , \\[6pt]
\nablasl_3 \overset{\text{\scalebox{.6}{$(1)$}}}{\mathfrak{\slashed{\mathfrak{f}}}}_{4} +\chib^{\sharp_2} \cdot\, \overset{\text{\scalebox{.6}{$(1)$}}}{\mathfrak{\slashed{\mathfrak{f}}}}_{4}    &=  \nablasl\overset{\text{\scalebox{.6}{$(1)$}}}{\mathfrak{\underline{\mathfrak{f}}}}_{4} +\frac{1}{2}\,(\slashed{\varepsilon}\cdot\chi)\, {}^{\star}\overset{\text{\scalebox{.6}{$(1)$}}}{\mathfrak{\underline{\mathfrak{f}}}} +\omegah \overset{\text{\scalebox{.6}{$(1)$}}}{\mathfrak{\slashed{\mathfrak{f}}}}_{3} + \etao -\zetao   
\end{align}
and the transport equations
\begin{align}
\nablasl_{4}\overset{\text{\scalebox{.6}{$(1)$}}}{\mathfrak{\underline{\mathfrak{f}}}}_{4} &= -2\,(\eta-\etab,\overset{\text{\scalebox{.6}{$(1)$}}}{\mathfrak{\slashed{\mathfrak{f}}}}_{4}) \, , \\[6pt]
\nablasl_4 \overset{\text{\scalebox{.6}{$(1)$}}}{\mathfrak{\underline{\mathfrak{f}}}}_{3}+\omegah \overset{\text{\scalebox{.6}{$(1)$}}}{\mathfrak{\underline{\mathfrak{f}}}}_{3}&=\omegabo   -2\,(\eta-\etab,\overset{\text{\scalebox{.6}{$(1)$}}}{\mathfrak{\slashed{\mathfrak{f}}}}_{3})-(\eta+\etab,\overset{\text{\scalebox{.6}{$(1)$}}}{\mathfrak{\underline{\mathfrak{f}}}}) \, , \\[6pt]
\nablasl_4\overset{\text{\scalebox{.6}{$(1)$}}}{\mathfrak{\underline{\mathfrak{f}}}}- \chi^{\sharp_1}\cdot \overset{\text{\scalebox{.6}{$(1)$}}}{\mathfrak{\underline{\mathfrak{f}}}}  +\omegah\overset{\text{\scalebox{.6}{$(1)$}}}{\mathfrak{\underline{\mathfrak{f}}}} &= -2\etao +2\,\overset{\text{\scalebox{.6}{$(1)$}}}{\mathfrak{\underline{\mathfrak{f}}}}_{4}(\eta-\etab)+ 2\,\overset{\text{\scalebox{.6}{$(1)$}}}{\widehat{\slashed{g}}}{}^{\sharp}\cdot(\eta-\etab)+(\text{tr}\overset{\text{\scalebox{.6}{$(1)$}}}{\slashed{g}})(\eta-\etab) \, .   
\end{align}
We have the elliptic equations
\begin{align}
\slashed{\text{curl}}\overset{\text{\scalebox{.6}{$(1)$}}}{\mathfrak{\slashed{\mathfrak{f}}}}_{4} &= \frac{1}{2} \,(\overset{\text{\scalebox{.6}{$(1)$}}}{\slashed{\varepsilon}\cdot\chi}) +\frac{1}{2}\,(\slashed{\varepsilon}\cdot\chi)\overset{\text{\scalebox{.6}{$(1)$}}}{\mathfrak{\underline{\mathfrak{f}}}}_4   +\frac{1}{4}\,(\slashed{\varepsilon}\cdot\chi)(\text{tr}\overset{\text{\scalebox{.6}{$(1)$}}}{\slashed{g}})   \, ,  \\[6pt] 
\slashed{\text{curl}}\overset{\text{\scalebox{.6}{$(1)$}}}{\mathfrak{\slashed{\mathfrak{f}}}}_{3}-(\eta+\etab) \wedge \overset{\text{\scalebox{.6}{$(1)$}}}{\mathfrak{\slashed{\mathfrak{f}}}}_{3} &= \frac{1}{2}\, (\overset{\text{\scalebox{.6}{$(1)$}}}{\slashed{\varepsilon}\cdot\chib}) +\frac{1}{2}\,(\slashed{\varepsilon}\cdot\chi)\overset{\text{\scalebox{.6}{$(1)$}}}{\mathfrak{\underline{\mathfrak{f}}}}_3 +\frac{1}{4}\,(\slashed{\varepsilon}\cdot\chib)(\text{tr}\overset{\text{\scalebox{.6}{$(1)$}}}{\slashed{g}})  \, .  
\end{align}

\subsection{Linearised null structure equations}  \label{sec_lin_eqns_null_structure}

The linearised null structure equations read as follows. We have the linearised first variational formulae
\begin{align}
\nablasl_4\overset{\text{\scalebox{.6}{$(1)$}}}{\widehat{\slashed{g}}}  +(\slashed{\varepsilon}\cdot\chi)\,{}^{\star}\overset{\text{\scalebox{.6}{$(1)$}}}{\widehat{\slashed{g}}} &= 2\,\chiho +2\,(\etab-\eta)\,\widehat{\otimes}\overset{\text{\scalebox{.6}{$(1)$}}}{\mathfrak{\slashed{\mathfrak{f}}}}_{4} \, , \\[6pt]
\nablasl_3\overset{\text{\scalebox{.6}{$(1)$}}}{\widehat{\slashed{g}}} +(\slashed{\varepsilon}\cdot\chib)\,{}^{\star}\overset{\text{\scalebox{.6}{$(1)$}}}{\widehat{\slashed{g}}}   &=2\,\chibho +2\,(\slashed{\mathcal{D}}_2^{\star}\overset{\text{\scalebox{.6}{$(1)$}}}{\underline{\mathfrak{f}}})-2\,(\etab-\eta)\,\widehat{\otimes}\overset{\text{\scalebox{.6}{$(1)$}}}{\mathfrak{\slashed{\mathfrak{f}}}}_{3} 
\end{align}
and
\begin{align}
\nablasl_4(\textup{tr}\overset{\text{\scalebox{.6}{$(1)$}}}{\slashed{g}})&= 2\,(\trchio) +4\,(\etab-\eta,\overset{\text{\scalebox{.6}{$(1)$}}}{\mathfrak{\slashed{\mathfrak{f}}}}_{4})   \, , \\[6pt]
\nablasl_3(\textup{tr}\overset{\text{\scalebox{.6}{$(1)$}}}{\slashed{g}})  &=  2\, (\trchibo)-4\,(\etab-\eta,\overset{\text{\scalebox{.6}{$(1)$}}}{\mathfrak{\slashed{\mathfrak{f}}}}_{3}) -2\,(\slashed{\textup{div}}\overset{\text{\scalebox{.6}{$(1)$}}}{\underline{\mathfrak{f}}}) -2\,(\textup{tr}\chi)\overset{\text{\scalebox{.6}{$(1)$}}}{\underline{\mathfrak{f}}}_3 - 2\,(\textup{tr}\chib)\overset{\text{\scalebox{.6}{$(1)$}}}{\underline{\mathfrak{f}}}_4  \, .
\end{align}
We have the linearised second variational formulae
\begin{align}
\nablasl_4 \chiho +(\text{tr}\chi)\chiho-\omegah \,\chiho &= -\alphao \, , \\[6pt]
\nablasl_3 \chibho+(\text{tr}\chib)\chibho &= -2\,\slashed{\mathcal{D}}_2^{\star} \ybo+(\etab-\eta)\,\widehat{\otimes}\ybo    -\alphabo \, .
\end{align}
We have the linearised Raychaudhuri equations
\begin{align}
\nablasl_4 (\trchio)+ (\text{tr}\chi)(\trchio)-\omegah \, (\trchio) =& \,  (\slashed{\varepsilon}\cdot\chi)(\overset{\text{\scalebox{.6}{$(1)$}}}{\slashed{\varepsilon}\cdot\chi}) \, , \\[6pt]
\nablasl_3 (\trchibo)+ (\text{tr}\chib)(\trchibo) =& \, (\slashed{\varepsilon}\cdot\chib)(\overset{\text{\scalebox{.6}{$(1)$}}}{\slashed{\varepsilon}\cdot\chib})+(\text{tr}\chib)\omegabo +2\, \slashed{\text{div}} \ybo   +2\,(\etab-\eta,\ybo) \\ &- (\nablasl(\text{tr}\chib),\overset{\text{\scalebox{.6}{$(1)$}}}{\mathfrak{\underline{\mathfrak{f}}}})-\nablasl_4(\text{tr}\chib)\overset{\text{\scalebox{.6}{$(1)$}}}{\mathfrak{\underline{\mathfrak{f}}}}_3  -\nablasl_3(\text{tr}\chib)\overset{\text{\scalebox{.6}{$(1)$}}}{\mathfrak{\underline{\mathfrak{f}}}}_4  \, . \nonumber
\end{align} 
We have the linearised mixed transport equations
\begin{align}
\nablasl_4 \chibho+\chi\times\chibho+\omegah\,\chibho =& \, -\frac{1}{2}\,(\text{tr}\chib)\chiho+\frac{1}{2}\,(\slashed{\varepsilon}\cdot\chib){^{\star}\chiho}  \\  &  - \widehat{(\nablasl\overset{\text{\scalebox{.6}{$(1)$}}}{\widehat{\slashed{g}}})^{\sharp_3}\cdot\etab}+ (\nablasl\overset{\text{\scalebox{.6}{$(1)$}}}{\widehat{\slashed{g}}})^{\sharp_1}\cdot\etab  -\frac{1}{2}\,\etab\,\widehat{\otimes}\,(\nablasl(\textup{tr}\overset{\text{\scalebox{.6}{$(1)$}}}{\slashed{g}}))-(\slashed{\textup{div}}\,\etab +(\etab,\etab))\overset{\text{\scalebox{.6}{$(1)$}}}{\widehat{\slashed{g}}} \nonumber\\
&   -\frac{1}{2}\,(\slashed{\varepsilon}\cdot\chi)\,({}^{\star}\etab)\,\widehat{\otimes}\,(-2\overset{\text{\scalebox{.6}{$(1)$}}}{\mathfrak{\slashed{\mathfrak{f}}}}_{3}+\overset{\text{\scalebox{.6}{$(1)$}}}{\mathfrak{\underline{\mathfrak{f}}}})  + (\slashed{\varepsilon}\cdot\chib)\,({}^{\star}\etab)\,\widehat{\otimes}\, \overset{\text{\scalebox{.6}{$(1)$}}}{\mathfrak{\slashed{\mathfrak{f}}}}_{4} \nonumber\\
&+(\nablasl_3\etab-\chib^{\sharp_2}\cdot\etab)\,\widehat{\otimes}\overset{\text{\scalebox{.6}{$(1)$}}}{\mathfrak{\slashed{\mathfrak{f}}}}_{4}+(\nablasl_4\etab-\chi^{\sharp_2}\cdot\etab)\,\widehat{\otimes}\overset{\text{\scalebox{.6}{$(1)$}}}{\mathfrak{\slashed{\mathfrak{f}}}}_{3} \, ,  \nonumber \\[6pt]
\nablasl_3 \chiho+\chib\times\chiho =& \, -\frac{1}{2}\,(\text{tr}\chi)\chibho+\frac{1}{2}\,(\slashed{\varepsilon}\cdot\chi){^{\star}\chibho} -2\, \slashed{\mathcal{D}}_2^{\star}\etao +2\, \eta \,\widehat{\otimes}\,\etao   \\ 
&- \widehat{(\nablasl\overset{\text{\scalebox{.6}{$(1)$}}}{\widehat{\slashed{g}}})^{\sharp_3}\cdot\eta}+ (\nablasl\overset{\text{\scalebox{.6}{$(1)$}}}{\widehat{\slashed{g}}})^{\sharp_1}\cdot\eta  -\frac{1}{2}\,\eta\,\widehat{\otimes}\,(\nablasl(\textup{tr}\overset{\text{\scalebox{.6}{$(1)$}}}{\slashed{g}}))-(\slashed{\textup{div}}\,\eta +(\eta,\eta))\overset{\text{\scalebox{.6}{$(1)$}}}{\widehat{\slashed{g}}}  \nonumber\\
& -\frac{1}{2}\,(\slashed{\varepsilon}\cdot\chi)\,({}^{\star}\eta)\,\widehat{\otimes}\,(-2\overset{\text{\scalebox{.6}{$(1)$}}}{\mathfrak{\slashed{\mathfrak{f}}}}_{3}+\overset{\text{\scalebox{.6}{$(1)$}}}{\mathfrak{\underline{\mathfrak{f}}}})  + (\slashed{\varepsilon}\cdot\chib)\,({}^{\star}\eta)\,\widehat{\otimes}\, \overset{\text{\scalebox{.6}{$(1)$}}}{\mathfrak{\slashed{\mathfrak{f}}}}_{4} \nonumber\\
&+(\nablasl_3\eta-\chib^{\sharp_2}\cdot\,\eta)\,\widehat{\otimes}\overset{\text{\scalebox{.6}{$(1)$}}}{\mathfrak{\slashed{\mathfrak{f}}}}_{4}+(\nablasl_4\eta-\chi^{\sharp_2}\cdot\eta)\,\widehat{\otimes}\overset{\text{\scalebox{.6}{$(1)$}}}{\mathfrak{\slashed{\mathfrak{f}}}}_{3}   \nonumber
\end{align}
and
\begin{align}
\nablasl_4( \trchibo)+ \frac{1}{2}\,(\text{tr}\chi)(\trchibo)+\omegah \, (\trchibo) =& \, - \frac{1}{2}\,(\text{tr}\chib)(\trchio)+\frac{1}{2}\,(\slashed{\varepsilon}\cdot\chib)(\overset{\text{\scalebox{.6}{$(1)$}}}{\slashed{\varepsilon}\cdot\chi})+\frac{1}{2}\,(\slashed{\varepsilon}\cdot\chi)(\overset{\text{\scalebox{.6}{$(1)$}}}{\slashed{\varepsilon}\cdot\chib}) \\ 
 &   - 2\,(\slashed{\textup{div}}\overset{\text{\scalebox{.6}{$(1)$}}}{\widehat{\slashed{g}}})^{\sharp}\cdot\etab -(\slashed{\textup{div}}\,\etab+(\etab,\etab))(\textup{tr}\overset{\text{\scalebox{.6}{$(1)$}}}{\slashed{g}}) \nonumber \\ & 
-(\textup{tr}\chi)\,(-2\overset{\text{\scalebox{.6}{$(1)$}}}{\mathfrak{\slashed{\mathfrak{f}}}}_{3}+\overset{\text{\scalebox{.6}{$(1)$}}}{\mathfrak{\underline{\mathfrak{f}}}},\etab) -(\slashed{\varepsilon}\cdot\chi)\,(-2\overset{\text{\scalebox{.6}{$(1)$}}}{\mathfrak{\slashed{\mathfrak{f}}}}_{3}+\overset{\text{\scalebox{.6}{$(1)$}}}{\mathfrak{\underline{\mathfrak{f}}}},{}^{\star}\etab)   \nonumber\\
&  +2\,(\textup{tr}\chib)(\overset{\text{\scalebox{.6}{$(1)$}}}{\mathfrak{\slashed{\mathfrak{f}}}}_{4},\etab)  
+2\,(\slashed{\varepsilon}\cdot\chib)(\overset{\text{\scalebox{.6}{$(1)$}}}{\mathfrak{\slashed{\mathfrak{f}}}}_{4}, {}^{\star}\etab) \nonumber \\
&  +2\,(\nablasl_3\etab-\chib^{\sharp_2}\cdot\etab,\overset{\text{\scalebox{.6}{$(1)$}}}{\mathfrak{\slashed{\mathfrak{f}}}}_{4}) +2\,(\nablasl_4\etab-\chi^{\sharp_2}\cdot\etab,\overset{\text{\scalebox{.6}{$(1)$}}}{\mathfrak{\slashed{\mathfrak{f}}}}_{3}) \nonumber\\ &  +(\text{tr}\chi)(\etab,\overset{\text{\scalebox{.6}{$(1)$}}}{\mathfrak{\underline{\mathfrak{f}}}})+2\rhoo \, , \nonumber \\[6pt]
\nablasl_3 (\trchio)+\frac{1}{2}\,(\text{tr}\chib)(\trchio) =& \, - \frac{1}{2}\,(\text{tr}\chi)(\trchibo)+\frac{1}{2}\,(\slashed{\varepsilon}\cdot\chi)(\overset{\text{\scalebox{.6}{$(1)$}}}{\slashed{\varepsilon}\cdot\chib})+\frac{1}{2}\,(\slashed{\varepsilon}\cdot\chib)(\overset{\text{\scalebox{.6}{$(1)$}}}{\slashed{\varepsilon}\cdot\chi})   \\ 
& -(\text{tr}\chi)\omegabo +2\,\slashed{\textup{div}}\etao+4\,(\eta,\etao) \nonumber\\
& -2 \, (\slashed{\textup{div}}\overset{\text{\scalebox{.6}{$(1)$}}}{\widehat{\slashed{g}}})^{\sharp}\cdot\eta - (\slashed{\textup{div}}\,\eta+(\eta,\eta))(\textup{tr}\overset{\text{\scalebox{.6}{$(1)$}}}{\slashed{g}}) \nonumber\\ & - (\textup{tr}\chi)\,(-2\overset{\text{\scalebox{.6}{$(1)$}}}{\mathfrak{\slashed{\mathfrak{f}}}}_{3}+\overset{\text{\scalebox{.6}{$(1)$}}}{\mathfrak{\underline{\mathfrak{f}}}},\eta) - (\slashed{\varepsilon}\cdot\chi)\,(-2\overset{\text{\scalebox{.6}{$(1)$}}}{\mathfrak{\slashed{\mathfrak{f}}}}_{3}+\overset{\text{\scalebox{.6}{$(1)$}}}{\mathfrak{\underline{\mathfrak{f}}}},{}^{\star}\eta)   \nonumber\\
&  +2\,(\textup{tr}\chib)(\overset{\text{\scalebox{.6}{$(1)$}}}{\mathfrak{\slashed{\mathfrak{f}}}}_{4},\eta)  
+2\,(\slashed{\varepsilon}\cdot\chib)(\overset{\text{\scalebox{.6}{$(1)$}}}{\mathfrak{\slashed{\mathfrak{f}}}}_{4}, {}^{\star}\eta) \nonumber\\
&+2\,(\nablasl_3\eta-\chib^{\sharp_2}\cdot\eta,\overset{\text{\scalebox{.6}{$(1)$}}}{\mathfrak{\slashed{\mathfrak{f}}}}_{4}) +2\,(\nablasl_4\eta-\chi^{\sharp_2}\cdot\eta,\overset{\text{\scalebox{.6}{$(1)$}}}{\mathfrak{\slashed{\mathfrak{f}}}}_{3}) \nonumber\\ 
&-(\nablasl(\text{tr}\chi),\overset{\text{\scalebox{.6}{$(1)$}}}{\mathfrak{\underline{\mathfrak{f}}}})-\nablasl_4(\text{tr}\chi)\overset{\text{\scalebox{.6}{$(1)$}}}{\mathfrak{\underline{\mathfrak{f}}}}_3  -\nablasl_3(\text{tr}\chi)\overset{\text{\scalebox{.6}{$(1)$}}}{\mathfrak{\underline{\mathfrak{f}}}}_4    \nonumber \\
& +(\text{tr}\chi)(\eta,\overset{\text{\scalebox{.6}{$(1)$}}}{\mathfrak{\underline{\mathfrak{f}}}})  +2\rhoo  \, . \nonumber
\end{align}
We have the linearised transport equations for the antitraces
\begin{align}
\nablasl_4 (\overset{\text{\scalebox{.6}{$(1)$}}}{\slashed{\varepsilon}\cdot\chi})+(\text{tr}\chi)(\overset{\text{\scalebox{.6}{$(1)$}}}{\slashed{\varepsilon}\cdot\chi})-\omegah \, (\overset{\text{\scalebox{.6}{$(1)$}}}{\slashed{\varepsilon}\cdot\chi})   =& \, -(\slashed{\varepsilon}\cdot\chi)(\trchio) \, , \\[6pt]
\nablasl_3 (\overset{\text{\scalebox{.6}{$(1)$}}}{\slashed{\varepsilon}\cdot\chib})+(\text{tr}\chib)(\overset{\text{\scalebox{.6}{$(1)$}}}{\slashed{\varepsilon}\cdot\chib})  =& \, -(\slashed{\varepsilon}\cdot\chib)(\trchibo)+ (\slashed{\varepsilon}\cdot\chib)\omegabo - 2\,(\eta+\etab)\wedge \ybo + 2\, \slashed{\text{curl}} \, \ybo \\ & -(\nablasl(\slashed{\varepsilon}\cdot\chib),\overset{\text{\scalebox{.6}{$(1)$}}}{\mathfrak{\underline{\mathfrak{f}}}})  -\nablasl_4(\slashed{\varepsilon}\cdot\chib)\overset{\text{\scalebox{.6}{$(1)$}}}{\mathfrak{\underline{\mathfrak{f}}}}_3  -\nablasl_3(\slashed{\varepsilon}\cdot\chib)\overset{\text{\scalebox{.6}{$(1)$}}}{\mathfrak{\underline{\mathfrak{f}}}}_4  \nonumber
\end{align}
and the mixed transport equations
\begin{align}
\nablasl_4 (\overset{\text{\scalebox{.6}{$(1)$}}}{\slashed{\varepsilon}\cdot\chib}) +\frac{1}{2}\,(\text{tr}\chi)(\overset{\text{\scalebox{.6}{$(1)$}}}{\slashed{\varepsilon}\cdot\chib})+\omegah \, (\overset{\text{\scalebox{.6}{$(1)$}}}{\slashed{\varepsilon}\cdot\chib})  =& \, -\frac{1}{2}\,(\slashed{\varepsilon}\cdot\chib)(\trchio)-\frac{1}{2}\,(\text{tr}\chib)(\overset{\text{\scalebox{.6}{$(1)$}}}{\slashed{\varepsilon}\cdot \chi})-\frac{1}{2}\,(\slashed{\varepsilon}\cdot \chi)(\trchibo) \\ 
&   -2\,(\nablasl_3\etab-\chib^{\sharp_2}\cdot\etab)\wedge\overset{\text{\scalebox{.6}{$(1)$}}}{\mathfrak{\slashed{\mathfrak{f}}}}_{4} -2\,(\nablasl_4\etab-\chi^{\sharp_2}\cdot\etab)\wedge\overset{\text{\scalebox{.6}{$(1)$}}}{\mathfrak{\slashed{\mathfrak{f}}}}_{3}  \nonumber \\
&+(\slashed{\varepsilon}\cdot\chi)(\etab,\overset{\text{\scalebox{.6}{$(1)$}}}{\mathfrak{\underline{\mathfrak{f}}}}) - (\slashed{\textup{curl}}\etab)(\textup{tr}\overset{\text{\scalebox{.6}{$(1)$}}}{\slashed{g}})+2\sigmao \, , \nonumber \\[6pt]
\nablasl_3 (\overset{\text{\scalebox{.6}{$(1)$}}}{\slashed{\varepsilon}\cdot\chi}) +\frac{1}{2}\,(\text{tr}\chib)(\overset{\text{\scalebox{.6}{$(1)$}}}{\slashed{\varepsilon}\cdot\chi}) =& \,-\frac{1}{2}\,(\slashed{\varepsilon}\cdot\chi)(\trchibo) -\frac{1}{2}\,(\text{tr}\chi)(\overset{\text{\scalebox{.6}{$(1)$}}}{\slashed{\varepsilon}\cdot \chib})-\frac{1}{2}\,(\slashed{\varepsilon}\cdot \chib)(\trchio)   \\
&-(\slashed{\varepsilon}\cdot\chi)\omegabo +2\,\slashed{\textup{curl}}\etao \nonumber \\
&-2\,(\nablasl_3\eta-\chib^{\sharp_2}\cdot\eta)\wedge\overset{\text{\scalebox{.6}{$(1)$}}}{\mathfrak{\slashed{\mathfrak{f}}}}_{4} -2\,(\nablasl_4\eta-\chi^{\sharp_2}\cdot\eta)\wedge\overset{\text{\scalebox{.6}{$(1)$}}}{\mathfrak{\slashed{\mathfrak{f}}}}_{3}  \nonumber \\
&-(\nablasl(\slashed{\varepsilon}\cdot\chi),\overset{\text{\scalebox{.6}{$(1)$}}}{\mathfrak{\underline{\mathfrak{f}}}})  -\nablasl_4(\slashed{\varepsilon}\cdot\chi)\overset{\text{\scalebox{.6}{$(1)$}}}{\mathfrak{\underline{\mathfrak{f}}}}_3  -\nablasl_3(\slashed{\varepsilon}\cdot\chi)\overset{\text{\scalebox{.6}{$(1)$}}}{\mathfrak{\underline{\mathfrak{f}}}}_4  \nonumber \\
&+(\slashed{\varepsilon}\cdot\chi)(\eta,\overset{\text{\scalebox{.6}{$(1)$}}}{\mathfrak{\underline{\mathfrak{f}}}})  - (\slashed{\textup{curl}}\,\eta)(\textup{tr}\overset{\text{\scalebox{.6}{$(1)$}}}{\slashed{g}})  -2\sigmao  \, . \nonumber
\end{align}
We have the linearised transport equations
\begin{align}
\nablasl_4 \etao+\frac{1}{2}\,(\text{tr}\chi)\etao -\frac{1}{2}\,(\slashed{\varepsilon}\cdot\chi){}^{\star}\etao =& \, \chiho{}^{\sharp}\cdot\etab+\frac{1}{2}\,(\trchio)\etab-\frac{1}{2}\,(\overset{\text{\scalebox{.6}{$(1)$}}}{\slashed{\varepsilon}\cdot\chi})({}^{\star}\etab-2\,{}^{\star}\eta)-\betao \\
&+2\,(\etab-\eta,\eta)\overset{\text{\scalebox{.6}{$(1)$}}}{\slashed{\mathfrak{f}}}_4+\frac{1}{2}\,(\slashed{\varepsilon}\cdot\chi)({}^{\star}\overset{\text{\scalebox{.6}{$(1)$}}}{\widehat{\slashed{g}}})^{\sharp_1}\cdot(\etab -2\,\eta)  \, , \nonumber \\[6pt]
\nablasl_4\ybo +2\,\omegah\ybo =& \,  -\chibho{}^{\sharp}\cdot\eta -\frac{1}{2}\,(\trchibo)\,\eta+\frac{1}{2}\,(\overset{\text{\scalebox{.6}{$(1)$}}}{\slashed{\varepsilon}\cdot\chib})({}^{\star}\eta-2\,{}^{\star}\etab)-\betabo \\
&-\frac{1}{2}\,(\text{tr}\chib)\etao  +\frac{1}{2}\,(\slashed{\varepsilon}\cdot\chib)\,{}^{\star}\etao -\frac{1}{2}\,(\slashed{\varepsilon}\cdot\chib)({}^{\star}\overset{\text{\scalebox{.6}{$(1)$}}}{\widehat{\slashed{g}}})^{\sharp_1}\cdot(\eta -2\,\etab) \nonumber\\
&+(\nablasl\etab)^{\sharp_1}\cdot \overset{\text{\scalebox{.6}{$(1)$}}}{\underline{\mathfrak{f}}}  -(\slashed{\mathcal{D}}_2^{\star}\overset{\text{\scalebox{.6}{$(1)$}}}{\underline{\mathfrak{f}}})^{\sharp_2}\cdot \etab+\frac{1}{2}\,(\slashed{\textup{div}}\overset{\text{\scalebox{.6}{$(1)$}}}{\underline{\mathfrak{f}}})\,\etab+\frac{1}{2}\,(\slashed{\textup{curl}}\overset{\text{\scalebox{.6}{$(1)$}}}{\underline{\mathfrak{f}}})\,{}^{\star}\etab \nonumber \\
&  -2\,(\eta-\etab,\etab)\overset{\text{\scalebox{.6}{$(1)$}}}{\slashed{\mathfrak{f}}}_3 +(\nablasl_4\etab+\chi^{\sharp_2}\cdot\etab)\overset{\text{\scalebox{.6}{$(1)$}}}{\underline{\mathfrak{f}}}_3+ (\nablasl_3\etab+\chib^{\sharp_2}\cdot\etab)\overset{\text{\scalebox{.6}{$(1)$}}}{\underline{\mathfrak{f}}}_4 \, , \nonumber  
\end{align}
\begin{align}
\nablasl_4\omegabo+2\,\omegah\omegabo  =& \,   -2\,(\eta-\etab,\etao)-2\,(\eta-\etab,\zetao) -2\rhoo   \\
&+((\eta-2\,\etab)\,\widehat{\otimes}\,\eta, \overset{\text{\scalebox{.6}{$(1)$}}}{\widehat{\slashed{g}}})+(\textup{tr}\overset{\text{\scalebox{.6}{$(1)$}}}{\slashed{g}})\,(\eta-2\,\etab,\eta) \nonumber \\
& -(\nablasl\omegah,\overset{\text{\scalebox{.6}{$(1)$}}}{\mathfrak{\underline{\mathfrak{f}}}})  -(\nablasl_4\omegah)\overset{\text{\scalebox{.6}{$(1)$}}}{\mathfrak{\underline{\mathfrak{f}}}}_3  -(\nablasl_3\omegah)\overset{\text{\scalebox{.6}{$(1)$}}}{\mathfrak{\underline{\mathfrak{f}}}}_4  \, , \nonumber
\end{align}
\begin{align}
\nablasl_4 \zetao +\chi^{\sharp_2}\cdot \zetao +\omegah \zetao    =& \,  \chiho {}^{\sharp}\cdot \etab +\frac{1}{2}\,(\trchio) \, \etab +\frac{1}{2}\,(\overset{\text{\scalebox{.6}{$(1)$}}}{\slashed{\varepsilon}\cdot\chi}) \, {^{\star}\etab} -\betao  \\
& +2\,(\etab-\eta,\eta)\overset{\text{\scalebox{.6}{$(1)$}}}{\slashed{\mathfrak{f}}}_4-(\nablasl_3\omegah)\overset{\text{\scalebox{.6}{$(1)$}}}{\slashed{\mathfrak{f}}}_4-(\nablasl_4\omegah)\overset{\text{\scalebox{.6}{$(1)$}}}{\slashed{\mathfrak{f}}}_3 \nonumber \\
&-\frac{1}{2}\,(\slashed{\varepsilon}\cdot\chi)({}^{\star}\overset{\text{\scalebox{.6}{$(1)$}}}{\widehat{\slashed{g}}})^{\sharp_1}\cdot\etab \, , \nonumber \\[6pt]
\nablasl_3\zetao +\chib^{\sharp_2}\cdot\zetao  =& \,  \nablasl \omegabo  +\chi^{\sharp_2}\cdot \ybo-\omegah \ybo -\chib^{\sharp_2}\cdot \etao -\betabo \\
&-  \,\chibho {}^{\sharp}\cdot \eta -\frac{1}{2}\,(\trchibo)\,\eta  -\frac{1}{2}\,(\overset{\text{\scalebox{.6}{$(1)$}}}{\slashed{\varepsilon}\cdot\chib}){^{\star}\eta}     \nonumber\\
&  +2\,(\eta-\etab,\eta)\overset{\text{\scalebox{.6}{$(1)$}}}{\slashed{\mathfrak{f}}}_3 -(\nablasl_4\eta+\chi^{\sharp_2}\cdot\eta)\overset{\text{\scalebox{.6}{$(1)$}}}{\underline{\mathfrak{f}}}_3 - (\nablasl_3\eta+\chib^{\sharp_2}\cdot\eta)\overset{\text{\scalebox{.6}{$(1)$}}}{\underline{\mathfrak{f}}}_4 \nonumber \\
&-(\nablasl\eta)^{\sharp_1}\cdot \overset{\text{\scalebox{.6}{$(1)$}}}{\underline{\mathfrak{f}}}+(\slashed{\mathcal{D}}_2^{\star}\overset{\text{\scalebox{.6}{$(1)$}}}{\underline{\mathfrak{f}}})^{\sharp_2}\cdot \eta -\frac{1}{2}\,(\slashed{\textup{div}}\overset{\text{\scalebox{.6}{$(1)$}}}{\underline{\mathfrak{f}}})\,\eta -\frac{1}{2}\,(\slashed{\textup{curl}}\overset{\text{\scalebox{.6}{$(1)$}}}{\underline{\mathfrak{f}}})\,{}^{\star}\eta  \nonumber \\
&+\frac{1}{2}\,(\slashed{\varepsilon}\cdot\chib)({}^{\star}\overset{\text{\scalebox{.6}{$(1)$}}}{\widehat{\slashed{g}}})^{\sharp_1}\cdot\eta   \nonumber
\end{align}
and the linearised elliptic equation
\begin{align}
\slashed{\textup{curl}}\zetao   =& \,  \frac{1}{4}\,(\slashed{\varepsilon}\cdot\chib)(\trchio)-\frac{1}{4}\,(\text{tr}\chib)(\overset{\text{\scalebox{.6}{$(1)$}}}{\slashed{\varepsilon}\cdot\chi})+\frac{1}{4}\,((\text{tr}\chi)-2\,\omegah)(\overset{\text{\scalebox{.6}{$(1)$}}}{\slashed{\varepsilon}\cdot\chib}) -\frac{1}{4}\,(\slashed{\varepsilon}\cdot\chi)(\trchibo)  \\ & +\frac{1}{2}\,(\slashed{\varepsilon}\cdot\chi)\omegabo  +\sigmao \nonumber \\
&+(\nablasl_3\eta-\chib^{\sharp_2}\cdot\eta)\wedge \overset{\text{\scalebox{.6}{$(1)$}}}{\slashed{\mathfrak{f}}}_4 +(\nablasl_4\eta-\chi^{\sharp_2}\cdot\eta)\wedge \overset{\text{\scalebox{.6}{$(1)$}}}{\slashed{\mathfrak{f}}}_3-\frac{1}{2}\,(\slashed{\varepsilon}\cdot\chi) (\eta,\overset{\text{\scalebox{.6}{$(1)$}}}{\underline{\mathfrak{f}}}) \nonumber \\
&-\frac{1}{2}\,(\slashed{\textup{curl}}\,\eta)(\textup{tr}\overset{\text{\scalebox{.6}{$(1)$}}}{\slashed{g}})    \, . \nonumber
\end{align}
We have the linearised Codazzi equations
\begin{align}
\slashed{\text{div}}\chiho+\chiho{}^{\sharp}\cdot\eta =& \, \frac{1}{2}\,\nablasl(\trchio)-\frac{1}{2}\,{}^{\star}\nablasl(\overset{\text{\scalebox{.6}{$(1)$}}}{\slashed{\varepsilon}\cdot\chi}) -\betao \\
& +\frac{1}{2}\,(\trchio)\,\eta -\frac{3}{2}\,(\overset{\text{\scalebox{.6}{$(1)$}}}{\slashed{\varepsilon}\cdot\chi})\,{}^{\star}\eta-\frac{1}{2}\,(\slashed{\varepsilon}\cdot\chi)\,{}^{\star}\zetao  +\frac{1}{2}\,(\text{tr}\chi)\zetao -(\slashed{\varepsilon}\cdot\chi)\,{}^{\star}\etao  \nonumber\\
& +\frac{3}{2}\,(\slashed{\varepsilon}\cdot\chi)({}^{\star}\overset{\text{\scalebox{.6}{$(1)$}}}{\widehat{\slashed{g}}})^{\sharp_1}\cdot\eta+\frac{1}{2}\,({}^{\star}\overset{\text{\scalebox{.6}{$(1)$}}}{\widehat{\slashed{g}}})^{\sharp_1}\cdot(\nablasl(\slashed{\varepsilon}\cdot\chi))  \nonumber\\
&+\frac{1}{2}\,(\nablasl_3(\text{tr}\chi))\overset{\text{\scalebox{.6}{$(1)$}}}{\slashed{\mathfrak{f}}}_4+\frac{1}{2}\,(\nablasl_4(\text{tr}\chi))\overset{\text{\scalebox{.6}{$(1)$}}}{\slashed{\mathfrak{f}}}_3-\frac{1}{2}\,(\nablasl_3(\slashed{\varepsilon}\cdot\chi))\,{}^{\star}\overset{\text{\scalebox{.6}{$(1)$}}}{\slashed{\mathfrak{f}}}_4-\frac{1}{2}\,(\nablasl_4(\slashed{\varepsilon}\cdot\chi))\,{}^{\star}\overset{\text{\scalebox{.6}{$(1)$}}}{\slashed{\mathfrak{f}}}_3 \, , \nonumber \\[6pt]
\slashed{\text{div}}\chibho-\chibho{}^{\sharp}\cdot\eta =& \,\frac{1}{2}\,\nablasl(\trchibo)-\frac{1}{2}\,{}^{\star}\nablasl(\overset{\text{\scalebox{.6}{$(1)$}}}{\slashed{\varepsilon}\cdot\chib}) +\betabo \\
&-\frac{1}{2}\,(\trchibo)\,\eta +\frac{1}{2}\,(\overset{\text{\scalebox{.6}{$(1)$}}}{\slashed{\varepsilon}\cdot\chib})({}^{\star}\eta-2\,{}^{\star}\etab)+\frac{1}{2}\,(\slashed{\varepsilon}\cdot\chib)\,{}^{\star}\zetao   -\frac{1}{2}\,(\text{tr}\chib)\zetao -(\slashed{\varepsilon}\cdot\chi)\,{}^{\star}\ybo  \nonumber \\
& -\frac{1}{2}\,(\slashed{\varepsilon}\cdot\chib)({}^{\star}\overset{\text{\scalebox{.6}{$(1)$}}}{\widehat{\slashed{g}}})^{\sharp_1}\cdot(\eta-2\,\etab)+\frac{1}{2}\,({}^{\star}\overset{\text{\scalebox{.6}{$(1)$}}}{\widehat{\slashed{g}}})^{\sharp_1}\cdot(\nablasl(\slashed{\varepsilon}\cdot\chib)) \nonumber \\
&+\frac{1}{2}\,(\nablasl_3(\text{tr}\chib))\overset{\text{\scalebox{.6}{$(1)$}}}{\slashed{\mathfrak{f}}}_4+\frac{1}{2}\,(\nablasl_4(\text{tr}\chib))\overset{\text{\scalebox{.6}{$(1)$}}}{\slashed{\mathfrak{f}}}_3-\frac{1}{2}\,(\nablasl_3(\slashed{\varepsilon}\cdot\chib))\,{}^{\star}\overset{\text{\scalebox{.6}{$(1)$}}}{\slashed{\mathfrak{f}}}_4-\frac{1}{2}\,(\nablasl_4(\slashed{\varepsilon}\cdot\chib))\,{}^{\star}\overset{\text{\scalebox{.6}{$(1)$}}}{\slashed{\mathfrak{f}}}_3  \nonumber
\end{align}
and the linearised Gauss equation
\begin{align}
\overset{\text{\scalebox{.6}{$(1)$}}}{\widetilde{\slashed{K}}} &= -\frac{1}{4}\,(\text{tr}\chib)(\trchio)-\frac{1}{4}\,(\text{tr}\chi)(\trchibo)-\frac{1}{4}\,(\slashed{\varepsilon}\cdot\chib)(\overset{\text{\scalebox{.6}{$(1)$}}}{\slashed{\varepsilon}\cdot\chi})-\frac{1}{4}\,(\slashed{\varepsilon}\cdot\chi)(\overset{\text{\scalebox{.6}{$(1)$}}}{\slashed{\varepsilon}\cdot\chib})-\rhoo \, .
\end{align}

\subsection{Linearised Bianchi equations}  \label{sec_lin_eqns_Bianchi}

The linearised Bianchi equations read
\begin{gather}
\nablasl_3\alphao+\frac{1}{2}\,(\text{tr}\chib)\alphao +\frac{1}{2}(\slashed{\varepsilon}\cdot\chib){}^{\star}\alphao =   -2\,\slashed{\mathcal{D}}_2^{\star}\betao-3\,\rho\,\chiho -3\,\sigma\,{}^{\star}\chiho+5\,\eta\,\widehat{\otimes}\betao  \, ,\\[6pt]
\nablasl_4\betao+2 \, (\text{tr}\chi)\betao-2\, (\slashed{\varepsilon}\cdot\chi){}^{\star}\betao-\omegah\betao =  \slashed{\text{div}}\alphao+(\etab^{\sharp}+2\,\eta^{\sharp})\cdot\alphao \, , 
\end{gather}
\begin{align}
\nablasl_3\betao+(\text{tr} \chib)\betao+(\slashed{\varepsilon}\cdot\chi){}^{\star}\betao  =& \,\slashed{\mathcal{D}}_1^{\star}(-\rhoo,\sigmao)+3\, \rho\etao +3\rhoo\eta  +3\,\sigma\,{}^{\star}\etao +3\sigmao {}^{\star}\eta \\
&+(\nablasl_3\rho)\overset{\text{\scalebox{.6}{$(1)$}}}{\mathfrak{\slashed{\mathfrak{f}}}}_{4}+(\nablasl_3\sigma){}^{\star}\overset{\text{\scalebox{.6}{$(1)$}}}{\mathfrak{\slashed{\mathfrak{f}}}}_{4}+ (\nablasl_4\rho)\overset{\text{\scalebox{.6}{$(1)$}}}{\mathfrak{\slashed{\mathfrak{f}}}}_{3}+ (\nablasl_4\sigma){}^{\star}\overset{\text{\scalebox{.6}{$(1)$}}}{\mathfrak{\slashed{\mathfrak{f}}}}_{3}  \nonumber \\
&-({}^{\star}\overset{\text{\scalebox{.6}{$(1)$}}}{\widehat{\slashed{g}}})^{\sharp_1}\cdot(\nablasl\sigma +3\,\sigma\,\eta) \nonumber \, ,
\end{align}
\begin{gather}
\nablasl_4 \rhoo+\frac{3}{2}\,(\text{tr}\chi)\rhoo  = \slashed{\text{div}}\betao+(2\etab+\eta,\betao)-\frac{3}{2}\,\rho\,(\trchio)-\frac{3}{2}\,\sigma\,(\overset{\text{\scalebox{.6}{$(1)$}}}{\slashed{\varepsilon}\cdot\chi}) -\frac{3}{2}\,(\slashed{\varepsilon}\cdot\chi)\sigmao \, , \\[6pt]
\nablasl_4 \sigmao+\frac{3}{2}\,(\text{tr}\chi)\sigmao =  -\slashed{\text{curl}}\betao-(2\etab+\eta)\wedge\betao-\frac{3}{2}\,\sigma\,(\trchio) +\frac{3}{2}\,\rho\,(\overset{\text{\scalebox{.6}{$(1)$}}}{\slashed{\varepsilon}\cdot\chi})+\frac{3}{2}\,(\slashed{\varepsilon}\cdot\chi)\rhoo \, ,  
\end{gather}
\begin{align}
\nablasl_3 \rhoo+\frac{3}{2}\,(\text{tr}\chib)\rhoo  =& \, -\slashed{\text{div}}\betabo-(\eta,\betabo)-\frac{3}{2}\,\rho\,(\trchibo)+\frac{3}{2}\,\sigma\,(\overset{\text{\scalebox{.6}{$(1)$}}}{\slashed{\varepsilon}\cdot\chib}) +\frac{3}{2}\,(\slashed{\varepsilon}\cdot\chib)\sigmao \\
& -(\nablasl \rho,\overset{\text{\scalebox{.6}{$(1)$}}}{\mathfrak{\underline{\mathfrak{f}}}}) -(\nablasl_4\rho)\overset{\text{\scalebox{.6}{$(1)$}}}{\mathfrak{\underline{\mathfrak{f}}}}_3  -(\nablasl_3\rho)\overset{\text{\scalebox{.6}{$(1)$}}}{\mathfrak{\underline{\mathfrak{f}}}}_4 \, , \nonumber \\[6pt]
\nablasl_3 \sigmao+\frac{3}{2}\,(\text{tr}\chib)\sigmao  =& \, -\slashed{\text{curl}}\betabo-\eta\wedge\betabo-\frac{3}{2}\,\sigma\,(\trchibo)-\frac{3}{2}\,\rho\,(\overset{\text{\scalebox{.6}{$(1)$}}}{\slashed{\varepsilon}\cdot\chib})-\frac{3}{2}\,(\slashed{\varepsilon}\cdot\chib)\rhoo \\
& -(\nablasl \sigma,\overset{\text{\scalebox{.6}{$(1)$}}}{\mathfrak{\underline{\mathfrak{f}}}})  -(\nablasl_4\sigma)\overset{\text{\scalebox{.6}{$(1)$}}}{\mathfrak{\underline{\mathfrak{f}}}}_3  -(\nablasl_3\sigma)\overset{\text{\scalebox{.6}{$(1)$}}}{\mathfrak{\underline{\mathfrak{f}}}}_4 \, , \nonumber
\end{align}
\begin{align}
\nablasl_4\betabo+(\text{tr} \chi)\betabo+(\slashed{\varepsilon}\cdot\chib){}^{\star}\betabo+\omegah\betabo  =& \,\slashed{\mathcal{D}}_1^{\star}(\rhoo,\sigmao)-3\,\rhoo\etab  +3\,\sigmao {}^{\star}\etab  \\
&-(\nablasl_3\rho)\overset{\text{\scalebox{.6}{$(1)$}}}{\mathfrak{\slashed{\mathfrak{f}}}}_{4}+ (\nablasl_3\sigma){}^{\star}\overset{\text{\scalebox{.6}{$(1)$}}}{\mathfrak{\slashed{\mathfrak{f}}}}_{4}- (\nablasl_4\rho)\overset{\text{\scalebox{.6}{$(1)$}}}{\mathfrak{\slashed{\mathfrak{f}}}}_{3}+ (\nablasl_4\sigma){}^{\star}\overset{\text{\scalebox{.6}{$(1)$}}}{\mathfrak{\slashed{\mathfrak{f}}}}_{3} \nonumber \\
&-({}^{\star}\overset{\text{\scalebox{.6}{$(1)$}}}{\widehat{\slashed{g}}})^{\sharp_1}\cdot(\nablasl\sigma +3\,\sigma\,\etab)  \, , \nonumber 
\end{align}
\begin{gather}
\nablasl_3\betabo+2\,(\text{tr}\chib)\betabo-2\,(\slashed{\varepsilon}\cdot\chib){}^{\star}\betabo =  -\slashed{\text{div}}\alphabo+\eta^{\sharp}\cdot\alphabo -3\,\rho\ybo+3\,\sigma\, {}^{\star}\ybo \, , \\[6pt] 
\nablasl_4\alphabo+\frac{1}{2}\,(\text{tr}\chi)\alphabo-\frac{1}{2}\,(\slashed{\varepsilon}\cdot\chi){}^{\star}\alphabo+2\,\omegah\alphabo =    2\,\slashed{\mathcal{D}}_2^{\star}\betabo-3\,\rho\chibho +3\,\sigma\, {}^{\star}\chibho-(4\etab-\eta)\,\widehat{\otimes}\betabo \, .  
\end{gather}

\appendix

\section{Non-integrable frames and the algebraically special frame of Kerr} \label{sec_intro_appendix}

This section serves as an appendix to the introduction of the paper.

\medskip

We give an informal definition of non-integrable frames tailored to our problem. 

\medskip

\begin{integrprop}
A \emph{non-integrable} null frame $\mathcal{N}$ of a Lorentzian manifold $(\mathcal{M},g)$ is a null frame whose null frame vector fields $e_3$ and $e_4$ induce an orthogonal (relative to $g$) distribution $$\mathfrak{D}_{\mathcal{N}}:=\left\langle e_3,e_4 \right\rangle^{\perp}$$ which is a \emph{non-integrable} distribution.\footnote{In differential geometry, a \emph{distribution} is a sub-bundle $\mathfrak{D} \subset T\mathcal{M}$ of the tangent bundle $T\mathcal{M}$ of $\mathcal{M}$.~A distribution $\mathfrak{D}$ is \emph{integrable} if, for any vector fields $X,Y\in \Gamma(\mathfrak{D})$, one has $[X,Y]\in\Gamma(\mathfrak{D})$.}
\end{integrprop}

\medskip

For the algebraically special frame $\mathcal{N}_{\text{as}}$ of the Kerr exterior manifold $(\mathcal{M},g_{a,M})$, one can choose the frame vector fields $(e^{\text{as}}_1,e^{\text{as}}_2)$ such that\footnote{To give the coordinate form of the frame vector fields $(e^{\text{as}}_1,e^{\text{as}}_2)$, we adopt the standard Boyer--Lindquist differentiable structure.~We point out that, although in the paper the spacelike frame vector fields of a null frame are always assumed to be orthonormal, the frame vector fields \eqref{intro_horizontal_vf} are not orthonormal.~This allows a more convenient choice of $(e^{\text{as}}_1,e^{\text{as}}_2)$ for the arguments of the present appendix, which can be however (less neatly) repeated for any choice of orthonormal frame vector fields $(e^{\text{as}}_1,e^{\text{as}}_2)$.}
\begin{align}
e^{\text{as}}_1&= \partial_{\theta} \, , &  e^{\text{as}}_2&= |a|\,\sin^2\theta\,\partial_t+\partial_{\phi} \, .  \label{intro_horizontal_vf}
\end{align}
The vector field commutator 
\begin{equation}
[e^{\text{as}}_1,e^{\text{as}}_2]=|a|\,\sin(2\theta)\,\partial_t \label{intro_commutator_horizontal_vf}
\end{equation}
is, for $|a|>0$, such that $[e^{\text{as}}_1,e^{\text{as}}_2]\not\in \left\langle e^{\text{as}}_1,e^{\text{as}}_2 \right\rangle $, and thus $[e^{\text{as}}_1,e^{\text{as}}_2]\not\in\Gamma(\mathfrak{D}_{\mathcal{N}_{\text{as}}})$.~The distribution $\mathfrak{D}_{\mathcal{N}_{\text{as}}}$ is therefore manifestly \emph{non-integrable} for $|a|>0$.~For $|a|=0$, the commutator \eqref{intro_commutator_horizontal_vf} identically vanishes and the frame $\mathcal{N}_{\text{as}}$ is \emph{integrable}.

\medskip

\begin{figure}[H]
\centering
\begin{tikzpicture}[scale=1.5]

\draw[->, ultra thin] (0,0)--(6.5,0);
\draw[->, ultra thin] (0,0)--(0,2.8);
\draw[->, ultra thin] (0,0)--(2,2);

\draw[ultra thin] (5,-0.1)--(5,0.1);

\draw[ultra thin] (0.8,-0.3)--(1.75,1.0);
\draw[ultra thin] (2.8,-0.5)--(3.45,1.2);

\draw[ultra thin] (1.8,-0.3)--(2.75,1.0);
\draw[ultra thin] (3.8,-0.5)--(4.45,1.2);

\draw[ultra thin] (1.75,1.0)--(2.75,1.0);
\draw[ultra thin] (3.45,1.2)--(4.45,1.2);

\draw[ultra thin] (0.8,-0.3)--(1.8,-0.3);
\draw[ultra thin] (2.8,-0.5)--(3.8,-0.5);

\draw[->, ultra thick] (1,0)--(2,0);
\draw[->, ultra thick] (3,0)--(4,0);

\draw[->, ultra thick] (1,0)--(1.3,0.35);
\draw[->, ultra thick] (3,0)--(3.2,0.5);

\draw[dashed, ultra thin] (3.75,1.0)--(4.75,1.0);
\draw[dashed, ultra thin] (2.8,-0.3)--(3.8,-0.3);

\draw[dashed, ultra thin] (2.8,-0.3)--(3.75,1.0);
\draw[dashed, ultra thin] (3.8,-0.3)--(4.75,1.0);

\node at (6.5,-0.3) {$\theta$};
\node at (-0.2,2.8) {$t$};
\node at (1.8,2.2) {$\phi$};
\node at (-0.15,-0.1) {$0$};
\node at (5,-0.3) {$\pi/2$};

\node at (0.8,-0.125) {$p$};

\node at (1.8,0.2) {$e_1^{\text{as}}$};
\node at (3.8,0.2) {$e_1^{\text{as}}$};
\node at (1,0.4) {$e_2^{\text{as}}$};
\node at (2.85,0.4) {$e_2^{\text{as}}$};

\draw (1,0) circle (1pt);
\draw (3,0) circle (1pt);

\end{tikzpicture}

\caption{The distribution of planes generated by $(e^{\text{as}}_1,e^{\text{as}}_2)$ cannot be possibly thought (even locally) as the tangent bundle to the two-dimensional leaves of \emph{any} foliation of $\mathcal{M}$.~Indeed, if one assumes that there exists a two-dimensional sub-manifold $\mathcal{S}\subset\mathcal{M}$ with $p\in\mathcal{S}$ and $T\mathcal{S}=\left\langle e^{\text{as}}_1,e^{\text{as}}_2 \right\rangle$ in an open neighbourhood $\Omega_p\subset\mathcal{S}$, then, by the invariance of $(e^{\text{as}}_1,e^{\text{as}}_2)$ along the $e^{\text{as}}_2$-direction, one can immediately conclude that $\Omega_p$ has to contain a line segment generated by $e^{\text{as}}_2$ and passing through $p$.~By $\mathcal{S}$ being a manifold, the fact that $\Omega_p$ contains a line segment though $p$ implies that $\mathcal{S}$ contains an open neighbourhood $\Omega^{\prime}_p\subset \left\langle e^{\text{as}}_1,e^{\text{as}}_2 \right\rangle_p$, i.e.~a portion of the two-plane generated by $(e^{\text{as}}_1,e^{\text{as}}_2)$ at $p$.~This latter statement contradicts the fact that the rotation planes generated by $(e^{\text{as}}_1,e^{\text{as}}_2)$ along the line segment $\left\lbrace \phi=\phi(p),t=t(p)\right\rbrace \cap\Omega^{\prime}_p$ are tangent to $\mathcal{S}$.}

\label{fig:rotating_vf}

\end{figure}
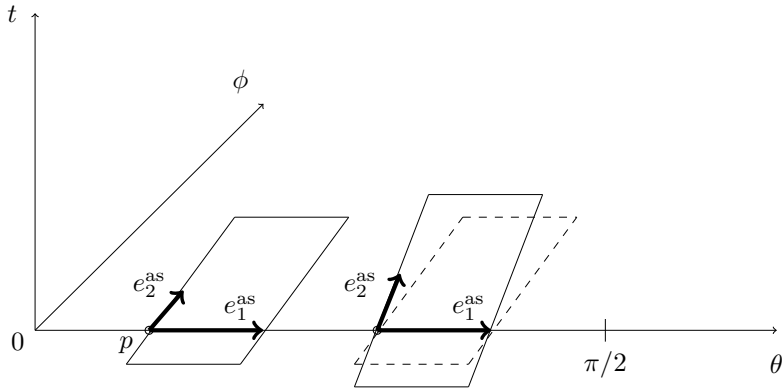

The (non-)integrability of null frames has a meaningful relation with the geometry of spacetime foliations.~This fact can already be elucidated by considering the non-integrable frame $\mathcal{N}_{\text{as}}$ of $(\mathcal{M},g_{a,M})$.~By depicting the frame vector fields \eqref{intro_horizontal_vf} in $(\theta,\phi,t)$ coordinates (see Figure \ref{fig:rotating_vf}), one realises that the vector fields $(e^{\text{as}}_1,e^{\text{as}}_2)$ generate \emph{rotating} two-dimensional planes along any line of constant $(\phi,t)$.~While it is already evident from \eqref{intro_horizontal_vf} that the vector fields $(e^{\text{as}}_1,e^{\text{as}}_2)$ are not tangent to the two-spheres of constant $(t,r)$, Figure \ref{fig:rotating_vf} suggests that the distribution of planes generated by $(e^{\text{as}}_1,e^{\text{as}}_2)$ cannot be possibly thought (even locally) as the tangent bundle to the two-dimensional leaves of \emph{any} foliation of $\mathcal{M}$. 

\medskip

We remark that the rotating behaviour of the planes generated by $(e^{\text{as}}_1,e^{\text{as}}_2)$, which is crucial to exclude the existence of foliations by two-dimensional leaves whose tangent bundle is (at least locally) generated by $(e^{\text{as}}_1,e^{\text{as}}_2)$, is indeed captured by the failure of the vector field commutator \eqref{intro_commutator_horizontal_vf} to be in the span of $(e^{\text{as}}_1,e^{\text{as}}_2)$.~In this sense, the non-integrability of the frame $\mathcal{N}_{\text{as}}$ is intimately related to the non-existence of foliations of $\mathcal{M}$ with a certain geometry.\footnote{One can equivalently think that the non-integrability of the frame $\mathcal{N}_{\text{as}}$ excludes the existence of foliations by two-dimensional leaves to which the vector fields $(e^{\text{as}}_3,e^{\text{as}}_4)$ are (at least locally) orthogonal.~A related problem is treated in Appendix B.3 of \cite{WaldBook}, where given a spacetime vector field $X$, one wants to determine whether there exists a family of hypersurfaces foliating the spacetime and (at least locally) orthogonal to the vector field $X$.~In this case, the (non-)existence of such hypersurfaces is related to the (non-)integrability of the distribution $\left\langle X \right\rangle^{\perp}$.}

\medskip

\medskip

It turns out that Figure \ref{fig:rotating_vf} is suggestive of a more general fact, which is the manifestation of a classical result by Frobenius:  

\medskip

\begin{frobprop}
A null frame $\mathcal{N}$ of a Lorentzian manifold $(\mathcal{M},g)$ is integrable \emph{if and only if} the frame vector fields $(e_1,e_2)$ generate, at least locally, the tangent bundle to the two-dimensional leaves of a foliation of $\mathcal{M}$.
\end{frobprop}

\medskip

The definition and properties of non-integrable frames are rigorously discussed in Section \ref{sec_preliminary_defns}.

\section{The hierarchical structure of the system of linearised gravity} \label{sec_hierarchy_system}

We present the hierarchical structure of the system of linearised gravity of Section \ref{sec_system_linearised_gravity}.~The gauge dependent part of the hierarchy only includes outgoing transport equations.~All linearised quantities (see Table \ref{table:linearised_unknowns}) appear in the hierarchy of equations.

\medskip

At each level of the hierarchy, the equations only couple with linearised quantities appearing either higher or at the same level in the hierarchy.~The equations which couple at the same level of the hierarchy form pairs of equations of the schematic form
\begin{align*}
\nablasl_4 \overset{\text{\scalebox{.6}{$(1)$}}}{q}_1&=f\cdot\overset{\text{\scalebox{.6}{$(1)$}}}{q}_2 + \text{lin.~quantities higher in the hierarchy} \, , \\
\nablasl_4 \overset{\text{\scalebox{.6}{$(1)$}}}{q}_2&=h\cdot\overset{\text{\scalebox{.6}{$(1)$}}}{q}_1 + \text{lin.~quantities higher in the hierarchy}
\end{align*}
and are denoted by $(\dagger)$.~We note that, for all such pairs of coupled equations, at least one of the coupling background quantities $f$ and $h$ identically vanishes along the event horizon.

\medskip

The hierarchy reads as follows:
\begin{itemize}
\item $\alphao$ and $\alphabo$
\begin{equation*}
\text{Spin $\pm 2$ Teukolsky equations,}
\end{equation*}
$(\trchio)$ and $(\overset{\text{\scalebox{.6}{$(1)$}}}{\slashed{\varepsilon}\cdot\chi})$
\begin{gather*}
\nablasl_4 (\trchio)+ (\text{tr}\chi)(\trchio)-\omegah \, (\trchio) =  (\slashed{\varepsilon}\cdot\chi)(\overset{\text{\scalebox{.6}{$(1)$}}}{\slashed{\varepsilon}\cdot\chi}) \, , \tag{$\dagger$} \\
\nablasl_4 (\overset{\text{\scalebox{.6}{$(1)$}}}{\slashed{\varepsilon}\cdot\chi})+(\text{tr}\chi)(\overset{\text{\scalebox{.6}{$(1)$}}}{\slashed{\varepsilon}\cdot\chi})-\omegah \, (\overset{\text{\scalebox{.6}{$(1)$}}}{\slashed{\varepsilon}\cdot\chi})   = -(\slashed{\varepsilon}\cdot\chi)(\trchio) \, , \tag{$\dagger$}
\end{gather*}
\item $\overset{\text{\scalebox{.6}{$(1)$}}}{\mathfrak{\slashed{\mathfrak{f}}}}_{4}$, $\chiho$ and $\betao$
\begin{gather*}
\nablasl_4 \overset{\text{\scalebox{.6}{$(1)$}}}{\mathfrak{\slashed{\mathfrak{f}}}}_{4}+\chi^{\sharp_2}\cdot \overset{\text{\scalebox{.6}{$(1)$}}}{\mathfrak{\slashed{\mathfrak{f}}}}_{4} -\omegah \overset{\text{\scalebox{.6}{$(1)$}}}{\mathfrak{\slashed{\mathfrak{f}}}}_{4}  = 0 \, , \\
\nablasl_4 \chiho +(\text{tr}\chi)\chiho-\omegah \,\chiho = -\alphao \, , \\
\nablasl_4\betao+2 \, (\text{tr}\chi)\betao-2\, (\slashed{\varepsilon}\cdot\chi){}^{\star}\betao-\omegah\betao =  \slashed{\text{div}}\alphao+(\etab^{\sharp}+2\,\eta^{\sharp})\cdot\alphao \, ,
\end{gather*}
\item $\overset{\text{\scalebox{.6}{$(1)$}}}{\widehat{\slashed{g}}}$, $(\textup{tr}\overset{\text{\scalebox{.6}{$(1)$}}}{\slashed{g}})$ and $\overset{\text{\scalebox{.6}{$(1)$}}}{\mathfrak{\underline{\mathfrak{f}}}}_{4}$
\begin{align*}
\nablasl_4\overset{\text{\scalebox{.6}{$(1)$}}}{\widehat{\slashed{g}}}  +(\slashed{\varepsilon}\cdot\chi)\,{}^{\star}\overset{\text{\scalebox{.6}{$(1)$}}}{\widehat{\slashed{g}}} =& \, 2\,\chiho +2\,(\etab-\eta)\,\widehat{\otimes}\overset{\text{\scalebox{.6}{$(1)$}}}{\mathfrak{\slashed{\mathfrak{f}}}}_{4} \, , \\
\nablasl_4(\textup{tr}\overset{\text{\scalebox{.6}{$(1)$}}}{\slashed{g}})=& \, 2\,(\trchio) +4\,(\etab-\eta,\overset{\text{\scalebox{.6}{$(1)$}}}{\mathfrak{\slashed{\mathfrak{f}}}}_{4})   \, , \\
\nablasl_{4}\overset{\text{\scalebox{.6}{$(1)$}}}{\mathfrak{\underline{\mathfrak{f}}}}_{4} =& \, -2\,(\eta-\etab,\overset{\text{\scalebox{.6}{$(1)$}}}{\mathfrak{\slashed{\mathfrak{f}}}}_{4}) \, , 
\end{align*}
$\rhoo$ and $\sigmao$
\begin{gather*}
\nablasl_4 \rhoo+\frac{3}{2}\,(\text{tr}\chi)\rhoo  = \slashed{\text{div}}\betao+(2\etab+\eta,\betao)-\frac{3}{2}\,\rho\,(\trchio)-\frac{3}{2}\,\sigma\,(\overset{\text{\scalebox{.6}{$(1)$}}}{\slashed{\varepsilon}\cdot\chi}) -\frac{3}{2}\,(\slashed{\varepsilon}\cdot\chi)\sigmao \, , \tag{$\dagger$} \\
\nablasl_4 \sigmao+\frac{3}{2}\,(\text{tr}\chi)\sigmao =  -\slashed{\text{curl}}\betao-(2\etab+\eta)\wedge\betao-\frac{3}{2}\,\sigma\,(\trchio) +\frac{3}{2}\,\rho\,(\overset{\text{\scalebox{.6}{$(1)$}}}{\slashed{\varepsilon}\cdot\chi})+\frac{3}{2}\,(\slashed{\varepsilon}\cdot\chi)\rhoo \, , \tag{$\dagger$}
\end{gather*}
\item $\overset{\text{\scalebox{.6}{$(1)$}}}{\mathfrak{\underline{\mathfrak{f}}}}$ and $\etao$
\begin{align*}
\nablasl_4\overset{\text{\scalebox{.6}{$(1)$}}}{\mathfrak{\underline{\mathfrak{f}}}}- \chi^{\sharp_1}\cdot \overset{\text{\scalebox{.6}{$(1)$}}}{\mathfrak{\underline{\mathfrak{f}}}}  +\omegah\overset{\text{\scalebox{.6}{$(1)$}}}{\mathfrak{\underline{\mathfrak{f}}}} =& \,  -2\etao +2\,\overset{\text{\scalebox{.6}{$(1)$}}}{\mathfrak{\underline{\mathfrak{f}}}}_{4}(\eta-\etab)+ 2\,\overset{\text{\scalebox{.6}{$(1)$}}}{\widehat{\slashed{g}}}{}^{\sharp}\cdot(\eta-\etab)+(\text{tr}\overset{\text{\scalebox{.6}{$(1)$}}}{\slashed{g}})(\eta-\etab) \, ,   \\
\nablasl_4 \etao+\frac{1}{2}\,(\text{tr}\chi)\etao -\frac{1}{2}\,(\slashed{\varepsilon}\cdot\chi){}^{\star}\etao =& \, \chiho{}^{\sharp}\cdot\etab+\frac{1}{2}\,(\trchio)\etab-\frac{1}{2}\,(\overset{\text{\scalebox{.6}{$(1)$}}}{\slashed{\varepsilon}\cdot\chi})({}^{\star}\etab-2\,{}^{\star}\eta)-\betao \\
&+2\,(\etab-\eta,\eta)\overset{\text{\scalebox{.6}{$(1)$}}}{\slashed{\mathfrak{f}}}_4+\frac{1}{2}\,(\slashed{\varepsilon}\cdot\chi)({}^{\star}\overset{\text{\scalebox{.6}{$(1)$}}}{\widehat{\slashed{g}}})^{\sharp_1}\cdot(\etab -2\,\eta)  \, , \nonumber
\end{align*}
$\overset{\text{\scalebox{.6}{$(1)$}}}{\mathfrak{\slashed{\mathfrak{f}}}}_{3}$ and $\zetao$
\begin{align*}
\nablasl_4 \overset{\text{\scalebox{.6}{$(1)$}}}{\mathfrak{\slashed{\mathfrak{f}}}}_{3} +\chi^{\sharp_2} \cdot \overset{\text{\scalebox{.6}{$(1)$}}}{\mathfrak{\slashed{\mathfrak{f}}}}_{3}  =& \,   \zetao \, , \tag{$\dagger$} \\
\nablasl_4 \zetao +\chi^{\sharp_2}\cdot \zetao +\omegah \zetao    =& \,  \chiho {}^{\sharp}\cdot \etab +\frac{1}{2}\,(\trchio) \, \etab +\frac{1}{2}\,(\overset{\text{\scalebox{.6}{$(1)$}}}{\slashed{\varepsilon}\cdot\chi}) \, {^{\star}\etab} -\betao  \tag{$\dagger$} \\
& +2\,(\etab-\eta,\eta)\overset{\text{\scalebox{.6}{$(1)$}}}{\slashed{\mathfrak{f}}}_4-(\nablasl_3\omegah)\overset{\text{\scalebox{.6}{$(1)$}}}{\slashed{\mathfrak{f}}}_4-(\nablasl_4\omegah)\overset{\text{\scalebox{.6}{$(1)$}}}{\slashed{\mathfrak{f}}}_3 \nonumber \\
&-\frac{1}{2}\,(\slashed{\varepsilon}\cdot\chi)({}^{\star}\overset{\text{\scalebox{.6}{$(1)$}}}{\widehat{\slashed{g}}})^{\sharp_1}\cdot\etab \, , \nonumber 
\end{align*}
\item $\chibho$ and $\betabo$
\begin{align*}
\nablasl_4 \chibho+\chi\times\chibho+\omegah\,\chibho =& \, -\frac{1}{2}\,(\text{tr}\chib)\chiho+\frac{1}{2}\,(\slashed{\varepsilon}\cdot\chib){^{\star}\chiho}  \\  &  - \widehat{(\nablasl\overset{\text{\scalebox{.6}{$(1)$}}}{\widehat{\slashed{g}}})^{\sharp_3}\cdot\etab}+ (\nablasl\overset{\text{\scalebox{.6}{$(1)$}}}{\widehat{\slashed{g}}})^{\sharp_1}\cdot\etab  -\frac{1}{2}\,\etab\,\widehat{\otimes}\,(\nablasl(\textup{tr}\overset{\text{\scalebox{.6}{$(1)$}}}{\slashed{g}}))-(\slashed{\textup{div}}\,\etab +(\etab,\etab))\overset{\text{\scalebox{.6}{$(1)$}}}{\widehat{\slashed{g}}} \nonumber\\
&   -\frac{1}{2}\,(\slashed{\varepsilon}\cdot\chi)\,({}^{\star}\etab)\,\widehat{\otimes}\,(-2\overset{\text{\scalebox{.6}{$(1)$}}}{\mathfrak{\slashed{\mathfrak{f}}}}_{3}+\overset{\text{\scalebox{.6}{$(1)$}}}{\mathfrak{\underline{\mathfrak{f}}}})  + (\slashed{\varepsilon}\cdot\chib)\,({}^{\star}\etab)\,\widehat{\otimes}\, \overset{\text{\scalebox{.6}{$(1)$}}}{\mathfrak{\slashed{\mathfrak{f}}}}_{4} \nonumber\\
&+(\nablasl_3\etab-\chib^{\sharp_2}\cdot\etab)\,\widehat{\otimes}\overset{\text{\scalebox{.6}{$(1)$}}}{\mathfrak{\slashed{\mathfrak{f}}}}_{4}+(\nablasl_4\etab-\chi^{\sharp_2}\cdot\etab)\,\widehat{\otimes}\overset{\text{\scalebox{.6}{$(1)$}}}{\mathfrak{\slashed{\mathfrak{f}}}}_{3} \, , \nonumber \\
\nablasl_4\betabo+(\text{tr} \chi)\betabo+(\slashed{\varepsilon}\cdot\chib){}^{\star}\betabo+\omegah\betabo  =& \,\slashed{\mathcal{D}}_1^{\star}(\rhoo,\sigmao)-3\,\rhoo\etab  +3\,\sigmao {}^{\star}\etab  \\
&-(\nablasl_3\rho)\overset{\text{\scalebox{.6}{$(1)$}}}{\mathfrak{\slashed{\mathfrak{f}}}}_{4}+ (\nablasl_3\sigma){}^{\star}\overset{\text{\scalebox{.6}{$(1)$}}}{\mathfrak{\slashed{\mathfrak{f}}}}_{4}- (\nablasl_4\rho)\overset{\text{\scalebox{.6}{$(1)$}}}{\mathfrak{\slashed{\mathfrak{f}}}}_{3}+ (\nablasl_4\sigma){}^{\star}\overset{\text{\scalebox{.6}{$(1)$}}}{\mathfrak{\slashed{\mathfrak{f}}}}_{3} \nonumber \\
&-({}^{\star}\overset{\text{\scalebox{.6}{$(1)$}}}{\widehat{\slashed{g}}})^{\sharp_1}\cdot(\nablasl\sigma +3\,\sigma\,\etab)  \, , \nonumber 
\end{align*}
$(\trchibo)$ and $(\overset{\text{\scalebox{.6}{$(1)$}}}{\slashed{\varepsilon}\cdot\chib})$
\begin{align*}
\nablasl_4( \trchibo)+ \frac{1}{2}\,(\text{tr}\chi)(\trchibo)+\omegah \, (\trchibo) =& \, - \frac{1}{2}\,(\text{tr}\chib)(\trchio)+\frac{1}{2}\,(\slashed{\varepsilon}\cdot\chib)(\overset{\text{\scalebox{.6}{$(1)$}}}{\slashed{\varepsilon}\cdot\chi})+\frac{1}{2}\,(\slashed{\varepsilon}\cdot\chi)(\overset{\text{\scalebox{.6}{$(1)$}}}{\slashed{\varepsilon}\cdot\chib}) \tag{$\dagger$} \\ 
 &   - 2\,(\slashed{\textup{div}}\overset{\text{\scalebox{.6}{$(1)$}}}{\widehat{\slashed{g}}})^{\sharp}\cdot\etab -(\slashed{\textup{div}}\,\etab+(\etab,\etab))(\textup{tr}\overset{\text{\scalebox{.6}{$(1)$}}}{\slashed{g}}) \nonumber \\ & 
-(\textup{tr}\chi)\,(-2\overset{\text{\scalebox{.6}{$(1)$}}}{\mathfrak{\slashed{\mathfrak{f}}}}_{3}+\overset{\text{\scalebox{.6}{$(1)$}}}{\mathfrak{\underline{\mathfrak{f}}}},\etab) -(\slashed{\varepsilon}\cdot\chi)\,(-2\overset{\text{\scalebox{.6}{$(1)$}}}{\mathfrak{\slashed{\mathfrak{f}}}}_{3}+\overset{\text{\scalebox{.6}{$(1)$}}}{\mathfrak{\underline{\mathfrak{f}}}},{}^{\star}\etab)   \nonumber\\
&  +2\,(\textup{tr}\chib)(\overset{\text{\scalebox{.6}{$(1)$}}}{\mathfrak{\slashed{\mathfrak{f}}}}_{4},\etab)  
+2\,(\slashed{\varepsilon}\cdot\chib)(\overset{\text{\scalebox{.6}{$(1)$}}}{\mathfrak{\slashed{\mathfrak{f}}}}_{4}, {}^{\star}\etab) \nonumber \\
&  +2\,(\nablasl_3\etab-\chib^{\sharp_2}\cdot\etab,\overset{\text{\scalebox{.6}{$(1)$}}}{\mathfrak{\slashed{\mathfrak{f}}}}_{4}) +2\,(\nablasl_4\etab-\chi^{\sharp_2}\cdot\etab,\overset{\text{\scalebox{.6}{$(1)$}}}{\mathfrak{\slashed{\mathfrak{f}}}}_{3}) \nonumber\\ &  +(\text{tr}\chi)(\etab,\overset{\text{\scalebox{.6}{$(1)$}}}{\mathfrak{\underline{\mathfrak{f}}}})+2\rhoo \, , \nonumber \\
\nablasl_4 (\overset{\text{\scalebox{.6}{$(1)$}}}{\slashed{\varepsilon}\cdot\chib}) +\frac{1}{2}\,(\text{tr}\chi)(\overset{\text{\scalebox{.6}{$(1)$}}}{\slashed{\varepsilon}\cdot\chib})+\omegah \, (\overset{\text{\scalebox{.6}{$(1)$}}}{\slashed{\varepsilon}\cdot\chib})  =& \, -\frac{1}{2}\,(\slashed{\varepsilon}\cdot\chib)(\trchio)-\frac{1}{2}\,(\text{tr}\chib)(\overset{\text{\scalebox{.6}{$(1)$}}}{\slashed{\varepsilon}\cdot \chi})-\frac{1}{2}\,(\slashed{\varepsilon}\cdot \chi)(\trchibo) \tag{$\dagger$} \\ 
&   -2\,(\nablasl_3\etab-\chib^{\sharp_2}\cdot\etab)\wedge\overset{\text{\scalebox{.6}{$(1)$}}}{\mathfrak{\slashed{\mathfrak{f}}}}_{4} -2\,(\nablasl_4\etab-\chi^{\sharp_2}\cdot\etab)\wedge\overset{\text{\scalebox{.6}{$(1)$}}}{\mathfrak{\slashed{\mathfrak{f}}}}_{3}  \nonumber \\
&+(\slashed{\varepsilon}\cdot\chi)(\etab,\overset{\text{\scalebox{.6}{$(1)$}}}{\mathfrak{\underline{\mathfrak{f}}}}) - (\slashed{\textup{curl}}\etab)(\textup{tr}\overset{\text{\scalebox{.6}{$(1)$}}}{\slashed{g}})+2\sigmao \, , \nonumber 
\end{align*}
$\overset{\text{\scalebox{.6}{$(1)$}}}{\mathfrak{\underline{\mathfrak{f}}}}_3$ and $\omegabo$
\begin{align*}
\nablasl_4 \overset{\text{\scalebox{.6}{$(1)$}}}{\mathfrak{\underline{\mathfrak{f}}}}_{3}+\omegah \overset{\text{\scalebox{.6}{$(1)$}}}{\mathfrak{\underline{\mathfrak{f}}}}_{3} =& \, \omegabo   -2\,(\eta-\etab,\overset{\text{\scalebox{.6}{$(1)$}}}{\mathfrak{\slashed{\mathfrak{f}}}}_{3})-(\eta+\etab,\overset{\text{\scalebox{.6}{$(1)$}}}{\mathfrak{\underline{\mathfrak{f}}}}) \, , \tag{$\dagger$} \\
\nablasl_4\omegabo+2\,\omegah\omegabo  =& \,   -2\,(\eta-\etab,\etao)-2\,(\eta-\etab,\zetao) -2\rhoo  \tag{$\dagger$}  \\
&+((\eta-2\,\etab)\,\widehat{\otimes}\,\eta, \overset{\text{\scalebox{.6}{$(1)$}}}{\widehat{\slashed{g}}})+(\textup{tr}\overset{\text{\scalebox{.6}{$(1)$}}}{\slashed{g}})\,(\eta-2\,\etab,\eta) \nonumber \\
& -(\nablasl\omegah,\overset{\text{\scalebox{.6}{$(1)$}}}{\mathfrak{\underline{\mathfrak{f}}}})  -(\nablasl_4\omegah)\overset{\text{\scalebox{.6}{$(1)$}}}{\mathfrak{\underline{\mathfrak{f}}}}_3  -(\nablasl_3\omegah)\overset{\text{\scalebox{.6}{$(1)$}}}{\mathfrak{\underline{\mathfrak{f}}}}_4  \, , \nonumber
\end{align*}
\item $\ybo$
\begin{align*}
\nablasl_4\ybo +2\,\omegah\ybo =& \,  -\chibho{}^{\sharp}\cdot\eta -\frac{1}{2}\,(\trchibo)\,\eta+\frac{1}{2}\,(\overset{\text{\scalebox{.6}{$(1)$}}}{\slashed{\varepsilon}\cdot\chib})({}^{\star}\eta-2\,{}^{\star}\etab)-\betabo \\
&-\frac{1}{2}\,(\text{tr}\chib)\etao  +\frac{1}{2}\,(\slashed{\varepsilon}\cdot\chib)\,{}^{\star}\etao -\frac{1}{2}\,(\slashed{\varepsilon}\cdot\chib)({}^{\star}\overset{\text{\scalebox{.6}{$(1)$}}}{\widehat{\slashed{g}}})^{\sharp_1}\cdot(\eta -2\,\etab) \nonumber\\
&+(\nablasl\etab)^{\sharp_1}\cdot \overset{\text{\scalebox{.6}{$(1)$}}}{\underline{\mathfrak{f}}}  -(\slashed{\mathcal{D}}_2^{\star}\overset{\text{\scalebox{.6}{$(1)$}}}{\underline{\mathfrak{f}}})^{\sharp_2}\cdot \etab+\frac{1}{2}\,(\slashed{\textup{div}}\overset{\text{\scalebox{.6}{$(1)$}}}{\underline{\mathfrak{f}}})\,\etab+\frac{1}{2}\,(\slashed{\textup{curl}}\overset{\text{\scalebox{.6}{$(1)$}}}{\underline{\mathfrak{f}}})\,{}^{\star}\etab \nonumber \\
&  -2\,(\eta-\etab,\etab)\overset{\text{\scalebox{.6}{$(1)$}}}{\slashed{\mathfrak{f}}}_3 +(\nablasl_4\etab+\chi^{\sharp_2}\cdot\etab)\overset{\text{\scalebox{.6}{$(1)$}}}{\underline{\mathfrak{f}}}_3+ (\nablasl_3\etab+\chib^{\sharp_2}\cdot\etab)\overset{\text{\scalebox{.6}{$(1)$}}}{\underline{\mathfrak{f}}}_4 \, . \nonumber 
\end{align*}
\end{itemize}

The transport equations in the hierarchy have the schematic form
\begin{align}
\nablasl_4\overset{\text{\scalebox{.6}{$(1)$}}}{\widetilde{\mathfrak{f}}}+\frac{p_1}{2}\,(\text{tr}\chi) \,\overset{\text{\scalebox{.6}{$(1)$}}}{\widetilde{\mathfrak{f}}}+\frac{p_2}{2}\,(\slashed{\varepsilon}\cdot\chi)\, {}^{\star}\overset{\text{\scalebox{.6}{$(1)$}}}{\widetilde{\mathfrak{f}}}+k\,\omegah \overset{\text{\scalebox{.6}{$(1)$}}}{\mathfrak{f}} &= \Gamma\cdot \overset{\text{\scalebox{.6}{$(1)$}}}{\mathfrak{f}}+\overset{\text{\scalebox{.6}{$(1)$}}}{\Gamma}  \, , \label{intro_schematic_form_outgoing_eqns_1}\\
\nablasl_4\overset{\text{\scalebox{.6}{$(1)$}}}{\widetilde{\Gamma}}+\frac{p_1}{2}\,(\text{tr}\chi) \,\overset{\text{\scalebox{.6}{$(1)$}}}{\widetilde{\Gamma}}+\frac{p_2}{2}\,(\slashed{\varepsilon}\cdot\chi)\, {}^{\star}\overset{\text{\scalebox{.6}{$(1)$}}}{\widetilde{\Gamma}}+k\,\omegah\, \overset{\text{\scalebox{.6}{$(1)$}}}{\widetilde{\Gamma}} &=\Gamma\cdot \overset{\text{\scalebox{.6}{$(1)$}}}{\Gamma}+\slashed{D}\,\Gamma\cdot \overset{\text{\scalebox{.6}{$(1)$}}}{\mathfrak{f}}+\Gamma\cdot \slashed{D}\,\overset{\text{\scalebox{.6}{$(1)$}}}{\mathfrak{f}}+\overset{\text{\scalebox{.6}{$(1)$}}}{\psi} \, ,  \label{intro_schematic_form_outgoing_eqns_2}\\
\nablasl_4\overset{\text{\scalebox{.6}{$(1)$}}}{\widetilde{\psi}}+\frac{p_1}{2}\,(\text{tr}\chi) \,\overset{\text{\scalebox{.6}{$(1)$}}}{\widetilde{\psi}}+\frac{p_2}{2}\,(\slashed{\varepsilon}\cdot\chi)\, {}^{\star}\overset{\text{\scalebox{.6}{$(1)$}}}{\widetilde{\psi}}+k\,\omegah\, \overset{\text{\scalebox{.6}{$(1)$}}}{\widetilde{\psi}} &=\nablasl\overset{\text{\scalebox{.6}{$(1)$}}}{\psi}+\Gamma\cdot\overset{\text{\scalebox{.6}{$(1)$}}}{\psi}+\psi\cdot \overset{\text{\scalebox{.6}{$(1)$}}}{\Gamma}+\slashed{D}\,\psi\cdot \overset{\text{\scalebox{.6}{$(1)$}}}{\mathfrak{f}} \label{intro_schematic_form_outgoing_eqns_3}
\end{align}
with $p_1,p_2,k\in\mathbb{Z}_0$, where the quantities $\overset{\text{\scalebox{.6}{$(1)$}}}{\widetilde{\mathfrak{f}}}$, $\overset{\text{\scalebox{.6}{$(1)$}}}{\mathfrak{f}}$ denote distinct general linearised induced metric or frame coefficients, the quantities $\overset{\text{\scalebox{.6}{$(1)$}}}{\widetilde{\Gamma}}$, $\overset{\text{\scalebox{.6}{$(1)$}}}{\Gamma}$ distinct general linearised connection coefficients, the quantity $\overset{\text{\scalebox{.6}{$(1)$}}}{\widetilde{\psi}}$, $\overset{\text{\scalebox{.6}{$(1)$}}}{\psi}$ distinct general linearised curvature components and $\slashed{D}=\left\lbrace \nablasl, \nablasl_4, \nablasl_3 \right\rbrace$.~The schematic forms \eqref{intro_schematic_form_outgoing_eqns_1}-\eqref{intro_schematic_form_outgoing_eqns_3} suppress terms for which the linearised quantity has the same structure of one of the terms already appearing in the equations, but multiplied by lower order derivatives of a background quantity or higher powers of a background quantity. This means, for instance, that the equations \eqref{intro_schematic_form_outgoing_eqns_2} may contain terms of the form $\Gamma\cdot \overset{\text{\scalebox{.6}{$(1)$}}}{\mathfrak{f}}$ or $\Gamma^2\cdot \slashed{D}\overset{\text{\scalebox{.6}{$(1)$}}}{\mathfrak{f}}$ on the right hand side. 

\medskip

All the equations for the linearised induced metric and frame coefficients do not contain any derivatives of linearised induced metric or frame coefficients on the right hand side.~Similarly, all the equations for the linearised connection coefficients do not contain any derivatives of linearised connection coefficients on the right hand side.

\medskip

The transport equations \eqref{intro_schematic_form_outgoing_eqns_1}-\eqref{intro_schematic_form_outgoing_eqns_3} are referred to as \emph{red-shifted}, \emph{no-shifted} or \emph{blue-shifted} if $k$ is positive, zero or negative respectively.~The equations for the linearised quantities
\begin{align*}
\overset{\text{\scalebox{.6}{$(1)$}}}{\widetilde{\mathfrak{f}}}&= \left\lbrace \overset{\text{\scalebox{.6}{$(1)$}}}{\underline{\mathfrak{f}}} \, , \, \overset{\text{\scalebox{.6}{$(1)$}}}{\underline{\mathfrak{f}}}_3 \right\rbrace \, , & \overset{\text{\scalebox{.6}{$(1)$}}}{\widetilde{\Gamma}}&=\left\lbrace \overset{\text{\scalebox{.6}{$(1)$}}}{\widehat{\underline{\chi}}} \, , \, (\overset{\text{\scalebox{.6}{$(1)$}}}{\text{tr}\underline{\chi}}) \, , \, (\overset{\text{\scalebox{.6}{$(1)$}}}{\slashed{\varepsilon}\cdot\chib}) \, , \,  \overset{\text{\scalebox{.6}{$(1)$}}}{\underline{\omegah}} \, , \,\overset{\text{\scalebox{.6}{$(1)$}}}{\underline{\xi}}   \, , \,  \zetao \right\rbrace \, , & \overset{\text{\scalebox{.6}{$(1)$}}}{\widetilde{\psi}}&= \betabo
\end{align*}
are red-shifted, the equations for the linearised quantities 
\begin{align*}
\overset{\text{\scalebox{.6}{$(1)$}}}{\widetilde{\mathfrak{f}}}&=\left\lbrace \overset{\text{\scalebox{.6}{$(1)$}}}{\widehat{\slashed{g}}} \, , \, (\text{tr}\overset{\text{\scalebox{.6}{$(1)$}}}{\slashed{g}}) \, , \, \overset{\text{\scalebox{.6}{$(1)$}}}{\underline{\mathfrak{f}}}_4 \, , \, \overset{\text{\scalebox{.6}{$(1)$}}}{\slashed{\mathfrak{f}}}_3 \right\rbrace \, , & \overset{\text{\scalebox{.6}{$(1)$}}}{\widetilde{\Gamma}}&= \overset{\text{\scalebox{.6}{$(1)$}}}{\eta} \, , & \overset{\text{\scalebox{.6}{$(1)$}}}{\widetilde{\psi}}&=\left\lbrace  \rhoo \, , \, \sigmao \right\rbrace
\end{align*}
are no-shifted and the equations for the linearised quantities
\begin{align*}
\overset{\text{\scalebox{.6}{$(1)$}}}{\widetilde{\mathfrak{f}}}&= \overset{\text{\scalebox{.6}{$(1)$}}}{\slashed{\mathfrak{f}}}_4   \, , & \overset{\text{\scalebox{.6}{$(1)$}}}{\widetilde{\Gamma}}&=\left\lbrace  \overset{\text{\scalebox{.6}{$(1)$}}}{\widehat{\chi}} \, , \, (\overset{\text{\scalebox{.6}{$(1)$}}}{\text{tr}\chi}) \, , \, (\overset{\text{\scalebox{.6}{$(1)$}}}{\slashed{\varepsilon}\cdot\chi})  \right\rbrace \, , & \overset{\text{\scalebox{.6}{$(1)$}}}{\widetilde{\psi}}&= \betao 
\end{align*}
are blue-shifted.

\bibliography{Stream_kerr} 
\bibliographystyle{hsiam}

\end{document}